\documentclass[11pt,headings=big,numbers=noenddot,DIV=14,a4paper]{article}%

\pdfoutput=1

\usepackage[margin=2.5 cm]{geometry}
\usepackage{multirow}

\usepackage{graphicx}  
\usepackage{dcolumn}   
\usepackage{bm,relsize}        
\usepackage{amsfonts,amsmath,amssymb,amsthm,mathtools}
\usepackage{slashed}
\usepackage[normalem]{ulem}
\usepackage{placeins}
\usepackage{framed}
\usepackage{lscape}
\usepackage{cancel}
\usepackage[sort&compress,numbers,merge]{natbib}

\usepackage{fancyhdr}
\fancypagestyle{specialfooter}{%
  \fancyhf{}
  
  \fancyfoot[L]{{\noindent\rule{4cm}{0.5pt}\\
  \footnotesize E-mail: \textcolor{PineGreen}{\texttt{yann.gouttenoire@gmail.com, geraldine.servant@desy.de, peera.simakachorn@desy.de}}}}
}

\usepackage{fourier}

\usepackage{lipsum, color}
\usepackage[usenames,dvipsnames,svgnames,table]{xcolor}
\usepackage[linktoc=page,bookmarks=false,colorlinks=true,linkbordercolor=PineGreen,citebordercolor=PineGreen,urlbordercolor=PineGreen]{hyperref}
\hypersetup{
    colorlinks=true,
    linkcolor=PineGreen,
    filecolor=Mahogany,      
    urlcolor=PineGreen,
    citecolor=PineGreen,
}

\renewcommand{\over}[2]{\left(\frac{#1}{#2}\right)}
\newcommand{\nor}[1]{\textrm{#1}}
\usepackage{chngcntr}
\counterwithin*{equation}{section}

\def\beq{\begin{equation}}
\def\eeq{\end{equation}}
\def\bea{\begin{eqnarray}}
\def\eea{\end{eqnarray}}

\def\MPl		{M_{\mathsmaller{\rm Pl}}}

\def\mreff        {m_{r,\textrm{eff}}}
\def\osc        {\textrm{osc}}

\begin{document}

\color{Black}

\thispagestyle{specialfooter}

\begin{flushright}
\footnotesize
DESY 21-134\\
\end{flushright}
\color{black}

\begin{center}

{\huge \bf 
{\fontfamily{put}\selectfont 
Kination cosmology from scalar fields\\[0.5em]
and gravitational-wave signatures
}
}

\vspace{0.25cm}

\bigskip

{
{\Large Yann Gouttenoire}$^{a}$,
{\Large G\'eraldine Servant}$^{b,c}$,
{\Large Peera Simakachorn}$^{b,c}$
}
\\[0.75em]

{\it  $^a$ School of Physics and Astronomy, Tel-Aviv University, Tel-Aviv 69978, Israel}\\
{\it  $^b$ Deutsches Elektronen-Synchrotron DESY, Notkestra{\ss}e 85, 22607, Hamburg, Germany}\\
{\it  $^c$ II. Institute of Theoretical Physics, Universit\"{a}t  Hamburg, 22761, Hamburg, Germany}\\[0.75em]
\end{center}

\bigskip

\centerline{\large \bf Abstract}

\bigskip

Kination denotes an era in the cosmological history corresponding to an equation of state  $\omega=+1$ such that the total energy density of the universe redshifts as the  sixth inverse power of the scale factor. This arises if the universe is dominated by the kinetic energy of a scalar field.
It has often been motivated in the literature  as an era following inflation, taking place before the radiation era.
In this paper, we review instead the possibility that kination is disconnected from primordial inflation and  occurs much later, inside the Standard Model radiation era. 
We study the implications on all main sources of primordial gravitational waves.
We show how this leads to very distinctive peaked spectra in the stochastic background of long-lasting cosmological sources of gravitational waves, namely the irreducible gravitational waves from inflation, and  gravitational waves from cosmic strings, both local and global, with promising observational prospects. We present model-independent signatures and detectability predictions at SKA, LIGO, LISA, ET, CE, BBO, as a function of the energy scale and duration of the kination era.
We then argue that such intermediate kination era is in fact symptomatic in a large class of axion models. 
We analyse in details the scalar field dynamics, the working conditions and constraints in the underlying models. We  present the gravitational-wave predictions as a function of particle physics parameters. We derive the general relation between the gravitational-wave signal and the axion dark matter abundance as well as the baryon asymmetry. We investigate the predictions for the special case of the QCD axion.
The key message is that gravitational-waves of primordial origin represent an alternative experimental probe of axion models.

\bigskip

\clearpage
\noindent\makebox[\linewidth]{\rule{\textwidth}{1pt}} 
\\[-2.5em]
\tableofcontents
\noindent\makebox[\linewidth]{\rule{\textwidth}{1pt}} 	

\newpage
\section{Introduction}

The measurement  of the abundances of the  light elements as predicted by the theory of  Big-Bang Nucleosynthesis (BBN) constrains the universe to be dominated by radiation when the temperature was $1~$MeV. The  smoothness and flatness of the universe, and the temperature anisotropies in the Cosmic Microwave Background (CMB), support the idea that much earlier than BBN, the universe was inflating exponentially, dominated by the energy density of a slowly-rolling scalar field. The non-detection of the fundamental B-mode polarization patterns in the CMB suggest that the maximal Hubble rate during inflation $H_{\rm inf}$ is  $5 \times 10^{13}$~GeV, which corresponds to a maximal energy scale of $10^{16}$~GeV \cite{Ade:2018gkx, Akrami:2018odb,BICEP:2021xfz}. 

The equation of state (EOS) of the universe between the end of inflation and the onset of BBN, encoded by the parameter $\omega = p/\rho$, where $p$ and $\rho$ are the local pressure and energy densities, is currently unconstrained \cite{Allahverdi:2020bys}. While the standard paradigm assumes that the energy density of the post-inflationary universe is radiation-dominated, $\omega = 1/3$, alternative cosmological histories are not unlikely.  For instance, the dynamics of the inflaton at the end of inflation can trigger  a {\it stiff }  EOS, $\omega > 1/3$ such that the total energy density of the universe redshifts faster than radiation. In this scenario,  the universe can be dominated by the kinetic energy of a fast-rolling scalar field, with  $\omega = 1$ \cite{Spokoiny:1993kt,Joyce:1996cp,Ferreira:1997hj,Joyce:thesis}.

The possibility that the universe has had a {\it stiff }  EOS, $\omega > 1/3$, is particularly relevant for the observation of a Stochastic Background of Gravitational Waves (SGWB) of primordial origin. The main sources of gravitational waves (GW) in the early universe are inflation, reheating/preheating, first-order phase transitions, and cosmic strings \cite{Caprini:2018mtu}. 
The observation of such GW in future interferometers or at pulsar-timing arrays would not only offer a unique probe of the high-energy particle physics phenomena responsible for their production but also of  the cosmological history, as the GW spectra encode information about the EOS of the universe between GW production at very early times and GW detection today.

The interesting aspect of an era with a stiff EOS is that it leads to an amplification of the GW energy density $\rho_\textrm{GW}$ produced by long-lasting sources compared to the predicted value in standard cosmology. This can be understood as follows. 
A given GW frequency $f$ in the primordial GW spectrum measured today corresponds to a GW emitted at time $t_*$ with frequency of order $H_* \propto \sqrt{\rho_*}$ when  the total energy density of the universe $\rho_*$. Reading the frequency spectrum from low to high frequencies is like going back to earlier emission times. 
The emitted energy density in GW  is proportional to $\rho_*$, and therefore to $H_*^2$.
For a given GW frequency today, the corresponding total energy density of the universe is necessarily higher in the scenario with a kination era than in the standard cosmology, see Fig.~\ref{diagram_intro}.
If a stiff era occurred,  the GW energy density today is larger than the value obtained assuming  standard cosmology.
If a stiff era lasts too long, this leads to a substantial amplification of the primordial GW signal, violating the bounds on the number of massless degrees of freedom $N_{\rm eff}$ from BBN.
 There is a large literature on the impact of a stiff era on the nearly-scale invariant primordial GW spectrum generated during inflation in the case where kination happens right after inflation~\cite{ Giovannini:1998bp, Giovannini:2009kg, Riazuelo:2000fc, Sahni:2001qp, Seto:2003kc, Tashiro:2003qp, Nakayama:2008ip, Nakayama:2008wy, Durrer:2011bi, Kuroyanagi:2011fy, Kuroyanagi:2018csn, Jinno:2012xb, Lasky:2015lej, Li:2016mmc, Saikawa:2018rcs, Caldwell:2018giq, Bernal:2019lpc, Figueroa:2019paj, DEramo:2019tit,Li:2013nal, Li:2016mmc, Li:2021htg}. The effect of a kination era following inflation on the SGWB generated by cosmic strings was also discussed in \cite{Cui:2017ufi, Cui:2018rwi, Auclair:2019wcv, Ramberg:2019dgi, Chang:2019mza, Gouttenoire:2019kij, Gouttenoire:2019rtn}. 
  
 On the other hand, so far (up to the suggestions in \cite{Co:2019wyp,Co:2020jtv} and the coincident studies \cite{Gouttenoire:2021wzu,Co:2021lkc}), there has been no investigation of the scenario where kination is disconnected from inflation and happens much later after reheating,  inside the radiation era. In this situation, BBN-$N_{\rm eff}$ bounds are easily evaded, while the observational prospects  at future gravitational-wave observatories are excellent.
 This is the main topic of this article.
Such scenario can be realised only if the kination era is preceeded by a matter era. 
The main task of this paper is to motivate, from particle physics, such a scenario, to derive the GW signatures and the prospects for their detectability. 
We identify main classes of models where this happens naturally. They are linked to axion models where the axion acquires a large kinetic energy before its low-energy potential develops, therefore leading to a spinning stage along the circular orbit of the axion potential.  
The interplayed dynamics between the radial mode and the angular mode of a complex scalar field generates the desired sequence of events.  A letter version of this work was presented in  \cite{Gouttenoire:2021wzu}. We provide many details and a thorough discussion in this paper, in particular on the damping  of radial motion.
  
 The plan of this paper is the following. We start  with the phenomenology and the observational implications of a kination era. We then present the particle physics implementation.
 In Sec.~\ref{sec:kination}, we review the status of kination in a broad sense, we sketch the three possible cosmological scenarios that can lead to a kination era.
We first discuss the universal experimental prospects for probing {\it a kination era following inflation}: We derive in Fig.~\ref{BBNbound_kination}  the values of the inflation scale and of the reheating temperature that are already constrained by BBN and by the scalar fluctuation, concluding that  LIGO, LISA, ET and BBO do not have sensitivity to probe the allowed region of parameter space and only ultra-high-frequency experiments could do so.  We then turn to our main topic: {\it a kination era inside the radiation era}. 
Gravitational-wave signatures are first  studied  in a model-independent way in Sec.~\ref{sec:modelindependent}. We predict the effect of kination on the irreducible GW spectrum from inflation  as well as on the GW from (both local and global) cosmic strings. We present the constraints on the duration and energy scale of kination. Prospects for detection at future experiments are derived in detail.
Having motivated an intermediate kination era with axion models, we determine the relation between the relic abundance of the axion and  the GW energy density today in Sec.~\ref{sec:darkmatter}. We also comment on the relation between the  GW energy density today and the baryonic energy density predicted through the  axiogenesis mechanism in Sec.~\ref{sec:baryon_asymmetry}.
We then discuss the particle physics realisations in Sec.~\ref{sec:trapped_mis} 
, \ref{sec:PQ:exampleII}, \ref{sec:scenario_I_non_thermal_damping}, \ref{sec:complex_field_thermal_potential} and
\ref{sec:complex_field_low_reh_temp}.
The damping of the radial mode energy density is a crucial aspect in this story and we discuss this extensively in Sec.~\ref{sec:scenario_I_non_thermal_damping} and \ref{sec:complex_field_thermal_potential}
and \ref{sec:complex_field_low_reh_temp}. Thermal effects are investigated in details in Sec.
 \ref{sec:complex_field_thermal_potential} and \ref{sec:complex_field_low_reh_temp}.
The conditions that lead to kination are analysed in terms of model parameters for both classes of models. 
Precise predictions for the GW signal and the prospects for detection for each class of models are given.  
We conclude in Sec.~\ref{sec:conclusion}.

A number of technical details are presented in the appendices.
App.~\ref{sec:quintessence} reviews the occurrence of a kination era following inflation before reheating. App.~\ref{sec:otherstiff} shows the constraints and observability prospects of a generic stiff era with EOS  $1/3<\omega<1$. App.~\ref{app:NKD_max} discusses different limitations on the duration of a kination era. App.~\ref{app:scalar_pot_SUSY} explains in details the origin of the complex scalar potential in the  UV completion and the role of each term in the dynamics.
In App.~\ref{app:issus_inflation_fluct_model_B}, we discuss the usual issues of adiabatic and isocurvature perturbations in the axion models  and the solutions.  App.~\ref{app:radial_damping} provides more details on the thermal and non-thermal damping mechanisms for the radial mode. App.~\ref{app:field_evolution_model_B} reports the detailed solution of the equation of motion for the complex scalar field model,  discussing the various steps over the full cosmological evolution.

\section{What is kination?}
\label{sec:kination}

\begin{figure}[h!]
\centering
\raisebox{0cm}{\makebox{\includegraphics[width=0.9\textwidth, scale=1]{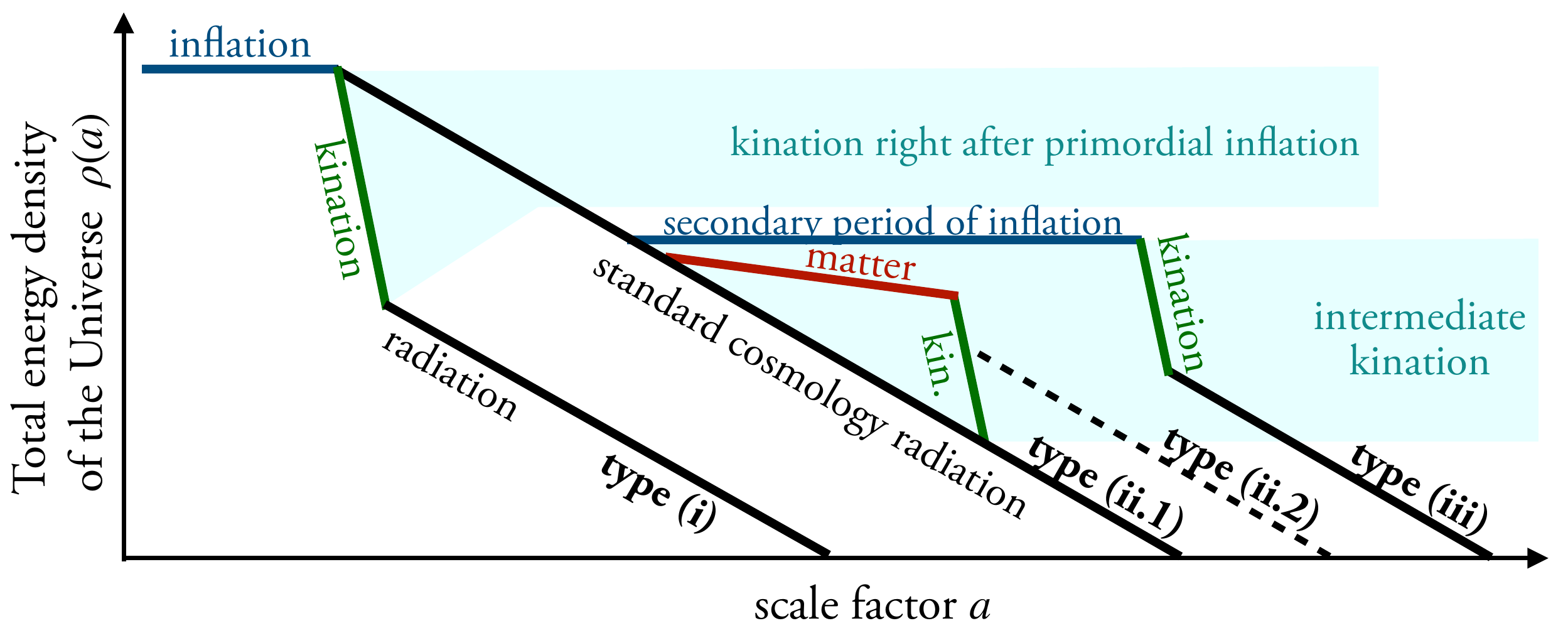}}}
\caption{\textit{ \small Possible cosmological histories involving a period of kination.}}
\label{diagram_intro}
\end{figure}

The term \emph{kination} was introduced for the first-time in \cite{Joyce:1996cp} and describes a scalar field whose kinetic energy dominates the dynamics. The corresponding   EOS is
$ \omega ~ = ~ \frac{p}{\rho} ~ = ~ \frac{KE ~ - ~  PE}{KE ~ + ~ PE} ~ \simeq ~ 1$, where $KE$ and $PE$ denote kinetic and potential energy density, respectively.
For example,  inflation can end when the inflaton potential becomes steep and the inflaton fast-rolls, inducing a kination period \cite{Spokoiny:1993kt}. This is the scenario of type (i) represented in Fig.~\ref{diagram_intro}.

We can generalise the definition of a kination era to  an  epoch when the universe is dominated by a fluid with EOS  $\omega \equiv p/\rho = 1$, where $p$ and $\rho$ are pressure and energy density.
According to this definition,  kination dates back to the exotic cosmological model by Zel'dovich \cite{l1962equation}.
Kination has the maximum EOS allowed by causality, i.e. the sound speed is the speed of light.
Its energy density has the fastest redshift $\rho \propto a^{-6}$, where $a$ is the scale factor of the universe, and the universe has the slowest expansion, $a \propto t^{1/3}$.
Hence, a kination era at early times will end by becoming subdominant to the Standard Model radiation without the need of decay\footnote{Though the kination-decaying scenario can be considered as in \cite{Visinelli:2017qga}.}.
Such slowest rate of expansion can affect for example reheating after inflation \cite{Ford:1986sy,Chun:2009yu, Dimopoulos:2018wfg,Nakama:2018gll,Opferkuch:2019zbd, Bettoni:2019dcw, Bettoni:2021zhq}, electroweak baryogenesis \cite{Joyce:1996cp,Joyce:1997fc}, the enhancement of Dark Matter (DM) relics \cite{Salati:2002md, Profumo:2003hq, Chung:2007vz, Chung:2007cn, Visinelli:2009kt, Redmond:2017tja, DEramo:2017gpl, DEramo:2017ecx, Visinelli:2017qga}, matter perturbations and small-scale structure formation \cite{Redmond:2018xty, Visinelli:2018wza},  GW signals from inflation \cite{Tashiro:2003qp, Figueroa:2018twl, Figueroa:2019paj, Bernal:2019lpc},  GW from both local and global cosmic strings \cite{Cui:2017ufi, Cui:2018rwi, Bettoni:2018pbl, Chang:2019mza, Gouttenoire:2019kij, Chang:2021afa}, and GW from phase transitions \cite{Chung:2010cb, Allahverdi:2020bys}.

The kination EOS can also arise outside of the fast-rolling scalar field context.
The small-scale anisotropic stress in the coarse-grained homogenous expanding background has the energy density $\propto a^{-6}$ \cite{Barrow:1981pa, Turner:1986tc}.
Recently, it was pointed-out that the cosmic fluid after a first-order phase transition can also produce the kination-liked anisotropy \cite{Niedermann:2019olb, Niedermann:2020dwg}.
A late intermediate kination era could thus occur after a second inflation stage arising  for instance due to a supercooled phase transition.  Such case is denoted type (iii) in Fig.~\ref{diagram_intro}.
The EOS evolution after bubble collision would require a dedicate study. We do not consider kination after a secondary inflation in this work. 
We instead focus on the cases where  kination occurs right after inflation (type (i)) and the intermediate  kination following a matter era (type (ii.1) \& (ii.2)). As we will explain below, a post-reheating kination era cannot happen inside the standard radiation era, it has to be preceeded by a matter era. 
In summary, the following  cosmological histories involving a period of kination are possible:
\begin{itemize}
\item \textbf{Type (i)}: Inflation $\rightarrow$ Kination $\rightarrow$ Radiation
\item \textbf{Type (ii.1)}: Inflation $\rightarrow$ Radiation $\rightarrow$ Matter $\rightarrow$  Kination $\rightarrow$ Radiation\\(without entropy injection)
\item \textbf{Type (ii.2)}:  Inflation $\rightarrow$ Radiation $\rightarrow$ Matter $\rightarrow$  Kination $\rightarrow$ Radiation\\(with entropy injection)
\item \textbf{Type (iii)}:  Inflation $\rightarrow$ Radiation $\rightarrow$ Inflation $\rightarrow$  Kination $\rightarrow$ Radiation
\end{itemize}
They are compared on Fig.~\ref{diagram_intro} and we discuss them in turn below.
\FloatBarrier
\FloatBarrier

\subsection{Kination right after inflation}
\label{subsec:kinafterinf}
In the literature,  two classes of models predict a stiff EOS,  $\omega > 1/3$, following primordial inflation.

\paragraph{Steep oscillatory potential.}
There are models where the inflaton ends up oscillating in a steep potential, $V(\phi) \propto \phi^{p}$ with $p>4$ \cite{Turner:1983he,Spokoiny:1993kt,Joyce:1996cp}. The stiff equation of state, however, hardly lasts longer than a few e-folds \cite{Lozanov:2016hid, vandeVis:2020qcp,Saha:2020bis, Antusch:2020iyq}. 

\paragraph{Steep non-oscillatory potential.}
There are non-oscillatory or quintessential inflation models \cite{Peebles:1998qn,Dimopoulos:2001ix} where the inflaton potential has a sudden drop responsible for the fast-roll of the scalar field after the end of inflation. On both sides of the drop, the inflaton potential features two asymptotically flat regions, the inflationary plateau and the quintessence tail, in which the scalar field can slow-roll and generate both primordial inflation and the late dark energy with a unified description.  
The first problem of quintessence inflation model is the need for super-Planckian field excusion. Indeed, during a period of kinetic energy domination, 
\begin{equation}
\ddot{\phi} + 3H\dot{\phi} =0 \quad \text{with} ~ H= \sqrt{ \frac{\dot{\phi}^2/2}{3M_{\rm pl}^2} }
\qquad \implies \qquad \Delta \phi \simeq \sqrt{6} M_{\rm pl} N_{\rm KD}, \qquad \textrm{with} \quad N_{\rm KD} \equiv \Delta \log{a}, \label{eq:kination_eom}
\end{equation}
a canonically-normalized scalar field $\phi$ varies over $\mathcal{O}(M_{\rm Pl})$ during each e-fold of kination which is a no-go if one takes seriously the swampland distance conjecture \cite{Ooguri:2006in, Brennan:2017rbf,Palti:2019pca}. This problem can be circumvented  by considering non-canonical kinetic terms as in $\alpha$-attractor models, see App.~\ref{sec:quintessence}.
The second problem is how to reheat the universe as 
kination does not feature  the coherent oscillations that can lead to the usual reheating or preheating mechanism.
Ways out require extra ingredients, either additional non-minimal couplings or extra fields.
We do not discuss this further as a myriad of models have been discussed in the literature and defer to App.~\ref{sec:quintessence}, a report status with references, on model-building related to the scenario of kination following inflation.
In the next subsection, we present model-independent constraints on this scenario.

\subsection{BBN, CMB \& and scalar fluctuation bounds}
\label{subsec:afterinflation}
A kination era is constrained by BBN and CMB for two reasons.
\paragraph{Kination cannot end after BBN.} The universe must be in the radiation era at the time of BBN, so in this paper we will impose that the  kination era ends before BBN, at a temperature $T_{\Delta}$ higher than, cf. App.~\ref{sec:CMB_BBN_bound_NKD}  and Ref. \cite{DESYfriendpaper}
\begin{equation}
T_{\Delta} \gtrsim T_{\rm BBN} = 1 ~{\rm MeV}.
\end{equation}
The possibility that a kination era starts after BBN and ends before matter-radiation equality was considered in \cite{Co:2021lkc} and is constrained by CMB.
In App.~\ref{sec:CMB_BBN_bound_NKD}, we show consistency between BBN constraints and the CMB upper bound on the inflationary scale leads to 
\begin{align}
\label{eq:max_NKD_CMB_BBN_main}
&\textrm{Type (i):}\,\,\qquad \qquad N_{\rm KD} ~\lesssim ~29 + \frac{2}{3} \log\left({\frac{E_{\rm inf}}{1.4 \times 10^{16}~\rm GeV}}\right)\\
&\textrm{Type (ii):}\qquad\qquad N_{\rm KD}~  \lesssim~ 14.6 + \frac{1}{3}\log\left({\frac{E_{\rm reh}}{1.4 \times 10^{16}~\rm GeV}}\right).
\end{align}
where  $E_{\rm inf}$ is the inflation energy scale, $E_{\rm reh}$ is the energy scale when the universe becomes radiation dominated right after inflation, and where the scenarios of Type (i) and Type (ii) are defined in Fig.~\ref{diagram_intro}.

\paragraph{BBN constraints on inflationary GW.} The pre-BBN kination enhances the GW signal $\Omega_\textrm{GW}(f)=\rho_\textrm{GW}/\rho_c$ from primordial times, as discussed in the next  section. 
If the duration of kination is too long, the enhanced GW energy density can impact the expansion rate at the time of BBN as it acts as an effective number of neutrino relics.
\begin{equation}
N_{\rm eff} = \frac{8}{7}\left( \frac{\rho_{\rm tot}-\rho_{\gamma}}{\rho_\gamma} \right)\left( \frac{11}{4} \right)^{4/3},
\end{equation}
which is constrained by CMB measurements \cite{Aghanim:2018eyx} to $N_{\mathsmaller{\rm eff}} = 2.99_{-0.33}^{+0.34}$ and by BBN predictions \cite{Mangano:2011ar, Peimbert:2016bdg} to $N_{\mathsmaller{\rm eff}} = 2.90_{-0.22}^{+0.22}$ whereas the Standard Model (SM) prediction \cite{Mangano:2005cc, deSalas:2016ztq} is $N_{\mathsmaller{\rm eff}}  \simeq 3.045$.
Using $\Omega_{\gamma} h^2 \simeq 2.47\times 10^{-5}$ \cite{Tanabashi:2018oca}, we obtain the following bound on the GW spectrum 
\begin{equation}
\int^{f_{\rm max}}_{f_\textrm{BBN}}\frac{df}{f}h^2 \Omega_\textrm{GW}(f) \leq 5.6 \times 10^{-6} ~ \Delta N_\nu, \label{eq:BBN_bound_inflation_GW}
\end{equation}
where we set $\Delta N_\nu \leq 0.2$ \cite{Tanabashi:2018oca}, $f_{\rm BBN}$ is the characteristic frequency corresponding to the BBN time ($f_{\rm BBN} \simeq 0.18 ~ {\rm nHz}$ for inflationary GW, see next section) and $f_{\rm max}$ is the cutoff frequency (associated with the end of inflation).
This bound applies to all sources of primordial GW. A conservative bound can be derived by considering the irreducible inflationary GW background\footnote{To use the GW background from cosmic strings, one needs a mechanism to generate cosmic strings during a kination era. We leave this issue for future work.}.

We show now that the kination right after inflation is strongly constrained by BBN which prevents the possibility that GW observatories such as LIGO, LISA, ET and BBO  could probe such  kination era. A  similar  analysis was performed for an arbitrary stiff era in \cite{Figueroa:2019paj}, considering LIGO and LISA prospects. The conclusion was that only $\omega \sim 0.5$ could still lead to signals at LISA while not being excluded by BBN, but they would correspond to a very low-energy stiff era, below a GeV.

\begin{figure}[h!]
\centering
\raisebox{0cm}{\makebox{\includegraphics[width=0.475\textwidth, scale=1]{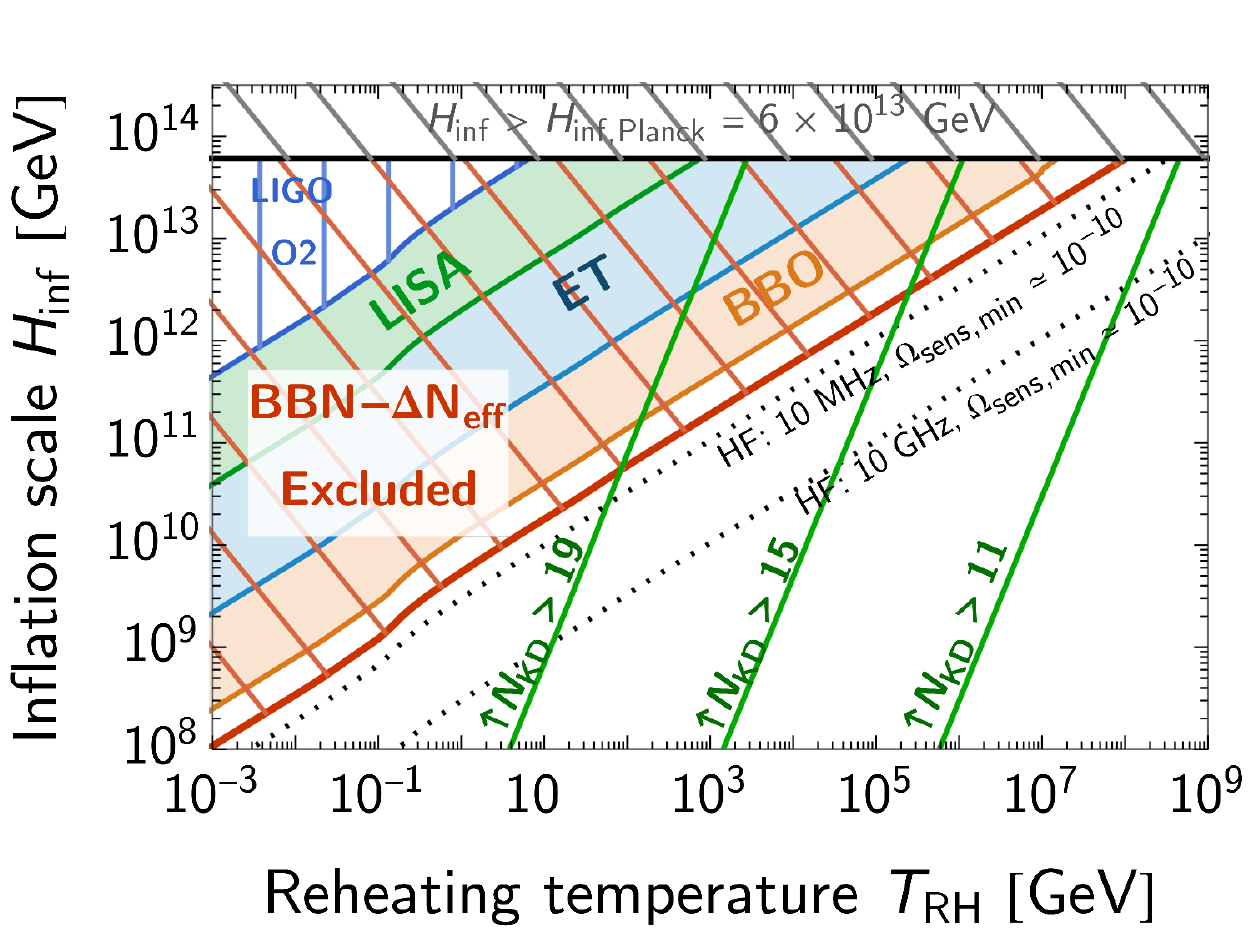}}}
\quad
\raisebox{0cm}{\makebox{\includegraphics[width=0.48\textwidth, scale=1]{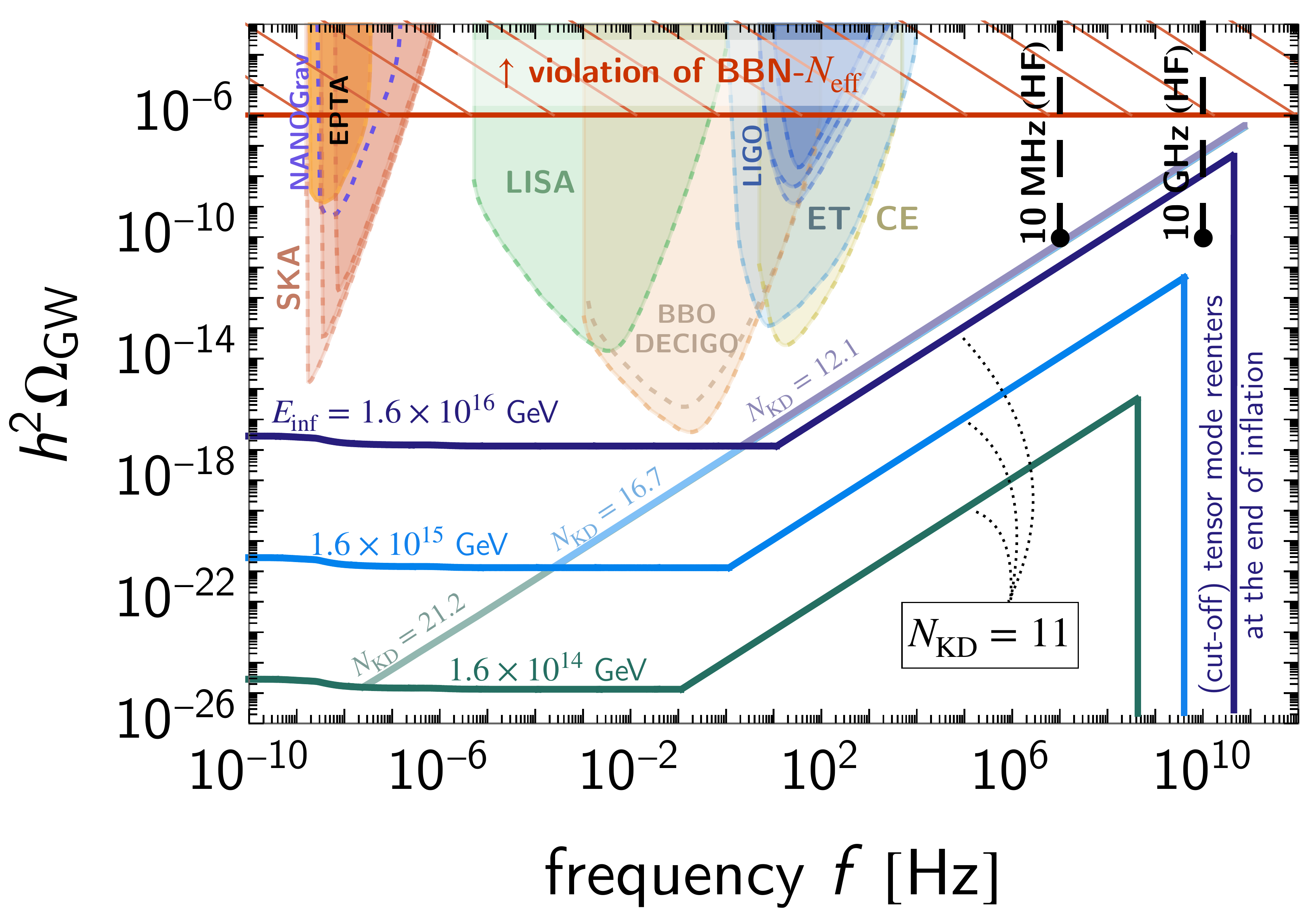}}}
\caption{\textit{ \small Ability of future GW experiments to probe a kination era taking place right after inflation (Scenario Type (i) in Fig.~\ref{diagram_intro}) via the imprint on the irreducible inflationary SGWB, assuming an inflation scale $H_{\rm inf} = E_{\rm inf}^2/\sqrt{3}\MPl$. 
Given the BBN-$N_{\rm eff}$ bound shown in red, cf. Eq.~\eqref{eq:bound_NKD_GW_inf_main}, none of the planned experiments can observe the kination-induced GW enhancement, even for the highest allowed inflationary scale $H_{\rm inf,Planck}$.
Above the black dotted lines of the left panel, the kination enhancement could be probed by  high-frequency (HF) experiments, operating at a given frequency and at the required sensitivity $\Omega_{\rm sens,min}$ shown on the right panel. 
Alternatively, the pale spectra on the right panel -- with a specific value of the number of  kination-efolds $N_{\rm KD}$-- could violate the BBN bound for a given $H_{\rm inf}$, but are still undetectable with future experiments. 
However, these spectra are  unphysical due to the scalar fluctuation bound, $N_{\rm KD} \lesssim 11$ for $\rho/\delta\rho \, (H_{\rm inf}) \lesssim 10^{9 \div 10}$ cf. Eq.~\eqref{eq:NKD_upper_bound_Cem}, as shown as the green lines on the left panel. By imposing $N_{\rm KD} \lesssim 11$ (could be relaxed in some inflation model), the spectra in the right panel can never violate the BBN-$N_{\rm eff}$ bound.
}}
\label{BBNbound_kination}
\end{figure}

Consider the scenario where kination occurs after inflation characterised by the Hubble scale $H_{\rm inf}$ and ends at the reheating temperature $T_{\rm RH}$. From Eqs.~\eqref{kination_end_frequency} and \eqref{inflation_peak_frequency}, the GW from inflation gets enhanced between the frequency corresponding to reheating $f_\Delta$ and the cut-off frequency corresponding to the end of inflation
\begin{align}
f_{\rm max}  ~ &\simeq ~ 9.64 \times 10^{11} ~ \textrm{Hz} ~ \left(\frac{g_*(T_{\rm RH})}{106.75}\right)^{1/6} \left(\frac{g_{*,s}(T_{\rm RH})}{106.75}\right)^{-1/3} \left(\frac{H_{\rm inf}}{\mathrm{10^{13} ~ GeV}}\right)^{2/3} \left(\frac{\textrm{1 TeV}}{T_{\rm RH}}\right)^{1/3}.
\label{max_freuquency_inflation}
\end{align}
We provide in Fig.~\ref{BBNbound_kination} model-independent bounds  on a kination era ($\omega=1$) happening just after inflation as a function of the inflationary scale and the reheating temperature. In App.~\ref{sec:inflation_bound_NKD}, we show that BBN-$N_{\rm eff}$ bound on inflationary GW leads to the following upper bound on the duration of kination
\begin{equation}
\label{eq:bound_NKD_GW_inf_main}
\textrm{Scenarios of Type (i) and (ii) in Fig.~\ref{diagram_intro}:} \qquad \qquad N_{\rm KD} ~\lesssim ~11.9 + \log\left({\frac{5 \times 10^{13}~\rm GeV}{H_{\rm inf}} }\right).
\end{equation}
The BBN-$N_{\rm eff}$ bound excludes the region where kination is too long and leads to a too large GW signal. All future planned experiments cannot beat this bound so we expect no discovery of the enhanced signal from a kination era right after inflation.
A way-out would be  to use high-frequency (HF) experiments as discussed in \cite{Aggarwal:2020olq}.
In Fig.~\ref{BBNbound_kination}, we show how  HF experiments operating at  10 MHz and 10 GHz with sensitivity $h^2\Omega_{\rm sens}^{\rm min} = 10^{-10}$ can potentially probe the parameter space beyond the BBN bound.
We also find that HF experiments operating at 1 kHz, 1 MHz, and 1 GHz need at least $h^2\Omega_{\rm sens}^{\rm min} \lesssim 10^{-14}, 10^{-11}, {\rm ~and ~}10^{-8}$ respectively,  to make a discovery.
For experiments operating at $\gtrsim$ THz range, the  cut-off frequency in Eq.~\eqref{max_freuquency_inflation} is smaller such that they cannot probe the GW signal.

We show in App.~\ref{sec:otherstiff} the analogues of  Fig.~\ref{BBNbound_kination} for a stiff era corresponding to $\omega=1/2$ and $\omega=2/3$.
For $\omega=1/2$, there is no BBN bound, LISA, ET and BBO can probe the enhanced GW signal from inflation while high-frequency experiments would not bring additional insight due to the gentle slope of the signal. For $\omega=2/3$, the BBN bounds prevents LISA's sensitivity while there is a potential for ET and BBO. In the rest of this paper, we only consider the maximal stiff era known as kination and we fix $\omega=1$.

\paragraph{Possible constraint from scalar fluctuation.}
Kination is triggered by a freely-rolling scalar field, whose fluctuation behaves as a hot gas of massless particles, therefore red-shifts as radiation. 
The energy density of the fluctuation of the scalar field can eventually dominate that of the zero-mode which  red-shifts as the inverse sixth power of the scale factor of the universe.
Assuming that the fluctuation with energy density $\delta \rho$ is generated at the end of inflation, it dominates the zero-mode of energy density $\rho$ after the kination era expands by $N_{\rm KD} \simeq \log(\rho_{\rm inf}/\delta \rho_{\rm inf})$. For instance, the scalar might fluctuate as the same order as the curvature fluctuation -- $\rho/\delta \rho \sim 10^{9\div 10}$ which leads to $N_{\rm KD}^{\rm max} \sim 11$, cf. also App.~\ref{sec:curvature_perturbation_kination}. 
For a suppressed fluctuation, the bound on kination duration can be relaxed for some particular inflation models.
Fig.~\ref{BBNbound_kination} shows that the theoretical kination-constraint from fluctuation is  stronger than the usually-considered BBN-$N_{\rm eff}$ bound.  The very interesting implications of the fluctuation from the kination-like field will be discussed further in \cite{DESYfriendpaper2}.

\FloatBarrier
\FloatBarrier

We now move to the new scenario investigated in this paper: \emph{an intermediate matter-kination era,} corresponding to Scenario Type (ii) in Fig.~\ref{diagram_intro}.

\subsection{Matter-Kination inside radiation}
\label{sec:intermediate_kination}

As we will motivate from particle physics in Secs.~\ref{sec:trapped_mis}, \ref{sec:PQ:exampleII}, \ref{sec:scenario_I_non_thermal_damping}, \ref{sec:complex_field_thermal_potential} and \ref{sec:complex_field_low_reh_temp},   kination can occur at lower energy scales well after reheating and for a short period. Therefore, it could enhance GW produced either during inflation, at preheating or much later in the post-reheating era by a network of cosmic strings, within the observable ranges of future-planned experiments, while the BBN-$N_{\rm eff}$ bound is not violated.

\paragraph{A matter-kination era.}
A period when the total energy density redshifts slower than radiation is needed, for a kination era inside the radiation era.
As we will see, the UV completions we present naturally generate a kination after a matter era.
The matter era brings the energy density of the universe above the radiation energy density.
This enables a period of kination that redshifts faster than radiation afterwards.
The longer the matter era dominates, the longer kination lasts.
The cosmological history with the intermediate matter-kination era is described by 
the total energy density of the universe
\begin{equation}
\rho(a)=\rho_{r,0} \, G[T(a),T_0] \left(\frac{a}{a_0}\right)^4+\rho_{m,0}\left(\frac{a}{a_0}\right)^3+\rho_{\Lambda,0} + \rho_\phi(a),
\label{friedmann_eq}
\end{equation}
where the function
\begin{equation}
G(T,\,T_0)=\left[\frac{g_*(T)}{g_*(T_0)}\right]\left[\frac{g_{*s}(T_0)}{g_{*s}(T)}\right]^{4/3},
\label{eq:DeltaR}
\end{equation}
accounts for the change in the number of relativistic degrees of freedom, assuming the conservation of the comoving entropy $g_{*s}\,T^3\,a^3$. We take the functions $g_{*}$ and $g_{*s}$ from App.~C of \cite{Saikawa:2018rcs}.
The first three terms of Eq.~\eqref{friedmann_eq} follow from  the $\Lambda$CDM assumption, while $\rho_\phi$ is the scalar field energy density that generates the non-standard matter and kination eras. 

\begin{figure}[h!]
\centering
\raisebox{0cm}{\makebox{\includegraphics[width=0.65\textwidth, scale=1]{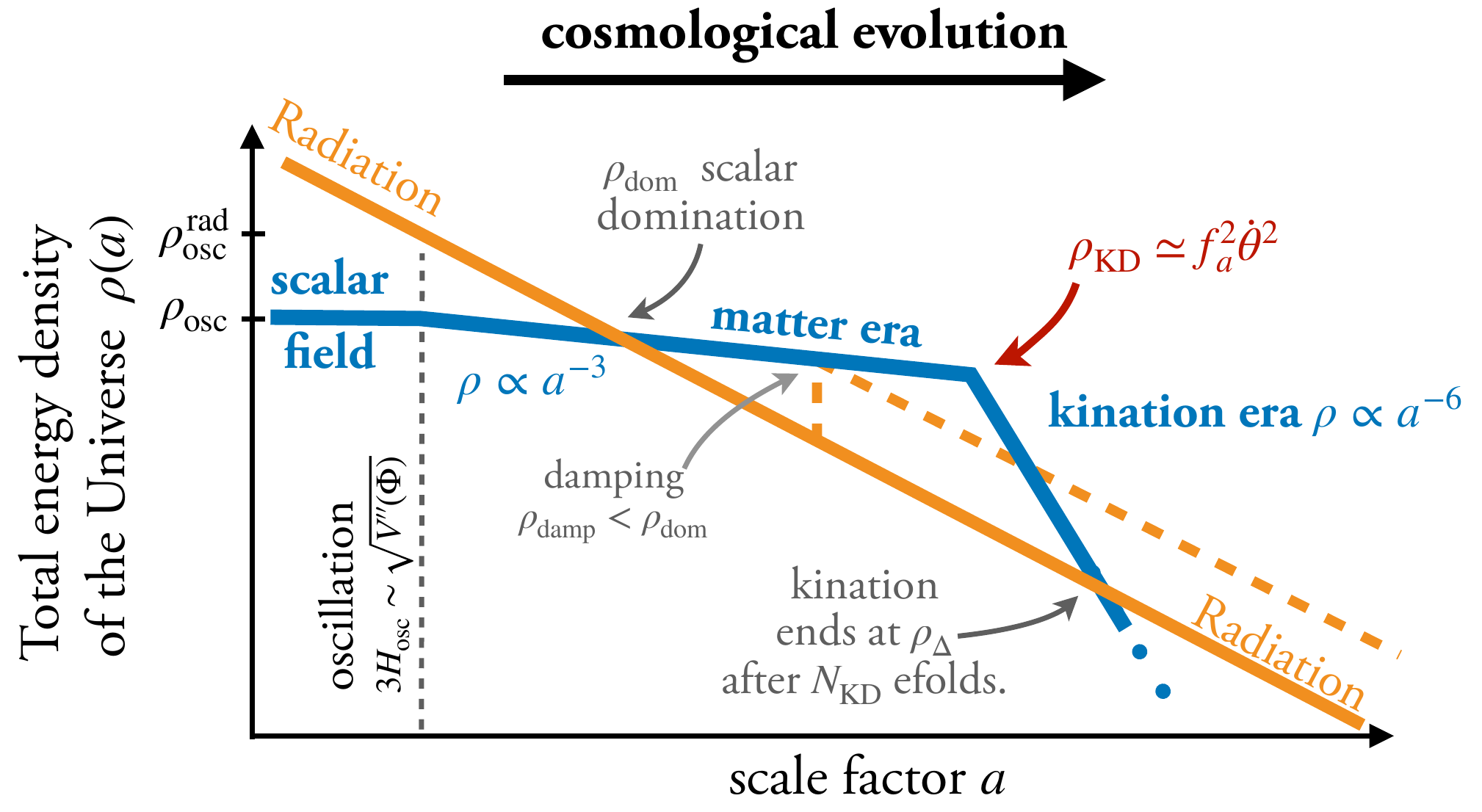}}}
\caption{\textit{ \small Scalar field dynamics that generates a matter-kination era inside the radiation epoch. The solid orange line shows Scenario of Type (ii.1) where the entropy injection is absent or happens before the scalar domination. For Scenario of Type (ii.2), the entropy injection occurs after scalar domination and suppresses the kination duration, as shown in dashed orange line. $f_a$ denotes the radius of the circular orbit of the field spinning with velocity $\dot{\theta}$.}}
\label{diagram_intermediate_kination}
\end{figure}

The cosmological evolution is sketched in Fig.~\ref{diagram_intermediate_kination}.
We start  when the Standard Model radiation dominates, while the scalar field $\phi$ is frozen and contributes to a subdominant cosmological constant. When the scalar field mass becomes larger than the expansion rate, it can start to move and oscillate.
Its coherent motion then behaves as pressure-less matter and leads to the matter era. Later, its kinetic energy dominates its dynamics, the kination era starts and lasts until the SM radiation dominates again.

\FloatBarrier
\FloatBarrier

The cosmological history with the intermediate-scale matter era followed by the kination era can be described in a model-independent way  by the following quantities: 
\begin{enumerate}
\item $\rho_\mathrm{osc}^\mathrm{rad}$ -- Energy density of the background radiation when the scalar field starts oscillating at $a_{\rm osc}$,
\item $\rho_\mathrm{osc}$ -- Energy density of the scalar field at oscillation, 
\item $\rho_{\mathrm{KD}}$ -- Energy density of the scalar field when the kination era starts,
\end{enumerate}
All of them can be related to the model-dependent quantities, see later sections.
For convenience, we define the energy scale at each event by 
\begin{equation}
\rho_i \equiv E_i^4 .
\end{equation}
The non-standard matter era starts at the so-called time of scalar domination, $a=a_{\rm dom}$, when the scalar field energy density is
\begin{align}
\label{eq:rho_M_def}
\rho_{\rm dom} ~ = ~ \rho_\mathrm{osc} \left(\frac{a_\mathrm{osc}}{a_\mathrm{dom}}\right)^3 ~ = ~  \rho_\mathrm{osc} \left(\frac{\rho_\mathrm{osc}}{\rho_\mathrm{osc}^\mathrm{rad}}\right)^3.
\end{align}
It lasts until kination starts at $a=a_{\rm KD}$ with
\begin{equation}
\frac{a_{\rm KD}}{a_{\rm dom}} = \left(\frac{\rho_{\rm dom}}{\rho_{\rm KD}}\right)^{1/3},
\end{equation}
and kination ends when the radiation bath dominates again at $a=a_\Delta$
\begin{align}
\label{eq:rho_Delta_def}
\rho_\Delta ~ = ~ \frac{\rho_{\mathrm{KD}}^2}{\max(\rho_{\rm dom}, \rho_{\rm damp})}.
\end{align}
The duration of kination is given by the e-folding number
\begin{align}
e^{N_{\rm KD}} ~ \equiv ~ \frac{a_{\Delta}}{a_{\rm KD}} = \left[\frac{\max(\rho_{\rm dom}, \rho_{\rm damp})}{\rho_{\rm KD}}\right]^{1/6}.
\label{model_independent_kination_end}
\end{align}

\paragraph{Absence of entropy injection.}
We have introduced the quantity $ \rho_{\rm damp}$, that will enter  in the particle physics implementations where the radial mode of the complex scalar field plays a role, see  Sec.~\ref{sec:PQ:exampleII}.
It is crucial for the duration of kination as the increase in the thermal bath energy density from radial mode damping shortens the duration of  kination, see orange dashed line in Fig.~\ref{diagram_intermediate_kination}.
The longest kination era is obtained when the universe evolves adiabatically during the whole matter-kination era. This implies no entropy injection during the matter-kination era and therefore $\rho_{\rm damp} > \rho_{\rm dom}$. In that case $\rho_\Delta = \rho_{\rm dom} \left( a_{\rm dom}/a_{\Delta}\right)^4$  together with Eq.~\eqref{eq:rho_M_def} and \eqref{eq:rho_Delta_def} imply
\begin{equation}
\label{eq:NKD_vs_NMD}
N_{\rm KD} = N_{\rm MD}/2,
\end{equation}
where $ N_{\rm MD} \equiv \log{(a_{\rm KD}/a_{\rm dom})}$ is the duration of the matter era. Except when explicitly specified, we assume Eq.~\eqref{eq:NKD_vs_NMD} to hold in our plots.

\paragraph{Impossibility of a kination era inside radiation.}
We now comment on the impossibility in our opinion of the radiation-kination-radiation scenario (adopted in \cite{Chang:2021afa}).

\begin{figure}[h!]
\centering
\raisebox{0cm}{\makebox{\includegraphics[width=0.6\textwidth, scale=1]{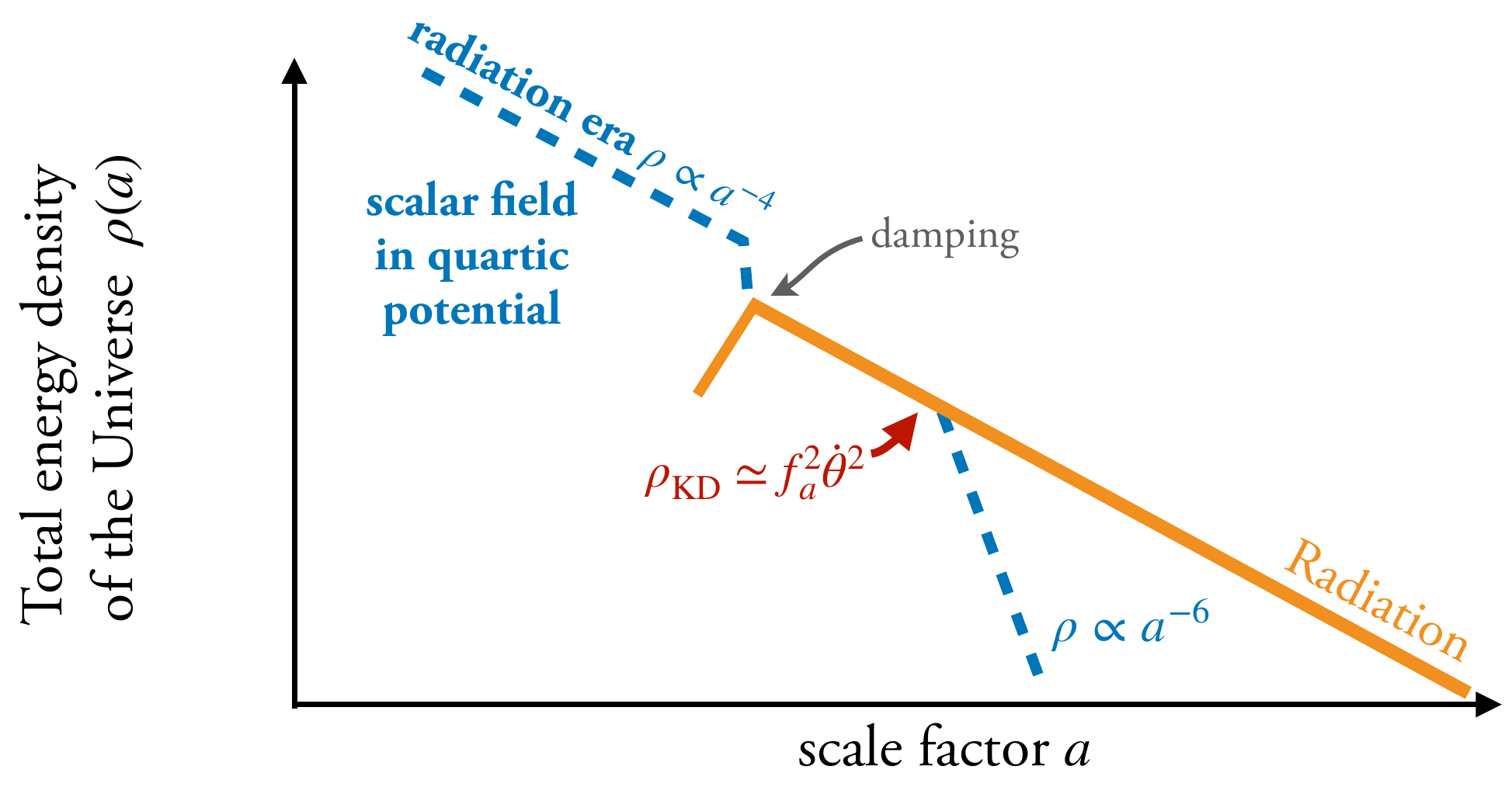}}}
\caption{\textit{ \small A complex scalar field orbiting in a quartic potential has the same EOS as radiation $\omega = 1/3$ (see App.~\ref{app:virial_th} and App.~\ref{paragraph:quartic_potential}). The radial mode must be damped for kination to occur. However, the damping mechanism transfers the radial mode energy to the thermal bath, preventing a kination era. An intermediate kination era must therefore be preceeded by  a matter era as shown in Fig.~\ref{diagram_intermediate_kination}.}}
\label{diagram_quartic_not_work}
\end{figure}

A spinning field inside a quartic potential can lead to the same EOS as radiation.
As we show in Fig.~\ref{afterdecayplot} of App.~\ref{paragraph:quartic_potential}, if the trajectory is circular, then the EOS becomes kination-like once the scalar field reaches the bottom of the potential. Nevertheless, as we illustrate in Fig.~\ref{diagram_quartic_not_work}, this scenario appears unfeasible. The damping of the radial motion responsible for the circular trajectory is expected to produce particles redshifting as radiation (or worse as matter), which prevents the universe to enter a kination stage.

\FloatBarrier
\FloatBarrier

\subsection{Preview: UV completions with intermediate kination cosmology }
\begin{figure}[h!]
\centering
\raisebox{0cm}{\makebox{\includegraphics[width=0.45\textwidth, scale=1]{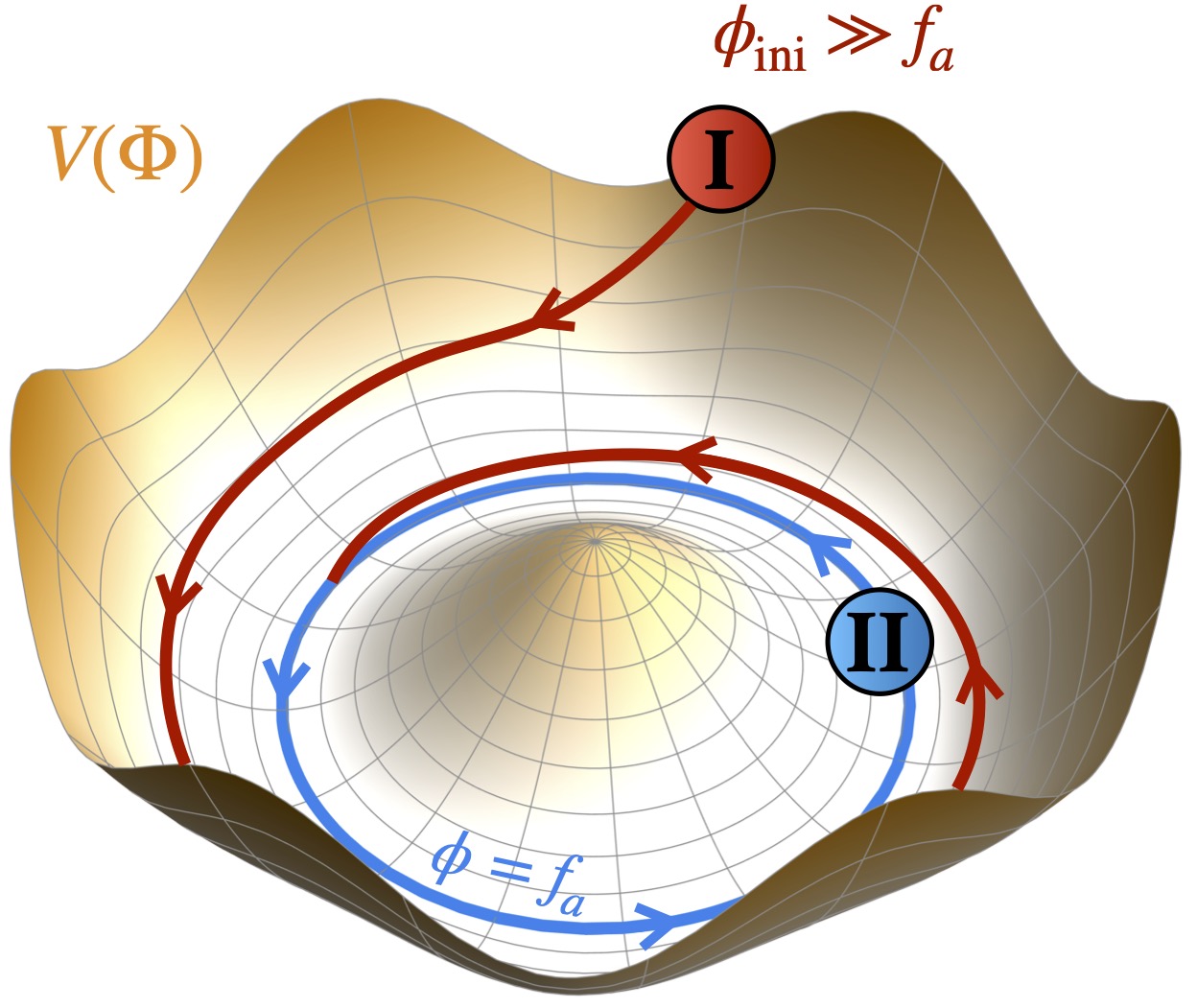}}}
\caption{\textit{ \small Axion models naturally provide a kination era preceded by a matter era. 
The field starts at large radius (red trajectory "I") before reaching stage "II".  }}
\label{cartoon_models}
\end{figure}

The sequence of events presented in the previous subsection and in Fig.~\ref{diagram_intermediate_kination} requires a non-trivial dynamics which can nevertheless occur naturally in well-motivated particle physics models. This will be the subject of a thorough analysis in Secs.~\ref{sec:trapped_mis}, \ref{sec:PQ:exampleII}, \ref{sec:scenario_I_non_thermal_damping}, \ref{sec:complex_field_thermal_potential}
and \ref{sec:complex_field_low_reh_temp}.
Here we give a preview of  the main properties. Generally, the kination era arises from a stage when the energy density of the universe is dominated by the spinning of an axion field around a circular orbit with  vanishing potential energy (at the bottom of the $U(1)$-symmetric potential). The key questions are: What imprinted the initial velocity of the axion? How was the axion kicked in the first place?  
We discuss the  main  mechanism illustrated in Fig.~\ref{cartoon_models}.

In the first attempt in Sec.~\ref{sec:trapped_mis}, we show why it is not enough to invoke only the axion degree of freedom (angular direction of the complex scalar field). 
In fact, the radial component of the complex scalar field is the key feature. The dynamics of the radial mode will trigger a motion in the angular direction. The interplayed dynamics induces a matter era. Eventually the field will reach the bottom of the potential. There is still an obstacle for a kination EOS to follow: the energy density in the radial mode must be damped. The optimal case happens when this damping occurs before the scalar field energy density dominates. A kination era may still happen otherwise but its duration will be reduced by the entropy injection. Two damping mechanisms can be invoked: through parametric resonance of the radial mode at early times or through thermal effects. The latter case relies on the interaction of the radial mode with  particles in the thermal bath. 

In the following Sec.~\ref{sec:modelindependent}, we present the implications for the GW signals of the intermediate matter+kination era in the most general way, that does not rely on any assumptions about the particle physics realisation. 
In Sec.~\ref{sec:darkmatter} and Sec.~\ref{sec:baryon_asymmetry}, we connect the GW signal to the axion dark matter abundance and the baryon asymmetry.
We will present the specific model parameter dependences of the GW signal in Secs.~\ref{sec:trapped_mis}, \ref{sec:PQ:exampleII}, \ref{sec:scenario_I_non_thermal_damping}, \ref{sec:complex_field_thermal_potential},  and \ref{sec:complex_field_low_reh_temp}.

\FloatBarrier
\FloatBarrier

\newpage
\addcontentsline{toc}{part}{Model-independent predictions}
\begin{center}
{\LARGE \bf \noindent Model-independent predictions}
\end{center}

\section{Gravitational-wave peaked signature}\label{sec:modelindependent}

\paragraph{Cosmic archeology. }
We will discuss the effect of an intermediate matter-kination era on primordial GW. There are four main sources of GW of primordial origin: from inflation \cite{Bartolo:2016ami}, from reheating/preheating \cite{Caprini:2018mtu}, from first-order phase transitions \cite{Caprini:2015zlo,Caprini:2019egz} and from networks of cosmic strings \cite{Gouttenoire:2019kij}. The first and last can be considered long-lasting sources while the others are typically short-lasting (meaning only active for a Hubble time or so).

As GW from long-lasting sources are produced at different times, they encode information about the cosmological history.
Their GW spectrum spans a wide range of frequencies. GW from the earlier times are produced when the horizon size was smaller and, thus, have higher frequencies.
Different  cosmological histories lead to different amounts of GW today. A non-standard cosmological history  imprints a GW spectral distortion which enables to trace the early-to-late history of our universe in the direction of high-to-low frequencies.

Examples of the cosmic-archeology works using long-lasting GW are \cite{Cui:2017ufi, Saikawa:2018rcs, Cui:2018rwi, DEramo:2019tit, Figueroa:2019paj, Bernal:2019lpc, Chang:2019mza, Gouttenoire:2019kij, Gouttenoire:2019rtn, Blasi:2020wpy, Ringwald:2020vei, Co:2021lkc, Gouttenoire:2021wzu}, see \cite{Allahverdi:2020bys} for a review.
Developing from the idea of cosmic archeology, this section focuses on the intermediate kination following the matter era and its smoking-gun signature, the peak\footnote{A peak signature also arises from some inflationary models when the tensor perturbations leave the horizon during the non-slow-rolling phase. The peak has a log-normal shape and also incorporates the oscillation feature \cite{Chen:2014cwa,Braglia:2020eai, Fumagalli:2020nvq,Braglia:2020taf, Fumagalli:2021cel, Witkowski:2021raz}. This feature can be distinguished from the broken power-law arising in the post-reheating dynamics discussed in this paper.} in GW spectrum from primordial inflation (Sec.~\ref{sec:inflationaryGW}), cosmic strings (Sec.~\ref{sec:GW_from_CS}), or both (Sec.~\ref{sec:multiple_peak}).
For a short-lasting source, GW are produced at a specific time and the spectrum localizes at a specific frequency. The effect of the non-standard cosmological history shifts the spectrum as a whole, at the exception of the causality tail \cite{Hook:2020phx}. As an example of the short-lasting GW source, the effect of matter-kination era on the GW from first-order phase transitions is discussed in Sec.~\ref{sec:phase_transition}.

\paragraph{Why a peaked spectrum?}
Before providing the mathematical formulation of the GW spectrum, let us first  illustrate the origin of the peak signal from the matter-kination era.
For simplicity, the following argument assumes the absence of entropy injection into the thermal bath.
The spectrum observed today of GW produced at $H_*$ is
\begin{align}
\Omega_{\rm GW} ~ = ~ \left(\frac{\rho_{\rm GW,*}}{\rho_{\rm tot,*}}\right) \left(\frac{H_*}{H_0}\right)^2 \left(\frac{a_*}{a_0}\right)^4.
\label{eq:GW_spectrum_general_relation}
\end{align}
 
GW inherit a fraction of the total energy density of the universe  at the time of production.
The amplitude of the GW spectrum from long-lasting GW sources (such as GW from inflation or cosmic strings) is therefore a probe of the  cosmological history.

\begin{figure}[h!]
\centering
\raisebox{0cm}{\makebox{\includegraphics[width=0.975\textwidth, scale=1]{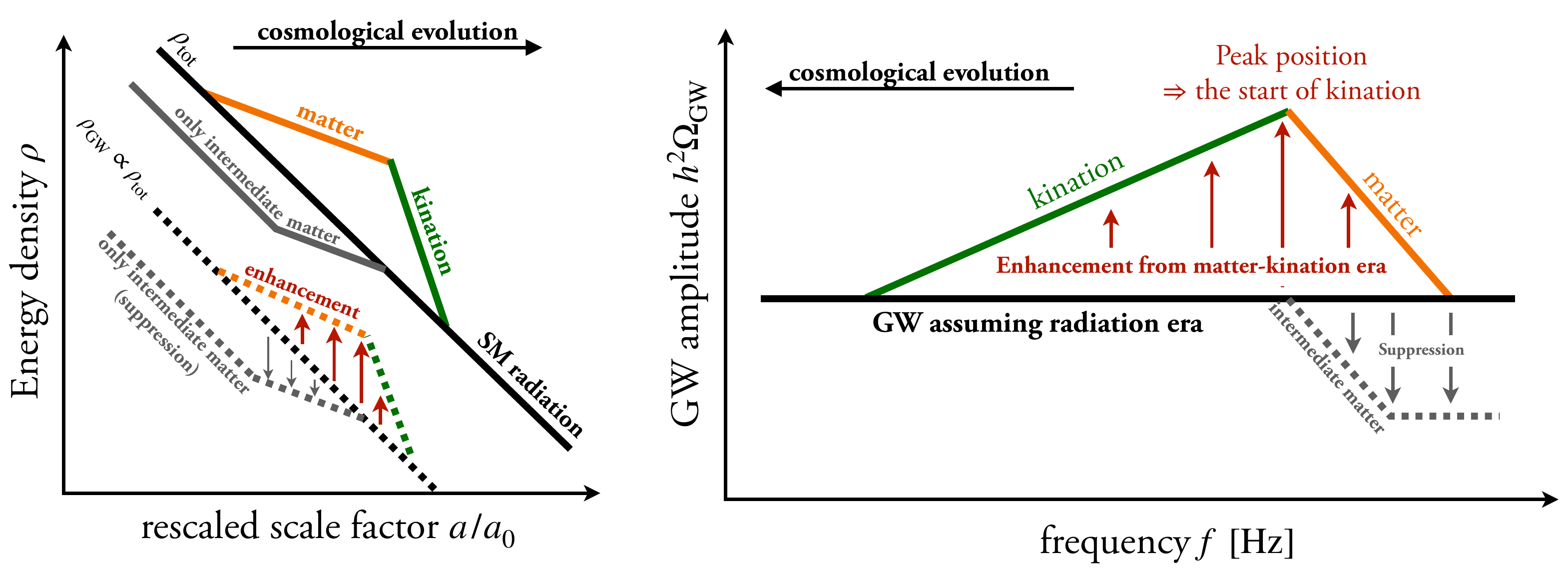}}}
\caption{\textit{ \small The left panel shows that the energy density in GW is enhanced by an intermediate matter-kination era. This is true for long-lasting sources that produce GW as a fraction of the total energy density. The right panel shows the enhanced GW spectrum whose peak position directly relates to the start of the kination era. By rescaling the scale factor of the universe, a matter era without kination leads to a universe that has smaller energy density at earlier times and thus leads to a suppression of the GW signal.}}
\label{fig:peak_cartoon}
\end{figure}

The inflationary GW energy density is sourced by the scale-invariant tensor perturbation from the primordial inflation, and Fourier modes remain frozen until they re-enter the Hubble horizon $f = H/2\pi \propto \rho_{\rm tot}^{1/2}$. Modes continuously re-enter the horizon in the post-inflationary cosmological history. In this sense, inflation can be understood as a long-lasting source of GW.
In the case of cosmic strings, loops are continuously produced and  radiate GW with amplitude $\rho_{\rm GW} \propto \rho_{\rm tot}$ as the network of long strings in the scaling regime has an energy density proportional to $\rho_{\rm tot}$.
For both sources, GW from a matter-kination era is enhanced because of the larger $\rho_{\rm tot}$, compared to the standard cosmology,
\begin{align}
\frac{\rho_{\rm GW}^{\rm NS}}{\rho_{\rm GW}^{\rm ST}} ~ = ~ \frac{\rho_{\rm tot}^{\rm NS}}{\rho_{\rm tot}^{\rm ST}} ~ \geq ~ 1.
\end{align}
During the matter era, the above ratio increases with times so we expect the GW spectrum to increase in the direction of high-to-low frequencies.
This ratio decreases during the kination era over time so the GW amplitude decreases for lower frequencies.
GW produced during the transition between the matter and the kination era are maximally enhanced, and correspond to the peak illustrated in Fig.~\ref{fig:peak_cartoon}.

On the other hand, a matter era without kination leads to a suppression of the SGWB from long-lasting sources.
 The above argument using Eq.~\eqref{eq:GW_spectrum_general_relation} applies for the rescaled scale factor. Since a matter era without kination leads to a horizon size today  which is larger than the one predicted in standard cosmology, the total energy density before the matter era is smaller than in the standard case after rescaling, cf. Fig.~\ref{fig:peak_cartoon}. Hence, it induces the step-liked suppression which might be an observable signature for a large GW signal such as cosmic-string SGWB \cite{Gouttenoire:2019rtn}.

\FloatBarrier
\FloatBarrier

\subsection{Inflationary Gravitational Waves}

The content of Sec.~\ref{sec:inflationaryGW} is reported as well in \cite{Gouttenoire:2021wzu}.
\label{sec:inflationaryGW}
\subsubsection{Standard cosmology}
Today, the irreducible stochastic GW background from inflationary tensor perturbations, denoted by its fraction of the total energy density, reads \cite{Caprini:2018mtu}
\begin{equation}
\Omega_{\rm GW}~=~ \frac{k^2a_k^2}{24H_0^2} \Omega_{\rm GW,inf}
\label{eq:Omega}
\end{equation}
and arises from modes with comoving wave number 
\begin{equation}
k=a_k H_k 
\end{equation}
which re-entered the cosmic horizon when the scale factor of the universe was $a_k$
and the Hubble rate was $H_k$. $H_0$ is the Hubble rate today.
The spectrum contains a high-frequency cut-off corresponding to the inflationary scale $k_{\rm inf} = a_{\rm inf}H_{\rm inf}$, where  $a_{\rm inf}$ is the scale factor at the end of inflation \cite{Nakayama:2008wy}.
After being generated by quantum fluctuations during inflation, the metric tensor perturbation are stretched outside the Hubble horizon and are well-known to lead to a nearly scale-invariant power spectrum at the horizon re-entry,
\begin{equation}
\Omega_{\rm GW,inf} \simeq \frac{2}{\pi^2} \left( \frac{H_{\rm inf}}{M_{\rm pl}} \right)^2 \left(  \frac{k}{k_p} \right)^{n_t},\label{eq:primordial_inf_spectrum}
\end{equation}
where $H_{\rm inf}$ the Hubble rate during inflation, and $k_p$ is the pivot scale used for CMB observation $k_p/a_0 \simeq 0.002~\rm Mpc^{-1}$  \cite{Akrami:2018odb} (equivalent to the GW frequency $f_p = 3.1 \times 10^{-18} ~ \mathrm{Hz}$).
In slow-roll inflation, the spectral index $n_t$ is expected to be only slightly red-tilted
\begin{equation}
n_t \simeq -2\epsilon \simeq - \frac{r}{8} \gtrsim -0.0045,
\end{equation}
since the non-observation of primordial B-modes by BICEP/Keck Collaboration  constrains the tensor-to-scalar ratio to be $r\lesssim 0.036$ \cite{BICEP:2021xfz}. The presence of this red-tilt suppresses the GW energy density by $\mathcal{O}$(10\%) correction in the ranges of Pulsar-Timing-Arrays (PTA) and Earth-based interferometers.
In the rest of the paper, we neglect this suppression and assume $n_t =0$ for simplicity.

Tensor modes that enter during the radiation era have the standard flat spectrum 
\begin{equation}
\Omega^\mathrm{st}_{\rm GW} h^2  \simeq  (1.3  \times 10^{-17}) G(T_k)  \left(\frac{E_\mathrm{inf}}{10^{16}\textrm{ GeV}}\right)^4,
\label{st_GW_inflation}
\end{equation}
where $E_{\rm inf}$ is the inflationary energy scale, $G(T_k)=  ({g_*(T_k)}/{106.75}) ({g_{*,s}(T_k)}/{106.75})^{-4/3}$ and $T_k$ is the temperature when a given mode enters the Hubble horizon.
This GW background is beyond the sensitivity of future GW observatories: LISA \cite{Audley:2017drz} and Einstein Telescope \cite{Hild:2010id, Punturo:2010zz}. Only Big Bang Observer \cite{Yagi:2011wg} could be sensitive  to if we assume  the largest inflation energy scale allowed by CMB data \cite{Akrami:2018odb}.

\subsubsection{In the presence of a matter-kination era}
\paragraph{Spectral index.}
 The inflationary GW which are produced with the horizon-size wavelength $H_k^{-1}$ have the frequency today
\begin{equation}
f =  \frac{H_k}{2 \pi} \frac{a_k }{a_0}.
\end{equation}
Using the Friedmann equation $H={\sqrt{\rho/3}} M_{\rm Pl}$
where $\rho\propto a^{-3(1+\omega)}$, $\omega$ being the EOS of the universe, we have $f \propto a_k^{-{(1+3\omega)}/{2}}$ and Eq.~(\ref{eq:Omega}) gives
\begin{equation}
\Omega_{\rm GW}
 \propto ~ f^{\beta}, ~ ~ \textrm{with} ~ ~ \beta ~ \equiv ~ -2\left(\frac{1-3\omega}{1+3\omega}\right).
 \label{eq:spectralindex}
\end{equation}
Note that the non-trivial scaling comes from the factor $a_k^2$ in Eq.~\eqref{eq:Omega} which arises from the transfer function of GW after re-entry \cite{Boyle:2005se}.
Therefore, modes entering the horizon during radiation ($\omega=1/3$), matter ($\omega=0$) and kination ($\omega=1$)
eras have spectral indices $\beta = 0, -2$, and 1, respectively\footnote{For a more realistic power spectrum, i.e., by solving the full GW EOM, the non-standard cosmology alters the behavior of the GW transfer function.
The effect on the amplitude is  of order  $\mathcal{O}(1)$ \cite{Figueroa:2019paj}, while the transition between eras could feature a spectral oscillation from the change of Bessel function's orders.}.
\paragraph{Triangular shape.}
For the intermediate matter-kination era as illustrated in Fig.~\ref{diagram_intermediate_kination}, 
the sign-change in the spectral index occurs at the transition between these eras and leads to a peaked GW signature.  The high-frequency slope -2 is associated to the matter era while the low-frequency slope  +1  is associated to the kination era.
The GW spectrum in the presence of the matter-kination era reads
\begin{eqnarray}
\Omega_\mathrm{GW,0} h^2 (f) =  \Omega^\mathrm{st}_{\rm GW}(f_\Delta)h^2 \times 
\begin{cases}
1 &;  f < f_\Delta ~ ~ ~ \textrm{(late-time radiation)},\\
\left(f/f_\Delta\right)  &; f_\Delta < f < f_\mathrm{KD} ~ ~ ~ \textrm{(kination)},\\
\left(f_\mathrm{KD}/f_\Delta\right) \left(f_\mathrm{KD}/f\right)^2  &; f_\mathrm{KD} < f < f_\textrm{dom} ~ ~ ~ \textrm{(matter)},\\
\left(f_\mathrm{KD}/f_\Delta\right) \left(f_\mathrm{KD}/f_\textrm{dom}\right)^2 &; f_\textrm{dom} < f ~ ~ ~ \textrm{(early-time radiation)},
\end{cases}
\label{inflation_GW_master}
\end{eqnarray}
where the GW abundance assuming the standard radiation-dominated cosmology $  \Omega^\mathrm{st}_{\rm GW}$ is given by Eq.~(\ref{st_GW_inflation}), and
$f_\Delta$, $f_\mathrm{KD}$ (peak frequency), $f_\textrm{dom}$ are the characteristic frequencies corresponding to the modes re-entering the horizon right after the end of the
kination era, at the beginning of the kination era, and at the end of the matter era, respectively.
 They are defined as:
\begin{align}
\label{kination_end_frequency}
&f_\Delta  =  \frac{H_\Delta a_\Delta}{2 \pi a_0}   \simeq   (2.7 \times 10^{-6}  \textrm{ Hz}) \, \left(\frac{g_*(T_{\Delta})}{106.75}\right)^{1/2} \left(\frac{g_{*,s}(T_{\Delta})}{106.75}\right)^{-1/3} \left(\frac{T_{\Delta}}{10^{2}\textrm{ GeV}}\right),\\
\label{inflation_peak_frequency}
&f_\textrm{KD} =  \frac{H_\textrm{KD} a_\textrm{KD}}{2 \pi a_0}  
=  f_\Delta  \left(\frac{\rho_\textrm{KD}}{\rho_\Delta} \right)^{1/3}
=  f_\Delta e^{2 N_\mathrm{KD}}
\simeq (1.1 \times 10^{-3}  \textrm{Hz}) \, G^{1/4}(T_\Delta)
\left(\frac{\rho_\textrm{KD}^{1/4}}{\textrm{10 TeV}}\right) \left(\frac{e^{N_\textrm{KD}/2}}{10}\right),
\end{align}
where the e-folding of the kination era is $e^{N_\textrm{KD}} \equiv (\rho_\mathrm{KD}/\rho_\Delta)^{1/6}$.
This peak frequency $f_\textrm{KD} $ thus contains information about the energy scale and the duration of the kination era.
The peak amplitude at $f_\mathrm{KD}$ is
\begin{eqnarray}
\nonumber
\Omega_\mathrm{GW,KD} & =& \Omega^{\mathrm{st}}_{\rm GW} h^2 (f_\Delta) \left(\frac{f_\mathrm{KD}}{f_\Delta}\right)   
 ~ =  ~ \Omega^{\mathrm{st}}_{\rm GW} h^2 (f_\Delta) e^{2 N_\mathrm{KD}}\\
 &\simeq & 2.8 \times 10^{-13}  \left(\frac{g_*(T_\Delta)}{106.75}\right) \left(\frac{g_{*,s}(T_
\Delta)}{106.75}\right)^{-4/3} \left(\frac{E_{\inf}}{10^{16}\textrm{ GeV}}\right)^4 \left( \frac{\exp(2 N_\mathrm{KD})}{e^{10}} \right),
\label{inflation_peak_amplitude}
\end{eqnarray}
and is enhanced by the duration of kination era.
Finally, the frequency corresponding to the start of the matter era is
\begin{align}
f_\textrm{dom}  =  \frac{H_\textrm{dom} a_{\rm dom}}{2 \pi a_0}  
=  f_{\rm KD}  \left(\frac{\rho_{\rm dom}} {\rho_\textrm{KD}}\right)^{1/6}.
\end{align}
The amplitude difference between flat parts from radiation eras before and after the matter-kination era is
\begin{equation}
\frac{\Omega_\mathrm{GW}(f > f_\textrm{dom})}{\Omega_\mathrm{GW} (f<f_\Delta) }  =  \left( \frac{f_\mathrm{KD}}{f_\Delta}\right) \left( \frac{f_\mathrm{KD}}{f_\textrm{dom}}\right)^2
=  \left( \frac{1}{\rho_\Delta} \cdot \frac {\rho_\textrm{KD}^2} {\rho_{\rm dom}}\right)^{1/3} ~ \leq ~ 1.
\end{equation}
The above equality is satisfied when no entropy dilution occurs during the whole completion of the matter domination era and $\rho_\Delta = \rho_\mathrm{KD}^2/ \rho_{\rm dom}$, cf. Eq.~\eqref{eq:rho_Delta_def}.
Otherwise, the amount of dilution is imprinted in the difference between the amplitudes of the two flat parts of the spectrum.

\begin{figure}[h!]
\centering
{\bf Gravitational waves from primordial inflation}\\
\raisebox{0cm}{\makebox{\includegraphics[width=0.47\textwidth, scale=1]{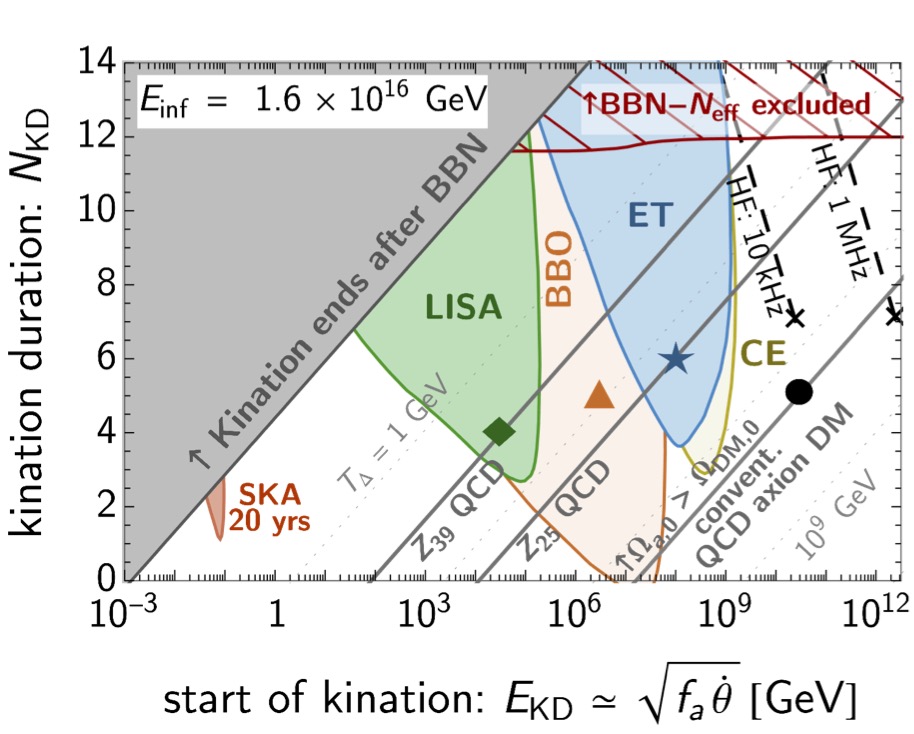}}}
\raisebox{0cm}{\makebox{\includegraphics[width=0.475\textwidth, scale=1]{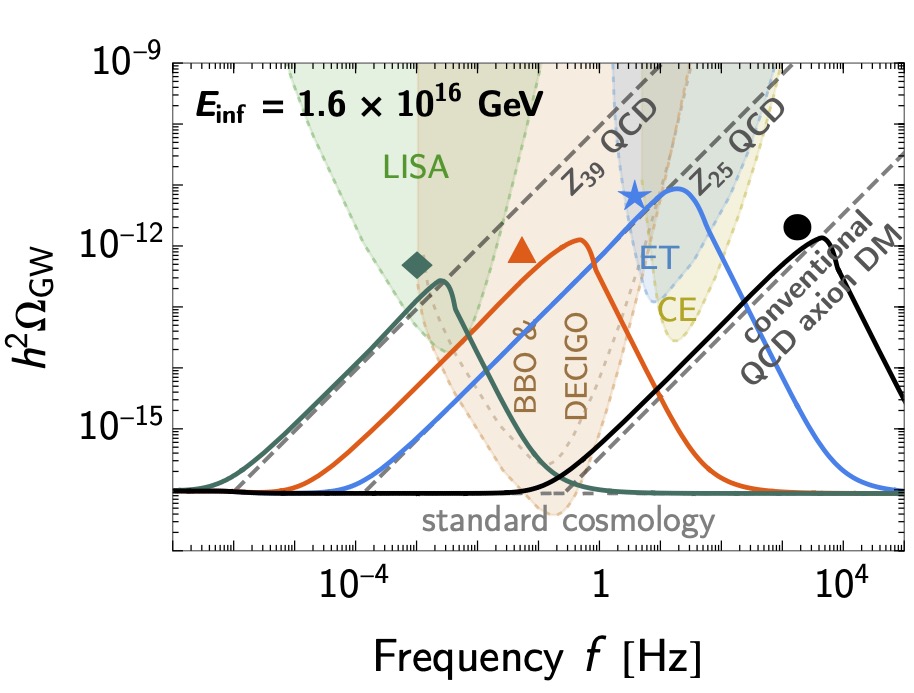}}}\\[0.1em]
\raisebox{0cm}{\makebox{\includegraphics[width=0.47\textwidth, scale=1]{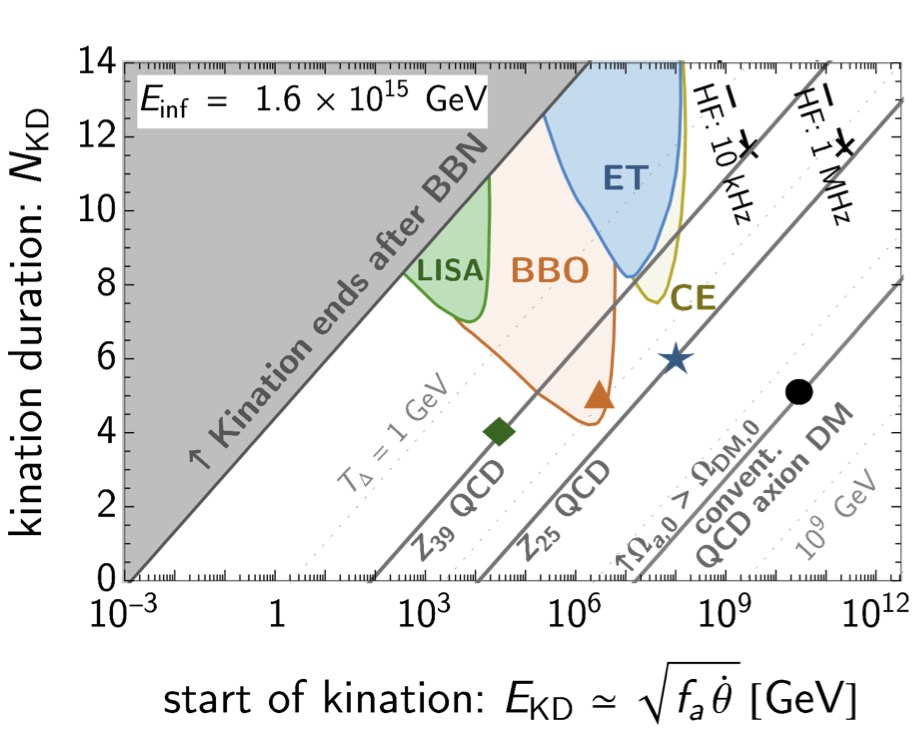}}}
\raisebox{0cm}{\makebox{\includegraphics[width=0.475\textwidth, scale=1]{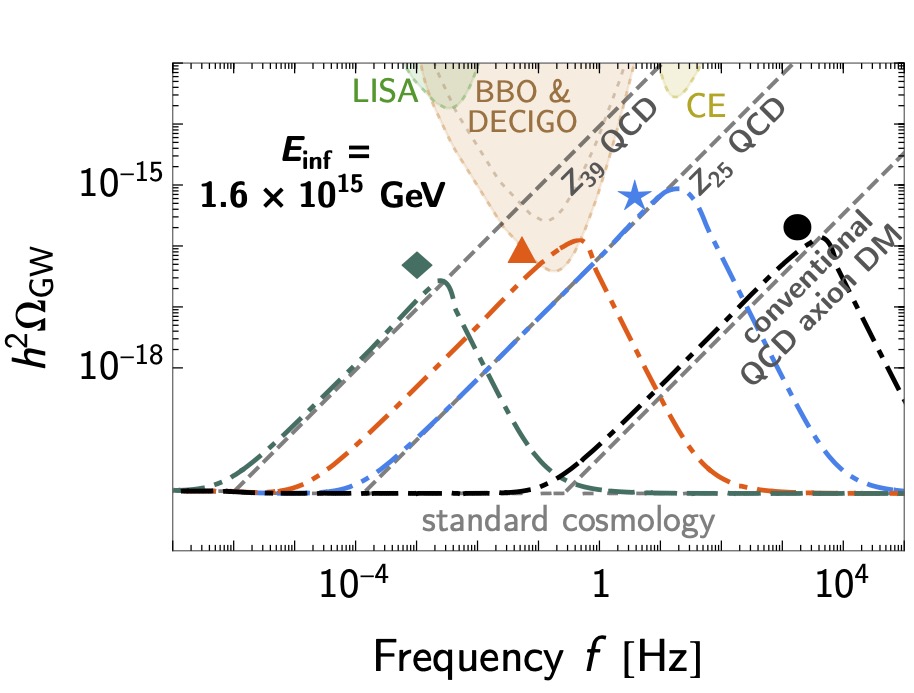}}}
\caption{\textit{ \small \textbf{Left}: Detectability of the irreducible GW background from inflation with energy scale $E_{\rm inf}$, enhanced by a period of matter-kination lasting for $(2N_\mathrm{KD}+N_\mathrm{KD})$ efolds with no entropy injection, cf. Eq.~\eqref{eq:NKD_vs_NMD}. The kination era starts at energy scale $E_\mathrm{KD}$ and ends when the temperature of the universe is $T_\Delta$ (gray dotted lines).  The observable windows of the peak signal are shown in the colored regions. BBN constrains the late kination eras (gray) and the duration of kination (red-hatched) (see Sec.~\ref{subsec:afterinflation}).
The QCD axion that allows a kination era could be DM along the solid-gray lines for the conventional and $\mathbb{Z}_\mathcal{N}$ QCD-axion models, assuming kinetic misalignment (see Sec.~\ref{sec:darkmatter}).
The smaller the inflation scale, the weaker the GW amplitude. The black dashed lines show the prospects for the detectability by hypothetical high-frequency experiments operating at 10 kHz and 1 MHz with sensitivity $h^2\Omega_{\rm sens} = 10^{-10}$.
\textbf{Right:} GW spectra corresponding to the benchmark points shown in the left panel. Dashed lines show the peak position for the matter-kination era from the QCD axion DM. }}
\label{detect_peak_inf}
\end{figure}

\begin{figure}[h!]
\centering
\raisebox{0cm}{\makebox{\includegraphics[width=0.47\textwidth, scale=1]{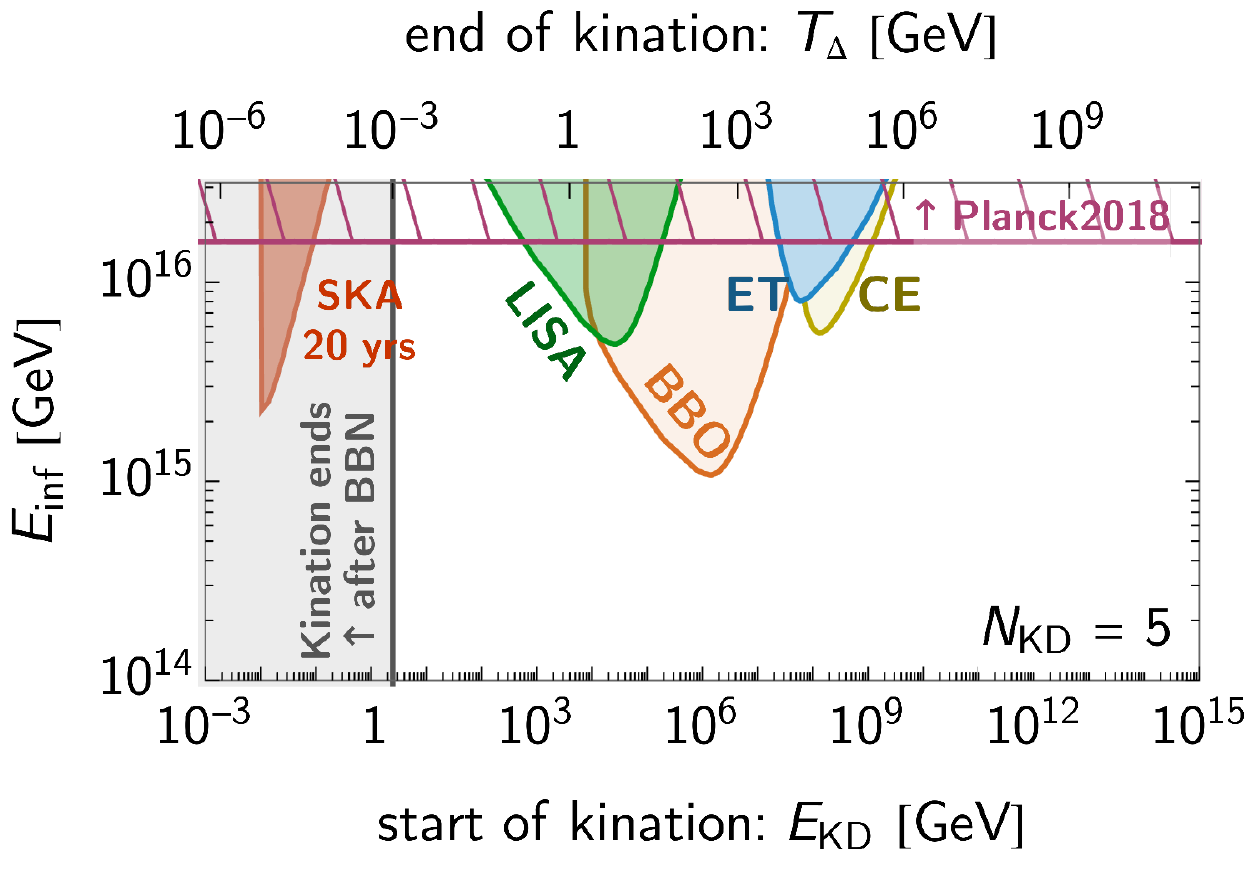}}}
\qquad
\raisebox{0cm}{\makebox{\includegraphics[width=0.47\textwidth, scale=1]{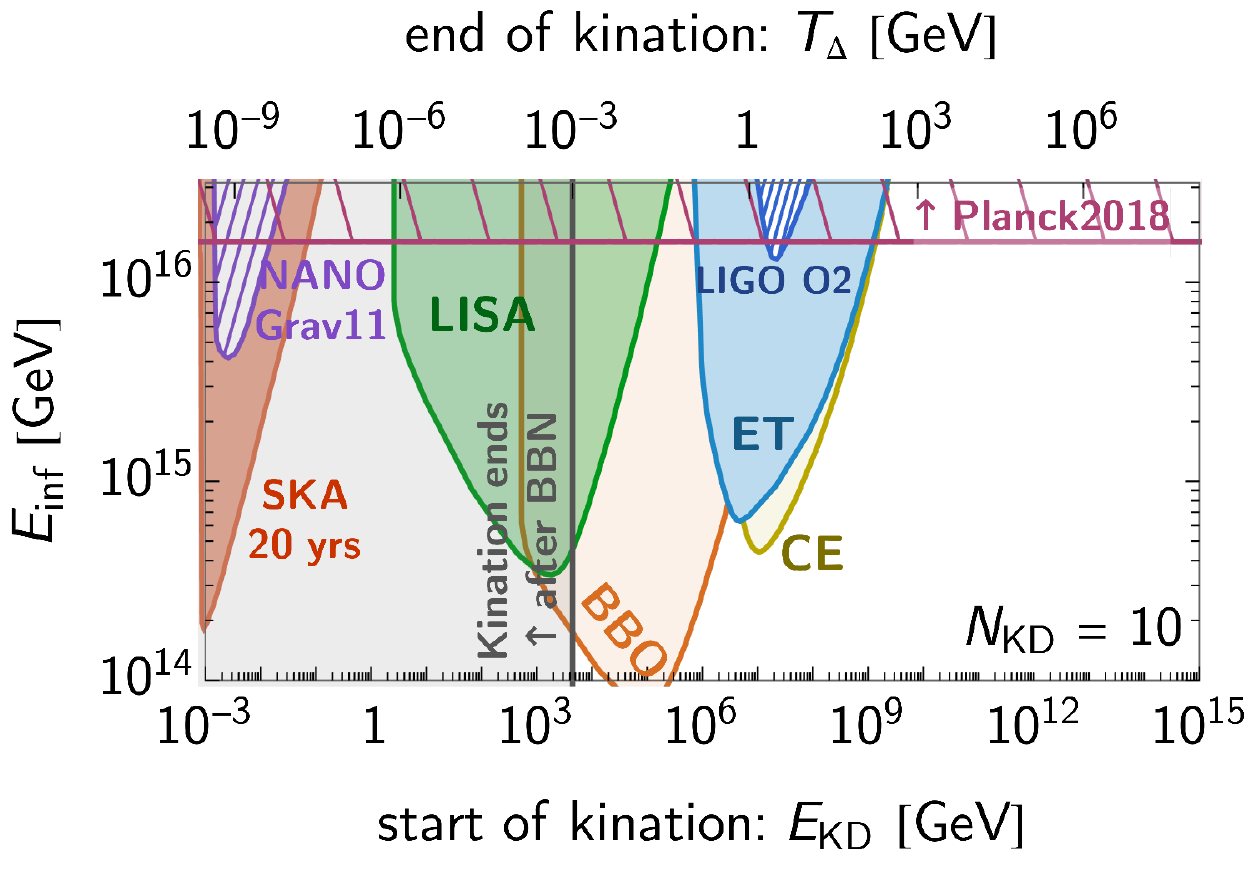}}}  
\caption{\textit{ \small Observability regions when varying the inflation scale for two kination durations, $N_\mathrm{KD}=5$ (\textbf{left}) and $N_\mathrm{KD}=10$ (\textbf{right}).
Planck2018 data \cite{Ade:2018gkx, Akrami:2018odb} implies the upper bound $E_\mathrm{inf} \lesssim 1.6 \times 10^{16} ~ {\rm GeV}$ (purple-hatched). }}
\label{detect_peak_Einf}
\end{figure}

\paragraph{Detectability}
The resulting typical spectra are plotted in the right panel of Fig.~\ref{detect_peak_inf} for three benchmark points reported in 
the left panel and corresponding to different choices of kination energy scales and kination durations. 
A large parameter space allows the peak from the matter-kination era to be  probed by LISA \cite{Audley:2017drz}, BBO \cite{Yagi:2011wg}, ET \cite{Hild:2010id, Punturo:2010zz}, CE \cite{Evans:2016mbw}  and SKA \cite{Janssen:2014dka}, where we have used the integrated power-law sensitivity curves\footnote{We denote a signal to be detectable when its amplitude surpasses the power-law sensitivity curve for a given signal-to-noise ratio (SNR). We note that the SNR formula given in \cite{Gouttenoire:2019kij} is an approximated one which works in the limit of a large detector noise. The generic formula can be found in \cite{Allen:1997ad, Kudoh:2005as, Brzeminski:2022haa}. We have checked that the two formulae agree for the power-law sensitivity curves with ${\rm SNR} \lesssim 10$ used in this paper.}  of \cite{Gouttenoire:2019kij}. 
Note that a kination era lasting more than $\sim 12$ e-folds is not viable as a too large energy density in GW violates the BBN-$N_{
\rm eff}$ bound, see Eq.~\eqref{eq:BBN_bound_inflation_GW}.
Fig.~\ref{detect_peak_Einf} shows that the longer duration of the kination era enhances the detectability of the peak signature.

The peak signature which we are exploring should be distinguished from GW peaked signals produced by cosmological first-order phase transitions, e.g. \cite{Grojean:2006bp,Caprini:2019egz}, or by network of cosmic strings \cite{Ramberg:2019dgi,Gouttenoire:2019kij}.
Another scenario with large primordial GW from inflation is axion inflation \cite{Barnaby:2010vf,Barnaby:2011qe}. The spectral shape of this signal is however very different from what we predict from an intermediate matter-kination era.

\FloatBarrier

\FloatBarrier
\FloatBarrier

\subsection{Gravitational waves from cosmic strings}
\label{sec:GW_from_CS}

\subsubsection{Short review}
A network of cosmic strings (CS) arises from a $U(1)$-symmetry-breaking phase transition in the early universe \cite{Kibble:1976sj}. We refer to \cite{Gouttenoire:2019kij,Auclair:2019wcv} for reviews of their GW emission and \cite{Vilenkin:2000jqa} for a textbook.
\paragraph{Scaling regime.}
It is well-known that the network of topological defects reaches the so-called scaling regime where the correlation length of the network grows linearly with the cosmic time \cite{Vilenkin:2000jqa, Martins:2000cs, Martins:2016ois}.
Unlike other types of defects, only cosmic strings attain a constant fraction of the total energy density of the universe, regardless of the cosmology.
Loops are being produced, which later decay either into GW or into particles.

\paragraph{Nambu-Goto string.}
As in \cite{Gouttenoire:2019kij, Gouttenoire:2019rtn}, we first assume the Nambu-Goto approximation where CS are infinitely thin classical objects which are described by their tension.
\begin{equation}
\mu = 2\pi n \, \eta^2 \times \begin{cases}
1 ~ ~ ~ & {\rm for ~ local ~ strings},\\
\log(\eta t)  ~ ~ ~ & {\rm for ~ global ~ strings},
\end{cases}
\label{string_tension}
\end{equation}
where $\eta$ is the VEV of the scalar field that forms strings, $n$ is the winding number assumed to be $n=1$, and the global strings have a massless Goldstone mode with logarithmically-divergent gradient energy \cite{Vilenkin:2000jqa}.

\paragraph{GW emission from loops.}
The SGWB from CS is dominated by the emission of loops with the emission frequency $\tilde{f}$ corresponding to the mode $m \in \mathbb{Z}^+$ of the loop oscillation, $\tilde{f} = 2m/l$,
 where $l$ is the loop length. 
The frequency today is $f = \tilde{f}[a(\tilde{t})/a(t_0)].$
Loops are formed at time $t_i$ with a size $\alpha t_i$ at a rate
\begin{equation}
\frac{dn_{\rm loop}}{dt_i} = 0.1 \frac{C_{\rm eff}(t_i)}{\alpha t_i^4},
\end{equation}
where the local-string loop-formation efficiency reaches the asymptotic value $C_{\rm eff} \simeq 0.39$, $5.4$, $29.6$ during matter, radiation and kination respectively \cite{Gouttenoire:2019kij}. For global strings, the long strings lose energy through particle production in addition to loop formation. This suppresses the loop-production efficiency with log-dependence, which is approximated to be $C_{\rm eff} \sim \mathcal{O}(1)$ for all cosmologies \cite{Chang:2019mza, Gouttenoire:2019kij, Chang:2021afa}. However, for the plots and analysis of this paper, we solve $C_{\rm eff}(t)$ as a solution of the velocity-dependent one-scale (VOS) equations governing the string network evolution, see \cite{Gouttenoire:2019kij} for more details on VOS equations of both local and global strings.
As pointed-out in \cite{Gouttenoire:2019kij}, the VOS evolution should capture the network's behavior more realistically than a constant $C_{\rm eff}$ for each era.

We suppose that the GW spectrum is dominated by the largest loops formed with size equal to $10\%$ of the horizon \cite{Blanco-Pillado:2013qja}. Hence we impose the following monochromatic loop size distribution
\begin{equation}
P_\alpha(\alpha) = \delta(\alpha - 0.1).
\end{equation}
After loops are formed, their oscillation of the $m^{\rm th}$ mode emit GW of power 
\begin{equation}
P_{\rm GW}^{(m)} = \Gamma^{(m)} G\mu^2, \qquad \text{with} \quad \Gamma^{(m)} = \frac{\Gamma m^{-{4/3}}}{\sum_{p=1}^\infty p^{-4/3}},
\end{equation}
where $\Gamma = 50$ \cite{Blanco-Pillado:2017oxo} and where, without too much loss of generality, we have assumed the small-scale structure to be cusp-dominated \cite{Olmez:2010bi}.
Due to GW and particle emissions, the loops length $l$ shrinks as
\begin{equation}
l(\tilde{t}) =  \alpha t_i - (\Gamma G \mu + \kappa) (\tilde{t} - t_i). \label{eq:length_shrink}
\end{equation}
where $\Gamma G \mu$ and $\kappa$ are shrinking rates due to GW and particle emissions, respectively.
Local-string loops only decay via GW emission ($\kappa = 0$), while global-string loops dominantly decay into particles ($\kappa \gg \Gamma G\mu$).
Moreover, the particle production rate $\kappa$ contains the logarithmic dependence which follows from the string tension $\kappa = \Gamma_{\rm Gold}/2 \pi \log(\eta t)$, where $\Gamma_{\rm Gold} \approx 65 $ \cite{Vilenkin:1986ku}.
The master formula for SGWB from CS can be written pedagogically as
\begin{equation}
\Omega_{\rm GW}(f) = \sum_m \frac{1}{\rho_c} \int_{t_{\rm osc}}^{t_0} d\tilde{t} \int_0^1 d\alpha \,\Theta\left(t_i - \frac{l_*}{\alpha}\right)\cdot \Theta(t_i - t_{\rm osc})\cdot\left[\frac{a(\tilde{t})}{a(t_0)} \right]^4 \cdot P_{\rm GW}^{(m)} \cdot\left[\frac{a(t_i)}{a(\tilde{t})} \right]^3 \cdot P_\alpha(\alpha)\cdot \frac{dt_i}{df}\cdot \frac{d n_{\rm loop}}{d t_i} , \label{eq:master_eq_pedagogical}
\end{equation}
such that the chronology of the involved processes can be understood from right to left.
Loops are formed at rate $dn_{\rm loop}/dt_i$ at $t_i$ with a size distribution $P_\alpha(\alpha)$. They dilute like $a^{-3}$ due to Hubble expansion, before emitting GW with power spectrum $P_{\rm GW}^{(m)}$ which subsequently redshifts like $a^{-4}$. The two Heaviside functions represent high-frequency cut-offs that might appear on the GW spectrum and will be discussed in the next paragraphs. We integrate over all loop sizes $\alpha$, all emission times $\tilde{t}$ and we sum over all loop modes $m$.  Using the previous equations, we can reshuffle Eq.~\eqref{eq:master_eq_pedagogical} into the ready-to-use form
	\begin{equation}
\Omega_{\rm GW}(f) =\sum_m\frac{1}{\rho_c}\cdot\frac{2m}{f}\cdot\frac{(0.1) \,\Gamma^{(m)}G\mu^2}{\alpha(\alpha+\Gamma G \mu + \kappa)}\int^{t_0}_{t_{\rm osc}}d\tilde{t} \, \frac{C_{\rm{eff}}(t_i)}{t_i^4}\left[\frac{a(\tilde{t})}{a(t_0)}\right]^5\left[\frac{a(t_i)}{a(\tilde{t})}\right]^3\Theta\left(t_i-\frac{l_*}{\alpha}\right)\Theta(t_i-t_{\rm osc}).
	\label{eq:master_eq_ready_to_use}
	\end{equation}
We refer to \cite{Gouttenoire:2019kij} for more details. 

\paragraph{GW amplitude for the standard cosmology}
The general formula -- Eq.~\eqref{eq:master_eq_ready_to_use} -- can be evaluated in two limits: $\kappa = 0, \, \alpha \gg \Gamma G\mu$ for local strings and $\kappa \gg \alpha \gg \Gamma G\mu$ for global strings.
Taking the analytic estimates in \cite{Gouttenoire:2019kij}, the local-string GW spectrum during the standard radiation era is estimated to be flat (scale-invariant) with an amplitude
\begin{align}
\Omega_\textrm{std}^\textrm{local} h^2 ~ &\simeq ~ 1.5 \pi \, \Omega_r h^2 \, C_{\rm eff}^{\rm rad} \left(\frac{\alpha G \mu}{\Gamma}\right)^{1/2} ~ \simeq ~ (1.5 \times  10^{-10}) \left(\frac{G\mu}{10^{-11}}\right)^{1/2} ~ \simeq ~ (1.0 \times  10^{-10}) \left(\frac{\eta}{10^{14} ~ \rm GeV}\right),
\label{cs_flat_amp_local}
\end{align}
where $\Omega_{r}h^2 \simeq 2.473 \times 10^{-5}$ \cite{ParticleDataGroup:2020ssz} is the radiation density today, and here the summation to large harmonics has been done.
The GW spectrum from global strings in a radiation era can be obtained using 
the formula from the fundamental mode  \cite{Gouttenoire:2019kij} 
\begin{align}
\Omega_\textrm{std,1st}^\textrm{global} h^2 ~ &\simeq ~ 2.5 \, \Omega_r h^2 \, C_{\rm eff}^{\rm rad} \left(\frac{\Gamma}{\Gamma_{\rm Gold}}\right) \log^3(\eta \tilde{t}_M) \left( \frac{\eta}{\MPl} \right)^4,\\ 
\Omega_\textrm{std,sum}^\textrm{global}  ~ &\simeq ~ 1.2 \times 10^{-17} \, \log^3\left[ (5.6 \times 10^{30}) \left(\frac{\eta}{10^{15} ~ {\rm GeV}}\right) \left( \frac{1 ~ {\rm mHz}}{f} \right)^2 \right]  \left( \frac{\eta}{10^{15} ~ {\rm GeV}} \right)^4,
\label{cs_flat_amp_global}
\end{align}
where the last line uses Eq.~\eqref{eq:t_M_expression} and \eqref{eq:f_GW_strings_general}, and we have also multiplied the factor 3.6 coming from the inclusion of higher-modes (cusp) \cite{Cui:2017ufi}.
Here we obtain the extra $\log^3$-dependence compared to the local-string case. From Eq.~\eqref{eq:master_eq_ready_to_use}, two of the $\log$-factors come from the string tension $\mu$ and another one from the particle production rate $\kappa$.
There is also another mild $\log$-dependence from the loop formation efficiency $C_{\rm eff}$ \cite{Gouttenoire:2019kij}, which could lead to the $\log^4$-dependence as observed in the recent field-theoretic simulations \cite{Gorghetto:2021fsn}.

\paragraph{Spectral index and the non-standard cosmology}
First we consider the local-string case. From the master equation Eq.~\eqref{eq:master_eq_ready_to_use}, the GW amplitude depends on frequency as
\begin{align}
\Omega_{\rm GW}^{\rm  local} ~ \propto ~ f^{-1} \, \tilde{t} \, t_i^{-4} \,a^2(\tilde{t}) \, a^3(t_i) ~ \propto ~ f^\beta, ~ ~ {\rm with } ~ ~ \beta ~ = ~ = 2\left[\frac{3m + n - nm}{n(2-m)}\right],
\label{spectral_index_local_string}
\end{align}
where the emission time and the loop formation time are related to frequency via Eq.~\eqref{eq:t_M_expression} and \eqref{eq:f_GW_strings_general}.
In the second step, we assume that the formation and the emission happen when the universe is dominated by the energy densities $\rho \propto a^{-n}$ and $a^{-m}$, respectively.
For loops that are produced and emit from the same era, i.e., $n=m$, the spectral index simplifies to $2(n-4)/(n-2)$.
For example, the GW spectrum has slope $0$, $-2$, and $1$ for loops from radiation, matter, and kination eras, respectively.
This scaling is the same as the inflationary GW spectrum. However, their degeneracy is broken by other effects of the cosmic-string network.
For example, the VOS evolution smoothens the GW around the transition between eras \cite{Gouttenoire:2019kij} and the effect of the higher harmonics makes the scaling during matter era dropped slower with the spectral index $-1/3$, $-2/3$, and $-2$ for cusps, kinks, and kink-kink collision \cite{Blasi:2020wpy}.\footnote{In addition to enhancing GW signals, the presence of a matter-kination era could probe the string small-scale structure by measuring the spectral index of the matter slope.}
In contrast, as shown in \cite{Gouttenoire:2019kij} the kination slope does not change after summing higher harmonics.

The spectral index for the global strings in the presence of a non-standard EOS reads
\begin{align}
\Omega_{\rm GW}^{\rm  global} ~  &\propto ~ f^{2\left(\frac{n-4}{n-2}\right)} \log^3\left[ \eta t_\Delta \left(\frac{\tilde{t}_M}{t_\Delta}\right) \right],\\
&\propto ~ f^{2\left(\frac{n-4}{n-2}\right)} \log^3\left[ (5.6 \times 10^{30}) \left(\frac{\eta}{10^{15} ~ {\rm GeV}}\right) \left( \frac{1 ~ {\rm mHz}}{f_\Delta}  \right)^2 \left(\frac{f_\Delta}{f}\right)^{n/(n-2)} \right],
\label{spectral_index_global_string}
\end{align}
where we took $n=m$ as global-string loops decay right after their formation. The scaling inside the $\log$ term comes from Eq.~\eqref{eq:f_GW_strings_general} and slightly reduces the spectral index of the kination slope.
We recover the scaling which is found in \cite{Chang:2021afa}. 

\paragraph{Energy-frequency relation}

The energy-frequency relations were derived for the first-time in \cite{Cui:2017ufi, Cui:2018rwi} for local strings and in \cite{Chang:2019mza} for global strings, and were later corrected through the numerical computation of the string network evolution in \cite{Gouttenoire:2019kij}. 
The time $\tilde{t}_M$ when loops maximally contribution to GW emission is defined to be when loop size in Eq.~\eqref{eq:length_shrink} is half of its initial size $\alpha t_i$,
\begin{align}
\tilde{t}_M ~ = ~ \frac{\left( \frac{1}{2} \alpha + \Gamma G \mu + \kappa \right)}{(\Gamma G \mu + \kappa)} t_i.
\end{align}
A difference between local and global strings is the particle-production efficiency
\begin{align}
\tilde{t}_M ~ \simeq ~  \begin{cases}
\frac{\alpha}{2 \Gamma G \mu} t_i ~ ~ &\mathrm{for} ~ \kappa = 0, ~ \alpha \gg \Gamma G \mu ~ \mathrm{(local ~ strings)},\\
t_i ~ ~ &\mathrm{for} ~ \kappa \gg \alpha \gg \Gamma G \mu ~ \mathrm{(global ~ strings)}.\\
\end{cases}
\label{eq:t_M_expression}
\end{align}
The global-string loops decay fast after their production, while the local-string loops live much longer.
For the emitted GW, the frequency today $f$  relates to the loop length $l_M = l(\tilde{t}_M)$
\begin{align}
\left(\frac{2}{f}\right) \left[\frac{a(\tilde{t}_M)}{a(t_0)}\right] ~ = ~ l_M ~ = ~ \frac{\alpha t_i}{2} ~ ~ ~ \Rightarrow ~ ~ ~ f ~ = ~ \frac{4}{\alpha t_i} \left[\frac{a(\tilde{t}_M)}{a(t_0)}\right],
\label{eq:f_GW_strings_general}
\end{align}
where the fundamental mode is considered here, $f_M = 2/l_M$.
The above equation shows how the  GW spectrum traces the cosmic-string evolution across  the cosmological history.
In Ref.~\cite{Gouttenoire:2019kij}, we reported the frequencies in GW spectrum that correspond to loops production during radiation era at energy scale $\rho_\Delta^{1/4}$
\begin{align}
f_\Delta ~ \simeq ~ \begin{cases}
(1.3 \times 10^{-2} ~ \mathrm{Hz}) \left(\frac{0.1 \times 50 \times 10^{-11} }{\alpha \times \Gamma G\mu}\right)^{1/2}\left(\frac{\rho_\Delta^{1/4}}{\rm GeV}\right) ~ ~ ~ &\mathrm{(local ~ strings)},\\[1em]
(9.0 \times 10^{-7} ~ \mathrm{Hz}) \left(\frac{0.1}{\alpha}\right) \left(\frac{\rho_\Delta^{1/4}}{\rm GeV}\right) ~ ~ ~ &\mathrm{(global ~ strings)},
\end{cases}
\label{turning_point_CS}
\end{align}
which are obtained from evaluating Eq.~\eqref{eq:f_GW_strings_general}, after assuming that $\tilde{t}_M$ and $t_i$ are in radiation era.
We have multiplied the numerical-fitted of 0.2 for local and global strings, justified by numerical simulations, in order to account for VOS evolution \cite{Gouttenoire:2019kij}.
Note that the above relation works for all loops which both decay and are produced during the radiation era. For example, the BBN temperature $T_{\rm BBN} \simeq 1~\rm MeV$ corresponds to the GW frequency  \cite{Gouttenoire:2019kij} 
\begin{equation}
f_\textrm{BBN} ~ \simeq ~ \begin{cases}
8.6 \times 10^{-5}~\textrm{Hz}~\left(\frac{0.1\times 50 \times 10^{-11}}{\alpha \Gamma G \mu}\right)^{1/2}, ~ ~ ~ &\mathrm{(local ~ strings)},\\[1em]
(5.9 \times 10^{-9} ~ \mathrm{Hz}) \left(\frac{0.1}{\alpha}\right), ~ ~ ~ &\mathrm{(global ~ strings)},
\end{cases}
\end{equation}
and where the GW amplitude should not exceed the BBN-$N_{\rm eff}$ bound in Eq.~\eqref{eq:BBN_bound_inflation_GW}.

The peak frequency from matter-kination era is obtained in a similar manner, but now $t_i$ is in the kination era.
The lifetime of global-string loops is short, such that we can safely assume $\tilde{t}_M$ in the kination era.
On the other hand, the time $\tilde{t}_M$ for local strings could reside in either kination or radiation era, depending on the kination duration and the loop lifetime.

\paragraph{High-frequency cut-offs.}
The string network forms around the energy scale $\eta$ defined in Eq.~\eqref{string_tension}, 
\begin{align}
T_\textrm{form} ~ &\simeq ~ (10^{11} \textrm{ GeV})\left(\frac{G\mu}{10^{-15}}\right)^{1/2}.
\label{string_formation_cutoff}
\end{align}
The above temperature, when plugged into the energy-frequency relation~\eqref{turning_point_CS}, corresponds to a UV cut-off on the GW spectrum, assuming a standard cosmology\footnote{If network formation takes place in a kination era, the corresponding cut-off frequency $f_\textrm{form} $ is obtained from Eq.~\eqref{peak_kination_freq_local} and \eqref{peak_kination_freq_global}.}
\begin{align}
f_\textrm{form} ~ &\simeq ~ 206 ~ {\rm GHz} \left(\frac{0.1\times 50}{\alpha \Gamma}\right)^{1/2} \left[\frac{g_*(T_{\rm form})}{g_*(T_0)}\right]^{1/4},
\label{string_formation_cutoff_frequency}
\end{align}
which is interestingly independent of $G\mu$. Indeed, string networks with smaller $G\mu$ are formed at later times, but the associated loops decay much slower, cf. Eq. \eqref{eq:length_shrink}. By varying $G\mu$, the GW frequency today remains constant by the compensation of smaller red-shift.

The GW spectrum from cosmic strings can experience other high-frequency cut-offs due to some UV physics or  to the dynamics of strings at early times.
The first Heaviside function in Eq.~\eqref{eq:master_eq_ready_to_use}  $\Theta(t_i- l_*/\alpha)$ discards loops whose size $\alpha t_i$ is smaller than a critical length $l_*$ below which massive particle production are responsible for the loop decay \cite{Auclair:2019jip}.

With the second Heaviside function in Eq.~\eqref{eq:master_eq_ready_to_use}, $\Theta(t_i-t_{\rm osc})$, we eliminate loops which are formed when loop oscillation is frozen due to thermal friction, i.e., strings motion is damped by interaction with the thermal plasma \cite{Vilenkin:1991zk} and hence the GW is suppressed. String oscillation can start when thermal friction becomes smaller than Hubble friction.

As we show in \cite{Gouttenoire:2019kij}, the presence of these high-frequency cut-offs can lift the BBN bounds on kination-enhanced GW from cosmic strings, however we expect them to lie at frequency higher than the windows of current and future GW interferometers. 
We leave the study of high-frequency cut-offs in the presence of kination for future work.

\paragraph{Local vs. global strings.}
Parametrically, the GW spectra from local and global CS scale as, cf. Eq.~\eqref{cs_flat_amp_local} and \eqref{cs_flat_amp_global}
\begin{equation}
\Omega_{\rm GW}^{\rm local} \simeq \Omega_r \frac{\eta}{M_{\rm pl}}, \qquad \textrm{and} \qquad \Omega_{\rm GW}^{\rm global} \simeq \Omega_r  \left( \frac{\eta}{M_{\rm pl}} \right)^4\log^3{\left( \eta t_{i} \right)}.
\end{equation}
In order to understand the scaling difference, let us consider the contribution to the GW spectrum coming from loops produced at time $t_i$. For local strings, the corresponding GW are dominantly emitted at time $\tilde{t}_M^{\rm local} \simeq \alpha t_i / (2 \Gamma G \mu_{\rm local}) $, see Eq.~\eqref{eq:t_M_expression}, which means that GW emission occurs $\left(M_{\rm pl}/\eta\right)^2$ Hubble times after loop production. Instead, global loops decay at $\tilde{t}_M^{\rm global} \simeq t_i $, so within one Hubble time after production, even though their tension is logarithmically enhanced. Therefore, with respect to local strings, the GW spectrum from global strings in standard radiation cosmology is:
\begin{itemize}
\item
suppressed by the shorter Hubble time $\tilde{t}_{M}$ at the time of GW emission: factor $\tilde{t}_{\rm M}^{\rm global}/\tilde{t}_{\rm M}^{\rm local} \propto G\mu_{\rm local} \propto \left( \eta/M_{\rm pl}\right)^2 $,
\item
suppressed by the larger GW redshift factor since emission occurs earlier: factor $\left[\frac{a\left(\tilde{t}_{\rm M}^{\rm global}\right)}{a\left(\tilde{t}_{\rm M}^{\rm local}\right)}\right]^4 \propto \left( \eta/M_{\rm pl}\right)^4$,
\item
enhanced by the lower loop redshift factor since GW emission occurs right after loop production: factor $\left(a\left(\tilde{t}_{\rm M}^{\rm local}\right)/a\left(\tilde{t}_{\rm M}^{\rm global}\right)\right)^3\propto \left( \eta/M_{\rm pl}\right)^{-3}$,
\item
increased by the logarithmically-enhanced GW power emission rate: factor $\log^2{\left( \eta t_{i} \right)} $,
\item
increased by the logarithmically-enhanced loop lifetime: factor $\log{\left( \eta t_{i} \right)} $.
\end{itemize}
The GW spectrum from global strings could be further enhanced by a fourth power of logarithmic factor due to a deviation from scaling regime in the loop production rate $C_{\rm eff}$ \cite{Gouttenoire:2019kij,Gorghetto:2021fsn}.
Additionally, for a given loop-production time $t_i$, the earlier GW emission for global loops implies that the associated frequency today, assuming a standard radiation cosmology, is lowered by a factor
\begin{equation}
\label{eq:global_vs_local_frequency}
\textrm{fixed loop formation time }~ t_i \qquad\implies \qquad \frac{f_{\rm global}}{f_{\rm local}}~\simeq~\frac{a\left(\tilde{t}_{\rm M}^{\rm global}\right)}{a\left(\tilde{t}_{\rm M}^{\rm local}\right)}~\simeq~\frac{\eta}{M_{\rm pl}},
\end{equation}
which indeed coincides with Eq.~\eqref{turning_point_CS}.
The next subsections provide expressions for the peak position for both local and global strings, and the GW detectability at current and future-planned detectors is discussed.

\subsubsection{Local strings}
\paragraph{Peak frequency.}
Local-string loops that are produced at the start of kination $t_\mathrm{KD}$ could decay long after the end of a short kination era at $t_\Delta$.
The condition for the GW emission at $\tilde{t}_M^\mathrm{KD}$ to take place during the late-radiation era is
\begin{align}
1 ~ < ~ \frac{\tilde{t}_M^\mathrm{KD}}{t_\Delta} ~ \simeq ~ \left(\frac{\alpha}{2 \Gamma G\mu}\right) \left(\frac{t_\mathrm{KD}}{t_\Delta}\right) ~ = ~ \left(\frac{\alpha}{2 \Gamma G\mu}\right) \left(\frac{a_\mathrm{KD}}{a_\Delta}\right)^3 ~ ~ \Rightarrow ~ ~  N_\mathrm{KD} < \frac{1}{3} \log\left(\frac{\alpha}{2\Gamma G\mu}\right),
\end{align}
where we used Eq.~\eqref{eq:t_M_expression} to relate GW emission times $\tilde{t}_M^{x}$ and loop production times $t_x$.
For $\tilde{t}_M^\mathrm{KD} > t_\Delta$, the peak frequency $f_\mathrm{KD}$ follows from Eq.~\eqref{eq:f_GW_strings_general} 
\begin{align}
f_\mathrm{KD} ~ &= ~ f_\Delta \left[\frac{a(\tilde{t}_M^\mathrm{KD})}{a(\tilde{t}_M^\Delta)}\right] \left(\frac{t_\Delta}{t_\mathrm{KD}}\right) ~ = ~ f_\Delta  \left(\frac{t_\Delta}{t_\mathrm{KD}}\right)^{1/2} ~ = ~ f_\Delta  \left(\frac{\rho_\mathrm{KD}}{\rho_\Delta}\right)^{1/4},
\end{align}
where we used Eq.~\eqref{eq:t_M_expression} once again.
For $\tilde{t}_M^\mathrm{KD} < t_\Delta$, the peak frequency $f_\mathrm{KD}$ is
\begin{align}
f_\mathrm{KD} ~ &= ~ f_\Delta \left[\frac{a(\tilde{t}_M^\mathrm{KD})}{a(t_\Delta)}\right] \left[\frac{a(t_\Delta)}{a(\tilde{t}_M^\Delta)}\right] \left(\frac{t_\Delta}{t_\mathrm{KD}}\right) ~ = ~ f_\Delta  \left(\frac{2 \Gamma G\mu}{\alpha}\right)^{1/6} \left(\frac{t_\Delta}{t_\mathrm{KD}}\right)^{2/3} ~ = ~ f_\Delta \left(\frac{2 \Gamma G\mu}{\alpha}\right)^{1/6} \left(\frac{\rho_\mathrm{KD}}{\rho_\Delta}\right)^{1/3}.
\end{align}
From the expression for $f_\Delta$ in Eq.~\eqref{turning_point_CS}, we deduce the  frequency of the peak signature of the presence of a matter-kination era in the  GW spectrum from local strings 
\begin{align}
f_\mathrm{KD} ~ \simeq ~ \begin{cases}
(1.8 \times 10^{3} ~ \mathrm{Hz}) \left( \frac{0.1 \times 50 \times 10^{-11}}{\alpha \Gamma G \mu}\right)^{1/2}  \left( \frac{E_\mathrm{KD}}{10^5 ~ \mathrm{GeV}}\right) ~ ~ &\mathrm{for} ~  N_\mathrm{KD} < \frac{1}{3} \log\left(\frac{\alpha}{2\Gamma G\mu}\right),\\[1em]
(6.1 \times 10^{2} ~ \mathrm{Hz}) \left( \frac{0.1}{\alpha}\right)^{2/3} \left( \frac{50 \times 10^{-11}}{\Gamma G \mu}\right)^{1/3}  \left( \frac{E_\mathrm{KD}}{10^5 ~ \mathrm{GeV}}\right)  \left[ \frac{\exp(N_\mathrm{KD}/2)}{10}\right] ~ ~ &\mathrm{for} ~  N_\mathrm{KD} > \frac{1}{3} \log\left(\frac{\alpha}{2\Gamma G\mu}\right).
\end{cases}
\label{peak_kination_freq_local}
\end{align}

\paragraph{Peak amplitude.}
The amplitude at the  peak is obtained from Eq.~\eqref{spectral_index_local_string}
\begin{align}
\Omega_\textrm{GW,KD} ~ \simeq ~ \frac{1}{2.5}\Omega_\textrm{GW,st}(10 f_\Delta) \left(\frac{f_\textrm{KD}}{10 f_\Delta}\right)  ,
\label{bump_peak_kination_amp_local}
\end{align}
where the factor 2.5 accounts for the change of relativistic degrees of freedom\footnote{Precisely, Eq.~\eqref{spectral_index_local_string} has a factor of $\left(\frac{g_*(T)}{g_*(T_0)}\right) \left(\frac{g_{*s}(T_0)}{g_{*s}(T)}\right)^{4/3}$, which goes to $2.5^{-1}$ in high-temperature limit.}, and we multiply the factor 10 to $f_\Delta$, which is fitted well with the peak from numerical simulations, in order to account for VOS evolution and mode summation. An analytical estimate of the GW amplitude $\Omega_\textrm{GW,st}(f)$ emitted by local strings in standard cosmology is given by Eq.~\eqref{cs_flat_amp_local}.

\begin{figure}[h!]
\centering
{\bf Gravitational waves from local cosmic strings}\\
\raisebox{0cm}{\makebox{\includegraphics[width=0.47\textwidth, scale=1]{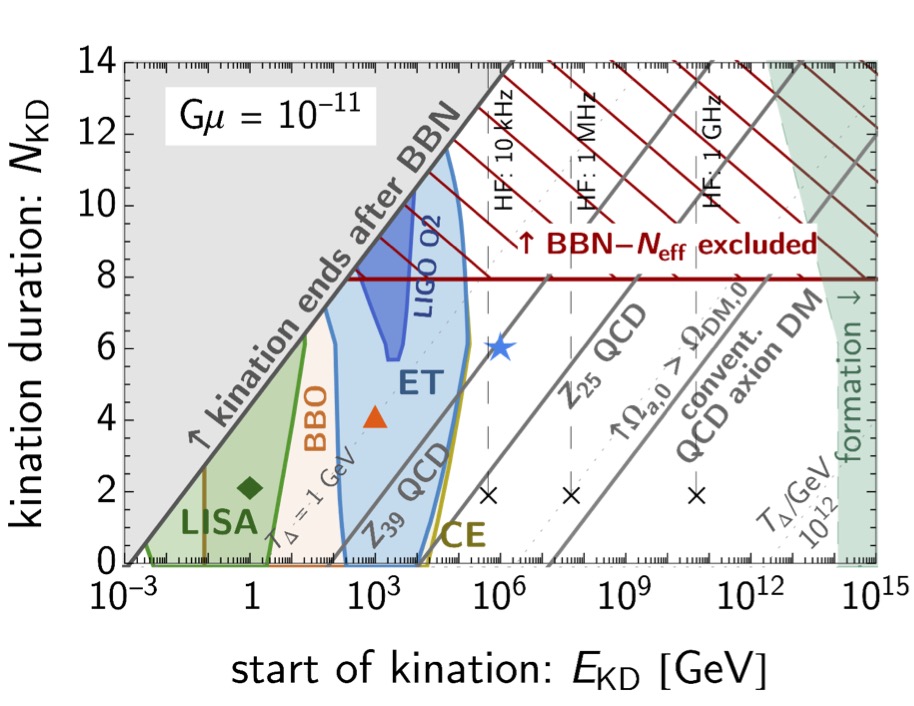}}}
\raisebox{0cm}{\makebox{\includegraphics[width=0.475\textwidth, scale=1]{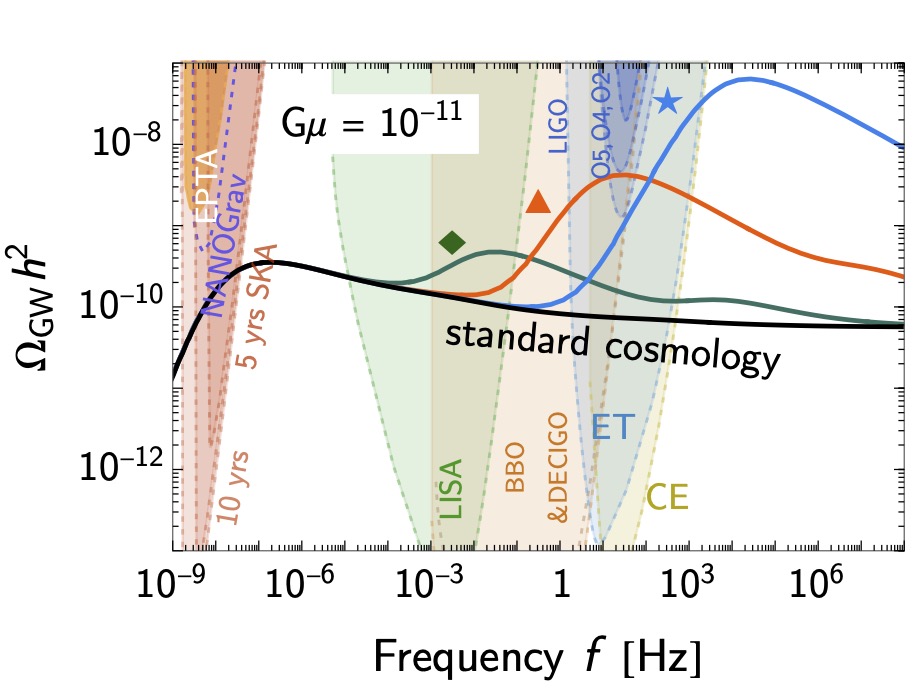}}}\\[1em]
\raisebox{0cm}{\makebox{\includegraphics[width=0.47\textwidth, scale=1]{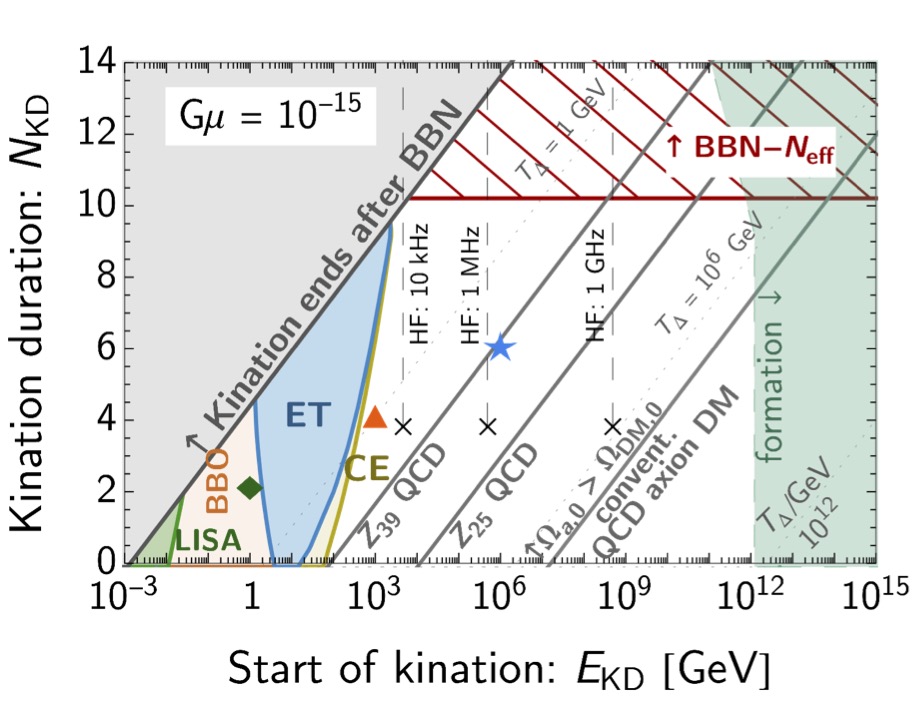}}}
\raisebox{0cm}{\makebox{\includegraphics[width=0.475\textwidth, scale=1]{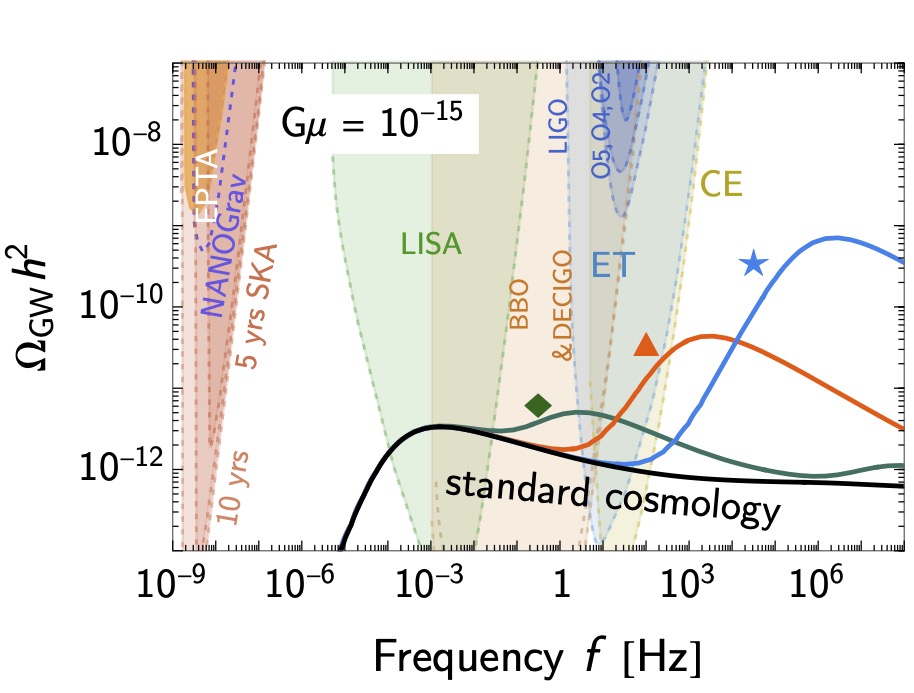}}}
\caption{\textit{ \small The GW background from local strings with tension $G\mu$ is enhanced by a period of matter-kination lasting for $(2N_\mathrm{KD}+N_\mathrm{KD})$ efolds, cf. Eq.~\eqref{eq:NKD_vs_NMD}. The kination era starts at energy scale $E_\mathrm{KD}$ and ends when the temperature of the universe is $T_\Delta$ (dashed lines).
\textbf{Left panel:}  In the coloured regions, the peak  is observable. BBN constrains late kination eras (gray) and long  kination eras (red-hatched) (see Sec.~\ref{subsec:afterinflation}). 
 The black dashed lines show the detectability prospects of hypothetical  HF experiments operating at 10 kHz, 1 MHz, 1 GHz frequencies with sensitivity $h^2\Omega_{\rm sens} = 10^{-10}$. The QCD axion that allows a kination era could be DM along the solid-gray lines for the conventional and $\mathbb{Z}_\mathcal{N}$ QCD-axion models, assuming kinetic misalignment (see Sec.~\ref{sec:darkmatter}).
\textbf{Right panel:} The GW spectra correspond to benchmark points in the left panel.
Note the second peak at high-frequency for the green line, that comes  from loops produced during the radiation era and decaying at the start of kination, cf. Eq.~\eqref{eq:2-peak_separation}. 
}}
\label{detect_peak_cs_local}
\end{figure}

\paragraph{Detectability.} 
The peak signature of a matter-kination era in the GW spectrum from local cosmic strings is potentially observable by future observatories, as shown in Fig.~\ref{detect_peak_cs_local}. 
As the GW spectrum from local CS assuming standard cosmology is already at the observable level,
even a few e-folds of kination era can induce the smoking-gun peak signature.
On the top panel, we show the string tension of $G\mu \sim 10^{-11}$ \cite{Ellis:2020ena,Blasi:2020mfx}, which could explain hints from NANOGrav 12.5 years \cite{NANOGrav:2020bcs}, PPTA 15 years \cite{Goncharov:2021oub} and EPTA 24 years \cite{Chen:2021rqp}.

A large parameter space (white) cannot be probed by the planned observatories, but ultra-high frequency experiments could do so.
We expect the ability to probe such high-energy kination era to get reduced by other cut-offs that  we have not discussed so far, i.e., friction and particle production. We leave the dedicated study for further work.

\begin{figure}[h!]
\centering
{\bf Gravitational waves from local cosmic strings}\\

\raisebox{0cm}{\makebox{\includegraphics[width=0.47\textwidth, scale=1]{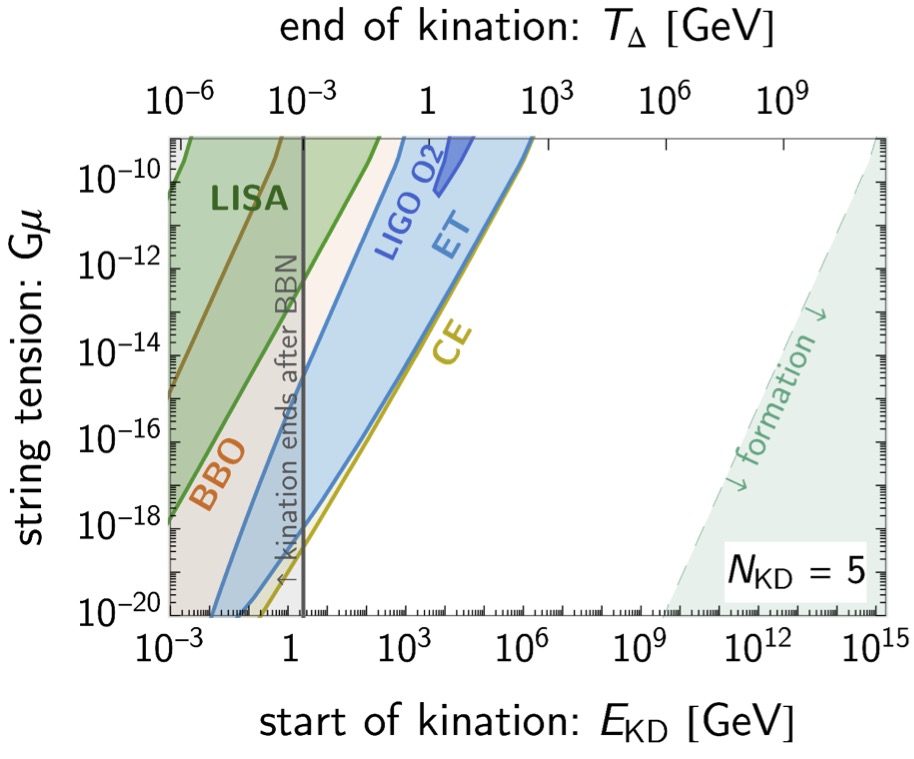}}}
\quad
\raisebox{0cm}{\makebox{\includegraphics[width=0.47\textwidth, scale=1]{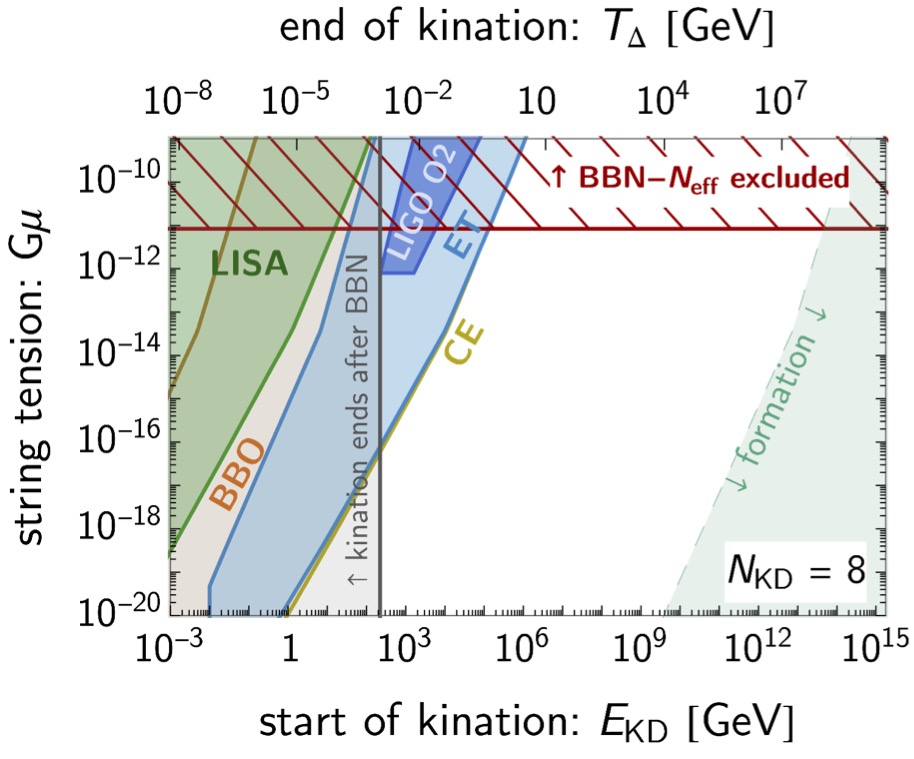}}}  
\caption{\textit{ \small Detectability of the GW peak for varying string tension $G\mu$. The longer the matter-kination era,  the higher the peak signature, allowing strings with smaller tension $G\mu$  to be probed.}}
\label{detect_peak_cs_local_gmu}
\end{figure}

In Fig.~\ref{detect_peak_cs_local_gmu}, a few e-folds of kination render GW signal from strings of tension $G\mu \simeq 10^{-19}$ observable, but at the price of having kination ending after BBN.\\

\paragraph{Second kination peak at high-frequency.}
The delayed decay of local string loops introduces another spectral enhancement at high-frequency, see top left panel in Fig.~\ref{detect_peak_cs_local}. In contrast to the main peak from loops produced at the start of kination, the smaller second peak corresponds to loops created deep inside the earlier radiation era.
Eq.~\eqref{spectral_index_local_string} tells us that the spectral index changes sign between loops from radiation which decay in matter era, $\beta = -1/2$, and those decay during kination era, $\beta = 1/4$. So the second peak corresponds to loops produced during radiation era and decay right at the start of kination era, i.e.
\begin{align}
\tilde{t}_M^{\rm KDII} ~ = ~  t_{\rm KD} ~ ~ \Rightarrow ~ ~ t_{\rm KDII} ~ = ~ \frac{2\Gamma G\mu}{\alpha} t_{\rm KD},
\end{align}
where we applied Eq. \eqref{eq:t_M_expression} to relate GW emission times $\tilde{t}_M^{x}$ and loop production times $t_x$.
The visibility of the second peak depends on the separation with respect to the first biggest kination peak, which we can derive from Eq. \eqref{eq:f_GW_strings_general}
\begin{align}
\frac{f_\mathrm{KDII}}{f_{\rm KD}} ~ &= ~ \left[\frac{a(\tilde{t}_M^{\rm KDII})}{a(\tilde{t}_M^{\rm KD})}\right] \left(\frac{t_{\rm KD}}{t_\mathrm{KDII}}\right) ~ = ~ \left[\frac{a(t_{\rm KD})}{a(\tilde{t}_M^{\rm KD})}\right] \left(\frac{\alpha}{2\Gamma G\mu}\right),
\end{align}
where the first bracket depends on the two limits of $N_{\rm KD}$, as in Eq. \eqref{peak_kination_freq_local}.
Here we report the separation between two kination peaks
\begin{align}
\frac{f_\mathrm{KDII}}{f_{\rm KD}} ~ &= ~ \begin{cases}
10^4 \, \left(\frac{\alpha \, \times \, 50 \, \times \, 10^{-11}}{0.1 \, \times \, \Gamma G\mu}\right)^{1/2} \exp(N_{\rm KD}/2)~ ~ &\mathrm{for} ~  N_\mathrm{KD} < \frac{1}{3} \log\left(\frac{\alpha}{2\Gamma G\mu}\right),\\[1em]
2.15 \times 10^{5} \,  \left(\frac{\alpha \, \times \, 50 \, \times \, 10^{-11}}{0.1 \, \times \,  \Gamma G\mu}\right)^{2/3}  ~ ~ &\mathrm{for} ~  N_\mathrm{KD} > \frac{1}{3} \log\left(\frac{\alpha}{2\Gamma G\mu}\right). \label{eq:2-peak_separation}
\end{cases}
\end{align}
 Eq.~\eqref{eq:2-peak_separation} underestimates the two-peak separation in Fig. \ref{detect_peak_cs_local} by approximately one order of magnitude because the EOS change only impacts the network evolution which results in correcting the loop formation and, thus,  the position of the biggest peak. It takes a few e-folds for the network to adapt to the change of cosmology which moves the first biggest peak to lower frequency than naively expected, same as the factor $\mathcal{O}(0.1)$ in Eq. \eqref{turning_point_CS}. On the other hand, the position of the smallest peak depends only on the loops' emission time and is not affected by the time that the network adapts to a change of cosmological era.

Before moving to the global strings, let us emphasize that the second peak will not be seen in the global string spectrum. 
This peak is linked to GW emission occurring for local strings at a time $\tilde{t}_M$ much later than the {loop-formation} time $t_i$, while global loops decay almost instantaneously after their formation.

\FloatBarrier
\FloatBarrier

\FloatBarrier
\FloatBarrier

\subsubsection{Global strings}

\paragraph{Short-lived global strings.}
The short-lived global strings emit GW right after loop production, much earlier than for local strings. As a consequence, GW redshift for a longer time, at fixed loop-formation time $t_i$, and the frequency observed today is lowered by a factor equal $\eta/M_{\rm pl}$, see Eq.~\eqref{eq:global_vs_local_frequency}. Hence, for fixed string scale $\eta$, the peak signature from the matter-kination era from global strings sits at a lower frequency than in the local-string case. 

Due to PTA constraints $G\mu \lesssim 10^{-11}$ \cite{Lentati:2015qwp,NANOGRAV:2018hou}, we only consider local strings with scale $\eta \lesssim 3 \times  10^{12}~\rm GeV$. In contrast the scale of global strings is only bounded by the largest inflationary scale $\eta \lesssim 10^{16}~\rm GeV$. Hence, in our plots the peak frequencies  from global and local strings can only be compared if we keep in mind that the string scales $\eta$ which we consider for both are different.

\paragraph{Peak frequency.}
The peak GW frequency $f_\mathrm{KD}$ from loops that formed at the start of the kination era $t_\mathrm{KD}$  is written, via Eq.~\eqref{eq:f_GW_strings_general}, in terms of the frequency $f_\Delta$ corresponding to  the end of kination  at time $t_\Delta$, 
\begin{align}
f_\mathrm{KD} ~ = ~ f_\Delta \left[\frac{a(\tilde{t}_M^\mathrm{KD})}{a(\tilde{t}_M^\Delta)}\right] \left(\frac{t_\Delta}{t_\mathrm{KD}}\right),
\label{}
\end{align}
where  $\tilde{t}_M^x$ and $t_x$ are the emission and loop-production times, respectively.
Applying $a \propto t^{1/3} \propto \rho^{1/6}$ during kination and $a \propto t^{1/2} \propto \rho^{1/4}$  during radiation era, the peaked frequency for global strings is
\begin{align}
f_\mathrm{KD} ~ &= ~ f_\Delta \left[\frac{a(\tilde{t}_M^\mathrm{KD})}{a(t_\Delta)}\right] \left[\frac{a(t_\Delta)}{a(\tilde{t}_M^\Delta)}\right] \left(\frac{t_\Delta}{t_\mathrm{KD}}\right) ~ = ~ f_\Delta \left(\frac{t_\Delta}{t_\mathrm{KD}}\right)^{2/3} ~ = ~ f_\Delta \left(\frac{\rho_\mathrm{KD}}{\rho_\Delta}\right)^{1/3},
\end{align}
where we apply $\tilde{t}_M^x \simeq t_x$ for global strings, see Eq.~\eqref{eq:t_M_expression}. Acquiring $f_\Delta$ from Eq.~\eqref{turning_point_CS}, we obtain the peak frequency from the presence of a matter-kination era in the global-string GW spectrum 
\begin{align}
f_\mathrm{KD} ~ \simeq ~ (0.9 ~ \mathrm{Hz}) \left( \frac{0.1}{\alpha}\right) \left( \frac{E_\mathrm{KD}}{10^5 ~ \mathrm{GeV}}\right)  \left[ \frac{\exp(N_\mathrm{KD}/2)}{10}\right],
\label{peak_kination_freq_global}
\end{align}
where the numerically-fitted $f_\Delta$, Eq.~\eqref{turning_point_CS},  is used to account for the transition between two scaling regimes of the string network.

\paragraph{Peak amplitude.}
From the relation in Eq.~\eqref{spectral_index_global_string}, the GW amplitude at the peak is 
\begin{align}
\Omega_\textrm{GW,KD} ~ \simeq ~ \Omega_\textrm{GW,st}(10 f_\Delta) \left(\frac{f_\mathrm{KD}}{10 f_\Delta}\right) ~L ,
\label{bump_peak_kination_amp_global}
\end{align}
where the factor $10 f_\Delta$  fits well with the peak from numerical simulations, accounting for VOS evolution and mode summation and where $L= \mathcal{O}(1)$ is a ratio of log factors.\footnote{$L ~\equiv~\left\{ \frac{\log^3\left[ (5.6 \times 10^{30}) \left(\frac{\eta}{10^{15} ~ {\rm GeV}}\right) \left( \frac{1 ~ {\rm mHz}}{10 f_\Delta}  \right)^2 \left(\frac{10 f_\Delta}{f_{\rm KD}}\right)^{3/2} \right]}{\log^3\left[ (5.6 \times 10^{30}) \left(\frac{\eta}{10^{15} ~ {\rm GeV}}\right) \left( \frac{1 ~ {\rm mHz}}{10 f_\Delta}  \right)^2 \right]} \right\} $.} An analytical estimate of the GW amplitude $\Omega_\textrm{GW,st}(f)$ emitted by global strings in standard cosmology is given by Eq.~\eqref{cs_flat_amp_global}.
In terms of the kination parameters, we have
\begin{align}
\Omega_{\rm GW,KD} ~ \simeq ~ (1.2 \, \times \, 10^{-18}) \left(\frac{\eta}{10^{15} ~ {\rm GeV}}\right)^4 \exp(2 N_{\rm KD}) \log^3\left[ (2.2 \, \times \, 10^{18}) \left(\frac{\eta}{10^{15} ~ {\rm GeV}}\right) \left(\frac{\alpha}{0.1}\right)^2 \left(\frac{10^9 ~ {\rm GeV}}{E_{\rm KD}}\right)^2 \right]
\label{bump_peak_kination_amp_global_simp}
\end{align}

\begin{figure}[h!]
\centering
{\bf Gravitational waves from global cosmic strings}\\
\raisebox{0cm}{\makebox{\includegraphics[width=0.46\textwidth, scale=1]{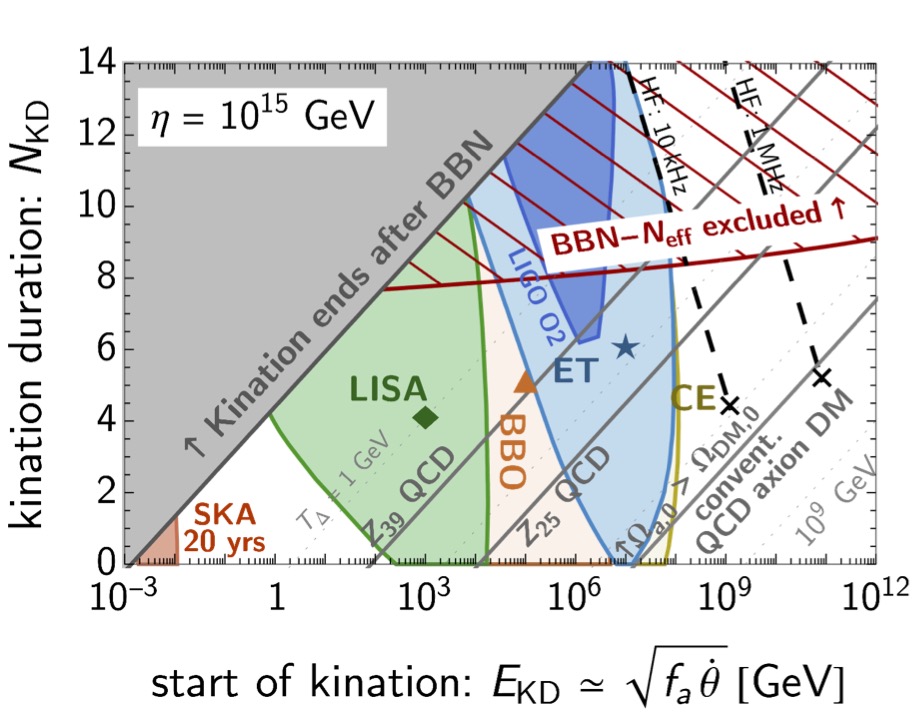}}}  
\qquad
\raisebox{0cm}{\makebox{\includegraphics[width=0.47\textwidth, scale=1]{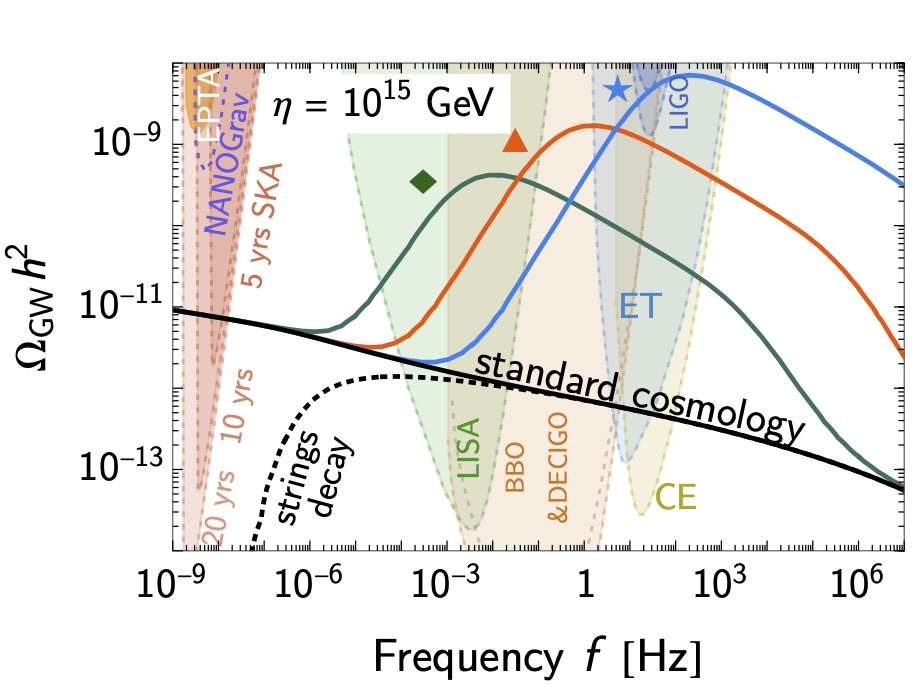}}}\\[0.1em]
\raisebox{0cm}{\makebox{\includegraphics[width=0.46\textwidth, scale=1]{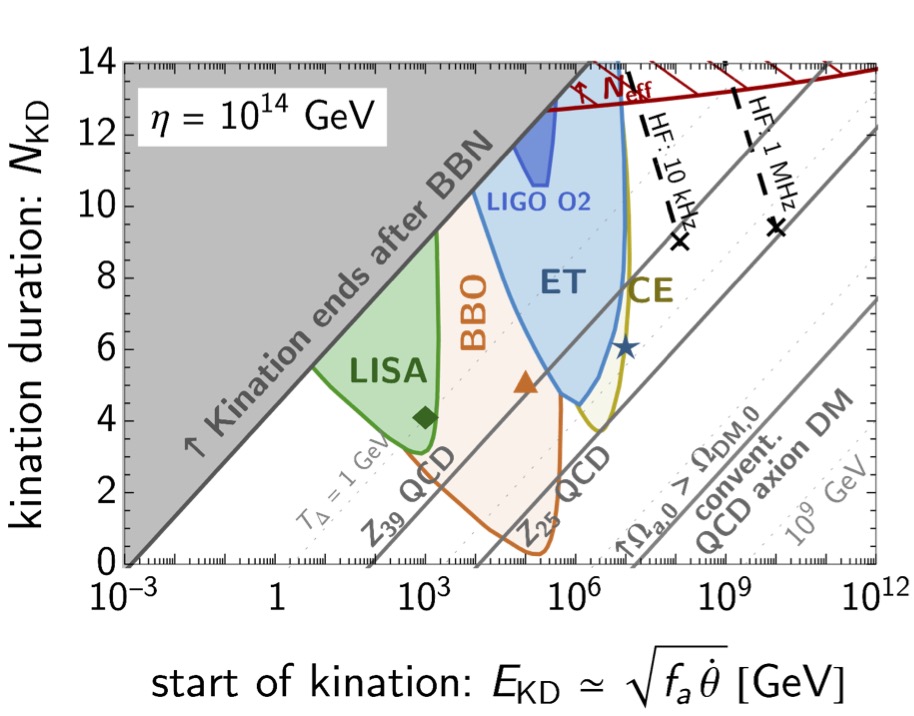}}}
\qquad
\raisebox{0cm}{\makebox{\includegraphics[width=0.47\textwidth, scale=1]{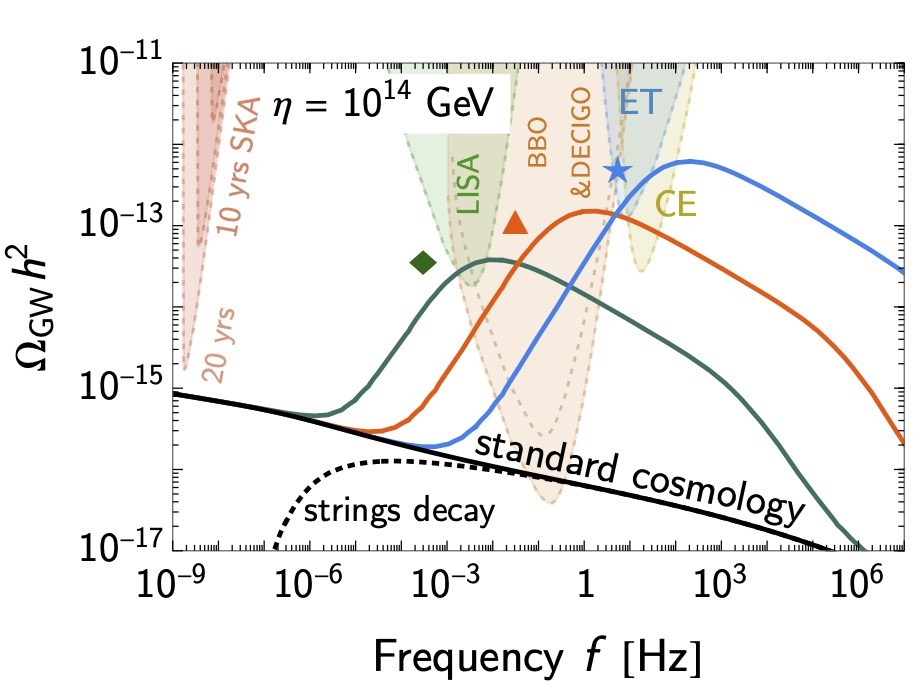}}}
\caption{\textit{ \small The GW spectrum from global strings with tension $\eta$ is enhanced by a period of matter-kination lasting for $(2N_\mathrm{KD}+N_\mathrm{KD})$ efolds, cf. Eq.~\eqref{eq:NKD_vs_NMD}. The kination era starts at energy scale $E_\mathrm{KD}$ and ends when the temperature of the universe is $T_\Delta$ (dashed lines). \textbf{Left panel:}  In the coloured regions, the peak  is observable. BBN constrains late kination eras (gray) and long  kination eras (red-hatched). 
The peak is described by Eq.~\eqref{peak_kination_freq_global} and \eqref{bump_peak_kination_amp_global_simp}. 
 The black dashed lines show the detectability prospects of hypothetical  HF experiments operating at 10 kHz, 1 MHz, 1 GHz frequencies with sensitivity $h^2\Omega_{\rm sens} = 10^{-10}$.
 The QCD axion that allows a kination era could be DM along the solid-gray lines for the conventional and $\mathbb{Z}_\mathcal{N}$ QCD-axion models, assuming kinetic misalignment (see Sec.~\ref{sec:darkmatter}).
\textbf{Right panel:} The GW spectra correspond to benchmark points in the left panel.
The effect of metastable strings cut the spectrum at a low-frequency, as shown by the black-dashed line for a network decay at $T \sim 100 ~ {\rm MeV}$.}}
\label{detect_peak_global_string}
\end{figure}

\paragraph{Detectability.}
Fig.~\ref{detect_peak_global_string} shows the detectability of GW produced by global strings and enhanced by the intermediate matter-kination era. The spectra shown on the right panel correspond to the benchmark points in the contour plot on the left panel.
The GW amplitude scales as $\eta^4$ up to  the log-suppression, therefore the string tension $\eta$ is required to be large for detectability.
The spectral index corresponding to loops formed during the matter era goes like $f^{-1/3}$ due to the summation of higher harmonics, instead of $f^{-1}$ in the spectrum of the only fundamental Fourier mode \cite{Blasi:2020wpy, Gouttenoire:2019kij,Chang:2021afa}. The drop at some high frequency is an artefact bacause we only sum up to $5 \times 10^{5}$ modes. 

GW from strings could experience a high-frequency cut-off due to friction effect. This could shift the spectral peak if the friction cut-off has frequency lower than the peak from the matter-kination era. We leave the dedicated study for future work. 
On the other hand, the spectrum could exhibit the low-frequency cut-off (black dotted lines in Fig.~\ref{detect_peak_global_string}) if the CS network manifests the metastability similar to \cite{Buchmuller:2019gfy,Gouttenoire:2019kij,Buchmuller:2021mbb} in the context of local strings.
The contour plot in Fig.~\ref{detect_peak_global_string}-left shows the compromise between the enhancement of the GW signal and the BBN bound when the kination duration is increased.

\paragraph{A common origin for matter-kination era and GW source: axion strings.} 
An intriguing possibility is if the physics responsible for kination induced by a spinning axion and the physics responsible for the cosmic strings have a common origin. A $U(1)$-breaking phase transition generates cosmic strings at early times, and the dynamics of the axion at later times generates a kination era.
In this paper, we consider models (Sec.~\ref{sec:PQ:exampleII}) where  the radial mode of the complex scalar field obtains a large VEV at early times during inflation so all topological defects are diluted away. 
However, in alternative constructions \cite{DESYfriendpaper3},
the $U(1)$ could be broken after inflation. This can lead to formation of a cosmic string network.  A few efolds of kination for large $f_a$ values  would then be compatible with global strings with large tension and a detectable GW signal.
For this class of models, the axion could generate the multiple-peak GW signals from both inflation and cosmic strings. We discuss the detectability of the axion-string GW enhanced by kination from spinning axion in Fig.~\ref{ma_fa_plot_axion_string} in Sec.~\ref{axionic_string_DM}.

\FloatBarrier
\FloatBarrier

\FloatBarrier
\begin{figure}[h!]
\centering
\centering
\raisebox{0cm}{\makebox{\includegraphics[width=0.44\textwidth, scale=1]{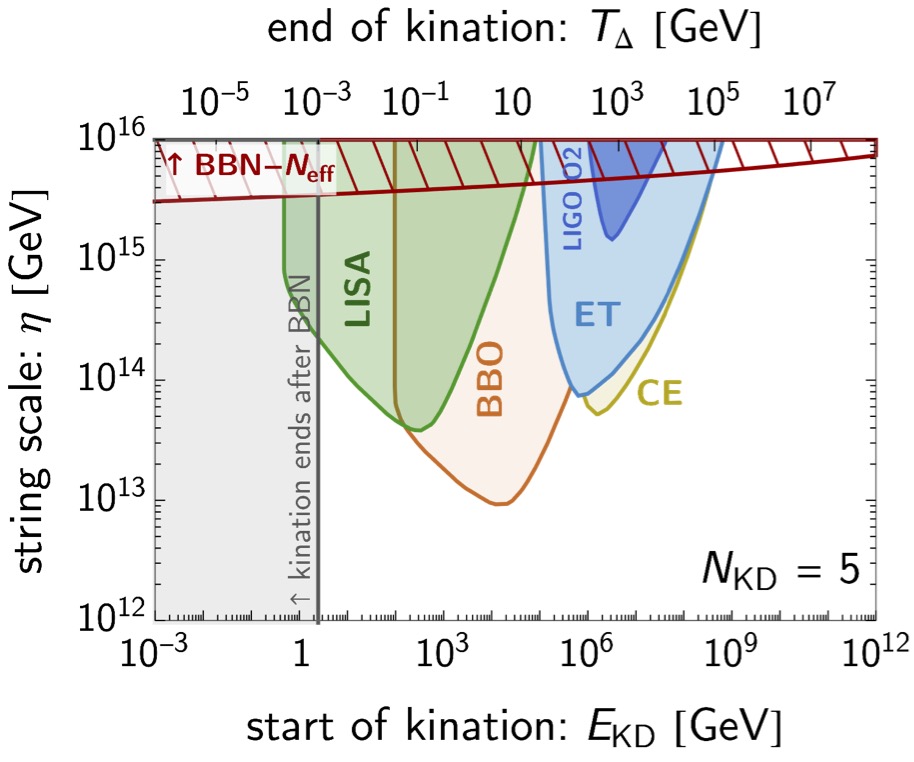}}}
\quad
\raisebox{0cm}{\makebox{\includegraphics[width=0.44\textwidth, scale=1]{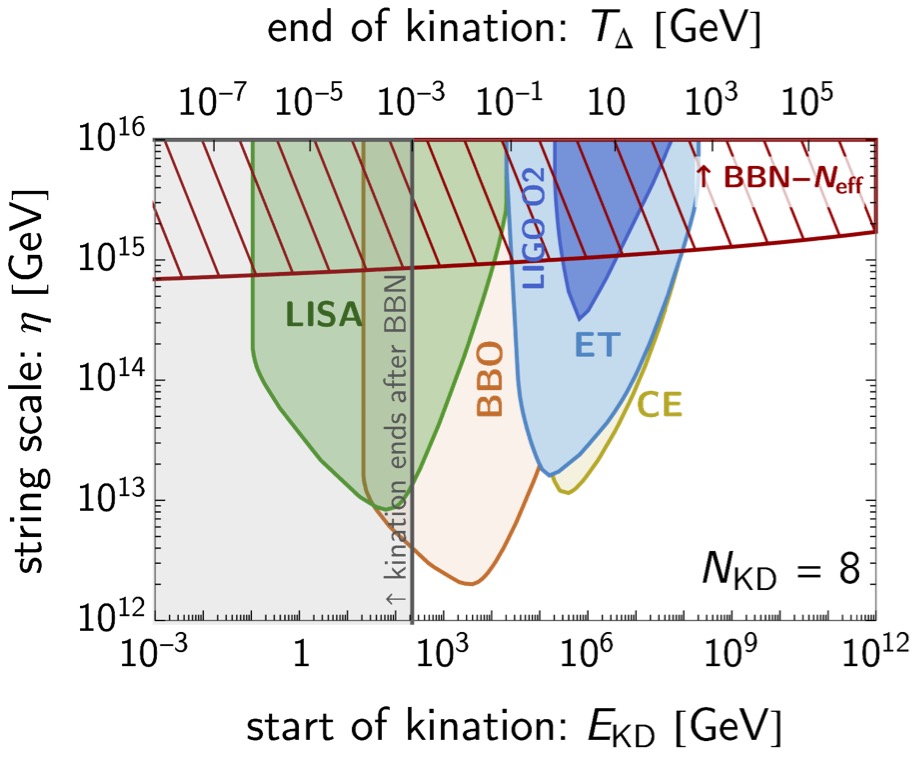}}}  
\caption{\textit{ \small 
The longer kination era enhances the peak, allowing strings with smaller string scale $\eta$  to be probed.}}
\label{detect_peak_eta_global}
\end{figure}
\FloatBarrier

\subsection{Multiple-peak signature}
\label{sec:multiple_peak}
\subsubsection{Inflation + local cosmic strings}
\paragraph{Three types of peaks.}
The physics explaining the presence of the cosmic strings is generally unrelated to the inflationary sector.
In the presence of multiple SGWB, the intermediate matter-kination era can lead to a multiple-peak GW signal which could be probed by the synergy of future detectors.
\begin{enumerate}
\item
Peak signature of matter-kination era in inflationary GW, cf. Eq.~\eqref{inflation_peak_frequency}.
\item
Peak signature of matter-kination era in SGWB from local CS, cf. Eq.~\eqref{peak_kination_freq_local}.
\item
Peak in SGWB from local CS due to the transition between radiation and later matter era around the temperature $0.75$~eV, and whose frequency reads \cite{Gouttenoire:2019kij}
\begin{align}
f_\textrm{low}^\textrm{cs} ~ \simeq ~ 1.48 \times 10^{-7} ~ \textrm{ Hz} \left(\frac{50\times10^{-11}}{\Gamma G\mu}\right).
\label{cs_low_peak_freq}
\end{align}
\end{enumerate}
The inflationary peak (1) can be easily distinguished from the CS peaks (2 and 3) which are broader because the CS network takes time to react to the change of cosmology \cite{Gouttenoire:2019kij}.
In this section, we point-out the possibility of a two-peak spectrum (two matter-kination peaks) and a three-peak spectrum (two matter-kination peaks + one radiation-matter peak at lower frequency, Eq.~\eqref{cs_low_peak_freq}).

\FloatBarrier
\begin{figure}[h!]
\centering
{\bf Gravitational waves from inflation and local cosmic strings}\\
\raisebox{0cm}{\makebox{\includegraphics[width=0.47\textwidth, scale=1]{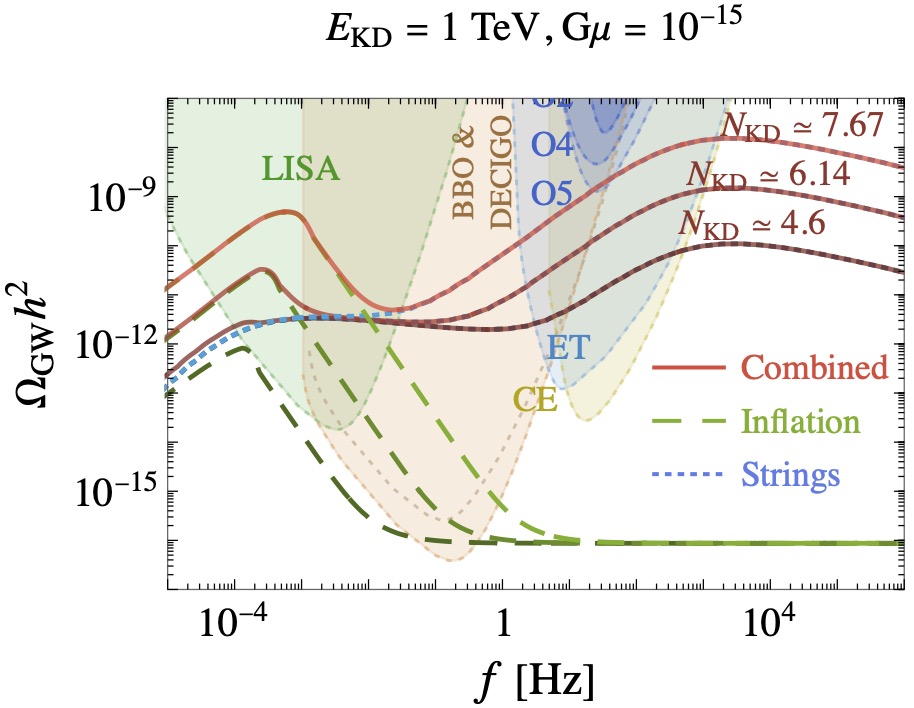}}}
\quad
\raisebox{0cm}{\makebox{\includegraphics[width=0.47\textwidth, scale=1]{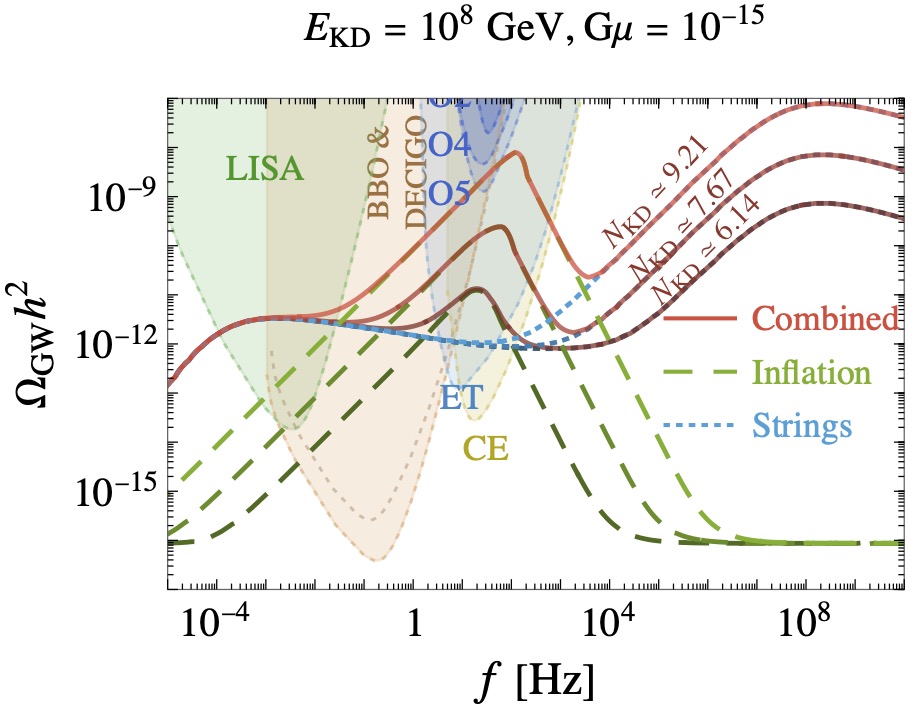}}}
\caption{\textit{ \small Two-peak (\textbf{left}) and three-peak (\textbf{right}) GW spectra from inflation and local CS network.
We assume the maximum inflationary scale allowed by CMB data $E_\mathrm{inf}= 1.6 \times 10^{16} ~ \mathrm{GeV}$ \cite{Ade:2018gkx, Akrami:2018odb}. 
}}
\label{fig:combined_spectra_local}
\end{figure}
\FloatBarrier

\paragraph{Peaks separation.}
We could observe either two (left panel) or three peaks (right panel) depending on the separation between each peak, which are estimated from Eqs.~\eqref{inflation_peak_frequency}, \eqref{peak_kination_freq_local}, and \eqref{cs_low_peak_freq}
\begin{align}
\frac{f_\textrm{peak}^\textrm{cs}}{{f_\textrm{low}^\textrm{cs}}} ~ &\approx ~ 1.2 \times 10^{9} \left(\frac{E_\textrm{KD}}{\textrm{10 TeV}}\right)\left( \frac{\Gamma G\mu }{50\times 10^{-11}}\right)^{1/2}\left( \frac{0.1 }{\alpha}\right)^{1/2},
\label{separation_cs_inf_cs_low}\\[1em]
\frac{f_\textrm{peak}^\textrm{cs}}{{f_\textrm{peak}^\textrm{inf}}} ~ &\approx ~ 1.6\times 10^{5} \left[\frac{10}{\exp(N_\textrm{KD}/2)}\right] \left(\frac{0.1\times50\times10^{-11}}{\alpha\Gamma G\mu}\right)^{1/2},
\label{separation_cs_inf_kd_peak}\\[1em]
\frac{f_\textrm{peak}^\textrm{inf}}{{f_\textrm{low}^\textrm{cs}}} ~ &\approx ~ 0.7 \times 10^{4} \left(\frac{E_\textrm{KD}}{\textrm{10 TeV}}\right)  \left[\frac{\exp(N_\textrm{KD}/2)}{10}\right] \left(\frac{\Gamma G\mu}{50\times10^{-11}}\right),
\label{separation_cs_inf_kd_low}
\end{align}
where we have assumed for simplicity that loops from kination era decay in the radiation era, $N_{\rm KD} < \log(\alpha/2\Gamma G\mu)/3$.
For observable multiple peaks, the separations should be small  but not overlapping.

\paragraph{Detectability of two peaks.}
The combined GW spectra are shown in Fig.~\ref{fig:combined_spectra_local}. The two-peak spectrum can be observed in synergy by LISA and ET/CE. 

\paragraph{Detectability of three peaks.}

The lowest-frequency peak in CS spectrum receives no boost from kination, but requires a large $G\mu$ for its observability in PTA range. However, the flat part of CS in Eq.~\eqref{cs_flat_amp_local} could dominate over the boosted inflationary peak, Eq.~\eqref{inflation_peak_amplitude}. The ratio between them is
\begin{align}
\frac{\Omega_\textrm{peak}^\textrm{inf}}{\Omega_\textrm{flat}^\textrm{cs}} ~ \approx ~0.8 \times 10^{-3}  \left(\frac{E_{\rm inf}}{10^{16}\textrm{ GeV}}\right)^4 \left(\frac{0.1 \times 10^{-11}}{\alpha G\mu}\right)^{1/2}\left(\frac{\Gamma}{50}\right)^{1/2}   \left[\frac{\exp(N_\textrm{KD}/2)}{10}\right]^4.
\label{compare_peak_flat}
\end{align}
For $N_\mathrm{KD} = 5$ and $E_\mathrm{inf}= 1.6 \times 10^{16} ~ \mathrm{GeV}$, the string network with tension $G\mu \lesssim 10^{-15}$ allows the inflationary peak to emerge. 
However, as shown in Fig.~\ref{fig:combined_spectra_local}-right, the simultaneous observation of the three peaks could be possible with the help  of HF experiments \cite{Aggarwal:2020olq}.

\subsubsection{Inflation + global cosmic strings}
\begin{figure}[h!]
\centering
{\bf Gravitational waves from inflation and global cosmic strings}\\
\raisebox{0.75em}{\makebox{\includegraphics[width=0.485\textwidth, scale=1]{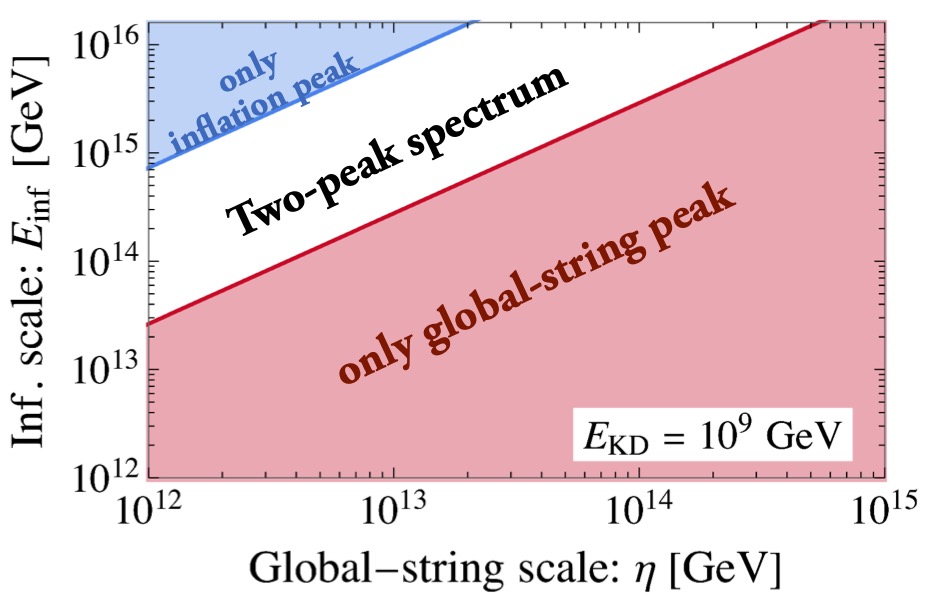}}}
\raisebox{0cm}{\makebox{\includegraphics[width=0.495\textwidth, scale=1]{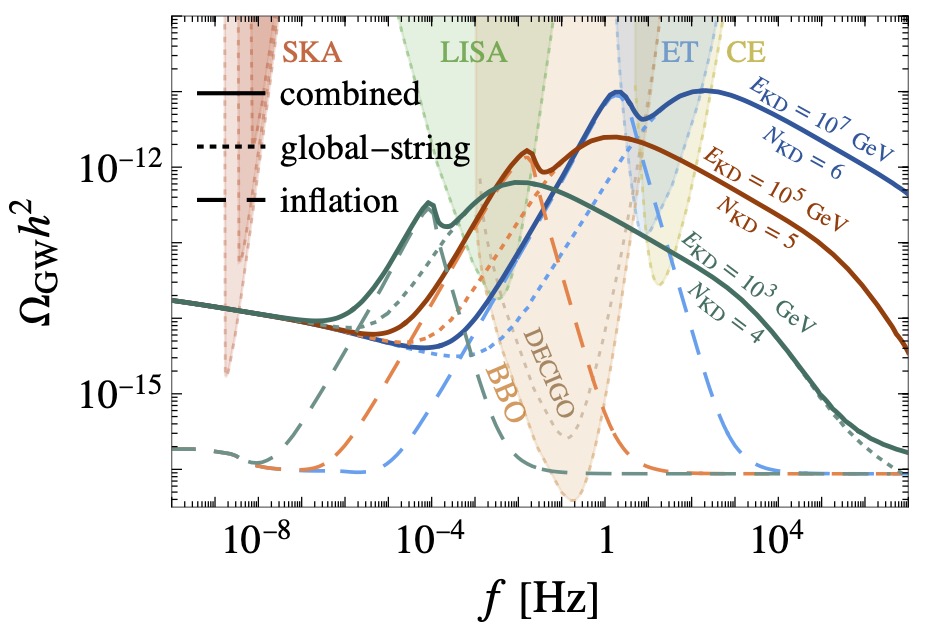}}}
\caption{\textit{ \small
\textbf{Left}: In the presence of GW from both inflation and global strings, as well as a matter-kination era, the signal can have either one or two peaks.
\textbf{Right}: Two-peak GW spectra from inflation at the maximum inflation scale allowed by CMB data $E_{\rm inf} = 1.6 \times 10^{16} ~ {\rm GeV}$ \cite{Ade:2018gkx, Akrami:2018odb} and  from global strings with  energy scale $\eta = 2 \times 10^{14} ~ {\rm GeV}$.
}}
\label{two_peak_spectrum_inf_global}
\end{figure}

\paragraph{Peaks separation.}
The separation between the matter-kination peak in SGWB from global string and primordial inflation can be read out from Eqs.~\eqref{inflation_peak_frequency} and \eqref{peak_kination_freq_local}
\begin{align}
\frac{f_{\rm KD}^{\rm inf}}{f_{\rm KD}^{\rm glob}} ~ \simeq ~ (1.2 \, \times \, 10^{-2}) G^{1/4}(T_\Delta) \left(\frac{\alpha}{0.1}\right).
\label{eq:two_peak_separation_freq}
\end{align}
Interestingly, the peak separation is independent of the matter-kination parameters. The reason is that global-string loops decay right after their production. So the GW frequency reflects directly the horizon size at that time, similar to the inflationary GW.
\paragraph{Detectability of two peaks.}
The visibility of each peak depends on their respective height, determined by the string scale $\eta$ and the inflationary scale $E_{\rm inf}$. The matter-kination peak in the global-string spectrum, Eq.~\eqref{bump_peak_kination_amp_global_simp}, is visible if its amplitude exceeds the inflation shoulder, Eq.~\eqref{inflation_peak_amplitude},
\begin{align}
1 ~ < ~ \left.\frac{\Omega_{\rm GW}^{\rm glob}}{\Omega_{\rm GW}^{\rm inf}}\right|_{f_{\rm KD}^{\rm glob}} ~ = ~ \frac{\Omega_{\rm GW,KD}^{\rm glob}}{\Omega_{\rm GW,KD}^{\rm inf} \left(\frac{f_{\rm KD}^{\rm inf}}{f_{\rm KD}^{\rm glob}}\right)^2} ~ = ~ & (6.4 \, \times \, 10^{2})  \left(\frac{\eta}{10^{15} ~ {\rm GeV}}\right)^4  G^{-3/2}(T_\Delta) \left(\frac{0.1}{\alpha}\right)^2   \left(\frac{10^{16} ~ {\rm GeV}}{E_{\rm inf}}\right)^4 \, \times\nonumber\\
&~ ~ \times \,  \log^3\left[ (2.2 \, \times \, 10^{18}) \left(\frac{\eta}{10^{15} ~ {\rm GeV}}\right) \left(\frac{\alpha}{0.1}\right)^2 \left(\frac{10^9 ~ {\rm GeV}}{E_{\rm KD}}\right)^2 \right].
\label{eq:two_peak_condition1}
\end{align}
Conversely, the matter-kination peak signature in the primordial inflationary GW is visible if its amplitude exceed the global-string blue-tilted part,
\begin{align}
1 ~ > ~ \left.\frac{\Omega_{\rm GW}^{\rm glob}}{\Omega_{\rm GW}^{\rm inf}}\right|_{f_{\rm KD}^{\rm inf}} ~ = ~ \frac{\Omega_{\rm GW,KD}^{\rm glob} \left(\frac{f_{\rm KD}^{\rm inf}}{f_{\rm KD}^{\rm glob}}\right)}{\Omega_{\rm GW,KD}^{\rm inf}} ~ = ~ & (1.1 \, \times \, 10^{-3})  \left(\frac{\eta}{10^{15} ~ {\rm GeV}}\right)^4  G^{-3/4}(T_\Delta) \left(\frac{\alpha}{0.1}\right)   \left(\frac{10^{16} ~ {\rm GeV}}{E_{\rm inf}}\right)^4 \, \times\nonumber\\
&~ ~ \times \,  \log^3\left[ (2.2 \, \times \, 10^{18}) \left(\frac{\eta}{10^{15} ~ {\rm GeV}}\right) \left(\frac{\alpha}{0.1}\right)^2 \left(\frac{10^9 ~ {\rm GeV}}{E_{\rm KD}}\right)^2 \right].
\label{eq:two_peak_condition2}
\end{align}
Both conditions in Eq.~\eqref{eq:two_peak_condition1} and \eqref{eq:two_peak_condition2} must be satisfied  for a visible two-peak signature, as illustrated in the white region of Fig.~\ref{two_peak_spectrum_inf_global}. Otherwise, only one peak is visible, either the one from global strings (red region) or the one from inflation (blue region).
Fig.~\ref{two_peak_spectrum_inf_global} only depends logarithmically on $E_{\rm KD}$, the white region moving to lower $E_{\rm inf}$ by only 10\% when $E_{\rm KD}$ increases by three orders-of-magnitude.

\FloatBarrier
\FloatBarrier

\subsection{GW from first-order phase transitions}
\label{sec:phase_transition}

In the previous subsections, we have shown that the presence of a matter-kination era leads to a peak shape in the GW spectrum produced by primordial inflation or cosmic strings. More generally, any GW signal whose production period lasts longer than the duration of the matter-kination era itself, will receive a triangular shaped spectral distortion. In Sec.~\ref{sec:spectral_distortion_1stOPT}, we show that this is also the case for super-horizon Fourier modes of GW from short-lasting sources such as a cosmological first-order phase transition (1stOPT). Moreover,  Sec.~\ref{sec:uniform_shift_1stOPT} shows that whenever the 1stOPT is produced during the non-standard era, the amplitude of the GW peak is reduced and its frequency is blue-shifted.

\subsubsection{Spectral distortion}
\label{sec:spectral_distortion_1stOPT}

\paragraph{GW from 1stOPT.}
We consider a 1stOPT driven by a scalar field initially at thermal equilibrium with the radiation component. Depending on the amount of supercooling, GW are either sourced by the collision of bubble walls of by fluids motions, e.g. \cite{Caprini:2015zlo,Caprini:2019egz,Ellis:2020awk}.
The peak amplitude of the GW can be formulated as
\begin{align}
\left.h^2\Omega_{\rm GW}(k)\right|_{t_0} \simeq h^2 \over{a_p}{a_0}^4 \over{\rho_{\nor{tot}, p}}{\rho_{\rm tot,0}}\over{H_p}{\beta}^m\over{\kappa \alpha}{1+\alpha + \gamma}^2 \Delta(k,\beta),
\label{eq:pt_modified_amplitude}
\end{align}
where $\rho_{\nor{tot}, i}$ is the total energy density of the universe at time $i$, $T_p$ and $H_p$ are the temperature and Hubble scale at the time of GW production, $\beta^{-1}$ is the duration of the transition, $\alpha$ is the ratio of the vacuum energy difference over the radiation energy density, $\kappa$ is the conversion coefficient and $ \Delta(k,\beta)$ is the spectral shape. We expect $m = 1$ for GW from long-lived fluid motion and $m=2$ for GW from short-lived fluid motion or bubble wall collisions. Since our focus in on the effects from the matter-kination era, we have neglected factors involving the wall velocity $v_{\rm w}$.
The factor $\gamma$ is the ratio of the energy density of the new sector, the spinning axion in our case, to that of radiation 
\begin{equation}
\gamma \equiv \rho_{\rm NS}/\rho_{\rm rad}.\label{eq:def_gamma_1stOPT}
\end{equation}
The case $\gamma = 0$ corresponds to the 1stOPT occuring during the standard radiation era.
Additionally, the peak frequency is shifted with respect to the standard scenario by
\begin{align}
f_{\rm NS}=f_{\rm ST}\left(\frac{a_0}{a_p H_p}\right)_{\rm ST}\left(\frac{a_p H_p}{a_0}\right)_{\rm NS}.
\label{eq:pt_modified_frequency}
\end{align}

\begin{figure}[h!]
\centering
\raisebox{0cm}{\makebox{\includegraphics[width=0.47\textwidth, scale=1]{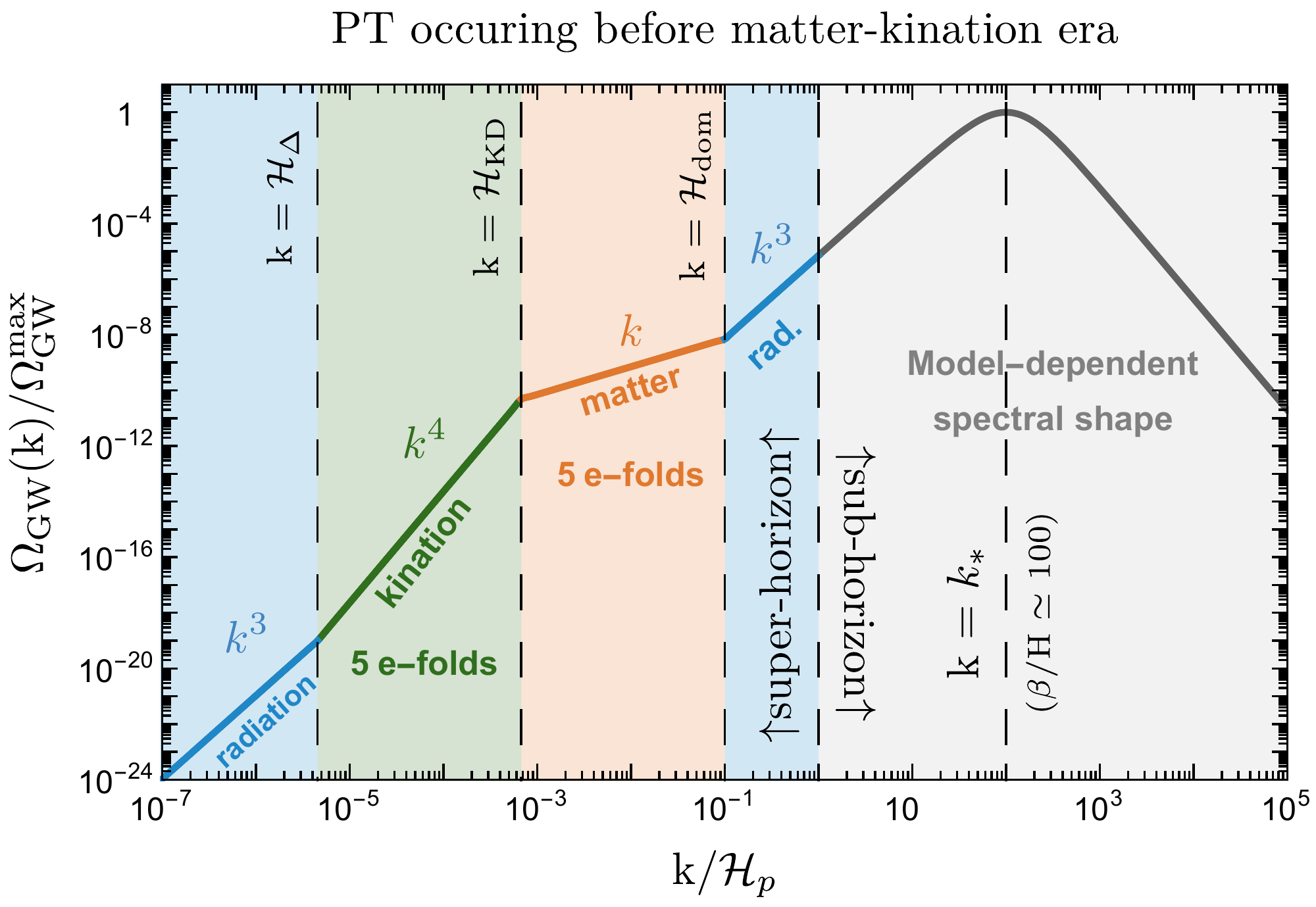}}}
\raisebox{0cm}{\makebox{\includegraphics[width=0.47\textwidth, scale=1]{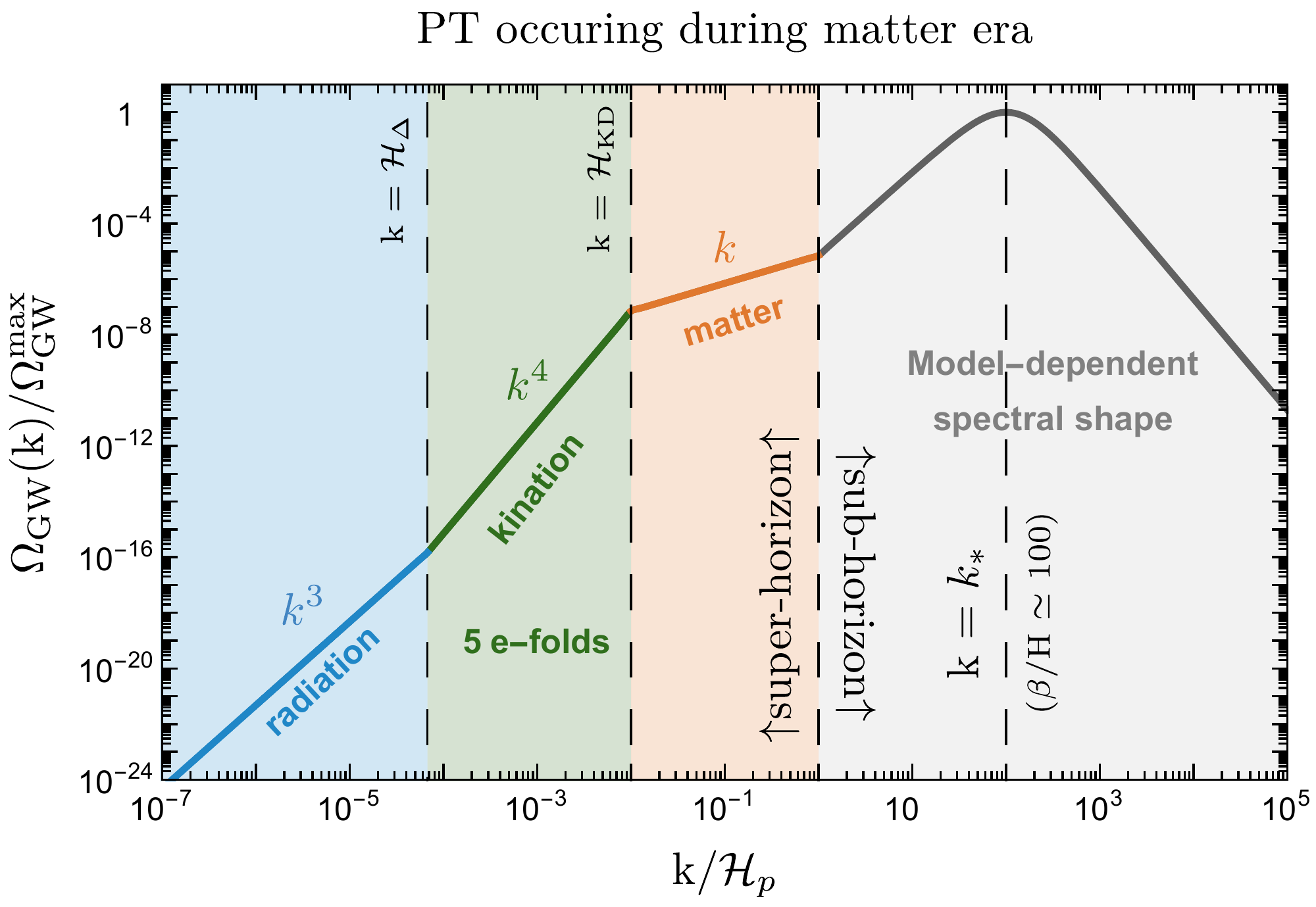}}}
\raisebox{0cm}{\makebox{\includegraphics[width=0.47\textwidth, scale=1]{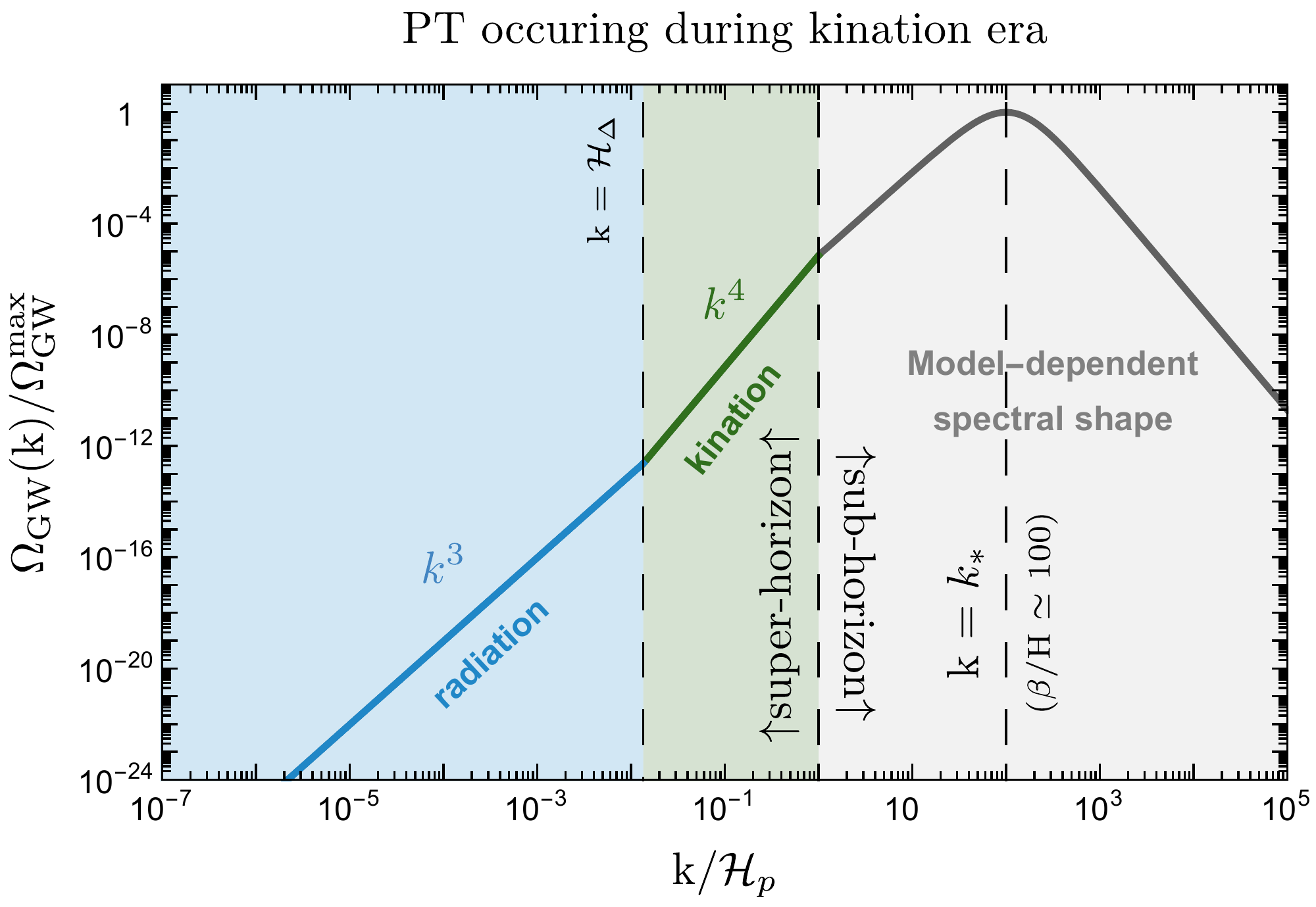}}}
\caption{\textit{ \small Spectral distortion of GW produced from a 1stOPT occurring before the matter-kination era (\textbf{Top left}), during a matter era which is followed by a kination era (\textbf{Top right}), and during a kination era (\textbf{Bottom}). To observe a triangular shape would be a smoking-gun of the scenarios studied in this work. The present figure does not show the peak suppression and the overall frequency blue-shift when the PT takes place during the non-standard era (e.g. middle and bottom panel here), and which we reserve for Fig.~\ref{fig:pt_cartoon} and Fig.~\ref{fig:spectral_distortion_GW_1stOPT_frequency}. $\mathcal{H}_{\rm dom}$, $\mathcal{H}_{\rm KD}$, and $\mathcal{H}_\Delta$ are the comoving Hubble scales ($\mathcal{H} \equiv aH$) at the beginning of the matter era, at the transition to the kination era, and at  the end of the kination era, respectively. The angular wave number $k$ is related to the linear frequency $f$ by $k= 2\pi f$.}}
\label{fig:spectral_distortion_GW_1stOPT}
\end{figure}

\paragraph{Super-horizon modes are sensitive to the EOS.}
Due to causality, the IR slope of GW spectrum from 1stOPT is expected to scale as $\Omega_{\rm GW} \propto k^3$ during radiation domination \cite{Durrer:2003ja, Caprini:2009fx,Cai:2019cdl}.
However, in generic background with EOS $\omega$, we expect the spectral index of super-Hubble Fourier modes to be \cite{Hook:2020phx} (see also \cite{Barenboim:2016mjm, Domenech:2020kqm,Ellis:2020nnr})
\begin{equation}
\label{eq:super_hor_mode_GW_1stOPT_eos}
\Omega_{\rm GW}(k) \propto 
 \left\{
                \begin{array}{ll}
                  k^3 ,\qquad \quad ~~\textrm{for} \quad k \gtrsim \mathcal{H}_p, \vspace{0.25cm}\\
                  k^{\frac{1+15\omega}{1+3\omega}},\qquad \textrm{for} \quad k \lesssim \mathcal{H}_p, 
                \end{array}
              \right. 
\end{equation}
where $\mathcal{H}_p = \frac{a_p}{a_0}H_p$ is the comoving Hubble radius at the time of the PT.  Therefore during matter and kination era the slopes become $k^1$ and $k^4$ for superhorizon modes. The resulting spectral shape is shown in Fig.~\ref{fig:spectral_distortion_GW_1stOPT}. We recognize the same triangular shape as the imprint in GW from primordial inflation and cosmic strings, cf. Sec.~\ref{sec:modelindependent}. The potential detection of such a feature with future interferometers would be a smoking-gun of the scenarios presented in this paper. 

\subsubsection{Uniform shift of the spectrum}
\label{sec:uniform_shift_1stOPT}
\begin{figure}[h!]
\centering
\raisebox{0cm}{\makebox{\includegraphics[width=\textwidth, scale=1]{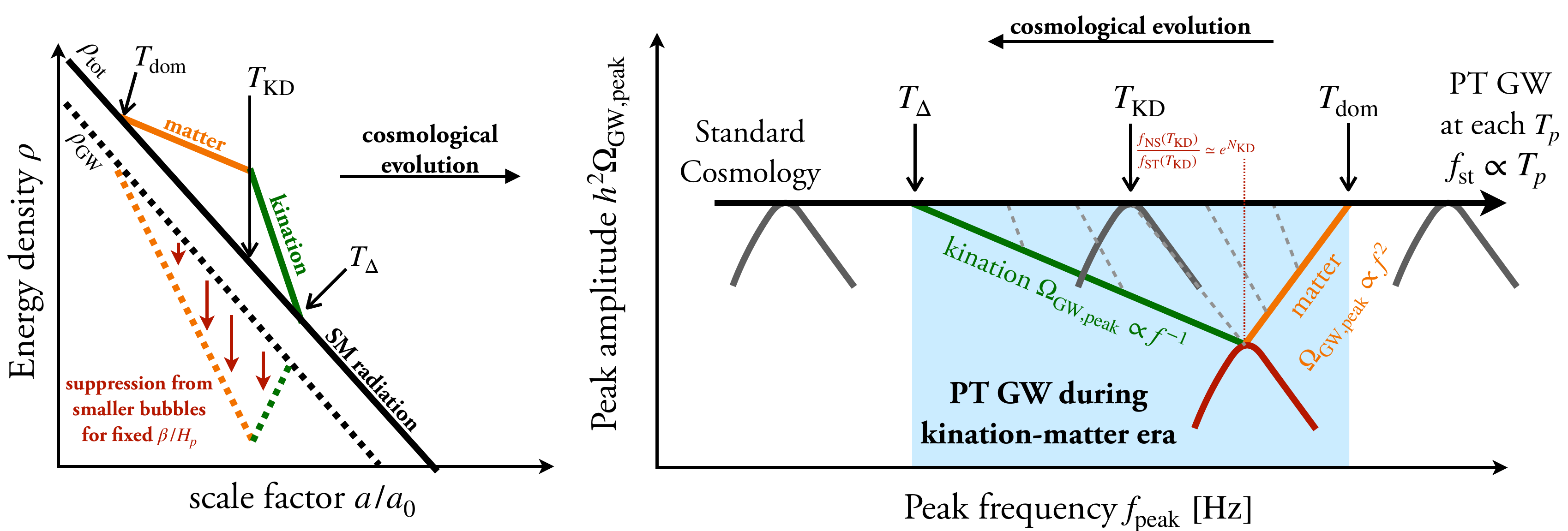}}}
\caption{\textit{ \small The \textbf{left} panel compares the evolution of the total energy density of the universe in the kination-matter scenario (colored) and the one in the standard cosmological history (black). The black dashed line is the expected energy density in GW produced during a thermal 1stOPT taking place at some temperature $T_p$.
Since $\beta/H_p$ is independent of the EOS of the universe, the colliding bubbles during the new-sector domination are smaller in size and the GW production is suppressed.
The \textbf{right} panel shows the amplitude suppression (green and orange lines) and blue-shift (gray dashed lines) of the GW peak when the 1stOPT takes place during the matter-kination era, cf. Eq. \eqref{eq:pt_modified_amplitude} and \eqref{eq:pt_modified_frequency}, with respect to the standard radiation-dominated history (black line).
The maximum suppression, shown in red, occurs when the PT takes place at the onset of the kination era. }}
\label{fig:pt_cartoon}
\end{figure}

\paragraph{Amplitude suppression and blue-shifting of the GW peak.}
Usually, a matter era is followed by a heated radiation era which implies a violated of entropy conservation, see e.g. \cite{Cirelli:2018iax}.
Instead, if the matter era is followed by a kination era, as considered in this paper, cf. Fig.~\ref{fig:pt_cartoon}, there is no entropy injection, which implies
\begin{equation}
\left(\frac{a_p}{a_0}\right)_{\rm NS} =\left(\frac{a_p}{a_0}\right)_{\rm ST} . \label{eq:no_entropy_inj_scale_fac}
\end{equation}
As we will see later in Eq.~\eqref{eq:pt_modified_amplitude}, \eqref{eq:pt_modified_frequency} and \eqref{eq:no_entropy_inj_scale_fac}, we deduce the displacement of the GW peak amplitude and frequency if emission occurs during the matter-kination era
\begin{align}
\frac{\Omega^{\rm NS}_{\rm GW}}{\Omega^{\rm ST}_{\rm GW}} ~ &= ~ \left(\frac{\rho_{\rm p,tot}^{\rm NS}}{\rho_{\rm p,tot}^{\rm ST}}\right) \left[\frac{(1+\alpha)_{\rm ST}}{(1+\alpha + \gamma)_{\rm NS}}\right]^2 ~ = ~ \left(\frac{\rho_{\rm p,tot}^{\rm ST}}{\rho_{\rm p,tot}^{\rm NS}}\right),\label{eq:pt_ratio_amplitude}\\
\frac{f_{\rm NS}}{f_{\rm ST}} ~ &= ~ \left(\frac{H_{\rm p}^{\rm NS}}{H_{\rm p}^{\rm ST}}\right) 
~ = ~ \left(\frac{\rho_{\rm p,tot}^{\rm NS}}{\rho_{\rm p,tot}^{\rm ST}}\right)^{1/2},\label{eq:pt_ratio_frequency}
\end{align}
where we have assumed unchanged $\alpha$, $\kappa$, and $\beta/H_p$.\footnote{$\alpha \equiv \frac{\Delta V}{\rho_{\rm rad}(T_p)}$, where $\Delta V$ is the vacuum energy difference, is left unchanged if $T_p$ is unchanged. $\kappa \equiv \frac{\rho_{\rm source}}{\rho_{\rm rad}}$ where $\rho_{\rm source}$ is the energy density of the GW source, is intrinsically independent of the background. $\beta /H_p \equiv \frac{T}{\Gamma} \frac{\partial \Gamma}{\partial T}\Big|_{T_p}$ where $\Gamma$ is the tunneling rate, is left unchanged if $\Gamma(T)$ is unchanged.}
We see that if the PT occurs during the non-standard era, $\rho^{\rm NS}_{\rm p,tot}  > \rho^{\rm ST}_{\rm p,tot} $,  the amplitude of the GW peak is suppressed and its frequency is blue-shifted, with respect to the one assuming a standard cosmological history, whch is in agreement with previous litterature \cite{Chung:2010cb, simakachorn2019probing, Allahverdi:2020bys}.
\paragraph{Case where PT occurs before non-standard era.}
In contrast if the spinning axion energy density is sub-dominant at the time of GW production, $\gamma \to 0$ in Eq.~\eqref{eq:def_gamma_1stOPT}, then there is no modification of the GW peak position with respect to the standard scenario.
\paragraph{Comparison with standard matter era.}
Due to the absence of entropy injection, cf. Eq.~\eqref{eq:no_entropy_inj_scale_fac}, the amplitude $\Omega^{\rm peak}_{\rm GW}$ and frequency $f_{p}$ of the peak are dispensed from the additional suppression factor $1/D^{4/3}$ and redshift factor $1/D$, respectively, where $D \equiv \frac{S_{\rm after}}{S_{\rm before}}   \geq 1$ is the usual dilution factor, e.g. \cite{Ertas:2021xeh}.

\paragraph{The change of the GW peak position, more precisely.}
We consider a kination-matter era with energy scale at the onset of kination $E_{\rm KD} \equiv \rho_{\rm KD}^{1/4}$ and with $N_{\rm KD}$ efolds of kinations, as in Fig.~\ref{fig:pt_cartoon}. Using Eq.~\eqref{eq:rho_M_def} and \eqref{eq:rho_Delta_def}, we obtain the corresponding temperatures of the radiation bath at the onset of matter, at the onset of kination, and at the end of kination, respectively
\begin{align}
&T_{\rm dom} ~ = ~ \left(\frac{30}{\pi^2 g_*(T_{\rm dom})}\right)^{1/4} E_{\rm KD} \exp\left(\frac{3}{2} N_{\rm KD}\right),\\
&T_{\rm KD} ~ = ~ \left(\frac{30}{\pi^2 g_*(T_{\rm KD})}\right)^{1/4} E_{\rm KD} \exp\left(- \frac{1}{2} N_{\rm KD}\right),\\
&T_{\Delta} ~ = ~ \left(\frac{30}{\pi^2 g_*(T_{\Delta})}\right)^{1/4} E_{\rm KD} \exp\left(- \frac{3}{2} N_{\rm KD}\right).
\end{align}
The amplitude of the GW peak in the presence of a kination-matter era reads
\begin{align}
\frac{\Omega_{\rm GW}^{\rm NS}}{\Omega_{\rm GW}^{\rm ST}} ~ = ~ \begin{cases}
1  ~ ~ ~ & {\rm for} ~ ~ T_p < T_\Delta,\\
\left(\frac{g_*(T_{\rm KD}) g_{*}(T_{\Delta})}{g_*^2(T_p)}\right) \left(\frac{T_\Delta}{T_p}\right)^{2}  ~ ~ ~ & {\rm for} ~ ~ T_{\rm KD} > T_p \geq T_\Delta,\\
\left(\frac{T_p}{T_{\rm KD}}\right) \exp(-2 N_{\rm KD}) ~ ~ ~ & {\rm for} ~ ~ T_{\rm dom} > T_p \geq T_{\rm KD},\\
1  ~ ~ ~ & {\rm for} ~ ~ T_p \geq T_{\rm dom},
\end{cases}
\label{eq:pt_modified_amplitude}
\end{align}
while its frequency is
\begin{align}
\frac{f_{\rm NS}}{f_{\rm ST}} ~ = ~ \begin{cases}
1  ~ ~ ~ & {\rm for} ~ ~ T_p < T_\Delta,\\
\left(\frac{g_*^2(T_p)}{g_*(T_{\rm KD}) g_{*}(T_{\Delta})}\right)^{1/2} \left(\frac{T_p}{T_\Delta}\right)  ~ ~ ~ & {\rm for} ~ ~ T_{\rm KD} > T_p \geq T_\Delta,\\
 \left(\frac{T_{\rm KD}}{T_p}\right)^{1/2} \exp( N_{\rm KD}) ~ ~ ~ & {\rm for} ~ ~ T_{\rm dom} > T_p \geq T_{\rm KD},\\
1  ~ ~ ~ & {\rm for} ~ ~ T_p \geq T_{\rm dom}.
\end{cases}
\label{eq:pt_modified_frequency}
\end{align}
The largest modification occurs when the PT takes place at the start of kination era, $T_p = T_{\rm KD}$, for which the peak amplitude and frequency are given by
\begin{align}
\left. \frac{\Omega_{\rm GW}^{\rm NS}}{\Omega_{\rm GW}^{\rm ST}}  \right|_{\rm KD} ~ = ~ \exp(-2 N_{\rm KD}), ~ {\rm and }~ ~ 
\left. \frac{f_{\rm NS}}{f_{\rm ST}}  \right|_{\rm KD} ~ = ~\exp( N_{\rm KD}),
\label{eq:pt_modify_largest}
\end{align}
The right panel of Fig.~\ref{fig:pt_cartoon} shows the peak position of the modified GW spectrum in the presence of the kination-matter era, compared to the one assuming a standard cosmological history.
The GW amplitude in the standard cosmological history (black line) is approximately constant with varying $T_p$, i.e. $\Omega_{\rm GW,ST} \propto \left(a_{p}/a_0\right)^4 \rho_{\rm p,tot} \propto {\rm constant}$, while the peak frequency grows linearly with the temperature $f_{\rm ST} \propto a_p H_p \propto T_p$.
In contrast, during the kination and matter eras, the peak amplitude $\Omega_{\rm GW}^{\rm NS}$ scales with the peak frequency as $f_{\rm NS}^{-1}$ and $f_{\rm NS}^2$, respectively.

\begin{figure}[h!]
\centering
\raisebox{0cm}{\makebox{\includegraphics[width=0.6\textwidth, scale=1]{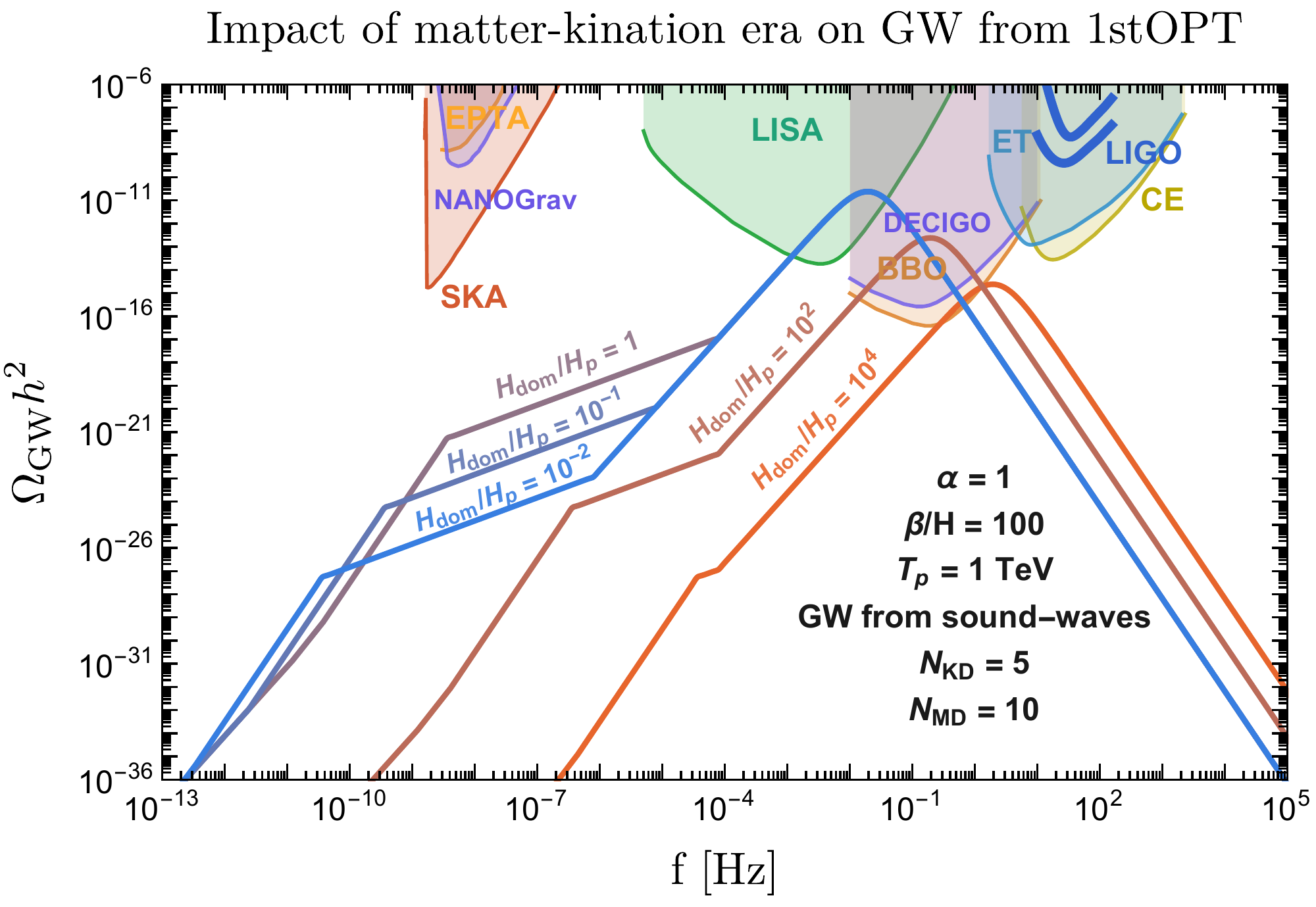}}}
\caption{\textit{ \small $H_p$ and $H_{\rm dom}$ are the Hubble scales at the time of GW production and at the onset of the matter-kination era, respectively. Super-horizon modes (i.e. emitted with a frequency $f_{\rm em.} < H/2\pi$), are sensitive to the EOS of the universe as stated in Eq.~\eqref{eq:super_hor_mode_GW_1stOPT_eos} (blue-ish lines). Here for $H_p< H_{\rm dom}$, the PT takes place during the non-standard era (either matter or kination) so its peak amplitude is suppressed and blue-shifted as stated by Eq.~\eqref{eq:pt_ratio_amplitude} and Eq.~\eqref{eq:pt_ratio_frequency} (red-ish lines).  The GW spectrum in standard radiation cosmology is computed according to \cite{Ellis:2020awk}. }}
\label{fig:spectral_distortion_GW_1stOPT_frequency}
\end{figure}
\paragraph{Origin of the peak suppression.}
The NS-to-standard GW density ratio in Eq.~\eqref{eq:pt_ratio_amplitude} can be rewritten as the ratio of Hubble horizon
\begin{align}
\frac{\Omega^{\rm NS}_{\rm GW}}{\Omega^{\rm ST}_{\rm GW}} ~ = ~ \left(\frac{\rho_{\rm p,tot}^{\rm ST}}{\rho_{\rm p,tot}^{\rm NS}}\right) ~=~ \left(\frac{H_{\rm p}^{\rm ST}}{H_{\rm p}^{\rm NS}}\right)^2.
\label{eq:GW_spectrum_general_relation}
\end{align}
At fixed $\beta/H_p$, the bubble size at collision is smaller during kination or matter era, which implies a smaller GW amplitude.
Finally, the overall impact of the matter-kination era on the short-lasting sources such as 1stOPT is shown in Fig.~\ref{fig:spectral_distortion_GW_1stOPT_frequency}.

\section{Axion Dark Matter } \label{sec:darkmatter}

The matter-kination scenario is well motivated by the spinning axion field, as we will see in the later section. We now discuss the relic abundances of the axion and axion-like particle (ALP). Then we study if this leads to extra constraints on the parameter space from overabundance. One of the motivation of our study is to demonstrate that spinning complex scalar fields can generate intermediate matter-kination era and enhance SGWB of cosmological origin.  By-products of such particle physics models are the baryon asymmetry discussed in Sec.~\ref{sec:baryon_asymmetry} and, in this section, DM,  whose {corresponding} production mechanism can either be kinetic misalignment or standard misalignment.
Let us stress  that this section only utilizes the generic setup of the spinning axion. The concrete realization of such model can put more constraints on the parameter space, cf. the `Axion model realizations' part of the paper.

\subsection{GW peak and axion abundance}
\subsubsection{GW peak and $U(1)$ charge}
\paragraph{$U(1)$ charge generation.}

The axion is the angular mode of a complex scalar field $\Phi = \phi \,e^{i \theta}$, the Peccei-Quinn (PQ) field, with a $U(1)_{\rm PQ}$ symmetry. At early times, it could receive a kick parametrized by the number density of Noether $U(1)_{\rm PQ}$ charge 
\begin{equation}
n_\theta = \phi^2 \dot{\theta},
\end{equation}
due to some $U(1)_{\rm PQ}$-breaking effect
\begin{align}
\frac{1}{a^3} \frac{\partial}{\partial t} (a^3 n_\theta) ~ = ~ - \frac{\partial V}{\partial \theta}.\label{eq:EoM_angular_AxionDM}
\end{align}
Assuming that the integral of Eq.~\eqref{eq:EoM_angular_AxionDM} is dominated by the latest time $t_r$, we obtain the resulting comoving number density of $U(1)$ charge
\begin{align}
Y_\theta(t) ~\equiv ~ \frac{n_\theta}{s}~ \simeq ~ \frac{a^3(t_r)}{a^3(t)} \cdot \frac{\partial V}{\partial \theta}(t_r) \cdot \frac{1}{H(t_r)},\label{eq:U1_charge_generation_AxionDM}
\end{align}
where $s$ is the entropy density of the universe and which is conserved through Hubble evolution, see Eq.~\eqref{PQ_charge_conservation}. 
See Sec.~\ref{sec:U1breaking} for more details, \cite{Affleck:1984fy} for the original work and e.g. Chap.~11.6 of \cite{Gorbunov:2011zz} for a review.

\paragraph{Relation between $U(1)$ charge and kination parameters.}

Assuming that the $U(1)_{\rm PQ}$ breaking effect decouples at later time $\partial V/\partial \theta \to 0$, the invariance of the interactions under the $U(1)$ symmetry of the complex scalar field, implies the conservation of the comoving $U(1)$ charge
\begin{equation}
Y_\theta ~ = ~ \frac{n_\theta}{s} ~ = ~ \frac{\phi^2 \dot{\theta}}{\frac{2 \pi^2}{45} g_{*s}(T) T^3} 
~ = ~ \left( \frac{2}{\dot{\theta}_\mathrm{KD}} \right) \left[\frac{E_\mathrm{KD}^4}{\frac{2 \pi^2}{45} g_{*s}(T_\mathrm{KD}) T_\mathrm{KD}^3} \right].
\label{eq:U1_charge_axion}
\end{equation}
 In the last equality of Eq.~\eqref{eq:U1_charge_axion}, we have evaluated $Y_\theta$ at the beginning of the kination era, when the complex scalar field sits in the circular minimum $\phi = f_a$ and the axion kinetic energy is given by $\rho_\mathrm{KD} = E_\mathrm{KD}^4 = f_a^2 \dot{\theta}_\mathrm{KD}^2/2$.
The temperature $T_\mathrm{KD}$ of the thermal bath when kination starts follows from
$
\rho^\mathrm{rad}_\mathrm{KD}  = E_\mathrm{KD}^4 \exp(-2 N_\mathrm{KD})$
and reads
$
T_\mathrm{KD}  = \left(30/\pi^2 g_*(T_\mathrm{KD}) \right)^{1/4} E_\mathrm{KD} \exp(-N_\mathrm{KD}/2).
$
It follows that the kination energy scale $E_{\rm KD}$ is directly related to the $U(1)$ charge $Y_\theta$
\begin{align}
E_\mathrm{KD}  ~ = ~ 0.436 \, G^{3/4}(T_\mathrm{KD}) \left( \frac{f_a}{Y_\theta } \right) \exp(3 N_\mathrm{KD}/2),
\label{Ekd_yield_DM_abundance}
\end{align}
where $G(T) \equiv  ({g_*(T)}/{106.75}) ({g_{*,s}(T)}/{106.75})^{-4/3}$.
As we will see below, this relation allows for a one-to-one relation between the frequency and amplitude of the GW peak induced by the kination era, and the $U(1)$ charge $Y_\theta$.
 
The $U(1)$ charge $Y_\theta$ can be partially transferred into baryon number and lead to successful baryogenesis as in the Affleck-Dine mechanism  \cite{Affleck:1984fy} or so-called axiogenesis mechanism \cite{Co:2019wyp}. We postpone the discussion of the baryon asymmetry to the next Sec.~\ref{sec:baryon_asymmetry}.
Instead in the present section, we discuss the transfer of $Y_\theta$ into axion coherent oscillations via the so-called kinetic misalignment mechanism, which can explain dark matter for smaller $f_a$ than in the standard misalignment mechanism  \cite{Co:2019jts,Chang:2019tvx}.

\subsubsection{Axion abundance}

\paragraph{Kinetic misalignment mechanism.}
After the kination starts, the energy density in the spinning axion  $\rho_\theta = f_a^2 \dot{\theta}^2/2$ decreases as $a^{-6}$.
Eventually, it drops below the axion potential at the top of the barrier $\rho_\mathrm{barrier} \simeq 2 m_a^2 f_a^2$ where $m_a$ is the axion mass, and the axion gets trapped and oscillates in one of the minima \cite{Co:2019jts,Chang:2019tvx}. This occurs when the axion spinning speed drops to
\begin{align}
\label{eq:T_trap}
\dot{\theta}(T_{\rm trap}) = 2 m_a(T_{\rm trap}),
\end{align}
where we neglect the temperature dependence of $m_a$.
Conservation of energy $\rho_{\theta} = \rho_a $ implies that the $U(1)$ charge yield  $Y_\theta = 2\rho_\theta/(\dot{\theta} s)$ is transferred to the yield of the axion oscillation $Y_a  = \rho_a/(m_a s)$
\begin{align}
 Y_a ~ = ~ Y_\theta. \label{eq:Ytheta_Ya}
\end{align}
However, Eq.~\eqref{eq:Ytheta_Ya} neglects correction due to non-linear effects which can enhance the axion abundance. Indeed, the fast-moving axion skipping the potential barrier is known to fragment into higher-momentum modes \cite{Fonseca:2019ypl}. It is found that the axion energy density in higher modes generated by the fragmentation is of the same order as the zero-mode component generated from Eq.~\eqref{eq:Ytheta_Ya} \cite{Fonseca:2019ypl,Morgante:2021bks,DESYfriendpaper}. Hence we replace Eq.~\eqref{eq:Ytheta_Ya} by
\begin{align}
 Y_a ~ = ~ 2Y_\theta, \label{eq:Ytheta_Ya_2}
\end{align}
as in the kinetic misalignment mechanism  \cite{Co:2019jts,Chang:2019tvx}.
We deduce the fraction of axion in the total DM energy density as a function of the $U(1)$ charge
 \begin{align}
\frac{\rho_{a,0}}{\rho_{{\rm DM,}0}} ~ = ~ \frac{\Omega_{a,0}h^2}{\Omega_{{\rm DM,}0}h^2} ~ \simeq ~ \frac{2 s_0 m_a Y_\theta h^2}{0.5745 ~ \mathrm{keV/cm^3}} ~ 
\simeq ~ 405.7~h^2~\left(\frac{m_a}{1 ~ \mathrm{eV}}\right) \left(\frac{Y_\theta}{40}\right).
\label{eq_axion_fraction}
\end{align}
where $2 s_0 m_a Y_\theta$ is the axion energy density today and $s_0 = 2\pi^2 g_{*s}(T_0) T_0^3/ 45$ is the entropy density today.
From using Eq.~\eqref{Ekd_yield_DM_abundance} and \eqref{eq_axion_fraction}, we deduce the kination of energy scale and duration as a function of the axion abundance 
\begin{framed}
\vspace{-1em}
\begin{align}
E_\mathrm{KD} ~  =  ~ (4.4\times 10^{9} ~h^2~ \mathrm{GeV}) G^{3/4}(T_\mathrm{KD}) \left( \frac{f_a}{10^9 ~ \mathrm{GeV}} \right) \left( \frac{m_a}{1 ~ \mathrm{eV}} \right) \left( \frac{\Omega_\mathrm{DM,0} h^2}{\Omega_\mathrm{a,0} h^2} \right)  \exp(3 N_\mathrm{KD}/2).
\label{eq:Ekd_axion_fraction}
\end{align}
\vspace{-1em}
\end{framed}\noindent
If the axion is the canonical Peccei-Quinn (PQ) QCD axion \cite{Peccei:1977ur, Peccei:1977hh,Wilczek:1977pj,Weinberg:1977ma}, the $m_a - f_a$ relation is fixed  \cite{GrillidiCortona:2015jxo}
\begin{equation}
m_a  \simeq  (5.70 ~ \mu \mathrm{eV}) \left(10^{12} ~ \mathrm{GeV}/f_a\right), \label{eq:QCD_axion_definition}
\end{equation}
and the kination energy scale in Eq.~\eqref{eq:Ekd_axion_fraction} becomes
\begin{align}
E_\mathrm{KD}^\mathrm{QCD} ~ &= ~ (2.5 \times 10^{7} ~h^2~ \mathrm{GeV}) G^{3/4}(T_\mathrm{KD}) \frac{\Omega_{{\rm DM},0}h^2}{\Omega_{a,0}h^2}  \exp(3 N_\mathrm{KD}/2).
\end{align}

\paragraph{Standard misalignment mechanism.}
The one-to-one relation in Eq.~\eqref{eq:Ekd_axion_fraction} between the kination energy scale and duration and the DM abundance is only valid if the axion abundance is set by the $U(1)$ charge $Y_\theta$ (kinetic misalignment mechanism).
Instead, in the limit of small $Y_\theta$, the axion can be trapped by the potential barrier at $T_{\rm trap}$ in Eq.~\eqref{eq:T_trap} before the conventional onset of axion oscillation at $T_{\rm mis}$
\begin{equation}
3 H(T_{\rm mis}) = m_a(T_{\rm mis}).
\end{equation}
In that case, for which $T_{\rm trap} > T_{\rm mis}$, the axion abundance is dominantly set by the standard misalignment mechanism with an axion number density \cite{Preskill:1982cy,Abbott:1982af,Dine:1982ah}
\begin{align}
Y_\mathrm{a}^{\rm mis} ~ = ~ \frac{2 m_a(T_{\rm mis}) f_a^2}{\frac{2 \pi^2}{45} g_{*s}(T_{\rm mis}) T_{\rm mis}^3} ~ \xrightarrow{\text{if}~m_a(T)~\rm ~is~T-independent} ~ \left(\frac{45}{30^{3/4} \pi^{1/2}} \right) \left(\frac{g_{*}^{3/4}(T_{\rm mis})}{g_{*s}(T_{\rm mis})} \right)  \left(\frac{f_a^2}{m_a^{1/2}\MPl^{3/2}} \right)\theta_{\rm mis}^2 , \label{eq:std_misalignment_mechanism}
\end{align}
where $\theta_{\rm mis}$ is the initial amplitude of the oscillation, which is expected to be order 1.
In the general case, the axion number density can be computed from
\begin{equation}
\label{eq:axion_abundance_general}
\frac{\Omega_{a,0}h^2}{\Omega_{{\rm DM,}0}h^2} ~
\simeq ~ 405.7~h^2~\left(\frac{m_a}{1 ~ \mathrm{eV}}\right) \left(\frac{Y_a}{40}\right), \qquad \textrm{with} \quad Y_\mathrm{a} ~=~\textrm{Max}\left[Y_\mathrm{a}^{\rm mis},~2Y_\theta\right],
\end{equation}
where we used Eq.~\eqref{eq_axion_fraction}, and where $Y_\theta$ and $Y_{\rm a}^{\rm mis} $ are defined in Eq.~\eqref{eq:U1_charge_axion} and \eqref{eq:std_misalignment_mechanism}.
We conclude that whenever
\begin{equation}
Y_\mathrm{a}^{\rm mis}~>~ 2Y_\theta,
\end{equation}
then we cannot relate the matter-kination parameters to the axion abundance using Eq.~\eqref{eq:Ekd_axion_fraction}, which assumes that the kinetic misalignment sets the axion abundance.

The temperature dependence of the axion mass $m_a(T)$ is model dependent. Considering the case of the canonical PQ QCD axion, the mass is supposed to vary as, e.g. \cite{Borsanyi:2016ksw}
\begin{equation}
\label{eq:QCD_axion_definition_finite_T}
m_a^2(T) ~=~ m_a^2 \frac{\chi(\rm GeV)}{\chi(0)} \left(\frac{1~\rm GeV}{T} \right)^\alpha = 0.12 m_a^2 \left(\frac{\rm \Lambda_{\rm QCD}}{T} \right)^\alpha,
\end{equation}
where\footnote{$\chi$ is the susceptibility of the topological charge, defined by $\chi(T) \equiv m_a(T)^2 f_a^2$ and $\Lambda_{\rm QCD}$ is the scale at which the perturbative QCD coupling constant diverges.} $\chi(0) \simeq (75,6 ~ \rm MeV)^{4}$, $\chi(1~\rm GeV) \simeq (2~\rm MeV)^4$, $\alpha = 8.16$ \cite{Borsanyi:2016ksw} and $\Lambda_{\rm QCD} \simeq 211 ~\rm MeV$ \cite{FlavourLatticeAveragingGroup:2019iem}.  
By comparing Eq.~\eqref{eq:std_misalignment_mechanism} with Eq.~\eqref{eq_axion_fraction}, we deduce that the kinetic misalignment is effective for 
\begin{align}
\sqrt{m_a f_a} ~ \lesssim ~ 1.1 \times 10^{-9} \left(\frac{g_{*s}(T_{\rm mis})}{g_{*}^{3/4}(T_{\rm mis})} \right) \left(\frac{\MPl}{f_a} \right)^{3/2},
\label{eq:condition_effective_misalign}
\end{align}
where we have taken $m_a(T_{\rm mis}) = m_a$. 
 
In the region of the parameter space leading to an observable peak in SGWB from primordial inflation, the DM abundance of conventional QCD axion DM is predicted to be too large, e.g. see Fig.~\ref{fig:peak_spectrum_amplitude} for kinetic misalignment. To prevent QCD axion DM overclosure, we can instead consider non-conventional QCD axion as discussed in Sec.~\ref{sec:non_conventional_QCD_axion} or a non-QCD generic ALP, see Fig.~\ref{fig:peak_spectrum_amplitude}. For the case where the DM abundance is set by standard misalignment, DM overclosure can be avoided if $\theta_{\rm mis}$ is tuned to be small. We leave this issue to future studies.

\subsection{Probing axion DM with inflationary GW}
\subsubsection{ALP DM}
\begin{figure}[h!]
\centering
{\bf Gravitational waves from primordial inflation}\\
\raisebox{0cm}{\makebox{\includegraphics[width=0.465\textwidth, scale=1]{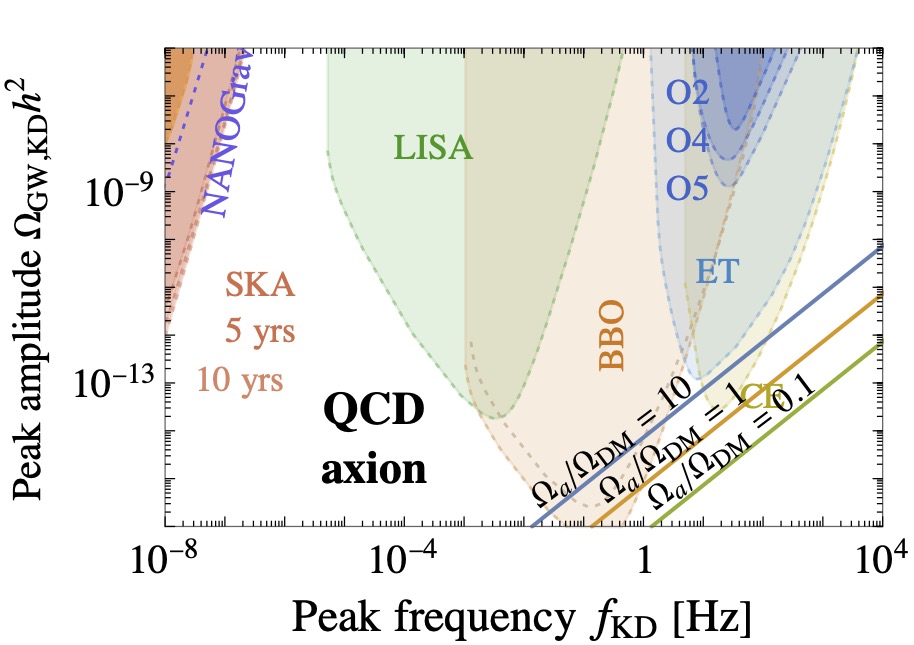}}}
\qquad
\raisebox{0cm}{\makebox{\includegraphics[width=0.465\textwidth, scale=1]{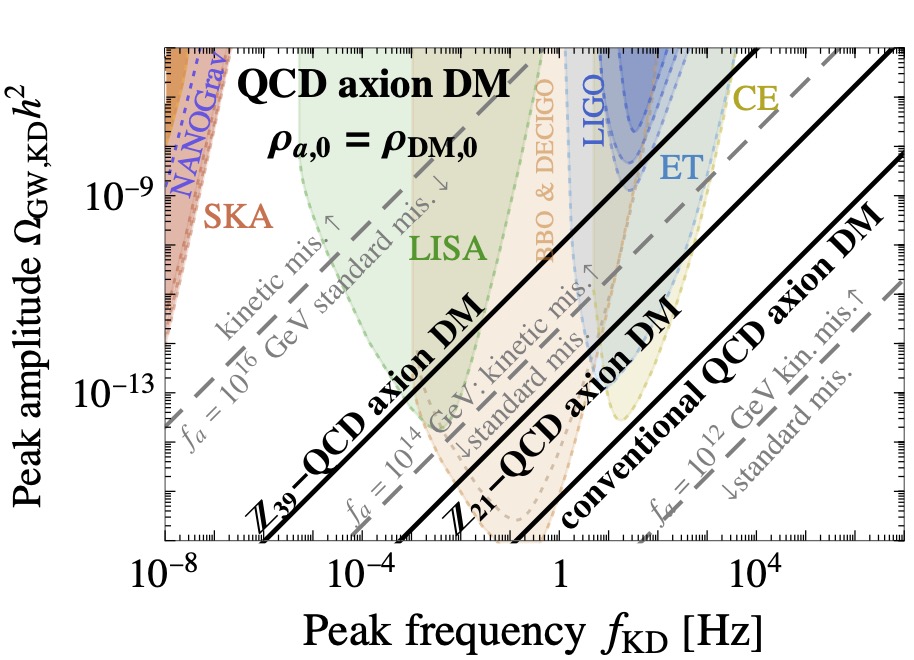}}}\\[1em]
\raisebox{0cm}{\makebox{\includegraphics[width=0.465\textwidth, scale=1]{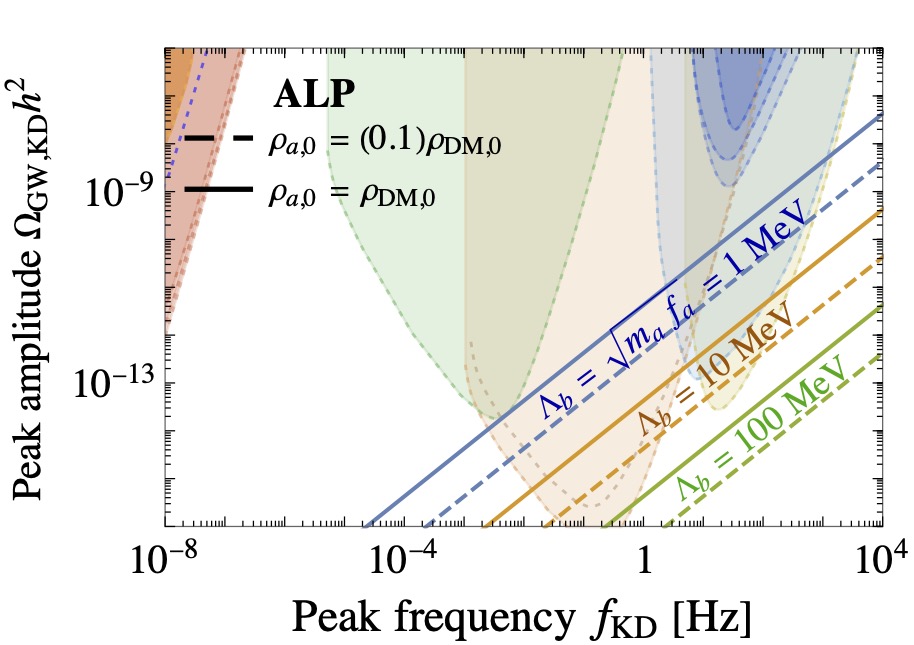}}}
\qquad
\raisebox{0cm}{\makebox{\includegraphics[width=0.465\textwidth, scale=1]{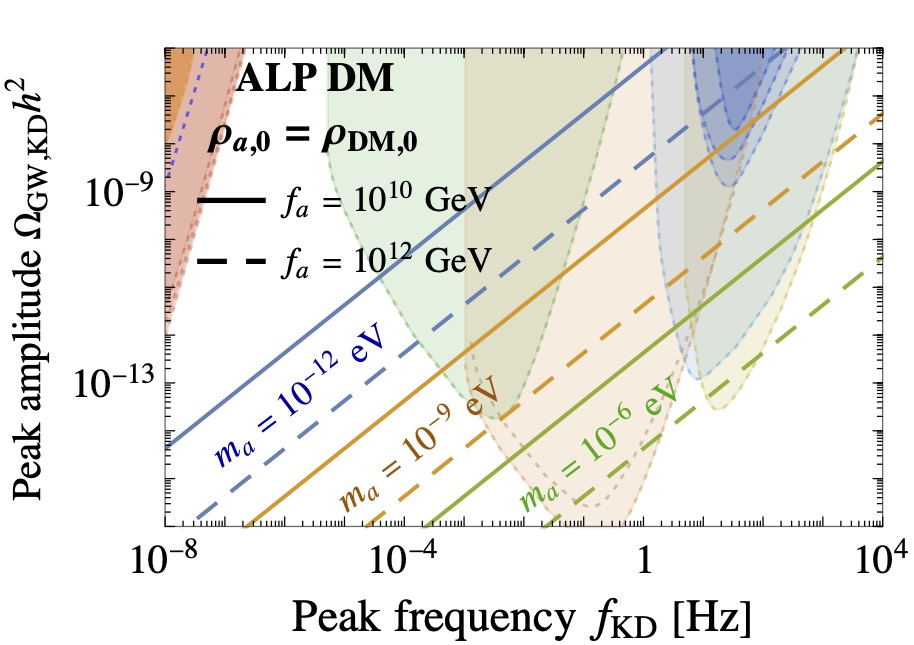}}}\\[0.5em]
\caption{\textit{ \small The solid lines indicate the position of the GW peak signature in the presence of a matter-kination generated by the spinning axion, for a fixed axion relic abundance, which is produced via kinetic misalignment for either the QCD axion (\textbf{Top}), see  Eq. \eqref{eq_peak_position_QCDaxion}, or a generic ALP (\textbf{Bottom}), see  Eq. \eqref{eq:peak_position_ALP2}. In the \textbf{top-right} plot, above the black solid lines, the axion is overabundant. Below the dashed gray lines, for a given $f_a$ value the axion abundance is set by the standard misalignment mechanism, such that the relation between the GW peak signature and the axion abundance in Eq.~\eqref{eq_peak_position_QCDaxion} and  \eqref{eq_peak_position_QCDaxion_freq} is not applicable.  The dependence on $m_a$ and $f_a$ is shown in Fig.~\ref{ma_fa_plot}.}}
\label{fig:peak_spectrum_amplitude}
\end{figure}

Let us first focus on the GW from inflation whose peaked signature simply depends on the inflationary scale.
From Eqs.~\eqref{inflation_peak_frequency}, \eqref{inflation_peak_amplitude}, and \eqref{Ekd_yield_DM_abundance}, the peaked frequency and amplitude read
\begin{align}
f_\mathrm{KD} ~ &= ~ (4.7\times 10^{-9} ~ \mathrm{Hz}) [G^{1/4}(T_\Delta) G^{3/4}(T_\mathrm{KD})] \left( \frac{f_a}{Y_\theta } \right) \exp(2 N_\mathrm{KD}), \label{eq:peak_yield_freq}\\
\Omega_\mathrm{GW,KD} h^2 ~ &= ~ (1.1 \times 10^{-16}) \left( \frac{G(T_\Delta)}{G(T_\mathrm{KD})} \right)^{3/4} \left( \frac{E_\mathrm{inf}}{10^{16} ~ \mathrm{GeV}} \right)^4 \left( \frac{f_\mathrm{KD}}{1 ~ \mathrm{Hz}} \right)  \left( \frac{10^{9} ~ \mathrm{GeV}}{f_a} \right)  \left( \frac{Y_\theta }{40} \right).
\label{eq:peak_yield_amp}
\end{align}
The GW peak position relates to the axion contribution to DM, set by the kinetic misalignment mechanism cf. Eq.~\eqref{eq_axion_fraction}
\begin{framed}
\vspace{-1em}
\begin{align}
f_\mathrm{KD} ~ &= ~ (21~{\rm Hz}) ~h^2~\left[\frac{G(T_\Delta)}{G(T_\mathrm{KD})}\right]^{1/4} \left(\frac{10^9 ~ \mathrm{GeV} }{f_a} \right)^{1/3} \left(\frac{1 ~ \mathrm{eV}}{m_a} \right)^{1/3} \left(\frac{\Omega_{a,0}}{\Omega_{DM,0}} \right)^{1/3} \left(\frac{E_\mathrm{KD}}{10^9 \mathrm{~ GeV}} \right)^{4/3}, \\
\Omega_\mathrm{GW,KD} h^2 ~ &= ~ (2.7 \times 10^{-19})~h^{-2}~ \left( \frac{G(T_\Delta)}{G(T_\mathrm{KD})} \right)^{3/4} \left( \frac{E_\mathrm{inf}}{10^{16} ~ \mathrm{GeV}} \right)^4 \left( \frac{f_\mathrm{KD}}{1 ~ \mathrm{Hz}} \right)  \left( \frac{10^{9} ~ \mathrm{GeV}}{f_a} \right) \left(\frac{1 ~ \mathrm{eV}}{{m_a}}\right) \left(\frac{\Omega_{a,0}}{\Omega_{DM,0}} \right).
\label{eq:peak_position_ALP2}
\end{align}
\vspace{-1em}
\end{framed}\noindent
For the QCD axion, Eq.~\eqref{eq:QCD_axion_definition} leads to simpler expressions
\begin{align}
\label{eq_peak_position_QCDaxion_freq}
f_\mathrm{KD}^{\rm QCD} ~ &= ~ (115 ~ \mathrm{Hz})~h^2~ \left[\frac{G(T_\Delta)}{G(T_\mathrm{KD})}\right]^{1/4} \left(\frac{\Omega_{a,0}}{\Omega_{DM,0}} \right)^{1/3} \left(\frac{E_\mathrm{KD}}{10^9 \mathrm{~ GeV}} \right)^{4/3},\\
\Omega_\mathrm{GW,KD}^{\rm QCD} h^2 ~ &= ~ (4.8 \times 10^{-17})~h^{-2}~ \left( \frac{G(T_\Delta)}{G(T_\mathrm{KD})} \right)^{3/4} \left( \frac{E_\mathrm{inf}}{10^{16} ~ \mathrm{GeV}} \right)^4 \left( \frac{f_\mathrm{KD}}{1 ~ \mathrm{Hz}} \right) \left(\frac{\Omega_{a,0}}{\Omega_{DM,0}} \right).
\label{eq_peak_position_QCDaxion}
\end{align}
The relation between observability of GW from primordial inflation and axion DM abundance is shown in Fig~\ref{fig:peak_spectrum_amplitude}. The matter-kination era generated by ALP DM with a mass $m_a \lesssim 10^{-6}~$eV can move the GW signal into observable windows of the future interferometers. In the specific case of the QCD axion DM, the GW signal is enhanced only at  frequencies larger than ET/CE, which motivates high-frequency GW searches \cite{Aggarwal:2020olq}. In the regions of observable GW signals, the conventional QCD axion is overabundant, as shown in Fig.~\ref{detect_peak_inf} and \ref{fig:peak_spectrum_amplitude}. As we show in the next Sec.~\ref{sec:non_conventional_QCD_axion}, only lighter (non-conventional) QCD axion can satisfy the correct DM abundance while leading to an observable GW peak signature. 
\FloatBarrier
\FloatBarrier

\subsubsection{Non-canonically lighter QCD axion dark matter}
\label{sec:non_conventional_QCD_axion}
In the previous section, we stressed that the QCD axion DM cannot induce  an observable matter-kination GW peak, except maybe at BBO.
Instead, for a given $f_a$ most of the interesting regions lies at a smaller $m_a$ region.
This motivates QCD axion models with a smaller mass than the one expected from the standard QCD relation in Eq.~\eqref{eq:QCD_axion_definition}.
We consider models where the QCD axion transforms non-linearly under a $\mathbb{Z}_\mathcal{N}$ symmetry \cite{Hook:2018jle, DiLuzio:2021pxd, DiLuzio:2021gos}.

The $\mathbb{Z}_\mathcal{N}$ symmetry suppresses the axion potential and the axion mass, see more details in the next section.
Ref. \cite{DiLuzio:2021pxd} improves the calculation of the axion mass in the large $\mathcal{N}$-limit
\begin{align}
\left( \frac{m_a}{m_a^\mathrm{QCD}} \right)^2 ~ \simeq ~ \frac{1}{\sqrt{\pi}}\sqrt{1-z^2} (1+z) \mathcal{N}^{3/2} z^{\mathcal{N}-1},
\end{align}
where $m_a^\mathrm{QCD}$ is the canonical QCD axion mass, and $z \equiv m_u/m_d \simeq 0.47$ \cite{Tanabashi:2018oca}.
Plugging Eq.~\eqref{eq:QCD_axion_definition}, the lighter QCD axion mass becomes
\begin{align}
m_a ~ \simeq ~ (4.9 ~ \mu \mathrm{eV}) \mathcal{N}^{3/4} z^{(\mathcal{N}-1)/2}\left(\frac{10^{12} ~ \mathrm{GeV}}{f_a}\right). \label{eq:lighter_axion_mass}
\end{align}
Assuming that the lighter QCD axion with mass in Eq.~\eqref{eq:lighter_axion_mass} forms DM, the kination energy scale $E_\mathrm{KD,\mathbb{Z}_\mathcal{N}}$ and duration $N_\mathrm{KD}$ become, cf. Eq.~\eqref{eq:Ekd_axion_fraction},
\begin{align}
E_\mathrm{KD,\mathbb{Z}_\mathcal{N}} ~ \simeq ~ (9.1 \times 10^{6} \mathrm{~ GeV}) G^{3/4}(T_\mathrm{KD}) \mathcal{N}^{3/4} z^{(\mathcal{N}-1)/2} \exp\left(\frac{3 N_\mathrm{KD}}{2}\right).
\end{align}
For values of $\mathcal{N}$ for which the axion mass $m_a$ in Eq.~\eqref{eq:lighter_axion_mass} corresponds to the benchmark points in Fig.~\ref{detect_peak_inf}, see Table~\ref{table:zN_value_benchmark}, the non-canonical QCD axion DM can induce a GW peak from primordial inflation that is observable by the future experiments.

\begin{table}[h!]
\centering
\begin{tabular}{c|ccc}
Observatories & $E_\mathrm{KD} ~ \mathrm{(GeV)}$ & $N_\mathrm{KD}$ & $\mathcal{N}$ \\[0.15em]  \hline\\[-0.85em] 
LISA          &   $3 \times 10^4$  &  4   & 39  \\[0.15em] \cline{1-1}\\[-0.85em] 
BBO           &  $3 \times 10^6$   & 5    &  31 \\[0.15em]  \cline{1-1}\\[-0.85em] 
ET            &   $10^8$  & 6    & 25  \\[0.15em]  \hline
\end{tabular}
\caption{\textit{ \small Benchmark points in Fig.~\ref{detect_peak_inf} and the corresponding $\mathcal{N}$-values of the non-canonical QCD axion DM models.}}
\label{table:zN_value_benchmark}
\end{table}

\subsubsection{Detectability of inflationary GW peak}

\begin{figure}[h!]
\centering
{\bf Gravitational waves from primordial inflation}\\[0.5em]
\raisebox{0cm}{\makebox{\includegraphics[width=0.85\textwidth, scale=1]{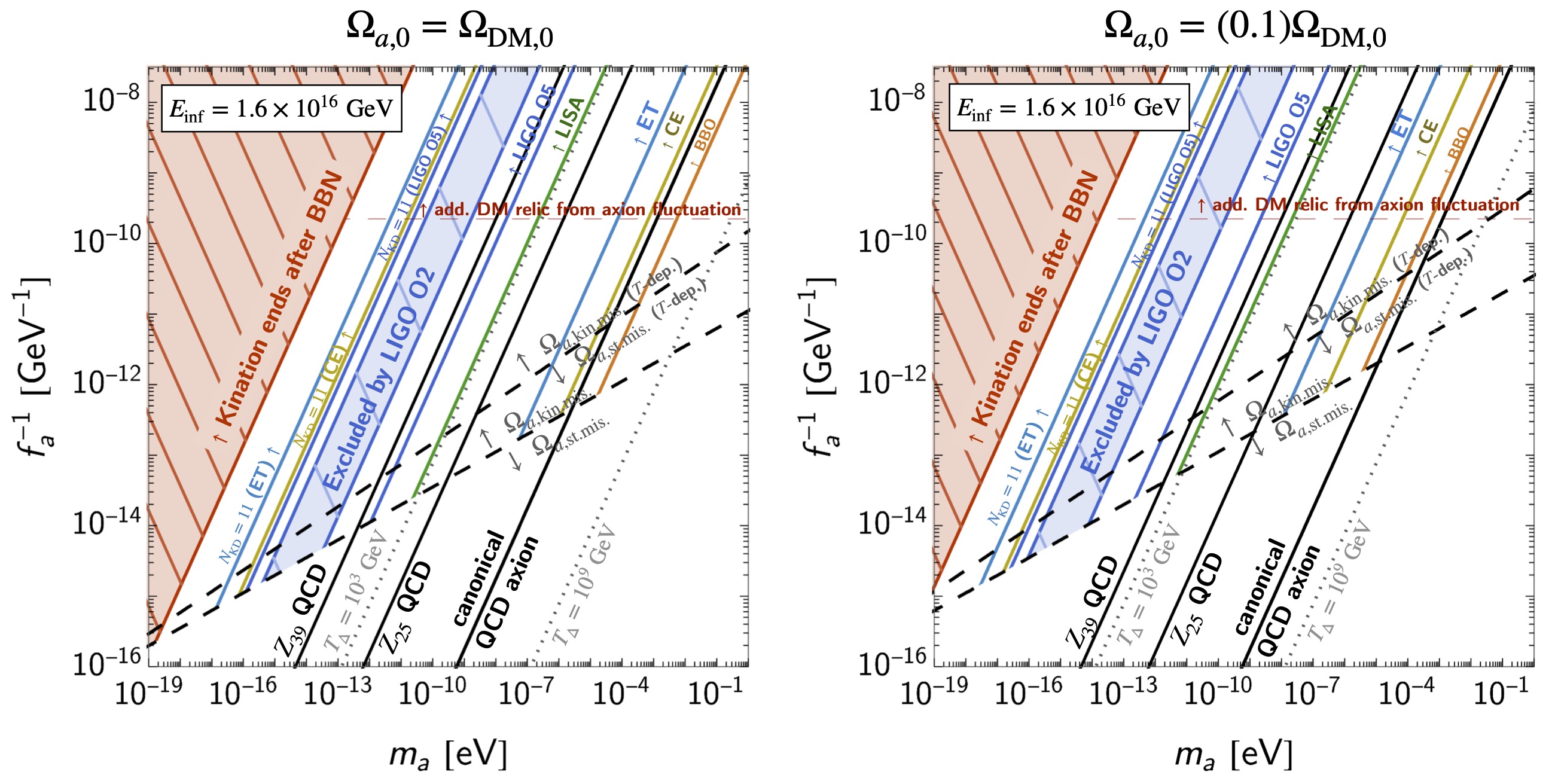}}}\\
\raisebox{0cm}{\makebox{\includegraphics[width=0.85\textwidth, scale=1]{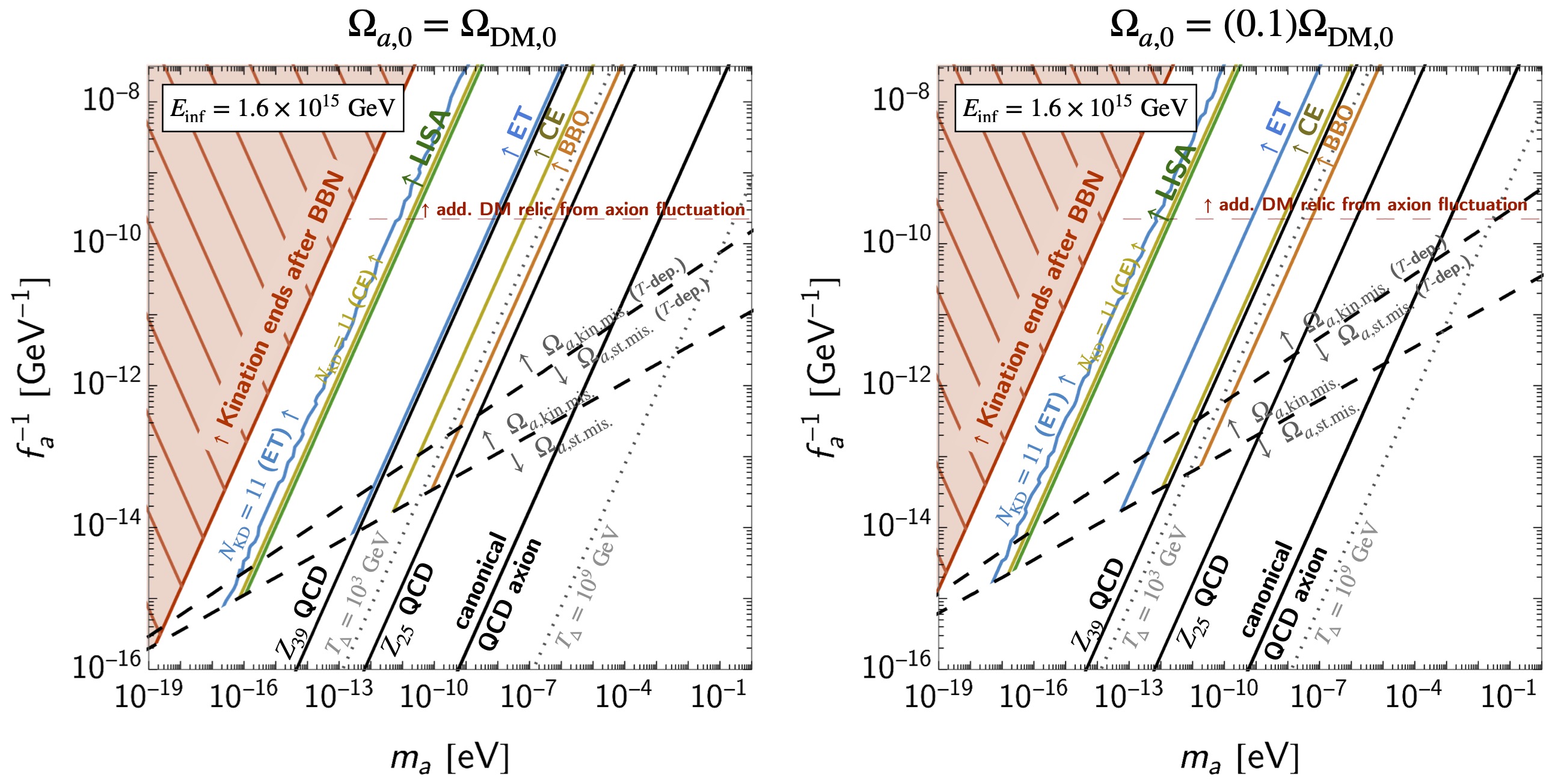}}}
\caption{\textit{ \small Ability of future planned GW experiments to probe axion DM through its matter-kination peak signature in inflationary SGWB, cf. Eq.~\eqref{model_independent_bound_fa_Ma}.  The BBN constraints, Eq.~\eqref{BBN_fa_Ma}, are shown in red-hashed. The black solid lines are different $m_a-f_a$ relations corresponding to either canonical, see Eq.~\eqref{eq:QCD_axion_definition_finite_T}, or non-canonical, see Eq.~\eqref{eq:lighter_axion_mass}, QCD axion. Below the black dashed lines, the axion abundance is set by standard misalignment such that the 1-to-1 connection between the matter-kination parameters $(E_{\rm KD}, \,f_{\rm KD})$ and the axion abundance $\Omega_{a,0}$ is lost, and the prediction of the GW peak signature via the kinetic misalignment, Eqs.~\eqref{eq_peak_position_QCDaxion} and \eqref{eq_peak_position_QCDaxion_freq},  is not applicable. We consider either a QCD axion mass dependence $m_a(T)$ or a constant mass $m_a$. The axion fluctuation mentioned in Sec.~\ref{subsec:afterinflation} allows the longest kination era to be $N_{\rm KD} \simeq 11$ and prevents the detectability bands to continue to smaller $m_a$. Moreover, the axion fluctuation can dominate the relic density from the zero-mode when $f_a \lesssim 10^{9 \div 10} ~ {\rm GeV}$ \cite{DESYfriendpaper2}. We only show $f_a$ values larger than $\sim 10^8$ GeV due to astrophysical constraints \cite{Ayala:2014pea,Li:2020pcn,Foster:2020pgt,Darling:2020uyo}.}}
\label{ma_fa_plot}
\end{figure}

\paragraph{Reach of GW interferometers. }
From Eq.~\eqref{eq:peak_position_ALP2} and Fig.~\ref{fig:peak_spectrum_amplitude}, we see that the GW peak amplitude scales as $\Omega_\mathrm{GW,KD} \propto f_{\rm KD}/m_a f_a$.
For a given observatory with the best sensitivity $\Omega_\mathrm{sens, min}$, there exists an maximal value of $m_a f_a$ below which the peak is observable. It depends on the frequency $f_\mathrm{sens, min}$ at which the signal-to-noise ratio is the largest\footnote{This estimation is valid only when the slope of the sensitivity curve is steeper than the scaling of $\Omega_\mathrm{GW}$. If not, the tip of sensitivity curve does not correspond to the largest $m_a f_a$.}.
Requiring $\Omega_\mathrm{GW,KD} > \Omega_\mathrm{sens,min}$ with $f_\mathrm{KD} = f_\mathrm{sens, min}$ in Eq.~\eqref{eq:peak_position_ALP2},  we deduce the maximal axion mass which leads to a detectable peak signature in the SGWB from primordial inflation
\begin{align}
m_a ~ & \lesssim  ~ (0.65 ~ \mathrm{\mu eV})  \left( \frac{E_\mathrm{inf}}{10^{16} ~ \mathrm{GeV}} \right)^4 \left( \frac{10^{-12}}{\Omega_\mathrm{sens, min}} \right) \left( \frac{f_\mathrm{sens, min}}{1 ~ \mathrm{Hz}} \right)  \left( \frac{10^{9} ~ \mathrm{GeV}}{f_a} \right) \left(\frac{\Omega_{a,0}}{\Omega_{\rm DM,0}} \right).
\label{model_independent_bound_fa_Ma}
\end{align}
We show the reach of future observatories in the $(m_a,~f_a)$ plane in Fig.~\ref{ma_fa_plot}.
For example, ET  ($f_\mathrm{sens, min} \simeq 1 \mathrm{~ Hz}$ and  $\Omega_\mathrm{sens, min} \simeq 10^{-13}$) can probe $m_a^{\rm ET} \lesssim 6.5 ~ \mu{\rm eV}(10^{9}~ {\rm GeV}/f_a)$ for the maximum inflationary scale.
Note that Eq.~\eqref{model_independent_bound_fa_Ma} is parallel to the $m_a(f_a)$ QCD axion mass relation.

\paragraph{BBN bound.} The energy scale $E_\Delta $ when the kination era ends follows from Eq.~\eqref{eq:Ekd_axion_fraction} 
\begin{align}
E_\Delta ~ & = ~ E_\mathrm{KD} \exp(-3 N_\mathrm{KD}/2) ~ =  ~ (1.9 \times 10^{9} ~ \mathrm{GeV}) G^{3/4}(T_\mathrm{KD}) \left( \frac{f_a}{10^9 ~ \mathrm{GeV}} \right) \left( \frac{m_a}{1 ~ \mathrm{eV}} \right) \left( \frac{\Omega_\mathrm{DM,0}}{\Omega_\mathrm{a,0} } \right).
\label{alp_DM_kination_ending}
\end{align}
The successful BBN requires that the universe is radiation dominated at the temperature $T_\mathrm{BBN} \sim 1 ~ \mathrm{MeV}$, which implies 
\begin{align}
E_\Delta \lesssim 1 ~ \mathrm{MeV}  \qquad \implies \qquad m_a ~ & \lesssim  ~ (0.5 \times 10^{-12}~ \mathrm{eV}) G^{-3/4}(T_\mathrm{KD}) \left( \frac{10^9 ~ \mathrm{GeV}}{f_a} \right) \left( \frac{\Omega_{a,0} }{\Omega_\mathrm{DM,0}} \right),
\label{BBN_fa_Ma}
\end{align}
which is shown as the red-hatched region in Fig.~\ref{ma_fa_plot}.

\paragraph{Scalar fluctuation bound.}
The presence of scalar fluctuation of the order $\rho/\delta\rho \sim 10^{9 \div 10}$, cf. Sec.~\ref{subsec:afterinflation}, puts an upper limit on the duration of kination $N_{\rm KD} \sim 11$. Therefore, the observable region cannot violate this bound for a given detector with the frequency $f_{\rm sens,min}$ with its best sensitivity if $\Omega_{\rm GW}(f_{\rm sens,min})$ in Eq.~\eqref{eq:peak_position_ALP2} is smaller than $\Omega_{\rm GW}(N_{\rm KD}^{\rm max} \sim 11)$ in Eq.~\eqref{inflation_peak_amplitude}. Equivalently, the scalar fluctuation bound excludes the observable region for
\begin{align}
m_a \lesssim (1.3 \times 10^{-11} ~ {\rm eV}) G^{-3/4}(T_{\rm KD}) G^{-1/4}(T_{\Delta}) \left(\frac{f_{\rm sens,min}}{\rm Hz}\right) \left(\frac{10^9 ~ {\rm GeV}}{f_a}\right) \left(\frac{e^{11}}{e^{N_{\rm KD}^{\rm max}}}\right)^2 \left( \frac{\Omega_{a,0} }{\Omega_\mathrm{DM,0}} \right),
\end{align}
corresponding to the region on the left of each ($N_{\rm KD} = 11$) line in Fig.~\ref{ma_fa_plot}.

\paragraph{Minimum inflationary scale.}
The amplitude of the GW spectrum from primordial inflation scales as $E_{\rm inf}^4$, see Eq.~\eqref{eq:peak_position_ALP2}. The discovery band of a particular detector,  Eq.~\eqref{model_independent_bound_fa_Ma}, becomes weaker than the BBN bound,  Eq.~\eqref{BBN_fa_Ma}, when the inflationary scale becomes lower than
\begin{align}
E_\mathrm{inf} ~ \lesssim ~ (3 \times 10^{14} ~ \mathrm{GeV} )  \left( \frac{\Omega_\mathrm{sens, min}}{10^{-12}} \right) \left( \frac{1 ~ \mathrm{Hz}}{f_\mathrm{sens, min}} \right).
\end{align}
For instance, ET can no longer probe the SGWB from primordial inflation enhanced by a period  of kination induced by ALP DM if $E_\mathrm{inf} \lesssim 10^{13} ~ \mathrm{GeV}$.

\paragraph{Trapping before kination ends.} After the start of kination, the axion speed $\dot{\theta}\propto a^{-3}$ should not drop below $m_a$.  Otherwise, kination stops earlier than expected and the universe is overclosed by the axion oscillation energy density.
We require that the energy scale at the end of kination must be larger than the scale of the potential barrier, see also App.~\ref{app:NKD_max}
\begin{align}
\rho_\Delta ~ = ~ \rho_{\rm KD}^2 / \rho_{\rm dom} ~ \gtrsim ~ m_a^2 f_a^2.
\label{condition_for_axion_moduli_problem}
\end{align}
Plugging Eq.~\eqref{alp_DM_kination_ending} in Eq.~\eqref{condition_for_axion_moduli_problem},  we obtain a lower bound on the axion mass
\begin{align}
m_a ~ & \lesssim  ~ (2.9  \times 10^{-19}~ \mathrm{eV}) G^{-3/2}(T_\mathrm{KD}) \left( \frac{10^9 ~ \mathrm{GeV}}{f_a} \right) \left( \frac{\Omega_\mathrm{a,0} }{\Omega_\mathrm{DM,0}} \right)^2. \label{eq:lower_bound_axion_mass_barrier}
\end{align}
For ALP DM, this bound is weaker than the BBN bound in Eq.~\eqref{BBN_fa_Ma}. 

\paragraph{Fragmentation before kination ends.}
From \cite{Fonseca:2019ypl} which was recently confirmed by lattice simulations \cite{Morgante:2021bks}, the condition for fragmention of the zero-mode axion is 
\begin{align}
2 H f_a \dot{\theta} ~ &< ~ \frac{\pi \Lambda_b^8}{2 f_a^3  \dot{\theta}^2}W_0^{-1},\\
\dot{\theta} ~ &< ~ \left(\frac{\pi}{4}\right)^{1/3} m_a \left(\frac{m_a}{H}\right)^{1/3} W_0^{-1/3},
\end{align}
where  $W_0$ is the 0th branch of the product log-function.  
We impose that kination does not end  because of  fragmentation  but due to the SM radiation overtaking the total energy density of the universe. In this case,  the spinning speed and the Hubble parameter at the end of kination is 
\begin{align}
\dot{\theta}_\Delta = \frac{E_{\rm KD}^2}{f_a} \exp(-3 N_{\rm KD}), ~ ~ H_{\Delta} = \frac{E_{\rm KD}^2}{\MPl \sqrt{3}}\exp(-3 N_{\rm KD}).
\end{align}
Plugging these expressions back into the fragmentation condition, the kination ends by fragmentation if 
\begin{align}
E_\Delta < \Lambda_b \left(\frac{\sqrt{3} \pi}{4}\right)^{1/8} \left(\frac{\MPl}{f_a}\right)^{1/8} W_0^{-1/8} \simeq \Lambda_b.
\label{eq:fragmentation_kination_ends}
\end{align}
This leads to a similar bound as the trapping condition before kination ends in Eq.~\eqref{eq:lower_bound_axion_mass_barrier}.

\FloatBarrier
\FloatBarrier

\subsection{Can axion DM produce its own GW ?}\label{axionic_string_DM}
\subsubsection{What if Axion DM generates both kination and GW}
An intriguing possibility would be if the $U(1)$-breaking producing the cosmic-string network is the same as the one leading to spinning axion and therefore to matter-kination. 
While it is not clear if this is a viable possibility\footnote{The emission of GW requires the existence of topological defects which imply the presence of inhomogeneities. In contrast, the generation of a matter-kination era assumes an homogeneous condensate. We leave to future work a precise investigation of how the presence of gradient terms in the energy density modifies the EOS of the complex scalar field.\label{footnote:axionDM_produce_own_GW}}, this section considers the special case where  the complex scalar field that sources the matter-kination era also generates the network of global strings, whose string scale matches the axion decay constant $\eta = f_a$.
SGWB produced by CS and their enhancement by a pre-BBN kination era are studied in Sec.~\ref{sec:modelindependent}.
Similar to the GW peak from primordial inflation in Eq.~\eqref{eq:peak_position_ALP2}, the peak position in the SGWB from CS can be related to the axion abundance.

Using Eqs.~\eqref{bump_peak_kination_amp_global_simp} and \eqref{peak_kination_freq_global}, the frequency and amplitude of the GW peak are 
\begin{align}
f_\mathrm{KD} ~ &\simeq ~ (3.96 h^2  ~ \mathrm{kHz}) G^{3/4}(T_\mathrm{KD}) \left(\frac{0.1}{\alpha}\right)   \left(\frac{f_a}{10^{15} ~ \mathrm{GeV} } \right) \left(\frac{m_a}{\mathrm{\mu eV}} \right)\left(\frac{\Omega_{{\rm DM},0}}{\Omega_{a,0}} \right) \exp(2 N_\mathrm{KD}),\\
\Omega_{\rm GW}^{\rm KD} ~ &\simeq ~ (3.03h^{-2} \, \times \, 10^{-22}) G^{-3/4}(T_{\rm KD}) \left(\frac{\alpha}{0.1}\right) \left(\frac{f_{\rm KD}}{{\rm Hz}}\right)   \left(\frac{f_a}{10^{15} ~ {\rm GeV}}\right)^3 \left(\frac{\mathrm{\mu eV}}{m_a} \right)\left(\frac{\Omega_{a,0}}{\Omega_{{\rm DM},0}} \right) \times \nonumber\\
& ~ ~ ~ \log^3\left[ (2.83h^{-1} \times 10^{22}) \left(\frac{\alpha}{0.1}\right)^{1/2} G^{3}(T_{\rm KD}) \left(\frac{{\rm Hz}}{f_{\rm KD}}\right)^{3/2} \left(\frac{f_a}{10^{15} ~ {\rm GeV}}\right)^{1/2}  \left(\frac{\mathrm{\mu eV}}{m_a} \right)^{1/2} \left(\frac{\Omega_{a,0}}{\Omega_{{\rm DM},0}} \right)^{1/2} \right].
\end{align}
Fig.~\ref{ma_fa_plot_axion_string} shows the axion parameter space where the peak signal from axionic strings is observable by future detectors. As expected, the higher the $f_a$ value, the larger the peak amplitude.
\subsubsection{Two-peak signature}
\begin{figure}[h!]
\centering
{\bf Gravitational waves from axionic strings: $\eta = f_a$}\\
\raisebox{0cm}{\makebox{\includegraphics[width=0.85\textwidth, scale=1]{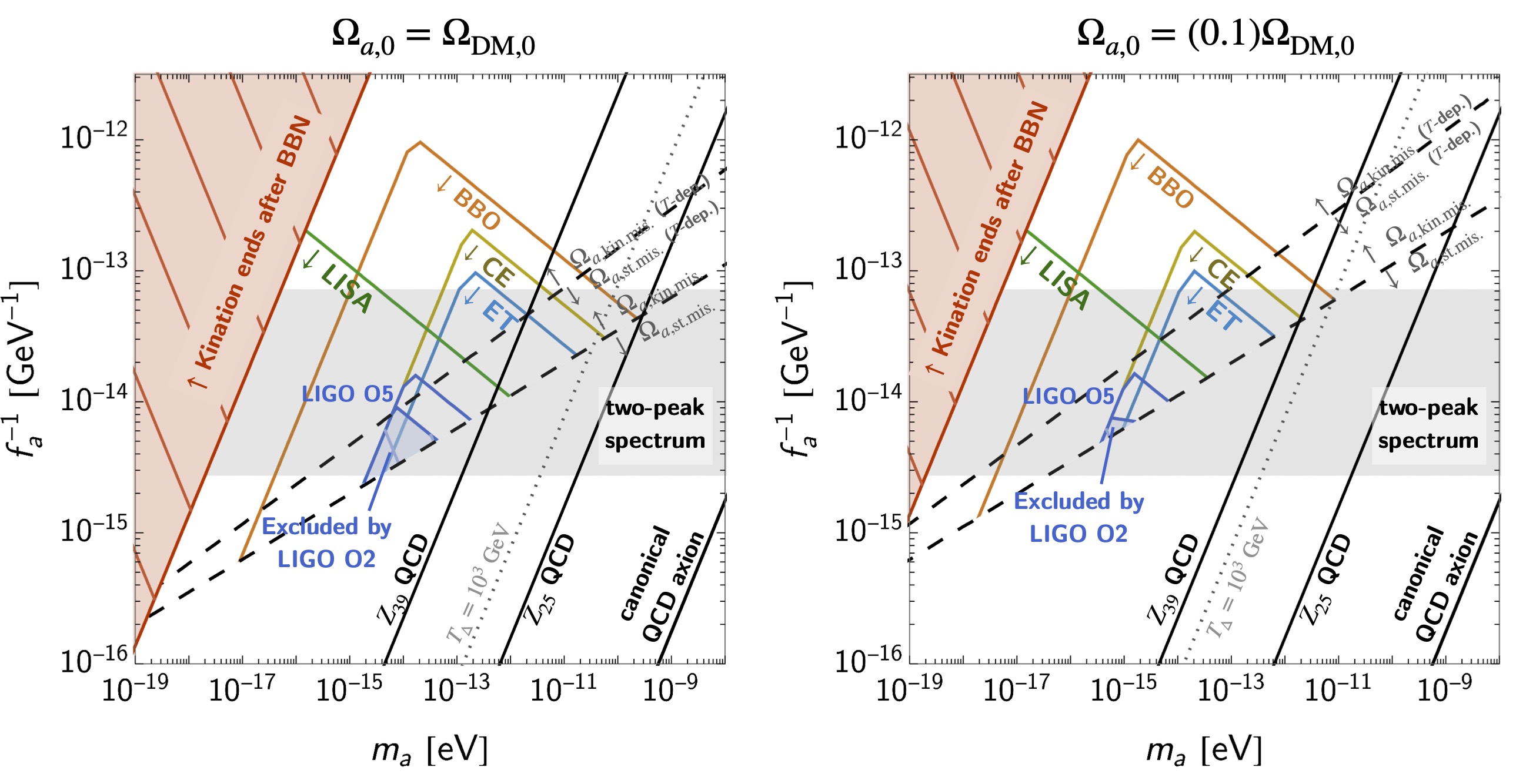}}}
\caption{\textit{ \small We consider the spinning axion-DM field generating both the matter-kination era and the cosmic strings sourcing the GW. The colored solid lines shows the ability of future observatories to probe the resulting matter-kination enhanced GW peak. The upper bounds on the duration of the kination era $N_{\rm KD}\lesssim 11$, discussed in Sec.~\ref{subsec:afterinflation}, prevents the detectability windows going to smaller $m_a$.
 Additionally, the gray band shows the region where the GW peak from the axionic string network could coexist with a second matter-kination peak from primordial inflation, assuming $E_{\rm inf} = 1.6 \times 10^{16} ~ {\rm GeV}$ (and $E_{\rm KD} = 10^{9} ~ {\rm GeV}$ even though the dependence on $E_{\rm KD}$ is only logarithmic, cf. Eq.~\eqref{eq:two_peak_condition1} and Eq.~\eqref{eq:two_peak_condition2}). See Fig.~\ref{ma_fa_plot} for a description of the other lines.}}
\label{ma_fa_plot_axion_string}
\end{figure}

In some regions of the parameter space, future experiments could observe two GW peaks, resulting from the imprint of the kination era, in the SGWB from the axionic string network, cf. Fig.~\ref{ma_fa_plot_axion_string}, and in the SGWB from primordial inflation, cf. Fig.~\ref{ma_fa_plot}, respectively.
In particular, Fig.~\ref{ma_fa_plot_axion_string} shows a region in gray band where the two peaks can be separated, i.e. where the inter-peak distance is more than two orders of magnitude in frequencies, cf. Eq.~\eqref{eq:two_peak_separation_freq}. This occurs when the two conditions~\eqref{eq:two_peak_condition1} and \eqref{eq:two_peak_condition2} are satisfied.

\FloatBarrier
\FloatBarrier

\section{Baryon asymmetry}
\label{sec:baryon_asymmetry}

In the presence of an interaction between the complex scalar field charged under the $U(1)$ and SM fields carrying baryon number, the $U(1)$ charge $Y_\theta$, defined in Eq.~\eqref{eq:U1_charge_generation_AxionDM}, can be transferred into a baryon asymmetry. This is the so-called Affleck-Dine mechanism \cite{Affleck:1984fy}. We can check that the three Sakharov conditions \cite{Sakharov:1967dj} are satisfied. $B$ number is violated by SM sphaleron, $CP$ is violated spontaneously by the rotation of the complex scalar field in one specific direction and the dynamic is far from equilibrium due to the presence of the condensate.
\subsection{Axiogenesis}
The $U(1)$ charge first transfers to the chiral asymmetry of SM quarks through $SU(3)_{\rm c}$ sphaleron transitions, and later this asymmetry is converted into the baryon asymmetry through the $SU(2)_L$ sphalerons \cite{Co:2019wyp, Co:2020xlh}.
As shown in the supplementary material of \cite{Co:2019wyp}, when the scalar field thermalizes with the plasma, most of the $U(1)$ charge remains in the condensates if the scalar field VEV is much larger than the temperature $\phi \gg T$. This implies that only a fraction $T_{\rm ws}^2/f_a^2$ of the $U(1)$ charge is converted into the baryon number
\begin{align}
Y_{B} ~ = ~ \frac{1}{s} \, \left(c_B \frac{T_{\rm ws}^2}{f_a^2} n_{\theta} \right) ~ = ~ 8 \times 10^{-11} \left(\frac{c_B}{0.1}\right) \left(\frac{T_{\rm ws}}{130 ~ {\rm GeV}}\right)^2 \left(\frac{10^{8} ~ {\rm GeV}}{f_a}\right)^2 \left(\frac{Y_\theta}{500}\right), \label{eq:Y_B_axiogenesis}
\end{align}
where $c_B \sim \mathcal{O}(0.1)$ stands for the interactions between baryons and the axion, $T_{\rm ws}$ is the temperature when electroweak spharelon becomes inefficient, which is around $\sim 130 ~ {\rm GeV}$ for SM \cite{DOnofrio:2014rug}.  
The current value for the yield of baryon asymmetry is given by $Y_B = n_{B,0}/s_0 \simeq 10^{-10}$, where the number density of baryon today is $n_B \simeq 2.515 \times 10^{-7} {\rm cm}^{-3}$ \cite{Tanabashi:2018oca}.
We deduce the value of the PQ charge which results in the baryon asymmetry $Y_B$
\begin{align}
Y_\theta ~ = ~ 692\left(\frac{0.1}{c_B}\right) \left(\frac{130 ~ {\rm GeV}}{T_{\rm ws}}\right)^2 \left(\frac{f_a}{10^{8} ~ {\rm GeV}}\right)^2 \left(\frac{Y_B}{10^{-10}}\right). \label{eq:Y_theta_Y_B}
\end{align}
From Eq.~\eqref{Ekd_yield_DM_abundance} and Eq.~\eqref{eq:Y_theta_Y_B}, we can relate the baryon asymmetry with the energy scale of kination
 \begin{align}
E_\mathrm{KD}  ~ = ~ (74~ {\rm TeV}) \, G^{3/4}(T_\mathrm{KD}) \left(\frac{c_B}{0.1}\right) \left(\frac{T_{\rm ws}}{130 ~ {\rm GeV}}\right)^2 \left(\frac{10^{8} ~ {\rm GeV}}{f_a}\right) \left(\frac{10^{-10}}{Y_B}\right) \exp\left(\frac{3 N_\mathrm{KD}}{2}\right).
\label{Ekd_yield_bau}
\end{align}
Using Eq.~\eqref{eq:peak_yield_freq}, \eqref{eq:peak_yield_amp} and \eqref{eq:Y_theta_Y_B}, the  frequency and amplitude of the associated peak in the SGWB from primordial inflation are

\begin{framed}
\vspace{-1em}
\begin{align}
f_\mathrm{KD} ~ &= ~ (0.79  ~ \mathrm{mHz}) [G^{1/4}(T_\Delta) G^{3/4}(T_\mathrm{KD})] \left(\frac{c_B}{0.1}\right) \left(\frac{T_{\rm ws}}{130 ~ {\rm GeV}}\right)^2 \left(\frac{10^{8} ~ {\rm GeV}}{f_a}\right) \left(\frac{10^{-10}}{Y_B}\right)   \exp(2 N_\mathrm{KD}), \label{eq:peak_position_bau_ALP1}\\
\Omega_\mathrm{GW,KD} h^2 ~ &= ~ (1.6 \times 10^{-14}) \left( \frac{G(T_\Delta)}{G(T_\mathrm{KD})} \right)^{3/4} \left( \frac{E_\mathrm{inf}}{10^{16} ~ \mathrm{GeV}} \right)^4 \left( \frac{f_\mathrm{KD}}{1 ~ \mathrm{Hz}} \right)  \left( \frac{f_a}{10^{8} ~ \mathrm{GeV}}\right) \left(\frac{0.1}{c_B}\right) \left(\frac{130 ~ {\rm GeV}}{T_{\rm ws}}\right)^2 \left(\frac{Y_B}{10^{-10}}\right).
\label{eq:peak_position_bau_ALP2}
\end{align}
\vspace{-1em}
\end{framed}\noindent

\begin{figure}[h!]
\centering
{\bf Gravitational waves from primordial inflation}\\
\raisebox{0cm}{\makebox{\includegraphics[width=0.65\textwidth, scale=1]{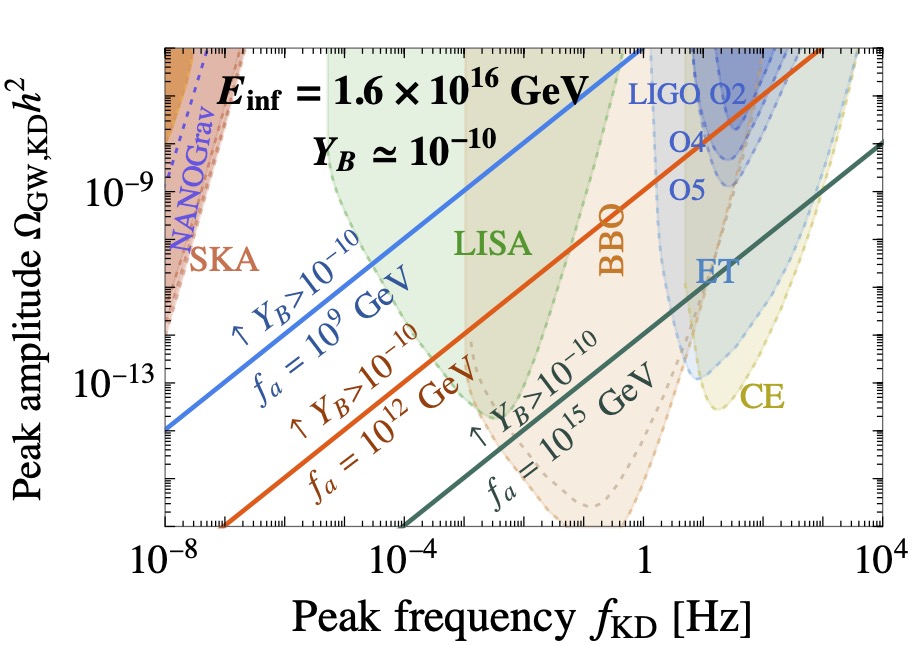}}}
\caption{\textit{ \small 
Reach of future interferometers to probe the GW peak signature in inflationary GW due to the presence of a matter-kination era induced by a spinning axion which also generates the correct baryonic asymmetry. We fixed $c_B = 0.1$ and  $T_{\rm ws} = 130~\rm GeV$ in Eq.~\eqref{eq:Y_theta_Y_B}.  }}
\label{bau_plot_spectrum}
\end{figure}

We show in Fig.~\ref{bau_plot_spectrum} the lines for the peak position of GW spectrum where the baryon asymmetry is explained.

\FloatBarrier
\FloatBarrier

\subsection{Connection to axion relic abundance}
Using Eq. \eqref{eq_axion_fraction} and \eqref{eq:Y_theta_Y_B}, the axion abundance relates to the baryon asymmetry
\begin{align}
\frac{\Omega_{a}}{\Omega_{\rm DM}} =  7006  \left(\frac{0.1}{c_B}\right) \left(\frac{130 ~ {\rm GeV}}{T_{\rm ws}}\right)^2 \left(\frac{m_a}{1 ~ {\rm eV}}\right)  \left(\frac{f_a}{10^{8} ~ {\rm GeV}}\right)^2 \left(\frac{Y_B}{10^{-10}}\right)
\end{align}
For the QCD axion, this translates into
\begin{align}
\frac{\Omega_{a}}{\Omega_{\rm DM}} =  399  \left(\frac{0.1}{c_B}\right) \left(\frac{130 ~ {\rm GeV}}{T_{\rm ws}}\right)^2  \left(\frac{f_a}{10^{8} ~ {\rm GeV}}\right) \left(\frac{Y_B}{10^{-10}}\right),
\end{align}
which clearly shows that the QCD axion overcloses the universe for a correct baryon asymmetry.

\subsection{More efficient charge transfer}
The transfer of the PQ charge to the baryon asymmetry can be made more efficient if the electroweak sphalerons freeze-out at a larger temperature $T_{\rm ws}$ \cite{Baldes:2018nel,Glioti:2018roy, Matsedonskyi:2020mlz,Harigaya:2021txz}, if the weak anomaly of $U(1)_{\rm PQ}$ is larger (larger $c_B$) \cite{Co:2019wyp},  in the presence of the dimension-5 Weinberg operator giving a Majorana mass to the SM neutrinos \cite{Weinberg:1979sa, Domcke:2020kcp,Co:2020jtv}, or in the presence of supersymmetric R-parity violating couplings \cite{Co:2021qgl}. It will be interesting to study the correlation between the presence of the matter-kination peak in SGWB from primordial origin and the successful baryogenesis in the case of the QCD axion.

After the general model-independent discussion of the previous sections, we are now ready to discuss specific model implementations of the matter-kination era induced by a spinning axion and work out  in details the conditions and particle physics parameter space.

\subsection{Axiogenesis and the scalar fluctuation}
The baryon asymmetry is mostly set at the time when the EW sphaleron becomes ineffective at temperature smaller than $T_{\rm ws}$. At this moment, the PQ charge in the axion zero-mode is transferred into the baryon asymmetry
\begin{align}
    Y_\theta ~ = ~ 692\left(\frac{0.1}{c_B}\right) \left(\frac{130 ~ {\rm GeV}}{T_{\rm ws}}\right)^2 \left(\frac{f_a}{10^{8} ~ {\rm GeV}}\right)^2 \left(\frac{Y_B}{10^{-10}}\right). 
\end{align}
\textbf{Assuming the transfer occurs after  the kination-like behavior starts.} Since the baryon asymmetry is induced from the zero mode, one needs the non-linear effect to occur below $T_{\rm ws}$ \cite{DESYfriendpaper2}
\begin{align}
    \frac{T_{\rm ws}}{T_{\rm KD}} = \frac{a_{\rm KD}}{a_{\rm ws}} > \frac{a_{\rm KD}}{a_{\rm nl}} = \xi^{1/2} ~ ~ \Rightarrow ~ ~ T_{\rm KD} < 10^7 ~ {\rm GeV} \left(\frac{T_{\rm ws}}{100 ~ {\rm GeV}}\right) \left(\frac{10^{-10}}{\xi}\right)^{1/2},
\end{align}
where $\xi = \delta\rho/ \rho$ describes the relative size of the scalar fluctuation.
Moreover, the PQ charge at the time of $T_{\rm ws}$ is bounded
\begin{align}
    Y_\theta = \frac{n_{\theta,{\rm ws}}}{s_{\rm ws}} \simeq \frac{f_a^2 m_r}{T_{\rm ws}^3} \left(\frac{a_{\rm KD}}{a_{\rm ws}}\right)^3 \gtrsim \frac{f_a^2 m_r}{T_{\rm ws}^3} \left(\frac{a_{\rm KD}}{a_{\rm nl}}\right)^3 = \frac{f_a^2 m_r}{T_{\rm ws}^3} \xi^{3/2},
\end{align}
where we use that the PQ charge red-shifts $\propto a^{-3}$ from the start of kination.
Applying this bound for the baryogesis prediction,
\begin{align}
    \frac{f_a^2 m_r}{T_{\rm ws}^3} \xi^{3/2} &< 692\left(\frac{0.1}{c_B}\right) \left(\frac{130 ~ {\rm GeV}}{T_{\rm ws}}\right)^2 \left(\frac{f_a}{10^{8} ~ {\rm GeV}}\right)^2 \left(\frac{Y_B}{10^{-10}}\right),\\
    \left(\frac{Y_B}{10^{-10}}\right) &> 8.55 \times 10^{8} \left(\frac{m_r}{T_{\rm ws}}\right) \left(\frac{c_B}{0.1}\right) \xi^{3/2}.
\end{align}
The baryogenesis from spinning axion w.r.t. the fluctuation bound reads
\begin{align}
    m_r < 1.52 \times 10^{8} ~ {\rm GeV} \left(\frac{Y_B}{10^{-10}}\right)  \left(\frac{T_{\rm ws}}{130 ~ {\rm GeV}}\right) \left(\frac{0.1}{c_B}\right) \left(\frac{10^{-10}}{\xi}\right)^{3/2}.
\end{align}
Note that this bound is independent of $f_a$.
It remains to be seen if the fluctuation plays any role in Axiogenesis.

\newpage

\addcontentsline{toc}{part}{Axion model realizations }
\begin{center}
{\LARGE \bf \noindent Axion model realizations }
\end{center}

\section{A first attempt: An axion-only realisation with spontaneous symmetry breaking scale $\phi = f_a$}\label{sec:trapped_mis}
In this section, we consider a class of models in which the axion acquires a large  kinetic energy due to a change in its mass at early times. Initially, the axion has a large mass and oscillates around the minimum of its potential. At some temperature $T_c$, the potential drops and the energy stored in the oscillations is converted into kinetic energy density.  Such dynamics does not involve at all the radial mode of the complex scalar field and we  investigate whether this simple model can lead to a kination era.
For this study, we will use the benchmark model   based on the work of \cite{DiLuzio:2021pxd,DiLuzio:2021gos}.
However, our results could be applied to other models in which the axion acquires a temporary large mass at early times.
The QCD axion potential is well-known to be generated by non-perturbative effects around the QCD scale.
At high temperature, the axion potential is unconstrained and can arise from a variety of PQ breaking effects \cite{Barr:2014vva, Takahashi:2015waa, Caputo:2019wsd, DiLuzio:2021pxd,DiLuzio:2021gos,Heurtier:2021rko}.
Based on the idea of \cite{Hook:2018jle}, the Refs.~\cite{DiLuzio:2021pxd,DiLuzio:2021gos} assume the existence of a discrete $\mathbb{Z}_\mathcal{N}$ shift symmetry for the axion field, where $\mathcal{N}$ is the number of mirror worlds to which the axion interacts with. As soon as the non-perturbative QCD effects from all worlds are effective, the axion mass receives an exponential suppression with respect to the usual prediction \cite{Kim:1979if, Shifman:1979if, Zhitnitsky:1980tq, Dine:1981rt, Kim:1984pt, Choi:1985cb}.
We extend the construction of \cite{DiLuzio:2021pxd,DiLuzio:2021gos} to any axion-like particles (ALP). We consider a toy model in which the ALP potential is a cosine whose barrier size suddenly drops to a much smaller value below a temperature $T_c$. 
The oscillation energy density of the axion in the high-temperature potential gets converted into kinetic energy density. 

\subsection{Trapped-misalignment setup}

\subsubsection{A two-stage axion potential}
We assume the potential of the axion $a \equiv \theta f_a$ to vary suddenly at a critical temperature $T_c$,
 \begin{align}
  V(T, ~\theta)~=~ \begin{cases}
M_{a}^{2} f_{a}^{2} \left(1- \cos  \frac{\theta-\delta}{N_{\rm DW}^{(1)}}\right) ~ ~ ~
      & \mathrm{ for ~} T>T_\mathrm{c} \\
m_{a}^{2} f_{a}^{2} \left(1- \cos  \frac{\theta}{{N_{\rm DW}^{(2)}}}\right) ~ ~ ~
      & \mathrm{ for ~} T<T_\mathrm{c} 
\end{cases},
\label{master_trappedmis}
\end{align}
where $\delta$ is the angular shift between potential minima. $N_{\rm DW}^{(1)}$ and $N_{\rm DW}^{(2)}$ are the domain wall (DW) numbers of the two potentials. We assume a large-scale separation between the high-temperature and low-temperature axion masses $M_a\gg m_a$. In \cite{DiLuzio:2021gos}, such a sudden suppression of the axion mass arises when a symmetry $\mathbb{Z}_\mathcal{N}$ turns on.\footnote{We thank Pablo Quilez for a valuable discussion on the set-up of the  trapped misalignment scenario.}

\subsubsection{Two stages of oscillation} 
Intially, the axion is frozen on the high-$T$ potential with an energy density
\begin{align}
\rho_\mathrm{osc} ~ = ~ \frac{1}{2} M_a^2 f_a^2 \left(\frac{\theta_i - \delta}{N_{\rm DW}^{(1)}}\right)^2,
\end{align}
where we assumed a small initial misaligned angle $\theta_i \lesssim \pi$.
The axion starts to coherently oscillate with a pressure-less equation of state when the expansion rate drops to $3H \sim M_a$, or equivalently, when the energy density of the thermal bath reaches
\begin{align}
\rho_\mathrm{osc}^\mathrm{rad} ~ = ~ \frac{1}{3} \MPl^2 M_a^2.
\end{align}
As the axion energy density redshifts slower than radiation, it dominates the universe and gives rise to a matter-dominated era at
the energy density determined by Eq.~\eqref{eq:rho_M_def}
\begin{align}
\rho_{\rm dom}  ~ = ~ \frac{27}{16} M_a^2 f_a^2 \left(\frac{f_a}{\MPl}\right)^6 \left(\frac{\theta_a -\delta}{N_{\rm DW}^{(1)}}\right)^8.
\end{align}
Below $T < T_c$, the finite-temperature axion potential vanishes and the smaller potential turns on.
The redshifted axion oscillation energy density 
\begin{align}
\rho(T_c) ~ = ~ \rho_\mathrm{osc}  \left(\frac{a_\mathrm{osc}}{a_c}\right)^3 ~ = ~  \rho_\mathrm{osc}  \left(\frac{\frac{\pi^2}{10}g_*(T_c) T_c^4}{M_a^2 \MPl^2}\right)^{3/4},
\end{align}
gets converted in kinetic energy density and the axion gets kicked with the speed
\begin{align}
\frac{f_a^2\dot{\theta}^2_c}{2} ~ = ~\left<\rho(T_c)\right>~=~ \frac{\rho(T_c)}{2} ~ ~ \Rightarrow ~ ~ \dot{\theta}_c~=~ \frac{\sqrt{\rho(T_c)}}{f_a} ~ = ~ \frac{1}{\sqrt{2}} \left[\frac{\pi^2}{10} g_*(T_c)\right]^{3/8} \frac{T_c^{3/2} M_a^{1/4}}{\MPl^{3/4}} \left(\frac{\theta_i - \delta}{N_{\rm DW}^{(1)}}\right), \label{eq:theta_dot_modelA}
\end{align}
where we have averaged $\rho(T_c)$ over many oscillations.
We checked that the dynamics of the radial mode can be safely neglected.

A freely moving scalar $\chi$ has the EOM $\ddot{\chi} + 3H \dot{\chi} =0$ and its solution is $\dot{\chi} \propto a^{-3}$.
Hence, if the axion speed is large enough, $\dot{\theta} \gg m_a$, its energy density redshifts as $a^{-6}$ and if it were to dominate the total energy density, this would trigger a kination era. 

\subsection{Impossibility to generate a kination era}

{The main issue of the scenario above is that the energy density of the sector responsible for the varying axion mass is not sub-dominant compared to the energy density of the axion rotation  after the axion mass is turned off, leading to the impossibility of a kination era\footnote{We thank Raymond Co, Nicolas Fernandez, Akshay Ghalsasi, and Keisuke Harigaya for a discussion on this point.}.
This can be seen in Eq. \ref{eq:theta_dot_modelA} where the kinetic energy of the axion after rotation begins is half of the energy density of the axion oscillation. The other half of the energy goes to the sector responsible for the varying axion mass, which further prevents kination domination. 
The thermal bath of the confining sector always has more energy than the axion oscillations.
To circumvent this problem one has to involve a second scalar field.  The simplest and most natural realisation is to consider the radial mode of the complex scalar field, and this is what will be done in the remaining sections of this paper.
Alternative UV-completions will be presented elsewhere \cite{DESYfriendpaper3}.

\newpage

\section{Interplayed dynamics between radial and angular modes of the complex scalar field with large spontaneous symmetry breaking scale $\phi \gg f_a$}
\label{sec:PQ:exampleII}

A main class of models that can lead to a kination era  is based 
on the work of \cite{Co:2019wyp,Co:2019jts,Co:2020dya,Co:2020jtv}. In this section, we review the evolution of a rotating complex scalar field in an expanding universe, from the time when it has large radial value  $\phi \gg f_a$  during inflation, until the time when it reaches the zero-temperature minimum in $\phi = f_a$.  We present what are the necessary conditions for generating a kination era.

Secs.~\ref{sec:scenario_I_non_thermal_damping}, \ref{sec:complex_field_thermal_potential}, \ref{sec:complex_field_low_reh_temp} present three possible scenarios for the mechanism of damping of the radial mode motion depending on whether  thermal effects play a role or not.
In each case, we determine the duration of the intermediate kination era and for scenario $I$ and $III$, we study the impact on the observability of the SGWB from primordial inflation, local or global cosmic strings.
Other works which rely on a similar set-up appear in \cite{Co:2019wyp,Co:2019jts,Chang:2019tvx,Co:2020dya,Co:2020jtv,Co:2021rhi,Harigaya:2021txz,Co:2021lkc}. 
Earlier works study the dynamics of a complex scalar field leading to a kination era following inflation,  by assuming the initial rotational velocity in the context of self-interacting dark matter \cite{Rindler-Daller:2013zxa,Li:2013nal,Li:2016mmc, Li:2021htg}. The initial rotational velocity can be predicted as a function of model parameters as we will now discuss.

\subsection{Requirements for a kination era}

\paragraph{Complex scalar field.}
We consider a complex scalar field $\Phi$ with a Lagrangian 
\begin{align}
\label{eq:complex_scalar_field_full_lag}
\mathcal{L} ~ = ~ (\partial_\mu \Phi)^\dagger \partial^\mu \Phi - V(\left| \Phi \right|) -V_{\rm th}(\left| \Phi \right|,\, T)  - V_{\cancel{U(1)}}(\Phi) - V_{H}(\Phi),
\end{align}
where $V(\left| \Phi \right|)$ is a global $U(1)$-symmetric potential with spontaneous symmetry breaking (SSB) vacuum, $V_{\rm th}(\left| \Phi \right|,\, T)$ are the finite-temperature corrections, $ V_{\cancel{U(1)}}$ is an explicit $U(1)$-breaking term, and  $ V_{H}$ is a Hubble-dependent term driving the field VEV to large values at early time.
The complex scalar field can be parameterized by two real fields describing radial $\phi$ and angular $\theta$ directions
\begin{align}
\Phi ~ = ~ \phi e^{i \theta},
\label{pfield_parametrize}
\end{align}
where the $U(1)$ symmetry acts as a shift symmetry for  $\theta$.
We consider only the homogeneous part of the field, such that the Lagrangian in the angular representation is 
\begin{framed}
\vspace{-1.5em}
\begin{align}
\mathcal{L} ~ = ~ \frac{1}{2}\dot{\phi}^2 + \frac{1}{2}\phi^2 \dot{\theta}^2  - V(\left| \Phi \right|) - V_{\rm th}(\left| \Phi \right|,\, T) - V_{\cancel{U(1)}}(\Phi)  - V_{H}(\Phi),
\label{lagragian}
\end{align}
\vspace{-1.5em}
\end{framed}\noindent
where the first and second terms denote the kinetic energy in the radial and angular modes, respectively. 

\paragraph{Ingredients for a kination era.}
First, let us chart the big picture and list the special features of the model required for generating a kination-dominated era.

\begin{itemize}
	\item \underline{a $U(1)$-conserving potential  $V(\left| \Phi \right|)$ with spontaneous breaking.}  In our scenario, the kination era occurs when a rotating scalar field, which dominates the energy density of the universe, rotates along the flat direction of its SSB minimum.
	\item \underline{an explicit $U(1)$-breaking potential $V_{\cancel{U(1)}}(\Phi)$.} The rotation of the field condensate is induced by an early kick in the angular direction due to the presence of an explicit breaking potential, similarly to the Affleck-Dine mechanism \cite{Affleck:1984fy}.
		\item \underline{a large initial radial field-value $\phi_\textrm{ini}$.} For the explicit breaking higher-order terms in the potential to play a role on the dynamics of the scalar field, we need a mechanism to drive the scalar field to large value in the early universe.
This is encoded in the term $V_{H}(\Phi)$.
	\item \underline{a mechanism for damping the radial mode.} After the kick, the field condensate undergoes an elliptic motion. A mechanism is necessary to damp the radial mode so that a circular trajectory is reached and the energy density will be  dominated by the kinetic energy of the angular mode when the field settles down to the SSB vacuum, resulting in  a kination era.
\end{itemize}

\subsection{$U(1)$-conserving potential with spontaneous breaking}
\label{sec:nearly-quadratic_potential}
\subsubsection{Zero-temperature potential}
In App.~\ref{app:field_traj_kick} and \ref{paragraph:quartic_potential}, we show that for the scalar field energy density to redshift slower than radiation and to dominate the energy density of the universe, we need to consider a potential shallower than quartic. Therefore, we consider a nearly-quadratic potential with a flat direction at the minimum
\begin{framed}
\vspace{-0.5em}
\begin{equation}
V(\left|\Phi \right|) ~=~ m_r^2 |\Phi|^2 \left(\ln \frac{|\Phi|^2}{f_a^2} -1 \right) + m_r^2 f_a^2~+~ \frac{\lambda^2}{M_{\rm pl}^{2l-6}} |\Phi|^{2l-2} ,\label{study_potential1}
\end{equation}
\vspace{-1.5em}
\end{framed}\noindent
where $f_a$ is the radial field value at the minimum. We can define an effective mass which is field dependent
\begin{equation}
 m_{r, \rm eff}^2 \equiv \frac{d^2 V}{d |\Phi|^2} = 4m_r^2 \left(1 + \ln \frac{|\Phi|}{f_a} \right). \label{eq:mr_eff}
\end{equation}
In App.~\ref{app:neglecting_H}, we show that the quadratic potential in Eq.~\eqref{study_potential1} can be generated in gravity-mediated SUSY-broken theories, with $m_r$ being equal to the gravitino mass 
\begin{equation}
m_r \simeq m_{32}.
\end{equation}
In the limit $\phi \lesssim M_{\rm pl}$ which we consider, we can reasonably neglect the quartic term, see App.~\ref{app:neglecting_H}. The logarithmic function is generated by the running of the soft mass \cite{Moxhay:1984am}, see App.~\ref{susy_potential} for a review.
The origin of the last term of Eq.~\eqref{study_potential1} is discussed in the next section, Sec.~\ref{sec:U1breaking}.

Before reaching $\phi \to f_a$, the equation of state of the scalar field in the potential in Eq.~\eqref{study_potential1} is matter-like, see App.~\ref{app:field_traj_kick} and \ref{paragraph:quartic_potential}, such that the scalar field energy density unavoidably overtakes the radiation energy density after some time.

\subsubsection{Finite-temperature corrections}
\label{sec:finite_T_corrections}
The interactions between the complex scalar and other fields at equilibrium with the thermal plasma can give rise to thermal corrections to the potential.
For definiteness, we assume that the complex scalar field $\phi$ is coupled to fermions $\psi$ charged under some (hidden or SM) gauge sector $A_\mu$
\begin{equation}
\mathcal{L} \supset y_\psi\phi \psi_{L}^{\dagger} \psi_{R}+h.c.+ g \, \bar{\psi} \gamma^\mu\psi\, A_\mu. \label{eq:KSVZ_lagrangian_2}
\end{equation}
Depending on whether the fermions of mass $y_\psi \phi$ are Boltzmann-suppressed or not, the thermal corrections read  \cite{Mukaida:2012qn,Mukaida:2012bz} 
\begin{align}
 V_{\rm th}(\phi,\, T) ~ = ~ \begin{cases}
\frac{1}{2} y_\psi^2 T^2 \phi^2,   & \mathrm{for} ~   y_\psi \phi \lesssim T, \\
\alpha^2 T^4 \ln(\frac{y_\psi^2 \phi^2}{T^2}) ,\qquad \qquad & \mathrm{for} ~   y_\psi \phi \gtrsim T, 
\end{cases}
\label{eq:thermal_correction_potential}
\end{align}
with $\alpha \equiv g^2/4\pi$ the gauge coupling constant.
\begin{enumerate}
\item  \textbf{Thermal mass} $y_\psi \phi < T$: fermions are relativistic and abundant in the thermal plasma.
\item  \textbf{Thermal-log} $y_\psi \phi \gg T$: the fermions abundance is Boltzmann suppressed. In that case the thermal corrections are obtained from the running of the gauge coupling constant $g$, after integrating out the heavy fermions \cite{Shifman:1979eb,Anisimov:2000wx,Bodeker:2006ij,Laine:2010cq,Carena:2012xa}.  
\end{enumerate}
A sketch of the thermal corrections to the zero-temperature potential is shown in Fig.~\ref{fig:finite_T_correction_sketch}.

\FloatBarrier
\begin{figure}[h!]
\centering
\raisebox{-0.5\height}{\makebox{\includegraphics[width=0.75\textwidth, scale=1]{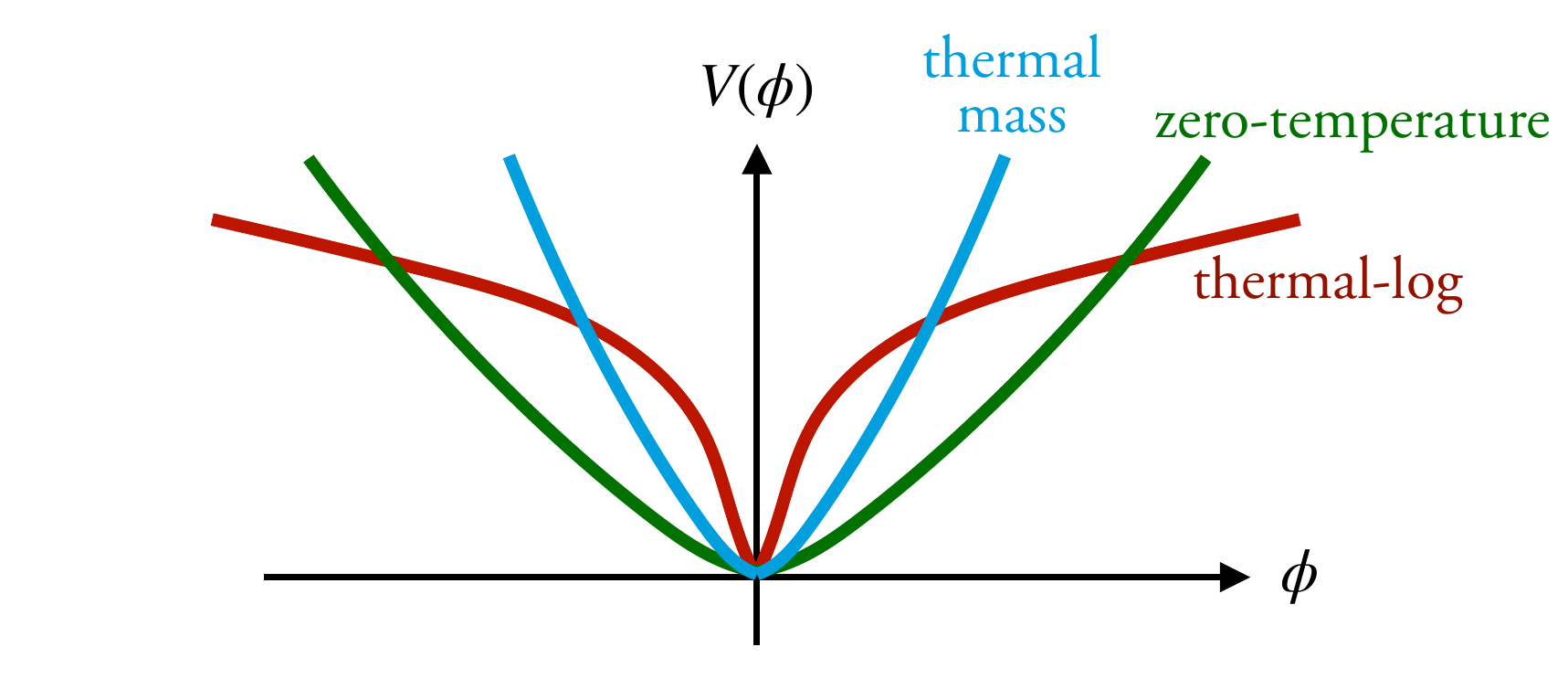}}}
\caption{\textit{ \small Sketch of the zero- and finite-temperature potentials.}}
\label{fig:finite_T_correction_sketch}
\end{figure}
\FloatBarrier

\subsubsection{$Q$-ball formation}
\label{sec:Q-balls}
Scalar fields moving in a potential $V(\phi)$ where $V(\phi)/\phi^2$ has a non-trivial minimum, have stable non-topological localized field configurations known has Q-balls \cite{Coleman:1985ki}. 
This is the case for instance of theories with negative radiative corrections to the mass term \cite{Kasuya:2000wx}
\begin{align}
m^2 ~ = ~ m_0^2(1+ \beta \ln (\phi)),\qquad \textrm{with}\quad \beta < 0.
\end{align}
The potential being flatter than quadratic, condensates with large rotational charge have a centrifugal force larger than the potential gradient and fragments into higher modes \cite{Coleman:1985ki}.
Thermal-log potentials, which are flatter than quadratic are also expected to form Q-balls  \cite{Kasuya:2001hg, Kasuya:2010vq, Mukaida:2012qn}. In order to preserve the condensate, the thermal-potential  in the second line of Eq.~\eqref{eq:thermal_correction_potential} must be sub-dominant meaning that one of the two following conditions must be satisfied
\begin{equation}
\label{eq:thermal_log_dominates}
\begin{cases}
y_\psi\, \phi~ \lesssim~ T , \qquad \\
2\alpha T^2 \ln^{1/2}\left(\frac{y_\psi \phi}{T}\right) ~ \lesssim ~\mreff \,\phi ,
\end{cases}
\quad \implies \quad
\begin{cases}
y_\psi ~\lesssim~ 3.4 \times 10^{-6} \left( \frac{T}{8.3 \times 10^{12}}\right) \left( \frac{M_{\rm pl}}{\phi}\right) , \\
T ~\lesssim ~8.3 \times 10^{12}\left(\frac{\mreff}{10^8~\rm GeV}\right)^{1/2}\left(\frac{\phi}{M_{\rm pl}}\right)^{1/2}\left(\frac{0.1}{\alpha}\right)^{1/2}\left(\frac{5}{\ln^{1/2}(\frac{y_\psi \phi}{T}) }\right)^{1/2}.
\end{cases}
\end{equation}
Q-ball formation is avoided either by choosing small Yukawa coupling $y_\psi$, cf. scenario I in Sec.~\ref{sec:scenario_I_non_thermal_damping}, or a small maximal plasma temperature $T_{\rm max}$, cf. scenario III in Sec.~\ref{sec:complex_field_low_reh_temp}, and in particular outside the red region in Fig.~\ref{fig:fa_MrOfa_TmaxOTreh}-top-left. In contrast, in scenario II presented in Sec.~\ref{sec:complex_field_thermal_potential} in which nothing is done to control neither Yukawa coupling nor the maximal plasma temperature, both conditions in Eq.~\eqref{eq:thermal_log_dominates} could be violated and Q-balls could possibly form \cite{Kasuya:2001hg, Kasuya:2010vq, Mukaida:2012qn}. We don't investigate this possibility further since scenario II does not lead to any kination era.

\subsection{Explicit $U(1)$-breaking potential}
\label{sec:U1breaking}

\subsubsection{Origin}
The general form of the explicit breaking term can be written as
\begin{align}
V_{\cancel{U(1)}}(\Phi) ~ = ~ \Lambda_b^4 \sum_l \sum_{k \neq l} \left[ \left(\frac{\Phi^\dagger}{M_{\rm pl}}\right)^l \left(\frac{\Phi}{M_{\rm pl}}\right)^k ~ + ~ \rm{h.c.}\right] ,
\label{breaking_potential_basic}
\end{align}
where $\Lambda_b$ is a mass scale, $M_{\rm pl}$ is the cut-off of the theory which we set equal to the Planck mass, and $l\neq k$ ensures the explicit breaking of the $U(1)$ symmetry.
In this work, we will assume a simpler form where $k = 0$ and only one value of $l$. As discussed in App.~\ref{app:scalar_pot_SUSY}, the origin of Eq.~\eqref{breaking_potential_basic} can be attributed to the interaction, in a SUSY theory, between soft-breaking terms and the higher-dimensional superpotential
\begin{equation}
\mathcal{L} \supset \int d^2\theta \,W(S_\Phi) ~+~h.c., \qquad \text{with}\quad W(S_\Phi) = \frac{\lambda}{l\,M_{\rm pl}^{l-3}} S_\Phi^l, \label{eq:superpotential_main_text}
\end{equation}
where $S_\Phi$ is the chiral superfield containing $\Phi$ and $\lambda = \mathcal{O}(1)$. The superpotential $W(S_\Phi) $ in Eq.~\eqref{eq:superpotential_main_text} also generates a positive $\lambda^2 \phi^{2l-2}$ term which insures stability at large field value and which we have already included in Eq.~\eqref{study_potential1}. In App.~\ref{app:neglecting_H}, we obtain
\begin{framed}
\vspace{-1.5em}
\begin{align}
V_{\cancel{U(1)}}(\phi,\theta) ~ = ~ \Lambda_b^4 \left[ \left(\frac{\Phi^\dagger}{M_{\rm pl}}\right)^l ~ + ~ \left(\frac{\Phi}{M_{\rm pl}}\right)^l \right] ,
\label{breaking_potential_thiswork}
\end{align}
\vspace{-1.5em}
\end{framed}\noindent
with
\begin{equation}
\Lambda_b^4 = \lambda\,m_{32} M_{\rm pl}^3. \label{eq:lambda_b_def_main_text}
\end{equation}
The integer $l$ corresponds both to the field order and to the number of wiggles along the angular direction.

\subsubsection{Equations of motion}
The evolution of the homogeneous field configuration is controlled by the Klein-Gordon equation in an expanding universe 
\begin{align}
\ddot{\Phi} + 3 H \dot{\Phi} + \frac{\partial}{\partial \Phi^\dagger}\left(V + V_{\cancel{U(1)}}\right) ~ &= ~ 0, \label{eq:KG_exp_U}
\end{align}
where $H$ is the Hubble rate.
Plugging Eq.~\eqref{breaking_potential_thiswork} in Eq.~\eqref{eq:KG_exp_U} and decomposing into radial and angular parts, we obtain the system of coupled equations of motion (EOM)
\begin{framed}
\vspace{-1.5em}
\begin{align}
\ddot{\phi} + 3 H \dot{\phi} + \frac{\partial V}{\partial \phi} + 2 l \cos\left[l\theta\right] \Lambda_b^3 \left(\frac{\Lambda_b}{M_{\rm pl}}\right) \left(\frac{\phi}{M_{\rm pl}}\right)^{l-1} ~ &= ~ \phi \dot{\theta}^2,\label{radial_break}\\
\phi \ddot{\theta} + 3 H \phi \dot{\theta} - 2 l\sin\left[l\theta\right] \Lambda_b^3 \left(\frac{\Lambda_b}{M_{\rm pl}}\right) \left(\frac{\phi}{M_{\rm pl}}\right)^{l-1} ~ &= ~ -2  \dot{\phi} \dot{\theta},
\label{angular_break}
\end{align}
\vspace{-1.5em}
\end{framed}\noindent
which is simply a Keplerian motion in a rotationally-invariant potential $V$, in the presence of small wiggles $V_{\cancel{U(1)}}$ and Hubble friction.
The equation of state of the universe is controlled by the Friedmann equation 
\begin{align}
H^2 ~ = ~ \frac{1}{3 \MPl^2} \left(\rho_\textrm{rad} + \dot{\phi}^2 + \phi^2 \dot{\theta}^2 + V + V_{\rm th} + V_{\cancel{U(1)}} + V_H\right), \label{eq:friedmann_eq}
\end{align}
where $\rho_\textrm{rad}$ is the radiation background energy density.
Note that the scalar field has three components in its energy density: radial and angular kinetic energies and potential energy.

\paragraph{$U(1)$-charge conservation.}
For $l>2$, the $U(1)$-breaking to $U(1)$-conserving ratio decreases at smaller field value $\phi \ll \phi_{\rm ini}$, such that the $U(1)$ symmetry is restored at later time, cf. Fig.~\ref{ratio_potential} in App.~\ref{app:scalar_pot_SUSY}. Therefore, after a few Hubble times of evolution during which $\phi$ has redshifted below $\phi \ll M_{\rm pl}$, the angular EOM, Eq.~\eqref{angular_break}, takes the form of a charge conservation equation
\begin{framed}
\vspace{-1.5em}
\begin{align}
 \frac{d}{dt}\left(a^3 n_\theta\right) ~=~ 0, \qquad \text{with} \quad n_\theta \equiv \phi^2 \dot{\theta},
\label{PQ_charge_conservation}
\end{align}
\vspace{-1.5em}
\end{framed}\noindent
where $n_\theta$ is the comoving Noether charge of the restored $U(1)$ symmetry.

\paragraph{Generation of the $U(1)$ charge.}
The angular EOM in Eq.~\eqref{angular_break}, can be written in the form of a Boltzmann equation for the $U(1)$ charge $n_\theta$
\begin{align}
\dot{n_\theta} + 3 H n_\theta ~ = ~ - \frac{\partial V_{\cancel{U(1)}}}{\partial \theta},\qquad \text{with} \quad n_\theta \equiv \phi^2 \dot{\theta}.
\label{charge_conservation_full}
\end{align}
In App.~\ref{app:kick_angular_direction}, we show that this implies that the field receives an angular kick at the onset of the radial mode oscillation, $t_{\rm osc} \sim \mreff^{-1}(\phi_{\rm ini})$ cf. Eq.~\eqref{eq:H_osc_def}, which for $V_{\cancel{U(1)}}(\theta)$ in Eq.~\eqref{breaking_potential_thiswork} and $l\geq 4$, is given by
\begin{framed}
\vspace{-1.5em}
\begin{align}
n_{\theta} ~= ~ \phi_\textrm{ini}^2 \dot{\theta}_\textrm{ini} \left( \frac{a_{\rm ini}}{a}\right)^3~ = ~ \left(\frac{12l}{6+q}\right) \Lambda_b^4 \left(\frac{\phi_\textrm{ini}}{M_{\rm pl}}\right)^l  \frac{\sin(l \theta_\textrm{ini})}{\mreff(\phi_{\rm ini})} \left( \frac{a_{\rm ini}}{a}\right)^3,
\label{epsilon_oscillation}
\end{align}
\vspace{-1.5em}
\end{framed}\noindent
where $q$ is related to the equation of state of the universe $H^2 \propto a^{-q}$ and where $m_{r, \rm eff}$ is defined in Eq.~\eqref{eq:mr_eff}.

\subsection{Motivation for operators of high dimension}

\paragraph{Quality problem of the Peccei-Quinn solution.}
If the global $U(1)$ symmetry is anomalous in a background of $SU(3)_c$ gluons, then a second $U(1)$-breaking potential is generated by QCD instantons around $T\simeq 100~$MeV \cite{Coleman:1985rnk,DiVecchia:1980yfw,GrillidiCortona:2015jxo}. In that case, the angular mode $\theta$ offers a solution to the strong CP problem known as the Peccei-Quinn QCD axion \cite{Peccei:1977ur, Peccei:1977hh,Wilczek:1977pj,Weinberg:1977ma}.
The non-detection of the electric dipole moment of the neutron (nEDM) implies the upper bound \cite{Baker:2006ts, Pendlebury:2015lrz,nEDM:2020crw}
\begin{equation}
\label{eq:EDM_bound}
\bar{\theta}_0 ~\lesssim ~10^{-10},
\end{equation}
where $\bar{\theta}_0$ is the angle value today with respect to one of the -- $CP$-preserving \cite{Vafa:1983tf} -- minimum of the QCD instanton potential. The presence of the higher dimentional $U(1)$-breaking potential in Eq.~\eqref{breaking_potential_thiswork} is expected to shift the $CP$-preserving minimum by\footnote{Here we calculate the shift in the axion's potential minimum and obtain a similar condition as provided in Ref.~\cite{Co:2020jtv} where the authors consider the mass contribution from the explicit breaking term.}
\begin{equation}
\Delta \theta~\simeq ~ l\frac{\Lambda_b^4}{\chi_0} \left(\frac{f_a}{M_{\rm pl}} \right)^l,
\end{equation}
where $\chi_0 \equiv m_a^2 f_a^2  \simeq (75.6 ~ \rm MeV)^{4}$ is the susceptibility of the topological charge at zero temperature for the canonical QCD axion \cite{Borsanyi:2016ksw}. Using Eq.~\eqref{eq:lambda_b_def_main_text} with $m_{32}\simeq m_r$, the nEDM bound in Eq.~\eqref{eq:EDM_bound} translates to
\begin{equation}
l\, \lambda\frac{m_r M_{\rm pl}^3}{\chi_0} \left(\frac{f_a}{M_{\rm pl}} \right)^l~\lesssim ~ 10^{-10}.
\label{eq:quality_problem_condition}
\end{equation}
For $m_r/f_a \simeq 10^{-2}$, $\lambda = 10^{-4}$ and $f_a \simeq (10^{8}~\rm GeV,\,10^{12}~\rm GeV,\,10^{16}~\rm GeV)$, the neutron EDM bound implies the following lower bounds on the order of the high-dimensional terms $l \geq (7,\, 12,\,34)$.

\paragraph{Long kination requires large $l$.}
Even though it is not restricted to the Peccei-Quinn QCD axion, our study also motivates large values of $l$ in order to maximize the amount of rotation $\epsilon$, defined in Eq.~\eqref{eq:epsilon_def}, resulting from the angular kick, see Eq.~\eqref{eq:epsilon_1_zero_temp}, and to have initial radial value $\phi_{\rm ini}$ as large as possible, see Eq.~\eqref{susy_phi_ini}. As we show in Eq.~\eqref{eq:longest_NKS_scenario_1_sketch} and Eq.~\eqref{eq:longest_NKS_scenario_3_sketch}, the duration of the kination era depends crucially on those quantities.
The impact of the value of $l$ on the detectability of the GW peak signature can be seen in Figs.~\ref{fig:complex_scenario1_inflation1}, \ref{fig:complex_scenario1_inflation2}, \ref{fig:complex_scenario1_inflation3}, \ref{fig:complex_scenario1_local1}. \ref{fig:complex_scenario1_local2}, \ref{fig:complex_scenario1_global1}, \ref{fig:complex_scenario1_global2} of Sec.~\ref{sec:scenario_I_non_thermal_damping} or  in Fig.~\ref{fig:complex_scenario3_inf1} of Sec.~\ref{sec:complex_field_low_reh_temp}.

\subsection{Large initial vacuum expectation field-value}
\label{sec:intitial_vev}
\subsubsection{Supersymmetric theory}
\label{sec:intitial_vev_SUSY}

During the early universe, the complex scalar field $\Phi$ can obtain a Hubble-induced negative mass and Hubble-induced higher dimensional terms \cite{Dine:1995uk,Dine:1995kz}
\begin{align}
V_H(\Phi) ~ \simeq ~ - c\,H^2 |\Phi|^2 ~ +~ a \frac{H }{m_{32}} \Lambda_b^4\left[\left( \frac{\Phi}{M_{\rm pl}}\right)^l +h.c.\right], \qquad  a,\,c = \mathcal{O}(1).
\label{eq:negative_hubble_mass}
\end{align}
As shown in App.~\ref{app:including_Hubble_curv}, these naturally arise in SUSY theories in the presence of a gravity-mediated interaction in the Kahler potential
\begin{equation}
\mathcal{L} = \int d\theta^2 d\bar{\theta}^2 \left( a \frac{S_I+S_I^*}{M_{\rm pl}}|S_\phi|^2 +c \frac{|S_I|^2}{M_{\rm pl}^2}|S_\Phi|^2 \right), \label{eq:gravi_int_X_Phi_main_text}
\end{equation}
where $S_\Phi$ is the chiral superfield containing $\Phi$ and $S_I$ is the chiral superfield whose $F$- or kinetic-term dominates the energy density of the universe. 
At early time, the radial VEV of the scalar field is governed by the $U(1)$-conserving, $U(1)$-breaking and Hubble-induced terms in Eq.~\eqref{study_potential1}, \eqref{breaking_potential_thiswork} and \eqref{eq:negative_hubble_mass}
\begin{align}
V(\Phi) ~ = ~ (m_{r,\rm eff}^2 - c H^2)|\Phi|^2 + \Lambda_b^4 \left(1 + a\frac{H }{m_r}\right) \left[\left( \frac{\Phi}{M_{\rm pl}}\right)^l +h.c.\right]+ \frac{\lambda^2}{M_{\rm pl}^{2n-6}} |\Phi|^{2l-2}.
\end{align}
With $c\sim \mathcal{O}(1)$ and $H \gtrsim  m_{r,\rm eff}$, the Hubble friction $3H$ and the Hubble-induced mass term $cH$ in the EOM have comparable size, such that the scalar field is nearly critically-damped \cite{Randall:1994fr}, and rolls exponentially fast (actually in $3/c$ e-folds of inflation) towards a non-trivial minimum at large field value, which for $a \lesssim c$, reads, cf. App.~\ref{app:phi_ini}
\begin{align}
\phi_\textrm{ini}~ = ~ M_{\rm pl}\left( \sqrt{c}\frac{\mreff(\phi_{\rm ini})}{\lambda \sqrt{2l-2}M_{\rm pl}} \right)^{\frac{1}{l-2}}.
\label{susy_phi_ini}
\end{align}
When the Hubble scale crosses its mass
\begin{equation}
H_{\rm osc} \simeq  m_{r,\rm eff}/3, \label{eq:H_osc_def}
\end{equation}
the field starts oscillating (under-damped motion) with an initial amplitude $\phi_\textrm{ini}$. 
An oscillation in the angular direction with initial amplitude $\theta_{\rm ini} \sim \mathcal{O}(1)$ is generated by the same dynamics thanks to the Hubble-dependent $U(1)$-breaking terms in Eq.~\eqref{eq:negative_hubble_mass}. We refer to App.~\ref{app:phi_ini} for more details on the evolution of $\Phi$ at early time.

\subsubsection{Random fluctuations during inflation}
\label{sec:stochastic_inflation}
Without any Hubble-induced mass term from the supersymmetry,
the field can be driven at large field value by the de Sitter (dS) fluctuations during inflation, namely the sub-horizon quantum modes in the dS background.
During each Hubble time $H_{\rm inf}^{-1}$, the dynamics of a real scalar field during inflation can be described as a superposition of quantum fluctuations $\delta \phi$ and classical motion $\Delta \phi$
\begin{equation}
\delta \phi  \simeq \frac{H_{\rm inf}}{2\pi}, \qquad
\Delta \phi  \simeq \frac{V'}{3H_{\rm inf}^2},
\end{equation}
where $H_{\rm inf}$ is the Hubble scale during inflation.
Out of this interplay between random walk and restoring force, the scalar field probability distribution, is a solution of the associated Fokker-Planck equation, and for a $U(1)$-preserving potential, spreads as  \cite{Linde:1984je, Starobinsky:1994bd}
\begin{equation}
\left< \phi^2  \right> = \frac{3H_{\rm inf}^4}{8\pi^2 m_{\rm \phi}^2} \left[ 1 - \exp \left[ -\frac{2 m_{\rm \phi}^2}{3H_{\rm inf}}(t-t_0)  \right] \right] \quad  \xrightarrow[]{N \gtrsim (H_{\rm inf}/m_{\rm \phi})^2}  \quad \frac{3H_{\rm inf}^4}{8\pi^2 m_{\phi}^2}, \label{eq:Bunch_Davies}
\end{equation}
where $m_{\rm \phi}$ is the mass of the real scalar field. The arrow shows when the so-called Bunch-Davies equilibrium distribution is reached.\footnote{Basically, we can interpret the quantum fluctuations as due to the temperature of the dS space-time $H_{\rm inf}/2\pi$ such that the scalar field gets a thermal distribution satisfying 
$
V(\phi) \simeq H_{\rm inf}^4.
$}
The correlation length $l$ of dS fluctuations \cite{Linde:1984je}
\begin{equation}
l \sim H^{-1} \exp(3H_{\rm inf}^2/2m_\phi^2),
\end{equation}
can be much larger than the Hubble horizon for relatively flat potential $m_\phi \ll H_{\rm inf}$, and therefore can give rise to an homogeneous condensate at later time. 
The case of a complex scalar field is treated in \cite{Wu:2020pej} where it is shown that the averaged radial value is equal to Eq.~\eqref{eq:Bunch_Davies}, up  to a factor 2
\begin{align}
\left<\phi_\textrm{ini}^2\right> ~ \simeq ~ \frac{3 H_{\rm inf}^4}{4 \pi^2 m_{r,\textrm{eff}}^2},
\label{ds_fluc}
\end{align}
where $m_{r,\textrm{eff}}$ is defined in Eq.~\eqref{eq:mr_eff}.
In the presence of an explicit $U(1)$-breaking, the averaged angle acquires a shift of order $\mathcal{O}(\epsilon)$, defined in Eq.~\eqref{eq:epsilon_def}, with respect to the values of $\theta$ along the valleys \cite{Wu:2020pej}.\footnote{The authors of \cite{Wu:2020pej} assume an explicit breaking of $U(1)$ of the form
\begin{align}
V_{\cancel{U(1)}}(\Phi) ~ = ~ \lambda \left(\Phi^3 \Phi^\dagger + \textrm{ h.c.}\right),\nonumber
\end{align}
where $\lambda$ can be either positive or negative.
As a result, they find that the averaged angular amplitude is 
\begin{align}
\left<\tan \theta_\textrm{ini}\right> ~ = ~ \frac{\pm \sqrt{1 - \frac{9 \lambda H^4}{8 \pi^2 m_{r,\textrm{eff}}^4}}}{\pm \sqrt{1 + \frac{9 \lambda H^4}{8 \pi^2 m_{r,\textrm{eff}}^4}}} ~ ~ \Rightarrow ~ ~ \left<\theta_\textrm{ini} \right>  ~ \simeq ~ (2n + 1) \frac{\pi}{4} \pm  \frac{9 H^4}{8 \pi^2 m_{r,\textrm{eff}}^4}\lambda,\nonumber
\end{align}
where $\lambda \ll m_{r,\textrm{eff}}^4/H^4$ by assumption, and $n$ is an integer.}
The scalar field remains frozen at $\left(\left<\phi_\textrm{ini}^2\right>,\,\left< \theta_\textrm{ini}  \right>\right) $ until the time of oscillation given by Eq.~\eqref{eq:H_osc_def}. Then, the $\mathcal{O}(\epsilon)$ shift of $\left< \theta_\textrm{ini}  \right>$ acts as a kick in the angular direction.

\paragraph{Issues with quantum fluctuations during inflation.}
In App.~\ref{app:issus_inflation_fluct_model_B}, we show that in the absence of Hubble-size mass terms for the radial and angular modes in Eq.~\eqref{eq:negative_hubble_mass}, quantum fluctuation during inflation leads to problematic adiabatic and isocurvature perturbations as well as to the formation of domain walls which are not bounded by cosmic strings. For those reasons, in this work we assume the presence of the Hubble-induced potential in Eq.~\eqref{eq:negative_hubble_mass} such that that the radial VEV $\left<\phi_\textrm{ini}^2\right>$ is set by Eq.~\eqref{susy_phi_ini} and not by Eq.~\eqref{ds_fluc}.

\FloatBarrier
\begin{figure}[h!]
\centering
\raisebox{0cm}{\makebox{\includegraphics[width=0.475\textwidth, scale=1]{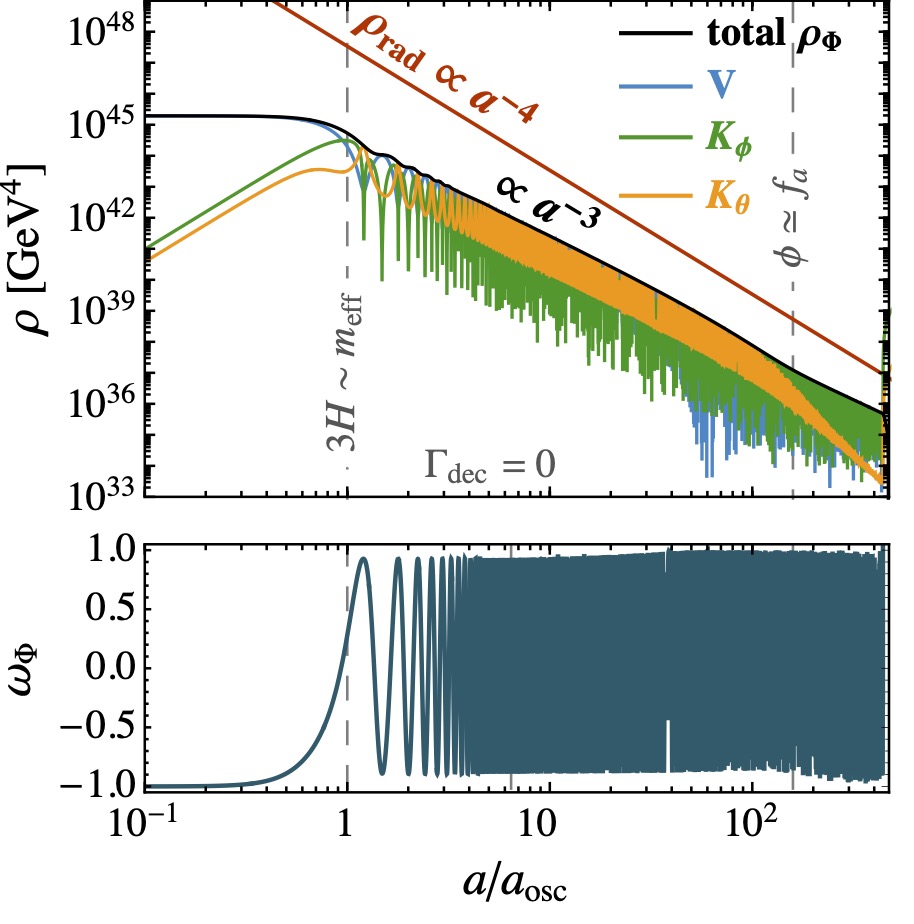}}}
\quad
\raisebox{0cm}{\makebox{\includegraphics[width=0.475\textwidth, scale=1]{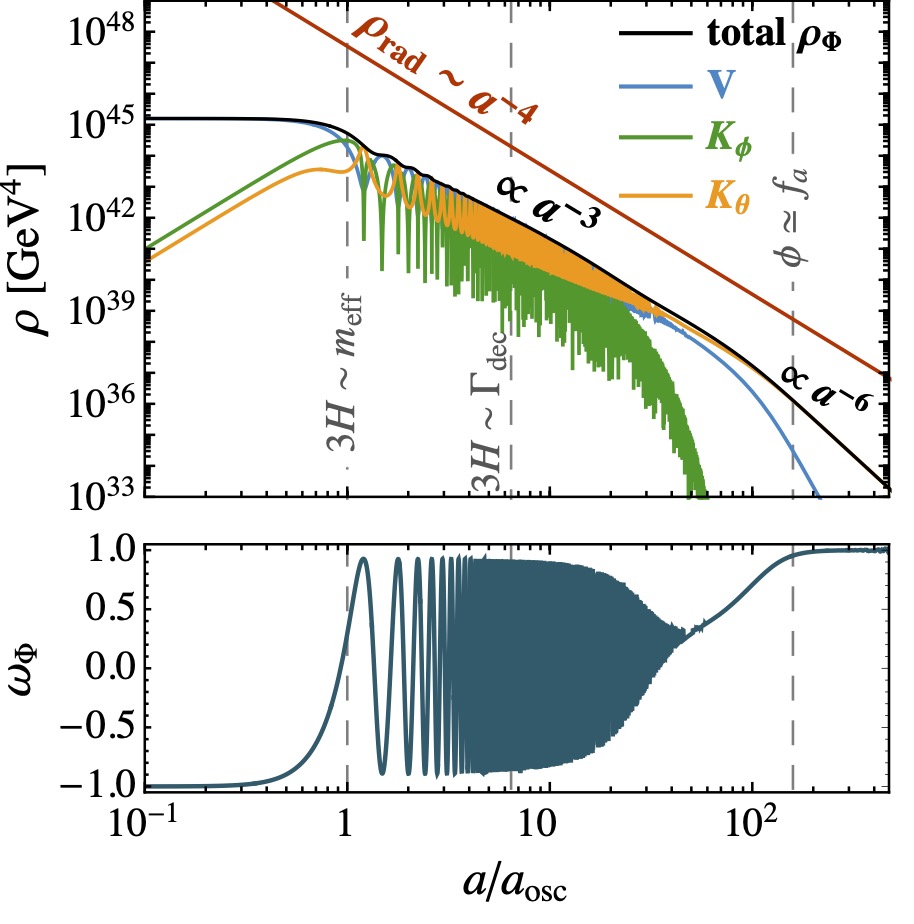}}}
\caption{\textit{ \small Evolution of complex field energy density without ({\textbf{left}}) and with ({\textbf{right}}) radial damping. Only the latter case gives rise to a kination equation of state when the complex scalar field reaches the degenerate minimum of its potential $\phi \to f_a$ (\textbf{bottom}). Obtained after numerically integrating the equations of motion in Eqs.~\eqref{radial_break}, \eqref{angular_break} and \eqref{eq:friedmann_eq}.}}
\label{fig:damping_nodamping}
\end{figure}
\FloatBarrier
\subsection{Damping of the radial mode}
\label{sec:themalization}

\subsubsection{Damping mechanisms}
After the complex fields starts oscillating, it accomplishes an elliptic orbit.  
In Fig.~\ref{fig:damping_nodamping}, we show that damping of the radial kinetic energy $K_\phi = \dot{\phi}^2/2$ (green lines - left vs right) is necessary for the complex scalar field to accomplish a circular orbit and to acquire a kination equation of state  when reaching the bottom of its potential $\phi \to f_a$.
In this work we consider two mechanisms for damping the radial mode.
\paragraph{1) Parametric resonance:} 
In App.~\ref{app:param_resonance}, we discuss in a qualitative manner the possibility that parametric resonance could damp the radial mode while preserving the $U(1)$ charge $n_\theta$. We leave however the quantitative study of the realistic efficiency of this damping mechanism for further studies. 
\paragraph{2) Thermalization:} 
We assume that the complex scalar field $\phi$ is coupled to fermions $\psi$ charged under some (hidden or SM) gauge sector $A_\mu$ (KSVZ-type interactions) 
\begin{equation}
\mathcal{L} \supset y_\psi\phi \psi_{L}^{\dagger} \psi_{R}+h.c.+ g \, \bar{\psi} \gamma^\mu\psi\, A_\mu. \label{eq:KSVZ_lagrangian}
\end{equation}
As a consequence, the scalar condensate thermalizes with the thermal plasma when the rate $\Gamma$ of decay into fermions, given in Eq.~\eqref{eq:fermion_damping_rate_YG} of App.~\ref{app:thermalization}, is larger than $H$. Higgs interactions are not sufficient enough as discussed in App.~\ref{par:Higgs_portal}, we therefore focus on thermal effects from the fermionic portal.
As shown in the supplementary material of \cite{Co:2019wyp}, as long as $\phi \gg T$, after thermalization it is energetically more favorable to keep the $U(1)$ charge in the condensate, which has an energy density $\rho_\phi=\phi^2 \dot{\theta}^2$, than in the plasma, which has an energy density $\rho_{\rm rad}\supset\phi^4\dot{\theta}^2/T^2$.
In the presence of this interaction, the scalar field dynamics can be dominated by its thermal mass
\begin{equation}
{m_{r,\textrm{eff}, T}^2} = \mreff^2 + y_\psi^2 T^2,
\end{equation}
after the onset of radial mode oscillation, if the Yukawa coupling $y_\psi $ is larger than
\begin{equation}
y_\psi T_{\rm osc} ~> ~\mreff \qquad \implies \qquad y_\psi ~\gtrsim ~ g_*^{1/4}\sqrt{\frac{\mreff}{M_{\rm pl}}}, \label{eq:yukawa_bound_no_thermal_mass}
\end{equation}
where $3H_{\rm osc}\simeq   \left(\frac{\pi^2 g_*}{10} \right)^{1/2} T_{\rm osc}^2/M_{\rm pl} \simeq \mreff$ and where we have replaced $\pi^2 \simeq 10$. 

\subsubsection{Three possible scenarios}
We consider three possible scenarios for the damping of the radial mode.
\begin{enumerate}
\item
\textbf{Scenario I.} The complex scalar field is assumed to evolve in the \textbf{zero-temperature} potential $V(\phi) = \mreff^2(\phi) \phi^2/2$, which implies a small Yukawa coupling $y_\psi \lesssim \sqrt{\mreff/M_{\rm pl}}$, cf. Eq.~\eqref{eq:yukawa_bound_no_thermal_mass}. As shown in gray regions of Fig.~\ref{fig:scenarioI_II_III_TmaxOTreh}, for such small Yukawa coupling, thermalization is not efficient enough to damp the radial mode early enough to generate kination. Instead, we assume that radial damping occurs \textbf{non-thermally} via some unknown mechanism, for which \textbf{parametric resonance} appears as a possible candidate, cf. App.~\ref{app:param_resonance}. This leads to the largest prediction for the number of kination e-folds $N_{\rm KD}$.  This is the scenario considered in the next Sec.~\ref{sec:scenario_I_non_thermal_damping}. The longest duration of kination is obtained in Eq.~\eqref{eq:longest_NKS_scenario_1}
\begin{align}
e^{N_{\rm KD}} 
~=~ e^{8.2}~\epsilon^{2/3}  \left( \frac{10^{9}~\rm GeV}{f_a}\right)^{1/3}   \left(\frac{\mreff(\phi_{\rm ini})}{5\mreff(f_a)} \right)^{1/3} \left( \frac{\phi_{\rm ini}}{M_{\rm pl}} \right)^{4/3}.\label{eq:longest_NKS_scenario_1_sketch}
\end{align}
\item
Radial damping occurs via \textbf{thermalization}. Large $N_{\rm KD}$ requires that thermalization occurs before scalar field domination in order to prevent entropy injection. This implies a large Yukawa coupling $y_\psi$ and potentially large thermal effects. We consider two possibilities.
\begin{enumerate}
\item
\textbf{Scenario II.} The dynamics of the complex scalar field is controlled by its \textbf{thermal mass} $V(\phi) =  \frac{1}{2}y_\psi^2 T^2 \phi^2$. In Sec.~\ref{sec:complex_field_thermal_potential} we show that the modification of the field dynamics in the presence of the thermal mass prevents the onset of a kinaton-dominated era. This is due to the initial angular kick $\epsilon$ being suppressed and also to the scalar field starting oscillating earlier.
\item
\textbf{Scenario III.} The fermions responsible for the thermal corrections of the potential are Boltzmann-suppressed when the scalar field starts oscillating. Therefore, the scalar field evolves in its zero-temperature potential and the efficiency of the angular kick is not spoiled by thermal effects. This scenario is presented in Sec.~\ref{sec:complex_field_low_reh_temp}. The longest period of kination is, cf. Eq.~\eqref{eq:NKD_scenario3_best}

\begin{equation}
\label{eq:longest_NKS_scenario_3_sketch}
e^{N_{\rm KD}}\big|_{y_\psi = y_{\psi,*}} = e^{5.1} \frac{\epsilon^{2/3}}{\alpha^{1/3}g_*^{1/4}}\left(\frac{10^8}{f_a} \right)^{1/6}\left(\frac{\mreff(f_a)}{f_a} \right)^{1/6}\left(\frac{\mreff(\phi_{\rm ini}}{\mreff(f_a)} \right)^{1/2} \left( \frac{0.1}{\phi_{\rm ini}/M_{\rm pl}} \right)^{2/3}.
\end{equation}

\end{enumerate}
\end{enumerate}

\section{Scenario I: non-thermal damping }
\label{sec:scenario_I_non_thermal_damping}

In the present section we study the first scenario (``Scenario I") for which the couplings between the complex scalar condensate $\Phi$ and the thermal plasma are suppressed such that we can ignore the thermal corrections to the potential in Eq.~\eqref{eq:thermal_correction_potential}. We impose the thermal mass to be negligible at the onset of the radial mode oscillation. Focusing on the case where $\Phi$ couples to thermal fermions through a Yukawa coupling $y_\psi$ as defined in Eq.~\eqref{eq:KSVZ_lagrangian_2}, this implies
\begin{equation}
y_\psi T_{\rm osc}~\lesssim ~ \mreff(\phi_{\rm ini}) \qquad \implies \quad y_\psi \lesssim g_*^{1/4} \frac{\mreff(\phi_{\rm ini})}{M_{\rm pl}}, \label{eq:thermal_mass_negligible_scenario_I}
\end{equation}
where we have used $y_\psi T_{\rm osc} \simeq 3H(T)$.

In App.~\ref{sec:impossibility_neglect_thermal_correction}, we show that it is impossible to efficiently damp the radial mode via thermalization while neglecting thermal corrections at the onset of the scalar field oscillation. Instead, in the present section we assume that radial damping takes place via a non-thermal mechanism (possibly parametric resonance cf. App.~\ref{app:param_resonance}), and we consider the radial damping rate $\Gamma$ to be a free parameter.

\subsection{Field  trajectory}\label{field_evol}
In Fig.~\ref{field_evolution}, we outline the main stages of evolution of the rotating complex scalar field.

First, thanks to peculiarities of the Hubble-induced potential in SUSY theories, the scalar field is initially frozen at a large field value $\phi_{\rm ini}$ and at a displaced angle $\theta_{\rm ini}$ with respect to the valleys of the potential, see Sec.~\ref{sec:intitial_vev}.

Second, the field starts oscillating when its dynamics become under-damped $3H_{\rm osc} \simeq \mreff$. Thanks to the initial kick induced by the $U(1)$ breaking potential, see Sec.~\ref{sec:U1breaking}, the complex scalar field accomplishes an elliptic motion, whose size reduces with time due to the cosmic expansion. For a nearly-quadratic potential, see Sec.~\ref{sec:nearly-quadratic_potential}, the scalar field energy density redshifts as matter $\rho_\phi \propto a^{-3}$ and starts dominating the energy density of the universe.

Third, the radial motion is damped due to either parametric resonance, cf. present Sec.~\ref{sec:scenario_I_non_thermal_damping}, or thermalization, cf. next Sec.~\ref{sec:complex_field_thermal_potential} and \ref{sec:complex_field_low_reh_temp}, while the angular motion remains, see Sec.~\ref{sec:themalization}. 

After this stage, the field rotates coherently in a circle with a continuously reducing size. When the orbiting field reaches the bottom of the potential, its kinetic energy dominates its potential energy and gives rise to a kination equation of state. 

The numerical computation of the full trajectory of the scalar field from the onset of the oscillation until the end of the kination era is shown in Fig.~\ref{fig:theta_dot_vary_epsilon_l}, Fig.~\ref{quad_field_evolution_whole} and in \href{https://www.youtube.com/watch?v=RdCAgcvfFy0}{our animation}.

\FloatBarrier
\begin{figure}[h!]
\centering
\raisebox{0cm}{\makebox{\includegraphics[width=0.8\textwidth, scale=1]{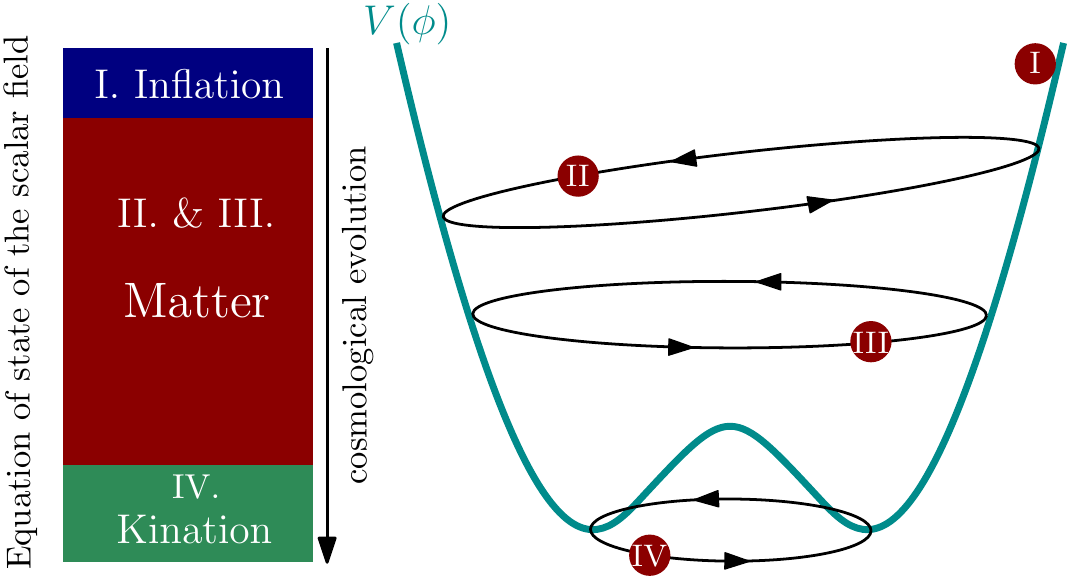}}}  \vspace{1cm}\\
\raisebox{0cm}{\makebox{\includegraphics[width=0.9\textwidth, scale=1]{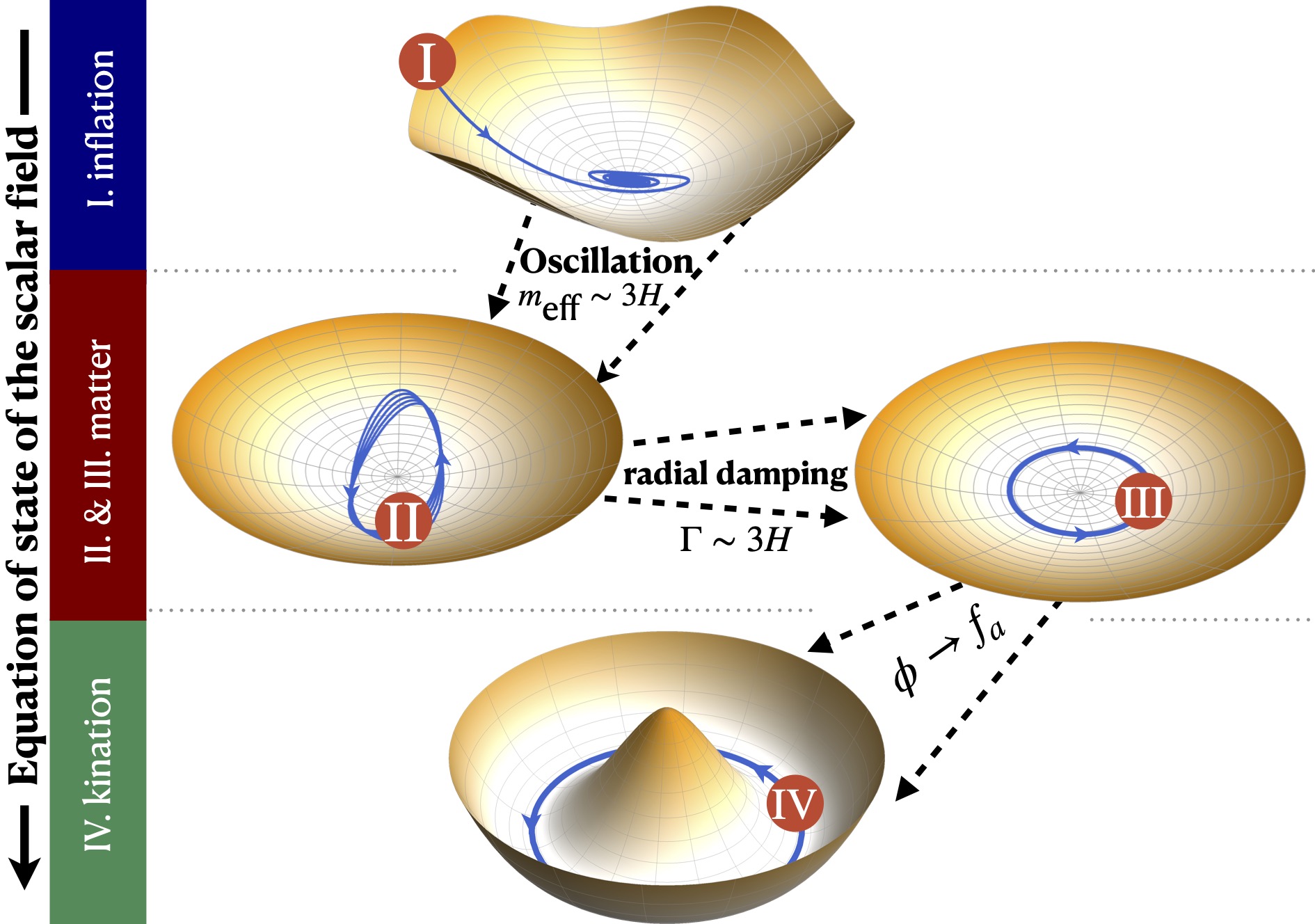}}}  \vspace{1cm}\\
\begin{tabular}{|cl|c|c|c|}
\hline
 & \qquad \qquad \qquad Stages                   &  Hubble factor                & Field value                  & Energy density
 \\ \hline
 \hline
I.   & Field frozen or track large-field minimum                   & $3 H > m_{r,\textrm{eff}}, ~\Gamma$                  & $\phi = \phi_\textrm{ini}$                        & $\rho_\Phi \propto a^0$                                                                                                        \\ \hline
II.  & Elliptic orbit: oscillation and rotation & $m_{r,\textrm{eff}} \geq 3 H > \Gamma$              & \multirow{2}{*}{$\phi_\textrm{ini} > \phi > f_a$} & \multirow{2}{*}{\begin{tabular}[c]{@{}c@{}}$\rho_\Phi \propto a^{-3}$\end{tabular}} \\ \cline{1-3}
III. & Circular orbit after radial damping            & \multirow{2}{*}{$m_{r,\textrm{eff}}, ~\Gamma > 3 H$} &                                                   &                                                                                                                           \\ \cline{1-2} \cline{4-5} 
IV.  & Rotation at potential minimum      &                                               & $\phi = f_a$                                      & $\rho \propto a^{-6}$                                                                                                     \\ \hline
\end{tabular}
\caption{\textit{ \small We show the different stages of evolution of the scalar field, leading to the matter-kination equation of state. The side bar on the left shows the EOS of the scalar field.
We invite the interested reader to visualize \href{https://www.youtube.com/watch?v=RdCAgcvfFy0}{our animation}.}}
\label{field_evolution}
\end{figure}
\FloatBarrier

\subsubsection{Kick in the angular direction.}

\paragraph{Amount of rotation.}
We introduce the ratio between number densities in the angular mode and in the radial mode\footnote{$\epsilon$ is related to $Y_\theta$ defined in Eq.~\eqref{eq:U1_charge_axion} through 
\begin{equation}
\epsilon ~= ~\frac{s}{2V(\phi)/m_{r,\textrm{eff}}(\phi)} Y_\theta,
\end{equation}
where $s = \frac{2\pi^2}{45}g_* T^3$. In successful setup, the quantity $\epsilon$ is bounded by above $\epsilon \lesssim 1$, where $\epsilon = 1$ corresponds to a field trajectory which would be already circular at the onset of the radial mode oscillation. In that sense, $\epsilon$ can be called the $U(1)$ charge fraction contained in the condensate, $1$ being the maximal value.
}
\begin{equation}
\epsilon ~ \equiv  ~ \frac{\phi^2 \dot{\theta}/2}{V(\phi)/m_{r,\textrm{eff}}(\phi)}~=~ \frac{\phi_{\rm ini}^2 \dot{\theta}_{\rm ini}/2}{V(\phi)/m_{r,\textrm{eff}}(\phi)} \left( \frac{a_{\rm ini}}{a} \right)^3.
 \label{eq:epsilon_def}
\end{equation}
From using $(\phi_{\rm ini}/M_{\rm pl})^{l-2}=\sqrt{c} \mreff(\phi_{\rm ini}) /  \lambda\sqrt{2l-2} M_{\rm pl}$, see Eq.~\eqref{susy_phi_ini}, we find that the term in the potential which dominates the dynamics of the scalar field at the onset of the radial-mode oscillation  depends on the value of $c$
\begin{equation}
V(\phi_{\rm ini}) = 
\begin{cases}
\lambda^2 \frac{\phi_{\rm ini}^{2l-2}}{M_{\rm pl}^{2l-6}},~ \qquad \qquad \qquad \qquad \quad \text{if}~ c ~> ~l-1 ,\\[0.75em]
\frac{1}{2}\mreff^2(\phi_{\rm ini}) \phi_{\rm ini}^2, \qquad  \qquad\quad ~~ \text{otherwise} .
\end{cases}
\end{equation}
In the first case of the previous equation, the scalar field redshift as, cf.~App.~\ref{app:virial_th}
\begin{equation}
\phi \propto a^{-\frac{6}{2+n}},\qquad\qquad \textrm{with}~n ~=~ 2l-2.
\label{eq:phi_redshift_higher_dim_pot}
\end{equation} 
until the nearly-quadratic term $\phi^2$ dominates around the field value $\phi_{\rm quad}$
\begin{equation}
\frac{1}{2}\mreff^2(\phi_{\rm quad})\phi_{\rm quad}^2~ =~ 
 \lambda^2 \frac{\phi_{\rm quad}^{2l-2}}{M_{\rm pl}^{2l-6}}.\label{eq:phi_quad_def}
\end{equation}
For $ \phi ~< ~\phi_{\rm quad}$, the quantity $\epsilon$ defined in Eq.~\eqref{eq:epsilon_def} becomes a conserved quantity since both numerator and denominator scale as $a^{-3}$.
 Plugging Eq.~\eqref{eq:phi_redshift_higher_dim_pot} into Eq.~\eqref{eq:phi_quad_def}, we obtain the scale factor $a_{\rm quad}$ below which $\epsilon$ becomes constant. From injecting $a_{\rm quad}$ into the generated $U(1)$ charge in Eq.~\eqref{epsilon_oscillation} and then back into the definition of $\epsilon$ in Eq.~\eqref{eq:epsilon_def}, we obtain
\begin{equation}
\epsilon~ =~ 
\begin{cases}
\frac{1}{\sqrt{2}}\frac{\mreff(f_a)}{\mreff(\phi_{\rm ini})} \,l\,\sin{l \theta_{\rm ini}},~ \qquad \qquad \qquad \qquad \quad~ \text{if} \quad c~> ~l-1,\\[0.75em]
\frac{1}{\sqrt{2}}\sqrt{\frac{c}{l-1}} \frac{\mreff(f_a)}{\mreff(\phi_{\rm ini})} \,l\,\sin{l \theta_{\rm ini}}, ~~~\quad\qquad\qquad \quad  ~  \text{otherwise} .
\end{cases} \label{eq:epsilon_1_zero_temp}
\end{equation}
For $c = \mathcal{O}(1)$ and $\sin{l \theta_{\rm ini}} = \mathcal{O}(1)$, a large value of $l$, as for instance $l \simeq 10$, can easily compensate for the ratio of $\mreff$, which is only $\log^{1/2}$ suppressed, cf. Eq.~\eqref{eq:mr_eff}, such that $\epsilon = \mathcal{O}(1)$ is realistic.\footnote{Values of $\epsilon$ larger than $1$ corresponds to the field climbing up the field potential due to the centrifugal force being larger than the mass and we expect them to be unphysical, therefore we should replace Eq.~\eqref{eq:epsilon_1_zero_temp} by $\textrm{Min}\left[\epsilon, ~1 \right]$.}
In Fig.~\ref{wiggly_potential}, we can see that the size of the wiggles increases with $\epsilon$. The larger the wiggles, the larger the potential gradient along the angular direction and the larger the initial kick.

\begin{figure}[h!]
\centering
\raisebox{0cm}{\makebox{\includegraphics[width=0.475\textwidth, scale=1]{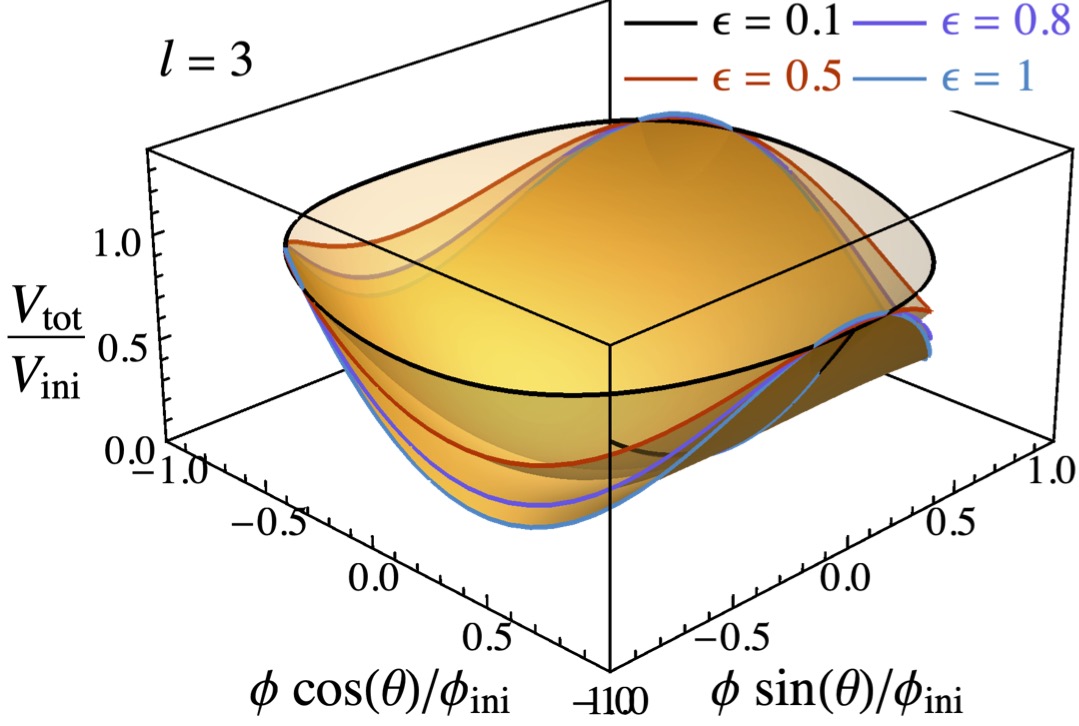}}}
\quad
\raisebox{0cm}{\makebox{\includegraphics[width=0.475\textwidth, scale=1]{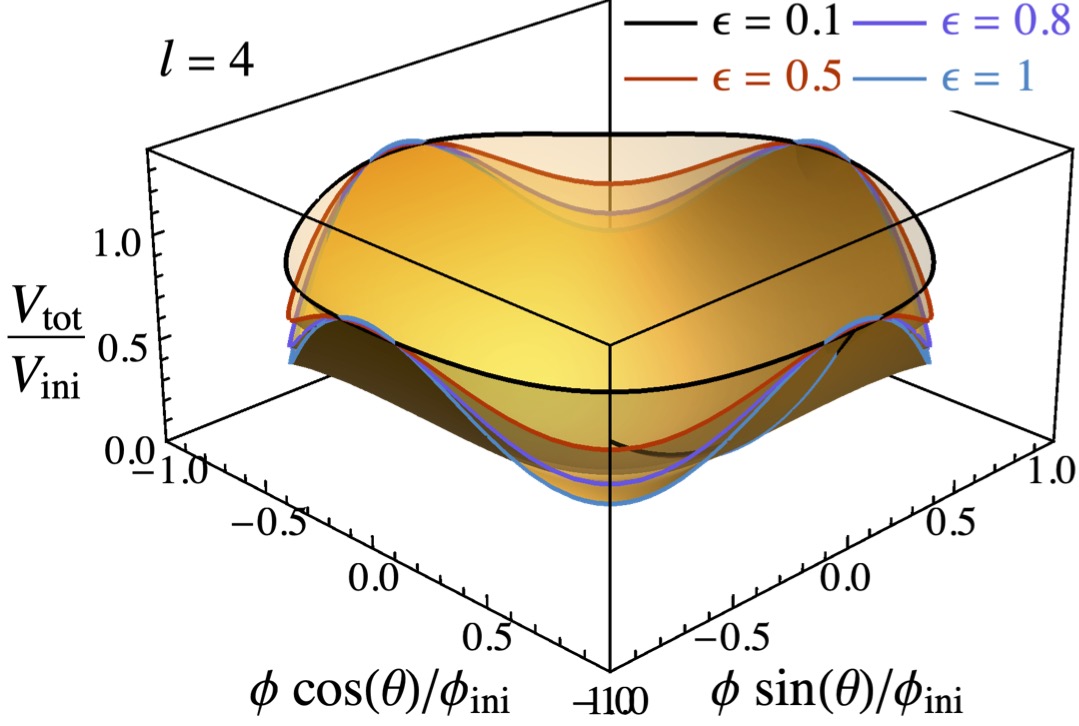}}}\\[0.5em]
\raisebox{0cm}{\makebox{\includegraphics[width=0.475\textwidth, scale=1]{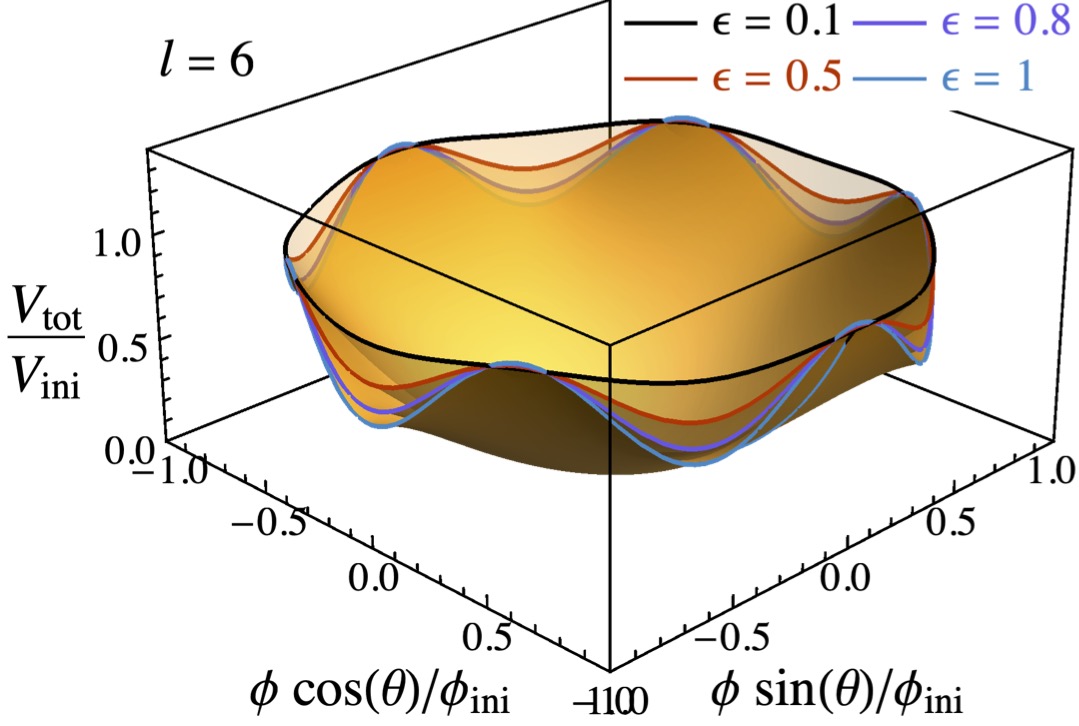}}}
\quad
\raisebox{0cm}{\makebox{\includegraphics[width=0.475\textwidth, scale=1]{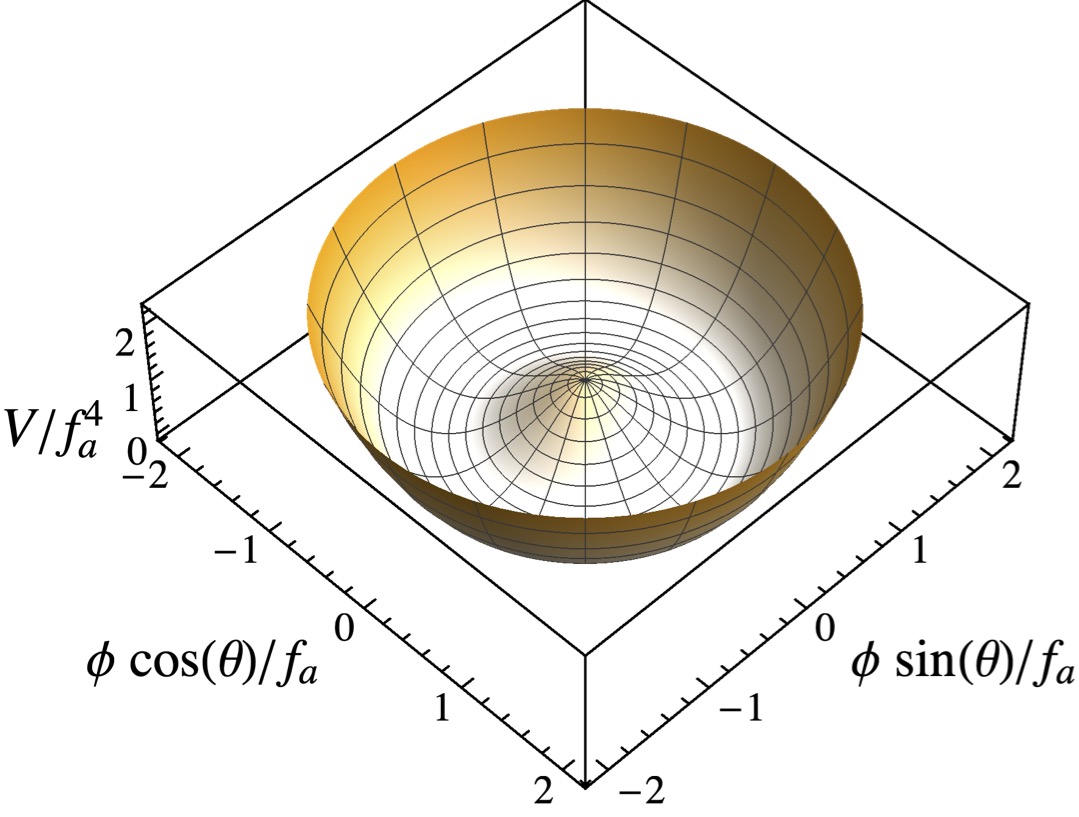}}}
\caption{\textit{ \small Nearly-quadratic potentials with the explicit-breaking term in Eq.~\eqref{breaking_potential_thiswork}. The integer $l$ corresponds to both the number of wiggles and the order of $\phi^l$.
The amplitude of the angular velocity kick $ \dot{\theta}_{\rm ini}$ is set by the explicit-breaking strength $\epsilon$, defined in Eq.~\eqref{eq:epsilon_def}.
The bottom-right figure shows a zoom of the circular vacuum at $\phi = f_a$.}}
\label{wiggly_potential}
\end{figure}

\subsubsection{Motion after the kick.}
In App.~\ref{app:exact_sol}, we show that after a few oscillations  in a nearly-quadratic potential, the angular velocity averaged over many oscillation periods $\left<\dot{\theta}\right>$ becomes independent of the initial kick $\dot{\theta}_{\rm ini}$ in Eq.~\eqref{epsilon_oscillation}. Instead, it converges to the attractor solution
\begin{equation}
\left< \dot{\theta}\right> ~ \equiv ~ \frac{1}{T}\int^T_{m_\textrm{eff}^{-1}} \dot{\theta}(t') dt' ~  =~ m_{r, \rm eff}, \label{eq:averaged_theta_dot_def}
\end{equation}
for which the quadratic potential $m_{r, \rm eff}^2 \phi^2$ is exactly compensated by the centrifugal potential $\phi^2 \dot{\theta}^2$. This is confirmed by the numerical integration of the equations of motion shown in Fig.~\ref{fig:theta_dot_vary_epsilon_l}. Notice that even if the stationary value of $\left< \dot{\theta}\right> $ is independent of $\dot{\theta}_{\rm ini}$, it is not the case of the fraction $\epsilon$ of $U(1)$ charge in the condensate, see Eq.~\eqref{eq:epsilon_1_zero_temp}.

In App.~\ref{app:virial_th}, using Virial theorem we show that the energy density $\rho_\Phi$ and the radial field value $\phi$ of a complex scalar field in a $U(1)$ conserving nearly quadratic potential, averaged over oscillations, scale like
\begin{equation}
\left< \rho_\Phi \right>  \propto a^{-3}, \qquad \left<\phi \right> \propto a^{-3/2}. \label{eq:redshift_laws_rho_phi}
\end{equation}

\begin{figure}[h!]
\centering
\raisebox{0cm}{\makebox{\includegraphics[width=0.56\textwidth, scale=1]{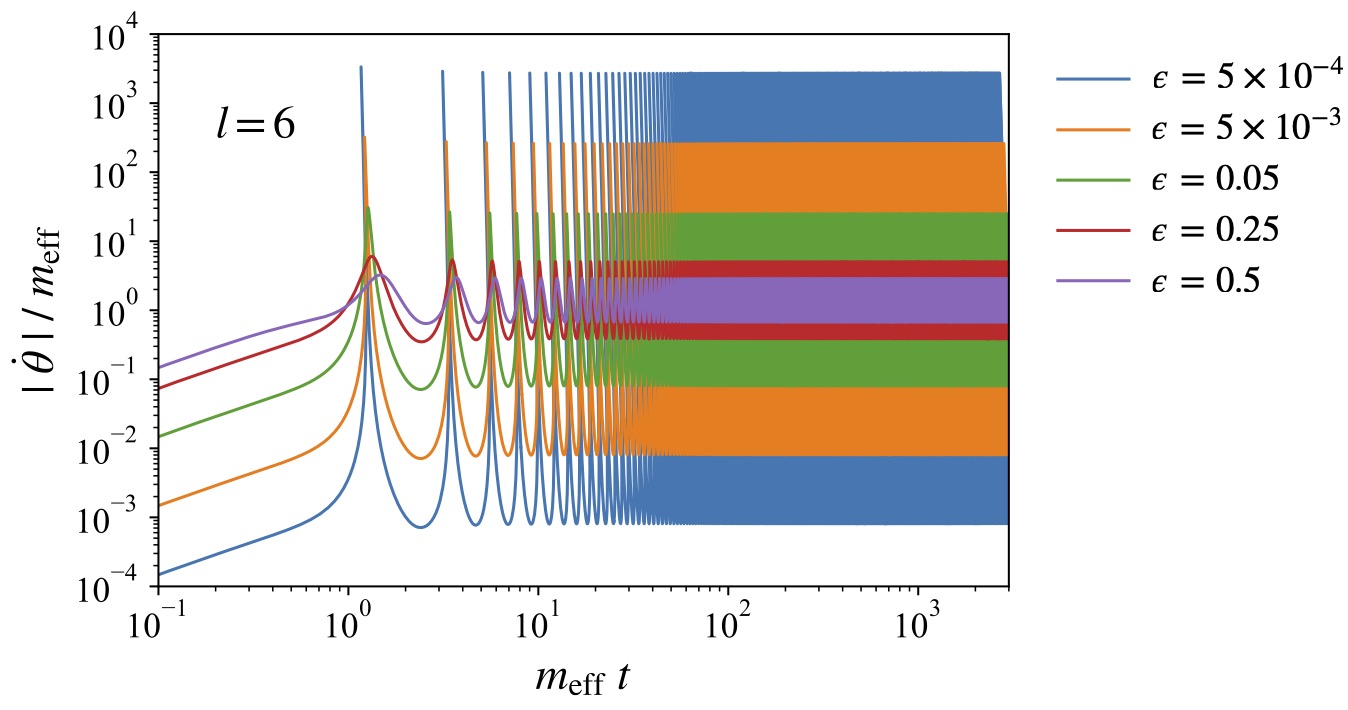}}}
\raisebox{0cm}{\makebox{\includegraphics[width=0.43\textwidth, scale=1]{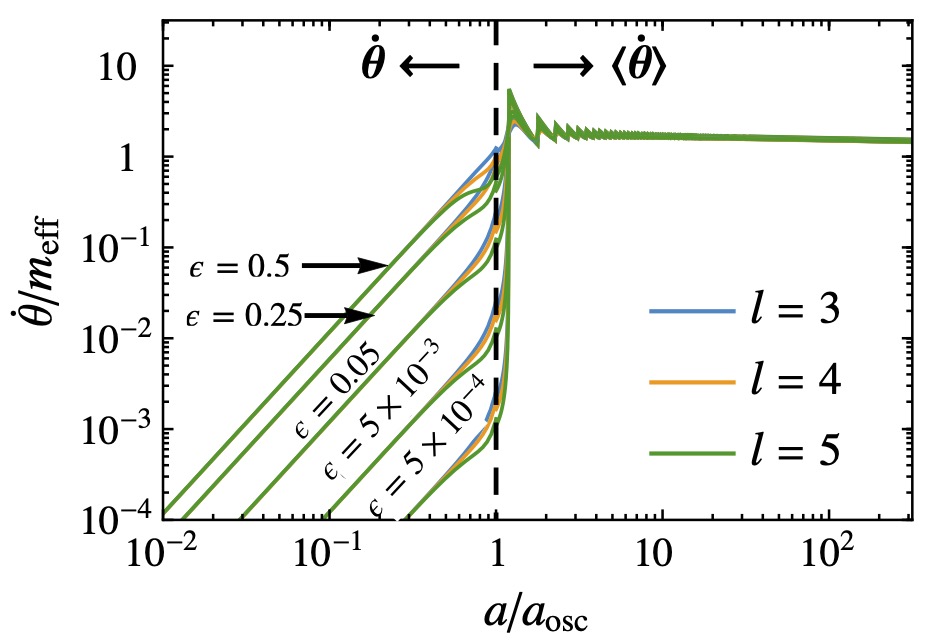}}}
\caption{\textit{ \small 
\textbf{Left:} Non-averaged angular velocity $\dot{\theta}(t)$, obtained after numerically integrating the equations of motion in Eqs.~\eqref{radial_break}, \eqref{angular_break} and \eqref{eq:friedmann_eq}. ($m_r = 10^5$ GeV, $M = \MPl$, $\phi_\textrm{ini} = 10^{17}$ GeV, $\theta_\textrm{ini} = \pi/2 l$, $\dot{\phi}_\textrm{ini} = \dot{\theta}_\textrm{ini}  = 0$). The trajectory is more and more circular as the $U(1)$ charge fraction $\epsilon \to 1$.
\textbf{Right:} Averaged angular velocity $\left< \dot{\theta} \right>$, defined in Eq.~\eqref{eq:averaged_theta_dot_def}.
After that the scalar field starts oscillating $a > a_{\rm osc}$, the trajectory becomes independent of the values of $\epsilon$ and $l$ and quickly converges to the attractor solution $\left< \dot{\theta} \right> = \mreff$.
}}
\label{fig:theta_dot_vary_epsilon_l}
\end{figure}
\FloatBarrier

\subsubsection{Motion due to radial damping.}
\label{sec:epsilon_suppression_kination}
The damping of the radial mode $\dot{\phi} \to 0$, see Sec.~\ref{sec:themalization}, converts the elliptic trajectory into a circular one. 
In App.~\ref{app:effect_epsilon_evolution}, we show that radial damping leads to a drop in total energy of the complex scalar field $\Phi$ equal to the $U(1)$ charge fraction $\epsilon$ defined in Eq.~\eqref{eq:epsilon_def}
\begin{align}
\rho_\Phi^\mathrm{after} ~ = ~ \epsilon \rho_\Phi^\mathrm{before} \qquad \implies \qquad \phi^2_\mathrm{after} ~ = ~ \epsilon \phi^2_\mathrm{before},
\label{epsilon_energy_transition_main}
\end{align}
where the label `before' and `after' denote the moments  just before and just after the time of damping. The suppression factor in Eq.~\eqref{epsilon_energy_transition_main} can be understood from the conservation of the rotational energy $\rho_\theta \equiv \dot{\theta}^2\phi^2/2$  during radial damping
\begin{align}
\rho_\theta ~ = ~ \epsilon V(\phi).
\label{epsilon_energy_main}
\end{align}
In App.~\ref{app:effect_epsilon_evolution}, we show that due to Eq.~\eqref{epsilon_energy_transition_main}, the number of e-folds of kination domination receives the suppression factor $-\frac{2}{3}\log{\epsilon}$.
The impact of $\epsilon$ on the energy density and the duration of kination can be understood from Fig.~\ref{diagram_epsilon_energy}.

\begin{figure}[h!]
\centering
\raisebox{0cm}{\makebox{\includegraphics[width=0.99\textwidth, scale=1]{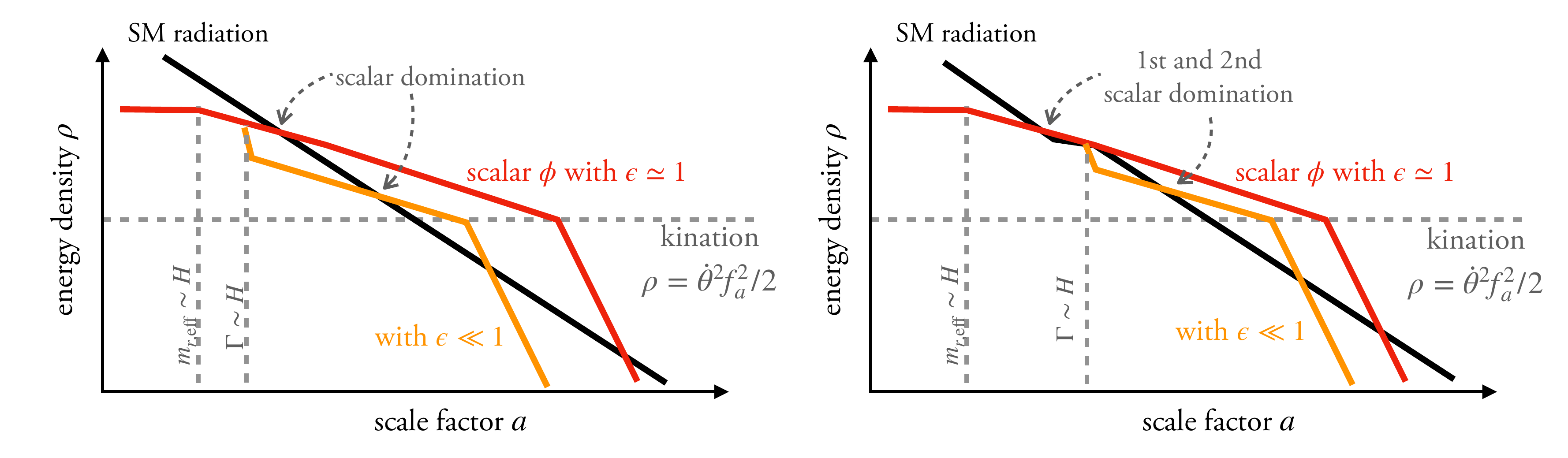}}}
\caption{\textit{ \small Impact of $\epsilon \ll 1$ on the evolution of the energy density of the complex scalar fields. We consider two cases according to whether the radial damping occurs before (\textbf{left}) or after (\textbf{right}) the complex scalar dominates the energy density of the universe. In both cases, the number of e-folds of kination is reduced by $-\frac{2}{3} \log{\epsilon}$. The best case scenario $\epsilon =1$ corresponds to a trajectory which is already circular even before radial damping. }}
\label{diagram_epsilon_energy}
\end{figure}

\begin{figure}[h!]
\centering
\raisebox{0cm}{\makebox{\includegraphics[width=0.48\textwidth, scale=1]{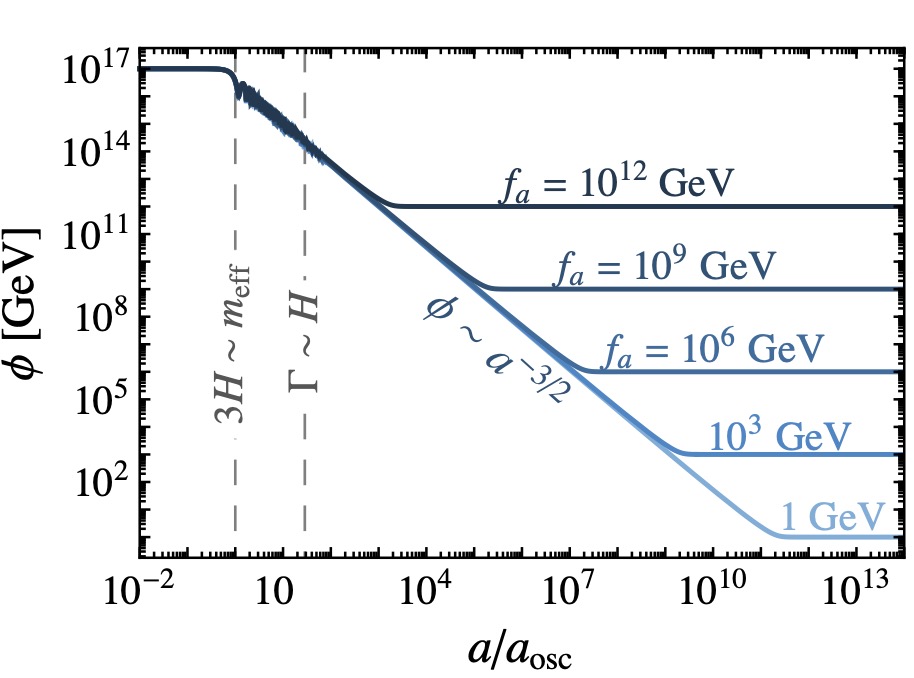}}}
\quad
\raisebox{0cm}{\makebox{\includegraphics[width=0.48\textwidth, scale=1]{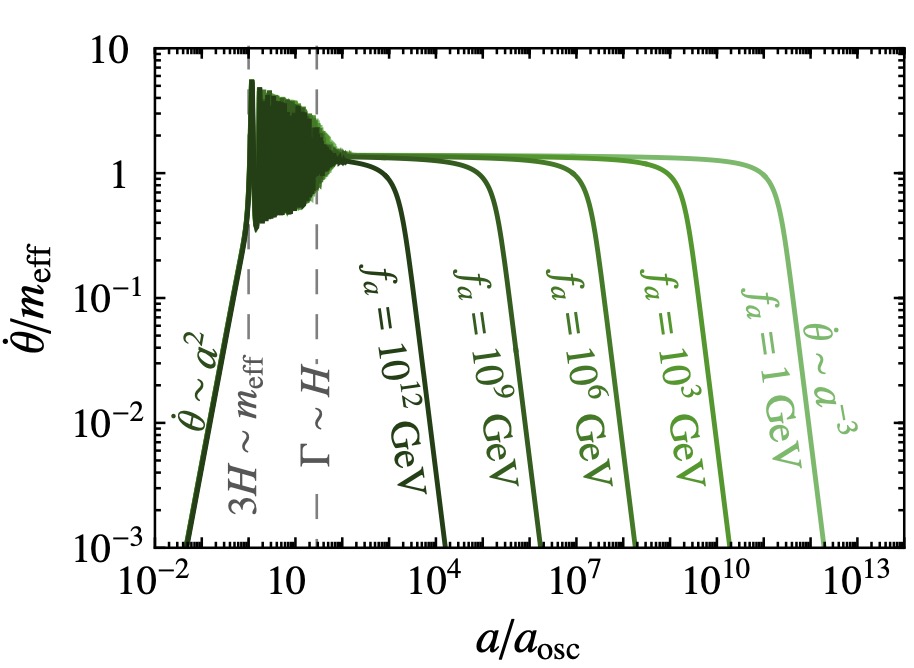}}}
\\
\raisebox{0cm}{\makebox{\includegraphics[width=0.48\textwidth, scale=1]{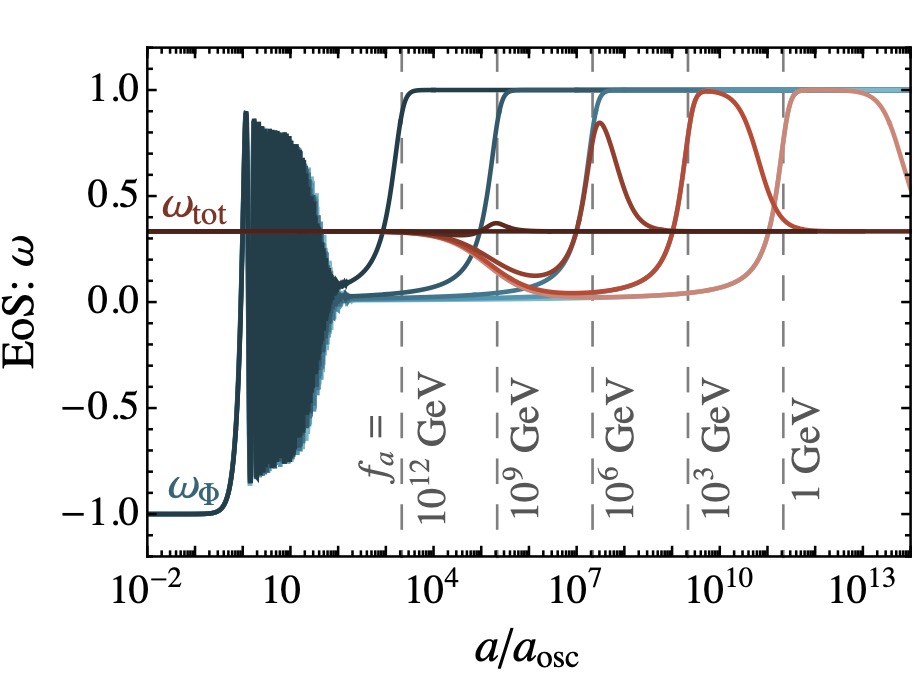}}}
\quad
\raisebox{0cm}{\makebox{\includegraphics[width=0.48\textwidth, scale=1]{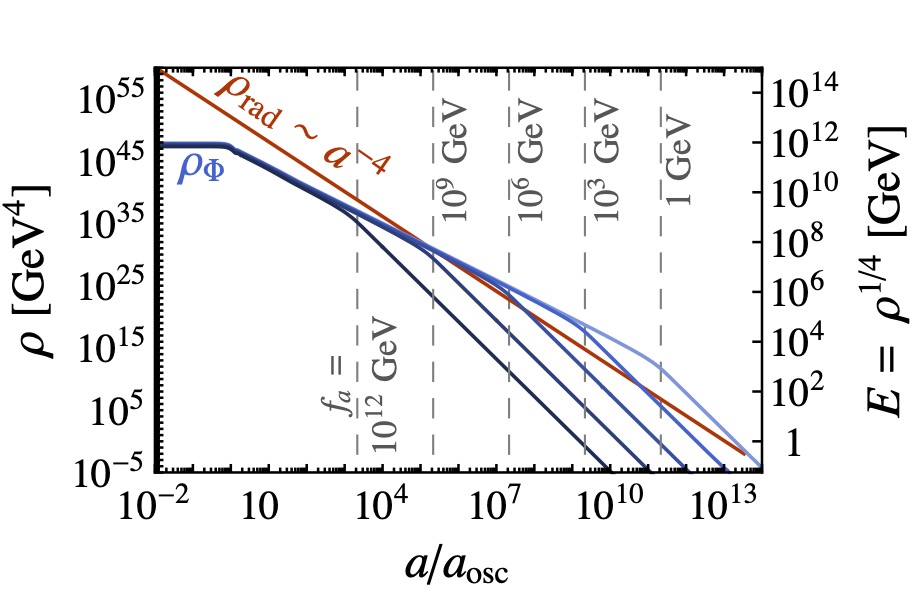}}}
\caption{\textit{ \small $m_r = 10^6$ GeV, $M = \MPl$, $\phi_\textrm{ini} = 10^{17}$ GeV, $\theta_\textrm{ini} = \pi/2 l$, $\dot{\phi}_\textrm{ini} = \dot{\theta}_\textrm{ini}  = 0$, $\epsilon = 0.4$, and $\Gamma = 10^4$ GeV.
Evolution of the radial field $\phi$, the angular velocity $\dot{\theta}$, the equation-of-state $\omega_\Phi$, and the energy density $\rho_\Phi$.
The complex scalar field has a matter EOS $\omega_\Phi=0$, when $\phi \gg f_a$ and reaches a kination EOS $\omega_\Phi = 1$, when $\phi \simeq f_a$.  Kination ends when the radiation energy density shown in red starts dominating the energy budget of the universe once again.
For a fixed $m_r$, the smaller $f_a$, the longer the matter era, the larger the domination of the energy density of the universe, and the longer the kination era. Obtained after numerically integrating the equations of motion in Eqs.~\eqref{radial_break}, \eqref{angular_break} and \eqref{eq:friedmann_eq}.
}}
\label{quad_field_evolution_whole}
\end{figure}

\subsubsection{Motion towards kination.}
After radial damping $\dot{\phi} \to 0$, the trajectory of the complex scalar field is reduced to a circular orbit whose radius decreases due to the Hubble friction.
From the conservation of the $U(1)$ charge in Eq.~\eqref{PQ_charge_conservation}
\begin{align}
a^3 \phi^2 \dot{\theta} ~ = ~ \textrm{constant},
\end{align}
 we see that once $\phi \to f_a$, the complex scalar field reaches a kination equation of state
\begin{equation}
\dot{\theta} \propto a^{-3} \qquad \text{and} \qquad \rho_{\Phi} = \frac{\phi^2 \dot{\theta}^2}{2} \propto a^{-6}.
\end{equation}
More precisely, in App.~\ref{app:traj_after_radial_damp} we compute the evolution of the complex scalar field $\Phi = \phi e^{i\theta}$ after radial damping $\dot{\phi} \to 0$
\begin{align}
\frac{d \ln \phi}{d \ln a} ~ = ~ \frac{-3 \log\left(\frac{\phi^2}{f^2}\right)}{2 \log\left(\frac{\phi^2}{f^2}\right) + 1} \qquad \text{and} \qquad \frac{d \ln \dot{\theta}^2}{d \ln a} ~ = ~ \frac{-6}{2 \log\left(\frac{\phi^2}{f^2}\right) + 1}, 
\end{align}
as well as its energy density $\rho_{\Phi}$ and equation of state $\omega_\Phi$
\begin{align}
\frac{d \ln \rho_{\Phi}}{d \ln a} ~ = ~ \frac{-6 \log\left(\frac{\phi^2}{f^2}\right)}{2 \log\left(\frac{\phi^2}{f^2}\right) - 1 + \frac{f^2}{\phi^2}} \qquad \text{and} \qquad  \omega_\Phi = \frac{\phi^2-f_a^2}{2\phi^2 \log{\frac{\phi^2}{f_a^2}} -f_a^2 + \phi^2}.
\end{align}
For $\phi \gg f_a$ we have
\begin{align}
\phi ~ \propto ~ a^{-3/2},\qquad \dot{\theta} ~ \propto ~ a^0, \qquad \rho_\Phi ~ \propto ~ a^{-3}, \qquad \omega_\Phi ~ \simeq ~ 0,
\end{align}
and for $\phi \simeq f_a$ we have
\begin{align}
\phi ~ \propto ~ a^0, \qquad \dot{\theta} ~ \propto ~ a^{-3}, \qquad \rho_\Phi ~ \propto ~ a^{-6}, \qquad \omega_\Phi ~ \simeq ~ 1.
\end{align}
Those analytical results agree with the numerical computation of the full trajectory of the complex scalar field in an expanding universe, from the onset of the oscillation until the end of the kination era, shown in Fig.~\ref{quad_field_evolution_whole}.

\subsection{Cosmological history}
\label{sec:cosmo_history}

In this section, we discuss the cosmological history of the universe in terms of the energy density of the scalar field $\rho_\Phi$. We defer the derivations of the expressions below to App.~\ref{derive_cosmo_history} (and App.~\ref{app:effect_epsilon_evolution} for the factor $\epsilon/2$). We show a sketch of the evolution of $\rho_\Phi$ in Fig.~\ref{diagram_kination}.

\subsubsection{Pre-kination stages} 
\paragraph{Onset of field oscillation.} 
We assume that the universe is initially radiation-dominated. The complex scalar field starts to roll when
\begin{equation}
3H = \mreff \qquad \implies \qquad T_{\rm osc} =g_*^{-1/4} \sqrt{\mreff M_{\rm pl}}, \label{eq:oscillation_temperarture}
\end{equation}
and 
\begin{equation}
\rho_{\rm osc} = V(\phi_{\rm ini}),
\end{equation}
with $\mreff$ and $\phi_{\rm ini}$ given in Eq.~\eqref{eq:mr_eff} and \eqref{susy_phi_ini}, respectively.

\paragraph{Oscillation after reheating.} 
In our framework, we assume that the scalar field starts oscillating during a radiation-dominated era  after reheating
\begin{equation}
T_{\rm osc}= g_*^{-1/4} \sqrt{\mreff M_{\rm pl}} \leq T_{\rm reh}.
\end{equation}
The maximum reheating temperature is of the order of the inflationary scale $E_{\rm inf}$. Hence in our plots we have the constraint
\begin{align}
T_{\rm osc} \leq E_{\rm inf},
\end{align}
which can be seen as the purple upper-right region in Fig.~\ref{fig:complex_scenario1_inflation1}, \ref{fig:complex_scenario1_inflation2}, \ref{fig:complex_scenario1_inflation3}, \ref{fig:complex_scenario1_local1}, \ref{fig:complex_scenario1_local2}, \ref{fig:complex_scenario1_global1}, and \ref{fig:complex_scenario1_global2}.

\paragraph{No second inflation.} 
In order for the scalar field to not induce a second period of slow-roll inflation, we must impose $\Phi$ to be sub-dominant at the onset of the oscillation
\begin{equation}
V(\phi_{\rm ini}) ~<  ~\frac{\mreff^2 M_{\rm pl}^2}{3}.
\label{eq:no_second_inflation}
\end{equation}
Note that this condition also guarantees that the initial radial field excursion is never superplanckian.

\begin{figure}[t]
\centering
\raisebox{0cm}{\makebox{\includegraphics[width=0.9\textwidth, scale=1]{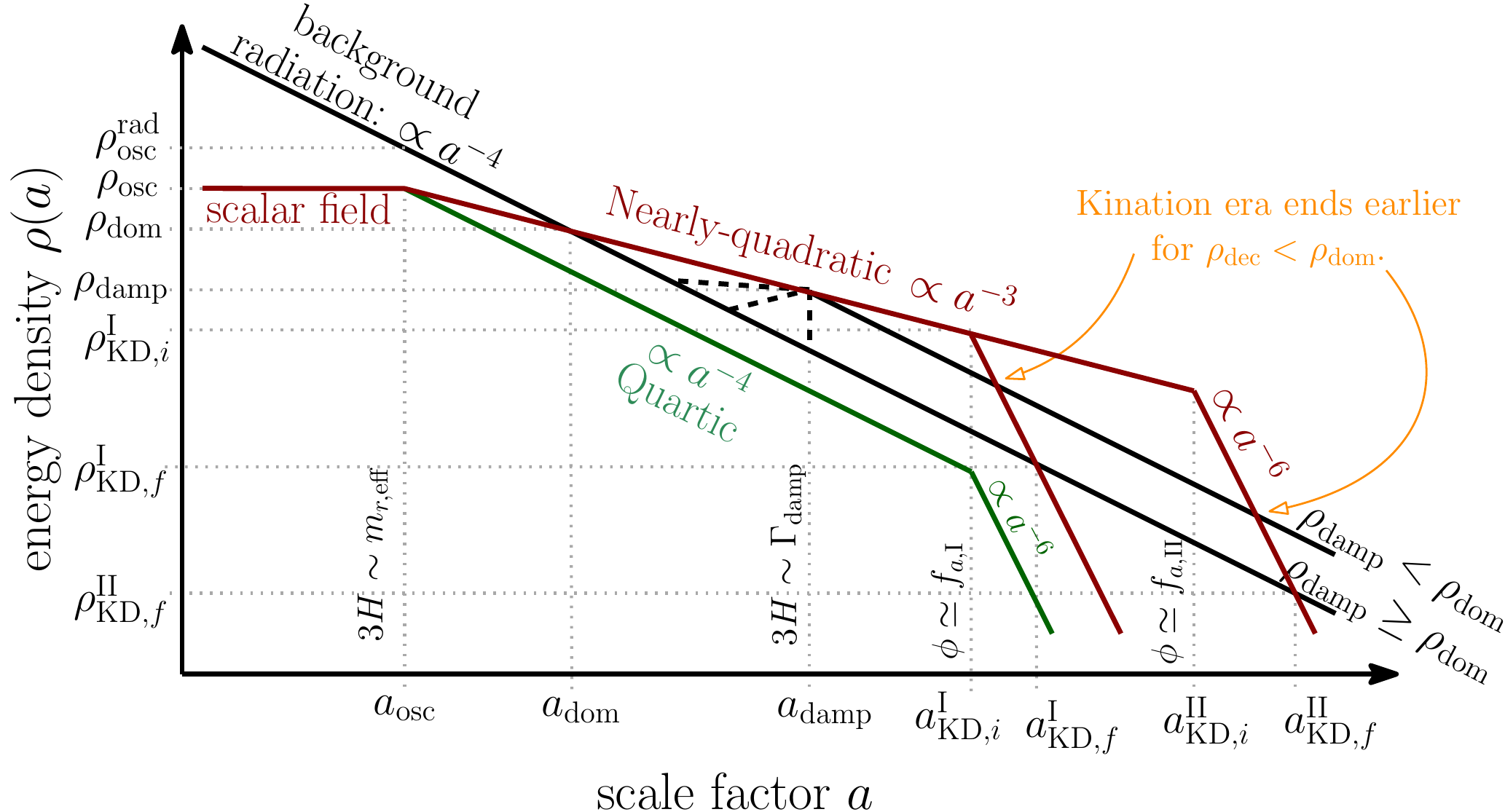}}}  
\caption{\textit{ \small Evolution of energy densities of SM radiation (black) and the complex scalar field in nearly-quadratic potential (red) and quartic potential (green).
A complex scalar field evolving in a quartic potential redshifts like radiation, see App.~\ref{app:field_traj_kick} and \ref{paragraph:quartic_potential} for the analytical justification, and can never generate a kination-dominated era.
Whenever it occurs after domination, $\rho_\mathrm{damp} <  \rho_{\rm dom}$, the radial damping heats the thermal bath (black dashed lines), which reduces the duration of the kination era. In contrast, the kination duration is optimized for $\rho_\mathrm{damp} >  \rho_{\rm dom}$.
We show two $f_a$-values, $f_{a,\mathrm{I}} > f_{a,\mathrm{II}}$, corresponding to  two durations of kination, $N_\mathrm{KD,I} < N_\mathrm{KD,II}$.
}}
\label{diagram_kination}
\end{figure}

\paragraph{Matter domination.} 
The scalar field redshifts like matter and dominates the energy density of the universe at
\begin{equation}
\rho_{\rm dom} ~ = ~  \frac{27\mreff(\phi_{\rm ini})^2 \phi_{\rm ini}^8}{16\MPl^6}  A_{\epsilon}^4
\label{cosmohist_zero_temp_matter_dom} \qquad \textrm{and} \qquad
\frac{a_{\rm dom}}{a_\textrm{osc}} ~ = ~ \frac{2\MPl^2 }{3 \phi_{\rm ini}^2}A_\epsilon^{-1}
\end{equation}
with
\begin{equation}
A_\epsilon = \begin{cases}
\epsilon , \qquad \qquad \text{if}~\rho_{\rm damp} > \rho_{\rm dom},\\[0.75em]
1, \qquad \qquad \text{if}~\rho_{\rm damp} < \rho_{\rm dom},
\end{cases}
\end{equation}
where $\epsilon$ is the amount of Noether charge, defined by Eq.~\eqref{eq:epsilon_def} and whose value is dynamically generated at the onset of oscillation, see Eq.~\eqref{eq:epsilon_1_zero_temp}. The impact of $\epsilon$ on the evolution of the scalar field energy density is discussed in  App.~\ref{app:effect_epsilon_evolution}.

\paragraph{Radial damping.} 
Denoting by $\Gamma$ the rate at which the radial motion is damped by some unspecified mechanism, cf.~Sec.~\ref{sec:themalization}, we obtain that the trajectory becomes circular when
\begin{equation}
\rho_{\rm damp}~ = ~ 3\MPl^2\Gamma^2\,B_{\epsilon}^4, \qquad \textrm{and} \qquad 
 \frac{a_{\rm damp}}{a_{\rm dom}} ~ = ~ \left(\frac{ \rho_{\rm dom}}{ \rho_{\rm damp}}\right)^{1/3},
\end{equation}
with
\begin{equation}
B_\epsilon = \begin{cases}
1 , \qquad \qquad \text{if}~\rho_{\rm damp} > \rho_{\rm dom},\\[0.75em]
\epsilon, \qquad \qquad \text{if}~\rho_{\rm damp} < \rho_{\rm dom}.
\end{cases}
\end{equation}
In this section, $\Gamma$ is considered as a free parameter and we assume that damping can occur before the onset of the matter domination, $\rho_{\rm damp} > \rho_{\rm dom}$.

\FloatBarrier
\begin{figure}[h!]
\centering
\raisebox{0cm}{\makebox{\includegraphics[width=0.6\textwidth, scale=1]{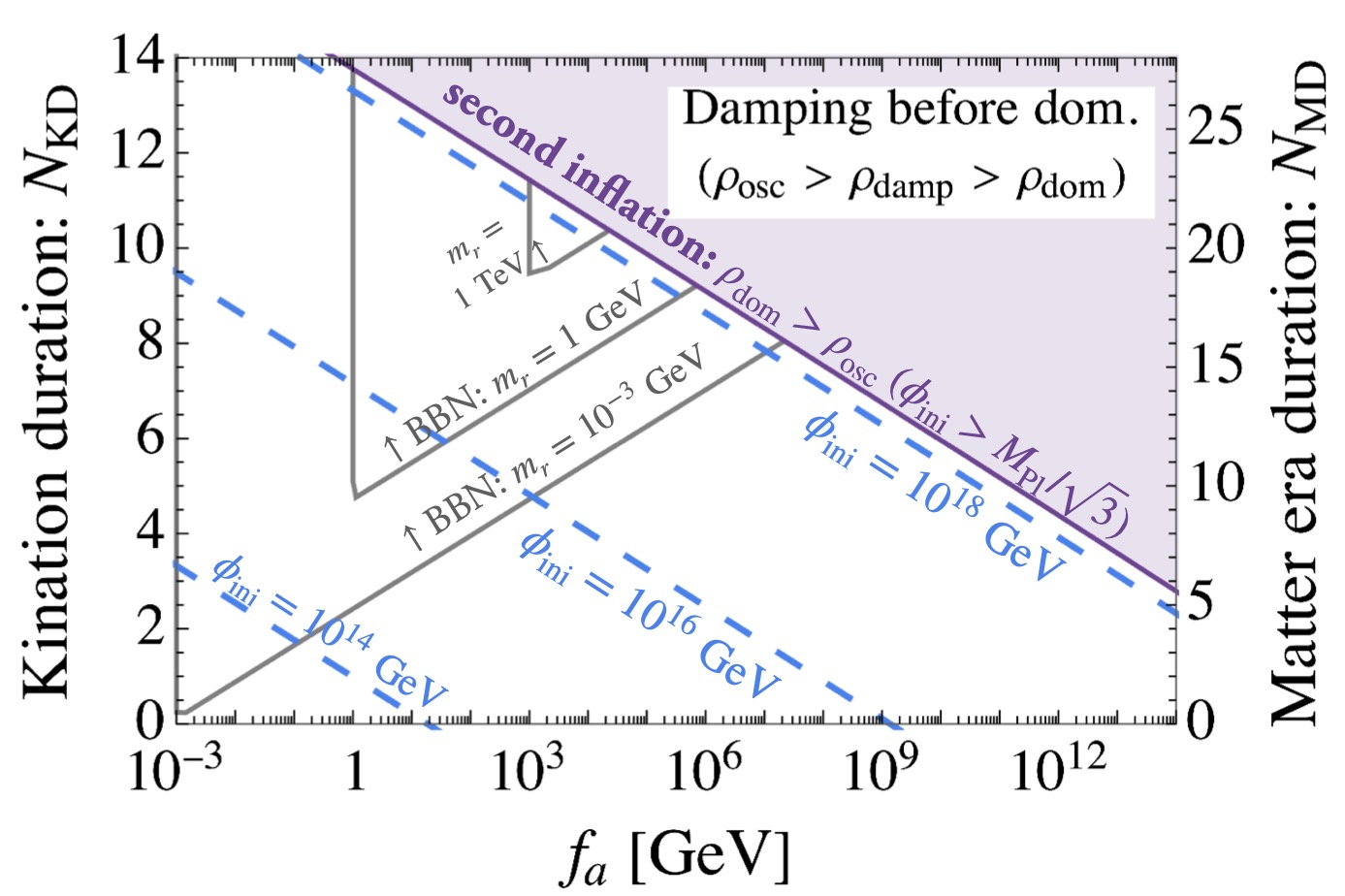}}} 
\caption{\textit{ \small 
Number of e-folds $N_{\rm MD}$ and $N_{\rm KD}$ of the matter and kination eras, for different values of the initial and final values of the radial mode $\phi_{\rm ini}$ and $f_a$, cf.  Eq.~\eqref{eq:longest_NKS_scenario_1} and \eqref{eq:NMD_duration}.
The purple region is excluded because the  potential energy of the scalar field dominates the universe before oscillation and leads to a second inflationary stage, cf. Eq.~\eqref{eq:no_second_inflation}.
For a given radial mass $m_r$, the region above the gray lines is excluded because the kination era ends after BBN.
The vertical cut is due to perturbativity violation $m_r > f_a$.
}}
\label{inter_duration}
\end{figure}
\FloatBarrier

\subsubsection{Duration of kination}

\paragraph{Start of kination.} 
The universe acquires a kination equation-of-state when the field reaches $\phi \to f_a$, corresponding to the energy density
\begin{equation}
 \rho_{\textrm{KD},i} ~ = ~  \frac{1}{2}f_a^2 \mreff ^2(f_a).
\end{equation}
Depending on whether radial damping occurs before or after the onset of matter domination, we obtain
\begin{equation}
\frac{a_{\textrm{KD},i}}{\max(a_{\rm dom},a_{\rm damp})} = \left(\frac{\text{min}(\rho_{\rm dom},\,\rho_{\rm damp})}{\rho_\textrm{KD,i}}\right)^{\!1/3}.
\label{eq:start_kination_scale_factor}
\end{equation} 
\paragraph{End of kination.} 
The kination era stops when the universe becomes radiation-dominated again. The energy scale at which it occurs depends on whether radial damping occurs before and after the onset of matter domination
\begin{equation}
\rho_{\textrm{KD},f} =\frac{\rho_{\textrm{KD},i}^2}{\text{min}(\rho_{\rm dom},\,\rho_{\rm damp})}= 
\begin{cases}
 \frac{4f_a^4 \mreff ^4(f_a) \MPl^6 }{27 \mreff^2(\phi_{\rm ini}) \phi_{\rm ini}^8} \left(\frac{1}{\epsilon} \right)^{4}, \qquad \qquad \qquad \quad\text{if}~\rho_{\rm damp} > \rho_{\rm dom},\\[0.75em]
 \frac{f_a^4 m_\textrm{eff} ^4(f_a)}{12 \MPl^2 \Gamma_{\rm damp}^2} \left(\frac{1}{\epsilon} \right)^{4}, ~~~\qquad \qquad \qquad \qquad \text{if}~\rho_{\rm damp} < \rho_{\rm dom}.\\
\end{cases}
\label{eq:modelB_rhof}
\end{equation}
The duration of the kination era $N_{\rm KD} \equiv \log{a_{\textrm{KD},f}/a_{\textrm{KD},i}}$ reads 
\begin{equation}
e^{N_{\rm KD}}  = \left(\frac{\text{min}(\rho_{\rm dom},\,\rho_{\rm damp})}{\rho_{\textrm{KD},i}}\right)^{\!1/6}=
\begin{cases}
\label{eq:end_kination_scale_factor}
\sqrt{\frac{3}{2}} \left(\frac{\mreff(\phi_{\rm ini})}{\mreff(f_a)} \frac{M_{\rm pl}}{f_a}\right)^{1/3}  \left( \frac{\phi_{\rm ini}}{M_{\rm pl}} \right)^{4/3}\epsilon^{2/3},   \qquad \text{if}~\rho_{\rm damp} > \rho_{\rm dom},\\[0.75em]
\left(\frac{6 \MPl^2 \Gamma_{\rm damp}^2}{f_a^2 \mreff ^2(f_a)}\right)^{1/6} \epsilon^{2/3},  ~~~\qquad \qquad\qquad \qquad \text{if}~\rho_{\rm damp} < \rho_{\rm dom}.
\end{cases}
\end{equation}
The first line of Eq.~\eqref{eq:end_kination_scale_factor}, corresponding to efficient radial damping before scalar field domination, gives the longest duration of kination for the complex scalar field model studied in this work
\begin{framed}
\begin{align}
\rho_{\rm damp} > \rho_{\rm dom}
\qquad
\implies \qquad e^{N_{\rm KD}} \simeq
& e^{8.2}~\epsilon^{2/3}  \left( \frac{10^{9}~\rm GeV}{f_a}\right)^{1/3}   \left(\frac{\mreff(\phi_{\rm ini})}{5\mreff(f_a)} \right)^{1/3} \left( \frac{\phi_{\rm ini}}{M_{\rm pl}} \right)^{4/3},\label{eq:longest_NKS_scenario_1}
\end{align}
\end{framed}\noindent
where $\epsilon$ can be $\mathcal{O}(1)$, see Eq.~\eqref{eq:epsilon_1_zero_temp}.

Note that for efficient thermalization, the number of matter e-folds $N_{\rm MD} \equiv \log{a_{\textrm{KD},i}/a_{\rm dom}}$, cf. first line of Eq.~\eqref{eq:start_kination_scale_factor}, verifies the property
\begin{equation}
\rho_{\rm damp} > \rho_{\rm dom} \qquad \implies \qquad N_{\rm MD}  = 2N_{\rm KD}. \label{eq:NMD_duration}
\end{equation}
The durations of the matter and kination eras are shown in Fig.~\ref{inter_duration}.

\FloatBarrier
\begin{figure}[h!]
\centering
\raisebox{0cm}{\makebox{\includegraphics[width=0.6\textwidth, scale=1]{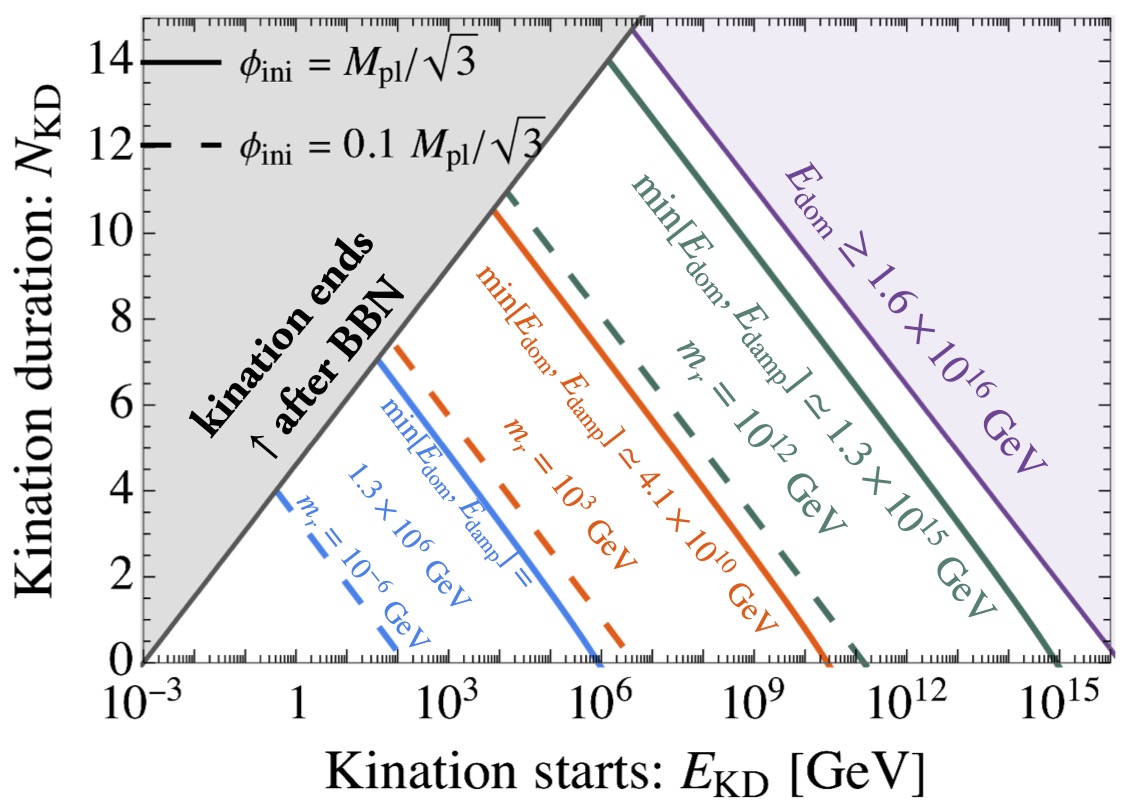}}} 
\caption{\textit{ \small 
The duration of kination $N_{\rm KD}$, cf. Eq.~\eqref{eq:end_kination_scale_factor}, depends on the energy scale at the `effective' start of the matter era $\textrm{Min}\left[E_{\rm dom},~E_{\rm damp}\right]$ and at the onset of the kination era $E_\textrm{KD} = (2 m_r^2 f_a^2)^{1/4}$. If $E_{\rm damp} < E_{\rm dom}$, entropy injection at the time of damping re-equilibrates the amount of matter vs radiation in the universe, hence the `\textrm{Min}' function. The larger $m_r$, the larger $E_{\rm dom}$ and the longer the duration of the matter-kination era. In the present Sec.~\ref{sec:scenario_I_non_thermal_damping}, we consider $E_{\rm damp}$ as a free parameter. The maximum duration of kination compatible with both BBN and the maximal inflationary scale allowed by CMB data, $N_{\rm KD} \lesssim 14.6$, is computed precisely in App.~\ref{sec:CMB_BBN_bound_NKD}.
}}
\label{inter_duration_2}
\end{figure}
\FloatBarrier

\subsection{Gravitational-wave signature and detectability}
 
 As discussed in Sec.~\ref{sec:modelindependent}, a main motivation for generating a kination era is the enhancement of the amplitude of SGWB produced beforehand.
The next figures show the detectability of SGWB produced by primordial inflation (Figs.~\ref{fig:complex_scenario1_inflation1}, \ref{fig:complex_scenario1_inflation2}, and \ref{fig:complex_scenario1_inflation3}), local strings (Figs.~\ref{fig:complex_scenario1_local1} and \ref{fig:complex_scenario1_local2}) and globals strings (Figs. \ref{fig:complex_scenario1_global1} and \ref{fig:complex_scenario1_global2}), in the presence of a kination era generated by the scenario I: a spinning complex scalar field assuming the existence of an efficient non-thermal damping mechanism that enables to neglect thermal corrections to the potential.

\FloatBarrier
\begin{figure}[h!]
\centering
{\bf Gravitational waves from primordial inflation: $E_{\rm inf} = 1.6 \times 10^{16} ~{\rm GeV}$}\\[0.25em]
\raisebox{0cm}{\makebox{\includegraphics[width=0.95\textwidth, scale=1]{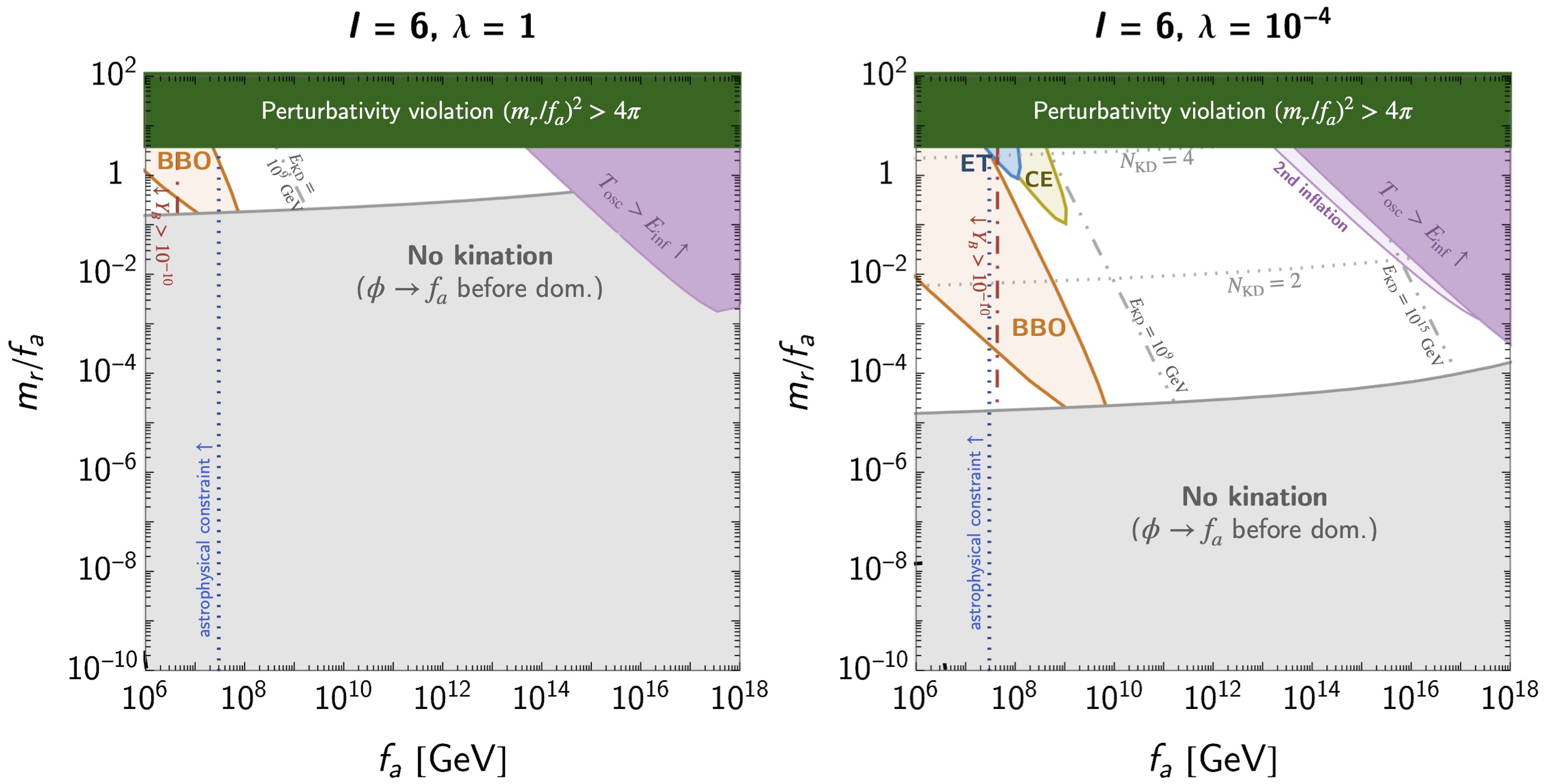}}}\\[0.25em]
\raisebox{0cm}{\makebox{\includegraphics[width=0.475\textwidth, scale=1]{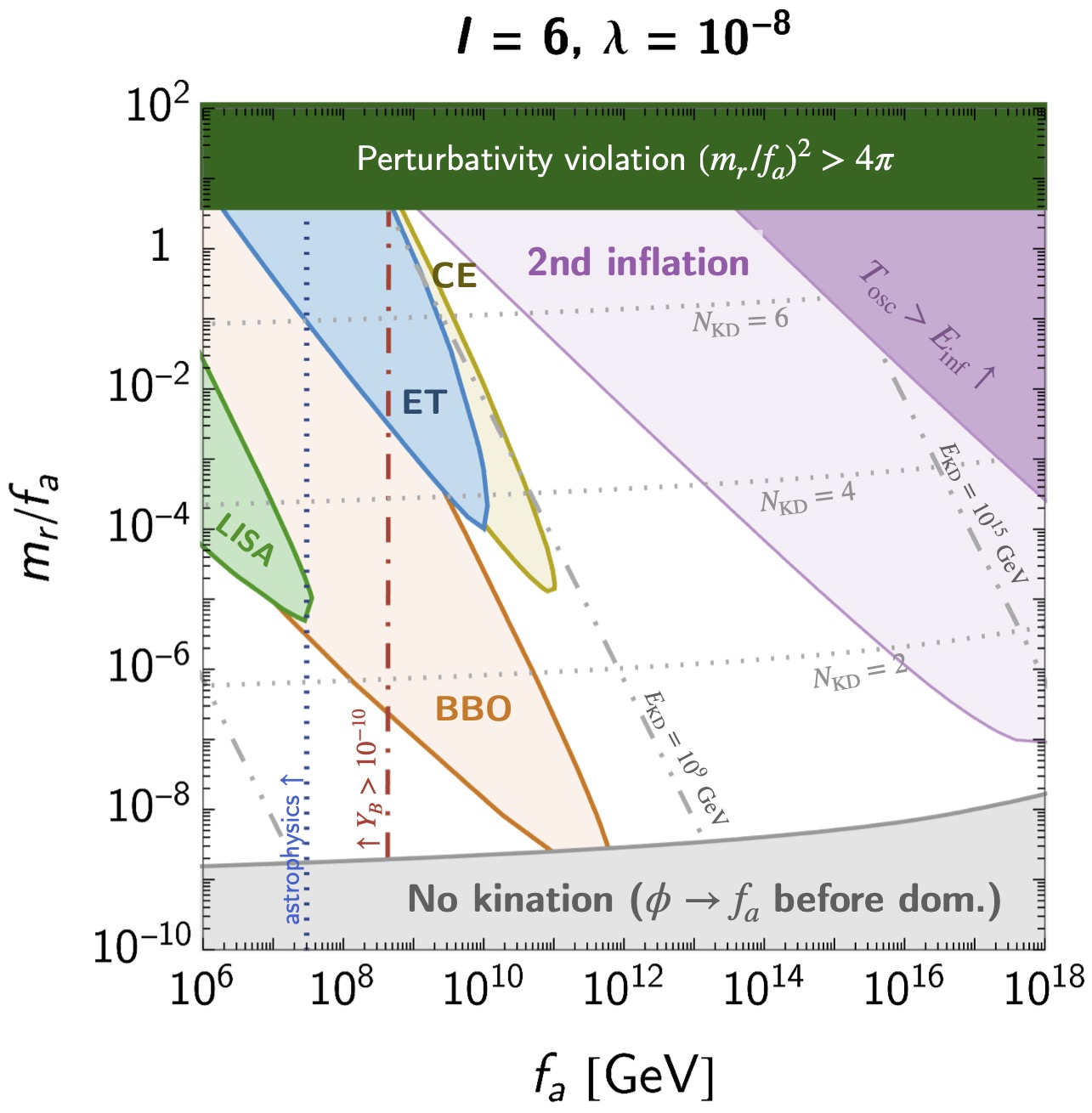}}}\\[0.25em]
\caption{\textit{ \small  
Ability of future-planned GW observatories to detect the peak signature  in the SGWB from primordial inflation  with energy scale $E_{\rm inf}$ of a matter-kination  era induced by \textbf{scenario I}. In this scenario,  a  kick in the angular direction of a complex scalar field is induced with a large radial value by operators of order $l=6$ and self-coupling $\lambda$, and the radial motion is assumed to be damped non-thermally.  For $l=6$, these plots only apply to non-QCD axions as the QCD axion is always overabundant, produced either from standard or kinetic misalignment mechanism, when imposing  \eqref{eq:quality_problem_condition} from the axion quality problem. A dotted-dashed red line denotes the parameter space where the spinning axion allows the correct baryon asymmetry, cf. Eq.~\eqref{Ekd_yield_bau}. Gray dotted and dot-dashed lines show contours of kination duration $N_{\rm KD}$ and energy scale $E_{\rm KD}$, respectively. Smaller $m_r$ and $\lambda$ implies larger initial scalar vev $\phi_{\rm ini}$, cf. Eq.~\eqref{susy_phi_ini}, and longer matter-kination. }}
\label{fig:complex_scenario1_inflation1}
\end{figure}
\FloatBarrier

\FloatBarrier
\begin{figure}[h!]
\centering
{\bf Gravitational waves from primordial inflation: $E_{\rm inf} = 1.6 \times 10^{16} ~{\rm GeV}$}\\[0.25em]
\raisebox{0cm}{\makebox{\includegraphics[width=0.95\textwidth, scale=1]{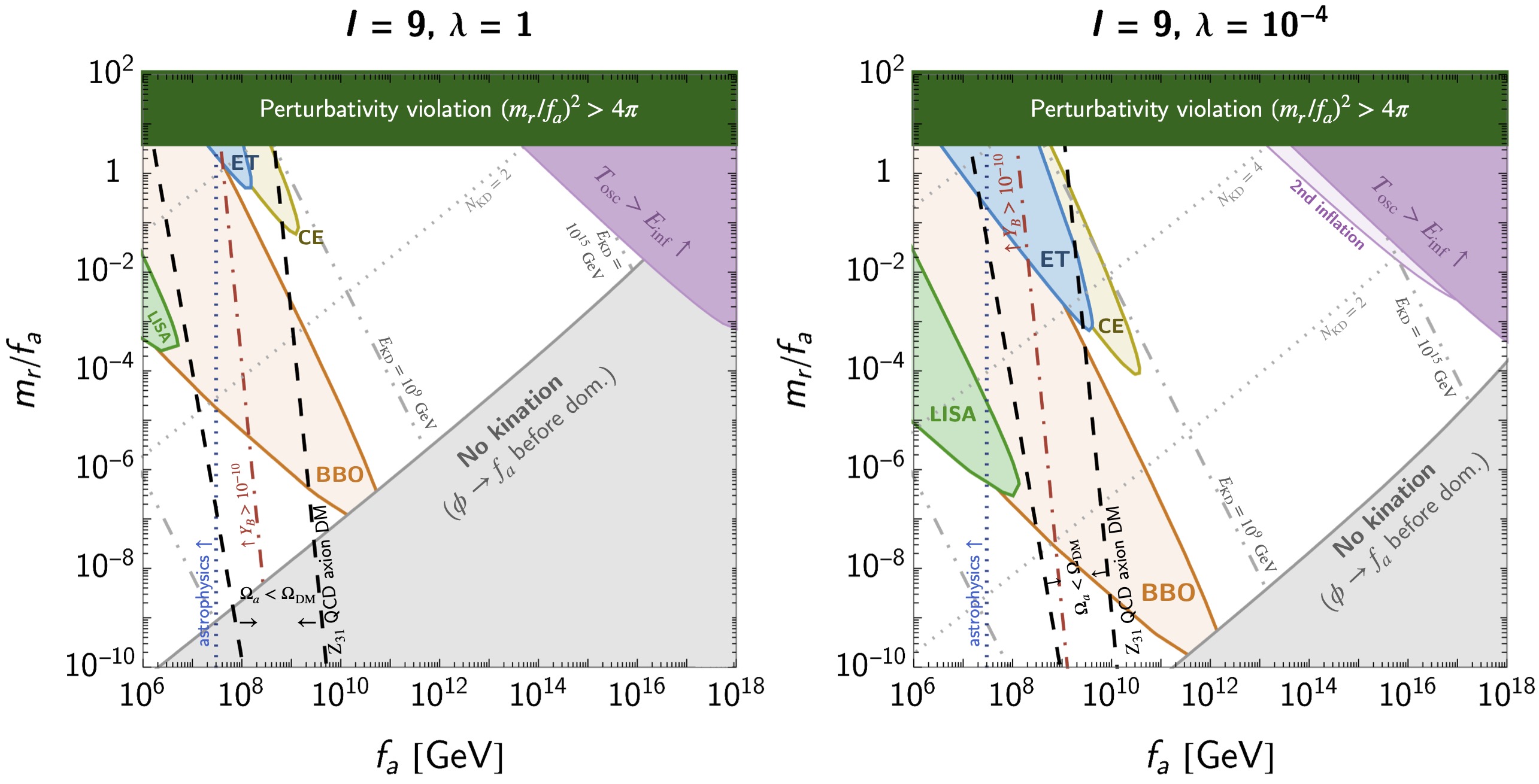}}}\\[0.25em]
\raisebox{0cm}{\makebox{\includegraphics[width=0.475\textwidth, scale=1]{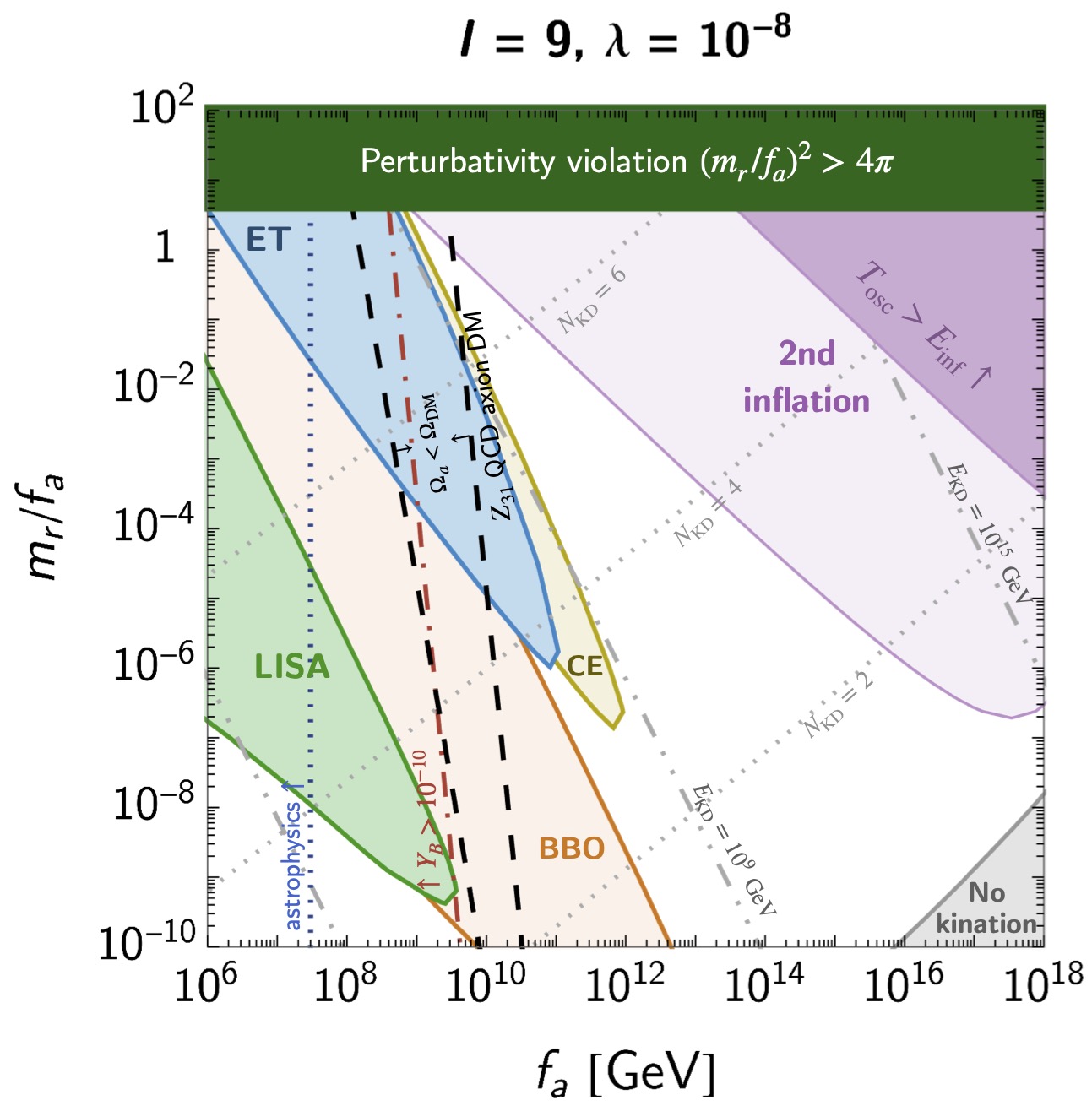}}}\\[0.25em]
\caption{\textit{ \small  Same as Fig.~\ref{fig:complex_scenario1_inflation1} for $l=9$. Black dashed lines indicate where the lighter non-canonical QCD axion abundance in  Eq.~\eqref{eq:lighter_axion_mass} is satisfied, cf. Eq.~\eqref{eq:axion_abundance_general}. The left boundary is set by the kinetic misalignment mechanism, while the right one is set by the axion quality problem (for larger $m_r$ depending on $f_a$), cf. Eq.~\eqref{eq:quality_problem_condition}. Only the region between these two lines does not over-produce DM.}}
\label{fig:complex_scenario1_inflation2}
\end{figure}
\FloatBarrier

\FloatBarrier
\begin{figure}[h!]
\centering
{\bf Gravitational waves from primordial inflation: $E_{\rm inf} = 1.6 \times 10^{16} ~{\rm GeV}$}\\[0.25em]
\raisebox{0cm}{\makebox{\includegraphics[width=0.95\textwidth, scale=1]{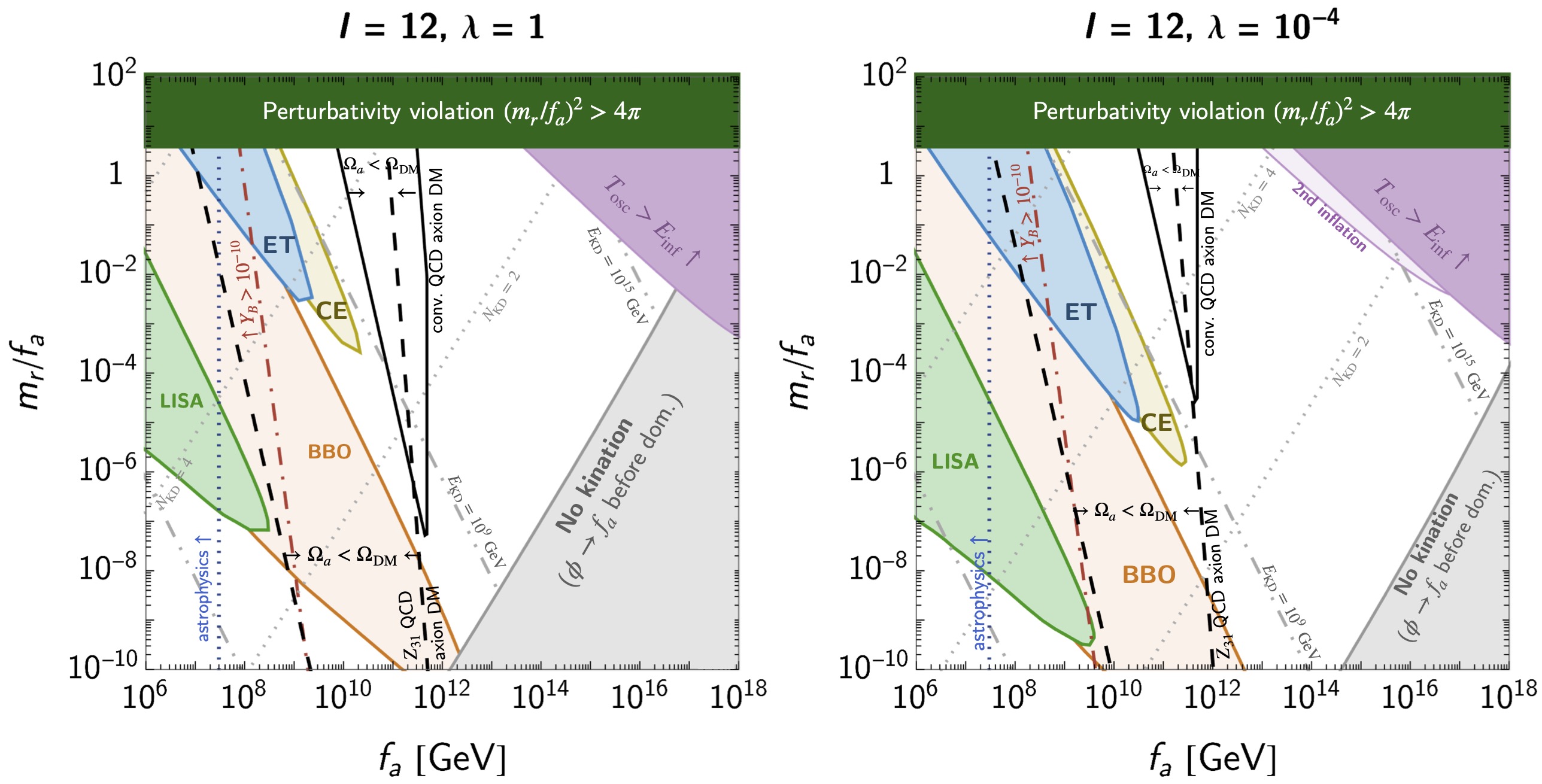}}}\\[0.25em]
\raisebox{0cm}{\makebox{\includegraphics[width=0.475\textwidth, scale=1]{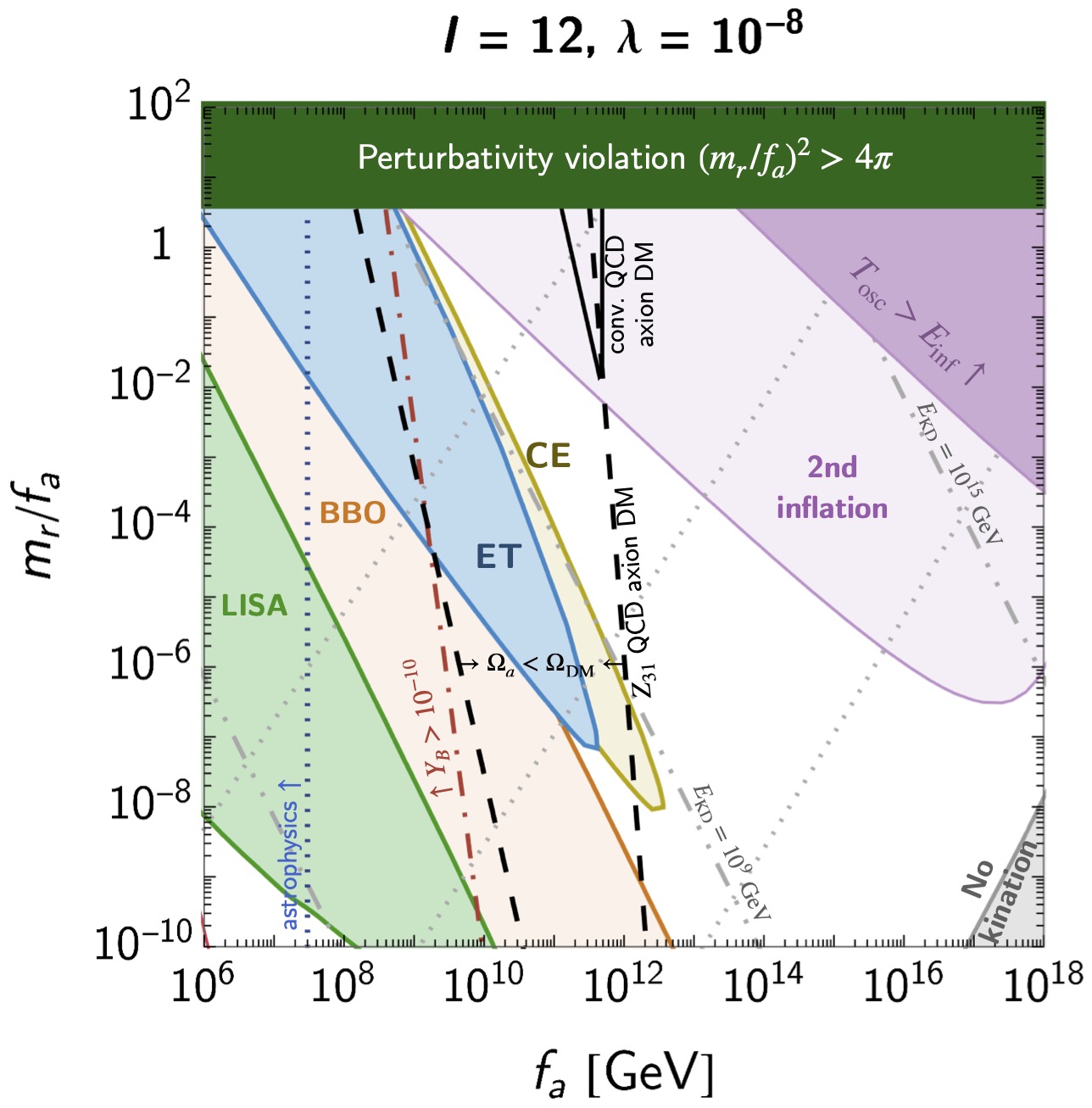}}}\\[0.25em]
\caption{\textit{ \small Same as Fig.~\ref{fig:complex_scenario1_inflation1} for $l=12$. 
Black solid lines indicate where the canonical QCD axion abundance is satisfied, cf. Eq.~\eqref{eq:axion_abundance_general}. The left boundary is set by the kinetic misalignment mechanism, while the right one is set by the standard misalignment (for small $m_r$ with a specific $f_a$) and by the axion quality problem (for larger $m_r$ depending on $f_a$), cf. Eq.~\eqref{eq:quality_problem_condition}. Only the region between the two lines does not over-produce DM. Dashed lines are the equivalent for lighter non-canonical QCD axion,  cf. Eq.~\eqref{eq:lighter_axion_mass}.}}
\label{fig:complex_scenario1_inflation3}
\end{figure}
\FloatBarrier

\FloatBarrier
\begin{figure}[h!]
\centering
{\bf Gravitational waves from local cosmic strings}\\[0.25em]
\raisebox{0cm}{\makebox{\includegraphics[width=0.95\textwidth, scale=1]{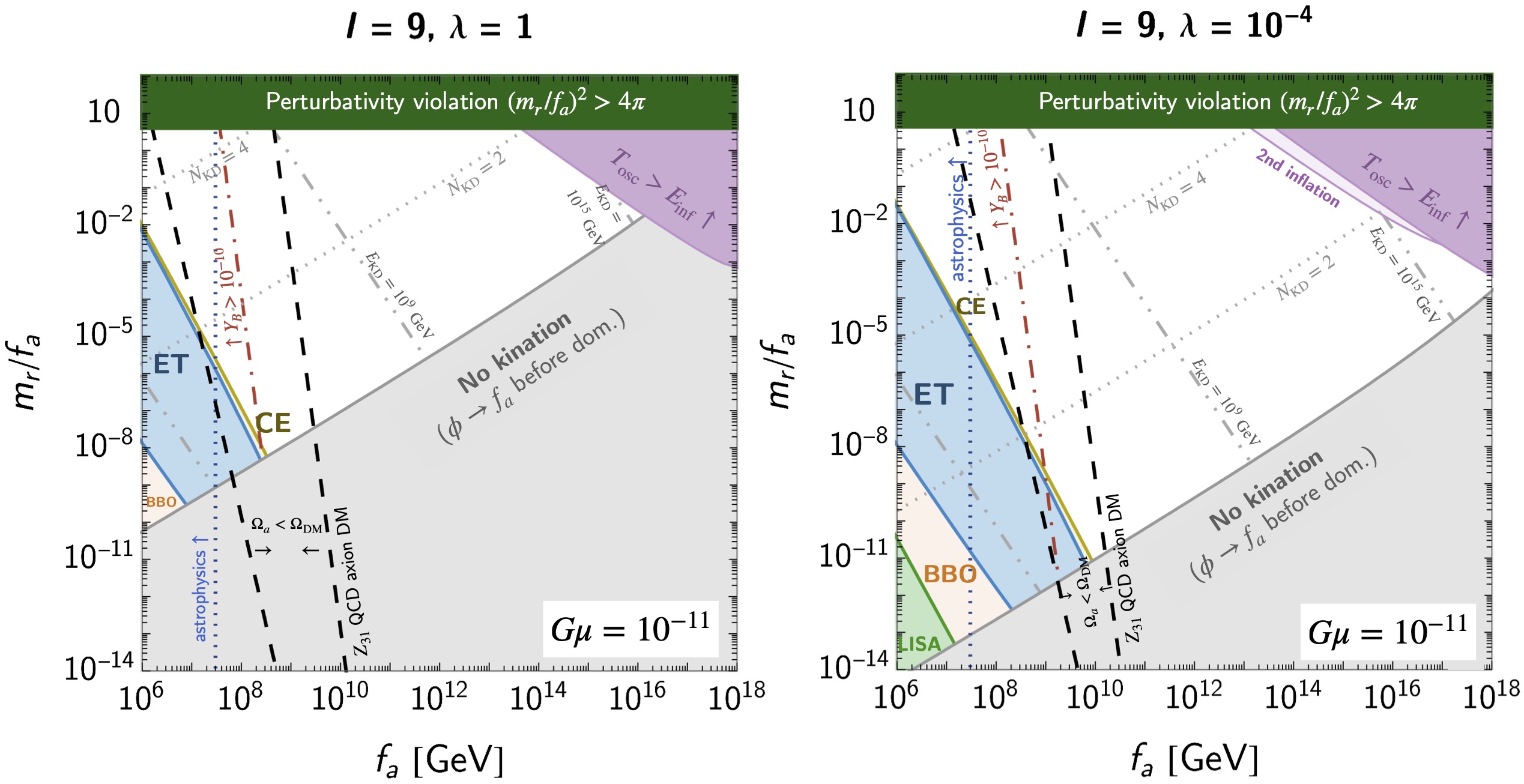}}}\\[0.25em]
\raisebox{0cm}{\makebox{\includegraphics[width=0.95\textwidth, scale=1]{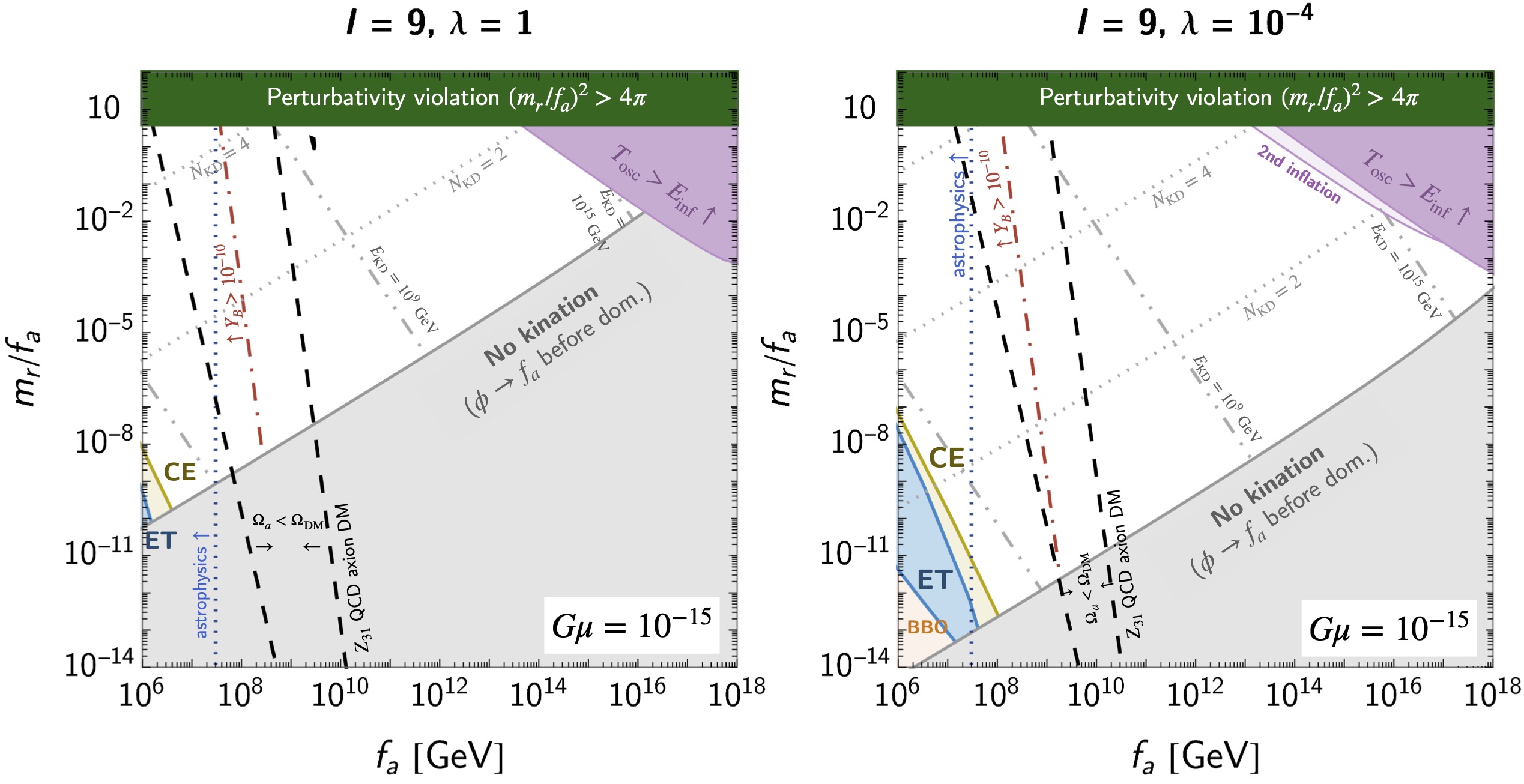}}}\\[0.25em]
\caption{\textit{ \small 
Ability of future-planned GW observatories to detect the peak signature  in the SGWB from local cosmic strings with tension $G\mu$ of a matter-kination  era induced by \textbf{scenario I}. In this scenario,  a  kick in the angular direction of a complex scalar field is induced with a large radial value by operators of order $l=9$ and self-coupling $\lambda$, and the radial motion is assumed to be damped non-thermally.
Black dashed lines indicate where the lighter non-canonical QCD axion abundance in  Eq.~\eqref{eq:lighter_axion_mass} is satisfied, cf. Eq.~\eqref{eq:axion_abundance_general}. The left boundary is set by the kinetic misalignment mechanism, while the right one is set by the axion quality problem (for larger $m_r$ depending on $f_a$), cf. Eq.~\eqref{eq:quality_problem_condition}. Only the region between the two lines does not over-produce DM. A dotted-dashed red line denotes the parameter space where the spinning axion allows the correct baryon asymmetry, cf. Eq.~\eqref{Ekd_yield_bau}. Gray dotted and dot-dashed lines show contours of kination duration $N_{\rm KD}$ and energy scale $E_{\rm KD}$, respectively. Smaller $m_r$ and $\lambda$ implies larger initial scalar vev $\phi_{\rm ini}$, cf. Eq.~\eqref{susy_phi_ini}, and longer matter-kination.}}
\label{fig:complex_scenario1_local1}
\end{figure}
\FloatBarrier

\FloatBarrier
\begin{figure}[h!]
\centering
{\bf Gravitational waves from local cosmic strings}\\[0.25em]
\raisebox{0cm}{\makebox{\includegraphics[width=0.95\textwidth, scale=1]{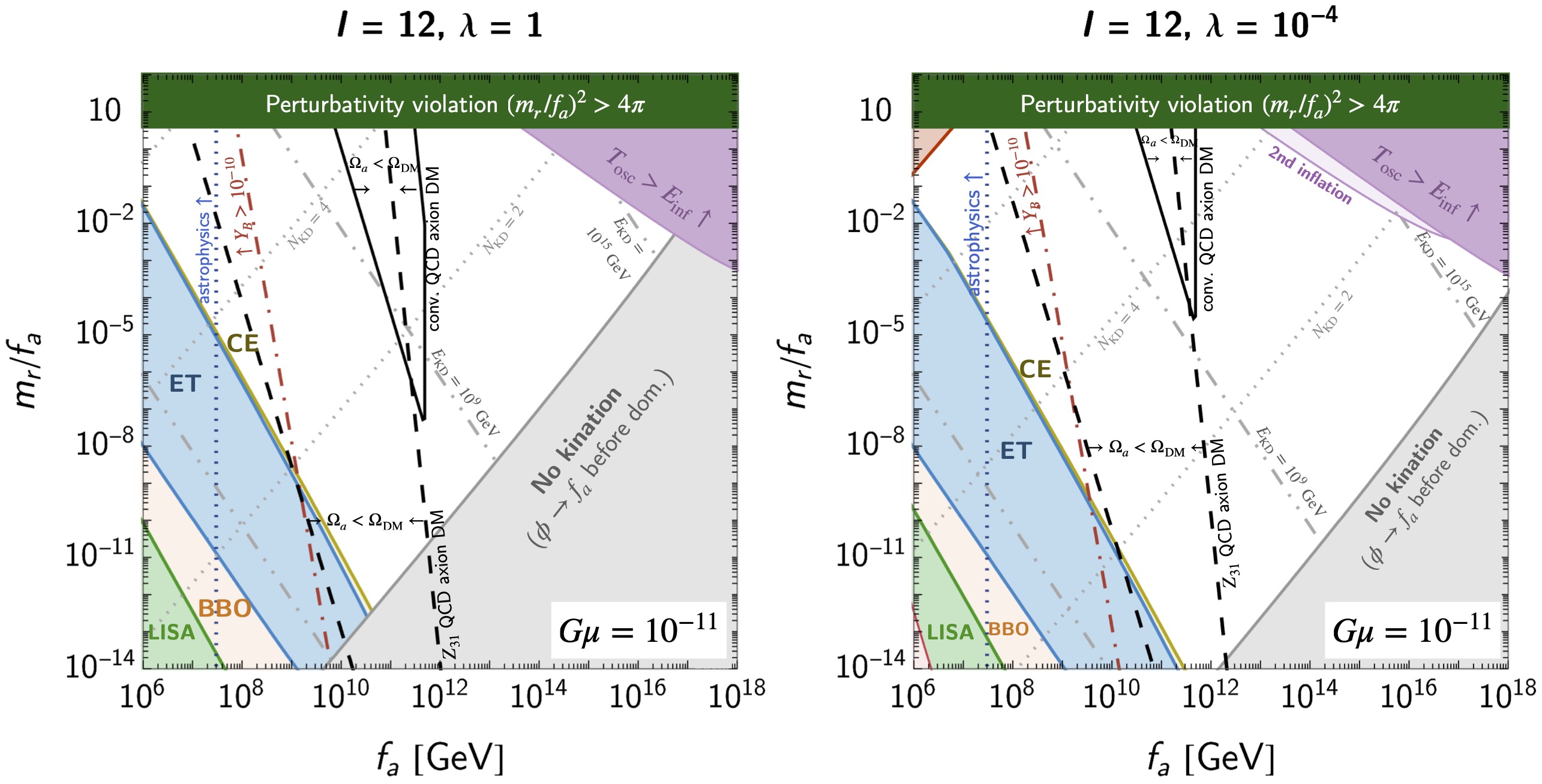}}}\\[0.25em]
\raisebox{0cm}{\makebox{\includegraphics[width=0.95\textwidth, scale=1]{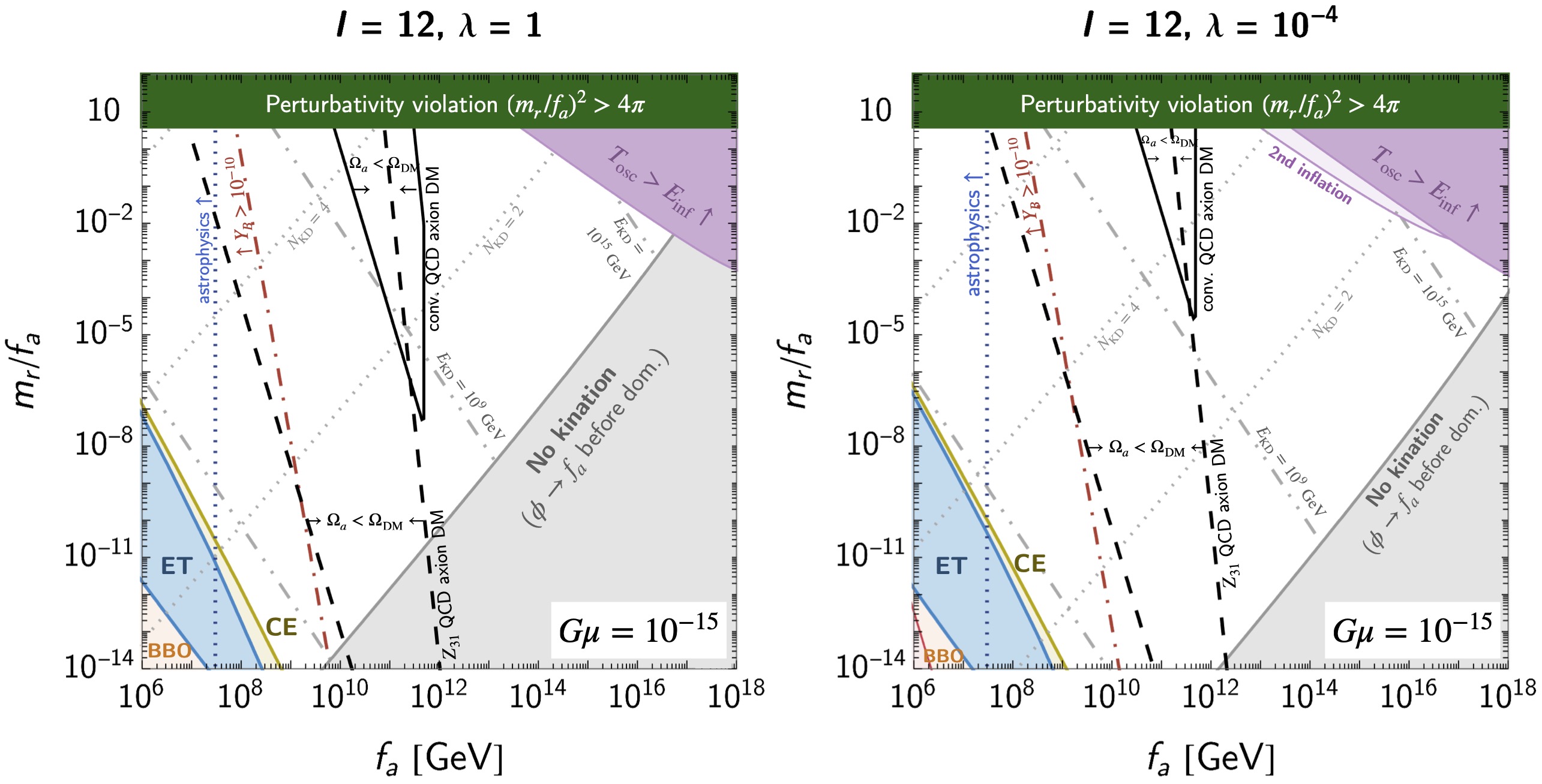}}}\\[0.25em]
\caption{\textit{ \small Same as Fig.~\ref{fig:complex_scenario1_local1} for $l=12$. Black solid lines indicate where the canonical QCD axion abundance is satisfied, cf. Eq.~\eqref{eq:axion_abundance_general}. The left boundary is set by the kinetic misalignment mechanism, while the right one is by the standard misalignment (for small $m_r$ with a specific $f_a$) and by the axion quality problem (for larger $m_r$ depending on $f_a$), cf. Eq.~\eqref{eq:quality_problem_condition}. Only the region between the two lines does not over-produce DM. Dashed lines are the equivalent for lighter non-canonical QCD axion,  cf. Eq.~\eqref{eq:lighter_axion_mass}.}}
\label{fig:complex_scenario1_local2}
\end{figure}
\FloatBarrier

\FloatBarrier
\begin{figure}[h!]
\centering
{\bf Gravitational waves from global cosmic strings}\\[0.25em]
\raisebox{0cm}{\makebox{\includegraphics[width=0.95\textwidth, scale=1]{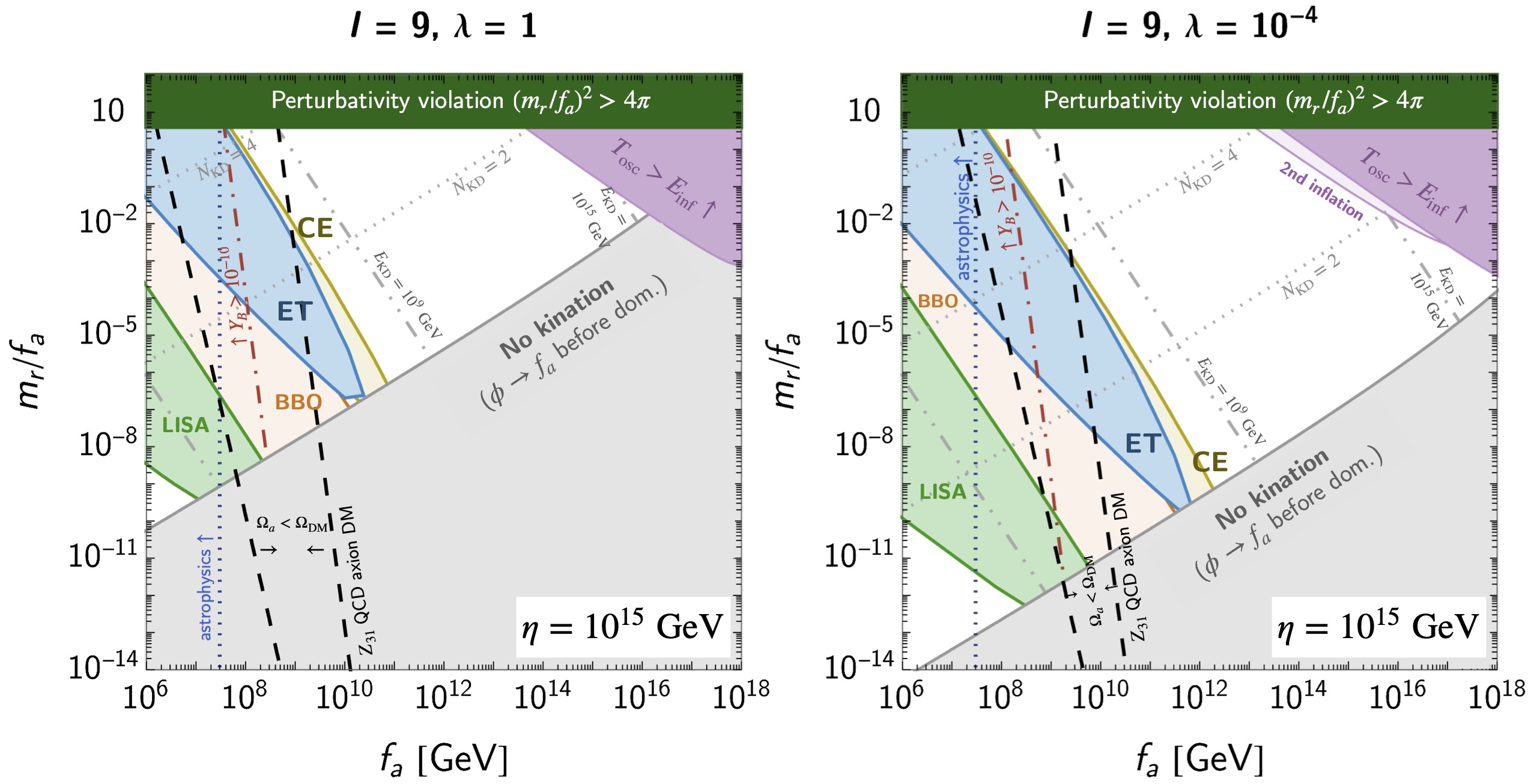}}}\\[0.25em]
\raisebox{0cm}{\makebox{\includegraphics[width=0.95\textwidth, scale=1]{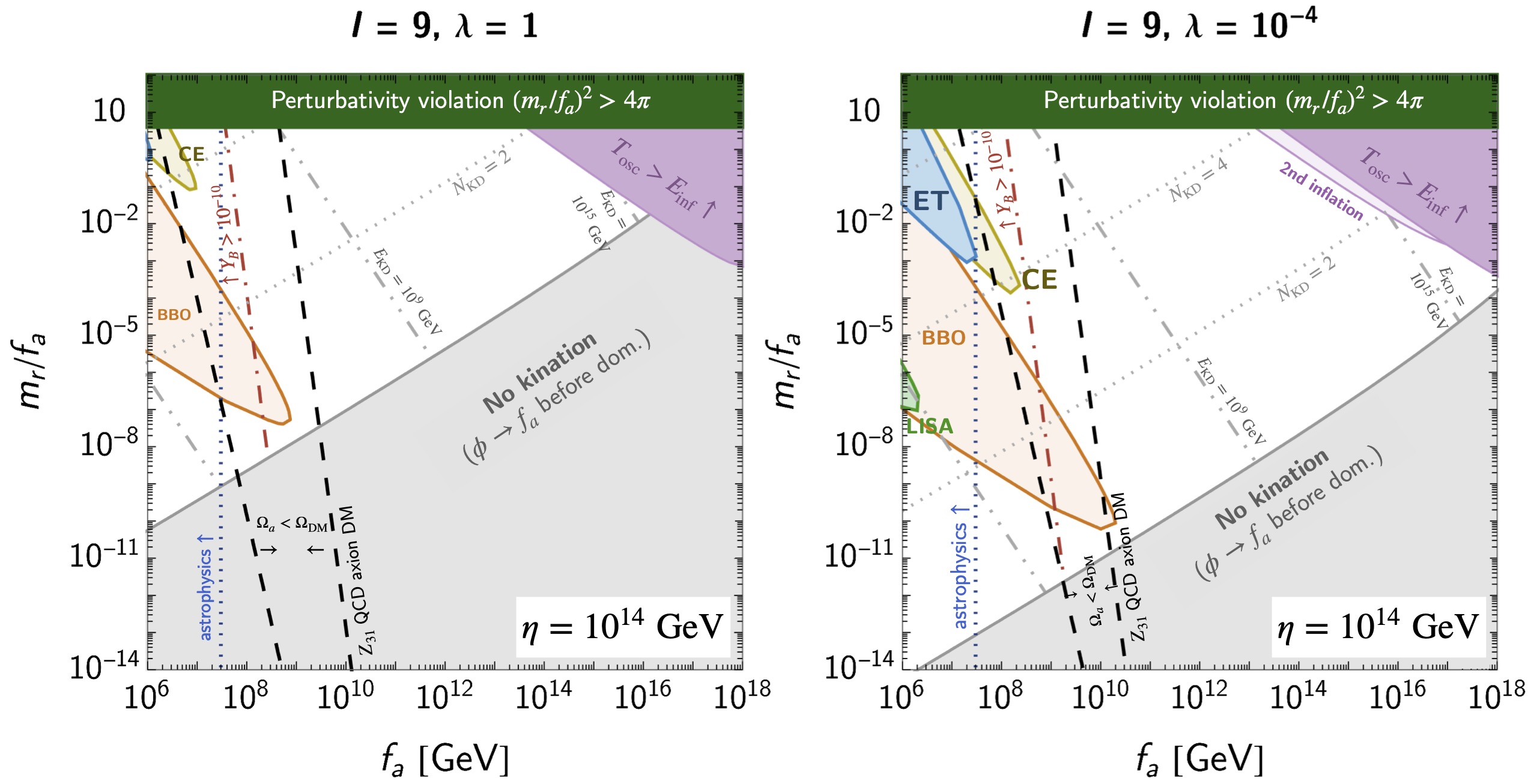}}}\\[0.25em]
\caption{\textit{ \small 
Ability of future-planned GW observatories to detect the peak signature  in the SGWB from global cosmic strings with string scale $\eta$ of a matter-kination  era induced by \textbf{scenario I}. In this scenario,  a  kick in the angular direction of a complex scalar field is induced with a large radial value by operators of order $l=9$ and self-coupling $\lambda$, and the radial motion is assumed to be damped non-thermally.
Black dashed lines indicate where the lighter non-canonical QCD axion abundance in  Eq.~\eqref{eq:lighter_axion_mass} is satisfied, cf. Eq.~\eqref{eq:axion_abundance_general}. The left boundary is set by the kinetic misalignment mechanism, while the right one is set by the axion quality problem (for larger $m_r$ depending on $f_a$), cf. Eq.~\eqref{eq:quality_problem_condition}. Only the region between the two lines does not over-produce DM. A dotted-dashed red line denotes the parameter space where the spinning axion allows the correct baryon asymmetry, cf. Eq.~\eqref{Ekd_yield_bau}. Gray dotted and dot-dashed lines show contours of kination duration $N_{\rm KD}$ and energy scale $E_{\rm KD}$, respectively. Smaller $m_r$ and $\lambda$ implies larger initial scalar vev $\phi_{\rm ini}$, cf. Eq.~\eqref{susy_phi_ini}, and longer matter-kination. }}
\label{fig:complex_scenario1_global1}
\end{figure}
\FloatBarrier

\FloatBarrier
\begin{figure}[h!]
\centering
{\bf Gravitational waves from global cosmic strings}\\[0.25em]
\raisebox{0cm}{\makebox{\includegraphics[width=0.95\textwidth, scale=1]{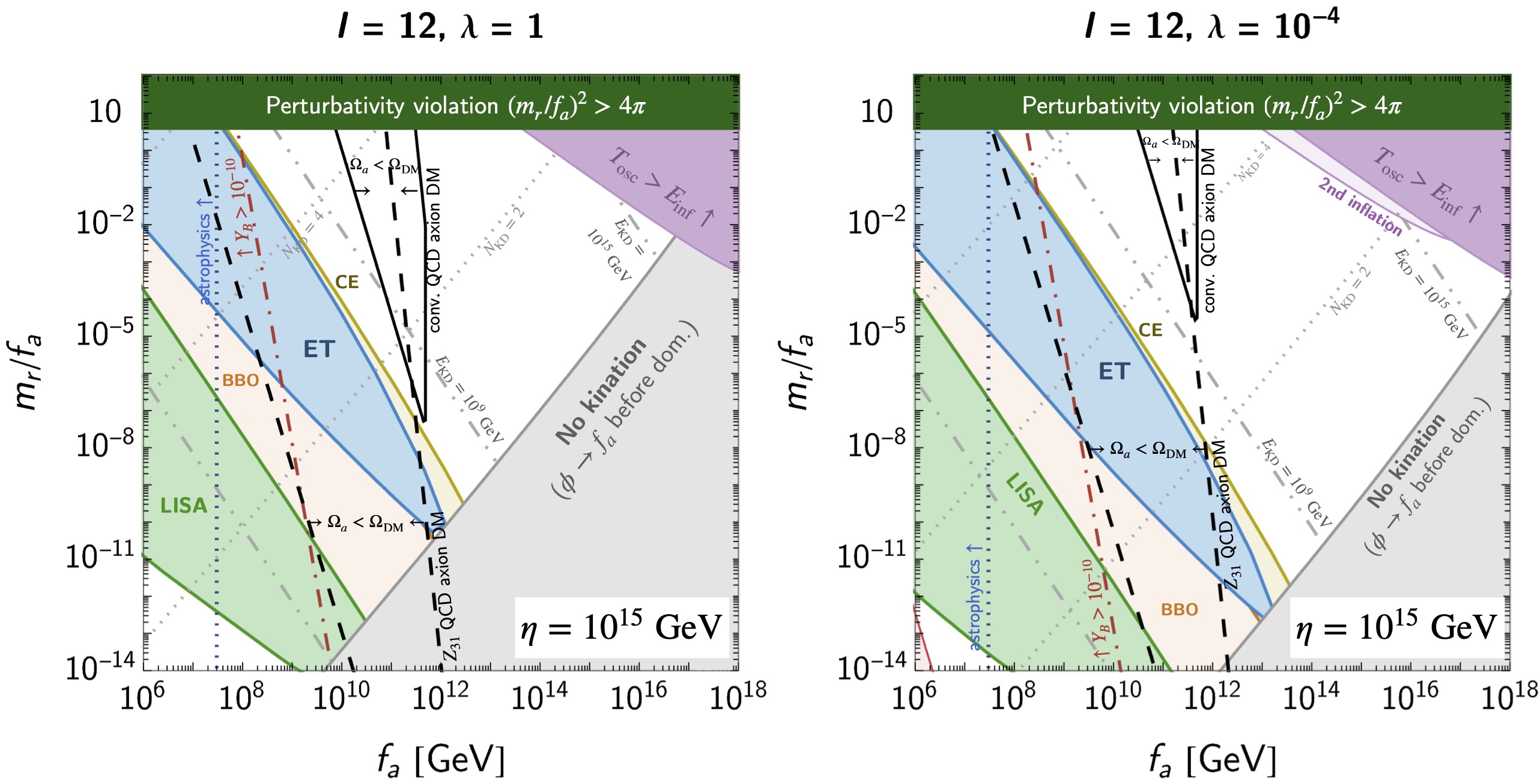}}}\\[0.25em]
\raisebox{0cm}{\makebox{\includegraphics[width=0.95\textwidth, scale=1]{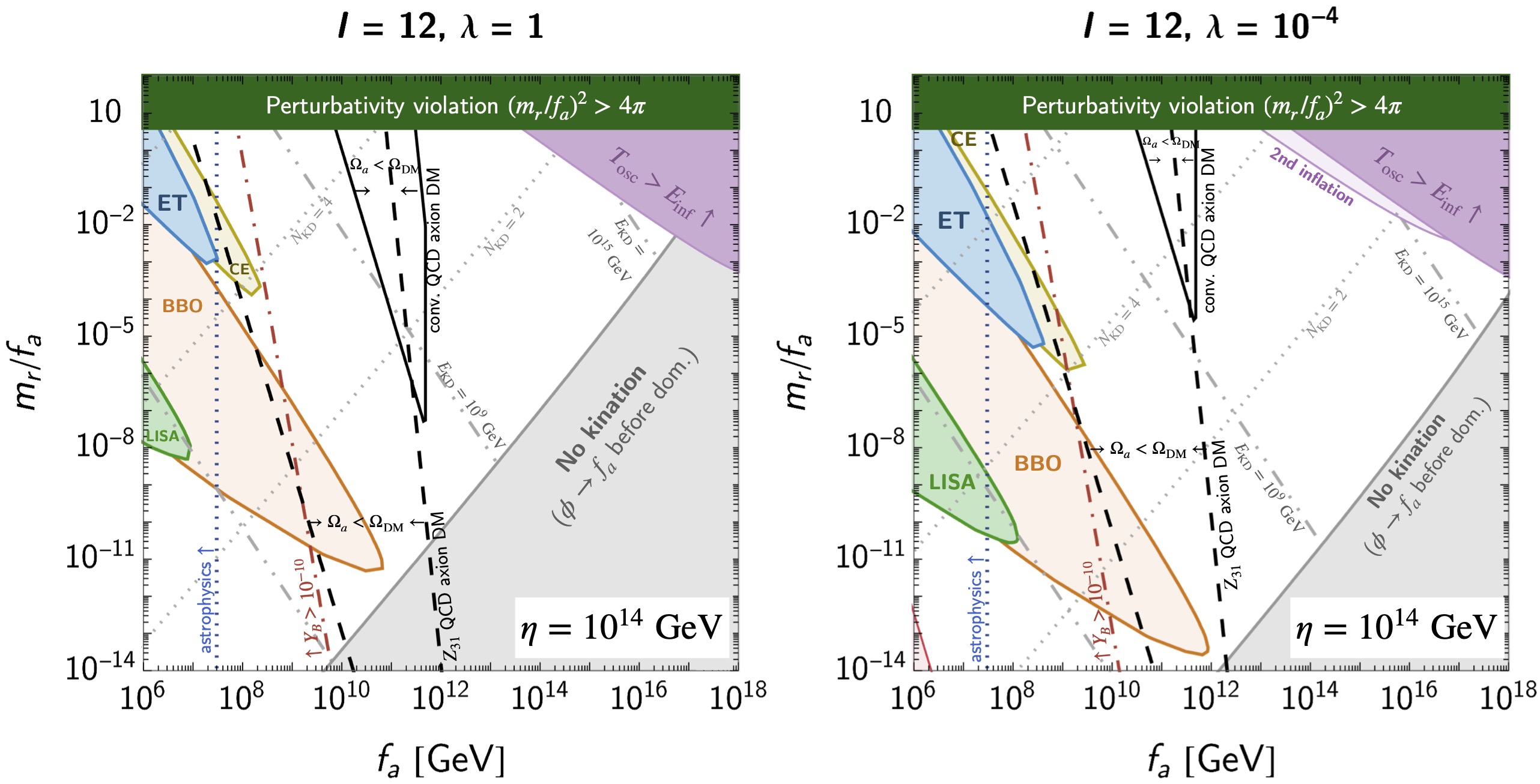}}}\\[0.25em]
\caption{\textit{ \small Same as Fig.~\ref{fig:complex_scenario1_global1} for $l=12$.  Black solid lines indicate where the canonical QCD axion abundance is satisfied, cf. Eq.~\eqref{eq:axion_abundance_general}. The left boundary is set by the kinetic misalignment mechanism, while the right one is by the standard misalignment (for small $m_r$ with a specific $f_a$) and by the axion quality problem (for larger $m_r$ depending on $f_a$), cf. Eq.~\eqref{eq:quality_problem_condition}. Only the region between the two lines does not over-produce DM. Dashed lines are the equivalent for lighter non-canonical QCD axion,  cf. Eq.~\eqref{eq:lighter_axion_mass}.}}
\label{fig:complex_scenario1_global2}
\end{figure}
\FloatBarrier

\subsubsection{Axion dark matter}
One interesting point in this paper is that the axion generating the matter-kination era and the peak GW signature could also explain DM. In Sec.~\ref{sec:darkmatter}, the nature of kination -- parametrized by its energy scale $E_{\rm KD}$ and duration $N_{\rm KD}$ -- is related model-independently to the axion parameters, namely axion mass $m_a$ and its decay constant $f_a$.
This means that the axion parameter space could be probed easily by reading the peak position of the GW background that gets boosted by this kination era.

This section provides the kination nature ($E_{\rm KD}, N_{\rm KD}$) in terms of UV-completion parameters, i.e. the radial-mode mass $m_r$ and $f_a$.
Therefore we translate the detectability plots in the UV-completion parameters into the usual axion parameter space.
Let us compare the model-independent expression of the energy scale at the end of kination, Eq.~\eqref{eq:Ekd_axion_fraction},
\begin{align}
E_\Delta = 4.4 \times 10^{9} h^2 ~ {\rm GeV} \left(\frac{f_a m_a}{\rm GeV^2}\right) \left(\frac{\Omega_{\rm DM}}{\Omega_{a}}\right),
\end{align}
and the model-dependent expression, Eq.~\eqref{eq:modelB_rhof},
\begin{align}
    E_\Delta^4 \simeq \frac{4 f_a^4 m_r^2 \MPl^6}{27 \phi_{\rm ini}^8} ~ ~ \Rightarrow ~ ~  E_\Delta \simeq \left(\frac{4}{27}\right)^{1/4} f_a \left(\frac{m_r}{M_{\rm Pl}}\right)^{\frac{l-6}{2(l-2)}},
\end{align}
where the log-dependence in $m_r$ introduces a factor $\mathcal{O}(1)$ to $E_f$ and thus is omitted for simplicity, and we use the initial VEV in Eq.~\eqref{susy_phi_ini}.
Equating these two expressions, the mass of radial mode is fixed when axion of mass $m_a$ contributes to the total DM density,
\begin{align}
   m_r &\simeq M_{\rm Pl} \left[\frac{7.1 \times 10^{9}h^2}{\left[\lambda^2 (2l-2)\right]^{\frac{1}{l-2}}} \left(\frac{m_a}{\rm GeV}\right)\left(\frac{\Omega_{\rm DM}}{\Omega_{a}}\right)\right]^{\frac{2(l-2)}{l-6}}.
\end{align}
Here we see a non-trivial result, namely $m_r$ grows with $m_a$. One might naively expect a larger $m_r$ for smaller $m_a$ because the larger PQ-charge yield is required for the correct $\Omega_{\rm a} = m_a Y_{\theta} s_0$. However, a larger charge does not mean larger $m_r$. A yield is proportional to $m_r/T_{\rm KD}^3$. For larger $m_r$, the kination energy scale also increases and hence enhances $T_{\rm KD}$ in the process.
Applying the new $m_r$ expression, the duration of kination in Eq.~\eqref{eq:end_kination_scale_factor} transforms into the axion DM parameters,
\begin{align}
    e^{N_{\rm KD}} \simeq \sqrt{\frac{3}{2}} \left(\frac{M_{\rm Pl}}{f_a}\right)^{\frac{1}{3}} \left(\frac{\phi_{\rm ini}}{M_{\rm Pl}}\right)^{\frac{4}{3}} \simeq \sqrt{\frac{3}{2}} \left(\frac{M_{\rm Pl}}{f_a}\right)^{\frac{1}{3}} \left[\frac{7.1 \times 10^{9}h^2}{\left(\lambda \sqrt{2l-2}\right)^{1/2}}\left(\frac{m_a}{\rm GeV}\right)\left(\frac{\Omega_{\rm DM}}{\Omega_{a}}\right)\right]^{\frac{8}{3(l-6)}}.
\end{align}

\section{Scenario II: thermal damping and relativistic fermions}
\label{sec:complex_field_thermal_potential}

\subsection{Effects of the thermal corrections}
\label{sec:effect_thermal_corrections}
In the scenario I in the previous Sec.~\ref{sec:scenario_I_non_thermal_damping}, we have assumed the existence of an unspecified non-thermal mechanism (maybe parametric resonance) responsible for efficiently damping the radial mode of the scalar condensate.
The advantage was that we could neglect the effect of the thermal corrections to the potential, presented in Sec.~\ref{sec:finite_T_corrections}, on the dynamics of the scalar field.
In the current section, we switch on the interaction with the thermal plasma by relaxing the condition of small Yukawa coupling $y_\psi$ in Eq.~\eqref{eq:thermal_mass_negligible_scenario_I}.
The advantage is that it leads to an early thermalization of the condensate with the plasma which is a well understood and efficient mechanism \cite{Co:2019wyp} for damping the radial mode.
The difficulty relies in the presence of the thermal mass which leads to a modification of the cosmological history of the scalar field:\footnote{Another difficulty inherent to the scenario II studied in the present section is the fragmentation of the scalar condensate into Q-balls whenever the thermal-log dominate, see Sec.~\ref{sec:Q-balls}. We do not investigate further this possibility as the present scenario II does not lead to any matter-kination era.}
\begin{itemize}
\item
the scalar field starts oscillating earlier,
 \item
the initial angular kick $\epsilon$ is substantially reduced,
\item
 the scalar field redshifts like radiation until the thermal mass becomes negligible below $T_{\rm zero} = \mreff/y_\psi$. This leads to a delay of the onset of the matter era.
\end{itemize}
We discuss those effects below and we show that they prevent the onset of the matter-kination era. A sketch of the cosmological history in the  presence of thermal effects is shown in Fig.~\ref{thermal_figur2}.

\FloatBarrier
\begin{figure}[h!]
\centering
\includegraphics[width=0.65\textwidth, scale=1]{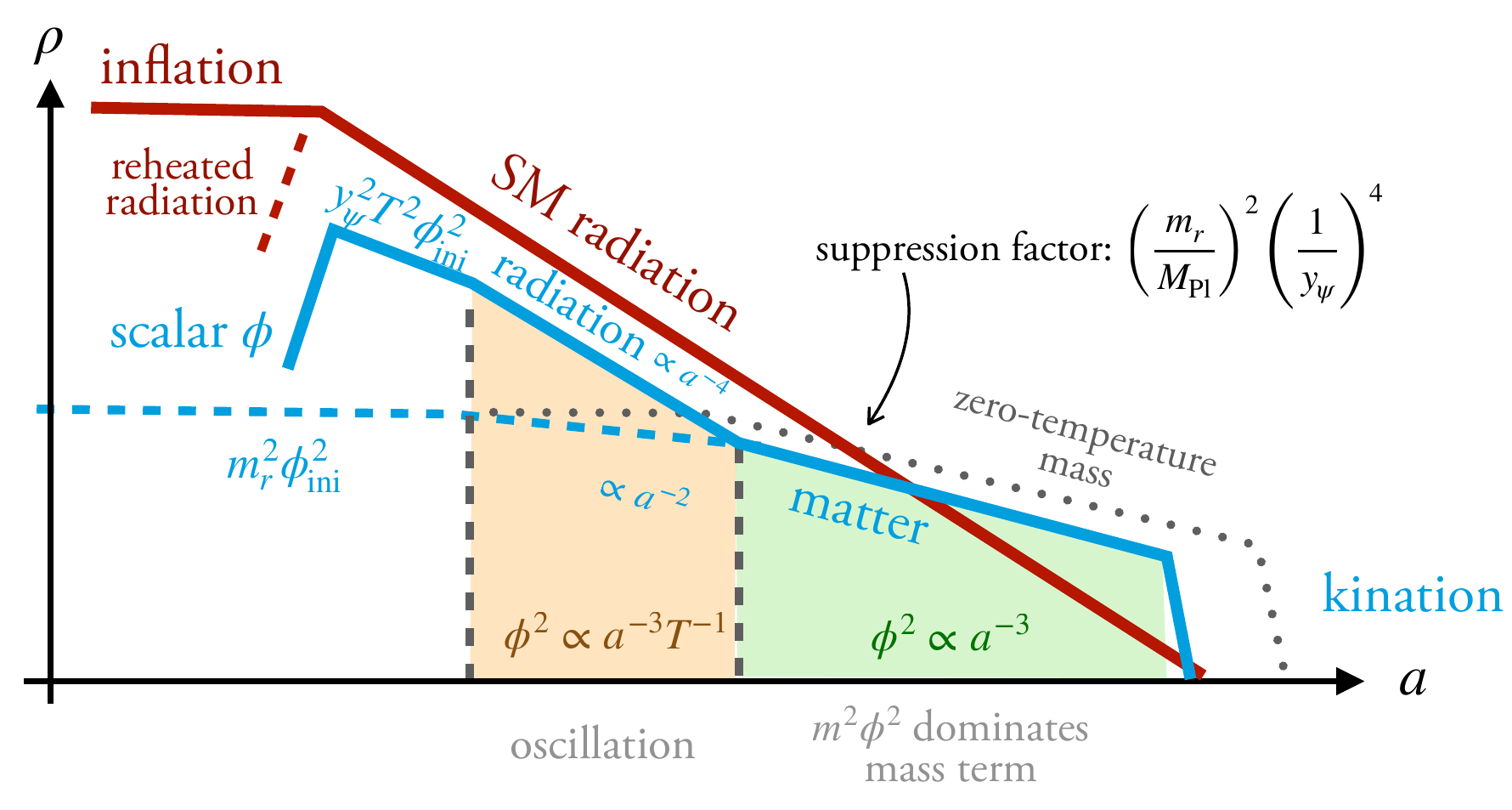}
\caption{\textit{ \small Evolution of the energy density of the complex scalar field with (blue) and without (gray) thermal mass. Due to the earlier oscillation, the thermal mass suppresses the energy density at the time of scalar domination by a factor proportional to $(m_r/\MPl)^2(1/y_\psi)^4$, and so is the duration of the kination era, see text for more details. The corresponding numerical trajectory is shown in Fig.~\ref{thermal_numerical2}. Not shown on this figure is the suppression of the angular kick by the thermal mass, see Eq.~\eqref{eq:epsilon_thermal_vs_non_thermal}. }}
\label{thermal_figur2}
\end{figure}
\FloatBarrier

\subsubsection{An earlier oscillation}
The earlier oscillation of the condensate due to the thermal effect has been pointed-out in \cite{Anisimov:2000wx,Allahverdi:2000zd}.
The field starts rolling when $3 H \sim m_\mathrm{eff,T}$ where the effective mass, in a radiation era is 
\begin{align}
\label{eq:def_thermal_mass}
m_\mathrm{eff,T}^2 ~ = ~ m_\mathrm{eff}^2 + y_\psi^2 T^2.
\end{align}
Assuming that the thermal mass dominates, we obtain that the field starts oscillating at the temperature
\begin{align}
3 H_\mathrm{osc} ~ = ~ y_\psi T_\mathrm{osc}, \qquad \implies \qquad T_\mathrm{osc} ~ = ~ g_*^{-1/2} \MPl y_\psi , \label{eq:oscillation_thermal_mass}
\end{align}
which is larger than the oscillation temperature in the zero-temperature potential, see Eq.~\eqref{eq:oscillation_temperarture}, for 
\begin{equation}
y_\psi~ \gtrsim~ y_{\psi,T=0} \equiv g_*^{1/4}  \sqrt{\frac{\mreff(\phi_{\rm ini})}{M_{\rm pl}}}. \label{eq:not_thermal_mass}
\end{equation}
An earlier oscillation due to thermal effects can be visualized in Fig.~\ref{thermal_numerical2}.

\FloatBarrier
\begin{figure}[h!]
\centering
\raisebox{-0.5\height}{\makebox{\includegraphics[width=0.5\textwidth, scale=1]{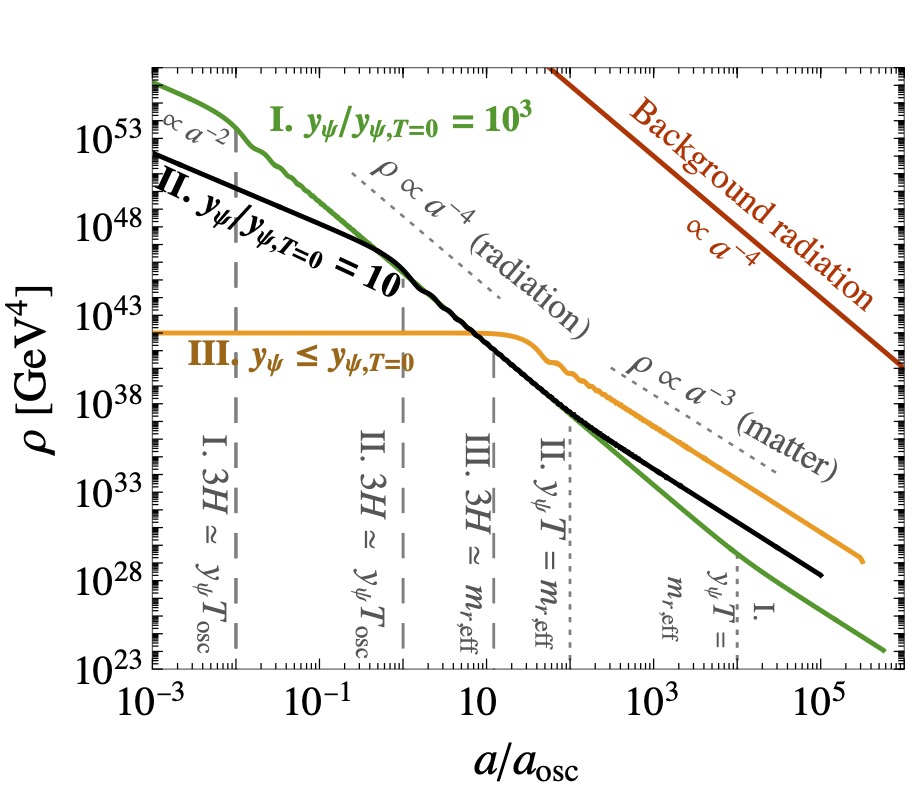}}}
\caption{\textit{ \small  Evolution of scalar field energy density with thermal mass $y_\psi T$ (case I in green and II in black) and without thermal mass (case III in yellow).  After the onset of oscillation (vertical dashed), the field redshifts as radiation until the zero-temperature mass dominates (vertical dotted), where it starts redshifting as matter. The smaller $y_\psi$, the later the start of oscillation and the earlier the start of the matter behavior.
Below the value $y_\psi < y_{\psi, T=0}$, where $y_{\psi, T=0}$ is defined in Eq.~\eqref{eq:not_thermal_mass}, the thermal mass never plays any role in the evolution, and the field redshifts as matter already at the onset of oscillation. 
We normalize the scale factor $a$ with $a_\mathrm{osc}$ of case II. Figure obtained after numerically integrating the EOM of the scalar field in quadratic potential in the presence of the thermal mass in Eq.~\eqref{eq:def_thermal_mass}.}}
\label{thermal_numerical2}
\end{figure}
\FloatBarrier

\subsubsection{A different redshift law}
When dominated by its thermal mass, the complex scalar field redshifts like radiation, see App.~\ref{app:virial_th}
\begin{align}
\phi^2 ~ \propto ~ a^{-3} T^{-1} \propto a^{-2}, \qquad \implies \qquad \rho_{\Phi} \simeq \frac{1}{2}y_\psi^2 T^2 \phi^2 \propto a^{-4}. \label{eq:thermal_mass_redshift_radiation}
\end{align}
On the other hand, the zero-temperature contribution redshifts slower than radiation
\begin{equation}
V(\phi,\,T=0) = \frac{1}{2}\mreff^2 \phi^2 \propto a^{-2}.
\label{eq:zero_temp_redshift_a2}
\end{equation}
The thermal potential becomes sub-dominant at the temperature, scale factor and field value
\begin{equation}
T_{\rm zero} = \frac{\mreff(\phi_{\rm zero})}{y_\psi}, \qquad \frac{a_{\rm zero}}{a_\mathrm{osc}} ~ = ~g_*^{-1/2} \frac{y_\psi^2\MPl}{\mreff(\phi_{\rm zero})} \qquad \textrm{and}\qquad  \phi_{\rm zero} = \phi_{\rm ini} \frac{T_{\rm zero}}{T_{\rm osc}} . \label{eq:Tzero_def}
\end{equation}
Changes in the redshift law are also seen in our numerical simulations, cf. Fig.~\ref{thermal_numerical1}.

\paragraph{No symmetry restoration when $\phi \to f_a$.}
To generate a kination equation of state, we must check that the complex scalar field reaches the minimum of the potential $\phi \to f_a$ after the thermal mass becomes sub-dominant, i.e.
\begin{equation}
\phi(T_{\rm zero} ) ~\gtrsim~ f_a.
\end{equation}
We checked that in the parameter space of interest, this never occurs.

\FloatBarrier
\begin{figure}[h!]
\centering
\raisebox{-0.5\height}{\makebox{\includegraphics[width=0.45\textwidth, scale=1]{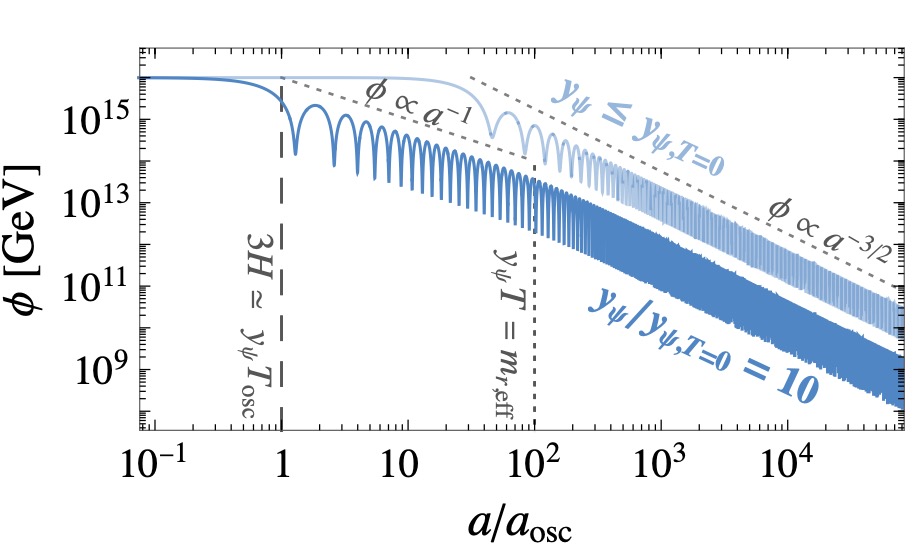}}}
\qquad
\raisebox{-0.5\height}{\makebox{\includegraphics[width=0.45\textwidth, scale=1]{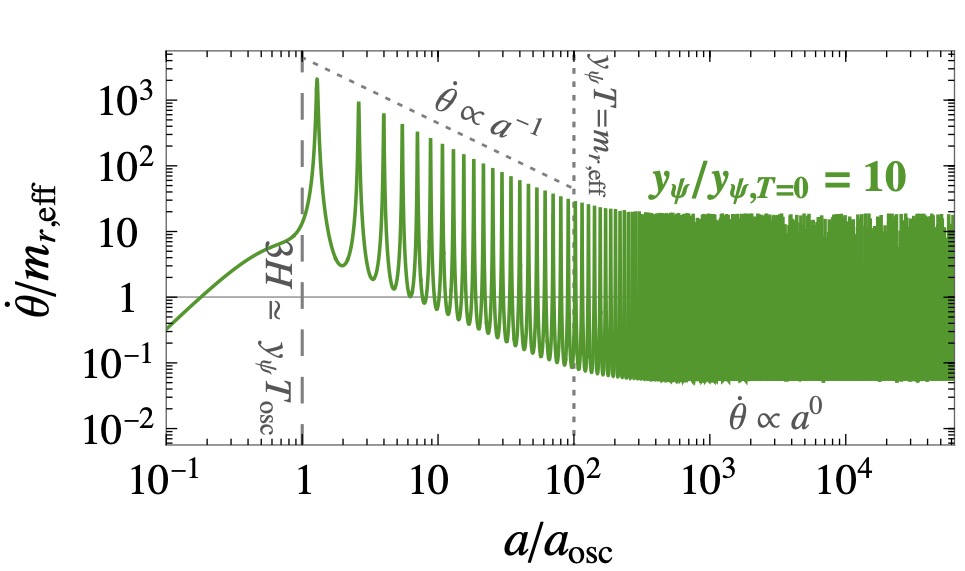}}}
\caption{\textit{ \small \textbf{Left:} In the presence of the thermal mass (dark blue), at the onset of oscillation (vertical dashed line), the radial mode $\phi$ starts evolving as $\left<\phi  \right> \propto a^{-3/2}T^{-1/2}  \propto a^{-1}$ in radiation era, until the zero-temperature mass dominates (vertical dotted line), after which the field starts redshifting as $\left<\phi\right> \propto a^{-3/2}$. Instead, when the potential is always dominated by its zero-temperature contribution (light blue), i.e. $y_\psi < y_{\psi, T=0}$ where $y_{\psi, T=0}$ is defined in Eq.~\eqref{eq:not_thermal_mass}, the field starts redshifting as $\left<\phi\right >\propto a^{-3/2}$ immediately after the start of oscillation.
\textbf{Right:} During the period of thermal mass domination, the angular velocity $\dot{\theta}$ redshifts as $\left<\dot{\theta}\right> \propto T \propto a^{-1}$ during radiation, and then oscillates around $\mreff$ when the zero-temperature mass dominates.}}
\label{thermal_numerical1}
\end{figure}
\FloatBarrier

\subsubsection{A smaller angular kick}
\label{sec:smaller_kick_thermal_mass}

We calculate the $U(1)$ charge fraction in Eq.~\eqref{eq:epsilon_def}
\begin{equation}
\epsilon ~ \equiv  ~ \frac{\phi^2 \dot{\theta}/2}{V(\phi)/m_{r,\textrm{eff}}(\phi)}~=~ \frac{\phi_{\rm ini}^2 \dot{\theta}_{\rm ini}/2}{V(\phi)/m_{r,\textrm{eff}}(\phi)} \left( \frac{a_{\rm ini}}{a} \right)^3.
 \label{eq:epsilon_def_thermal}
\end{equation}
From using $(\phi_{\rm ini}/M_{\rm pl})^{l-2}=\sqrt{c} \mreff(\phi_{\rm ini}) / \lambda\sqrt{2l-2}  M_{\rm pl}$, see Eq.~\eqref{susy_phi_ini}, we find that the term in the potential which dominates the dynamics of the scalar field at the onset of the radial mode oscillation  depends on the value of $c$
\begin{equation}
V(\phi_{\rm ini}) = 
\begin{cases}
\lambda^2 \frac{\phi_{\rm ini}^{2l-2}}{M_{\rm pl}^{2l-6}},~ \qquad \qquad \qquad \quad \text{if}~ c ~> ~l-1 ,\\[0.75em]
\frac{1}{2} y_\psi^2 T_{\rm osc}^2 \phi_{\rm ini}^2, \qquad  \qquad\quad ~~ \text{otherwise} .\\[0.75em]
\end{cases}
\end{equation}
In the first case of the previous equation, the scalar field redshifts as, cf.~App.~\ref{app:virial_th}
\begin{equation}
\phi \propto a^{-\frac{6}{2+n}},\qquad\qquad \textrm{with}~n ~=~ 2l-2.
\label{eq:phi_redshift_higher_dim_pot_thermal}
\end{equation} 
until the nearly-quadratic term $\phi^2$ dominates around the field value $\phi_{\rm quad}$
\begin{equation}
\frac{1}{2}y_\psi^2 T_{\rm quad}^2\phi_{\rm quad}^2~ =~ 
 \lambda^2 \frac{\phi_{\rm quad}^{2l-2}}{M_{\rm pl}^{2l-6}}. \label{eq:phi_quad_def_thermal}
\end{equation}
For $\phi < \phi_{\rm quad}$, the $U(1)$ charge fraction $\epsilon$ becomes constant since both numerator and denominator in Eq.~\eqref{eq:epsilon_def_thermal} scale like $a^{-3}$.
Plugging Eq.~\eqref{eq:phi_redshift_higher_dim_pot_thermal} into Eq.~\eqref{eq:phi_quad_def_thermal}, we obtain the scale factor $a_{\rm quad}$ below which the potential is dominated by the quadratic term. From injecting $a_{\rm quad}$ into the generated $U(1)$ charge in Eq.~\eqref{epsilon_oscillation} and the definition of $\epsilon$ in Eq.~\eqref{eq:epsilon_def_thermal}, we obtain
\begin{equation}
\epsilon~ =~ 
\begin{cases}
\sqrt{\frac{ g_*}{2}}\frac{\mreff(f_a)}{y_\psi^2 M_{\rm pl}} \,l\,\sin{l \theta_{\rm ini}},~ \qquad \qquad \qquad  \quad~ \text{if} \quad c~> ~l-1,\\[0.75em]
\sqrt{\frac{ g_* c}{2(l-1)}}\frac{\mreff(f_a)}{y_\psi^2 M_{\rm pl}} \,l\,\sin{l \theta_{\rm ini}}, ~~~\quad\qquad\qquad \quad  ~  \text{otherwise} .
\end{cases} \label{eq:epsilon_1_zero_temp_thermal}
\end{equation}
We conclude that for a fix value of $c$, $l$ and $\theta_{\rm ini}$, the value of $\epsilon$ in the presence of a thermal mass is suppressed by a factor  $\frac{g_*^{1/2} \mreff(\phi_{\rm ini})}{y_{\psi}^2M_{\rm pl}}$ with respect to the value of $\epsilon$ without thermal mass, cf. Eq.~\eqref{eq:epsilon_1_zero_temp}, which we denote $\epsilon_0$
\begin{equation}
\epsilon~=~\left(\frac{g_*^{1/2} \mreff(\phi_{\rm ini})}{y_{\psi}^2M_{\rm pl}}\right)~\epsilon_0.
\label{eq:epsilon_thermal_vs_non_thermal}
\end{equation}
The reason of the suppression of $\epsilon$ in Eq.~\eqref{eq:epsilon_thermal_vs_non_thermal} can be understood from Eq.~\eqref{eq:epsilon_def_thermal} where the denominator $V(\phi)$ is enlarged by the presence of the thermal mass while the explicit $U(1)$ breaking potential generating the numerator $\dot{\theta}$ is not.

\subsection{Evolution of the field and its energy density}

\subsubsection{Delay of matter domination}
The scalar field dominates the energy density of the universe at the scale factor $a_{\rm dom}$ defined by
\begin{equation}
V(\phi_{\rm zero},\,T=0)\left(\frac{a_{\rm zero}}{a_{\rm dom}}\right)^3 = \rho_{\rm rad}(T_{\rm zero})\left(\frac{a_{\rm zero}}{a_{\rm dom}}\right)^4 \qquad \textrm{with} \quad \rho_{\rm rad}= \frac{\pi^2}{30}g_* T^4,
\end{equation}
where $T_{\rm zero}$ denotes the temperature, defined in Eq.~\eqref{eq:Tzero_def}, below which the zero-temperature mass dominates over the thermal mass. We obtain
\begin{equation}
\frac{a_{\rm dom}}{a_{\rm zero}} = \frac{2}{3}\left(\frac{M_{\rm pl}}{\phi_{\rm ini}} \right)^2A_\epsilon^{-1}, \quad \quad  T_{\rm dom} =  \frac{\mreff(\phi_{\rm zero})}{y_{\psi}} \frac{3}{2}\left( \frac{\phi_{\rm ini}}{M_{\rm pl}}\right)^2\,A_\epsilon,
\end{equation}
and
\begin{equation}
\rho_{\rm dom} = \frac{27g_*}{16} \left( \frac{\mreff(\phi_{\rm zero})}{y_{\psi}}\right)^4 \left( \frac{\phi_{\rm ini}}{M_{\rm pl}}\right)^8A_\epsilon^4, \label{eq:rho_dom_scenario_2}
\end{equation}
with $A_\epsilon$ defined by
\begin{equation}
A_\epsilon = \begin{cases}
\epsilon , \qquad \qquad \text{if}~\rho_{\rm damp} > \rho_{\rm dom},\\[0.75em]
1, \qquad \qquad \text{if}~\rho_{\rm damp} < \rho_{\rm dom},
\end{cases}
\end{equation}
stands for the $\epsilon$-suppression discussed in  App.~\ref{app:effect_epsilon_evolution}. The value of $\epsilon$ is given by Eq.~\eqref{eq:epsilon_1_zero_temp_thermal}.
Note that we can rewrite Eq.~\eqref{eq:rho_dom_scenario_2} as
\begin{equation}
\rho_{\rm dom} =  \left(g_*^{1/2} \frac{\mreff^2(\phi_{\rm zero})}{y_{\psi}^2\mreff(\phi_{\rm ini})M_{\rm pl}}\right)^2 \rho_{\rm dom,1}
\label{eq:rho_dom_thermal_nonthermal}
\end{equation}
with $\rho_{\rm dom,\,1}$ being the energy density at domination, defined in Eq.~\eqref{cosmohist_zero_temp_matter_dom}, in the case of the first scenario where the thermal mass can be neglected due to a small Yukawa coupling. From Eq.~\eqref{eq:rho_dom_thermal_nonthermal}, we can see that the effect of the thermal mass is to delay the onset of the matter domination by a factor $\propto y_{\psi}^{-4}$ in the energy density. As we will see in Eq.~\eqref{eq:end_kination_scale_factor_thermal_mass}, the delay of the matter domination in Eq.~\eqref{eq:rho_dom_thermal_nonthermal} together with the suppression of $\epsilon$ in Eq.~\eqref{eq:epsilon_thermal_vs_non_thermal} are responsible for preventing the kination era to take place in the presence of thermal effects.

\subsubsection{Radial damping}
\label{sec:radial_damping_scenario_2}
In the presence of the Yukawa interactions with fermions $\psi$, cf. Eq.~\eqref{eq:KSVZ_lagrangian_2}, the scalar condensate thermalizes with the thermal plasma with the rate, see App.~\ref{app:thermalization} 
\begin{align}
\Gamma_\phi \simeq  \begin{cases}
\textrm{ for} ~ y_\psi \phi < T :
\begin{cases}
 \textrm{for}~ \alpha T  > y_\psi \phi ,\qquad \frac{y_\psi^2 \alpha T}{2 \pi^2},\\[0.5em]
\textrm{for}~\alpha T < y_\psi \phi ,\qquad \frac{y_\psi^4 \phi^2}{\pi^2 \alpha T},
\end{cases} 
\\[2em]
\textrm{ for} ~ y_\psi \phi > T: 
\qquad 
b \alpha^2 \frac{\textrm{Max}\left[T,~m_\phi\right]^3}{\phi^2} ,
\\
\end{cases}
+\quad \frac{y_\psi^2 m_{\phi}}{8\pi} \Theta\left(m_\phi/2 - \textrm{Max}\left[y_\psi \phi,\,gT\right]\right).
\label{eq:fermion_damping_rate_YG_main}
\end{align}
with $b \simeq 0.01$.  As shown in the supplementary material of \cite{Co:2019wyp}, thermalization conserves the $U(1)$ charge $\phi^2 \dot{\theta}$.
Radial damping takes place at the energy scale and scale factor 
\begin{equation}
\rho_{\rm damp} = 3M_{\rm pl}^2 \Gamma^2 B_{\epsilon}^4 \qquad \textrm{and} \qquad
\frac{a_{\rm damp}}{a_{\rm dom}} = \left( \frac{\rho_{\rm dom}}{\rho_{\rm damp}} \right)^{1/3} B_{\epsilon}^{-1},
\end{equation}
with 
\begin{equation}
B_\epsilon = \begin{cases}
1 , \qquad \qquad \text{if}~\rho_{\rm damp} > \rho_{\rm dom},\\[0.75em]
\epsilon, \qquad \qquad \text{if}~\rho_{\rm damp} < \rho_{\rm dom}.
\end{cases} \label{eq:Bepsilon_scenario2}
\end{equation}

For the sake of simplicity, in the present Sec.~\ref{sec:complex_field_thermal_potential}, we suppose that the thermal width is larger than the fermion mass $\alpha T  > y_\psi \phi$. The decay rate is given by the first line in Eq.~\eqref{eq:fermion_damping_rate_YG_main} and we compute
\begin{equation}
\Gamma_{\rm damp} =
\begin{cases}
 \frac{3\alpha^2}{4\pi^4}g_*^{-1/2} y_\psi^4 M_{\rm pl} ,~  \qquad \qquad \qquad \qquad ~ \text{for}~ y_\psi > y_{\psi, 1},\\[0.75em]
 \frac{y_\psi^2 \mreff(f_a)}{8\pi},~~\quad \qquad  \qquad \qquad\qquad \qquad \text{otherwise},
\end{cases} \label{eq:Gamma_damp_thermal_mass}
\end{equation}
and 
\begin{equation}
T_{\rm damp} =
\begin{cases}
\frac{\alpha}{2\pi} \frac{y_\psi^2 M_{\rm pl}}{g_*^{1/2}} ,~ \qquad \qquad  \qquad \qquad \qquad \qquad~  \text{for}~ y_\psi > y_{\psi, 1},\\[0.75em]
\frac{y_{\psi}}{\pi g_*^{1/4}}\sqrt{\mreff(f_a)M_{\rm pl}},  \qquad \quad\qquad \qquad \text{otherwise},
\end{cases} \label{eq:Tdamp_thermal_mass}
\end{equation}
where
\begin{equation}
y_{\psi, 1} \equiv 2\frac{g_s^{1/4}}{\alpha} \left(\frac{\mreff(f_a)}{M_{\rm pl}}\right)^{1/2}. \label{eq:def_ypsi1_ypsi_2}
\end{equation}

\subsubsection{Duration of the kination era}

\paragraph{Start of kination.} 
The universe acquires a kination equation-of-state when the field reaches $\phi \to f_a$, corresponding to the energy density and scale factor
\begin{equation}
 \rho_{\textrm{KD},i} ~ = ~  \frac{1}{2}f_a^2 \mreff ^2(f_a) \qquad \textrm{and} \qquad \frac{a_{\textrm{KD},i}}{\max(a_{\rm dom},a_{\rm damp})} = \left(\frac{\text{min}(\rho_{\rm dom},\,\rho_{\rm damp})}{\rho_\textrm{KD,i}}\right)^{\!1/3}.
\end{equation}

\paragraph{End of kination.} 
The kination era stops when the universe becomes radiation-dominated again. The energy scale at which it occurs depends on whether radial damping occurs before and after the onset of matter domination
\begin{equation}
\rho_{\textrm{KD},f} =\frac{\rho_{\textrm{KD},i}^2}{\text{min}(\rho_{\rm dom},\,\rho_{\rm damp})}.
\end{equation}
The duration of the kination era $N_{\rm KD} \equiv \log{\frac{a_{\textrm{KD},f}}{a_{\textrm{KD},i}}}$ reads
\begin{equation}
e^{N_{\rm KD}}= \left(\frac{\text{min}(\rho_{\rm dom},\,\rho_{\rm damp})}{\rho_{\textrm{KD},i}}\right)^{\!1/6}=
\begin{cases}
\label{eq:end_kination_scale_factor_thermal_mass}
\sqrt{\frac{3}{2}} \left(\frac{g_*^{1/2}}{y_{\psi}^2}\frac{\mreff(\phi_{\rm zero})}{\mreff(f_a)} \frac{\mreff(\phi_{\rm zero})}{f_a}\right)^{1/3}  \left( \frac{\phi_{\rm ini}}{M_{\rm pl}} \right)^{4/3}\epsilon^{2/3}, \quad  \text{if}~\rho_{\rm damp} > \rho_{\rm dom},\\[0.75em]
\left(\frac{6 \MPl^2 \Gamma_{\rm damp}^2}{f_a^2 \mreff ^2(f_a)}\right)^{1/6}\epsilon^{2/3},  \qquad \qquad\qquad \qquad\qquad \qquad \quad  \text{if}~\rho_{\rm damp} < \rho_{\rm dom}.
\end{cases} 
\end{equation}
Using the first lines of Eq.~\eqref{eq:fermion_damping_rate_YG_main} and \eqref{eq:Tdamp_thermal_mass}, we obtain 
\begin{equation}
e^{N_{\rm KD}}=
0.74 g_*^{1/18}\alpha^{2/9}\epsilon^{2/3} \left( \frac{M_{\rm pl}}{f_a} \right)^{1/3} \left( \frac{\mreff(f_a)}{M_{\rm pl}} \right)^{1/9} \left( \frac{\mreff(\phi_{\rm zero})}{\mreff(f_a)} \right)^{4/9} \left( \frac{\phi_{\rm ini}}{M_{\rm pl}}\right)^{8/9} \times~
\begin{cases}
\label{eq:end_kination_scale_factor_thermal_mass}
\left( \frac{y_{\psi, \rm th}}{y_{\psi}} \right)^{2/3}, \qquad\quad \text{if}~y_{\psi} \gtrsim y_{\psi, \rm th},\\[0.75em]
\left( \frac{y_{\psi}}{y_{\psi, \rm th}} \right)^{4/3}, \qquad \quad~ \text{if}~y_{\psi, \rm th} \gtrsim y_{\psi} \gtrsim y_{\psi, 1},\\[0.75em]
\left( \frac{y_{\psi, 1}}{y_{\psi, \rm th}} \right)^{4/3} \left( \frac{y_{\psi}}{y_{\psi, 1}} \right)^{2/3}, ~\text{if}~y_{\psi} \lesssim y_{\psi, 1},
\end{cases} 
\end{equation}
where $y_{\psi, 1}$ is defined in Eq.~\eqref{eq:def_ypsi1_ypsi_2} and where $y_{\psi, \rm th}$ is the Yukawa coupling above which thermalization starts to occur before scalar field domination
\begin{equation}
\rho_{\rm damp} > \rho_{\rm dom} \qquad \implies \qquad y_{\psi} ~>~y_{\psi, \rm th} ~=~ \pi^{2/3}\frac{g_*^{1/6}}{\alpha^{1/3}} \left(\frac{\mreff(\phi_{\rm zero})}{M_{\rm pl}}\right)^{1/3} \left(\frac{\phi_{\rm ini}}{M_{\rm pl}}\right)^{2/3}.
 \label{eq:rho_dam_above_rho_dom_thermal_mass}
\end{equation}
From plugging in Eq.~\eqref{eq:end_kination_scale_factor_thermal_mass} the most optimistic value of $\epsilon$, cf. first line of Eq.~\eqref{eq:epsilon_1_zero_temp_thermal}, we obtain
\begin{equation}
e^{N_{\rm KD}}=
0.2 g_*^{1/6}\alpha^{2/3} \left(\frac{\mreff(f_a)}{f_a}\right)^{1/3} \left(\frac{\mreff(\phi_{\rm ini})}{\mreff(f_a)}\right)^{2/3}\,(l\,\sin{l\theta_{\rm ini}})^{2/3}  \times~
\begin{cases}
\left( \frac{y_{\psi, \rm th}}{y_{\psi}} \right)^{2}, \qquad \quad ~~\quad \text{if}~y_{\psi} \gtrsim y_{\psi, \rm th},\\[0.75em]
1, \qquad \quad \quad \quad \quad \quad~~\text{if}~y_{\psi, \rm th} \gtrsim y_{\psi} \gtrsim y_{\psi, 1},\\[0.75em]
\left( \frac{y_{\psi, 1}}{y_{\psi}} \right)^{2/3} , \qquad \quad \quad~\text{if}~y_{\psi, 1} \gtrsim y_{\psi} \gtrsim y_{\psi, 0} ,\\[0.75em]
\left( \frac{y_{\psi, 1}}{y_{\psi, 0}} \right)^{2/3}\left( \frac{y_{\psi}}{y_{\psi, 0}} \right)^{2/3}\quad\text{if}~y_{\psi, 0} \gtrsim y_{\psi},
\end{cases} 
\end{equation}
where $y_{\psi, 0}$, defined in Eq.~\eqref{eq:not_thermal_mass}, is the Yukawa coupling below which oscillation is induced by the zero-temperature potential.
The longest duration of kination occurs for 
\begin{framed}
\begin{equation}
y_{\psi}\simeq y_{\psi, 0} \qquad \implies \qquad e^{N_{\rm KD}}=0.4 g_*^{1/6} \left(\frac{\mreff(\phi_{\rm ini})}{f_a}\right)^{1/3}\,(l\,\sin{l\theta_{\rm ini}})^{2/3} ~\lesssim ~1. \label{eq:longest_NKS_scenario_2}
\end{equation}
\end{framed}\noindent
We conclude that a period of kination-domination can not be induced by the spinning complex scalar field, starting oscillating in a radiation-dominated universe, when the damping of the radial mode relies on thermalization. To reduce the Yukawa coupling $y_\psi$ would reduce the thermal mass but delay thermalization too much. To increase $y_\psi$ would make thermalization more efficient but would increase the thermal mass too much.
In the next section, we show how the inhibition of the initial rotation due to the thermal mass can be avoided when the fermion abundance is Boltzmann-suppressed at the onset of the oscillation due to either a large Yukawa coupling $y_\psi$ or a small reheating temperature $T_{\rm reh}$
\begin{equation}
y_\psi \phi_{\rm osc} \gtrsim T_{\rm osc}, \quad \textrm{or} \quad y_\psi \phi_{\rm reh} \gtrsim T_{\rm reh}.
\end{equation}

\section{{Scenario III: thermal damping and non-relativistic fermions}}
\label{sec:complex_field_low_reh_temp}

In the first scenario in Sec.~\ref{sec:scenario_I_non_thermal_damping}, we have considered a small Yukawa coupling such that thermal corrections could be neglected. However as discussed in App.~\ref{sec:impossibility_neglect_thermal_correction}, the difficulty is that thermalization takes place too late and we must rely on a different mechanism for damping the radial mode.

In the second scenario in Sec.~\ref{sec:complex_field_thermal_potential}, we have enforced radial damping through thermalization. However, the suppression of the initial angular kick $\epsilon$, cf. Sec.~\ref{sec:smaller_kick_thermal_mass}, and the delay of the onset of the matter era, cf. Eq.~\eqref{eq:rho_dom_thermal_nonthermal}, due to the large thermal mass at early time, prevent the scalar field to induce a kination era, cf. Eq.~\eqref{eq:longest_NKS_scenario_2}.

In the present section, we consider a third scenario, depicted in Fig.~\ref{fig:scenario3_diagram}, which turns out to be the promising one and where radial damping occurs through thermalization, but  the thermal mass is absent at the onset of the radial mode oscillation due to the Boltzmann-suppression of the fermion abundance.

\begin{figure}[t!]
\centering
\raisebox{0cm}{\makebox{\includegraphics[width=1\textwidth, scale=1]{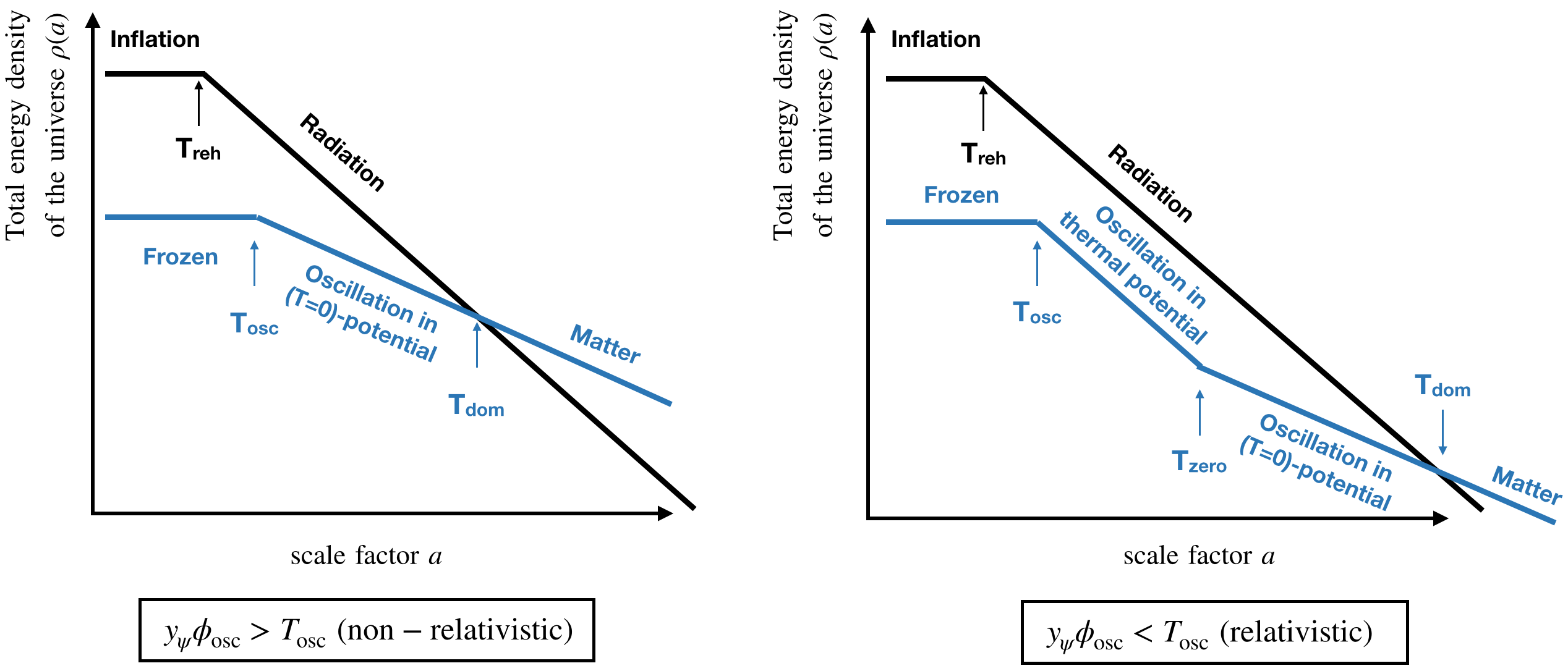}}}
\caption{\textit{ \small In scenario III of the present Sec.~\ref{sec:complex_field_low_reh_temp}, we consider the fermions responsible for the thermal mass to be Boltzmann-suppressed at the onset of the oscillation (\textbf{left} panel). Therefore, the thermal mass is turned off at the time of the angular kick and the $U(1)$ charge fraction $\epsilon$ in Eq.~\eqref{eq:epsilon_scenario_3} can be $\mathcal{O}(1)$ without any tuning of the Yukawa coupling to small values, in contrast to scenario II, cf. Sec~\ref{sec:complex_field_thermal_potential} (\textbf{right} panel).
Later, when the fermions become relativistic, Yukawa interactions are efficient enough to damp the radial motion before the onset of the matter era. }}
\label{fig:scenario3_diagram}
\end{figure}

\subsection{Boltzmann-suppression of the fermion abundance}
\label{sec:Boltzmann_suppression}

\subsubsection{Conditions} 
\label{sec:condition_boltzmann}
The presence of the thermal mass $ y_\psi T$ supposes the fermions $\psi$ to be abundant in the plasma. Instead, if their abundance is Boltzmann-suppressed, the thermal mass is set by the thermal log potential, cf. Eq.~\eqref{eq:thermal_correction_potential}
\begin{align}
 \mreff(\phi,\, T) ~ = ~ \begin{cases}
y_\psi T,   & \mathrm{for} ~   y_\psi \phi \lesssim T, \\
\sqrt{2}\alpha \frac{T^2}{\phi} ,\qquad \qquad & \mathrm{for} ~   y_\psi \phi \gtrsim T, 
\end{cases}
\label{eq:thermal_mass_2}
\end{align}
The Boltzmann-suppression of the fermion abundance can arise in two situations:
\begin{enumerate}
\item
 \textbf{At large Yukawa coupling} $y_\psi$
\begin{equation}
\label{eq:boltzmann_lower_bound_yukawa}
\begin{cases}
y_\psi \phi_{\rm osc} > T_{\rm osc},\\[0.75em]
\phi_{\rm osc} = M_{\rm pl}\left( \sqrt{c}\frac{y_{\psi}T_{\rm osc}}{\lambda \sqrt{2l-2}M_{\rm pl}} \right)^{\frac{1}{l-2}},\\[0.75em]
T_{\rm osc} = g_*^{-1/2}y_\psi M_{\rm pl},
\end{cases}
\qquad
\implies 
\qquad
y_\psi > 2^{1/4} \sqrt{\lambda}g_*^{\frac{(3-l)}{4}}  \left(\frac{l-1}{c} \right)^{1/4}.
\end{equation}
For $(\lambda,\, c,\,l,\,g_*) = (1,\, 1,\, 10, \, 100)$, we obtain $y_\psi \simeq 6.5 \times 10^{-4}$.
\item
\textbf{At small reheating temperature} $T_{\rm reh}$
\begin{equation}
\label{eq:T_reh_upper_bound}
\begin{cases}
y_\psi \phi_{\rm reh} > T_{\rm reh},\\[0.75em]
\phi_{\rm reh} = M_{\rm pl}\left( \sqrt{c}\frac{3H_{\rm reh}}{\lambda \sqrt{2l-2}M_{\rm pl}} \right)^{\frac{1}{l-2}},\\[0.75em]
T_{\rm reh} = g_*^{-1/4}\sqrt{3H_{\rm reh}M_{\rm pl}}.
\end{cases}
\qquad
\implies 
\qquad
T_{\rm reh}/M_{\rm pl} ~<~  y_\psi^{\frac{l-2}{l-4}} \left(\frac{g_*^{1/4}}{\sqrt{3}} \right)^{\frac{2}{l-4}}\left(\frac{\lambda \sqrt{2l-2}}{3\sqrt{c}} \right)^{\frac{1}{l-4}}.
\end{equation}
For $(\lambda,\, c,\,l,\,g_*,\,y_\psi) = (1,\, 1,\, 10, \, 100,\,10^{-4})$, we obtain $T_{\rm reh} < 1.5 \times 10^{13}~\rm GeV$. The upper bound \eqref{eq:T_reh_upper_bound} on the reheating temperature prevents the thermal mass $y_\psi T$ to be  active at the time of the kick. In our work we assume the universe to be radiation-dominated at the time of the kick. Therfore, for consistency we must also insure that the reheating temperature is larger than the temperature at the time of the kick
\begin{equation}
\label{eq:T_reh_lower_bound}
T_{\rm reh} ~>~T_{\rm osc}~ =~ g_*^{-1/4}\sqrt{\mreff(\phi_{\rm ini}) M_{\rm pl}} = 4.9 \times 10^{12}~\textrm{GeV} \left( \frac{100}{g_*}\right)^{1/4} \left( \frac{\mreff(\phi_{\rm ini})}{10^8~\rm GeV}\right)^{1/2} .
\end{equation}
Compatibility between Eq.~\eqref{eq:T_reh_upper_bound} and Eq.~\eqref{eq:T_reh_lower_bound} implies that the conditions for successful kination are lost as soon as
\begin{equation}
 g_*^{-1/4}\sqrt{\mreff(\phi_{\rm ini}) /M_{\rm pl}} ~>~y_\psi^{\frac{l-2}{l-4}} \left(\frac{g_*^{1/4}}{\sqrt{3}} \right)^{\frac{2}{l-4}}\left(\frac{\lambda \sqrt{2l-2}}{3\sqrt{c}} \right)^{\frac{1}{l-4}}.
\end{equation}
We find that the later condition does not add any constraint on our plots.

\end{enumerate}
\paragraph{No thermal-log domination.}
When fermions are Boltmann-suppressed, the thermal potential is given by the thermal log potential which is suppressed with respect to the quadratic thermal correction by $(\alpha T/y_\psi \phi)^2$. If the thermal-log dominates the potential at the onset of oscillation, we expect a suppression of the kick $\epsilon$ and a delay of the onset of the matter era as in scenario II, cf. Sec.~\ref{sec:effect_thermal_corrections}, but also the formation of Q-balls, see Sec.~\ref{sec:Q-balls}.
The thermal-log potential can be neglected whenever its associated thermal mass is smaller than the zero-temperature mass
\begin{equation}
\sqrt{2}\alpha \frac{T_{\rm osc}^2}{\phi_{\rm ini}} ~\lesssim ~\mreff(\phi_{\rm ini}) \quad \implies \quad \alpha~ \lesssim~ \frac{g_*^{1/2}}{\sqrt{2}}\left(\frac{\phi_{\rm ini}}{M_{\rm pl}}\right). \label{eq:no_thermal_log_at_osc}
\end{equation}
where we have used $\mreff(\phi_{\rm ini}) \simeq 3H_{\rm osc} \simeq g_*^{1/2}T_{\rm osc}^2/M_{\rm pl}$.

\subsubsection{Consequences} 
\paragraph{Non-suppressed angular kick.}
Since the thermal mass is absent at the onset of the radial mode oscillation, the onset of oscillation occurs when the Hubble scale crosses the zero-temperature mass
\begin{equation}
H_{\rm osc} \simeq \mreff(\phi_{\rm ini})/3.
\end{equation}
where $\phi_{\rm ini}$ is given by Eq.~\eqref{susy_phi_ini} and the fraction of $U(1)$ charge is the same as in the scenario I, cf. Eq.~\eqref{eq:epsilon_1_zero_temp}
\begin{equation}
\epsilon~ =~ 
\begin{cases}
\frac{1}{\sqrt{2}}\frac{\mreff(f_a)}{\mreff(\phi_{\rm ini})} \,l\,\sin{l \theta_{\rm ini}},~ \qquad \qquad \qquad \qquad \quad~ \text{if} \quad c~> ~l-1,\\[0.75em]
\frac{1}{\sqrt{2}}\sqrt{\frac{c}{l-1}} \frac{\mreff(f_a)}{\mreff(\phi_{\rm ini})} \,l\,\sin{l \theta_{\rm ini}}, ~~~\quad\qquad\qquad \quad  ~  \text{otherwise} .
\end{cases} \label{eq:epsilon_scenario_3}
\end{equation}
so that values $\epsilon \sim \mathcal{O}(1)$ are allowed.

\paragraph{No thermal mass domination at all.}
After the onset of oscillation, the scalar field $\phi \propto a^{-3/2}$ redshifts faster than the 
temperature $T \propto a^{-1}$.  This implies the existence of a temperature $T_{\rm rel}$ 
\begin{equation}
\label{eq:T_rel_def}
 T_{\rm rel} ~\equiv~ y_\psi \phi_{\rm rel},
\end{equation}
below which the fermion are relativistic and abundant in the plasma. 
Using that the scalar field $\phi$ reshifts as matter in a radiation-dominated universe, we get
 \begin{equation}
 \label{eq:phi_rel_sol}
 \phi_{\rm rel} =  \phi_{\rm ini} \left(\frac{H_{\rm rel}}{H_{\rm osc}} \right)^{3/4},
 \end{equation}
 where $H_{\rm osc}\simeq \mreff(\phi_{\rm ini})/3$, $H_{\rm rel} \simeq g_*^{1/2} T_{\rm damp}^2/3M_{\rm pl}$. From Eq.~\eqref{eq:T_rel_def} and Eq.~\eqref{eq:phi_rel_sol}, we obtain
\begin{equation}
\label{eq:Tref_def_jump_sol}
T_{\rm rel}\simeq \frac{1 }{g_*^{3/4}} \frac{\mreff^{3/2}(\phi_{\rm ini})}{y_{\psi}^2 M_{\rm pl}^{1/2}} \left(\frac{M_{\rm pl}}{\phi_{\rm ini}} \right)^{2}.
\end{equation}
When the fermions become relativistic, the thermal mass jumps according to, cf. Eq.~\eqref{eq:thermal_mass_2}
\begin{equation}
\mreff\big|_{\rm th} = \sqrt{2}\alpha \frac{T_{\rm rel}^2}{\phi_{\rm rel}} \quad \rightarrow \quad \mreff\big|_{\rm th} = y_\psi T_{\rm rel}.
\end{equation}
This jump in thermal mass has no impact on the dynamics if the thermal mass remains smaller than the zero-temperature mass
\begin{equation}
y_\psi T_{\rm rel}~ \lesssim~ \mreff(\phi_{\rm rel}).
\end{equation}
From plugging Eq.~\eqref{eq:Tref_def_jump_sol} in the previous equation, we obtain that the scalar field evolves in its zero-temperature potential during the whole time if
\begin{equation}
\label{eq:nothermalmassatall}
y_\psi  ~\gtrsim~ \frac{1}{g_*^{3/4}}\left(\frac{\mreff(\phi_{\rm ini})}{\mreff(\phi_{\rm rel})} \right) \left( \frac{\mreff(\phi_{\rm ini})}{M_{\rm pl}} \right)^{1/2} \left( \frac{M_{\rm pl}}{\phi_{\rm ini}} \right)^2.
\end{equation}

\subsection{Evolution of the field and its energy density}
\subsubsection{Delay of matter domination.}

If the condition in Sec.~\ref{sec:condition_boltzmann} and in Eq.~\eqref{eq:nothermalmassatall} are satisfied, then the scalar field redshifts like matter during its whole evolution. Therefore, the situation is similar to scenario I in Eq.~\eqref{cosmohist_zero_temp_matter_dom} and the scalar dominates the energy density of the universe at
\begin{equation}
\rho_{\rm dom} ~ = ~  \frac{27\mreff(\phi_{\rm ini})^2 \phi_{\rm ini}^8}{16\MPl^6}  A_{\epsilon}^4
\label{cosmohist_zero_temp_matter_dom_2} \qquad \textrm{and} \qquad
\frac{a_{\rm dom}}{a_\textrm{osc}} ~ = ~ \frac{2\MPl^2 }{3 \phi_{\rm ini}^2}A_\epsilon^{-1}
\end{equation}
with
\begin{equation}
A_\epsilon = \begin{cases}
\epsilon , \qquad \qquad \text{if}~\rho_{\rm damp} > \rho_{\rm dom},\\[0.75em]
1, \qquad \qquad \text{if}~\rho_{\rm damp} < \rho_{\rm dom}.
\end{cases}
\end{equation}
We recall that the impact of $\epsilon$ on the evolution of the scalar field energy density is discussed in  App.~\ref{app:effect_epsilon_evolution}.

\subsubsection{Radial damping.}
The scalar field decay rate is given by Eq.~\eqref{eq:fermion_damping_rate_YG_main} (see App.~\ref{app:thermalization} for more details) which we rewrite here
\begin{align}
\Gamma_\phi \simeq  \begin{cases}
\textrm{ for} ~ y_\psi \phi < T :
\begin{cases}
 \textrm{for}~ \alpha T  > y_\psi \phi ,\qquad \frac{y_\psi^2 \alpha T}{2 \pi^2},\\[0.5em]
\textrm{for}~\alpha T < y_\psi \phi ,\qquad \frac{y_\psi^4 \phi^2}{\pi^2 \alpha T},
\end{cases} 
\\[2em]
\textrm{ for} ~ y_\psi \phi > T: 
\qquad 
b \alpha^2 \frac{\textrm{Max}\left[T,~m_\phi\right]^3}{\phi^2} ,
\\
\end{cases}
+\quad \frac{y_\psi^2 m_{\phi}}{8\pi} \Theta\left(m_\phi/2 - \textrm{Max}\left[y_\psi \phi,\,gT\right]\right).
\label{eq:fermion_damping_rate_YG_main_2}
\end{align}
The different decay channels in Eq.~\eqref{eq:fermion_damping_rate_YG_main} are the ones induced by scattering with virtual fermions of the plasma $\phi\psi_{\rm th} \to \psi_{\rm th}$ $\Big(\Gamma = \frac{y_\psi^2 \alpha T}{2 \pi^2}$ or $\frac{y_\psi^4 \phi^2}{\pi^2 \alpha T}\Big)$, tree-level decay $\phi \to \psi \psi$ $\left(\Gamma =\frac{y_\psi^2 m_{\phi}}{8\pi}\right)$ or loop-induced decay into gauge boson pair $\phi \to AA$  $\left( \Gamma = b \alpha^2 \frac{\textrm{Max}\left[T,~m_\phi\right]^3}{\phi^2}\right)$. 
 In Fig.~\ref{fig:decay_rates}, we show the different decay rates and compute the decay temperature for three different values of the Yukawa coupling $y_\psi$.

\FloatBarrier
\begin{figure}[h!]
\centering
\raisebox{-0.5\height}{\makebox{\includegraphics[width=0.8\textwidth, scale=1]{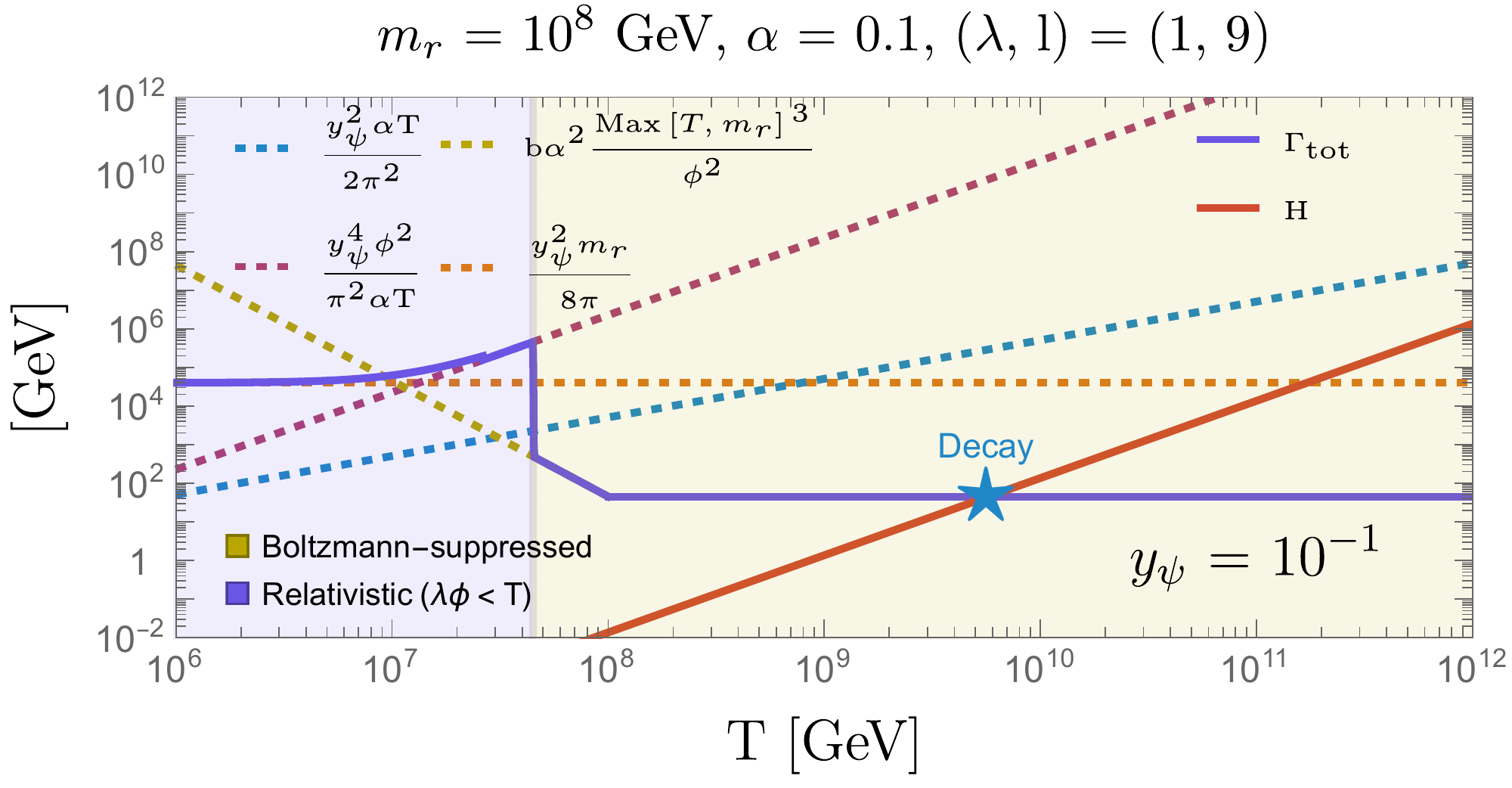}}}
\qquad
\raisebox{-0.5\height}{\makebox{\includegraphics[width=0.8\textwidth, scale=1]{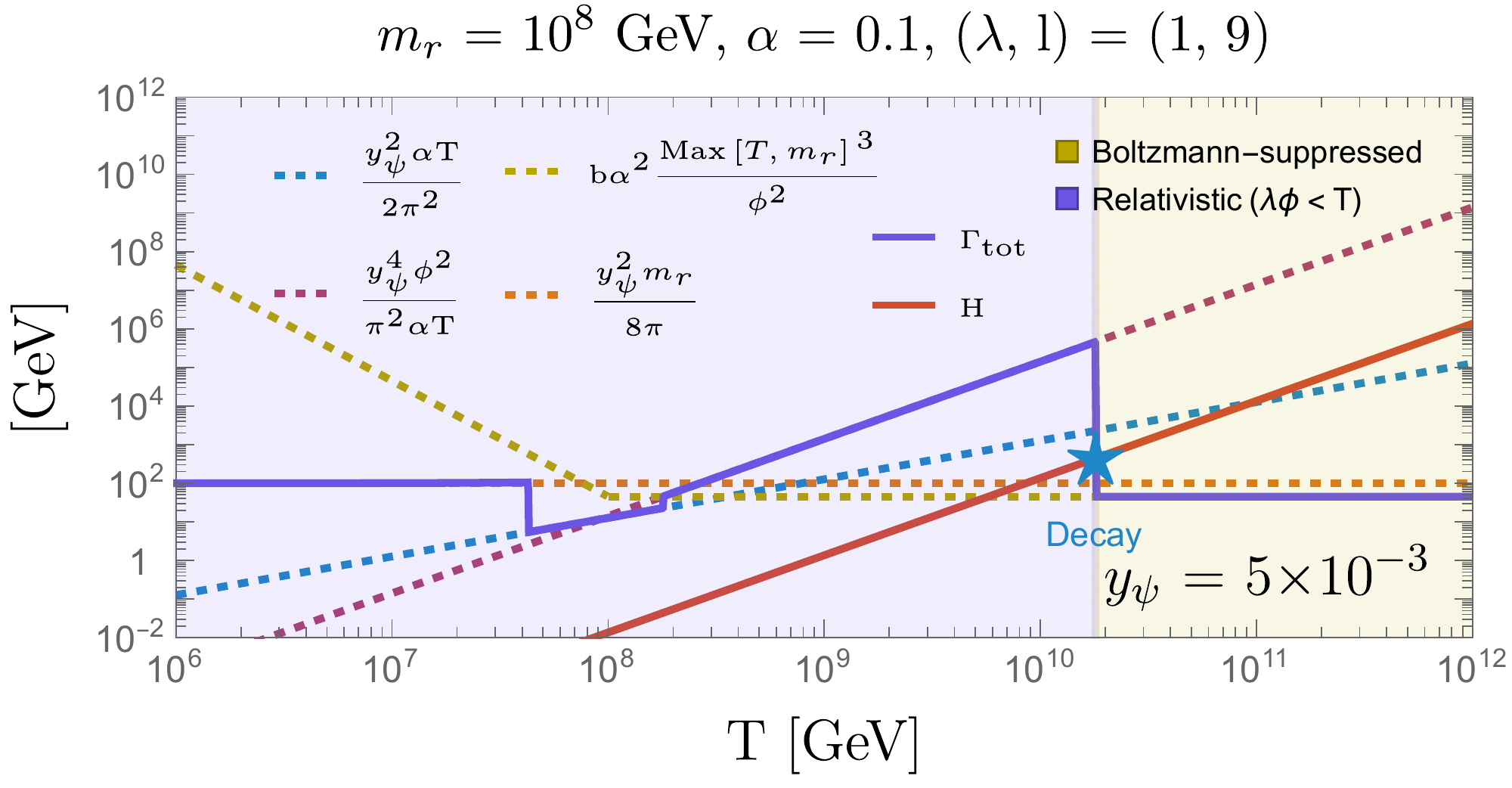}}}
\raisebox{-0.5\height}{\makebox{\includegraphics[width=0.8\textwidth, scale=1]{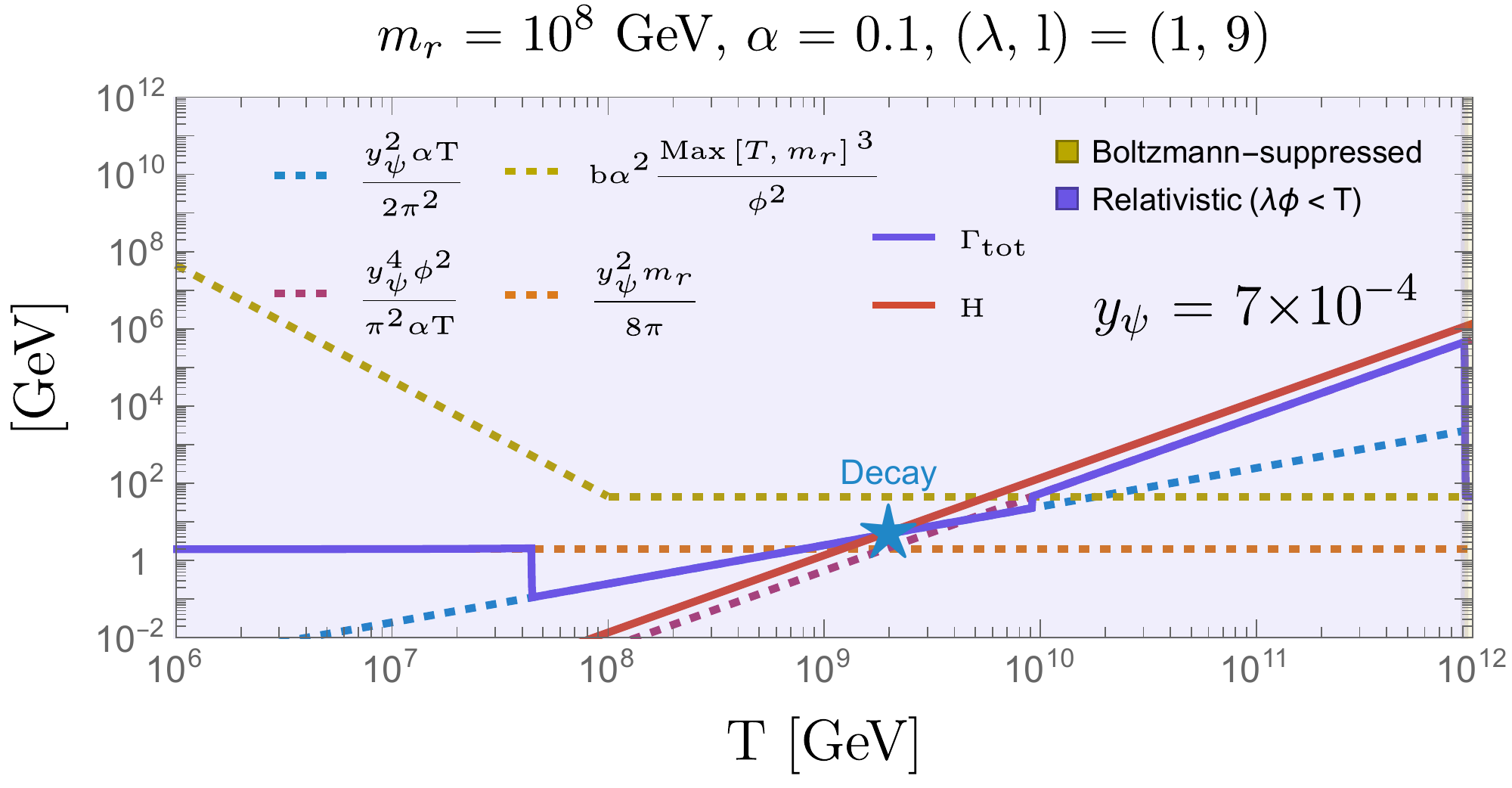}}}
\caption{\textit{ \small The total decay rate (solid purple) of the scalar condensate in the fermion portal  compared to the Hubble scale (solid red). The decay temperature is depicted by a blue star. We show three values of $y_\psi$, corresponding to the three decay channels given in Eq.~\eqref{eq:fermion_damping_rate_YG_main_2} and depicted in Fig.~\ref{fig:KSVZ_decay_diagrams} (see App.~\ref{app:thermalization} for more details). From \textbf{top} to \textbf{bottom}, the dominant channel is the decay into gauge boson pairs at one loop $\phi \to AA$ (dotted gold), scattering with virtual fermions of the plasma $\phi\psi_{\rm th} \to \psi_{\rm th}$ in the small thermal width $\alpha T \lesssim y_\psi \phi$ (dotted purple) and large thermal width limit $\alpha T \gtrsim y_\psi \phi$ (dotted blue). Visible on these plots is the suppression of the decay into fermions when the later are non-relativistic (yellow region).}}
\label{fig:decay_rates}
\end{figure}
\FloatBarrier

For the better understanding, we now derive analytically the expression for the maximum damping temperature.
We suppose that the conditions introduced in Sec.~\ref{sec:condition_boltzmann} are satisfied such that the fermion abundance is initially Boltzmann-suppressed at the onset of scalar field oscillation.
When the fermions become relativistic at the temperature $T_{\rm rel}\equiv y_{\psi}\phi_{\rm rel}$ given in Eq.~\eqref{eq:T_rel_def}, the damping rate suddenly increases due the scattering with virtual fermions of the plasma, see Fig.~\ref{fig:decay_rates}. We suppose $\alpha \lesssim 1$ in Eq.~\eqref{eq:fermion_damping_rate_YG_main_2}, such that the decay rate at $T_{\rm rel}$ is controlled by
\begin{equation}
\Gamma\simeq \Gamma_{\phi\psi_{\rm th} \to \psi_{\rm th}} = \frac{y_\psi^4 \phi^2}{\pi^2 \alpha T} \propto T^2.
\end{equation}
Since $\Gamma/H = \textrm{constant}$, the decay through that channel only depends on whether $\Gamma/H > 1$ or not at the relativistic threshold $T_{\rm rel}$. This defines the quantity $y_{\psi,*}$
\begin{equation}
\label{eq:y_psi_start_def_boost}
\Gamma/H\Big|_{T=T_{\rm rel}} > 1 \qquad \implies \qquad y_\psi >  y_{\psi,*} =\frac{1.35 \alpha^{1/4}}{g_*^{1/16}}\left(\frac{m_{\rm eff,ini}}{\MPl}\right)^{3/8}\left(\frac{M_{\rm pl}}{\phi_{\rm ini}}\right)^{1/2}.
\end{equation}

We deduce the maximum damping temperature
\begin{equation}
\label{eq:Tdamp_max}
\begin{cases}
\alpha \lesssim 1, \\
y_\psi >  y_{\psi,*},
\end{cases}
\qquad \implies 
\qquad
T_{\rm damp} = T_{\rm rel},
\end{equation}
with $T_{\rm rel}$ given in Eq.~\eqref{eq:T_rel_def}. 
The damping temperature peaks for $y_\psi =  y_{\psi,*}$, see maximum of the dashed yellow line in Fig.~\ref{fig:scenarioI_II_III_TmaxOTreh}-top.

The energy density just after damping is given by, see Eq.~\eqref{epsilon_case2_dom}
\begin{equation}
\rho_{\rm damp}~ = ~ \frac{\pi^2}{30}g_* T_{\rm damp}^4\,B_{\epsilon}^4,
\end{equation}
with
\begin{equation}
B_\epsilon = \begin{cases}
1 , \qquad \qquad \text{if}~\rho_{\rm damp} > \rho_{\rm dom},\\[0.75em]
\epsilon, \qquad \qquad \text{if}~\rho_{\rm damp} < \rho_{\rm dom}.
\end{cases}
\end{equation}

\paragraph{Preservation of the $U(1)$ charge in the condensate.}

The $U(1)$ charge carried by the condensate is preserved during thermalization if \cite{Co:2019wyp} (see also App.~\ref{app:thermalization})
\begin{equation}
f_a ~\gg~T_{\phi \to f_a}, \label{eq:cond_phi_damp_larger_Tdamp}
\end{equation}
where $T_{\phi \to f_a}$ is the temperature when $\phi$ reaches $f_a$.
Using that $\phi$ redshifts like matter, $\phi = \phi_{\rm ini} 
\left( \frac{T}{T_{\rm osc}}\right)^{3/2}$, we obtain 
\begin{equation}
\frac{f_a}{T_{\phi \to f_a}}~=~0.8 \times 10^5~ \left(\frac{f_a}{10^8~\rm GeV}\right)^{1/3}\left(\frac{10^{10}~\rm GeV}{T_{\rm osc}}\right)\left(\frac{\phi_{\rm ini}}{M_{\rm pl}}\right)^{2/3}~\gg~1. \label{eq:check_phi_damp_larger_Tdamp}
\end{equation}
We checked that all the parameter space shown in Figs.~\ref{fig:scenarioI_II_III_TmaxOTreh},  \ref{fig:fa_MrOfa_TmaxOTreh}, \ref{fig:complex_scenario3_inf1}, \ref{fig:complex_scenario3_local1} and \ref{fig:complex_scenario3_global1} satisfies that condition.

\subsubsection{Duration of the kination era}

\paragraph{Start of kination.} 
The universe acquires a kination equation-of-state when the field reaches $\phi \to f_a$, corresponding to the energy density and scale factor
\begin{equation}
 \rho_{\textrm{KD},i} ~ = ~  \frac{1}{2}f_a^2 \mreff ^2(f_a), \qquad \textrm{and} \qquad \frac{a_{\textrm{KD},i}}{\max(a_{\rm dom},a_{\rm damp})} = \left(\frac{\text{min}(\rho_{\rm dom},\,\rho_{\rm damp})}{\rho_\textrm{KD,i}}\right)^{\!1/3}.
\end{equation}

\paragraph{End of kination.} 
The kination era stops when the universe becomes radiation dominated. The energy scale at which it occurs depends on whether radial damping occurs before and after the onset of matter domination
\begin{equation}
\rho_{\textrm{KD},f} =\frac{\rho_{\textrm{KD},i}^2}{\text{min}(\rho_{\rm dom},\,\rho_{\rm damp})}.
\end{equation}

\paragraph{Duration of the kination era.} 
The duration of the kination era $N_{\rm KD} \equiv \log{\frac{a_{\textrm{KD},f}}{a_{\textrm{KD},i}}} = \frac{1}{6} \log  \left(\frac{\text{min}(\rho_{\rm dom},\,\rho_{\rm damp})}{\rho_{\textrm{KD},i}}\right)$ reads
\begin{equation}
e^{N_{\rm KD}}=
\begin{cases}
\label{eq:end_kination_scale_factor_scenario_3}
\sqrt{\frac{3}{2}} \left(\frac{\mreff(\phi_{\rm ini})}{\mreff(f_a)} \frac{M_{\rm pl}}{f_a}\right)^{1/3}  \left( \frac{\phi_{\rm ini}}{M_{\rm pl}} \right)^{4/3}\epsilon^{2/3},   \qquad  \qquad\qquad~~~~\text{if}~\rho_{\rm damp} > \rho_{\rm dom},\\[0.75em]
\left(\frac{\frac{\pi^2}{30}g_* T_{\rm damp}^4}{f_a^2 \mreff ^2(f_a)/2}\right)^{1/6}\epsilon^{2/3},  \qquad \qquad\qquad \qquad\qquad \qquad \quad  \text{if}~\rho_{\rm damp} < \rho_{\rm dom}.
\end{cases} 
\end{equation}
where $\epsilon$ can be $\mathcal{O}(1)$ and where $T_{\rm damp}$ is given by Eq.~\eqref{eq:Tdamp_max}. In the case where $y_{\psi} >y_{\psi,*}$ with $y_{\psi, *}$ given by Eq.~\eqref{eq:y_psi_start_def_boost}, we obtain 
\begin{equation}
e^{N_{\rm KD}}=
\begin{cases}
 e^{8.1} \epsilon^{2/3}\left(\frac{10^{8}\rm GeV}{f_a}\right)^{1/3}  \left(\frac{\mreff(\phi_{\rm ini})}{\mreff(f_a)} \right)\left( \frac{\phi_{\rm ini}}{M_{\rm pl}} \right)^{4/3},   \qquad\qquad \qquad\qquad\qquad\qquad ~~\quad\text{if}~\rho_{\rm damp} > \rho_{\rm dom},\\[0.75em]
e^{4.2}\frac{\epsilon^{2/3}}{g_*^{1/3}} \left( \frac{f_a}{10^{8}~\rm GeV}\right)^{1/3}\left( \frac{\mreff(f_a)}{f_a}\right)^{2/3}\left( \frac{\mreff(\phi_{\rm ini})}{\mreff(f_a)}\right)\left( \frac{10^{-4}}{y_{\psi}}\right)^{4/3} \left( \frac{M_{\rm pl}}{\phi_{\rm ini}}\right)^{4/3}\qquad   \qquad \text{if}~\rho_{\rm damp} < \rho_{\rm dom}.
\end{cases} 
\end{equation}
The maximal duration of kination is reached for $y_\psi = y_{\psi,*}$
\begin{framed}
\begin{equation}
\label{eq:NKD_scenario3_best}
e^{N_{\rm KD}}\big|_{y_\psi = y_{\psi,*}} = e^{5.1} \frac{\epsilon^{2/3}}{\alpha^{1/3}g_*^{1/4}}\left(\frac{10^8}{f_a} \right)^{1/6}\left(\frac{\mreff(f_a)}{f_a} \right)^{1/6}\left(\frac{\mreff(\phi_{\rm ini}}{\mreff(f_a)} \right)^{1/2} \left( \frac{0.1}{\phi_{\rm ini}/M_{\rm pl}} \right)^{2/3}.
\end{equation}
\end{framed}

\subsection{Gravitational-wave signature and detectability}\label{sec:GW_scenarioIII}
Next, we show the detectability of the SGWB produced by primordial inflation (Fig.~\ref{fig:complex_scenario3_inf1}), local strings (Fig.~\ref{fig:complex_scenario3_local1}) and globals strings (Figs.\ref{fig:complex_scenario3_global1}), in the presence of a kination era generated by the scenario III: a spinning complex scalar field with thermal damping and Boltzmann-suppression of the thermal corrections to the potential. 

The parameter space splits into two parts separated by the blue dashed line: the Boltzmann suppression by a large $y_\psi$ in Eq.~\eqref{eq:boltzmann_lower_bound_yukawa} and by a small reheating temperature in Eq.~\eqref{eq:T_reh_upper_bound}.
The kination duration in the former region is independent of $T_{\rm reh}$, and the inflationary or the string scales are not bounded. On the contrary, in the region with the small reheating temperature the  maximum energy scale is constrained if the universe reheats instantaneously.
In the figures below, some parts of the parameter space have the inflation scale and the string scale fixed at high-energy above the $T_{\rm reh,max}$. This is allowed when a period between the end of inflation and the completion of reheating exists. Its existence would induce the SGWB distortion at high-frequencies above the kination peak and might allow us to distinguish the large $y_\psi$ from the small $T_{\rm reh}$ cases.

Finally, we also show the constraints that apply for very low $m_r$, in which case the radial mode has a large thermal abundance and long lifetime, which is excluded either by overabundance if cosmologically stable or by late decay into photons after BBN  \cite{Cadamuro:2011fd}. The corresponding constraint $m_r > 10 ~\rm GeV (f_a / 10^9 ~\rm GeV)^{2/3}$  is reported in  the green hashed region in the plots.

These plots demonstrate that the concrete scenario III, where radial damping is not assumed but explicitly calculated via thermal effects, leads to observable signatures of an intermediate  matter-kination era.

\begin{figure}[h!]
\centering
~~~~~\raisebox{0cm}{\makebox{\includegraphics[width=0.775\textwidth, scale=1]{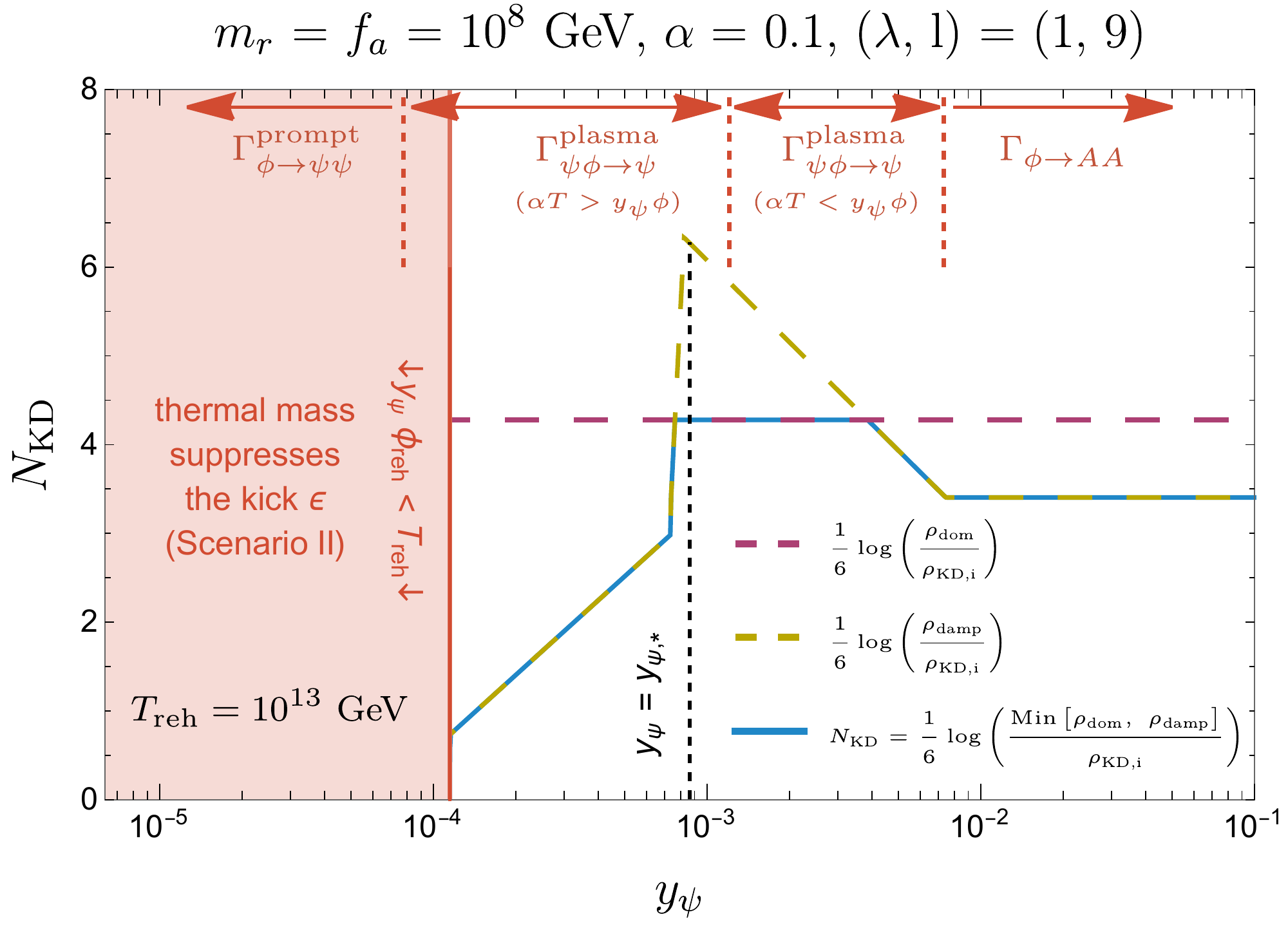}}}
\raisebox{0cm}{\makebox{\includegraphics[width=0.8\textwidth, scale=1]{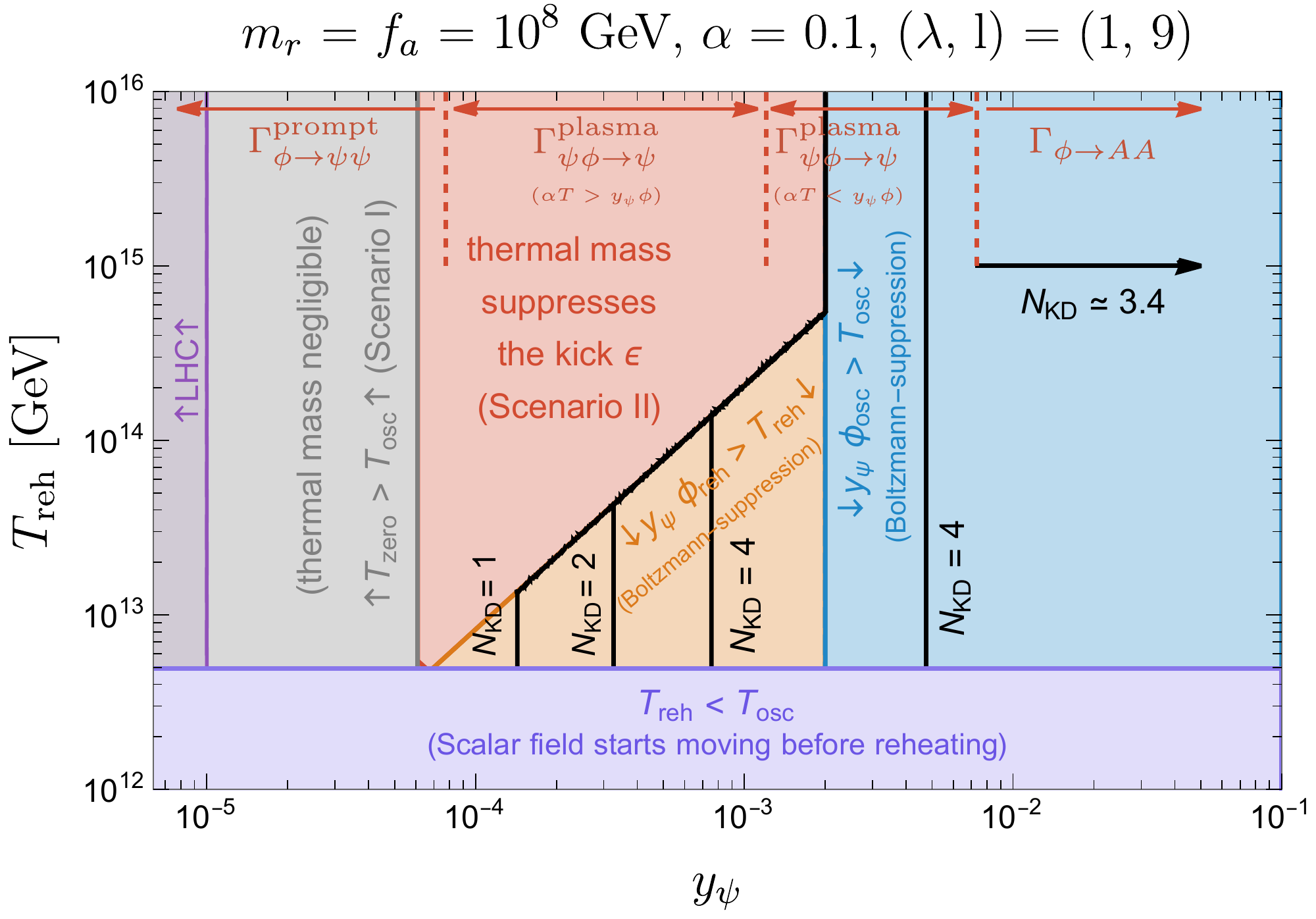}}}
\caption{\textit{ \small Number of kination e-folds when radial damping occurs via thermalization through fermion portal. The presence of the thermal mass at the onset of the oscillation (scenario II in Sec.~\ref{sec:PQ:exampleII}) suppresses the angular kick and prevents kination (red region). Instead, at large Yukawa coupling (blue region and condition~\eqref{eq:boltzmann_lower_bound_yukawa}) or at low reheating temperature (orange region and condition~\eqref{eq:T_reh_upper_bound}), the fermion abundance is Boltzmann-suppressed and the thermal mass is absent. 
 The $N_{\rm KD}$ black contour lines follow from Eq.~\eqref{eq:end_kination_scale_factor_scenario_3}, with the decay temperature $T_{\rm damp}$ being determined numerically as shown in Fig.~\ref{fig:decay_rates}. We write in red the dominant decay channel, based on Eq.~\eqref{eq:fermion_damping_rate_YG_main_2}. The maximal duration of kination is reached at $y_\psi = y_{\psi,*}$, cf. Eq.~\eqref{eq:y_psi_start_def_boost}, and is given by Eq.~\eqref{eq:NKD_scenario3_best}. The thermal mass is also negligible at small Yukawa (gray region and scenario I in Sec.~\ref{sec:scenario_I_non_thermal_damping}), but there, the thermalization rate is too small and a circular trajectory is not obtained before $\phi \to f_a$.  In the pale purple region, the kick at $T_{\rm osc}$ occurs before the universe is reheated which goes beyond the scope of our study, see Eq.~\eqref{eq:T_reh_lower_bound}. We show the LHC constraints on heavy colored fermions, $m_\psi = y_\psi f_a \gtrsim \rm TeV$. }}
\label{fig:scenarioI_II_III_TmaxOTreh}
\end{figure}

\FloatBarrier
\begin{figure}[h!]
\vspace{-1.2cm}
\centering
\raisebox{0cm}{\makebox{\includegraphics[width=0.48\textwidth, scale=1]{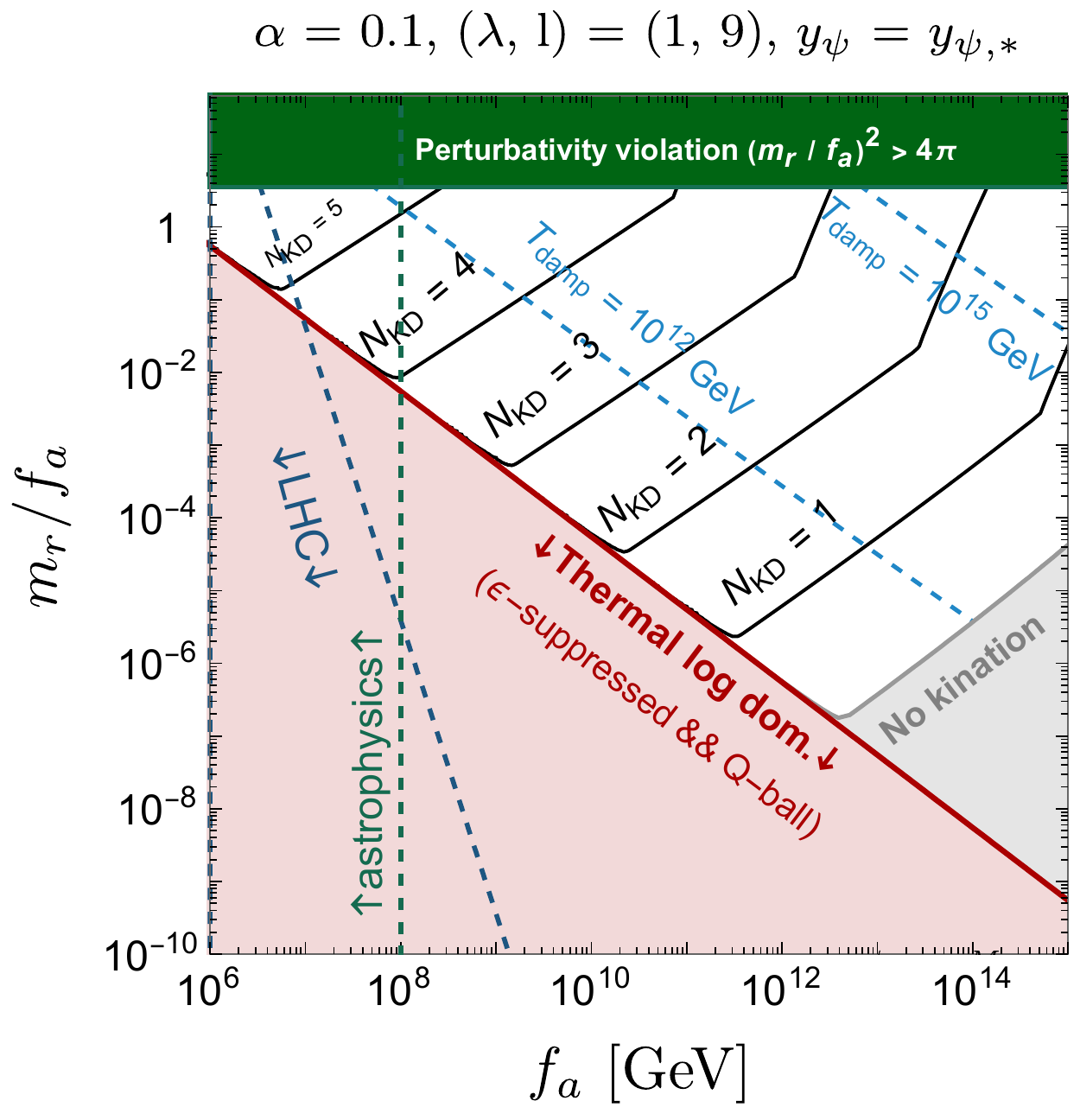}}}
\raisebox{0cm}{\makebox{\includegraphics[width=0.48\textwidth, scale=1]{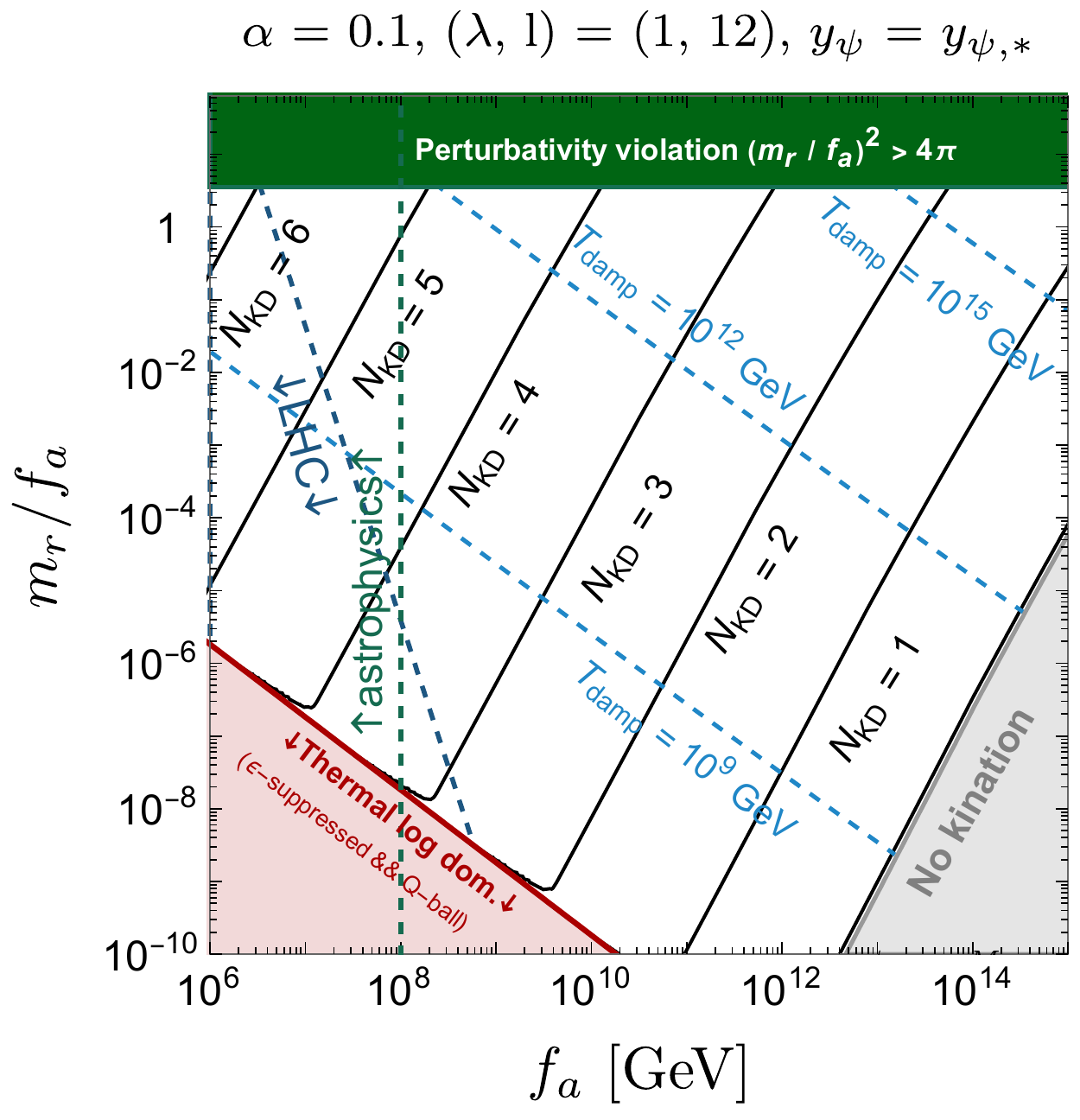}}}\\
\raisebox{0cm}{\makebox{\includegraphics[width=0.48\textwidth, scale=1]{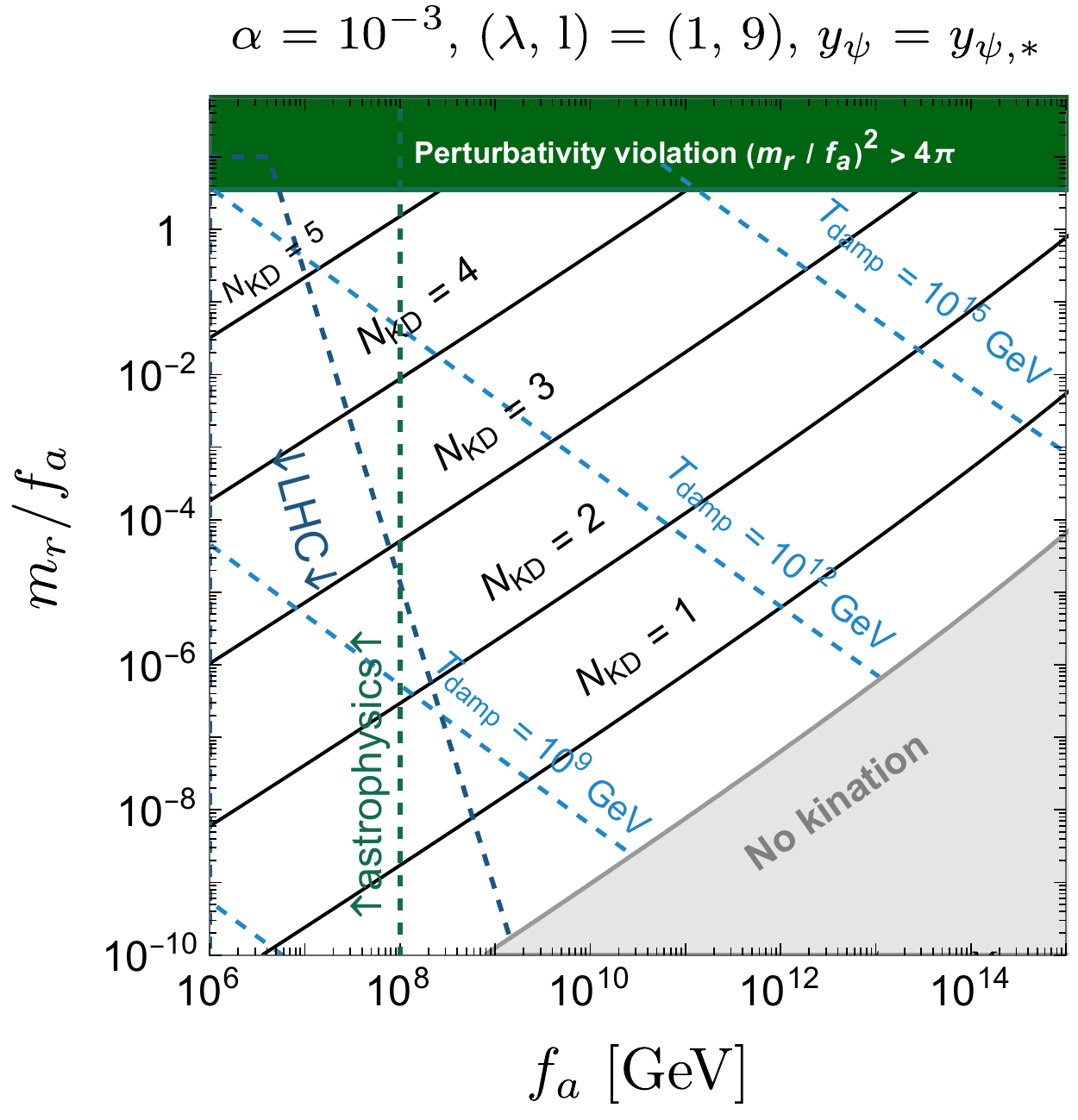}}}
\raisebox{0cm}{\makebox{\includegraphics[width=0.48\textwidth, scale=1]{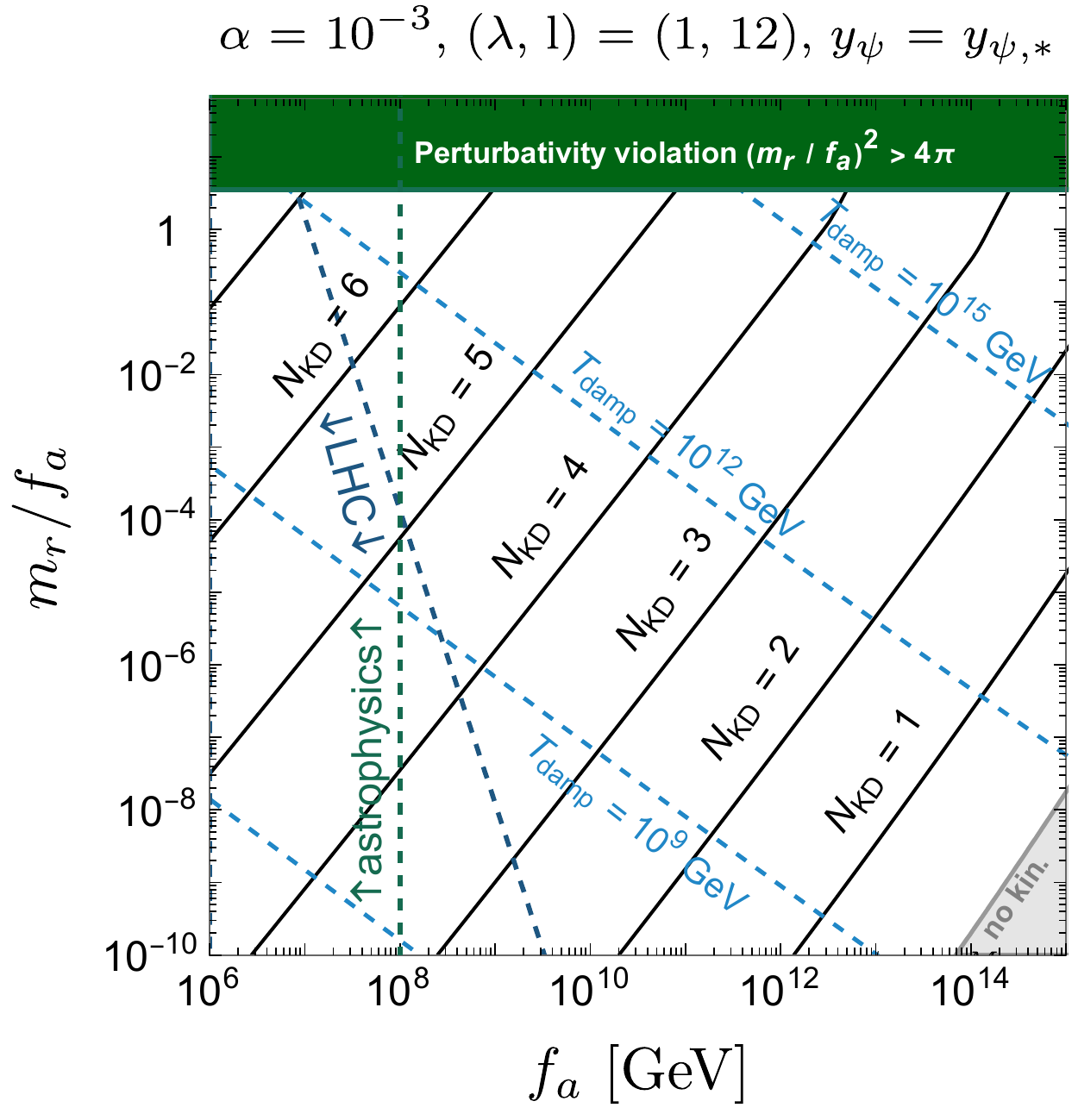}}}
\caption{\textit{ \small  Maximum number of kination e-folds in \textbf{scenario III}. In order to have matter-kination as long as possible, we chose values of the Yukawa coupling $y_\psi = y_{\psi,*}$ in Eq.~\eqref{eq:y_psi_start_def_boost} to maximize the duration of kination, see Eq.~\eqref{eq:NKD_scenario3_best}. The dashed blue lines show the temperature at which the radial mode is dampened by thermal effects. The dark blue region shows the LHC constraints $m_\psi \gtrsim \rm TeV$. The vertical green dashed line shows expected astrophysical constraints $f_a \gtrsim 10^8~\rm GeV$. The dark green region shows the limit of validity of the EFT. In the purple region, the thermal-log potential dominates the onset of oscillation, which suppresses the angular kick $\epsilon$, delays the onset of matter era, cf. Sec.~\ref{sec:effect_thermal_corrections}, and generates Q-balls, see Sec.~\ref{sec:Q-balls}.}}
\label{fig:fa_MrOfa_TmaxOTreh}
\end{figure}
\FloatBarrier

\FloatBarrier
\begin{figure}[h!]
\centering
{\bf Gravitational waves from primordial inflation}\\[0.5em]
\raisebox{0cm}{\makebox{\includegraphics[width=\textwidth, scale=1]{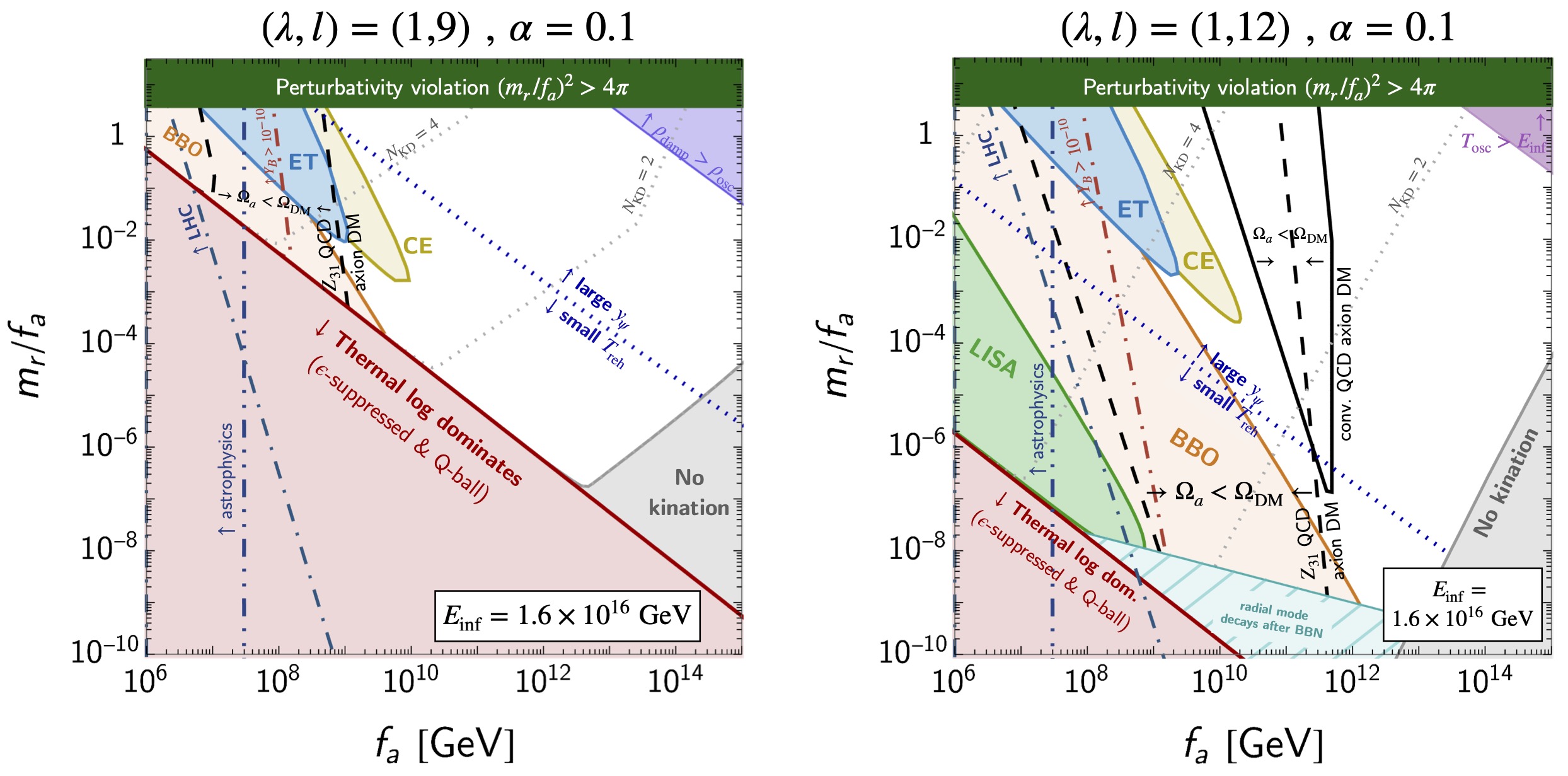}}}\\[0.25em]
\raisebox{0cm}{\makebox{\includegraphics[width=\textwidth, scale=1]{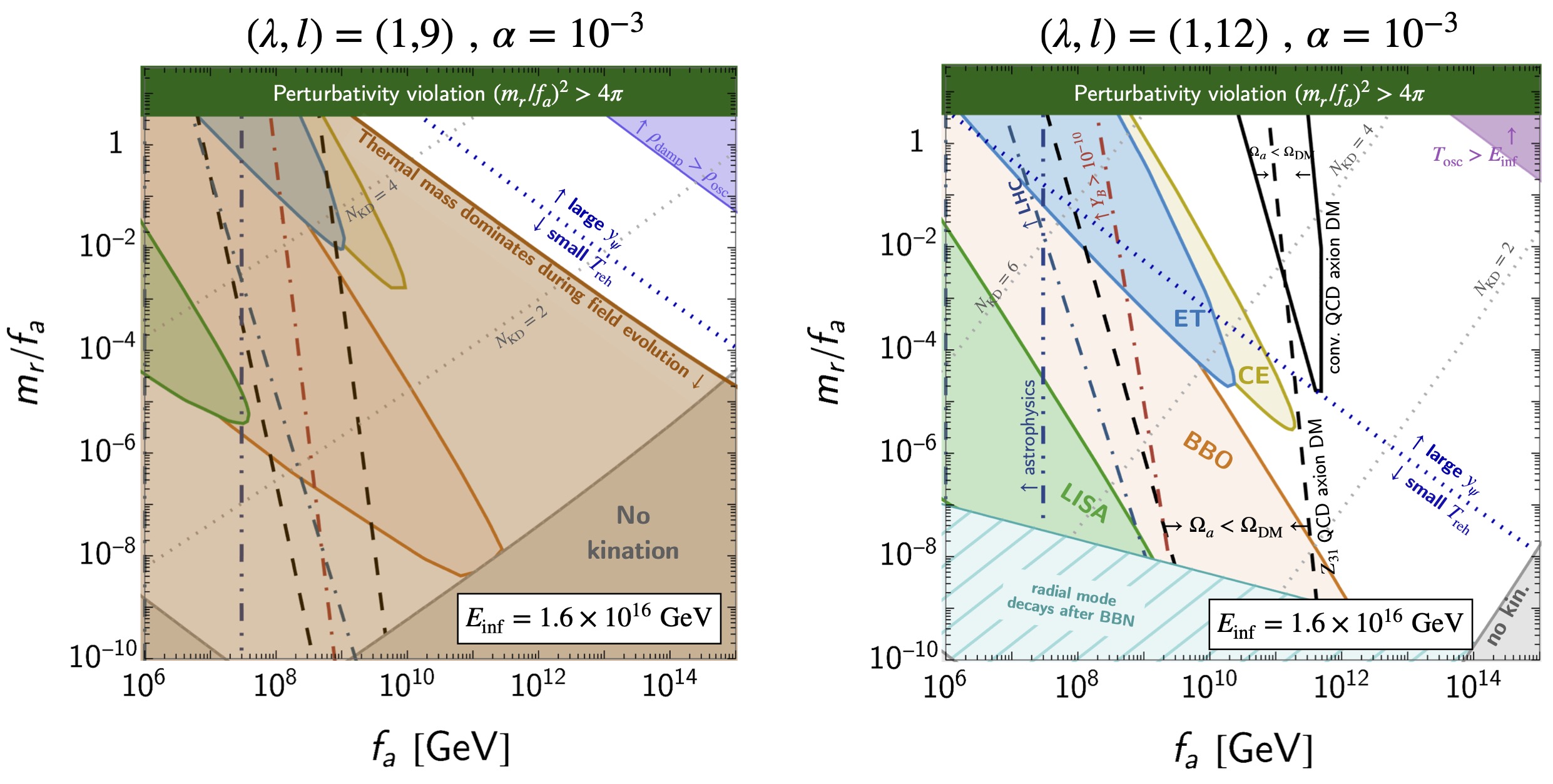}}}\\[0.25em]
\caption{\textit{ \small Ability of future-planned GW observatories to detect the peak signature of a matter-kination  era induced by \textbf{scenario III}  in the SGWB from inflation with energy scale $E_{\rm inf}$.  Black solid lines indicate where the canonical QCD axion DM abundance is satisfied, cf. Eq.~\eqref{eq:axion_abundance_general}. The left boundary is set by the kinetic misalignment mechanism, while the right one is by the standard misalignment (for small $m_r$ with a specific $f_a$) and by the axion quality problem (for larger $m_r$ depending on $f_a$), cf. Eq.~\eqref{eq:quality_problem_condition}. Only the region between the two lines does not over-produce DM. Dashed lines are the equivalent for lighter non-canonical QCD axion,  cf. Eq.~\eqref{eq:lighter_axion_mass}. A dotted-dashed red line denotes the parameter space where the spinning axion allows the correct baryon asymmetry, cf. Eq.~\eqref{Ekd_yield_bau}. Gray dotted lines show the kination duration contours. Smaller $m_r$ and $\lambda$ implies larger initial scalar vev $\phi_{\rm ini}$, cf. Eq.~\eqref{susy_phi_ini}, and longer matter-kination duration. For smaller $\alpha$, the coupling  $y_{\psi,*}$ decreases and cannot prevent the thermal mass to dominate during the field evolution as shown in bottom-left panel, cf. Eq.~\eqref{eq:nothermalmassatall}. The blue dashed line separates the large $y_\psi$ region where the kination duration is $T_{\rm reh}$-independent, from the small $T_{\rm reh}$ region. }}
\label{fig:complex_scenario3_inf1}
\end{figure}
\FloatBarrier

\FloatBarrier
\begin{figure}[h!]
\centering
{\bf Gravitational waves from local cosmic strings}\\[0.5em]
\raisebox{0cm}{\makebox{\includegraphics[width=0.9\textwidth, scale=1]{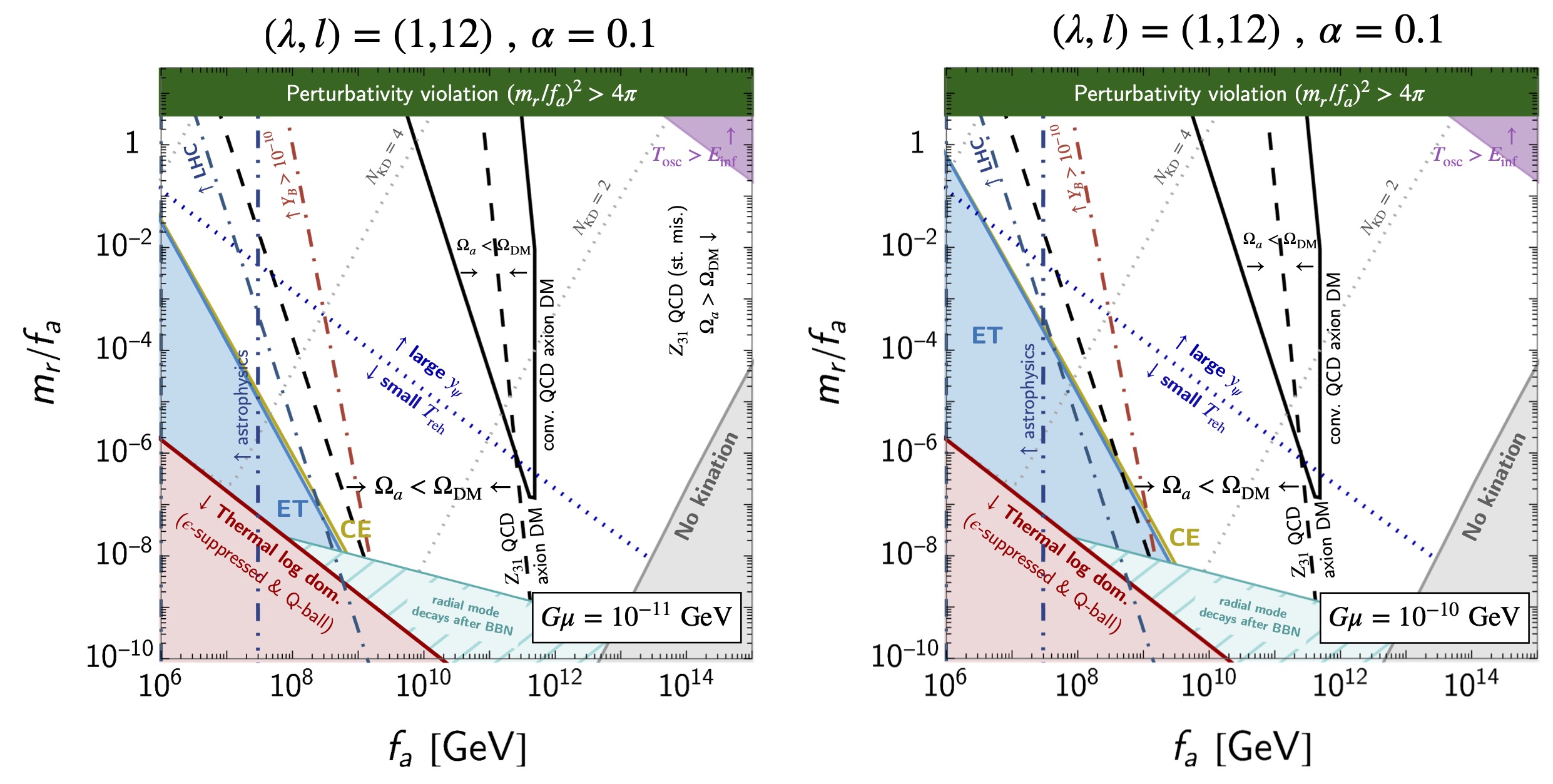}}}
\caption{\textit{ \small  Same as Fig.~\ref{fig:complex_scenario3_inf1} using the SGWB from local cosmic strings with tension $G\mu$. }}
\label{fig:complex_scenario3_local1}
\end{figure}
\FloatBarrier

\FloatBarrier
\begin{figure}[h!]
\centering
{\bf Gravitational waves from global cosmic strings}\\[0.5em]
\raisebox{0cm}{\makebox{\includegraphics[width=0.9\textwidth, scale=1]{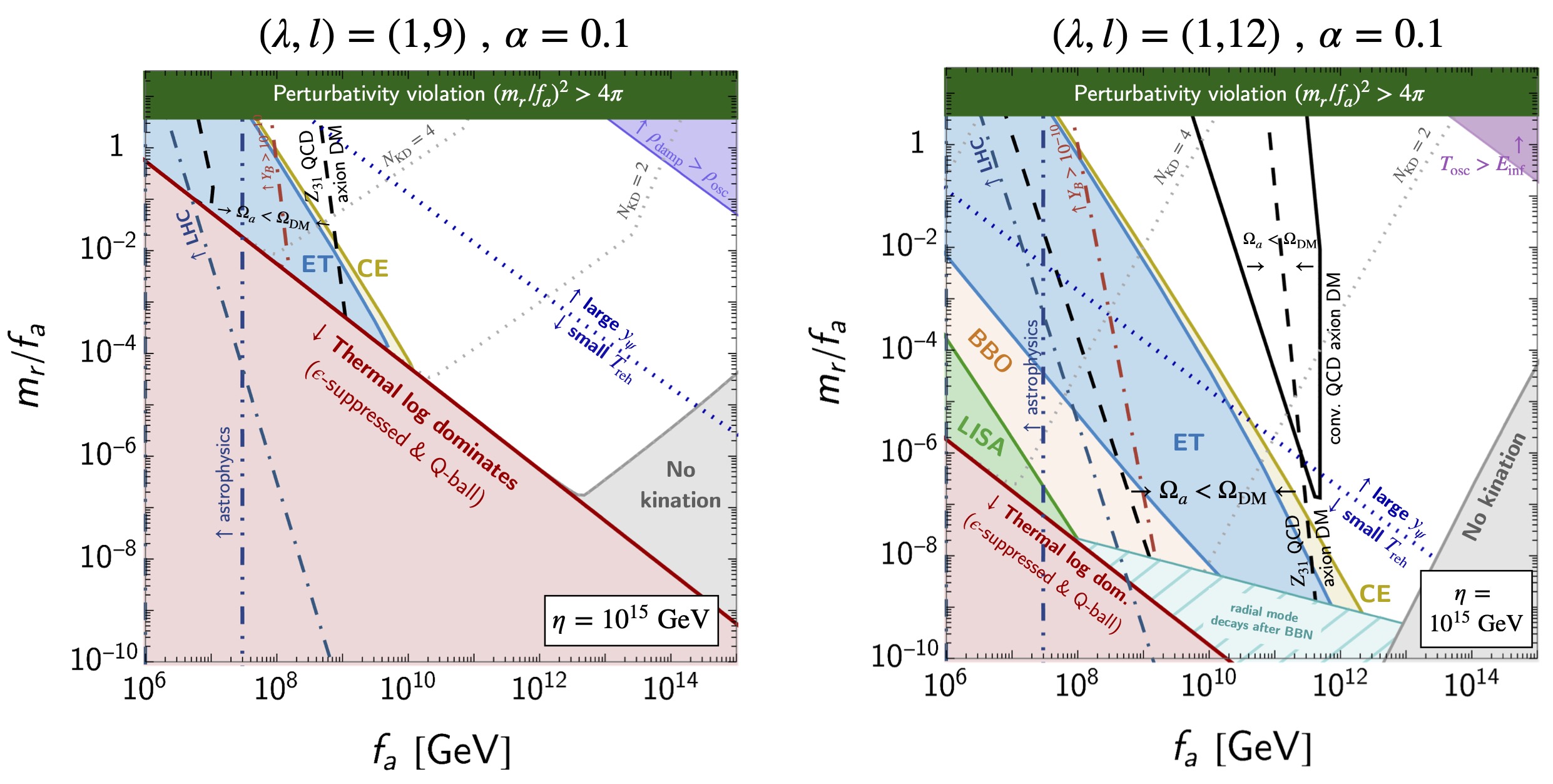}}}\\[0.25em]
\raisebox{0cm}{\makebox{\includegraphics[width=0.9\textwidth, scale=1]{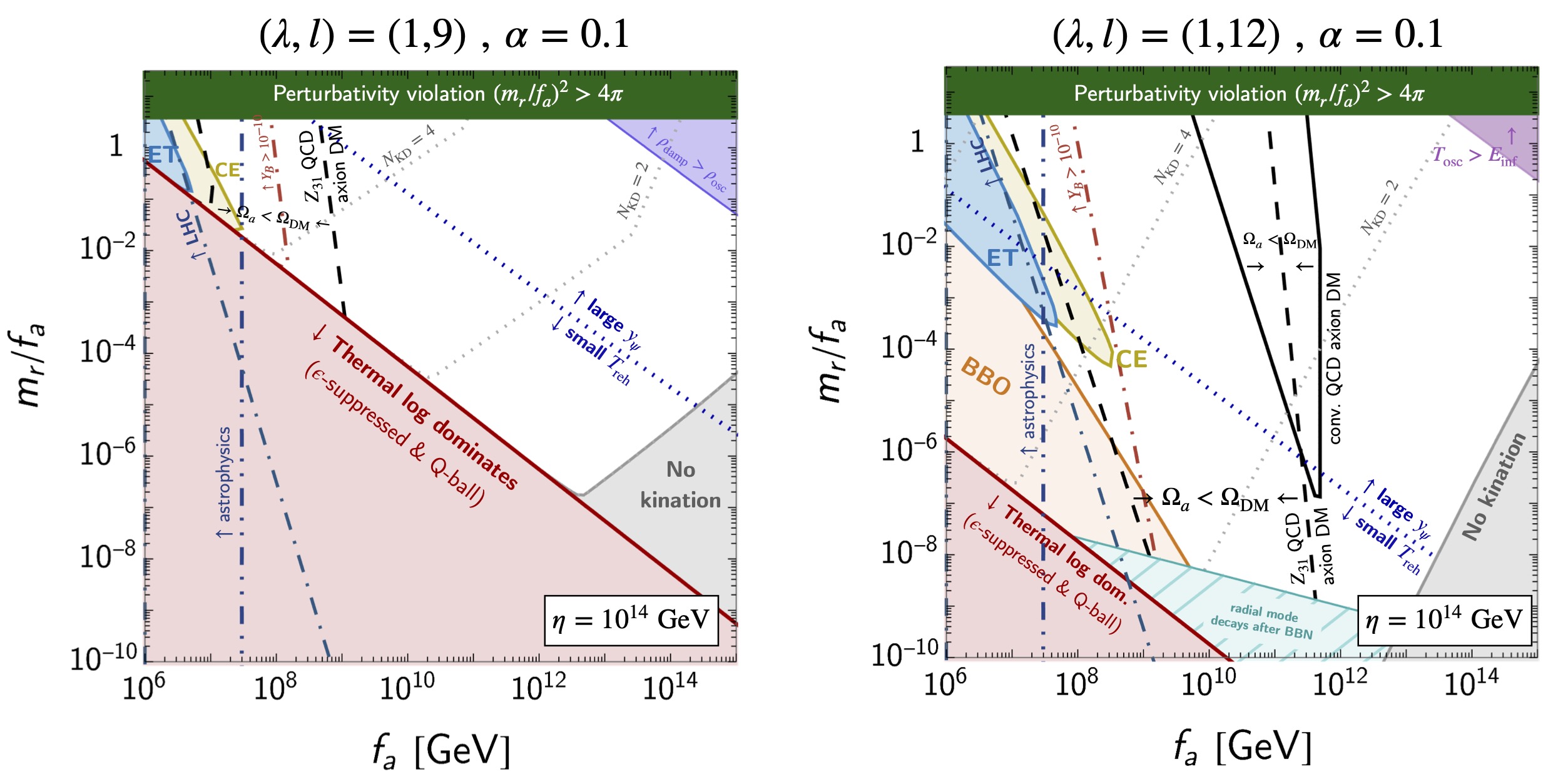}}}
\caption{\textit{ \small Same as Fig.~\ref{fig:complex_scenario3_inf1} using the SGWB from global cosmic strings with string scale $\eta$.  }}
\label{fig:complex_scenario3_global1}
\end{figure}
\FloatBarrier

\medskip

\section{Summary and conclusion}\label{sec:conclusion}

We have presented a comprehensive analysis of the possibility of a kination era taking place inside the standard radiation era in the cosmological history. So far, kination had been treated in the literature mainly as a phenomenon occurring at the end of inflation, before reheating to the Standard Model thermal bath, and therefore coming from the dynamics of the inflaton field itself. We have shown that a kination era can take place completely independently of the inflation physics, inside the standard radiation era, and rather generically  from the early dynamics of an axion field. In the first part of this paper, Sec.~\ref{sec:kination} and Sec.~\ref{sec:modelindependent}, we have shown model-independent results, displaying in detail  the consequences of this kination era for the main classes of gravitational-wave stochastic backgrounds of primordial origin, with emphasis on the long-lasting sources such as the irreducible one from inflation, and the signals from local and global cosmic strings.  The case of short-lasting primordial sources of gravitational waves such as first-order cosmological phase transitions is also discussed although it is not enhanced by kination.
GW from cosmic strings and inflation track the total energy density and therefore get enhanced during the matter+kination era. In contrast, GW from phase transitions come from an additional scalar-sector latent heat that is relatively suppressed in the presence of the axion-sector extra energy density responsible for the matter+kination era with respect to the case of standard cosmology.
For long-lasting sources, in Sec.~\ref{sec:darkmatter} we have  derived the relation between the amplitude and peak frequency of the GW signal to the relic abundance of axion DM. We have presented results for both the most general ALP particles as well as for the QCD axion. We discuss implications for the baryon asymmetry in Sec.~\ref{sec:baryon_asymmetry}.
In the second part of the paper, we have discussed the particle physics implementations of such intermediate kination era in one main class of axion models,  presented in Secs.~\ref{sec:PQ:exampleII}, \ref{sec:scenario_I_non_thermal_damping}, \ref{sec:complex_field_thermal_potential} and
\ref{sec:complex_field_low_reh_temp}, where the unavoidable partner of the axion, the radial mode of the complex scalar field, plays a crucial role in providing a kick to the axion. While the corresponding setup where the radial mode dynamics starts at large VEV $\gg f_a$  can be well-motivated in SUSY frameworks for instance, one non-trivial aspect  is that  the radial motion needs to be damped for successful kination.  

As this work was being completed, Refs.~\cite{Co:2021lkc, Gouttenoire:2021wzu} appeared which also discuss gravitational waves from a kination era triggered in axion models, and therefore overlap with our Secs.~\ref{sec:PQ:exampleII}, \ref{sec:scenario_I_non_thermal_damping}, \ref{sec:complex_field_thermal_potential} and
\ref{sec:complex_field_low_reh_temp}. As discussed extensively in these sections, the damping of the radial mode is absolutely crucial for kination to take place. Thermal damping of the radial mode requires to introduce new interactions of the Peccei-Quinn field. We have shown that these interactions induce a thermal mass for the Peccei Quinn field, thus delaying the time of the axion kick. This, together with a modified thermalisation temperature, prevents a kination era unless the reheating temperature of the universe is sufficiently low or the Yukawa coupling $y_\psi$ is sufficiently large.
We have derived  in detail the parameter space of the model that leads to a kination era and resulting observable gravitational waves.
We showed that $N_{\rm KD} < 6$ for most of the parameter space of the model. In our model-independent part we considered the energy scale and duration of kination as free parameters, with $N_{\rm KD}$ as high as $N_{\rm KD} < 11$. It would be interesting to investigate whether there are alternative models which can give longer kination eras.

The damping temperature is determined by the rate of thermalisation of the radial mode. We have used the thermalisation rate that follows from a yukawa interactions between the scalar field and new fermions, as derived in Refs.~\cite{Mukaida:2012qn,Mukaida:2012bz}, which we have relied on for our analysis.
While there is a large literature on axion cosmology, there is no study of the radial mode thermalisation beyond references \cite{Co:2019wyp,Co:2019jts,Chang:2019tvx,Co:2020dya,Co:2020jtv,Co:2021rhi,Harigaya:2021txz,Co:2021lkc}, on which we have extended by including the thermal mass effect in the equation of motion for the radial mode. We hope our work will motivate further investigations on these thermalisation effects as they are crucial for the early universe physics of axion, with far-reaching observable consequences. Another option is to assume non-thermal damping through parametric resonance. We have derived predictions in this case, treating the damping temperature as a free parameter. It remains to be checked whether this is indeed realizable. We have exposed the problematics in App.~\ref{app:param_resonance}. This will require a careful investigation of the coupled dynamics and of back-reaction effects.

While most of the literature on axion cosmology typically ignores the dynamics of its radial mode partner and only focuses on the angular mode dynamics, the analysis of our Secs.~\ref{sec:PQ:exampleII}, \ref{sec:scenario_I_non_thermal_damping}, \ref{sec:complex_field_thermal_potential} and
\ref{sec:complex_field_low_reh_temp} as well as references \cite{Co:2019wyp,Co:2019jts,Chang:2019tvx,Co:2020dya,Co:2020jtv,Co:2021rhi,Harigaya:2021txz,Co:2021lkc} show that the radial  mode may actually be a key to understand the early cosmological history in axion models, with important consequences for the experimental programme.
We note that if the complex scalar field belongs to a completely secluded sector, we expect the LHC and astrophysical bounds on $f_a$ in Fig.~\ref{fig:fa_MrOfa_TmaxOTreh}, \ref{fig:complex_scenario3_inf1}, \ref{fig:complex_scenario3_local1}, and \ref{fig:complex_scenario3_global1} to be relaxed. Additionally, the $N_{\rm eff}$ bounds can be evaded if the onset of matter domination takes place after thermalization. Low $f_a$ regions are interesting from the point of view of GW detection, see Fig.~\ref{detect_peak_inf}. In the case where the matter-kination era occurs after the end of BBN, below $6 ~ \rm keV$ \cite{Co:2021lkc}, the peak signature appears in the very low frequency region and could be probed by pulsar timing array experiments like SKA and NANOGrav \cite{Co:2021lkc}. We leave the dedicated study of the viability of this possibility for further works.

Fig.~\ref{fig:summary} shows the values of $f_a$ and of the kination energy scales that can be probed by GW experiments in the main class of axion model considered.  
We summarise the key expressions and figures of the paper in Tables~\ref{table:summary1} and \ref{table:summary2}.

Considering the particle physics implementations, Secs.~\ref{sec:PQ:exampleII}, \ref{sec:scenario_I_non_thermal_damping}, \ref{sec:complex_field_thermal_potential} and
\ref{sec:complex_field_low_reh_temp} improve the recent study \cite{Co:2021lkc} in different aspects which we can summarize as follows.
\begin{itemize}
\item
We compute the Noether $U(1)$ charge $\epsilon$ (or $Y_\theta$) explicitly from a particle physics model, and we show that it is suppressed when the thermal mass is present, see Sec.~\ref{sec:smaller_kick_thermal_mass}. The impact of $\epsilon$ on the kinaton duration is presented in Sec.~\ref{sec:epsilon_suppression_kination}.
\item
We show that the thermal mass is necessarily present at the time of the angular kick if we require efficient thermalization, and as a consequence, a period of matter-kination domination cannot be generated, see Sec.~\ref{sec:complex_field_thermal_potential}.
\item
A first solution is to consider radial damping through non-thermal processes as parametric resonance, see scenario I in Sec.~\ref{sec:scenario_I_non_thermal_damping}, even though a more quantitative analysis would be needed to assess this possibility.
\item
A second solution\footnote{We thank Keisuke Harigaya for fruitful discussions on this point.} is to consider the possibility that the fermions inducing the thermal mass are Boltzmann-suppressed at the time of the angular kick, when the value of $\epsilon$ is generated, see scenario III in Sec.~\ref{sec:complex_field_low_reh_temp}.
\item
We study the possibility of generating a matter-kination era from a spinning complex scalar field in all the possible values of the Yukawa coupling $y_\psi$ and reheating temperature $T_{\rm reh}$. We found four different regimes, pictured by colored regions in Fig.~\ref{fig:scenarioI_II_III_TmaxOTreh}.
\end{itemize}
As shown in Fig.~\ref{fig:fa_MrOfa_TmaxOTreh}, the parameter space associated with the longest matter-kination era lies in the region $f_a \lesssim 10^8$ GeV and $y_\psi f_a \lesssim \rm TeV$, potentially already excluded by astrophysics and LHC constraints. This motivates the study of an axion sector secluded from the SM.}

As the UV completion requires a number of assumptions (on the shape of the radial mode potential, on the absence of higher-dimensional operators with $l<6$, on the need for appropriate damping of the radial motion),  it will be interesting to motivate further constructions leading to high axion masses at early times compatible with a matter-kination era. We leave this for future work \cite{DESYfriendpaper2}.

Finally, a particularly intriguing scenario is the case where the same $U(1)$-breaking generating the axion leads to the global string network whose GW emission is enhanced by the matter-kination era induced by the spinning axion itself. It will be worth investigating such so-called axionic string framework in more details.

\FloatBarrier
\begin{figure}[h!]
\centering
\raisebox{0cm}{\makebox{\includegraphics[width=1\textwidth, scale=1]{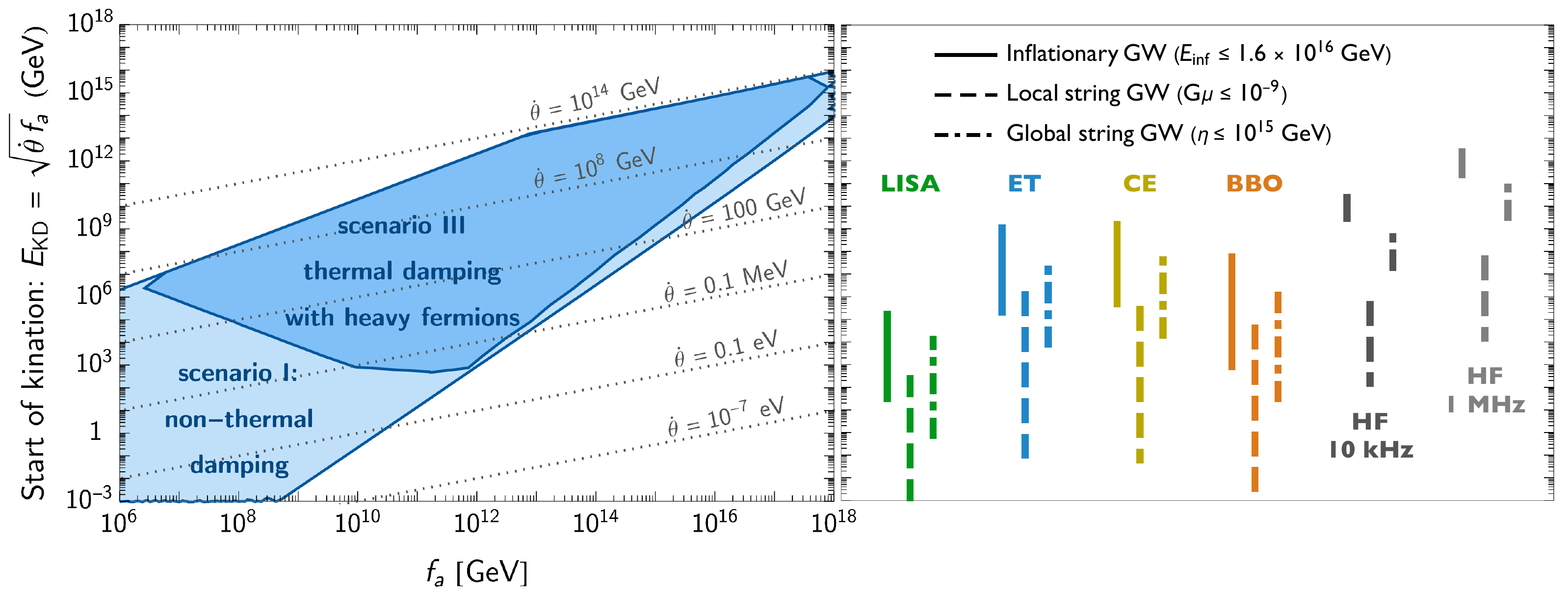}}}
\caption{\textit{ \small 
\textbf{Left}: Parameter space that leads to observable gravitational waves in the main class of model considered in this paper, scanning over $m_r$ and $l$ values and the corresponding values of the axion velocity at the start of kination. \textbf{Right}:  Detectable ranges of $E_{\rm KD}$ by the future planned GW experiments, for the three GW sources: primordial inflation, local and global cosmic strings. For the fictional high-frequency experiments, we assume detectors operating at $10 ~{\rm kHz}$ and $1 ~ {\rm MHz}$ with $\Omega_{\rm sens}h^2 \simeq 10^{-10}$.
}}
\label{fig:summary}
\end{figure}
\FloatBarrier

\begin{table}[h!]
\centering
{\bf Model-independent}\\[0.5em]
\begin{tabular}{|c|c|c|}
\hline
\multirow{2}{*}{\begin{tabular}[c]{@{}c@{}}GW from primordial inflation\\ Fig. \ref{detect_peak_inf}, \ref{detect_peak_Einf} \end{tabular}} & $f_{\rm peak}$ & Eq.~\eqref{inflation_peak_frequency}   \\ \cline{2-3} 
                  & $\Omega_{\rm peak}$ & Eq.~\eqref{inflation_peak_amplitude}  \\ \hline
\multirow{2}{*}{\begin{tabular}[c]{@{}c@{}}GW from local cosmic strings\\ Fig. \ref{detect_peak_cs_local}, \ref{detect_peak_cs_local_gmu} \end{tabular}} &   $f_{\rm peak}$    &  Eq.~\eqref{peak_kination_freq_local} \\ \cline{2-3} 
                  & $\Omega_{\rm peak}$ & Eq.~\eqref{bump_peak_kination_amp_local}  \\ \hline
\multirow{2}{*}{\begin{tabular}[c]{@{}c@{}}GW from global cosmic strings\\ Fig. \ref{detect_peak_global_string}, \ref{detect_peak_eta_global} \end{tabular}} &   $f_{\rm peak}$  & Eq.~\eqref{peak_kination_freq_global}  \\ \cline{2-3} 
                  & $\Omega_{\rm peak}$ & Eq.~\eqref{bump_peak_kination_amp_global_simp}  \\ \hline
\multirow{2}{*}{\begin{tabular}[c]{@{}c@{}}GW peak and axion relic abundance\\ Fig. \ref{fig:peak_spectrum_amplitude}, \ref{ma_fa_plot}, \ref{ma_fa_plot_axion_string} \end{tabular}} &   $E_{\rm KD} - \Omega_{a,0}$  & Eq.~\eqref{eq:Ekd_axion_fraction} \\ \cline{2-3} 
                  & $\Omega_{\rm peak} - \Omega_{a,0}$ & Eq.~\eqref{eq:peak_position_ALP2} \\ \hline
\multirow{2}{*}{\begin{tabular}[c]{@{}c@{}}GW peak and baryon asymmetry\\ Fig. \ref{bau_plot_spectrum} \end{tabular}} &  $E_{\rm KD} - Y_B$   & Eq.~\eqref{Ekd_yield_bau}  \\ \cline{2-3} 
                  & $\Omega_{\rm peak} - Y_B$ & Eq.~\eqref{eq:peak_position_bau_ALP2}  \\ \hline
\end{tabular}
\caption{List of key relations and figures in the model-independent analysis.}
\label{table:summary1}
\end{table}

\begin{table}[h!]
\centering
{\bf Model-dependent}\\[0.5em]
\begin{tabular}{|c|c|c|}
\hline
\multicolumn{1}{|c|}{Model: radial mode + axion ~ Fig. \ref{field_evolution}} & \multicolumn{2}{c|}{ \begin{tabular}[c]{@{}c@{}}  \\[-1.2em] $E_{\rm KD} = 2^{1/4} (f_a m_r)^{1/2}$\end{tabular}}                   \\ \hline
          \begin{tabular}[c]{@{}c@{}}Scenario I   (non-thermal damping)\\ Fig. \ref{inter_duration_2}, \ref{fig:complex_scenario1_inflation3}, \ref{fig:complex_scenario1_local2}, \ref{fig:complex_scenario1_global2} \end{tabular}           &             $N_{\rm KD}$          &    Eq.~\eqref{eq:longest_NKS_scenario_1_sketch}                   \\ \hline
              Scenario II    (thermal damping) Fig.~\ref{thermal_figur2}    &         $N_{\rm KD}$              &         Eq.~\eqref{eq:longest_NKS_scenario_2}              \\ \hline
\multicolumn{1}{|c|}{   \begin{tabular}[c]{@{}c@{}}Scenario III (thermal damping with non-relativistic fermions)\\ Fig. \ref{fig:scenario3_diagram}, \ref{fig:scenarioI_II_III_TmaxOTreh}, \ref{fig:fa_MrOfa_TmaxOTreh}, \ref{fig:complex_scenario3_inf1}, \ref{fig:complex_scenario3_local1}, \ref{fig:complex_scenario3_global1} \end{tabular}} & \multicolumn{1}{c|}{$N_{\rm KD}$} & \multicolumn{1}{c|}{Eq.~\eqref{eq:longest_NKS_scenario_3_sketch}} \\ \hline
\end{tabular}
\caption{List of key relations and figures in the model-dependent analysis.}
\label{table:summary2}
\end{table}

\section*{Acknowledgements}
Special thanks go to our DESY colleagues Cem Er{\"o}ncel, Pablo Quilez, Ryosuke Sato and Philip S{\o}rensen for important discussions. We are also grateful to Hyungjin Kim, Enrico Morgante, Juli\'an Rey,  Pedro Schwaller, Ivan Sobolev, and Alexander Westphal.
We thank Daniel Figueroa for his useful feedback during the DESY Theory Forum 2020 as well as Keisuke Harigaya and Kyohei Mukaida for clarifications and valuable insights on their papers. YG and GS are grateful to Galileo Galilei Institute for hospitality during the very final stage of this work while PS thanks Johannes Gutenberg Universit{\"a}t Mainz.  YG acknowledges the Azrieli Foundation for the award of an Azrieli Fellowship. This work is supported by the Deutsche Forschungsgemeinschaft under Germany Excellence Strategy - EXC 2121 ``Quantum Universe'' - 390833306.

\appendix

\section{Kination after inflation: quintessential inflation with $\alpha$-attractor}
\label{sec:quintessence}

Kination has been discussed in the literature as a follow-up of inflation, and before reheating. While  this paper instead investigates the possibility of kination inside the radiation era, except for the new constraints derived in Sec.~\ref{subsec:afterinflation},  we review in this appendix motivations for  models of kination following inflation for completeness. As already stated in Sec.~\ref{subsec:kinafterinf}, it is not possible to have a kination era from a canonical scalar field rolling down its potential without having superplanckian field excursions.  
On the other hand, kination arises in a popular class of inflation models called $\alpha$-attractor models which not only fit well CMB data but also link to quintessence at late time. In addition, they can be motivated in supergravity models.  We  summarise them in the following.

\subsection{A sharp transition between two plateaus}

\paragraph{$\alpha$-attractor.}
$\alpha$-attractor is a class of supergravity theory where kinetic terms have poles
\begin{equation}
\mathcal{L} = \frac{\frac{1}{2}(\partial \phi)^2}{(1- \frac{\phi^2}{6\alpha M_{\rm pl}^2})^2} - V(\phi) + \Lambda,   \qquad \alpha > 0.
\end{equation}
The use of $\alpha$-attractor for quintessential inflation has been studied in \cite{Linder:2015qxa,Mishra:2017ehw,Dimopoulos:2017zvq,Dimopoulos:2017tud,Akrami:2017cir,Dimopoulos:2019ogl}. 
Since the scalar field cannot cross the poles, the field range is limited to $-\sqrt{6 \alpha}M_{\rm pl} \lesssim \phi \lesssim +\sqrt{6 \alpha}M_{\rm pl}$. For $\alpha \lesssim 1/6$, it has the great advantage to prevent super-Planckian field excursion which plagues usual quintessential inflation scenarios. 

\paragraph{Scalar potential.}
Following \cite{Dimopoulos:2017zvq,Dimopoulos:2017tud,Dimopoulos:2019ogl}, we choose the following scalar potential
\begin{equation}
V(\phi) = V_0 \, e^{-\kappa \phi/M_{\rm pl}},
\end{equation}
where $\kappa > 0$, which can be motivated by supergravity \cite{Goncharov:1984qm,Ellis:1983sf}, brane inflation \cite{Burgess:2001vr}, string theory 
\cite{Cicoli:2008gp, Burgess:2016owb,Baumann:2014nda} or gaugino condensation \cite{Gorlich:2004qm,Lalak:2005hr,Haack:2006cy}.
At late time, the scalar field slows downs when reaching an infinite kinetic term at  $\phi \to +\sqrt{6 \alpha}M_{\rm pl}$. We suppose the existence of an unknown mechanism which set the cosmological constant (CC) to zero at $\phi \to +\sqrt{6 \alpha}M_{\rm pl}$
\begin{equation}
\Lambda = V(\sqrt{6 \alpha}M_{\rm pl}) = V_0\,e^{-\kappa \sqrt{6\alpha}},
\end{equation}
such that the scalar field potential energy + CC reads 
\begin{equation}
V(\phi) = V_0\,e^{-n} \left[ e^{n\left(1 - \frac{\phi}{\sqrt{6\alpha}M_{\rm pl}}\right) }- 1 \right], \qquad \text{with} \quad n \equiv \kappa \sqrt{6 \alpha}.
\end{equation}
\paragraph{Canonical normalization.}
Upon introducing the field transformation 
\begin{equation}
\frac{\partial \phi}{\partial \varphi} = 1 - \frac{\phi^2}{6\alpha M_{\rm pl}^2}, \qquad \leftrightarrow \qquad \phi = \sqrt{6 \alpha} M_{\rm pl} \, \textrm{tanh} \left( \frac{\varphi}{\sqrt{6 \alpha} M_{\rm pl}}  \right),
\end{equation}
we obtain a canonically normalized scalar field with potential energy
\begin{equation}
V(\varphi) = e^{-2n} M^4 \left[ e^{n\left(1 - \textrm{tanh} \frac{\varphi}{\sqrt{6 \alpha} M_{\rm pl}}\right)} - 1 \right], \qquad  \quad M^4 \equiv e^n V_0. \label{eq:pot_alpha_attractor} 
\end{equation}

\begin{figure}[t!]
\centering
\raisebox{0cm}{\makebox{\includegraphics[width=0.7\textwidth, scale=1]{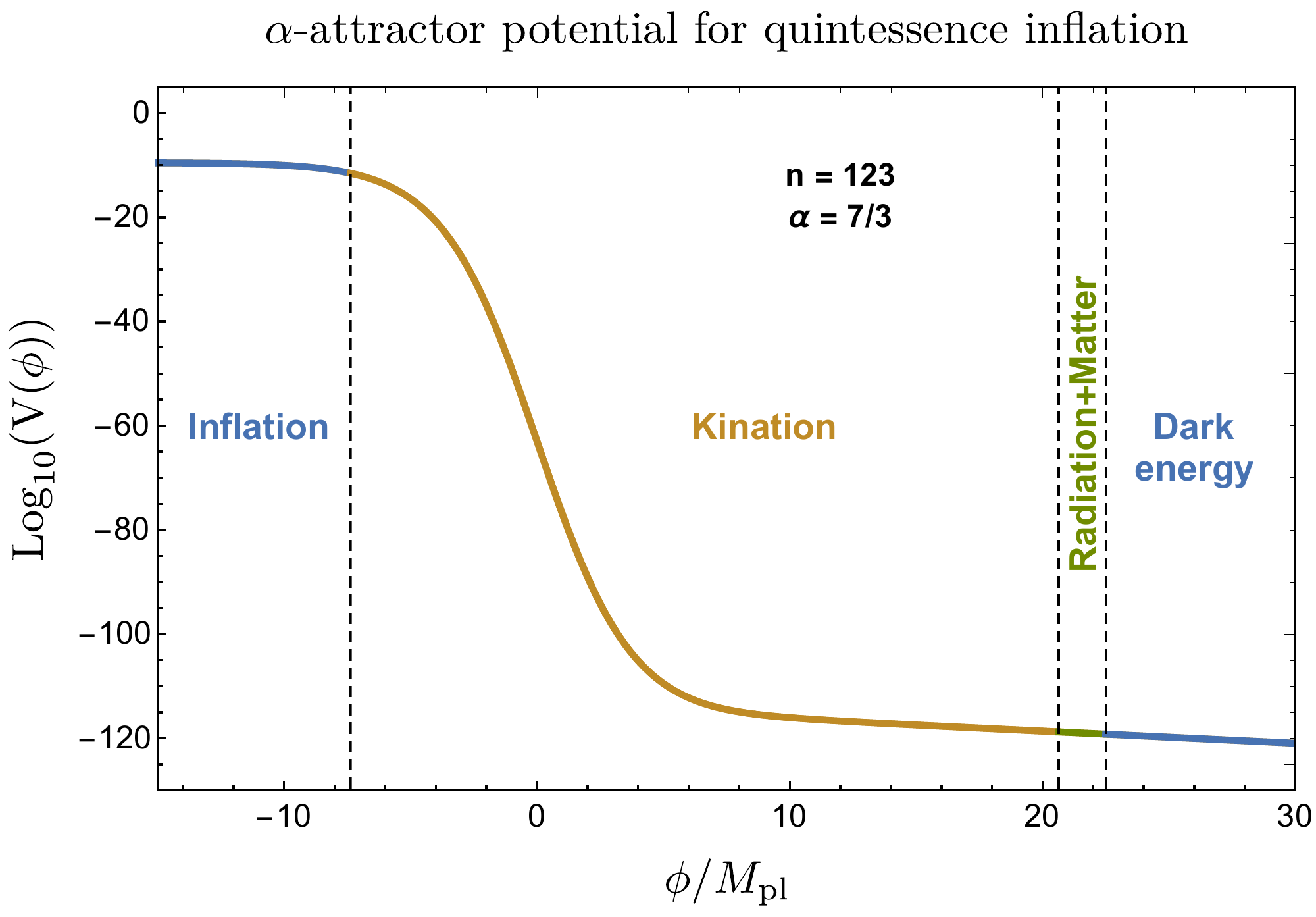}}}\hspace{0.2cm}
\caption{\textit{ \small  $\alpha$-attractor potential for quintessential inflation. We took the same parameters as in Fig.~\eqref{fig:quintessential_inf_alpha_attractor_energy}.}}
\label{fig:quintessential_inf_alpha_attractor_pot}
\end{figure}
\begin{figure}[ht!]
\centering
\raisebox{0cm}{\makebox{\includegraphics[width=0.7\textwidth, scale=1]{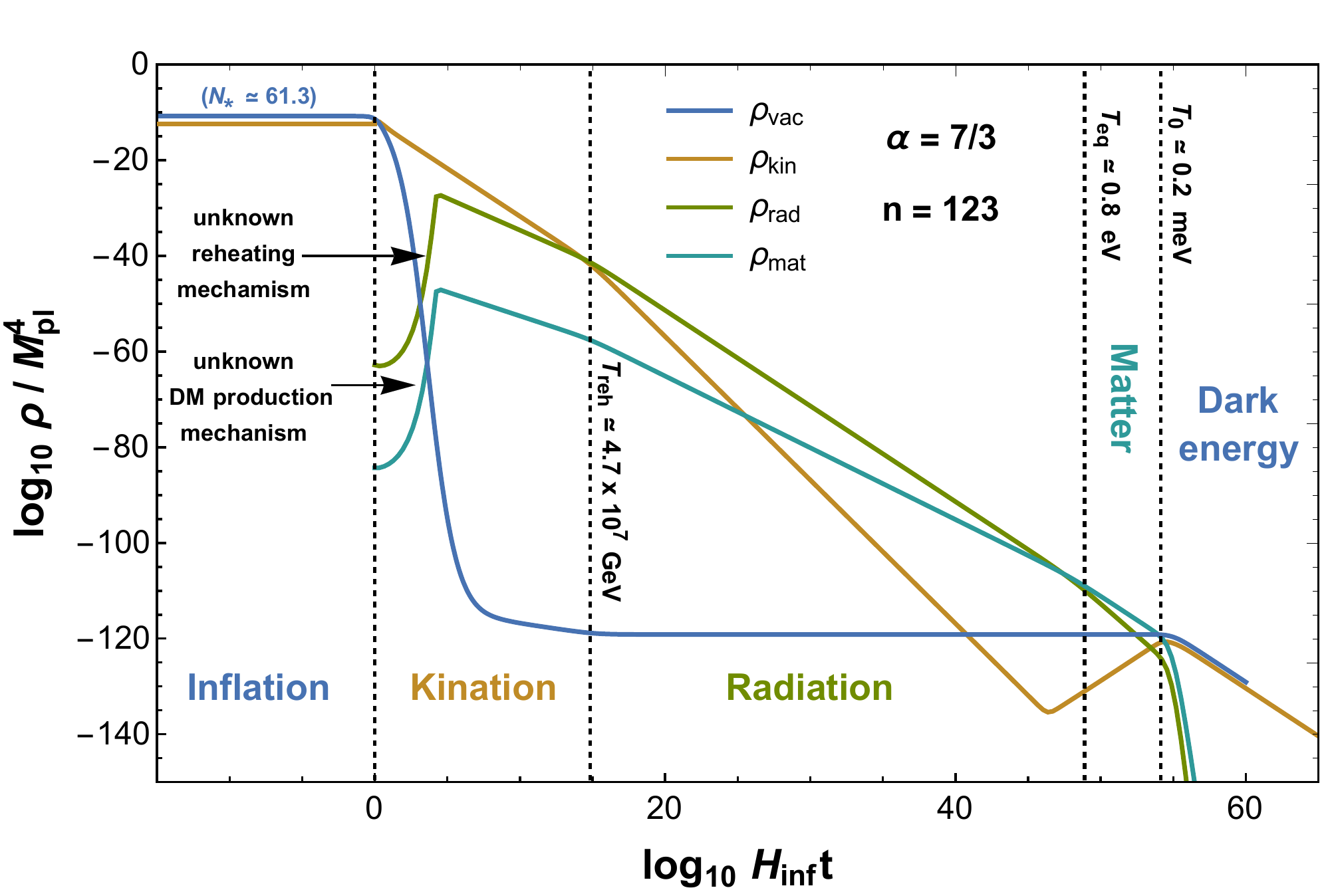}}}\hspace{0.2cm}
\raisebox{0cm}{\makebox{\includegraphics[width=0.68\textwidth, scale=1]{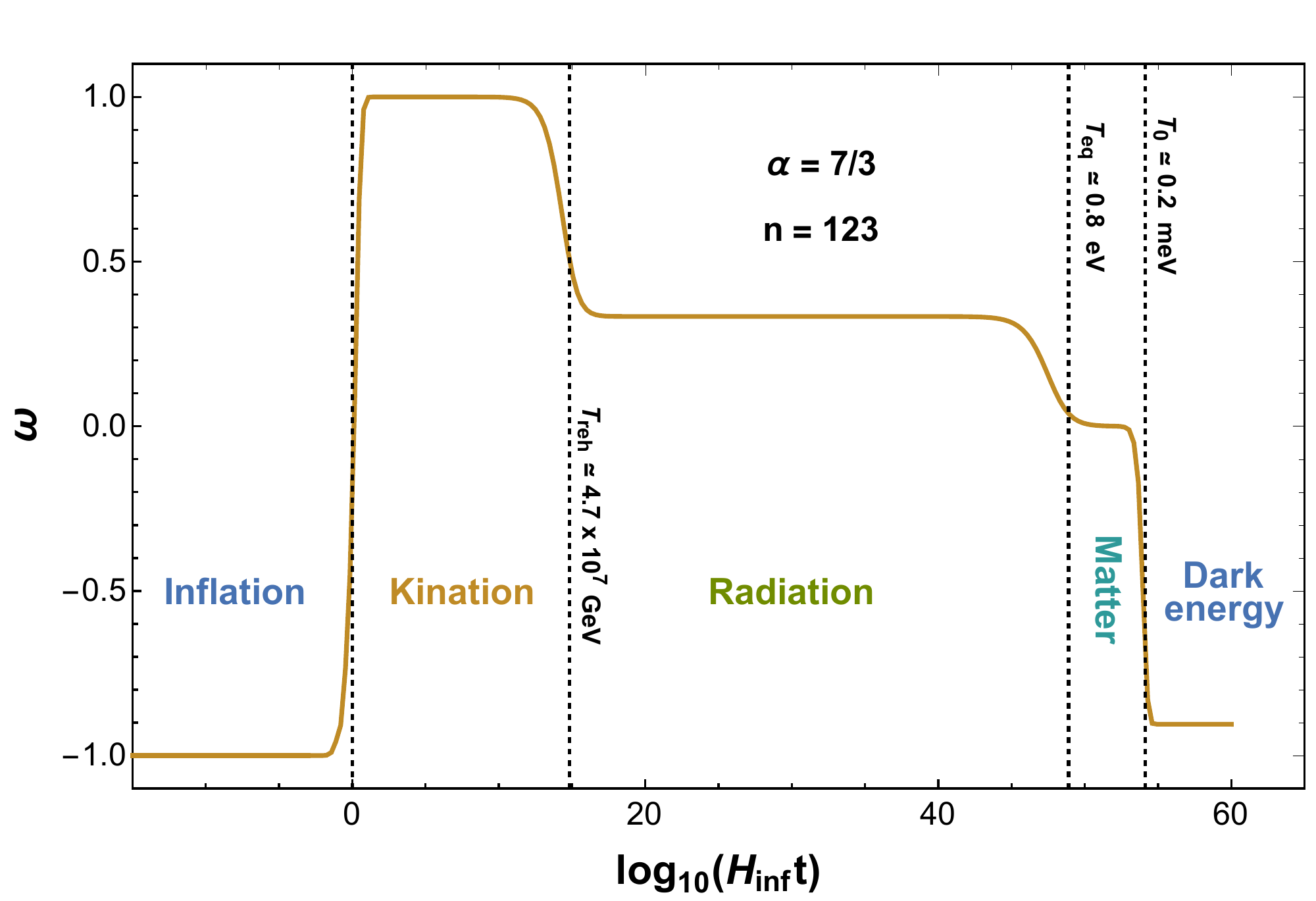}}}
\caption{\textit{ \small   \textbf{Top}: evolution of the different energy densities composing the universe in the quintessential inflation scenario based on the potential in Eq.~\eqref{eq:pot_alpha_attractor} for $\alpha = 7/3$ (motivated by string theory \cite{Akrami:2017cir}) and $n=123$. We solved numerically the equation of motion of the scalar field evolving in the potential in Eq.~\eqref{eq:pot_alpha_attractor} in the expanding universe.
\textbf{Bottom}: evolution of the equation of state $\omega = p/\rho$ of the universe.} }
\label{fig:quintessential_inf_alpha_attractor_energy}
\end{figure}
The poles at $\phi \to \pm \sqrt{6 \alpha}M_{\rm pl}$ have been sent to $\varphi \to \pm \infty$ and the potential $V(\varphi)$ features two plateaus
\begin{align}
&V(\varphi) \simeq M^4 \, \text{exp} \left(  -2n e^{\frac{2 \varphi}{\sqrt{6\alpha}M_{\rm pl}}} \right), \qquad \qquad \textrm{when}\quad \varphi \to -\infty,\label{eq:inf_pot_alpha_attractor} \\
&V(\varphi) \simeq 2ne^{-2n}M^4 \, \text{exp} \left(  -\frac{2\varphi}{\sqrt{6\alpha} M_{\rm pl}} \right), \qquad \textrm{when}\quad \varphi \to +\infty.\label{eq:quint_pot_alpha_attractor} 
\end{align}
This potential and the history of the universe in the quintessential inflation scenario are shown in Fig.~\ref{fig:quintessential_inf_alpha_attractor_pot} and Fig.~\ref{fig:quintessential_inf_alpha_attractor_energy} respectively.

\subsection{Kination followed by reheating}
\paragraph{Kination.}

After inflation, the universe is dominated by the kinetic energy of the scalar field which evolves according to Eq.~\eqref{eq:kination_eom}
\begin{equation}
\varphi = \varphi_{\rm end} + \sqrt{\frac{2}{3}}M_{\rm pl} \ln{\left(  \frac{t}{t_{\rm end}}  \right)} \label{eq:phi_kin}
\end{equation}
with the field position at the end of inflation estimated as
\begin{equation}
 \varphi_{\rm end} = \sqrt{\frac{3\alpha}{2}} M_{\rm pl} \ln \left( \frac{\sqrt{3\alpha}}{2n} \right).
 \label{eq:phi_end}
\end{equation}
Since the inflation potential is non-oscillatory, the standard reheating can not occur through the decay of the inflaton.

\paragraph{Gravitational reheating.}

The kination era ends when the universe becomes dominated by the energy density of the reheated plasma. An immediate possibility for producing radiation is gravitational reheating \cite{Ford:1986sy,Chun:2009yu} in which the reheated density reads
\begin{equation}
\rho_{\rm grav} \simeq \delta \times 10^{-2}\,H_{\rm end}^4, \label{eq:grav_reheating}
\end{equation}
where $\delta$ is an efficiency factor which depends on the number of fields, the nature of their coupling with gravity, their mass and their self-coupling, see \cite{Figueroa:2018twl} for a review.
However, unless we introduce a large number $\delta \gtrsim 50$ of self-interacting and/or non-minimally-coupled light fields, reheating through gravitational coupling only is inconsistent with the BBN bound on GW from inflation \cite{Sahni:2001qp, Tashiro:2003qp, Sami:2004xk, Artymowski:2017pua, AresteSalo:2017lkv,Figueroa:2018twl}. 

\paragraph{Other reheating mechanisms.}
A natural way-out, which may be the one realized in the SM \cite{Figueroa:2016dsc}, is to introduce a large non-minimal coupling to gravity to exploit the tachyonic instability generated by the change of sign of the Ricci scalar during kination \cite{Dimopoulos:2018wfg,Nakama:2018gll,Opferkuch:2019zbd, Bettoni:2019dcw, Bettoni:2021zhq}. The tachyonic instability can also be generated by a thermal phase transition \cite{Dimopoulos:2019ogl} or more generally when the inflaton crosses an enhanced symmetry point, see instant reheating \cite{Felder:1998vq,Felder:1999pv,Dimopoulos:2017tud,Haro:2018jtb} or trapping reheating \cite{BuenoSanchez:2006fhh, BuenoSanchez:2006epu}. Another well-known efficient reheating is curvaton reheating where a spectator field decay into SM \cite{Feng:2002nb,BuenoSanchez:2007jxm,Matsuda:2007ax}.

\section{Stiff-dominated era}
\label{sec:otherstiff}
\subsection{Example of stiff era}
The scalar Virial theorem, Eq.~\eqref{eq:final_solution_virial}, states that the averaged energy density of the scalar field in the power-law potential, whose motion follows the Klein-Gordon equation in the expanding universe, redshifts as
\begin{align}
\langle \rho_\Phi \rangle ~ \propto ~ a^{-6n/(2+n)} ~ ~ {\rm for} ~ ~ V(\Phi) ~ \propto ~ \Phi^{n},
\end{align}
which is equivalent to the EOS: $\omega = (n-2)/(n+2)$.

\subsection{BBN and interferometer probe}
\begin{figure}[h!]
\centering
\raisebox{0cm}{\makebox{\includegraphics[width=1\textwidth, scale=1]{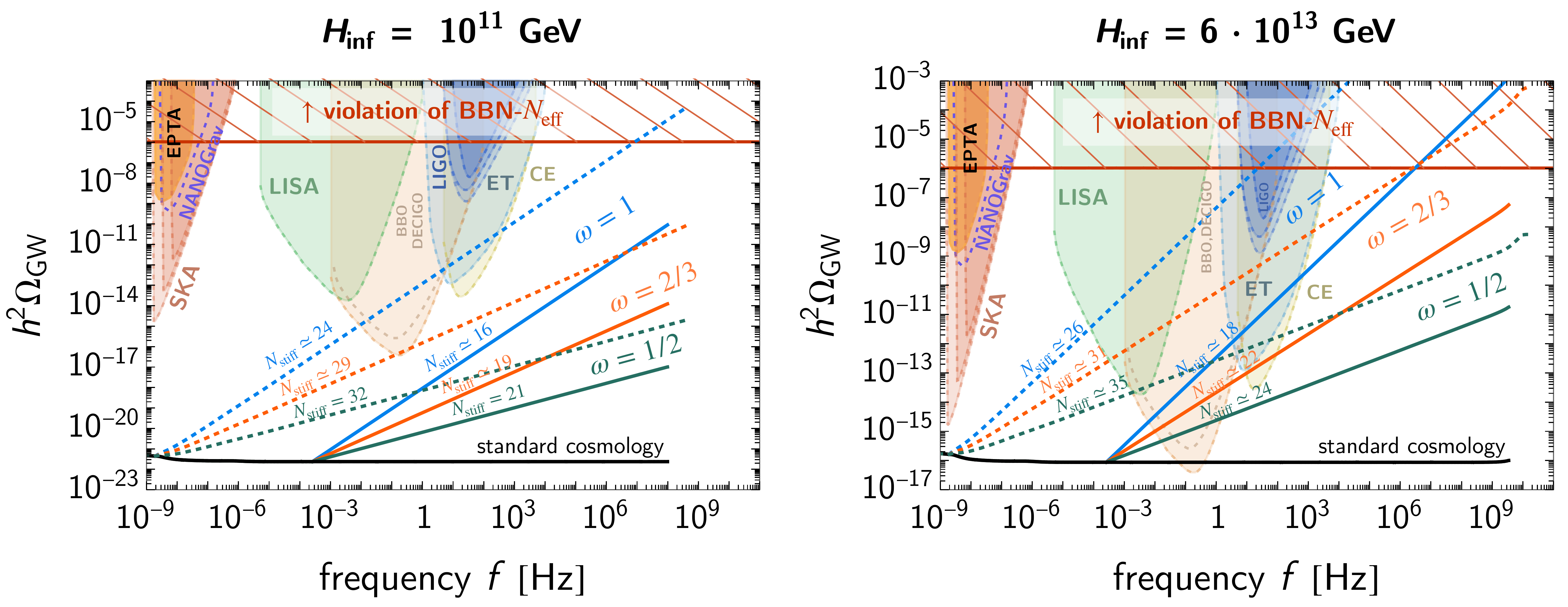}}}
\raisebox{0cm}{\makebox{\includegraphics[width=0.485\textwidth, scale=1]{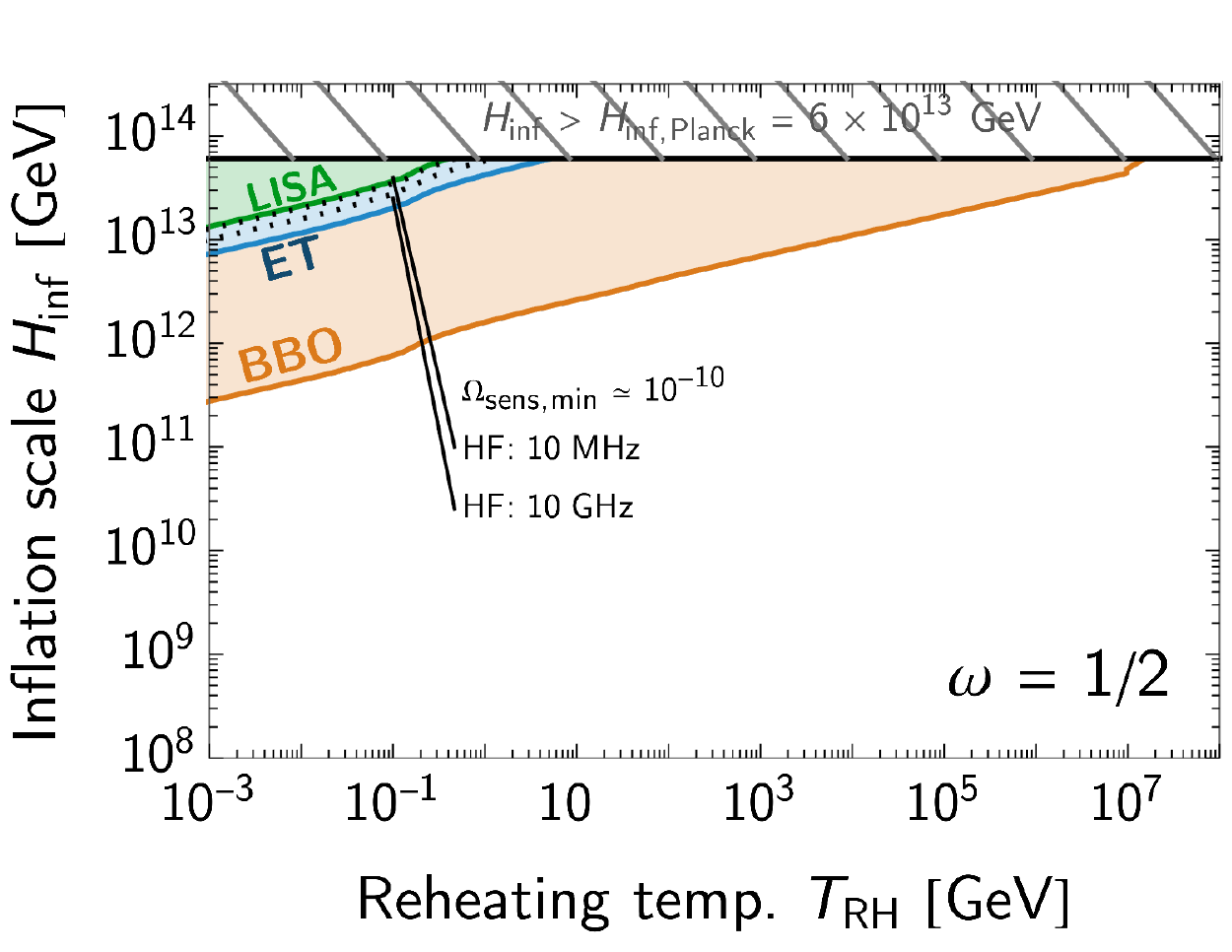}}}\quad
\raisebox{0cm}{\makebox{\includegraphics[width=0.485\textwidth, scale=1]{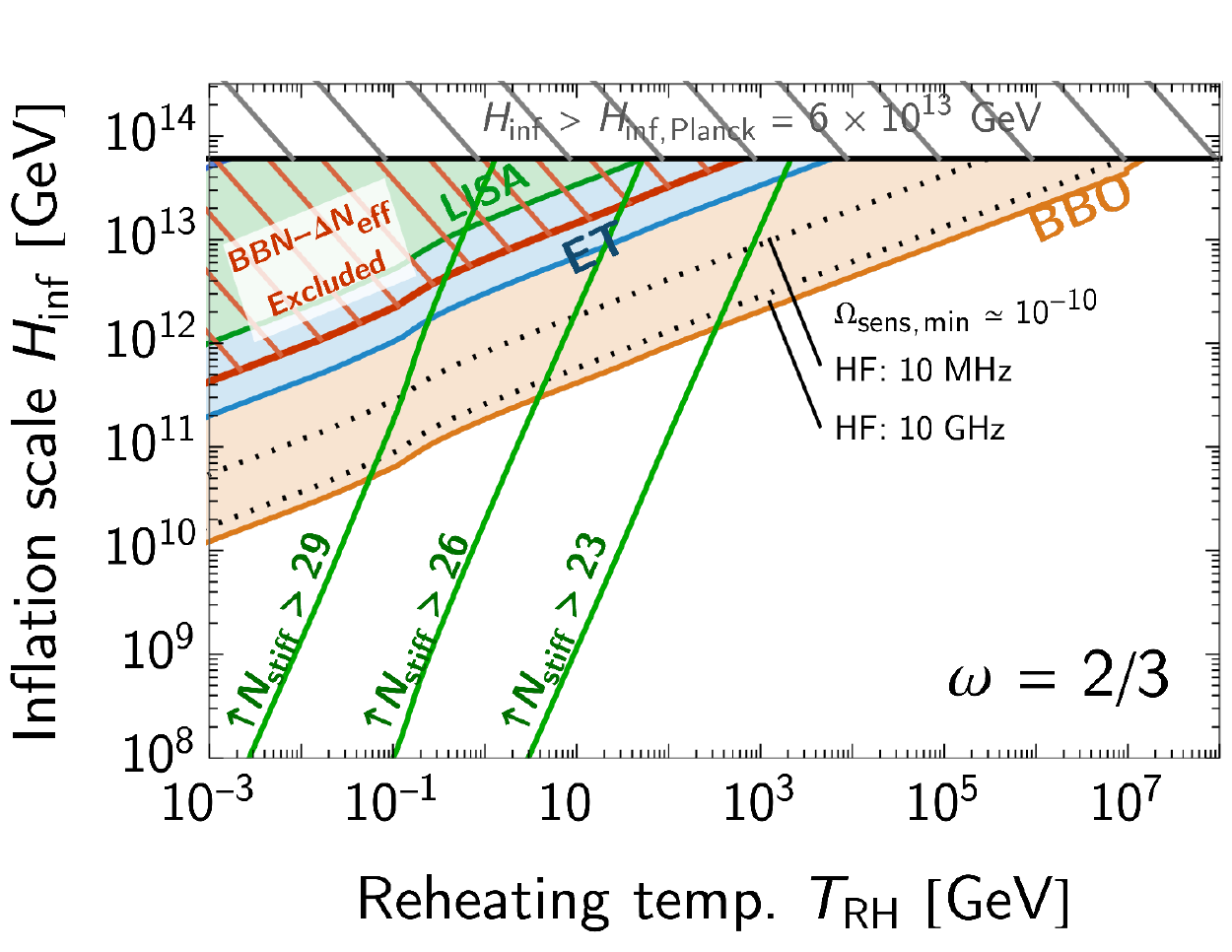}}}
\caption{\textit{ \small \textbf{Top:} In the presence of a stiff era with EOS $\omega$ occuring right after inflation, the GW spectrum from primordial inflation receives a blue-tilt, cf. Eq.~\eqref{eq:spectralindex}. The dashed and solid lines correspond to the reheating temperatures    $T_{\rm RH} = 10^{-1} ~ {\rm GeV}$ and $10^{4} ~ {\rm GeV}$, respectively. Some spectra associated with too long kination are subject to the fluctuation constraint Eq.~\eqref{eq:fluctuation_stiff}, which depends on the initial scalar fluctuation \cite{DESYfriendpaper2}.
\textbf{Bottom:} Ability of the future-planned experiments to compete with the BBN bound, for probing SGWB enhanced by post-inflation stiff era with EOS $\omega$ ending at the reheating temperature $T_{\rm RH}$. The lower $\omega$, the more competitive the GW interferometers with respect to BBN. The stiff era on the left of green solid line cannot be realized due to the radiation-like fluctuation, cf. Eq.~\eqref{eq:fluctuation_stiff}.}  An experiment operating at higher frequencies loses sensitivity for smaller $\omega$ because of the smaller enhancement.}
\label{BBNbound_stiff_contour}
\end{figure}

In Sec.~\ref{sec:inflationaryGW}, we have studied the spectrum of inflationary GW in presence of an arbitrary equation of state of the universe and we have found, see Eq.~\eqref{eq:spectralindex}
\begin{equation}
\Omega_{\rm GW}
\propto f^{-{2(1-3\omega)}/{1+3\omega}}
 \propto ~ f^{\beta}, ~ ~ \textrm{with} ~ ~ \beta ~ \equiv ~ -2\left(\frac{1-3\omega}{1+3\omega}\right)
 \label{eq:spectralindex_app}
\end{equation}
where $\omega$ is the equation of state of the universe. The energy-frequency relation is 
\begin{align}
f(\rho) = f_\Delta \left(\frac{\rho}{\rho_{\Delta}}\right)^\frac{1+3\omega}{6(1+\omega)},
\end{align}
where $f_\Delta$ and $\rho_{\Delta}$ are the GW frequency and the total energy density at the end of the stiff era.
Examples of inflationary GW spectra in the presence of a stiff era are shown in Fig.~\ref{BBNbound_stiff_contour}-top.
In Fig.~\ref{BBNbound_stiff_contour}-bottom, we show the reach of future-planned GW observatories.
In contrast to the case $\omega = 1$ shown in  Fig.~\ref{BBNbound_kination} where constraints from BBN were the strongest, here for $\omega=2/3$ and $\omega=1/2$, the ability of GW interferometers to probe a stiff era can be better than what BBN does.

\FloatBarrier
\FloatBarrier

\section{Maximal duration of kination}
\label{app:NKD_max}
In this appendix, we discuss what could lead to upper bounds on the number of e-folds of kination.
\begin{enumerate}
\item
The limited duration between end of inflation/reheating and BBN, cf. Sec.~\ref{sec:CMB_BBN_bound_NKD}
\begin{align}
\label{eq:max_NKD_CMB_BBN_intro}
&\textrm{Type (i):}\,\,\qquad \qquad N_{\rm KD} ~\lesssim ~29 + \frac{2}{3} \log{\frac{E_{\rm inf}}{1.4 \times 10^{16}~\rm GeV}}\\
&\textrm{Type (ii):}\qquad\qquad N_{\rm KD}~  \lesssim~ 14.6 + \frac{1}{3}\log{\frac{E_{\rm reh}}{1.4 \times 10^{16}~\rm GeV}}.
\end{align}
where the Types (i) and (ii) are defined in Fig.~\ref{diagram_intro}.
\item
The $N_{\rm eff}$ constraint of kination-enhanced inflationary GW, cf. Eq.~\eqref{sec:inflation_bound_NKD}
\begin{equation}
\label{eq:bound_NKD_GW_inf_intro}
\textrm{Types (i) and (ii):}\qquad \qquad N_{\rm KD} ~\lesssim ~11.9 + \log{\frac{5 \times 10^{13}~\rm GeV}{H_{\rm inf}}}.
\end{equation}
\item
The growth of scalar fluctuation during kination, cf. Sec.~\ref{sec:curvature_perturbation_kination}
\begin{equation}
\label{eq:NKD_upper_bound_Cem_intro}
\textrm{Types (ii):}\qquad \qquad N_{\rm KD}~\lesssim ~10~\textrm{or}~11.
\end{equation}

\item
The angular mode energy density drops below the wiggle step size of the axion potential, cf. Sec.~\ref{sec:axion_wiggles}
\begin{equation}
\qquad \qquad N_{\rm KD}~ \lesssim~ \frac{1}{3} \ln {\frac{\dot{\theta}_0}{2m_{a}}}, \label{eq:NKD_max_wiggles_intro}
\end{equation}
for  the axion models discussed in Secs.~\ref{sec:PQ:exampleII}, \ref{sec:scenario_I_non_thermal_damping}, \ref{sec:complex_field_thermal_potential},  and \ref{sec:complex_field_low_reh_temp}.
\item
The universe contains a non-vanishing vacuum energy $V_0$, cf. Sec.~\ref{sec:inflation_after_kination},
\begin{equation}
N_{\rm KD}~\lesssim~  \frac{1}{6}\log \left(\frac{\rho_{\rm KD}}{V_0}\right)
\end{equation}
\end{enumerate}
We now derive these bounds in turn in the following subsections.

\subsection{Duration between inflation/reheating and BBN}
\label{sec:CMB_BBN_bound_NKD}
\paragraph{Maximal reheating scale.}
The  non-detection of primordial B modes by 2021 BICEP/Keck Collaboration \cite{BICEP:2021xfz} constrains the tensor-to-scalar ratio to be smaller than
\begin{equation}
 r ~\equiv ~ \frac{A_s}{A_t}~\lesssim~ 0.036,
\end{equation} 
where $A_s \simeq 2.196 \times 10^{-9}$ \cite{ParticleDataGroup:2018ovx} and $A_t = \frac{2H_{\rm inf}^2}{\pi M_{\rm pl}^2}$ \cite{Baumann:2018muz}. This implies the maximal Hubble and energy scale at the end of inflation to be
\begin{equation}
\label{eq:max_inf_scale}
H_{\rm inf}~\lesssim ~ 5 \times 10^{-13}~\rm GeV, \qquad \textrm{and} \qquad E_{\rm inf} \equiv \left(3M_{\rm pl}^2 H_{\rm inf}^2\right)^{1/4} ~\lesssim ~ 1.4 \times 10^{16}~\rm GeV.
\end{equation}
The maximum reheating scale, corresponding to the extreme scenario where reheating takes place instantaneously at the very end of inflation, is given by 
\begin{equation}
E_{\rm reh}~\lesssim ~1.4 \times 10^{16}~\rm GeV.
\end{equation}
\paragraph{Successful BBN.}
By using the code AlterBBN \cite{Arbey:2011nf}, it can be shown\footnote{We thank Philip S{\o}rensen for this point.} that successful BBN requires the amount of kination energy density to be less than $92\%$ of the radiation energy density at the temperature $T=1$~MeV  \cite{DESYfriendpaper}
\begin{equation}
\rho_{\rm kin}~\lesssim~ 92\%~\rho_{\rm rad},\qquad \textrm{at}\quad T=1~\rm MeV.
\end{equation}
We conclude that kination must end before
\begin{equation}
\label{eq:bound_end_kination_BBN}
E_{\Delta}~\gtrsim ~E_{\rm BBN}\equiv 1.4~\rm MeV,
\end{equation}
where $E_{\Delta}$ is the energy scale at  the end of kination when $\rho_{\rm kin} = \rho_{\rm rad}$, and where we remind the reader of our notation $E_i \equiv \rho_i^{1/4}$.
 In the plots of our paper we actually use a slightly more aggressive constraint $E_{\Delta} \simeq 10 ~{\rm MeV}$. 

\paragraph{Maximal duration of kination after inflation (Scenario Type (i)).}
We consider the case of quintessential inflation, cf. Scenario Type (i) in Fig.~\ref{diagram_intro}. A simple constraint on $N_{\rm KD}$ comes from kination not being able to last more than the time between the end of inflation in Eq.~\eqref{eq:max_inf_scale}, and the end of BBN in Eq.~\eqref{eq:bound_end_kination_BBN}. We obtain the bound
\begin{equation}
\textrm{Type (i):}\qquad \qquad N_{\rm KD} ~\lesssim ~ \frac{1}{6}\log\frac{E_{\rm inf}^4}{\rho_{\rm BBN}} = 29 + \frac{2}{3} \log{\frac{E_{\rm inf}}{1.4 \times 10^{16}~\rm GeV}}
\end{equation}

\paragraph{Maximal duration of intermediate matter-kination (Scenario Type (ii)).}
We consider the case of an intermediate matter-kination era, corresponding to Scenario Type (ii) in Fig.~\ref{diagram_intro}.
As we already discussed along Eq.~\eqref{eq:NKD_vs_NMD}, the duration of the matter-kination is maximised when there is no entropy injection which would produce additional radiation component. For such an adiabatic evolution, the number of e-folds of matter and kination are related through 
\begin{equation}
N_{\rm MD} = 2N_{\rm KD}, \qquad  \textrm{with}\quad N_{\rm MD}\equiv \frac{1}{3}\log{\frac{\rho_{\rm dom}}{\rho_{\rm KD}}},\quad \textrm{and} \quad N_{\rm KD}\equiv \frac{1}{6}\log{\frac{\rho_{\rm KD}}{\rho_{\Delta}}}.
\end{equation}
In that scenario the energy scale $E_{\rm KD}$ at the onset of the kination era is given by the geometric mean of the energy scales $E_{\rm dom}$ and $E_{\Delta}$ at the onset of the matter era and at the end of the kination era
\begin{equation}
E_{\rm KD}~=~\sqrt{E_{\rm dom}E_{\Delta}},
\end{equation}
and the duration of the kination era alone can be written as
\begin{equation}
N_{\rm KD}  =\frac{1}{3}\log{\frac{E_{\rm dom}}{E_{\Delta}}}.
\end{equation}
The longest matter-kination era allowed by CMB and BBN is the one which starts right after the end of inflation $E_{\rm dom} = E_{\rm reh}$ and ends just before the onset of BBN $E_{\Delta} = E_{\rm BBN}$. We obtain the upper bound
\begin{equation}
\label{eq:max_NKD_CMB_BBN}
\textrm{Type (ii):}\qquad\qquad \qquad N_{\rm KD} ~ \lesssim ~14.6 + \frac{1}{3}\log{\frac{E_{\rm reh}}{1.4 \times 10^{16}~\rm GeV}}
\end{equation}
The constraint can be seen in Fig.~\ref{inter_duration_2}.

\subsection{$N_{\rm eff}$ bound on inflationary GW}
\label{sec:inflation_bound_NKD}

As we discuss in Sec.~\ref{subsec:afterinflation}, the GW energy density produced prior to the onset of BBN can contribute to the effective number of neutrino species, which leads to the bound, cf. Eq.~\eqref{eq:BBN_bound_inflation_GW}
\begin{equation}
\int^{f_{\rm max}}_{f_\textrm{BBN}}\frac{df}{f}h^2 \Omega_\textrm{GW}(f) \leq 5.6 \times 10^{-6} ~ \Delta N_\nu, \label{eq:BBN_bound_inflation_GW_app}
\end{equation}
where $\Delta N_\nu \leq 0.2$ \cite{Tanabashi:2018oca},
In Sec.~\ref{sec:inflationaryGW}, we have studied the impact of a kination EOS on the spectrum of primordial GW and we have found that a period of $N_{\rm KD}$ e-folds of kination leads to a blue-tilt $\Omega_\mathrm{GW} \propto f$ whose peak value is, cf. Eq.~\eqref{inflation_peak_amplitude}
\begin{equation}
\Omega_\mathrm{GW,KD} \simeq 2.8 \times 10^{-13}  \left(\frac{g_*(T_\Delta)}{106.75}\right) \left(\frac{g_{*,s}(T_
\Delta)}{106.75}\right)^{-4/3} \left(\frac{E_{\inf}}{10^{16}\textrm{ GeV}}\right)^4 \left( \frac{\exp(2 N_\mathrm{KD})}{e^{10}} \right),
\label{inflation_peak_amplitude_app}
\end{equation}
Upon injecting Eq.~\eqref{inflation_peak_amplitude_app} into Eq.~\eqref{eq:BBN_bound_inflation_GW_app}, we obtain the maximal duration of kination
\begin{equation}
\label{eq:bound_NKD_GW_inf}
\textrm{Scenarios Type (i) and (ii):} \qquad \qquad N_{\rm KD} ~\lesssim ~11.9 + \log{\frac{5 \times 10^{13}~\rm GeV}{H_{\rm inf}} }.
\end{equation}
where Types (i) and (ii) are defined in Fig.~\ref{diagram_intro}.
The constraint can be seen in Fig.~\ref{BBNbound_kination}.

\subsection{Growth of scalar fluctuation}
\label{sec:curvature_perturbation_kination}
So far the strongest bound on the kination duration comes from the requirement of radiation era at the time of BBN.
However, the scalar fluctuations in the early universe could lead to a stronger bound as the growth of fluctuation depends on the expansion history of the universe.
For example, a long matter era can lead to a non-linear behavior of fluctuations \cite{Erickcek:2011us,Musoke:2019ima,Eggemeier:2021smj}  or the enhanced formation of primordial black holes \cite{Harada:2016mhb}.
Similarly, the fluctuations of the massless field which behave as radiation grow during kination era and could eventually dominate over that of the zero-mode which scales as stiff fluid after some $N_{\rm stiff}$ efolds, determined by
\begin{align}
\left(\frac{\rho_{\rm inf}}{\rho_{\rm reh}}\right)^{\frac{1}{3(1+\omega)}} = \left(\frac{3 \MPl H_{\rm inf}^2}{\pi^2 g_*(T_{\rm reh}) T_{\rm reh}^4 / 30}\right)^{\frac{1}{3(1+\omega)}} = \frac{a_{\rm reh}}{a_{\rm inf}} = \exp\left(N_{\rm stiff}\right) < \xi^{-\frac{1}{3\omega-1}},
\label{eq:fluctuation_stiff}
\end{align}
where $\xi$ is the initial fluctuation's energy density at the end of inflation. 
To estimate the bound on kination duration, one can assume that the growth is sourced by the curvature perturbations.
The maximum energy density in the fluctuation would be the mode entering the horizon at the start of kination and could reach the level of the curvature perturbations observed in CMB $\sim 10^{-9\div-10}$.
From this we can estimate that the number of e-folds of kination era is bounded to\footnote{We thank Cem Er{\"o}ncel for discussing this point.}
\begin{equation}
\label{eq:NKD_upper_bound_Cem}
N_{\rm KD}~\lesssim ~10~\textrm{or}~11.
\end{equation}
This bound is further relaxed for less stiff era, e.g. $N_{\rm stiff}^{\rm max} \sim 23, 46$ for $\omega = 2/3, 1/2$, respectively.
We leave a more careful study of fluctuations during matter-kination era for future work \cite{DESYfriendpaper2}. We impose this constraint in Fig.~\ref{ma_fa_plot_axion_string}.

\subsection{Presence of axion wiggles}
\label{sec:axion_wiggles}

Assuming the presence of an axion potential
\begin{equation}
V(\theta)~ = ~f_a^2 m_a^2(1+ \cos{\theta}),
\end{equation}
where $m_a$ is the mass of the axion $a \equiv \theta f_a$.
The circular motion of the angular mode is only possible if its kinetic energy density is larger than the energy density on the top of the potential barrier
\begin{equation}
\frac{f_a^2 \dot{\theta}^2}{2} ~ > ~ 2f_a^2 m_a^2 \qquad \implies \qquad \dot{\theta}~>~2m_a.
\end{equation}
Since the axion velocity redshift as $\theta \propto a^{-3}$, we deduce the maximal number of e-fold of kination
\begin{equation}
N_{\rm KD}^{\rm max} = \frac{1}{3} \ln {\frac{\dot{\theta}_0}{2m_{a}}}. \label{eq:NKD_max_wiggles}
\end{equation}
We obtained a similar bound from fragmentation in Eq.~\eqref{eq:fragmentation_kination_ends}.
We also checked that the early wiggles from higher-dimensional $U(1)$-explicit breaking terms do not lead to fragmentation.

\FloatBarrier
\begin{figure}[h!]
\vspace{-1cm}
\centering
\raisebox{0cm}{\makebox{\includegraphics[width=0.45\textwidth, scale=1]{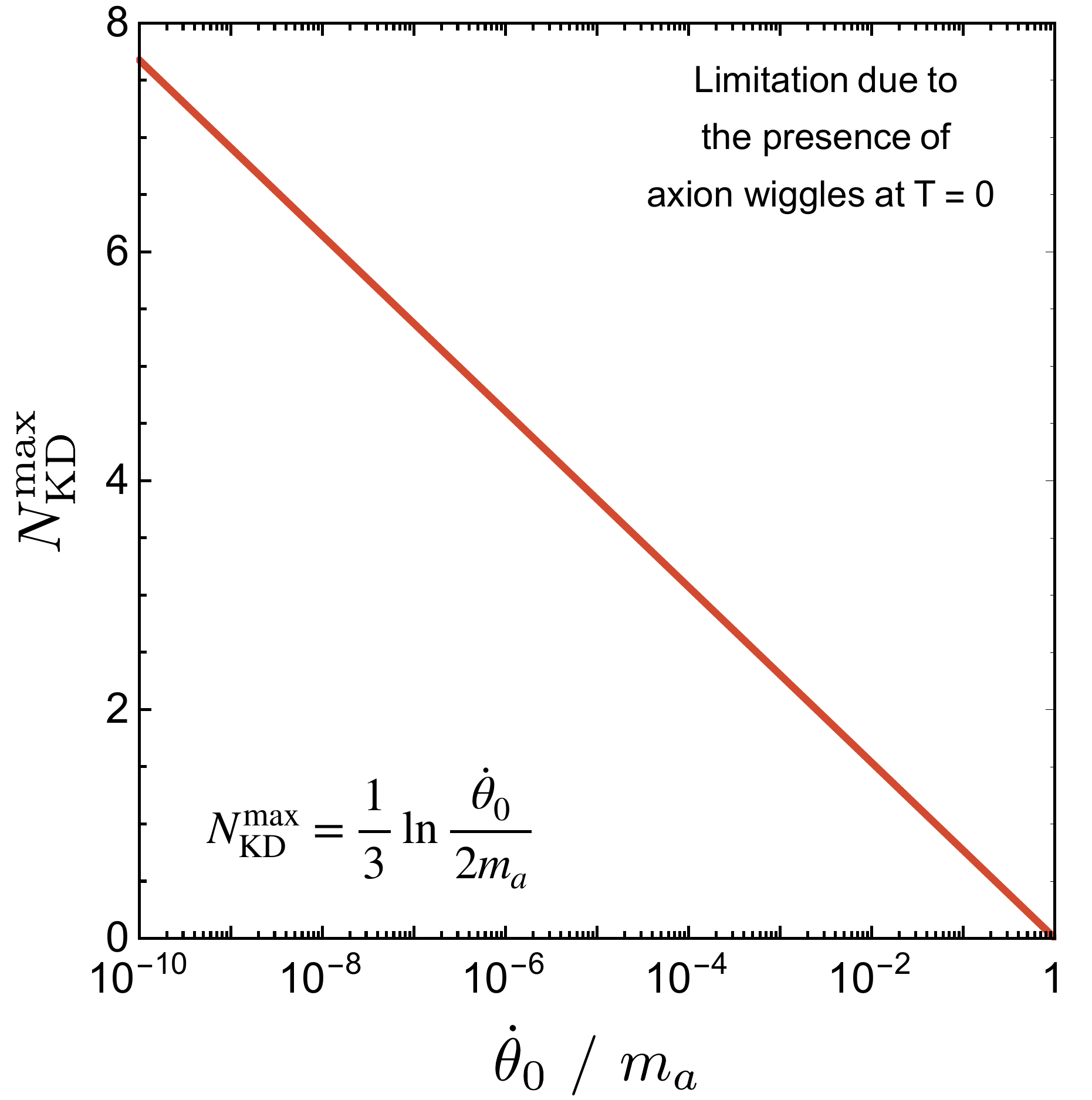}}}
\caption{\textit{ \small Maximal number of kination e-folds due to the presence of axion wiggles  cf. Eq.~\eqref{eq:NKD_max_wiggles}.  In the parameter space of interest, the number of e-folds $N_{\rm KD}$ is always smaller than $N_{\rm KD}^{\rm max}$ shown here, such that the bound (\ref{eq:NKD_max_wiggles}) is always satisfied. Instead, we expect the duration of kination to be dominantly constrained by the other bounds given in Eqs.~\eqref{eq:max_NKD_CMB_BBN_intro}, \eqref{eq:bound_NKD_GW_inf_intro} and \eqref{eq:NKD_upper_bound_Cem_intro}}}
\label{fig:Max_NKD_Pablo}
\end{figure}
\FloatBarrier

\subsection{Inflationary behavior after kination phase}
\label{sec:inflation_after_kination}
\begin{figure}[h!]
\centering
\raisebox{0cm}{\makebox{\includegraphics[width=0.55\textwidth, scale=1]{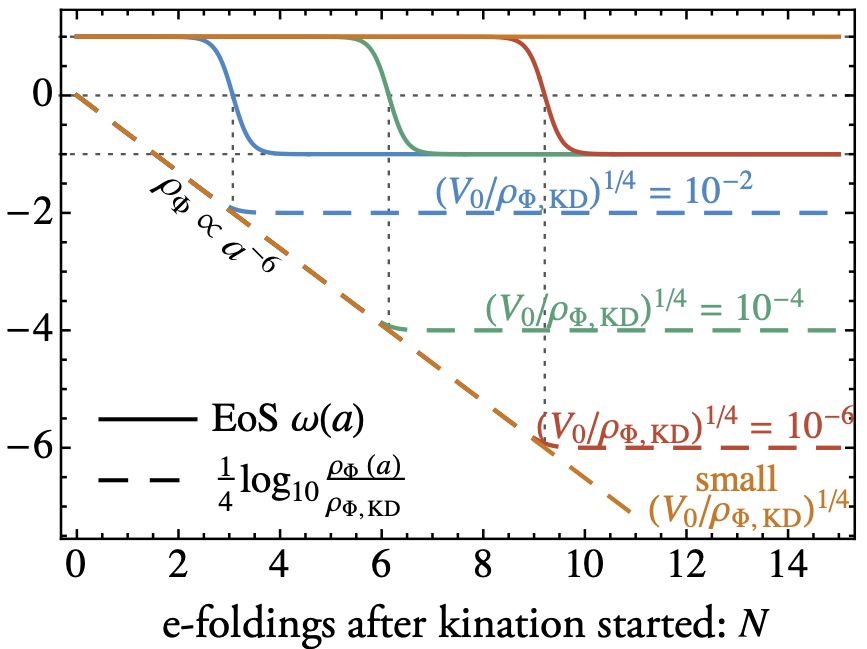}}}
\caption{\textit{ \small 
The kination EOS ends when the kinetic energy density $\rho_{\phi,\rm KD}$ drops below the non-zero vacuum energy $V_0$.
}}
\label{stiff_generation}
\end{figure}

\paragraph{Maximal number of kination e-folds.}
So far, we have neglected the vacuum energy at the potential minimum.
However, this non-zero potential energy can lead to an inflationary period at later times.
The kinetic energy density redshifts as $a^{-6}$ and eventually becomes as small as the vacuum energy
\begin{align}
\omega(a) ~ = ~ \frac{K(a) - V_0}{K(a) + V_0} ~ &= ~ \frac{\rho_\textrm{KD} (\frac{a_\textrm{KD}}{a})^6 - V_0}{\rho_\textrm{KD} (\frac{a_\textrm{KD}}{a})^6 + V_0} ~ = ~ \frac{\exp(-6 N_{\rm KD}) - V_0/\rho_\textrm{KD}}{\exp(-6 N_{\rm KD}) + V_0/\rho_\textrm{KD}},
\end{align}
where $N_{\rm KD}$ is the e-foldings of the cosmic expansion after the kination era started.
For a non-zero $V_0$, Fig.~\ref{stiff_generation} shows that after a number of kination e-folds given by
\begin{equation}
N_{\rm KD}^{\rm max} =  \frac{1}{6}\log \left(\frac{\rho_{\rm KD}}{V_0}\right),
\end{equation}
the EOS decreases abruptly to $\omega = -1$.

\paragraph{Physical motivations.}
The vacuum-energy domination could signal a cosmological supercooled first-order phase transition about to take place.

\FloatBarrier
\FloatBarrier

\section{Origin of the scalar potential in the rotating complex scalar field model}
\label{app:scalar_pot_SUSY}
\paragraph{The potential.}
We consider a complex scalar field $\Phi$ with the Lagrangian introduced in Eq.~\eqref{eq:complex_scalar_field_full_lag}
\begin{align}
\mathcal{L} ~ = ~ (\partial_\mu \Phi)^\dagger \partial^\mu \Phi - V (\left| \Phi \right|)  - V_{\rm th}(|\Phi|)- V_{\cancel{U(1)}}(\Phi) - V_{H}(\Phi),
\end{align}
with $V$ the global $U(1)$-symmetric potential with spontaneous symmetry breaking (SSB) vacuum, $V_{\rm th}$ the thermal corrections, $ V_{\cancel{U(1)}}$ the explicit $U(1)$-breaking term, and $V_H$ the Hubble-dependent potential
\begin{align}
&V (\left| \Phi \right|) = m_r^2 |\Phi|^2 \left(\ln \frac{|\Phi|^2}{f_a^2} -1 \right) + m_r^2 f_a^2~+ ~ \frac{\lambda^2}{M^{2l-6}} |\Phi|^{2l-2}, \label{eq:V_U1_app} \\
&V_{\cancel{U(1)}}(\Phi) = \Lambda_b^4 \left[ \left(\frac{\Phi^\dagger}{M}\right)^l ~ + ~ \left(\frac{\Phi}{M}\right)^l \right]  , \label{eq:V_break_U1_app}\\
&V_{H}(\Phi) =  - c H^2 |\Phi|^2 ~-~ a \frac{M}{M_{\rm pl}}  \frac{H}{m_{3/2}}\Lambda_b^4 \left[\left(\frac{\Phi^\dagger}{M}\right)^l ~ + ~ \left(\frac{\Phi}{M}\right)^l\right],\label{eq:V_H_app}
\end{align}
and
\begin{align}
V_{\rm th}(\phi,\, T) ~ = ~ \begin{cases}
\frac{1}{2} y_\psi^2 T^2 \phi^2,   & \mathrm{for} ~   y_\psi \phi \lesssim T, \\
\alpha^2 T^4 \ln(\frac{y_\psi^2 \phi^2}{T^2}) ,\qquad \qquad & \mathrm{for} ~   y_\psi \phi \gtrsim T
\end{cases}.
\end{align}
In this appendix, we derive the above scalar potential in SUSY framework.
In the main text, we take $M=M_{\rm pl}$.

\subsection{Neglecting Hubble curvature}
\label{app:neglecting_H}
The first two terms, in Eq.~\eqref{eq:V_U1_app} and Eq.~\eqref{eq:V_break_U1_app}, can be derived from the SUSY Lagrangian
\begin{equation}
\mathcal{L} \supset \int d^2\theta d^2\bar{\theta} \, K(S_\Phi,\,S_\Phi^*) + \int d^2\theta \, W(S_\Phi) + h.c. \label{eq:SUSY_lag}
\end{equation}
with the Kahler potential $K$ and the superpotential $W$ being given by
\begin{align}
&K(S_\Phi,\,S_\Phi^*)  = |S_\Phi|^2 - \frac{1}{M} |S_\Phi|^2\left(S_\chi + S_\chi^*\right) - \frac{1}{M^2}|S_\Phi|^2|S_\chi|^2  -  \frac{1}{M^4}|S_\Phi|^4|S_\chi|^2 , \label{eq:Kahler_pot}\\
&W(S_\Phi)  = \lambda \frac{S_\Phi^{l}}{l\,M^{l-3}} , \qquad l \in \mathbb{N}.\label{eq:superpotential}
\end{align}
$(\theta,\, \bar{\theta})$ are the Grassmanian coordinates in the superspace. $S_\Phi$ is the chiral superfield containing the complex scalar $\Phi$
\begin{equation}
S_\Phi = \Phi + \sqrt{2} \theta \psi + \theta^2 F. \label{eq:super_S_Phi}
\end{equation} 
$M$ is the messenger scale (e.g. the Planck scale for gravity mediation) and $S_\chi$ is a SUSY-breaking chiral superfield
\begin{equation}
S_\chi = \chi + \sqrt{2} \theta \psi_\chi + \theta^2 F_\chi, \label{eq:S_chi}
\end{equation}
which we assume to get a non-vanishing F-term VEV
\begin{equation}
\left< S_\chi \right> = \theta^2 \left<F_\chi\right> \qquad \text{with} ~\left<F_{\chi}\right> = m_{32} M_{\rm pl}, 
\end{equation}
where $m_{32}$ is the gravitino mass.
From using Eq.~\eqref{eq:S_chi}, together with $\int d\theta^2 \theta^2 = 1$, the last two terms of the Kahler potential in Eq.~\eqref{eq:Kahler_pot} contribute to the scalar potential
\begin{equation}
V(\Phi) ~\supset  ~\frac{\left<F_\chi\right>^2}{M^2}|\Phi|^2  +  \frac{\left<F_\chi\right>^2}{M^4} |\Phi|^4, \label{eq:scalar_pot_1}
\end{equation}
while the second term of the Kahler potential in Eq.~\eqref{eq:Kahler_pot} acts as a non-holomorphic effective superpotential
\begin{equation}
\int d^2\theta d^2\bar{\theta}  \frac{1}{M} |S_\Phi|^2\left(S_\chi + S_\chi^*\right)  \simeq  \int d^2\theta \frac{1}{M} |S_\Phi|^2 \left<F_\chi\right> + h.c.,
\end{equation}
and Eq.~\eqref{eq:superpotential} can be replaced by
\begin{equation}
W(S_\Phi)  = \frac{\left<F_\chi\right> }{M} |S_\Phi|^2 + \lambda \frac{S_\Phi^{l}}{l\,M^{l-3}} . \label{eq:superpotential_U1_breaking}
\end{equation}
In turn, from the equations of motion for the F-terms of $S_\Phi$ and $S_\Phi^*$
\begin{equation}
F = \left(\frac{\partial K}{\partial \Phi^\dagger \partial \Phi^\dagger }\right)^{-1} \frac{\partial W}{\partial \Phi^\dagger },\qquad F^* = \left(\frac{\partial K}{\partial \Phi\partial \Phi}\right)^{-1} \frac{\partial W}{\partial \Phi},
\end{equation}
we obtain the scalar potential
\begin{equation}
 V(\Phi)~ \supset~  \left(\frac{\partial K}{\partial \Phi \partial \Phi^\dagger }\right)^{-1} \left|\frac{\partial W}{\partial \Phi }\right|^2~ = ~ \frac{\left< F_{\chi} \right>^2}{M^2} |\Phi|^2 + \lambda \left< F_{\chi} \right> \frac{\Phi^{l}+h.c.}{M^{l-2}}  + |\lambda|^2 \frac{|\Phi|^{2l-2}}{M^{2l-6}}. \label{eq:scalar_pot_2}
\end{equation}
From Eq.~\eqref{eq:scalar_pot_1} and Eq.~\eqref{eq:scalar_pot_2}, we deduce $V_{U(1)}(\Phi ) $ and $V_{\cancel{U(1)}}(\Phi)$ in Eq.~\eqref{eq:V_U1_app} and Eq.~\eqref{eq:V_break_U1_app} 
\begin{align}
&V (\left| \Phi \right|) = m_r^2 |\Phi|^2 \left(\ln \frac{|\Phi|^2}{f_a^2} -1 \right) + m_r^2 f_a^2~+ ~ \frac{|\lambda|^2}{M^{2l-6}} |\Phi|^{2l-2}, \label{eq:V_U1_app_2} \\
&V_{\cancel{U(1)}}(\Phi) = \Lambda_b^4 \left[ \left(\frac{\Phi^\dagger}{M}\right)^l ~ + ~ \left(\frac{\Phi}{M}\right)^l \right]  , \label{eq:V_break_U1_app_2}
\end{align}
with 
\begin{align}
&m_r^2 = \frac{\left<F_\chi\right>^2}{M^2} = m_{32}^2 \frac{M_{\rm pl}^2}{M^2}, \\
&\Lambda_b^4 = \lambda \,\left<F_\chi\right> M^2 = \lambda \,m_{32} M_{\rm pl} M^2. \label{eq:Lambda_b_def}
\end{align}
The quartic term in Eq.~\eqref{eq:scalar_pot_1} is negligible as long as $|\Phi| \ll M$. The presence of the logarithmic function in Eq.~\eqref{eq:V_U1_app_2}, which is responsible for the spontaneous breaking of the $U(1)$ symmetry $\Phi \to e^{i\alpha}\Phi$, is generated radiatively \cite{Moxhay:1984am}, see App.~\ref{susy_potential} for a review. The constant term in the same equation is needed in order to tune the cosmological constant to zero.

\subsection{Including Hubble expansion}
\label{app:including_Hubble_curv}
Let's assume that the energy density is dominated by a complex scalar $I$, which can be the inflaton during inflation or matter-dominated preheating \cite{Dine:1995uk,Dine:1995kz} or a field in equilibrium with the thermal plasma during radiation domination \cite{Kawasaki:2011zi}
\begin{equation}
H^2 M_{\rm pl}^2 = \rho_I = \left\{
                \begin{array}{ll}
                  \left<F_I^* F_I \right>,~ ~\quad \text{during inflation or matter-domination,} 
                  \vspace{0.5cm}\\
                   \left< \partial I^* \partial I \right>, \quad \text{during radiation-domination,}
                \end{array}
              \right.
\end{equation}
where $F_I$ is the F-term of the chiral superpotential $S_I$ containing $I$.
Our complex scalar of interest $\Phi$, which is a sub-dominant fraction of the energy density of the universe $\rho_\Phi \ll H^2 M_{\rm pl}^2$, interacts with $I$ through gravity, which leads to the non-renormalizable Kahler potential
\begin{equation}
\mathcal{L} \supset \int d\theta^2 d\bar{\theta}^2 \left( a \frac{S_I+S_I^*}{M_{\rm pl}}|S_\phi|^2 +c \frac{|S_I|^2}{M_{\rm pl}^2}|S_\Phi|^2 \right), \qquad a,\,c = \mathcal{O}(1). \label{eq:gravi_int_I_Phi}
\end{equation}
Upon solving for the $S_\Phi$-F-term equations of motion from the total Lagrangian Eq.~\eqref{eq:Kahler_pot}, \eqref{eq:superpotential} and Eq.~\eqref{eq:gravi_int_I_Phi}, in which we replace\footnote{We remind the reader that chiral superfields in superspace read $S_I(y,\theta) =I(y) + \sqrt{2} \theta \psi(y) + \theta^2 F(y)$ with $y^\mu = x^\mu + i \theta \sigma^\mu \bar{\theta}$, which implies $S_I(y,\theta, \bar{\theta}) =I(x)   + i\bar{\theta} \bar{\sigma}^\mu \theta \partial_\mu I(x) + \frac{1}{4}\theta^2 \bar{\theta}^2 \Box I(x)+ \sqrt{2} \theta \psi(x) = \frac{i}{\sqrt{2}} \theta^2 \bar{\theta} \bar{\sigma}^\mu \partial_\mu \psi(x) + \theta^2 F(x) $. In Eq.~\eqref{eq:S_X_assumpt}, we have neglected the fermionic component $\psi$, the inflaton VEV $\phi$ and the mass term $ \Box I$.}
\begin{equation}
S_I = i\theta \sigma^\mu \bar{\theta} \partial_\mu I + \theta^2 F_I, \label{eq:S_X_assumpt}
\end{equation}
we find that the scalar potential receives the additional terms
\begin{equation}
V(\phi) \supset \frac{-c\partial I^\dagger \partial I + (a^2-c) F_I^* F_I + a \frac{M_{\rm pl}}{M} (F_I\,F_\chi^*+ F_I^*\,F_\chi) }{M_{\rm pl}^2} \left|\Phi\right|^2  - a \lambda \frac{F_I\,\Phi^l + h.c.}{M_{\rm pl}M^{l-3}}. \label{eq:curvature_induced_full_pot_0}
\end{equation}
If the dynamics occurs during radiation-domination, then $\left< \partial I^* \partial I \right>$ dominates and the scalar potential receives a negative curvature-induced mass whenever $c> 0$
\begin{equation}
V(\phi) \supset -c H^2 \Phi^2, \label{eq:curvature_induced_full_pot_radiation}
\end{equation}
If the dynamics occurs during inflation or matter-domination, then $\left< F_I^*F_I \right>$ dominates\footnote{The term $a H m_{32} \Phi^2$ in Eq.~\eqref{eq:curvature_induced_full_pot} is generated by the interaction between the three non trivial terms  $|S_\Phi|^2 S_\chi$,  $|S_\Phi|^2|S_\chi|^2$ and $|S_\Phi|^2 S_I$ in the Kahler potential.}
\begin{equation}
V(\phi) \supset \left( (a^2-c) H^2 + a m_{32} H\right)  \Phi^2 -  a \frac{M}{M_{\rm pl}} \frac{H}{m_{3/2}}\Lambda_b^4\left(\frac{\Phi}{M}\right)^l, \label{eq:curvature_induced_full_pot}
\end{equation}
with $\Lambda_b$ defined in Eq.~\eqref{eq:Lambda_b_def}.  The Hubble-induced mass is negative whenever $c> a^2$ and $H \gtrsim m_{32}$. For $M=M_{\rm pl}$, the latter condition is verified whenever the SUSY-breaking field $\chi$ does not dominate the energy density of the universe $F_\chi \lesssim F_I$.
We refer to \cite{Dine:1995uk,Dine:1995kz} for a discussion of supergravity corrections, which should become important during inflation whenever $I \gtrsim M_{\rm pl}$. 

In summary, we see that a large and negative Hubble-induced mass term is naturally generated in models where the complex scalar field $\Phi$ couples to a field that dominates the energy density of the universe. This is very important for justifying the initial conditions.

\subsection{Evolution of the scalar field in the negative Hubble-induced potential.}
\label{app:phi_ini}

\paragraph{Damped harmonic oscillator.}
We recall the damped harmonic oscillator 
\begin{align}
\ddot{\phi} + \Gamma \dot{\phi} + m^2 \phi ~ = ~ 0,
\end{align}
has three regimes according to the value of $\Gamma$
\begin{align}
\phi(t) ~ = ~ \begin{cases}
\phi_{1} \exp\left[\left( - \frac{\Gamma}{2} + \sqrt{\frac{\Gamma^2}{4}-m^2} \right) t\right] + \phi_{2} \exp\left[\left(- \frac{\Gamma}{2} - \sqrt{\frac{\Gamma^2}{4}-m^2} \right) t\right]&~ ~ \textrm{for} ~ ~ \Gamma^2 ~ > ~ 4 m^2 ~ ~ \textrm{(over-damped)},\\[0.75em]
\left(\phi_1 + \phi_2 \Gamma t\right) \exp\left(\frac{-\Gamma t}{2}\right) &~ ~ \textrm{for} ~ ~ \Gamma^2 ~ = ~ 4 m^2 ~ ~ \textrm{(critically-damped)},\\[0.75em]
\phi_1 \cos(\sqrt{m^2-\frac{\Gamma^2}{4}}t - \alpha) \exp\left(\frac{-\Gamma t}{2}\right) &~ ~ \textrm{for} ~ ~ \Gamma^2 ~ < ~ 4 m^2 ~ ~ \textrm{(under-damped)},
\end{cases}
\end{align}
where $\phi_1,\phi_2,\alpha$ are constants set by the initial conditions.
Taking $\Gamma = 3H$ and introducing $N_e\equiv Ht$, we obtain, e.g. \cite{Randall:1994fr}
\begin{align}
\phi(N_e) ~ \simeq ~ \begin{cases}
\phi_{1} e^{-\frac{m^2}{3H^2}N_e}&~ ~ \textrm{for} ~ ~ H ~ \gg ~  m ~ ~ \textrm{(over-damped)},\\[0.75em]
\left(\phi_1 + 3\phi_2 N_e\right) e^{-\frac{3}{2}N_e} &~ ~ \textrm{for} ~ ~ H~ = ~ \frac{2}{3}m ~ ~ \textrm{(critically-damped)},\\[0.75em]
\phi_1 \cos(m t - \alpha) e^{-\frac{3}{2}N_e} &~ ~ \textrm{for} ~ ~H ~ \ll ~m ~ ~ \textrm{(under-damped)}.
\end{cases}
\label{eq:damped_harmonic_osc}
\end{align}

\begin{figure}[h!]
\centering
\raisebox{-0.5\height}{\makebox{\includegraphics[width=0.465\textwidth, scale=1]{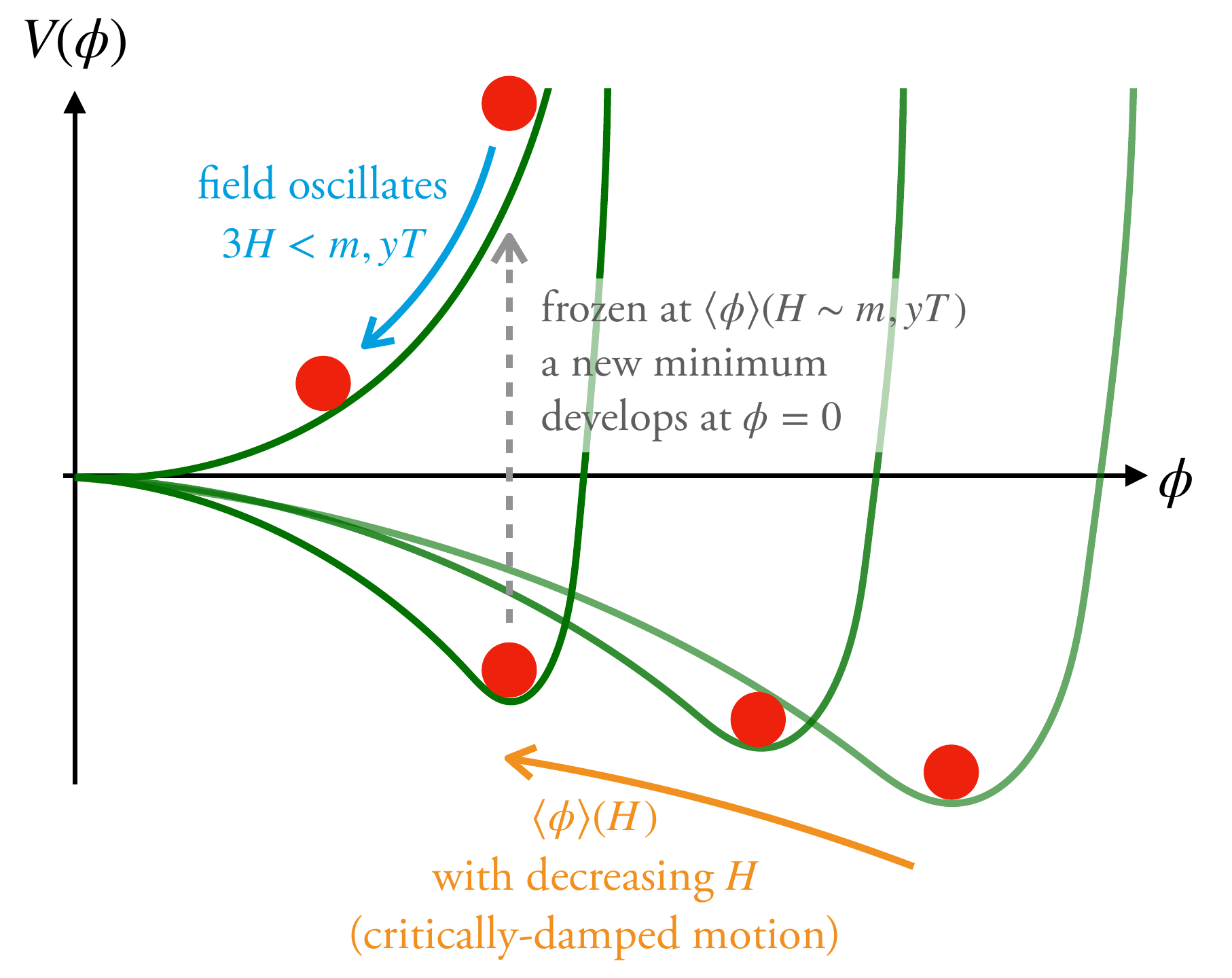}}}
\quad
\raisebox{-0.5\height}{\makebox{\includegraphics[width=0.495\textwidth, scale=1]{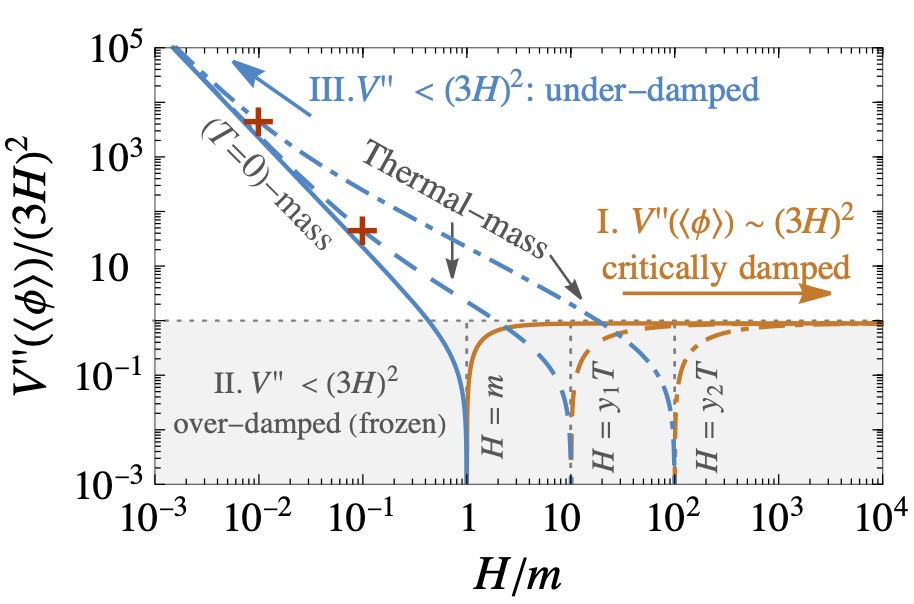}}}
\caption{\textit{ \small We numerically integrate the Klein-Gordon equation of motion in the presence of the negative Hubble-induced potential in Eq.~\eqref{eq:Hubble_induced_pot}, in radiation-dominated universe. There are three stages of field evolution, which depend on the value of $V''/H$: 1) critically-damped 2) over-damped 3) under-damped. 
Effects from the thermal mass become important for $y > \sqrt{m/\MPl}$. They lead to an earlier oscillation and a larger $\phi_\mathrm{ini}$, compared to the zero-temperature mass alone. 
The field later evolves in a zero-temperature potential at the points marked by the red crosses.
We chose the values $y_1 = \sqrt{10m/\MPl}$, and $y_2 = 10\sqrt{m/\MPl}$.}}
\label{fig:initial_VEV_traj}
\end{figure}

\paragraph{Initial radial value.}
Neglecting the $U(1)$-breaking term and assuming $a \ll \sqrt{c}$ in Eq.~\eqref{eq:curvature_induced_full_pot}, the potential at early time reads, cf. Eq.~\eqref{eq:V_U1_app_2}, Eq.~\eqref{eq:V_break_U1_app_2} and Eq.~\eqref{eq:curvature_induced_full_pot}
\begin{align}
V(\phi) ~ = ~ (\frac{1}{2}m^2_{\rm eff}(T) - c H^2)\phi^2 + \lambda^2\frac{\phi^{2l-2}}{\MPl^{2l-6}},\qquad m_{\rm eff}^2(T) \equiv m^2 + y^2 T^2, \label{eq:Hubble_induced_pot}
\end{align}
where we included a thermal mass $y^2 T^2$ coming from a possible interaction with the thermal bath. 
We study the dynamics of the scalar field at early time by numerically integrating the field equation of motion in the presence of the potential in Eq.~\eqref{eq:Hubble_induced_pot}.
In order to simplify the numerical study, we assume that the universe is radiation-dominated, so that the temperature $T$ is related to the Hubble parameter $H$ through
\begin{equation}
T^2 \simeq H \MPl.
\end{equation}
From Eq.~\eqref{eq:damped_harmonic_osc}, we deduce that there are three stages of field evolution after which the field starts oscillating with an amplitude $\phi_\mathrm{ini}$ in a potential with minimum $\phi_\mathrm{min} = 0$.\footnote{At early stage, we can forget about the existence of the SSB minimum at $\phi = f_a \ll \phi_\mathrm{ini}$.} The resulting field trajectory is plotted in Fig.~\ref{fig:initial_VEV_traj}.
\begin{enumerate}
\item $\sqrt{c} H \gg m_{\rm eff}$: the negative Hubble-mass dominates the mass term so the fields rolls within $N_e \simeq 3/c$ e-folds towards the non-trivial minimum at
\begin{align}
\phi_{\rm ini} (H) ~ \simeq ~ \left(\frac{\MPl^{l-3}\sqrt{c H^2 - m^2/2 -H \MPl y^2/2} }{\lambda\sqrt{2l-2}}\right)^{1/(l-2)}.
\end{align}

\item $m_{\rm eff} \gtrsim \sqrt{c} H$: the Hubble-mass becomes sub-dominant and the time-dependent minimum vanishes $\phi_{\rm min} = 0$. 

\item $m_{\rm eff} > 3H$: the Hubble friction drops, the field becomes under-damped, $V''(\left<\phi \right>) > 9H^2$, and starts to roll away from 
\begin{equation}
\phi_{\rm ini} (H \simeq m_{\rm eff}) ~ \simeq ~ M_{\rm pl}\left(\sqrt{c}\frac{m_{\rm eff}}{\lambda\sqrt{2l-2} M_{\rm pl}} \right)^{1/(l-2)}. \label{eq:initial_VEV_radiation}
\end{equation}
\end{enumerate}

\FloatBarrier
\FloatBarrier

The trajectories shown in Fig.~\eqref{fig:initial_VEV_traj} assume that the dynamics occurs during a radiation-dominated era.
If the Hubble-induced mass is generated from the inflaton sector, the Hubble-induced mass decreases instead during preheating and the field starts rolling away from the non trivial value in Eq.~\eqref{eq:initial_VEV_radiation} as soon as $\textrm{Max}\left[\sqrt{c},\,3\right]H$ drops below $m_{\rm eff}$.

\paragraph{Initial angle.}
For F-term dominated universe, cf. Eq.~\eqref{eq:curvature_induced_full_pot}, the $U(1)$ breaking term receives Hubble-dependent corrections
\begin{align}
V_{\cancel{U(1)}}(\Phi) ~ \simeq ~ ~\left(1 + a \frac{H }{m_{32}} \right)\Lambda_b^4\left[\left( \frac{\Phi}{M_{\rm pl}}\right)^l +h.c.\right], \qquad  a,\,c = \mathcal{O}(1).
\label{eq:Hubble_dependent_U(1)_breaking}
\end{align}
Therefore, the positions of the valleys of minimum potential $\theta_{\rm min}$ are time dependent. The same dynamics, just described above for the radial mode, also applies for the angular mode. The angular field rolls towards a temporary vallee $\theta_0 +\theta_H$ during inflation when $H_{\rm inf} \gtrsim m_{32}$, and starts rolling towards the Hubble-independent valley $\theta_0$ when $H$ drops below $m_{32}$.
In this work, we simply assume the initial angular amplitude\footnote{Note that for radiation-dominated universe, the A-term in Eq.~\eqref{eq:curvature_induced_full_pot_0}, at least in our analysis and in the ones of \cite{Anisimov:2000wx,Allahverdi:2000zd}, seems to be Hubble-independent. We leave for further study the question of whether such Hubble-dependent A-term is induced during radiation-domination. In any case, the angular kick could be generated by random fluctuations during inflation, see Sec.~\ref{sec:stochastic_inflation}.}
\begin{equation}
\theta_\textrm{ini} \sim \mathcal{O}(1).
\end{equation}

\subsection{Nearly-quadratic potentials}
\label{susy_potential}
In this appendix, we review the results presented in \cite{Moxhay:1984am} which show how spontaneous symmetry breaking can arise due to the running of the soft masses. The authors consider supersymmetric theories coupled to $\mathcal{N} = 1$ supergravity where $U(1)_\textrm{PQ}$ is a global symmetry of the superpotential. They show that $U(1)_\textrm{PQ}$ can be broken by the logarithmic running of renormalization group equations. Their results are not specific of the Peccei-Quinn symmetry and they can be generalized to other global $U(1)$. 
\paragraph{Introduce chiral superfields.}
The superpotential is given by 
\begin{align}
W ~ = ~ \lambda \bar{Q} Q \phi,
\end{align}
where the $SU(3)_c$ quantum numbers of the chiral superfields are
\begin{align}
Q(\textbf{3}), ~ \bar{Q}(\bar{\textbf{3}}), ~ \phi(\textbf{1}), 
\end{align}
and are singlets under $SU(2)_L$.
The fields transform under the $U(1)_\textrm{PQ}$ as
\begin{align}
(Q, ~ \bar{Q}) ~ \rightarrow ~ e^{i \alpha} (Q, ~ \bar{Q}),\qquad  ~ \phi ~ \rightarrow ~ e^{-2i\alpha} \phi.
\end{align}
There is a continuum of supersymmetric minima
\begin{align}
\phi ~ = ~ \textrm{undetermined}, ~ Q ~ = ~ \bar{Q} ~ = ~ 0.
\end{align}

\paragraph{SUGRA breaking.}
Supergravity couplings lift this flat direction by inducing soft supersymmetric breaking terms characterized by the gravitino mass $m_{3/2}$
\begin{align}
V_\textrm{soft} ~ = ~ m_{3/2} A \lambda \bar{Q} Q \phi + \textrm{h.c.} + m_{3/2}^2 (g_Q |Q|^2 + g_{\bar{Q}} |\bar{Q}|^2 +g_\phi |\phi|^2).
\end{align}
The initial conditions for the RGE are set at the Planck scale $M_{\rm pl}$,
\begin{align}
g_Q ~ = ~ g_{\bar{Q}} ~ = ~ g_\phi ~ = ~ 1.
\end{align}
The effective scalar potential is 
\begin{align}
V ~ = ~ \sum_i \left| \frac{\partial W}{\partial \phi_i} \right|^2 + \frac{1}{2} \Sigma_a D_a^2 + V_\textrm{soft},
\end{align}
and along the supersymmetric minimum $(Q ,~ \bar{Q}) = (0,\,0)$, it is reduced to
\begin{align}
V(\phi) ~ = ~ m_{3/2}^2 g_\phi |\phi|^2.
\end{align}

\paragraph{Renormalization group equations.}
Loop corrections to the couplings lead to the running \cite{Moxhay:1984am}
\begin{align}
\frac{d e}{d t} ~ &= ~ \frac{-2}{16 \pi^2} e^3,\\
\frac{d \lambda}{d t} ~ &= ~ \frac{\lambda}{16 \pi^2} \left( 5\lambda^2 -\frac{16}{3} e^2 \right),\\
\frac{d A}{d t} ~ &= ~ \frac{5 A \lambda^2}{8 \pi^2} e^3,\\
\frac{d g_Q}{d t} ~ &= ~ \frac{d g_{\bar{Q}}}{d t} ~ = ~ \frac{1}{3} \cdot \frac{d g_\phi}{d t} ~ = ~ \frac{\lambda^2}{8 \pi^2} \left( g_Q + g_{\bar{Q}} + g_\phi + A^2 \right),
\end{align}
where $t \equiv \ln(|\phi|/M_{\rm pl})$ and $e$ is the gauge-coupling of the $SU(3)_c$.
The equation for $e$ can be solved analytically with 
\begin{align}
e^2(t) ~ = ~ \frac{e_0^2}{1+\frac{e_0^2}{4 \pi^2} t}.
\end{align}
All the couplings $g$ are related $g_Q =  g_{\bar{Q}} = \frac{1}{3} g_\phi$, hence we have
\begin{align}
\frac{d g_\phi}{d t} ~ = ~ \frac{\lambda^2}{8 \pi^2} \left( 5 g_\phi + 3 A^2 \right).
\end{align}
The system of first-order differential equations above has a fixed point at
\begin{align}
\frac{\lambda^2}{e^2} ~ = ~ \frac{2}{3},
\end{align}
where $g_\phi$ is running as
\begin{align}
\frac{d g_\phi}{d t} ~ = ~ \frac{e_0^2}{12 \pi^2} \left( \frac{5 g_\phi + 3 A^2}{1+\frac{e_0^2}{4 \pi^2} t} \right),
\end{align}
and $A$ is approximately constant.

\paragraph{Dimensional transmutation.}
In the limit $\frac{e_0^2}{4 \pi^2} \ll 1$, the above equation is solved analytically by
\begin{align}
g_\phi(\phi) ~ \simeq ~ (3A + 5) \left(\frac{e_0^2}{24 \pi^2}\right) \ln\left(\frac{\phi^2}{M_{\rm pl}^2}\right) + 1,
\end{align}
where the approximation $\frac{e_0^2}{4 \pi^2} \ll 1$ is used.
Then we define scale $M_\textrm{PQ}$ by when the radiative correction changes the sign of effective potential, i.e. at
\begin{align}
g_\phi(\phi \rightarrow M_\textrm{PQ}) ~ = ~ 0 ~ \simeq ~  (3A + 5) \left(\frac{e_0^2}{24 \pi^2}\right) \ln\left(\frac{M_\textrm{PQ}^2}{M_{\rm pl}^2}\right) + 1,
\end{align}
where $M_\textrm{PQ}$ is related to $M_{\rm pl}$ by
\begin{align}
M_\textrm{PQ}^2 ~ = ~ M_{\rm pl}^2 \exp\left[\left(\frac{-1}{3A+5}\right)\frac{24\pi^2}{e_0^2}\right].
\end{align}
With the definition of $M_\textrm{PQ}$, the coupling $g_\phi$ is simplified to
\begin{align}
g_\phi(\phi) ~ = ~  (3A + 5) \left(\frac{e_0^2}{24 \pi^2}\right) \ln\left(\frac{\phi^2}{M_\textrm{PQ}^2}\right).
\end{align}
\paragraph{The effective potential.}
The effective potential reads
\begin{align}
V_\textrm{eff} ~ = ~ m_{3/2}^2 g_\phi \phi^2 ~ \simeq ~ m_{3/2}^2 \phi^2 \left[\ln\left(\frac{\phi^2}{f_a^2}\right) - 1\right],\qquad \text{and}\quad f_a ~ \equiv ~ M_\textrm{PQ} e^{-1/2}.
\end{align}
For $A = 2$ and $e_0 = 0.8$, we have $f_a \simeq 7  \times 10^{10} ~\rm GeV$.

\section{Inflationary fluctuations in the rotating complex scalar field model}
\label{app:issus_inflation_fluct_model_B}

\subsection{Adiabatic curvature perturbations}
Any light scalar present during inflation receives quantum fluctuations which classicalize upon horizon exit, e.g. \cite{Riotto:2002yw}. When such perturbations re-enter the horizon, they source either the adiabatic or the isocurvature  power spectrum according to whether the extra specie thermalizes with the SM or not. 
In our model, cf. Sec.~\ref{sec:PQ:exampleII}, the radial mode of the complex scalar field decays into thermal radiation and therefore contributes to the adiabatic part of the curvature perturbations.
Assuming a quadratic scalar potential with $m_r \ll H_{\rm inf}$, the perturbation from the decaying scalar field reads \cite{Torrado:2017qtr}
\begin{align}
\mathcal{P}_\zeta^\phi ~ = ~ r_\mathrm{dec}^2 \left(\frac{H_\mathrm{inf}}{3\pi \phi_\mathrm{ini}}\right)^2,\qquad \textrm{with} \quad r_\mathrm{dec}~ \simeq ~ \left(\frac{3 \rho_\phi}{3\rho_\phi + 4 \rho_\mathrm{rad}}\right)_\mathrm{dec}, \label{eq:r_dec_fun}
\end{align}
where $r_\mathrm{dec}$ is the energy density fraction carried by the scalar, evaluated at the time of radial damping.
The upper bound on $\mathcal{P}_\zeta^\phi$  corresponds to the value measured by CMB \cite{Akrami:2018odb}
\begin{align}
\mathcal{P}_\zeta^\phi ~ < ~ \mathcal{P}_\zeta^\mathrm{tot}  ~ \simeq ~ 2.2 \times 10^{-9}.\label{eq:planck_constraint_adiab}
\end{align}
To constrain the model, we consider two limits.\noindent

\begin{figure}[h!]
\centering
\raisebox{0cm}{\makebox{\includegraphics[width=0.475\textwidth, scale=1]{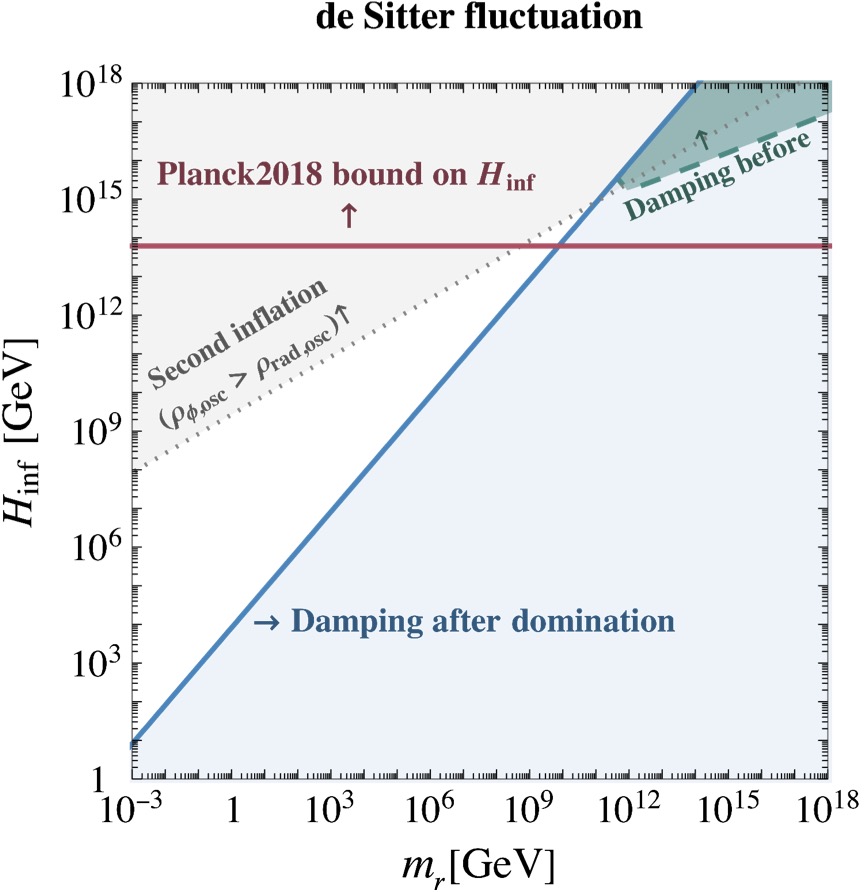}}}
\caption{\textit{ \small Parameter space of a thermalized complex scalar which generates too much adiabatic curvature perturbations. The weakest constraint (\textbf{green}) arises when damping occurs at the onset of oscillation. As damping takes place later, the constraint grows and reach \textbf{light blue} region when damping takes place at the onset of scalar domination. For the Hubble-induced mass scenario, the green and light blue constraints are avoided due to a large Hubble-sized mass during inflation, see Sec.~\ref{sec:intitial_vev}. The red line corresponds to Planck constraints from B modes non-observation \cite{Akrami:2018odb}. The gray region corresponds to the requirement of not generating a second inflation era, see Eq.~\eqref{eq:no_second_inflation}.}}
\label{adiabatic_radial_constraint}
\end{figure}

\textbf{1) Damping before domination.} Assuming the extreme case in which radial damping occurs right after the onset of radial mode oscillation, c.f. Eq.~\eqref{eq:H_osc_def}, in order to minimize $r_{\rm dec}$, the adiabatic curvature power spectrum reads
\begin{align}
\mathcal{P}_\zeta^\phi ~ = ~ \left(\frac{ \phi_\mathrm{osc}^2}{ \phi_\mathrm{osc}^2 + \frac{4}{9} \MPl^2}\right)^2 \left(\frac{H_\mathrm{inf}}{3\pi \phi_\mathrm{ini}}\right)^2,
\end{align}
where $\phi_\mathrm{osc}$ is the value at oscillation.
\begin{itemize}
\item
 If the initial field value is set by the stochastic-inflation process, i.e. $\phi_\mathrm{ini}^2 = 3H_\mathrm{ini}^4/(4\pi^2 m^2)$ in Eq.~\eqref{ds_fluc}, the above equation becomes
\begin{align}
\mathcal{P}_\zeta^\phi ~ = ~ \left(\frac{27}{64\pi^4}\right) \left(\frac{H_\mathrm{inf}^6}{m^2 \MPl^4}\right) \left(1 + \frac{27 H_\mathrm{inf}^4 }{16 \pi^2 m_r^2 \MPl^2}  \right)^{-2}. \label{eq:adiab_fluct_1_1}
\end{align}
In Fig.~\ref{adiabatic_radial_constraint}, we show that the constraints on $H_{\rm inf}$ from Eq.~\eqref{eq:planck_constraint_adiab} and Eq.~\eqref{eq:adiab_fluct_1_1} are weaker than the Planck constraints coming from the B-mode non-observation.
\item
If the initial field value is driven by the negative-Hubble mass and is stabilized by higher-order terms, the quantum fluctuation of the radial and angular fluctuation are suppressed, see Sec.~\ref{sec:solution_inflation_fluct_model_B}. This is the scenario which we assume in this paper.
\end{itemize}
\textbf{2) Damping when dominating.}
Assuming $\rho_\phi \gg \rho_\mathrm{rad}$ in Eq.~\eqref{eq:r_dec_fun} leads to $r_\mathrm{dec} ~ \simeq ~  1$ and to the adiabatic curvature perturbations $
\mathcal{P}_\zeta^\phi ~ = ~ \left(H_\mathrm{inf}/3\pi \phi_\mathrm{ini}\right)^2.
$
\begin{itemize}
\item
If the initial field value is set by the stochastic-inflation process, i.e. $\phi_\mathrm{ini}^2 = 3H_\mathrm{ini}^4/(4\pi^2 m_r^2)$, the above equation becomes
\begin{align}
\mathcal{P}_\zeta^\phi ~ = ~ \left(\frac{4}{27}\right) \left(\frac{m_r}{H_\mathrm{inf}}\right)^2.\label{eq:adiab_fluct_2_1}
\end{align}
As shown in light blue in Fig.~\ref{adiabatic_radial_constraint}, the constraint from Eq.~\eqref{eq:planck_constraint_adiab} and Eq.~\eqref{eq:adiab_fluct_2_1} can be quite limiting.
\item
As mentioned in the previous paragraph and as we will discuss more precisely in Sec.~\ref{sec:solution_inflation_fluct_model_B}, for scenarios with Hubble-size masses, quantum fluctuations are suppressed and the Planck constraints are avoided. 
\end{itemize}

\FloatBarrier
\FloatBarrier

\subsection{Domain wall problem}

Quantum fluctuations of the initial angular phase $\delta \theta_i$, and of the initial radial value $\delta \phi_i$  lead to a fluctuation of the final angular phase $\delta \theta$ at the time when the axion potential develops. This can lead to the formation of domain walls (DW). In the absence of CS, e.g. in the pre-inflationary PQ-breaking scenario, the DW are infinite and can not decay, so we must impose $\delta \theta \lesssim \mathcal{O}(1)$.

Assuming $V(\Phi) \propto \Phi^p$, from $U(1)$ charge conservation in Eq.~\eqref{PQ_charge_conservation} and from the scaling $\left<\phi \right> \propto a^{-\frac{6}{2+p}}$ in Eq.\eqref{eq:final_solution_virial}, we deduce the scaling of the angular velocity 
\begin{align}
Y_\theta ~ = ~ \phi^2 \dot{\theta} ~ \propto a^{-3} ~ ~ ~ ~  \Rightarrow ~\dot{\theta} ~ \propto ~ a^{-3(p-2)/(p+2)} ~ \propto ~ t^{-3(p-2)/2(p+2)} ,
\end{align}
where we assume a radiation-dominated universe.
The angular phase elapsed after the onset of radial mode oscillation which we denote by $t_i \simeq m_\mathrm{eff}^{-1}(\phi_i)$ reads
\begin{align}
\theta ~ = ~ \int_{t_i}^{t} \dot{\theta} dt' ~ \simeq ~\dot{\theta}_i t_i \left(\frac{\phi_i}{\phi}\right)^{(10-p)/6},
\end{align}

The fluctuation in the angular phase can be written as
\begin{align}
\frac{\delta \theta}{\theta} ~ = ~ \frac{\delta\dot{\theta}_i}{\dot{\theta}_i} + \frac{\delta t_i}{t_i} +  \left(\frac{10-p}{6}\right) \left(\frac{\phi_i}{\phi}\right)^{(10-p)/6} \left(\frac{\delta \phi_i}{\phi_i}\right). \label{eq:delta_theta_over_theta_ModelB}
\end{align}
From Eq.~\eqref{eq:theta_ini_app}, $t_i \simeq m_\mathrm{eff}^{-1}(\phi_i)$  and Eq.~\eqref{eq:mr_eff}, we can write
\begin{align}
\frac{\delta\dot{\theta}_i}{\dot{\theta}_i} ~ = ~ (l-p) \frac{\delta \phi_i}{\phi_i} + l \cot(l \theta_i) \delta \theta_i + \frac{\delta t_i}{t_i}, \qquad \textrm{and} \qquad \frac{\delta t_i}{t_i} ~ = ~ - \frac{\delta \phi_i / \phi_i}{2\left(1+ \log(\phi_i/f_a)\right)}. \label{eq:dot_theta_i_and_ti}
\end{align}

The late time phase fluctuation in Eq.~\eqref{eq:delta_theta_over_theta_ModelB} becomes
\begin{align}
\frac{\delta \theta}{\theta}  ~ \simeq ~ \left[(l-p) - \frac{1}{1+\log(\phi_i/f_a)} + \left(\frac{10-p}{6}\right) \left(\frac{\phi_i}{f_a}\right)^{(10-p)/6} \right]  \left(\frac{\delta \phi_i}{\phi_i}\right) + l \cot{l \theta_i} \delta \theta_i,
\end{align}
where the last term in the squared bracket dominates for $p < 10$ and $\phi_i \gg f_a$.
Plugging the typical standard deviation for a massless field during inflation, e.g. \cite{Riotto:2002yw}
\begin{equation}
\delta \phi_i = H_\mathrm{inf} / 2\pi = \phi_i \delta \theta_i, \label{eq:quantum_fluctuation_massless_field_app}
\end{equation}
we obtain the necessary condition for preventing DW formation once the axion potential switches on
\begin{align}
\delta \theta  ~ \simeq ~ \left[\left(\frac{10-p}{6}\right) \left(\frac{\phi_i}{f_a}\right)^{(10-p)/6} + l \cot(l \theta_i)\right] \left(\frac{H_\mathrm{inf}}{2 \pi \phi_i}\right) ~<~ 1,
\end{align}
where we replaced $\theta \sim \mathcal{O}(1)$. 
The flatter the potential, the slower the redshift of the angular velocity and the larger the final time fluctuation $\delta \theta $.

\subsection{Isocurvature perturbations} 
If the axion contributes to the DM abundance then the quantum fluctuation during inflation in Eq.~\eqref{eq:quantum_fluctuation_massless_field_app} generate isocurvature perturbations  \cite{Linde:1985yf, Lyth:1989pb, Kawasaki:1995ta, Beltran:2006sq}\footnote{This problem also arises in the context of the Affleck-Dine Baryogenesis, where the inflationary perturbation leads to the baryonic  isocurvature perturbations \cite{Enqvist:1998pf, Kasuya:2008xp, Kamada:2010yz, McDonald:2012pv, Harigaya:2014tla}.} whose amplitude is bounded by Planck data \cite{Akrami:2018odb}
\begin{align}
\mathcal{P}_{\rm iso} ~ = ~ \left\langle \left( \frac{\Omega_a}{\Omega_{\rm DM}} \cdot \frac{\delta \Omega_a}{\Omega_a}\right)^2 \right\rangle ~ < ~ 8.69 \times 10^{-11}.
\end{align}
For a spinning axion, the axion abundance is set by the kinetic misalignment mechanism $\Omega_a \propto \dot{\theta}_c$, cf. Eq.~\eqref{eq_axion_fraction}.  From Eq.~\eqref{eq:dot_theta_i_and_ti} and Eq.~\eqref{eq:quantum_fluctuation_massless_field_app}, we deduce
\begin{align}
\mathcal{P}_{\rm iso} ~ = ~ \left(\frac{\Omega_a}{\Omega_{\rm DM}}\right)^2 \left\langle \left(\frac{\delta \dot{\theta}_i}{\dot{\theta}_i}\right)^2 \right\rangle ~  = ~A \left(\frac{\Omega_a}{\Omega_{\rm DM}}\right)^2\left(\frac{H_{\rm inf}}{2 \pi \phi_i}\right)^2  ~ < ~ 8.69 \times 10^{-11}, \label{eq:isocurvature_bound_model_B}
\end{align}
where $A\equiv l-p - \frac{1}{1+\log(\phi_i/f_a)} + l \cot(l \theta_i) = \mathcal{O}(1)$.

\subsection{Solution}
\label{sec:solution_inflation_fluct_model_B}

A way to cure the three problems listed above - too large adiabatic and isocurvature perturbations and DW overclosure - is to suppress the initial quantum fluctuations by introducing a large mass for both the radial mode $\phi$ and the angular mode $\theta$, during inflation \cite{Dine:2004cq, Jeong:2013xta, Higaki:2014ooa, Dine:2014gba, Harigaya:2014tla, Choi:2015zra, Takahashi:2015waa}. 

As discussed along Eq.~\eqref{eq:negative_hubble_mass}, Hubble size masses for $\phi$ and $\theta$ arise naturally in SUSY scenario where we have
\begin{align}
m_\phi^2 ~ = ~ \partial^2 V_H/\partial \phi^2 \simeq ~ 4(l-2) H_{\rm inf}^2, ~ ~ {\rm and} ~ ~ m_\theta^2 ~ = ~ |\phi|^{-2} \partial^2 V_H/\partial \theta^2 \simeq ~ H_{\rm inf}^2 \sqrt{l^2/(l-1)}.
\end{align}
So, in fact, the solution to these problems is built-in in these models.
Quantum fluctuations of massive states are blue-tilted, e.g. \cite{Riotto:2002yw}, such that the amplitude of the associated curvature perturbations entering the Hubble horizon long after the end of inflation, are expected to be negligible. This is the scenario which we consider in this paper and therefore we assume that the initial field value $\phi_{\rm i}$ is set by the classical minimum of the Hubble-induced SUSY potential in Sec.~\ref{sec:intitial_vev_SUSY} and not by the Bunch-Davies quantum distribution in Sec.~\ref{sec:stochastic_inflation}.

\section{Damping of the radial motion in the rotating complex scalar field model}
\label{app:radial_damping}
\subsection{Thermalization}
\label{app:thermalization}

\paragraph{Fermion portal.}
We assume that the complex scalar field $\phi$ is coupled to heavy fermions $\psi$ charged under some (hidden or SM) gauge sector $A_\mu$ (KSVZ-type interactions)
\begin{equation}
\mathcal{L} \supset y_\psi\phi \psi_{L}^{\dagger} \psi_{R}+h.c.+ g \, \bar{\psi} \gamma^\mu\psi\, A_\mu. \label{eq:KSVZ_lagrangian}
\end{equation}

\begin{figure}[h!]
\centering
\includegraphics[width=1.0\columnwidth]{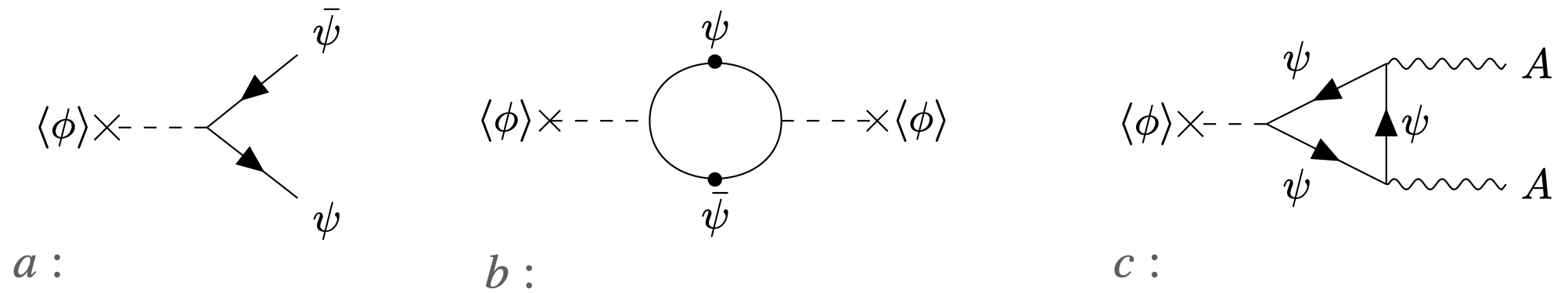}
\caption{\textit{ \small  Quantum processes responsible for the thermalization of the coherent scalar field in KSVZ type models. The scattering rates of a, b and c are respectively given by $\frac{y_\psi^2 m_{\phi}}{8\pi}$, $\frac{y_\psi^2 \alpha T}{2 \pi^2}$ and $\frac{b \alpha^2 T^3}{\phi^2}$, cf. Eq.~\eqref{eq:fermion_damping_rate_YG}. The black blobs in the middle diagram stand for the thermal field corrections to the fermion propagator accounting for plasma effects. The decay width is obtained from the finite-temperature analogue of Cutkosky's rule \cite{Laine:2016hma}.
}}
\label{fig:KSVZ_decay_diagrams}
\end{figure}

At zero temperature, the scalar condensate decays into fermions pairs with a rate, cf. diagram $a$ in Fig.~\ref{fig:KSVZ_decay_diagrams}
\begin{equation}
\Gamma_{\phi \to \psi \psi} \simeq \frac{y_\psi^2}{8\pi} m_{\psi},  \qquad \qquad\textrm{for}\quad  m_{\phi}/2 ~> ~\textrm{Max}[y_\psi \phi,\, gT].
\end{equation}
If the fermion mass, either from vacuum $y_\psi \phi$ or thermal origin $gT$, is larger than the scalar field mass, then the scalar field cannot excite thermal $\psi$ and it is dominantly depleted through scattering with thermally-dressed fermions of the plasma, cf. diagram $b$ in Fig.~\ref{fig:KSVZ_decay_diagrams}, with a rate \cite{Mukaida:2012qn,Mukaida:2012bz} 
\begin{equation}
 \label{eq:decay_width_into_fermion_app_Kyohei}
\Gamma_{\phi\psi_{\rm th} \rightarrow \psi_{\rm th}} \simeq 
\begin{cases}
\frac{y_\psi^2 \alpha T}{2 \pi^2},\qquad \textrm{for}~ \alpha T > y_\psi \phi ,\vspace{0.25cm} \\
\frac{y_\psi^4 \phi^2}{\pi^2 \alpha T},\qquad \textrm{for}~\alpha T < y_\psi \phi <  T,
\end{cases}
\qquad \quad 
\alpha \equiv \frac{g^2}{4\pi}.
\end{equation}
The two regimes in Eq.~\eqref{eq:decay_width_into_fermion_app_Kyohei} depend on whether the thermal width $\alpha T$ - the typical relaxation rate of the fermion density towards thermal equilibrium - is  larger or smaller than the fermion zero-temperature mass  $y_\psi \phi$.
If the temperature is smaller than the fermion mass, $T < y_\psi \phi$, the heavy fermions are absent of the thermal plasma. In that case, the scalar field can decay into gauge bosons through a loop of fermions \cite{Shifman:1979eb,Bodeker:2006ij,Laine:2010cq}, cf. diagram $c$ in Fig.~\ref{fig:KSVZ_decay_diagrams}
\begin{equation}
\Gamma_{\phi \to A A} \simeq \frac{b \alpha^2 T^3}{\phi^2}, \qquad b\simeq 0.01, \qquad \textrm{for}~y_\psi \phi > T. \label{eq:decay_width_into_gluons}
\end{equation}
For $T < m_\phi$, we replace $T$ in Eq.~\eqref{eq:decay_width_into_gluons} by $m_\phi$. The dependence on $y_\psi$ in Eq.~\eqref{eq:decay_width_into_gluons} appears in the logarithmic running of $\alpha$.
So we conclude 
\begin{align}
\Gamma_\phi \simeq  \begin{cases}
\textrm{ for} ~ y_\psi \phi < T :
\begin{cases}
 \textrm{for}~ \alpha T  > y_\psi \phi ,\qquad \frac{y_\psi^2 \alpha T}{2 \pi^2},\\[0.5em]
\textrm{for}~\alpha T < y_\psi \phi ,\qquad \frac{y_\psi^4 \phi^2}{\pi^2 \alpha T},
\end{cases} 
\\[2em]
\textrm{ for} ~ y_\psi \phi > T: 
\qquad 
b \alpha^2 \frac{\textrm{Max}\left[T,~m_\phi\right]^3}{\phi^2} ,
\\
\end{cases}
+\quad \frac{y_\psi^2 m_{\phi}}{8\pi} \Theta\left(m_\phi/2 - \textrm{Max}\left[y_\psi \phi,\,gT\right]\right).
\label{eq:fermion_damping_rate_YG}
\end{align}

\paragraph{The scalar field keeps spinning after thermalization.}
The coherent oscillation is a very ordered state and once the oscillation energy begins to be transferred into the thermal bath, the inverse process which creates coherent oscillation is unlikely to occur.
Therefore, the thermalization of the scalar field, when the universe reaches the energy density
\begin{equation}
\rho_{\rm damp} = 3 \Gamma^2 M_{\rm pl}^2,
\end{equation}
transfers all the kinetic energy of the radial mode to the plasma $\dot{\phi} \to 0$. 
The kinetic energy of the angular mode, aka the $U(1)$ charge, can also be transferred to the thermal bath, e.g. in the form of chiral charges of SM fermions. However, as shown by the supplementary material of \cite{Co:2019wyp}, the creation of an asymmetric fermion abundance costs chemical potential which leads to an energy density $\phi^4 \dot{\theta}^2/T^2$ and it turns out that it is thermodynamically favorable to keep the $U(1)$ charge in the condensate with energy density $\phi^2 \dot{\theta}^2$ as long as 
\begin{equation}
f_a ~\gg~T_{\phi \to f_a},
\end{equation}
where $T_{\phi \to f_a}$ is the temperature when $\phi$ reaches $f_a$.
We check that this condition is satisfied in Eq.~\eqref{eq:check_phi_damp_larger_Tdamp}.
The washout of $U(1)$ charge due to chirality flip mediated by Yukawa interactions is shown to be negligible if the temperature of thermalization is smaller than \cite{Co:2019wyp}
\begin{equation}
T_{\rm damp} \lesssim 10^{12}~\rm GeV \left(\frac{\left<\phi\right>}{10^9~\rm GeV}\right)^2.
\end{equation}
Using that the scalar field evolves in its quadratic potential as 
\begin{equation}
\phi = \phi_{\rm ini} (T/T_{\rm osc})^{3/2}, \qquad T_{\rm osc}=g_*^{-1/4}\sqrt{m_r(\phi_{\rm ini}) M_{\rm pl}},
\end{equation}
we obtain that the axion rotation is preserved from wash-out from the onset of oscillation until the final stage $\phi \to f_a$ as long as
\begin{equation}
f_a ~> ~10^{4}~\textrm{GeV}\left( \frac{m_r}{f_a}\right)^{3/5}\left( \frac{M_{\rm pl}}{\phi_{\rm ini}}\right)^{4/5}.
\end{equation}
The later condition is largely satisfied in the parameter space of our interest, and  wash-out never occurs.

\paragraph{Thermal corrections to the potential.}
The presence of the interactions in Eq.~\eqref{eq:KSVZ_lagrangian} generates thermal corrections to the scalar potential \cite{Mukaida:2012qn,Mukaida:2012bz} 
\begin{align}
&V(\phi) = m_r^2 \phi^2 \left(\ln \frac{\phi^2}{f_a^2} -1 \right) + m_r^2 f_a^2 + V_{\rm th}(\phi,\, T), \\
&V_{\rm th}(\phi,\, T) = \frac{1}{2}y_\psi^2 T^2 \phi^2 \Theta(T - y_\psi \phi) + a \alpha^2 T^4 \ln {\lambda_\phi^2 \phi^2 /T^2} \Theta(y_\psi \phi - T) \qquad a = \mathcal{O}(1), \quad \alpha \equiv \frac{g^2}{4\pi}. \label{eq:thermal_pot}
\end{align}
The first Heaviside function stands for the Boltzmann suppression of the fermion $\psi$ abundance in the thermal plasma in the large vev limit $y_\psi \phi \gg T$. In that case the thermal corrections are given by the thermal-log potential, obtained after integrating out the heavy fermions \cite{Shifman:1979eb,Carena:2012xa,Bodeker:2006ij,Laine:2010cq}. At small vev value  $\lambda \phi < T$, the running of the gauge coupling constant $g$ becomes independent of $\phi$, which explains the second Heaviside function.

\paragraph{Impossibility to generate efficient thermalization when neglecting thermal corrections.}
\label{sec:impossibility_neglect_thermal_correction}
Let's first assume that thermal corrections to the potential in Eq.~\eqref{eq:thermal_pot} do not impact the scalar field dynamics. This happens whenever the thermal mass of the scalar fields can be neglected at the time of oscillation $T_{\rm osc} \sim \sqrt{m_{\rm r,\, eff}M_{\rm pl}}$, see Eq.~\eqref{eq:thermal_mass_negligible_scenario_I}, \footnote{We backward check that this coincides with $ y_\psi \phi_{\rm ini} \lesssim T_{\rm osc}$ (the first Heaviside function in Eq.~\eqref{eq:thermal_pot}) whenever the initial field value $\phi_{\rm ini}$ is sub-planckian
\begin{equation}
\frac{T_{\rm osc} }{y_\psi} \gtrsim \frac{T_{\rm osc} ^2}{m_{\rm r, eff}} \simeq \frac{M_{\rm pl}}{\left(\frac{\pi^2 g_*}{10} \right)^{1/4}} \gtrsim \phi_{\rm ini}.
\end{equation}}
\begin{equation}
y_\psi T_{\rm osc}  \lesssim m_{\rm r, eff}, \quad \implies \quad y_\psi \lesssim \left(\frac{\pi^2 g_*}{10} \right)^{1/4} \sqrt{\frac{m_{\rm r, eff}}{M_{\rm, pl}}}. \label{eq:thermal_mass_negligible}
\end{equation}
The duration of the kination era reads, cf. Sec.~\ref{sec:cosmo_history}
\begin{equation}
e^{N_{\rm KD}}  \equiv\frac{a_{\rm Kf}}{a_{\rm Ki}} = \left( \frac{\textrm{Min}\left[\rho_{\rm dom},\, \rho_{\rm damp}\right]}{\rho_{\rm K, i}} \right)^{1/6} \left(\frac{\epsilon}{2}\right)^{2/3},
\end{equation}
with
\begin{align}
\rho_{\rm dom} = \frac{27V(\phi_{\rm ini})^4}{\mreff^6 M_{\rm pl}^6}, \qquad \rho_{\rm damp}= 3\Gamma^2 M_{\rm pl}^2 \qquad \textrm{and} \qquad  \rho_{\rm K, i} = \frac{1}{2}f_a^2 \mreff^2.
\end{align}
From plugging\footnote{We checked that $y_\psi \lesssim  \sqrt{\frac{m_{\rm r, eff}}{M_{\rm, pl}}}$ implies $\frac{y_\psi \phi_{\rm damp}}{T_{\rm damp}} \lesssim (b \alpha^2)^{1/3}\left( \frac{\mreff}{M_{\rm pl}}\right)^{1/6} $ and $\frac{T_{\rm damp}}{\mreff} \lesssim 1$, leading to $\Gamma \simeq \frac{y_\psi^2 m_{\phi}}{8\pi}$ in Eq.~\eqref{eq:fermion_damping_rate_YG}.} $\Gamma \simeq \frac{y_\psi^2 m_{\phi}}{8\pi}$, cf. Eq.~\eqref{eq:fermion_damping_rate_YG}, $\phi_{\rm ini} \simeq M_{\rm pl}$, which corresponds to the $l\to \infty$ limit of Eq.~\eqref{susy_phi_ini} and $y_\psi \lesssim  \sqrt{\frac{m_{\rm r, eff}}{M_{\rm, pl}}}$, cf. Eq.~\eqref{eq:thermal_mass_negligible} we obtain
\begin{equation}
e^{N_{\rm KD}} ~\lesssim ~0.3\, g_*^{1/6}\left( \frac{\mreff(f_a)}{f_a}  \right)^{1/3} \epsilon^{2/3}~<~1.
\end{equation}
So we conclude that whenever the conditions of neglecting the thermal mass in Eq.~\eqref{eq:thermal_mass_negligible} is satisfied, thermalization via fermion portal is not efficient enough to generate a kination era. 

For this reason, in scenario I of Sec.~\ref{sec:scenario_I_non_thermal_damping}, we must postulate the existence of an unknown mechanism, other than thermalization via fermion or Higgs portal, to damp the radial mode.

The effects from the thermal mass are considered in scenario II of Sec.~\ref{sec:complex_field_thermal_potential}. Unfortunately, the suppression of the angular kick and the delay of the matter domination prevents the onset of kination.

In scenario III in Sec.~\ref{sec:complex_field_low_reh_temp}, instead of choosing a low Yukawa coupling $y_\psi$, we get rid of the thermal effects at the onset of the radial mode oscillation by Boltzmann-suppressing the fermion abundance.

\paragraph{Higgs portal.}
\label{par:Higgs_portal}
In the previous paragraph, we have considered the fermion portal.
If instead the scalar condensate thermalizes through a Higgs mixing (DFSZ-type interactions)
\begin{equation}
\label{eq:higgs_portal}
\mathcal{L}~ \supset~ \lambda_{\rm H} |\phi|^2 H^{\dagger}H.
\end{equation}
The thermal mass correction to the condensate is
\begin{equation}
V_{\rm th}(\phi,\, T) = \lambda_{\rm H} T^2 \phi^2.
\end{equation}
The quantum processes responsible for the thermalization of the condensate are described in \cite{Mukaida:2013xxa}.
In contrast to the fermion portal (denoted by $\psi$), in the case of the Higgs portal (denoted by $H$) the thermalization rates scale as the the forth power of the thermal mass. Indeed
\begin{equation}
\Gamma_{\rm \psi} \propto m_{\rm th, \psi}^2,\qquad \text{and} \qquad \Gamma_{\rm H} \propto m_{\rm th, H}^4, \label{eq:fermion_vs_Higgs_portal}
\end{equation}
with
\begin{equation}
m_{\rm th,  \psi}^2 = y_\psi^2 T^2, \qquad \text{and} \qquad m_{\rm th, H}^2 = \lambda_H^2 T^2.
\end{equation}
 We conclude that the situation of the previous paragraph - to impose a small thermal mass prevents thermalization to occur before the start of the would-be kination era -  is even worse in the case of the Higgs portal.
 
 The larger dependence of the damping rate on the thermal mass in the Higgs portal, in Eq.~\eqref{eq:fermion_vs_Higgs_portal}, is the reason why we focus on the fermion portal in the main text, cf. Sec.~\ref{sec:complex_field_thermal_potential} and Sec.~\ref{sec:complex_field_low_reh_temp}.

\subsection{Parametric resonance}
\label{app:param_resonance}

In this paper we focus on the zero-mode dynamics. However, there may be circumstances where higher modes get produced. We now introduce the framework to study early parametric resonance, which may occur in the UV completion and may have important consequences on the whole dynamics.
The equation of motion of a complex scalar reads
\begin{equation}
\ddot{\Phi} - a^{-2}\nabla^2 \Phi + 3H \dot{\Phi} + \frac{\partial V}{\partial \Phi^{\dagger}}
\end{equation}
which after decomposing in polar coordinates $\Phi = \phi e^{i\theta}$, becomes
\begin{align}
&\ddot{\phi} -  a^{-2}\nabla^2 \phi+ 3 H \dot{\phi} + V'(\phi) =\phi \dot{\theta}^2-  a^{-2} \phi (\nabla {\theta})^2,\\
&\phi\ddot{\theta} -  a^{-2} \phi \nabla^2 \theta+ 3 H \phi \dot{\theta} = -2 \dot{\phi}\dot{\theta} + 2  a^{-2} \nabla{\phi}\nabla{\theta}.
\end{align}
We can decompose $(\phi(x,\,t),\, \theta(x,\,t))$ into the superposition of a classical homogeneous mode $(\phi(t),\, \theta(t))$ and small fluctuations $(\delta \phi(x,\,t),\, \delta \theta(x,\,t))$  around it
\begin{align}
&\phi(x,\,t) = \phi(t) + \delta\phi(x,\,t) = \phi(t) +  \left(\int \frac{d^3 k}{(2\pi)^3} a^\phi_k \, u^\phi_k(t) \, e^{ikx} + \textrm{h.c.} \right), \\
&\theta(x,\,t) = \theta(t) + \delta\theta(x,\,t) = \theta(t) +  \left(\int \frac{d^3 k}{(2\pi)^3} a_k^\theta \, u_k^\theta(t) \, e^{ikx} + \textrm{h.c.} \right),
\end{align}
with $a^i_k$, $a_k^{i\dagger}$ being the annihilation and creation operators of field $i$
\begin{equation}
\left[a_k^i,\, a_{k'}^{j\dagger}  \right] = (2\pi)^3 \delta_{ij}\delta^{(3)}(k - k^{'}),
\end{equation}
and the initial condition for the mode function $u_k^i$ at $t \to - \infty$ is given by
\begin{equation}
u_k^i(t) = \frac{e^{i(k/a)t}}{a\sqrt{2k}}.
\end{equation}
We now treat $(\delta \phi(x,\,t),\, \delta \theta(x,\,t))$  as perturbations. We expand the potential around the background solution 
\begin{equation}
V'(\phi(x,\,t)) = V'(\phi) + V''(\phi) \partial \phi + \frac{1}{2}V'''(\phi) \partial \phi \partial \phi,
\end{equation}
such that after spatial and quantum averaging $\left<\cdots \right>$ we get, e.g. \cite{Co:2020jtv,Fonseca:2019ypl,Dine:2003ax}.
\begin{align}
&\ddot{\phi} + 3 H \dot{\phi} + V'(\phi) + \frac{1}{2}V'''(\phi) \left<\delta \phi^2 \right> =  \phi \dot{\theta}^2 + \phi  \left<\delta \dot{\theta}^2 \right> + 2 \dot{\theta} \left<\delta \phi \delta \dot{\theta} \right> - 2  a^{-2} \phi \left<\nabla \delta \theta \nabla \delta \theta \right>, \label{eq:phi_zero_mode}\\
&\phi\ddot{\theta} + 3 H \phi \dot{\theta}  + 3 H \left<\delta \phi \delta \dot{\theta} \right>=  -2 \dot{\phi}\dot{\theta}  - 2 \left<\delta \dot{\phi}\delta \dot{\theta} \right> - \left< \delta \phi \delta \ddot{\theta} \right> +  2 a^{-2}\left<\nabla \delta \phi \nabla \delta \theta \right>, \label{eq:theta_zero_mode}
\end{align}
for the zero mode, and 
\begin{align}
&\delta \ddot{\phi} + 3 H \delta \dot{\phi} + V''(\phi) \delta \phi + \frac{k^2}{a^2} \delta \phi= \delta \phi \dot{\theta}^2 + 2\phi \dot{\theta} \delta \dot{\theta}, \label{eq:phi_higher_mode}\\
&\delta \phi\ddot{\theta}  +  \phi \delta\ddot{\theta}  + 3 H \delta \phi \dot{\theta} + 3 H \phi  \delta \dot{\theta} + \frac{k^2}{a^2} \delta \theta = -2 \delta \dot{\phi}\dot{\theta} - 2 \dot{\phi}\delta \dot{\theta}, \label{eq:theta_higher_mode}
\end{align}
for the higher modes, where
\begin{equation}
\left<\delta \phi^2 \right> = \int \frac{d^3 k}{(2\pi)^3} \left| u_k^\phi \right|^2, \quad \left<\delta \theta^2 \right> = \int \frac{d^3 k}{(2\pi)^3} \left| u_k^\theta \right|^2, \quad \left<\delta \theta \delta \phi \right> = 0. 
\end{equation}
Since $[\phi,\,\theta] = 0$, all the quantum average of cross terms in Eq.~\eqref{eq:phi_zero_mode} and \eqref{eq:theta_zero_mode} vanish and we get
\begin{align}
&\ddot{\phi} + 3 H \dot{\phi} + V'(\phi) + \frac{1}{2}V'''(\phi) \left<\delta \phi^2 \right> =  \phi \dot{\theta}^2 + \phi  \left<\delta \dot{\theta}^2 \right>  - 2  \phi \left<\frac{k^2}{a^2} \delta \theta^2 \right>, \label{eq:phi_zero_mode_2}\\
&\phi\ddot{\theta} + 3 H \phi \dot{\theta}  =  -2 \dot{\phi}\dot{\theta}, \label{eq:theta_zero_mode_2}
\end{align}
We conclude that the Noether charge 
\begin{equation}
n_\theta = \phi^2 \dot{\theta},
\end{equation}
is conserved during parametric resonance.
Note that from using Eq.~\eqref{eq:theta_zero_mode_2},  Eq.~\eqref{eq:theta_higher_mode} simplifies to \cite{Dine:2003ax}
\begin{equation}
 \delta\ddot{\theta}  + 3 H \delta \dot{\theta}  + \frac{k^2}{a^2} \delta \theta = -\frac{2 \delta \dot{\phi}\dot{\theta}}{\phi} - \frac{2 \dot{\phi}\delta \dot{\theta}}{\phi} + \frac{2 \dot{\phi} \dot{\theta}}{\phi^2}\delta \phi
\end{equation}
In order to address the question of whether parametric resonance is a successful mechanism for damping the radial mode $\dot{\phi} \to 0$ or not - an important question for scenario I in Sec.~\ref{sec:scenario_I_non_thermal_damping} - we need to solve the system of Eqs.~\eqref{eq:phi_higher_mode}, \eqref{eq:theta_higher_mode}, \eqref{eq:phi_zero_mode_2}  and \eqref{eq:theta_zero_mode_2}. We leave this question for further study.

\section{Detailed solution of field evolution in the rotating complex scalar field model}
\label{app:field_evolution_model_B}
\subsection{The angular kick}
\label{app:kick_angular_direction}

\begin{figure}[h!]
\centering
\raisebox{0cm}{\makebox{\includegraphics[width=0.425\textwidth, scale=1]{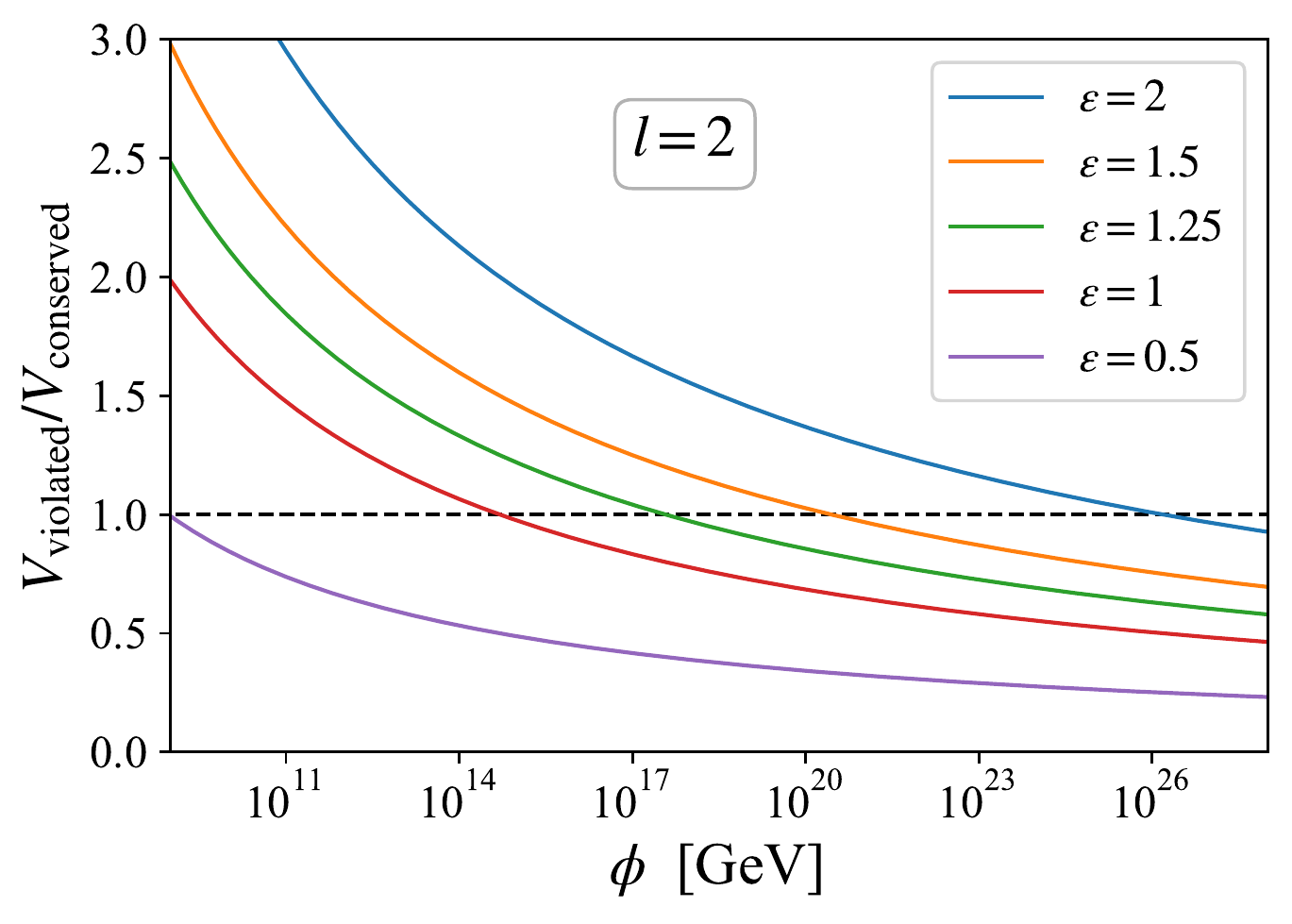}}}
\qquad
\raisebox{0cm}{\makebox{\includegraphics[width=0.45\textwidth, scale=1]{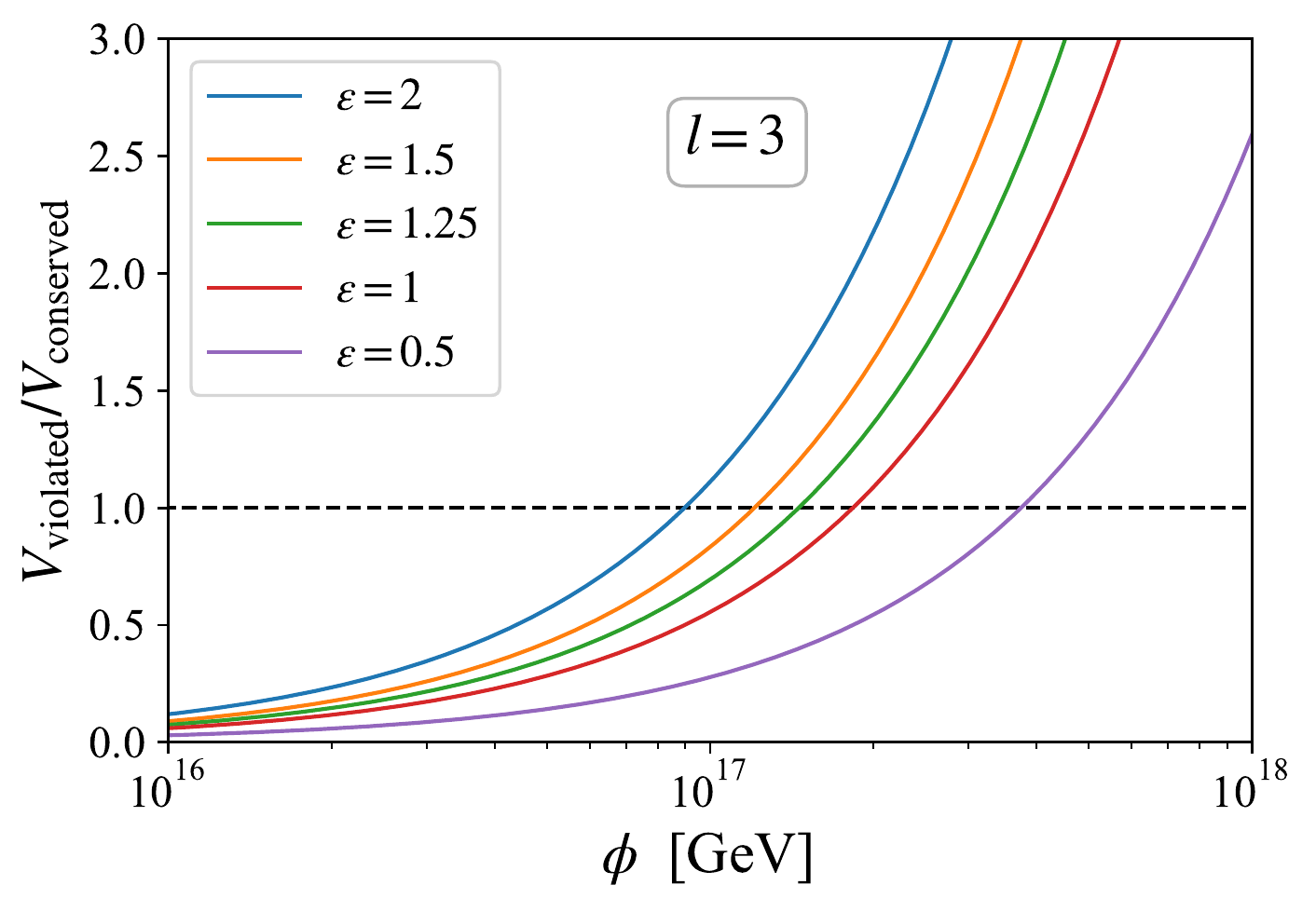}}}\\[0.5em]
\raisebox{0cm}{\makebox{\includegraphics[width=0.45\textwidth, scale=1]{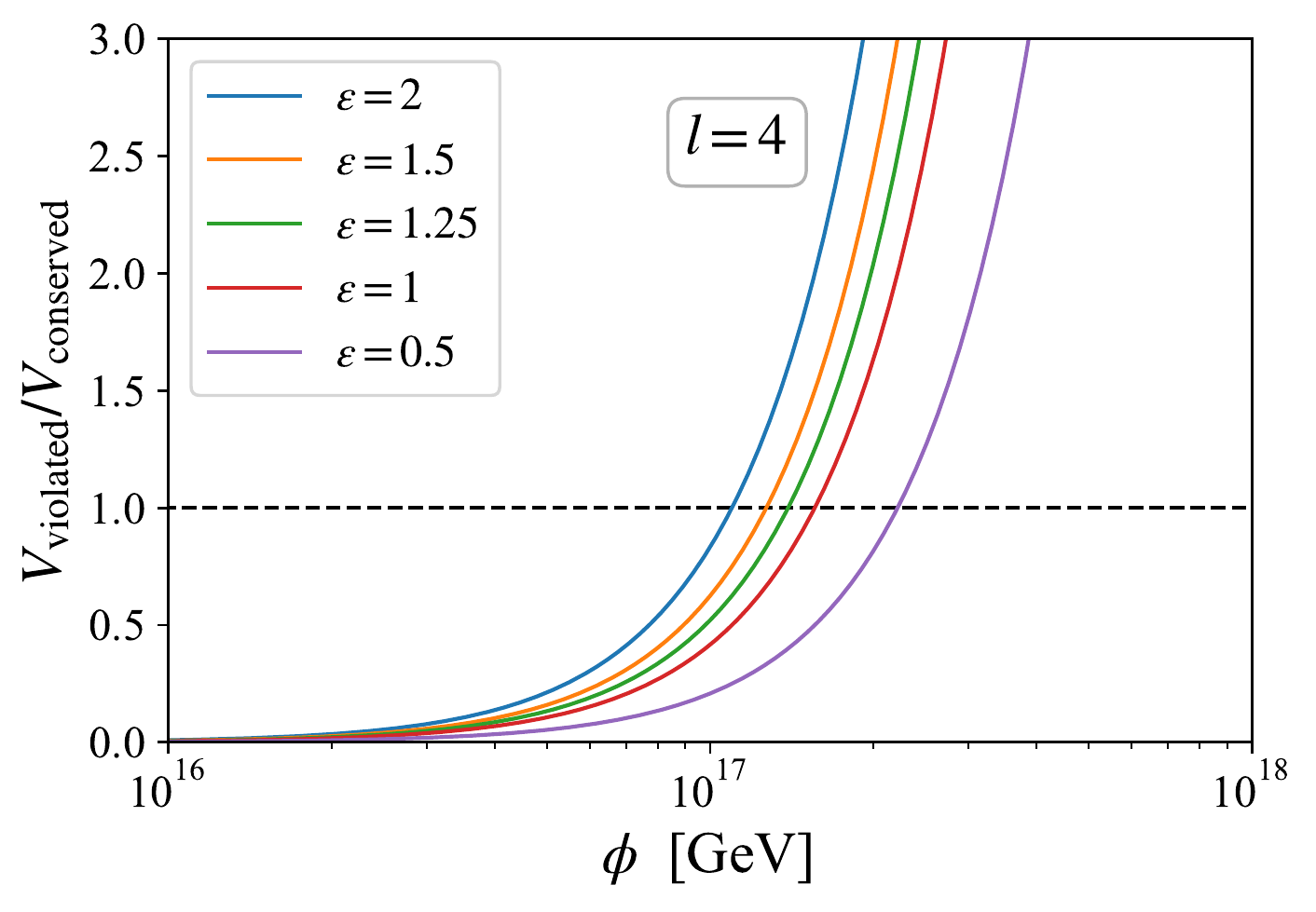}}}
\qquad
\raisebox{0cm}{\makebox{\includegraphics[width=0.45\textwidth, scale=1]{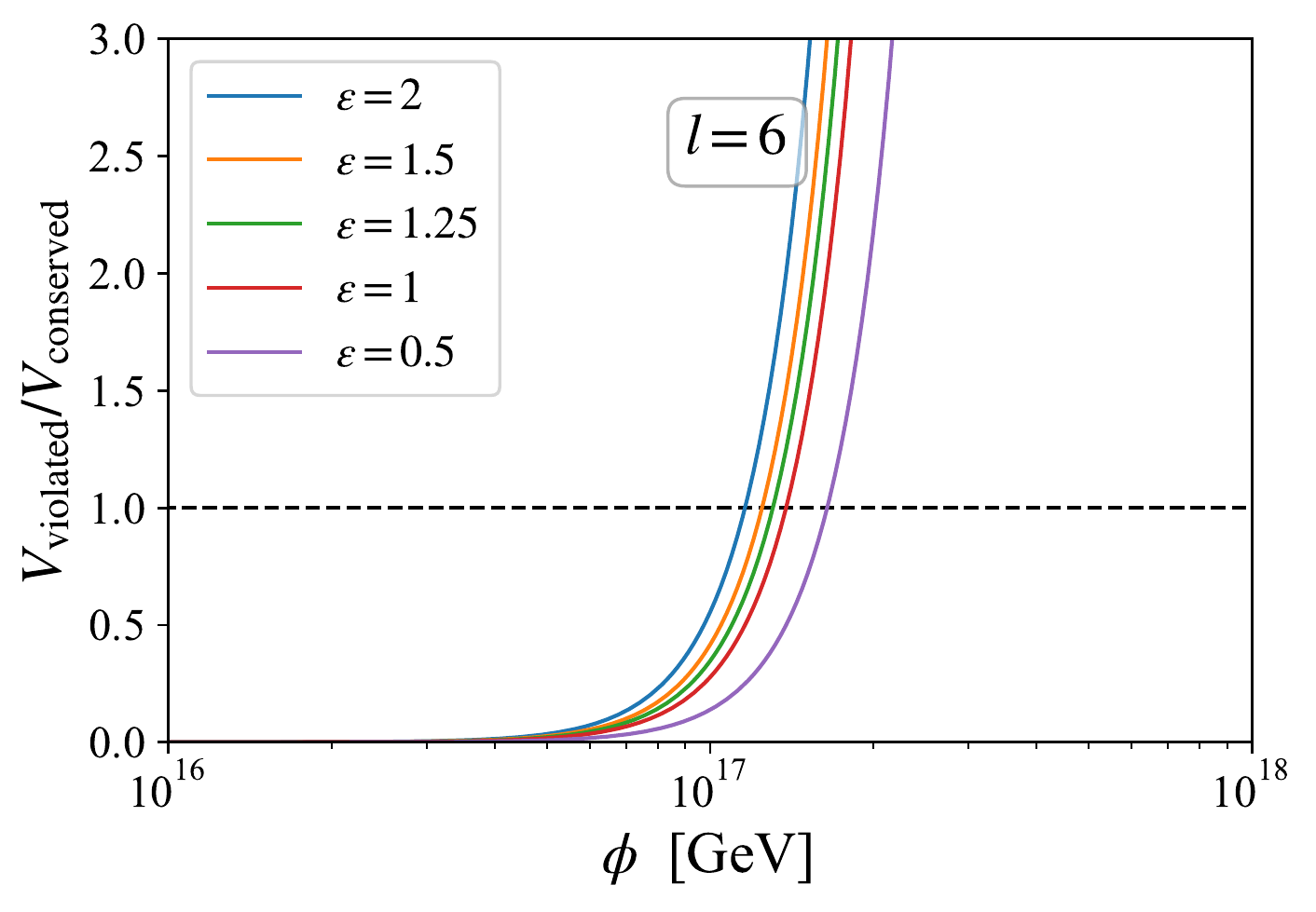}}}\\[0.5em]
\caption{\textit{ \small 
For each $l$, we show the ratio of the $U(1)$-violating potential to the $U(1)$-conserving one, assuming nearly-quadratic potential, along the $\theta=0$ direction which has the maximum $V_{\cancel{U(1)}}$.
For $l>2$, the conserving potential always dominates at small field values $\phi \ll M$, and the $U(1)$ charge, defined in Eq.~\eqref{PQ_charge_conservation}, is conserved in the subsequent evolution. 
}}
\label{ratio_potential}
\end{figure}

The angular EOM, Eq.~\eqref{angular_break}, can be rewritten as a Boltzmann equation for the $U(1)$ charge $n_\theta$
\begin{framed}
\vspace{-1.5em}
\begin{align}
\dot{n_\theta} + 3 H n_\theta ~ = ~ - \frac{\partial V_{\cancel{U(1)}}}{\partial \theta},\qquad \text{with} \quad n_\theta \equiv \phi^2 \dot{\theta}.
\label{charge_conservation_full_app}
\end{align}
\vspace{-1.5em}
\end{framed}\noindent
 which is equivalent to 
\begin{align}
\frac{d}{dt} \left(a^3 n_\theta\right) ~ = ~ - a^3 \frac{\partial V_{\cancel{U(1)}}}{\partial \theta} ~ ~ ~ \Rightarrow ~ ~ ~ \frac{d}{da} \left(a^3 n_\theta\right) ~ = ~ - \frac{a^2}{H^2} \frac{\partial V_{\cancel{U(1)}}}{\partial \theta}.
\label{charge_conservation_full_simplified}
\end{align}
This equation suggests that the $U(1)$ charge production rate is proportional to the potential gradient in the angular direction.
Numerical simulations plotted in Fig.~\ref{fig:theta_dot_vary_epsilon_l} show that the charge generation can possibly start even before that the field starts to roll in the radial direction at $3H \simeq \mreff$. 
Once the field value $\phi$ drops substantially, the explicit breaking term becomes negligible, see Fig.~\ref{ratio_potential}, and the $U(1)$ charge $n_\theta a^3$ becomes conserved.

We can determine the angular kick by integrating Eq.~\eqref{charge_conservation_full_simplified} 
\begin{align}
a^3 n_\theta (a) - a_i^3 n_\theta(a_i) ~ = ~ - \int_{a_i}^a d\tilde{a} ~ \frac{\tilde{a}^2}{H} \frac{\partial V_{\cancel{U(1)}}}{\partial \theta} ~ = ~ \int_{a_i}^a d\tilde{a} ~ \frac{\tilde{a}^2}{H}2 l \Lambda_b^4 \left(\frac{\phi}{M}\right)^l \sin(l \theta),
\label{charge_generation}
\end{align}
where $a_i$ is the scale factor of the universe in the far past,  where we plugged the explicit breaking potential in Eq.~\eqref{eq:V_break_U1_app}.
The production rate of the $U(1)$ charge $n_\theta$ behaves differently accordingly to whether the radial mode $\phi$ has started rolling $3H < \mreff$, or not.

\begin{itemize}
\item {\textbf{Before oscillation:}} the field initially stands at $\phi_\textrm{ini}$, see Sec.~\ref{sec:intitial_vev}.
Assuming the universe is dominated by the background energy density $\rho \propto a^{-q}$, the corresponding charge generation is 
\begin{align}
a^3 n_\theta (a) - a_i^3 n_\theta(a_i) ~ &= ~ 2 l \Lambda_b^4 \left(\frac{\phi_\textrm{ini}}{M}\right)^l \int_{a_i}^a d\tilde{a} ~ \frac{\tilde{a}^2}{H} \sin(l \theta),\\
&\simeq ~ 2 l \Lambda_b^4 \left(\frac{\phi_\textrm{ini}}{M}\right)^l \frac{\sin(l \theta_\textrm{ini})}{H_\textrm{ini} a_{\textrm{ini}}^{p/2}} \int_{a_i}^a d\tilde{a} ~ \tilde{a}^{2+q/2},\\
a^3 n_\theta (a) - a_i^3 n_\theta(a_i) ~ &\simeq ~ 2 l \Lambda_b^4 \left(\frac{\phi_\textrm{ini}}{M}\right)^l  \frac{\sin(l \theta_\textrm{ini})}{H_\textrm{ini} a_{\textrm{ini}}^{q/2}} \left(\frac{2}{6+q}\right) \left(a^{6+q}-a_i^{6+q}\right)^{1/2} ,
\end{align}
where we approximate $\sin(l \theta) \sim \mathcal{O}(1) \sim \sin(l \theta_\textrm{ini})$ and take it out of the integration.
The late-time contribution dominates the charge generation such that we have the $U(1)$ charge evolution
\begin{align}
n (a) ~ \simeq ~ \left(\frac{4l}{6+q}\right) \Lambda_b^4 \left(\frac{\phi_\textrm{ini}}{M}\right)^l  \frac{\sin(l \theta_\textrm{ini})}{H},
\end{align}
and we see that the field receives a kick even before the time of oscillation with $\dot{\theta} \propto a^{q/2}$.
The $U(1)$ charge is maximally generated at the onset of oscillation 
\begin{align}
n_\theta (a_\textrm{osc}) ~ \simeq ~ \left(\frac{4l}{6+q}\right) \Lambda_b^4 \left(\frac{\phi_\textrm{ini}}{M}\right)^l  \frac{\sin(l \theta_\textrm{ini})}{H_\textrm{osc}} ~ \simeq ~ \left(\frac{12l}{6+q}\right) \Lambda_b^4 \left(\frac{\phi_\textrm{ini}}{M}\right)^l  \frac{\sin(l \theta_\textrm{ini})}{\mreff},
\end{align}
and the corresponding angular velocity reads
\begin{align}
\dot{\theta}_\textrm{osc} ~ \simeq ~ \left(\frac{12l}{6+q}\right) \Lambda_b^4 \left(\frac{\phi_\textrm{ini}}{M}\right)^l  \frac{\sin(l \theta_\textrm{ini})}{\mreff \phi_\textrm{ini}^2}, \label{eq:theta_ini_app}
\end{align}

\item {\textbf{After the oscillation:}} gradients of the potential in angular and radial directions kick the field into an elliptic orbit whose size redshifts over time.
In App.~\ref{app:virial_th}, we determine the dynamics of the radial mode as a function of the shape of the $U(1)$-symmetric potential
\begin{align}
V(\phi) ~ \propto ~ \phi^p \qquad \implies \qquad \left<\phi\right> ~ \propto ~ a^{-6/(2+p)}.
\end{align}
Assuming that the Hubble factor evolves as $H \propto a^{-q/2}$, then the $U(1)$ charge generated  after oscillation $3H < \mreff$, in Eq.~\eqref{charge_generation}, reads
\begin{align}
a^3 n_\theta (a) - a_\osc^3 n_\theta(a_\osc) ~ &\sim ~ \int_{a_\osc}^a d\tilde{a} ~ \tilde{a}^{\frac{(6 + q)(2+p) - 12l}{2(2+p)} -1}.
\end{align}
Depending on the values of $q$, $p$, and $l$, we obtain three regimes 
\begin{align}
a^3 n_\theta (a) - a_\osc^3 n_\theta(a_\osc) ~ \sim ~ \begin{cases}
a^{\frac{(6 + q)(2+p) - 12l}{2(2+p)}} ~ ~ &\textrm{for} ~ ~ l < \frac{(6 + q)(2+p)}{12},\\[0.5em]
\log a~ ~ &\textrm{for} ~ ~ l = \frac{(6 + q)(2+p)}{12},\\[0.5em]
a^{-\frac{12l - (6 + q)(2+p)}{2(2+p)}} ~ ~ &\textrm{for} ~ ~ l > \frac{(6 + q)(2+p)}{12}.
\end{cases}
\end{align}
In the first and second cases, the $U(1)$ charge continues to increase after the oscillation, while in the third case at large $l$, the $U(1)$ charge stops being efficiently produced after a few Hubble times and the $U(1)$-symmetry is restored.
Assuming a radiation-dominated universe ($q=4$), the estimated $U(1)$-symmetry after the oscillation demands that $l$ is greater than $10/3$ and $5$ for quadratic  ($p=2$) and quartic  ($p=4$) potentials, respectively.
Hence, in this work we consider cases $l \geq 4$ in order to neglect explicit breaking terms at later time. 
\end{itemize}

\FloatBarrier

\subsection{After the kick}
\label{app:field_traj_kick}
The $U(1)$-conserving and the  $U(1)$-breaking potentials excite the field in the radial and angular directions, respectively. As argued in App.~\ref{app:kick_angular_direction}, after a few Hubble times of field evolution, for $l \geq 4$ we can neglect effects coming from the $U(1)$-breaking terms.

\paragraph{Exact solution.}
\label{app:exact_sol}
We now derive an exact analytical solution of the complex scalar equation of motion in a quadratic potential without Hubble friction, e.g. \cite{Arbey:2001jj, brizard2009primer}.
For $\phi \gg f_a$, the nearly-quadratic potential in Eq.~\eqref{study_potential1} is approximated to $V = m^2 \Phi^\dagger \Phi/2=m^2 \phi^2/2$, where $m$ is a constant or a slowly-changing variable.
The $U(1)$-charge conservation in Eq.~\eqref{PQ_charge_conservation}, and the radial EOM in Eq.~\eqref{radial_break}, give
\begin{align}
\ddot{\phi} + 3 H \dot{\phi} + \frac{\partial V}{\partial \phi} - \frac{Y^2}{\phi^3 a^6} ~ &= ~ 0,
\label{eom_exact_proof}
\end{align}
where $Y \equiv a^3 \phi^2 \dot{\theta} = $ is the conserved comoving $U(1)$ charge. 
We use a set of dimensionless parameters
\begin{align}
\tau ~ \equiv ~ m t, ~ ~ h~ \equiv ~ \frac{H}{m}, ~ ~ u ~ \equiv ~ \sqrt{\frac{m}{Y}} a^{3/2} \phi,
\end{align}
which simplifies the charge conservation equation to $\dot{\theta} u^2 ~ = ~ m$ and the EOM to
\begin{align}
0 ~ = ~ u'' + \left[1 - u^{-4} - \frac{3}{2}\left(h' + \frac{3}{2}h^2\right)\right]u ~ \simeq ~  u'' + u - \frac{1}{u^{3}},
\label{eq:eqdiff_eom_quadratic}
\end{align}
where $\cdots'$ denotes the derivative w.r.t. $\tau$, and where we considered the oscillation time scale to be much faster than a Hubble time $h \to 0$, which becomes true after oscillation has started $3H \simeq \mreff$.
The differential equation in Eq.~\eqref{eq:eqdiff_eom_quadratic} can be understood as the equation of motion of a scalar $u$ evolving in a potential  
\begin{align}
\mathcal{V}(u) ~ = ~ \frac{u^2 + u^{-2}}{2},
\end{align}
whose minimum is at $u = 1$,  or equivalently $\dot{\theta} = m$. Its corresponding constant-of-motion $E_u$ satisfies
\begin{align}
\frac{d E_u}{d \tau} ~ \equiv ~ \frac{d}{d \tau} \left( \frac{u'^2}{2} + \frac{u^2 + u^{-2}}{2}\right)  ~ = ~ 0,
\label{exact_sol_Eu}
\end{align}
with $E_u \geq 1$.
The parameter $E_u$ can be related to the energy density of the scalar field 
\begin{align}
\rho_\Phi ~ = ~ \frac{\dot{\phi}^2}{2} + \frac{\phi^2 \dot{\theta}^2}{2} + V ~ = ~ m Y E_u a^{-3}. \label{eq:rho_Phi_exact}
\end{align}
We just found that the energy density of a complex scalar field evolving in a quadratic potential, redshifts as matter $\rho_\Phi \propto a^{-3}$.  

To find the exact solution to the EOM, we rewrite Eq.~\eqref{exact_sol_Eu} as
\begin{align}
u'^2 ~ = ~ 2 E_u - u^2 - u^{-2} ~ ~ ~ \Rightarrow ~ ~ ~ U'^2 ~ = ~  -4 U^2 +8E_u U -4,
\end{align}
where $U \equiv u^2$.
Integrating this equation, we deduce
\begin{align}
\int dt ~ = ~  \int \frac{d U}{\sqrt{-4 U^2 +8E_u U -4}} ~ ~ ~ &\Rightarrow ~ ~ ~ 2 (\tau + \psi) ~ = ~ - \tan^{-1} \left(\frac{E_u - U}{\sqrt{- U^2 +2E_u U -1}}\right),\nonumber \\
&\Rightarrow ~ ~ u(\tau) ~ = ~ \sqrt{(E_u^2 - 1)^{1/2} \sin(2 \tau + \psi) + E_u}.\label{eq:exact_eom_quadratic}
\end{align}
where $\psi$ is a phase which depends on the initial conditions.
In $u$-space, the orbit is a fixed-sized ellipse, so the orbit in field space $\phi$ is an ellipse whose size scale as $a^{-3/2}$, i.e. an ellipse that spirals towards the origin as illustrated in Fig.~\ref{field_evolution} or by  \href{https://www.youtube.com/watch?v=RdCAgcvfFy0}{our animation.}. The parameter $E_u$ is related to the orbit eccentricity and to the previously defined parameter $0 < \epsilon \leq 2$ in Eq.~\eqref{eq:epsilon_def}
\begin{align}
m^2 \phi^2 ~ = ~ \rho_\Phi ~ = ~ m n_\theta E_u ~ ~ ~ ~ \Rightarrow ~ ~ ~ ~ 
\frac{1}{E_u} ~ = ~ \frac{n_\theta}{m \phi^2} ~ = ~ \frac{\epsilon}{2} ~ ~ ~ ~ \Rightarrow ~ ~ ~ ~ 
\epsilon ~ = ~ 2/E_u.
\end{align}
For $E_u = 1$ or $\epsilon = 2$, the orbit is essentially circular, while for large $E_u$ or $\epsilon \rightarrow 0$ the orbit has a large eccentricity and starts to resemble the trajectory of a  real scalar field.

We can calculate the average angular speed
\begin{align}
\dot{\theta} ~ = ~ \frac{m}{u^2}  ~ = ~ \frac{m}{(E_u^2 - 1)^{1/2} \sin(2 \tau + \psi) + E_u},
\end{align}
by averaging over a time $T = n\pi$, $n \in \mathbb{N}^+$, which is larger than the oscillation period but less than the Hubble time scale
\begin{align}
\left< \dot{\theta} \right> ~ = ~ \frac{1}{n\pi} \int^{n\pi}_0 d\tau \frac{m}{(E_u^2 - 1)^{1/2} \sin(2 \tau + \psi) + E_u} ~ = ~ m,
\end{align}
where we have integrated an elliptic integral\footnote{We can freely choose the phase $\psi$ such that the sine becomes a cosine and the elliptic integral reads
\begin{align}
\int^{2\pi}_0 \frac{dx}{a+b\cos x} ~ = ~ \frac{2 \pi}{\sqrt{a^2 - b^2}}.\nonumber
\end{align}}. 
We find that the averaged angular velocities becomes independent of the initial value $\theta_{\rm ini}$ in Eq.~\eqref{eq:theta_ini_app} and converges towards an attractor solution. This is confirmed by the numerical integration of the equations of motion in Fig.~\ref{fig:theta_dot_vary_epsilon_l}.

\paragraph{Exact solution (bis) : quartic potential.}
\label{exact_elliptic_quartic}
Even if we don't use it in our work, it may be useful for further studies (of axion dark matter for example \cite{DESYfriendpaper}) to give the analog of Eq.~\eqref{eq:exact_eom_quadratic} for a quartic potential.
When adding a quartic term to the potential  $V ~ = ~ \frac{1}{2}m^2 \phi^2 + \frac{1}{4}\lambda \phi^4$, the Eq.~\eqref{eq:eqdiff_eom_quadratic} becomes
\begin{align}
\Rightarrow ~ ~ ~ ~ u'' + u + \frac{4 \alpha}{a^3}u^3 - \frac{1}{u^{3}} ~ = ~ 0, \qquad \text{with}\quad \alpha \equiv \lambda Y/(4m^3).
\label{eq_quad2}
\end{align}
We now consider $\phi \gg  m/\sqrt{\lambda}$, such that the linear term $u$ corresponding to $ \frac{1}{2}m^2 \phi^2$ can be dropped and Eq. \eqref{eq_quad2} becomes
\begin{align}
 \nu'' + 4 \alpha \nu^3 - \frac{1}{\nu^3} ~ = ~ 0 \qquad \implies \qquad \frac{1}{2} \nu'^2 ~ = ~  E_\nu - V_\nu ~ = ~ 2 E_
\nu - 2 \alpha \nu^4 - \frac{1}{\nu^2},
\label{eq_quar}
\end{align}
with $\nu \equiv a^{-1/2} u$ and with $'$ denoting the derivative with respect to the conformal time $\eta$ ( $m dt = d \tau = a d \eta$) and where we have dropped Hubble depending terms. The motion can be interpreted as a scalar $\nu$ oscillating in a potential
$
V_\nu  =  \alpha \nu^4 + 1/(2 \nu^2),
$
with the constant of motion 
$
E_\nu ~ = ~ \nu'^2/2 + V_\nu.
$
Introducing the new variables $V \equiv - \nu^2/\sqrt{2}$ and $s \equiv \eta \sqrt{\alpha}$, we deduce
\begin{align}
\left( \frac{d V}{ds} \right)^2 ~ = ~  4 V^3 -\frac{4 E_\nu}{\alpha} V - \frac{2}{\alpha}.
\label{sim_de_weierstrass}
\end{align}
It is known that the \emph{Weierstrass elliptic function} $\wp(s; ~g_2, ~g_3)$ solves the differential equation of the form \cite{Gradshteyn:1702455}
\begin{align}
\left( \frac{d \wp(s)}{ds} \right)^2 ~ = ~  4 \wp^3(s) - g_2 \wp(s) - g_3.
\end{align}
Therefore, Eq. \eqref{sim_de_weierstrass} has a solution as an elliptic orbit described by $\wp(s; ~4E_\nu/\alpha,~ 2/\alpha)$.
Applying one of the properties of $\wp$, namely
$
\wp(u; g_2, g_3) ~ = ~ k^2 ~\wp (k u;~ g_2/k^4, ~g_3/k^6),
$
the exact solution to Eq. \eqref{eq_quar} reads
\begin{align}
\nu^2 ~ &= ~ - \sqrt{2} V ~ = ~ -\frac{\sqrt{2}}{\alpha} \wp(\eta; ~4\alpha E_\nu,~ 2 \alpha^2).
\end{align}

\paragraph{Virial theorem.}
\label{app:virial_th}
The Virial theorem in classical mechanics is a well-known tool, e.g. \cite{goldstein1980classical}, for studying the averaged behavior of a stable system. We now use it to study the dynamics of a complex scalar field in a central potential during the early universe.
The EOM in Eq.~\eqref{radial_break} and  \eqref{angular_break} can be written in terms of kinetic $K$ and potential $V$ energy densities as
\begin{align}
\frac{d}{d t} (K + V) ~ &= ~ \dot{\phi} \left(\ddot{\phi}  + \frac{\partial V}{\partial \phi} + \phi \dot{\theta}^2 + \frac{\phi^2}{\dot{\phi}}\dot{\theta}\ddot{\theta}\right) + \dot{T} \frac{\partial V}{\partial T},\\
&= ~ - (3H + \Gamma) \dot{\phi}^2 -3H \phi^2 \dot{\theta}^2 + \dot{T} \frac{\partial V}{\partial T},\\
&= ~ - (6H + 2\Gamma) K_\phi -6 H K_\theta + \dot{T} \frac{\partial V}{\partial T},
\end{align}
where $K_\phi \equiv \dot{\phi}^2/2$ and $K_\theta \equiv \phi^2 \dot{\theta}^2/2$ are the kinetic energy in radial and angular modes, respectively.
After $3H \simeq \mreff$, the field oscillates fast compared to the expansion and damping rate. The average over many field cycles but over a time shorter than the $H^{-1}$ reads
\begin{align}
\frac{d}{d t} \left< K + V \right> ~ &= ~ - (6H + 2\Gamma) \left< K_\phi \right> -6 H \left<K_\theta\right>+ \dot{T} \frac{\partial V}{\partial T}.
\label{average_EOM}
\end{align}
We introduce the \emph{Virial} parameter $G$, e.g. \cite{goldstein1980classical}
\begin{align}
G ~ \equiv ~ \sum_{i=\phi,\theta} p_i \cdot r_i ~ = ~ \dot{\phi}\phi + \phi^2 \dot{\theta} \theta,
\end{align}
where $r_\phi = \phi$ and $r_\theta = \theta$, and $p_i \equiv \partial \mathcal{L} / \partial r_i$.
By taking its time derivative, we deduce
\begin{align}
\frac{dG}{dt} ~ &= ~ \dot{\phi}^2 + \phi^2 \dot{\theta}^2 + \phi \ddot{\phi} + \phi^2 \theta \left(\ddot{\theta} + 2 \frac{\dot{\phi}}{\phi} \dot{\theta}\right),\\
&= ~  \dot{\phi}^2 + \phi^2 \dot{\theta}^2  - \phi \frac{\partial V}{\partial \phi} - (3 H + \Gamma) \phi \dot{\phi} -3H  \phi^2 \dot{\theta} \theta,\\
 &\simeq ~2 K - \phi \frac{\partial V}{\partial \phi},
\end{align}
where we used Eq.~\eqref{radial_break} and  \eqref{angular_break} in the second step, and where we neglected the damping terms in  the last step.
The Virial theorem states that whenever the system is stable, the averaged kinetic and potential energies are related 
\begin{align}
 \left<\frac{d G}{dt}\right> ~ = ~ \lim_{\Delta T \rightarrow \infty} \frac{G(T+\Delta T) - G(T)}{\Delta T} ~ = 0, \qquad \implies \qquad \left<2 K\right> ~  = ~  \left< \phi \frac{\partial V}{\partial \phi} \right>. \label{scalar_virial} 
\end{align}
which resembles the Virial theorem for a real scalar \cite{Mukaida:2012qn, Mukaida:2012bz}.
For convenience, we consider the field behavior before the damping term becomes effective and deduces  the average EOM from Eq.~\eqref{average_EOM} and \eqref{scalar_virial}
\begin{align}
\label{virial_eom}
\frac{d}{d t} \left< K + V \right> ~= ~ - 6H \left< K \right>+ \dot{T} \frac{\partial V}{\partial T} \qquad \implies \qquad
\frac{d}{d t} \left< \phi \frac{\partial V}{\partial \phi} + 2V \right> ~ = ~ - 6H \left< \phi \frac{\partial V}{\partial \phi} \right>+ \dot{T} \frac{\partial V}{\partial T}.
\end{align}
\paragraph{Monomial potential.}
From plugging the potential $V ~\propto ~ \phi^n $ into Eq.~\eqref{virial_eom}, we obtain
\begin{align}
(n+2) \frac{d}{d t} \left< V \right> ~ = ~ - 6H n \left< V \right>\quad \implies \quad \frac{d}{d t} \left< \rho_\Phi  \right> ~ = ~ - \frac{6n}{2+n}  H \left< \rho \right>\quad \implies \quad \frac{d\ln \left< \rho_\Phi \right>}{d \ln a}  ~ = ~ - \frac{6n}{2+n},
\label{eq:dede}
\end{align}
with $\left< \rho_\Phi \right>$ is the average total energy density of the field
\begin{equation}
\left< \rho_\Phi \right> = \left<K + V\right> = (2+n) \left< V \right>/2.
\end{equation}
We deduce the redshift laws of $\left< \rho_\Phi \right>$ and $\left<\phi \right>$
\begin{align}
 \left< \rho_\Phi  \right> ~ &\propto ~ a^{-\frac{6n}{2+n}}, \qquad \text{and} \quad \left< \phi \right> ~ \propto ~ a^{-\frac{6}{2+n}}.
 \label{eq:final_solution_virial}
\end{align}
For the quadratic and quartic potentials, the complex scalar field behaves like matter ($\rho_\Phi  \propto a^{-3}$, $\phi \propto a^{-3/2}$) and radiation ($\rho_\Phi  \propto a^{-4}$, $\phi \propto a^{-1}$), respectively. The scaling $\rho_\Phi  \propto a^{-3}$ is confirmed  by the exact solution in Eq.~\eqref{eq:rho_Phi_exact}.  We show that, cf.  Eq.~\eqref{eq:thetadot_after_damping}, that  Eq.~\eqref{eq:dede} holds without any average after radial damping $\dot{\phi} \to 0$.
\paragraph{Thermal mass.}
\label{sec:viriel_thermal_mass}
From plugging the potential $V = \lambda T^2\phi^2/2 $ into Eq.~\eqref{virial_eom}, we obtain
\begin{align}
\frac{d}{d t} \left< \phi^2 \right> ~ = ~ - 3H  \left<  \phi^2 \right> - \frac{\dot{T}}{T}\left<  \phi^2 \right>\quad \implies \quad \phi^2 \propto a^{-3}T^{-1} \qquad \textrm{and} \qquad V \propto a^{-3}T.
\label{eq:thermal_mass_redshift_law}
\end{align}
We conclude that in a radiation-dominated universe, a scalar field dominated by its thermal mass redshifts like radiation.

\subsection{The radial damping}
\label{app:effect_epsilon_evolution}
\paragraph{Amount of rotation.}
The parameter $\epsilon$ expresses the amount of rotation generated by the explicit breaking at the time of oscillation $t_{\rm osc}$, cf. Eq.~\eqref{eq:epsilon_def}
\begin{align}
\epsilon ~ = ~ \frac{\phi^2 \dot{\theta}/2}{V(\phi)/m_{r,\mathrm{eff}}} ,
\end{align}
From Eq.~\eqref{eq:redshift_laws_rho_phi}, we deduce that $\epsilon$ becomes a conserved quantity through Hubble expansion $d\epsilon/da = 0$, as soon as $V(\phi)$ become dominated by its quadratic term.
It can be rewritten in terms of the ratio between energy densities as 
\begin{align}
\epsilon ~ = ~ \left(\frac{ \rho_\theta}{V}\right)  \left(\frac{m_{r,\mathrm{eff}}}{\dot{\theta}}\right), \qquad \textrm{with}\quad \rho_\theta \equiv \dot{\theta^2} \phi^2 /2.
\end{align}
Since $\left<\dot{\theta} \right> = \mreff$, see Fig.~\ref{fig:theta_dot_vary_epsilon_l}, we deduce the rotational energy density to be an $\epsilon$ fraction of the potential energy density 
\begin{align}
\rho_\theta ~ = ~ \epsilon V(\phi).
\label{epsilon_energy}
\end{align}
Note that $\epsilon  \leq 1$ from preventing the field to spin upward.

\begin{equation}
\begin{cases}
\epsilon = \phi^2 \dot{\theta}/(2\rho_{r}(\phi)/\mreff), \vspace{0.5cm}\\
\phi^2 \dot{\theta}= \textrm{constant}, \vspace{0.5cm}\\
\dot{\theta}_{\rm after} = \mreff,
\end{cases}
\qquad \implies \qquad \rho_{\theta}\Big|_{\rm after} =  \epsilon \rho_{r}(\phi)
\end{equation}

\begin{figure}[h!]
\centering
\raisebox{0cm}{\makebox{\includegraphics[width=0.475\textwidth, scale=1]{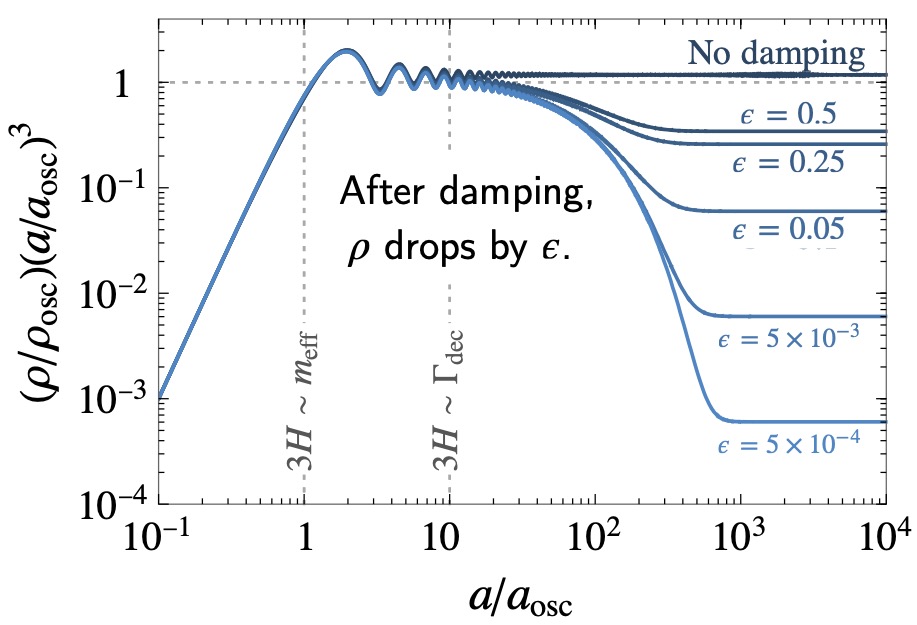}}}
\quad
{\makebox{\includegraphics[width=0.475\textwidth, scale=1]{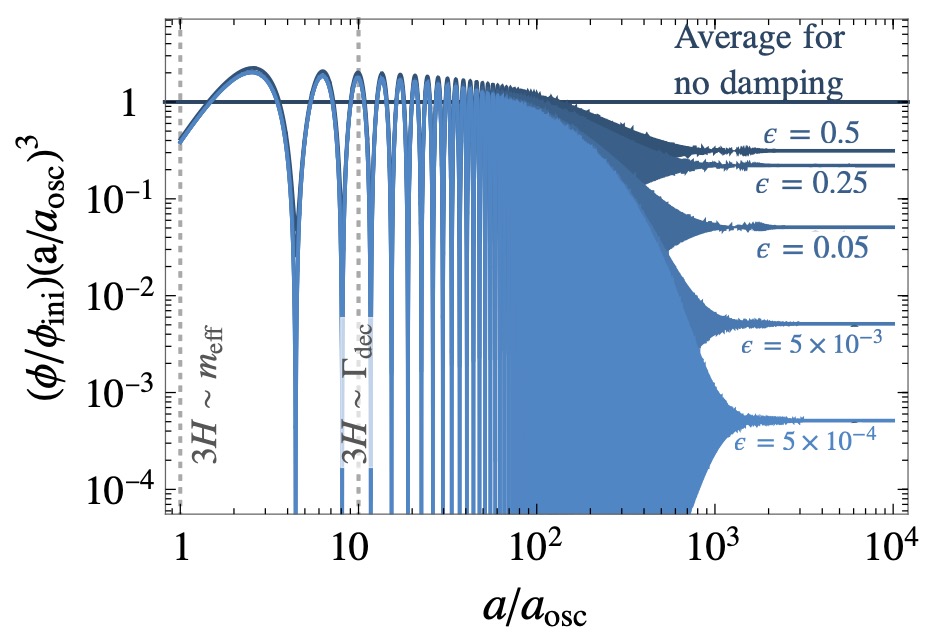}}}
\caption{\textit{ \small Numerical integration of the equations of motion in Eqs.~\eqref{radial_break}, \eqref{angular_break} and \eqref{eq:friedmann_eq}. The energy density (\textbf{left}) and the radial value (\textbf{right}) of the complex scalar field drops by a factor $\epsilon$ after the radial damping. The larger $\epsilon$, the larger the rotational energy density produced during the initial kick and the smaller the energy drop during the radial damping. The extreme case $\epsilon = 1$ corresponds to a trajectory which is already circular right after the initial angular kick. }}
\label{fig:rho_drop_epsi}
\end{figure}

\paragraph{Drop of energy density during radial damping.}
During the radial damping $\dot{\phi} \to 0$, see Sec.~\ref{sec:themalization}, the elliptic orbits becomes a circular one. During this process, the scalar field kinetic energy is damped but the rotational kinetic energy is preserved. Therefore, after the radial damping the total energy density of the complex scalar field drops by 
\begin{align}
\rho_\Phi^\mathrm{after} ~ = ~ \epsilon \,\rho_\Phi^\mathrm{before} \qquad \implies \qquad \phi^2_\mathrm{after} ~ = ~ \epsilon\,\phi^2_\mathrm{before},
\label{epsilon_energy_transition}
\end{align}
where the second equation assumes the quadratic potential. 
$\phi_\mathrm{before}$ is the field value just before thermalization, and it can be computed from Eq.~\eqref{eq:final_solution_virial}
\begin{align}
\frac{\rho_\Phi^\mathrm{osc}}{\rho_\Phi^\mathrm{before}} ~ = ~ \left(\frac{a_\textrm{damp}}{a_\textrm{osc}}\right)^3 ~ = ~ \left( \frac{\phi_\textrm{osc}}{\phi_\textrm{before}} \right)^2 \qquad \implies  \qquad \phi_\textrm{before} ~ = ~ \frac{(\rho_\Phi^\mathrm{before})^{1/2}}{\mreff(\phi_\textrm{osc})},
\end{align}
where we used that the energy density just after oscillation can be written as $\rho_\Phi^\mathrm{osc} = \phi_\textrm{osc}^2 m_\textrm{eff}(\phi_{\rm osc})^2$.
In Fig.~\ref{fig:rho_drop_epsi}, we show that the result in Eq.\eqref{epsilon_energy_transition} is confirmed by numerical integration of the equations of motion.

\FloatBarrier
\FloatBarrier

\paragraph{Impact on kination duration.}
We now study the impact of the energy drop in Eq.~\eqref{epsilon_energy_transition} on the duration of the kination era.
As shown in Fig.~\ref{diagram_epsilon_energy}, we must consider two possible scenarios, depending on whether the radial damping occurs before or after the scalar domination. For the sake of simplicity we assume an instantaneous drop in energy density at the time of radial damping.
\begin{itemize}
\item \textbf{Damping before domination.} The energy transferred to the thermal bath is negligible. However, the $\epsilon$ factor in the energy density in Eq.~\eqref{epsilon_energy_transition} delays the time when the scalar field dominates the energy density of the universe
\begin{align}
\rho'_\mathrm{dom} ~ = ~ \epsilon^4 \left.\frac{\rho_\Phi^4}{\rho_\mathrm{rad}^3}\right|_\mathrm{at ~ oscillation} ~ = ~ \epsilon^4 \rho_\mathrm{dom},
\label{epsilon_case1_dom}
\end{align}
where $\rho_\mathrm{dom}$ is the domination energy scale without energy drop during damping (case $\epsilon =1$) and where we have used Eq.~\eqref{eq:rho_dom_def_app}.

As a consequence of the energy drop, the kination era ends earlier and is then shorter
\begin{align}
\rho_{\mathrm{KD},f} ~ = ~ \frac{\rho_{\mathrm{KD},i}^2}{\rho'_\mathrm{dom}} ~ = ~ \frac{\rho_{\mathrm{KD},i}^2}{\rho_\mathrm{dom}}   \left(\frac{1}{\epsilon}\right)^4 \qquad \text{and} \qquad \frac{a_{\mathrm{KD},f}}{a_{\mathrm{KD},i}} ~ = ~ \left(\frac{\rho'_\mathrm{dom}}{\rho_{\mathrm{KD},i}}\right)^{1/6} ~ = ~  \left(\frac{\rho_\mathrm{dom}}{\rho_{\mathrm{KD},i}}\right)^{1/6} \epsilon^{2/3}.
\label{epsilon_case1_kd_duration}
\end{align}

\item \textbf{Damping after domination.} A substantial amount of energy is injected in the thermal bath. 
For $\epsilon = \mathcal{O}(0.1)$, the radial damping reduces the scalar energy by a factor $\mathcal{O}(10)$, so that the produced SM radiation dominates the energy density.
Later, the domination of the energy density by the rotating scalar generates a second matter era at
\begin{align}
\rho'_\mathrm{dom} ~ = ~ \epsilon^4 \rho_\mathrm{damp},
\label{epsilon_case2_dom}
\end{align}
The scale of kination ending and its duration are given by Eq.~\eqref{epsilon_case1_kd_duration} after replacing $\rho_\mathrm{dom}$ by $\rho_\mathrm{damp}$
\begin{align}
\rho_{\mathrm{KD},f} ~ =~ \frac{\rho_{\mathrm{KD},i}^2}{\rho_\mathrm{damp}}   \left(\frac{1}{\epsilon}\right)^4\qquad \text{and} \qquad \frac{a_{\mathrm{KD},f}}{a_{\mathrm{KD},i}} ~ =  ~  \left(\frac{\rho_\mathrm{damp}}{\rho_{\mathrm{KD},i}}\right)^{1/6} \epsilon^{2/3}.
\label{epsilon_case2_kd_f}
\end{align}
Note that the presence of the radiation era in between the two matter eras could leave a distinctive imprint in the SGWB.

\end{itemize}

\paragraph{Too small $\epsilon$ kills the kination era.}
In Eqs.~\eqref{epsilon_case1_kd_duration} and \eqref{epsilon_case2_kd_f}, we have seen that the kination duration receives a suppression factor $\epsilon^{2/3}$. This implies the existence of a lower bound on $\epsilon$ below which no kination era is generated
\begin{align}
\mathrm{Requiring} ~ ~ ~ ~ \frac{a_{\mathrm{KD},f}}{a_{\mathrm{KD},i}} \geq 1 ~ ~ ~ ~ \Longrightarrow  ~ ~ ~ ~ \epsilon \geq  \left[\frac{\rho_{\mathrm{KD},i}}{\max(\rho_\mathrm{dom},\rho_\mathrm{damp})}\right],
\label{epsilon_allow_kd}
\end{align}
for both damping before and after the scalar domination. In Sec.~\ref{sec:complex_field_thermal_potential}, we show that the scenario II, in which thermal corrections to the potential are present at the onset of the radial mode oscillation, necessarily predicts a value of $\epsilon $ smaller than Eq.~\eqref{epsilon_allow_kd}.

\subsection{After the radial damping} 
\label{app:traj_after_radial_damp}
In this section, we reproduce the results from \cite{Co:2019wyp} and derive the evolution of the radial $\phi$ and angular $\dot{\theta}$ field values, their energy density $\rho_\Phi$ and equation of state $\omega_\Phi$, in an arbitrary $U(1)$-symmetric potential $V(\phi)$, after that the radial mode has been damped $\dot{\phi}/\phi \ll \sqrt{V^{''}}$.

\paragraph{Radial evolution.}
In the limit $\dot{\phi} \to 0$, the radial EOM in Eq.~\eqref{radial_break} takes a simple form
\begin{align}
\dot{\theta}^2 \phi ~ = ~ \frac{\partial V}{\partial \phi} \qquad \implies \qquad \dot{\theta}^2 ~ = ~ 2 \frac{\partial V}{\partial \phi^2}, \label{eq:thetadot_after_damping}
\end{align}
We inject it in the equation of conservation of $n_\theta = \phi^2 \dot{\theta}$
\begin{align}
\frac{d(a^3 n_\theta)}{d a} ~ &= ~ 3 a^2 \phi^2 \dot{\theta} + 2 a^3 \phi \dot{\theta} \frac{d \phi}{d a}  + a^3 \phi^2 \frac{d \dot{\theta}}{d a},\\
\implies \qquad \qquad 0 ~ &= ~ 3 a^2 \phi^2 \dot{\theta} + 2 a^3 \phi \dot{\theta} \frac{d \phi}{d a}  + a^3 \phi^2 \frac{4\phi}{2\dot{\theta}} \cdot  \frac{d \phi}{d a} \cdot \frac{\partial^2 V}{(\partial \phi^2)^2},\\
\implies \qquad ~ a \frac{d \phi}{d a} ~ &= ~ \frac{-3 \phi \dot{\theta}^2}{2 \dot{\theta}^2 + 2 \phi^2 \frac{\partial^2 V}{(\partial \phi^2)^2}} ~ = ~ \frac{-3 \phi  \frac{\partial V}{\partial \phi^2}}{2  \frac{\partial V}{\partial \phi^2} + \phi^2 \frac{\partial^2 V}{(\partial \phi^2)^2}},
\end{align}
From which we obtain 
\begin{equation}
\frac{d \ln \phi}{d \ln a} ~ = ~ \frac{-3  \frac{\partial V}{\partial \phi^2}}{2  \frac{\partial V}{\partial \phi^2} + \phi^2 \frac{\partial^2 V}{(\partial \phi^2)^2}}. \label{eq:dlogphidloga}
\end{equation}

\paragraph{Angular evolution.}
Starting from Eq.\eqref{eq:thetadot_after_damping}, we can write
\begin{align}
\frac{d \ln \dot{\theta}^2}{d \ln a} ~ = ~ 2 \frac{a}{\dot{\theta}^2}\frac{d \phi^2}{d a} \cdot \frac{\partial^2 V}{(\partial \phi^2)^2} ~=~ 4 \frac{\phi^2}{\dot{\theta}^2} \frac{d \ln \phi}{d \ln a} \cdot \frac{\partial^2 V}{(\partial \phi^2)^2} ~=~ 2 \frac{\phi^2}{\frac{\partial V}{\partial \phi^2}} \frac{d \ln \phi}{d \ln a} \cdot \frac{\partial^2 V}{(\partial \phi^2)^2},\label{eq:dlogthetadloga}
\end{align}
where $\frac{d \ln \phi}{d \ln a}$ is given by Eq.~\eqref{eq:dlogphidloga}.

\paragraph{Energy density evolution.} 
After radial damping, the kinetic energy of the radial mode vanishes such that the energy density of the complex field reads
\begin{align}
\rho_\Phi ~ = ~ \frac{1}{2}\phi^2 \dot{\theta}^2 + V(\phi) ~ = ~ \phi^2 \frac{\partial V}{\partial \phi^2} + V(\phi).
\end{align}
Taking the derivative with respect to $\phi$, we get
\begin{align}
\frac{d \rho_\Phi }{d \phi} ~ = ~ 2 \phi \frac{\partial V}{\partial \phi^2} + 2 \phi^3 \frac{\partial^2 V}{(\partial \phi^2)^2} + \frac{\partial V}{\partial \phi} ~= ~ 2\phi \left(2 \frac{\partial V}{\partial \phi^2} + \phi^2 \frac{\partial^2 V}{(\partial \phi^2)^2} \right).
\end{align}
Now taking the derivative with respect to $a$ 
\begin{align}
a \frac{d \rho_\Phi }{d a} ~ = ~ \frac{d \rho_\Phi }{d \phi} \cdot a \frac{d \phi}{d a} ~= ~ \left[2\phi \left(2 \frac{\partial V}{\partial \phi^2} + \phi^2 \frac{\partial^2 V}{(\partial \phi^2)^2} \right)\right] \left[\frac{-3 \phi  \frac{\partial V}{\partial \phi^2}}{2  \frac{\partial V}{\partial \phi^2} + \phi^2 \frac{\partial^2 V}{(\partial \phi^2)^2}}\right] ~ = ~ -6 \phi^2 \frac{\partial V}{\partial \phi^2} ~ \equiv ~ -6\rho_{\theta},
\end{align}
where $\rho_\theta$ is the kinetic energy density of the angular field
\begin{align}
\rho_{\theta} \equiv \frac{1}{2} \phi^2 \dot{\theta}^2 ~ = ~ \phi^2 \frac{\partial V}{\partial \phi^2},
\end{align}
we deduce
\begin{align}
\frac{d \ln \rho_\Phi }{d \ln a} ~ = ~ \frac{-6 \phi^2 \frac{\partial V}{\partial \phi^2}}{\phi^2 \frac{\partial V}{\partial \phi^2} + V}.\label{eq:dlogrhodloga}
\end{align}

\paragraph{Equation of state.}
Using $\dot{\phi} \to 0$ and Eq.~\eqref{eq:thetadot_after_damping}, the equation of state of the complex scalar field becomes\footnote{As a sanity check, we can use
\begin{align}
\dot{\rho} + 3 H (\rho + P) ~ = ~ 0 ~\qquad \implies \qquad  \frac{d \ln \rho}{d \ln a} ~ = ~ -3(1 + \omega),
\end{align}
to show that Eq.~\eqref{eq:dlogrhodloga} and Eq.~\eqref{after_decay_eos} are consistent with each others. }
\begin{align}
\omega_\Phi ~ = ~ \frac{\frac{1}{2}\phi^2 \dot{\theta}^2 - V}{\frac{1}{2}\phi^2 \dot{\theta}^2 + V} ~ = ~ \frac{\phi^2 \frac{\partial V}{\partial \phi^2} - V}{\phi^2 \frac{\partial V}{\partial \phi^2} + V}.
\label{after_decay_eos}
\end{align}

\paragraph{Example: Nearly-quadratic potential.}
This is the scenario considered in this work. The potential and its derivatives read\footnote{
Note that is the minimum of potential is not vanishing $V_\textrm{min} \neq 0$, then the evolution of the energy density and EOS becomes
\begin{align}
\frac{d \ln \rho}{d \ln a} ~ = ~ \frac{-6 \log\left(\frac{\phi^2}{f^2}\right)}{2 \log\left(\frac{\phi^2}{f^2}\right) - 1 + \frac{f^2}{\phi^2} + \frac{V_\textrm{min}}{m^2 f^2}} \qquad \text{and} \qquad  \omega_\Phi = \frac{\phi^2-f_a^2}{2\phi^2 \log{\frac{\phi^2}{f_a^2}} -f_a^2 + \phi^2 - V_\textrm{min}/f_a^2}.
\end{align}
In order to generated a kination EOS, in this work we assume $V_\textrm{min} \ll m^2 f^2$.
}
\begin{align}
V ~ = ~ m^2 \phi^2 \left[ \log \left( \frac{\phi^2}{f^2}\right) -1 \right] + m^2 f^2,\qquad
\frac{\partial V}{\partial \phi^2} ~ = ~ m^2 \log \left(\frac{\phi^2}{f^2}\right), \qquad
\frac{\partial^2 V}{(\partial \phi^2)^2} ~ = ~ \frac{m^2}{\phi^2}.
\end{align}
\begin{framed}
From using Eq.~\eqref{eq:dlogphidloga} and \eqref{eq:dlogthetadloga}, we deduce the evolution of the radial and angular component of the scalar field after radial damping 
\begin{align}
\frac{d \ln \phi}{d \ln a} ~ = ~ \frac{-3 \log\left(\frac{\phi^2}{f^2}\right)}{2 \log\left(\frac{\phi^2}{f^2}\right) + 1} \qquad \text{and} \qquad \frac{d \ln \dot{\theta}^2}{d \ln a} ~ = ~ \frac{-6}{2 \log\left(\frac{\phi^2}{f^2}\right) + 1}.
\end{align}
Using Eq.~\eqref{eq:dlogrhodloga} and \eqref{after_decay_eos},  we deduce the evolution of the complex  scalar field energy density $\rho_{\Phi}$ and its equation of state $\omega_\Phi$
\begin{align}
\frac{d \ln \rho_{\Phi}}{d \ln a} ~ = ~ \frac{-6 \log\left(\frac{\phi^2}{f^2}\right)}{2 \log\left(\frac{\phi^2}{f^2}\right) - 1 + \frac{f^2}{\phi^2}} \qquad \text{and} \qquad  \omega_\Phi = \frac{\phi^2-f_a^2}{2\phi^2 \log{\frac{\phi^2}{f_a^2}} -f_a^2 + \phi^2}.
\end{align}
For $\phi \gg f_a$ we have
\begin{align}
\phi ~ \propto ~ a^{-3/2},\qquad \dot{\theta} ~ \propto ~ a^0, \qquad \rho_\Phi ~ \propto ~ a^{-3}, \qquad \omega_\Phi ~ \simeq ~ 0,
\end{align}
and for $\phi \simeq f_a$ we have
\begin{align}
\phi ~ \propto ~ a^0, \qquad \dot{\theta} ~ \propto ~ a^{-3}, \qquad \rho_\Phi ~ \propto ~ a^{-6}, \qquad \omega_\Phi ~ \simeq ~ 1.
\end{align}
\vspace{-1.5em}
\end{framed}

\begin{figure}[h!]
\centering
\raisebox{0cm}{\makebox{\includegraphics[width=0.495\textwidth, scale=1]{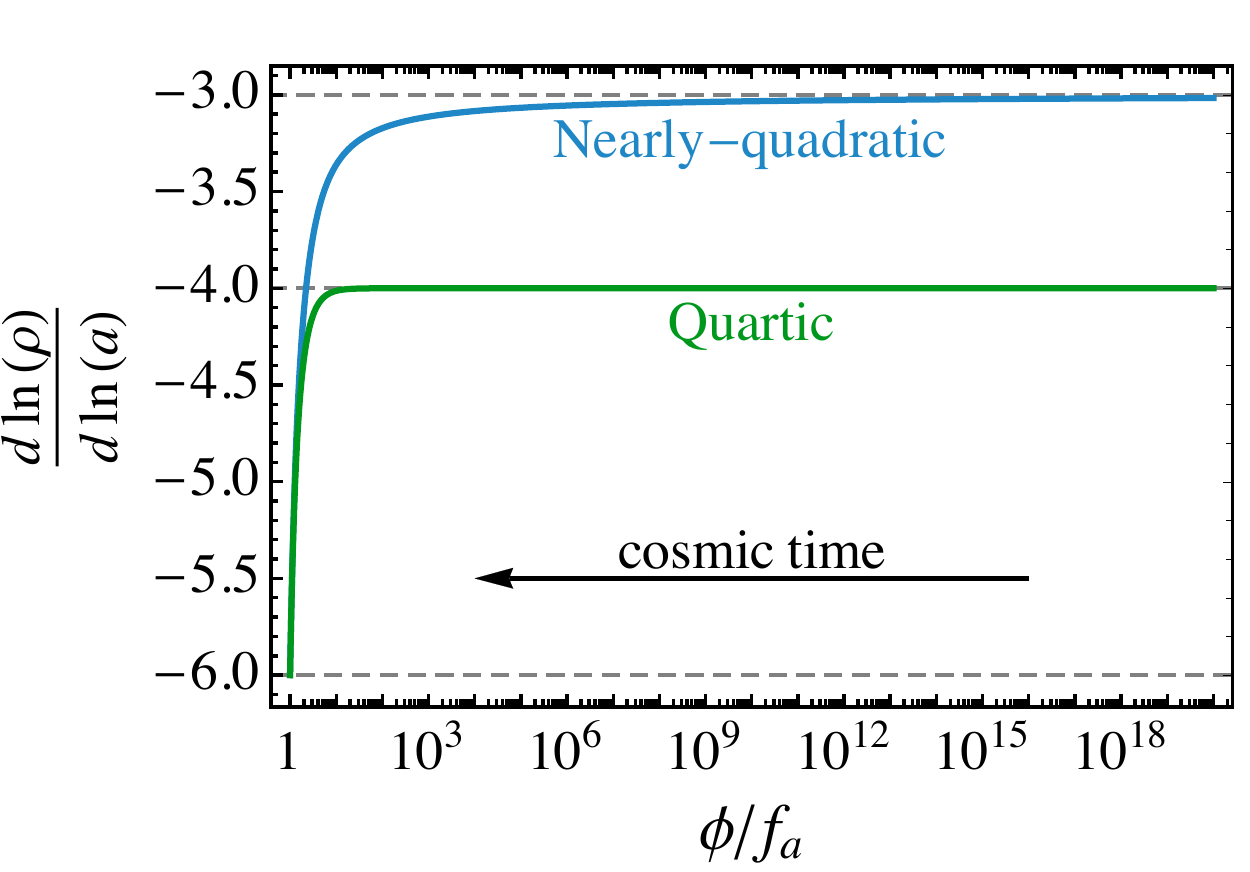}}}
\quad
\raisebox{0cm}{\makebox{\includegraphics[width=0.45\textwidth, scale=1]{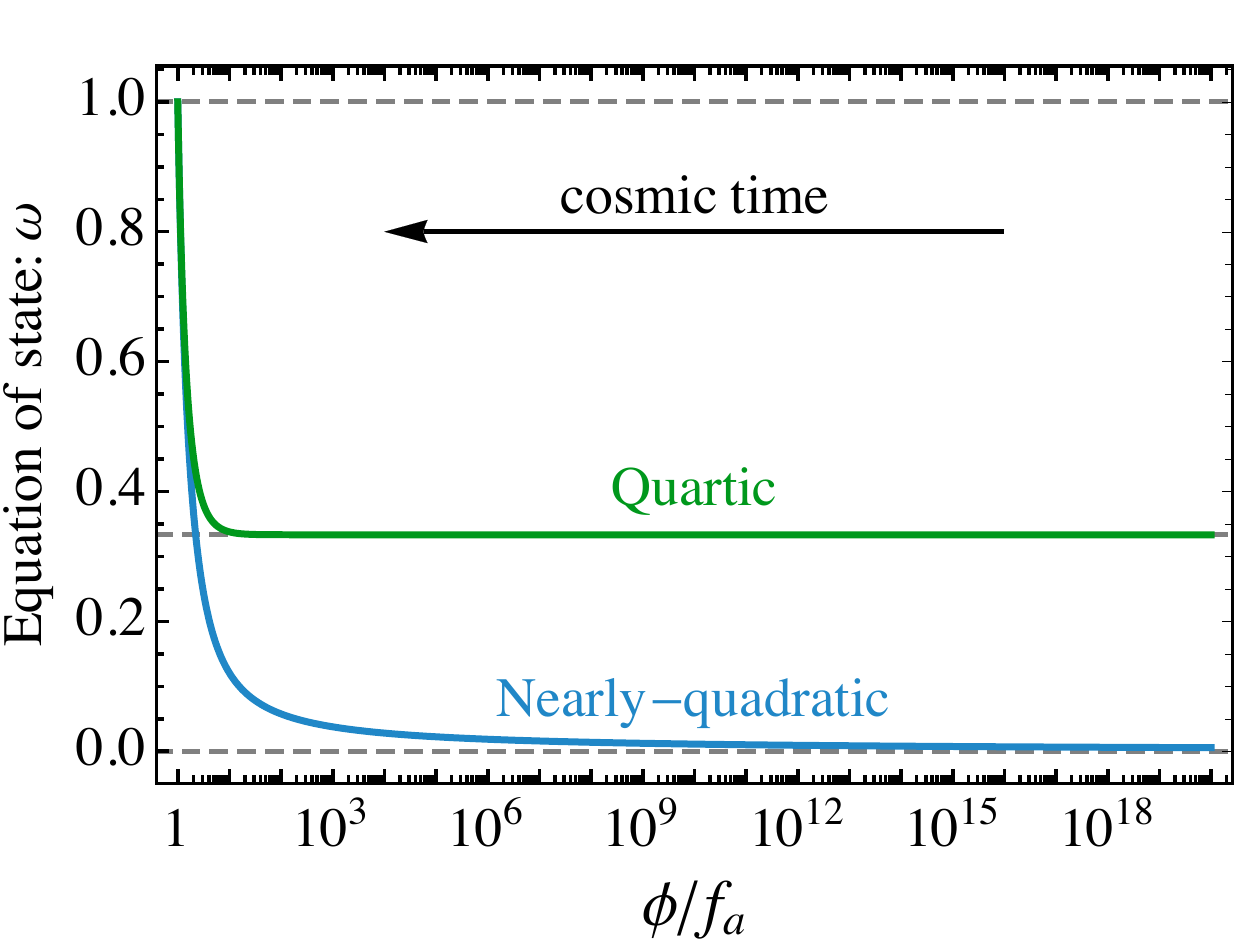}}}
\caption{\textit{ \small Evolutions of the energy density (\textbf{left}) and the equation-of-state (\textbf{right}) of the complex scalar field, rotating in a nearly-quadratic or quartic potential after its radial motion has been damped. Both cases lead to a kination EOS when $\phi \to f_a$. For $\phi \gg f_a$, the quartic potential gives a radiation EOS while the quadratic potential gives a matter EOS} }
\label{afterdecayplot}
\end{figure}

\paragraph{Example: quartic potential.}
\label{paragraph:quartic_potential}
For the sake of the comparison, we consider the potential 
\begin{align}
V ~ = ~ \lambda^2 \left( \phi^2 - f^2 \right)^2, \qquad \frac{\partial V}{\partial \phi^2} ~ = ~ 2 \lambda^2 \left( \phi^2 - f^2 \right), \qquad \frac{\partial^2 V}{(\partial \phi^2)^2} ~ &= ~ 2 \lambda^2.
\end{align}
\begin{framed}
Eq.~\eqref{eq:dlogphidloga} and \eqref{eq:dlogthetadloga}
From using Eq.~\eqref{eq:dlogphidloga} and \eqref{eq:dlogthetadloga}, we deduce the evolution of the radial and angular component of the scalar field after radial damping 
\begin{align}
\frac{d \ln \phi}{d \ln a} ~ = ~  \frac{-3 \left[  \frac{\phi^2}{f_a^2} -1\right]}{3 \frac{\phi^2}{f_a^2} -2} \qquad \text{and} \qquad \frac{d \ln \dot{\theta}^2}{d \ln a} ~ = ~ \frac{-6 \frac{\phi^2}{f_a^2}}{3 \frac{\phi^2}{f_a^2} - 1}.
\end{align}
Using Eq.~\eqref{eq:dlogrhodloga} and \eqref{after_decay_eos},  we deduce the evolution of the complex  scalar field energy density $\rho_{\Phi}$ and equation of state $\omega_\Phi$
\begin{align}
\frac{d \ln \rho_{\Phi}}{d \ln a} ~ = ~ \frac{- 12 \frac{\phi^2}{f_a^2}}{3 \frac{\phi^2}{f_a^2} - 1} \qquad \text{and} \qquad  \omega_\Phi = \frac{\phi^2+f_a^2}{3\phi^2-f_a^2}.
\end{align}
For $\phi \gg f_a$ we have
\begin{align}
\phi ~ \propto ~ a^{-1},\qquad \dot{\theta} ~ \propto ~ a^{-1}, \qquad \rho_\Phi ~ \propto ~ a^{-4}, \qquad \omega_\Phi ~ \simeq ~ \frac{1}{3},
\end{align}
and for $\phi \simeq f_a$ we have
\begin{align}
\phi ~ \propto ~ a^0, \qquad \dot{\theta} ~ \propto ~ a^{-3}, \qquad \rho_\Phi ~ \propto ~ a^{-6}, \qquad \omega_\Phi ~ \simeq ~ 1.
\end{align}
\vspace{-1.5em}
\end{framed}
The evolution of $\rho_{\Phi}$ and $\omega_\Phi$ in nearly-quadratic and quartic potentials are shown in Fig.~\ref{afterdecayplot}.  Only the matter phase induced by the nearly-quadratic potential can allow the initially sub-dominant $\Phi$ to dominate the energy density of the universe and, later, generate a kination-dominated era. For this reason, in this work we focus on a nearly-quadratic potential. In Fig.~\ref{quad_field_evolution_whole}, we show the evolution of $\phi$, $\dot{\theta}$, $\rho_\Phi$ and $\omega_\Phi$ in the nearly-quadratic potential, obtained after numerically integrating the equations of motion in Eqs.~\eqref{radial_break}, \eqref{angular_break} and \eqref{eq:friedmann_eq}.

\FloatBarrier
\FloatBarrier

\subsection{Derivation of the cosmological-history relations}
\label{derive_cosmo_history}
In this appendix, we derive the expression used in Sec.~\ref{sec:cosmo_history}, neglecting the factor $(\epsilon/2)$, and for which we refer to App.~\ref{app:effect_epsilon_evolution}.

\paragraph{Matter domination.}
After the field starts oscillating at $3H_{\rm osc} = \mreff$, it redshifts like matter and induces a matter-domination era as soon as it dominates the energy density of the universe when
\begin{align}
 \rho_{\rm dom} ~ \equiv ~V(\phi_{\rm ini})  \left(\frac{a_\textrm{osc}}{a_{\rm dom}}\right)^3~ \simeq ~ 3 H_{\rm osc}^2 M_{\rm pl}^2 \left(\frac{a_\textrm{osc}}{a_{\rm dom}}\right)^4,
\end{align}
where we have neglected the change of relativistic degrees of freedom in the thermal bath.
We obtain
\begin{align}
\frac{a_{\rm dom}}{a_\textrm{osc}} ~ \simeq  ~ \frac{\MPl^2 m_\textrm{eff}^2(\phi_\textrm{ini})}{3 V(\phi_\textrm{ini})} \qquad \textrm{and}\qquad
 \rho_{\rm dom} ~ = ~  \frac{27 V^4(\phi_\textrm{ini})}{\MPl^6 m_\textrm{eff}^6(\phi_\textrm{ini})}.  \label{eq:rho_dom_def_app}
\end{align}

\paragraph{Radial damping.}
The decay of the radial mode, occuring at the rate $\Gamma$, dissipates the radial kinetic energy when
\begin{equation}
\Gamma ~ \simeq ~  H \qquad \implies \qquad \rho_{\rm damp}~ = ~ 3\MPl^2\Gamma^2,
\end{equation}
after that the universe has expanded by
\begin{align}
\frac{a_{\rm damp}}{a_{\rm dom}} ~ = ~ \left(\frac{ \rho_{\rm dom}}{ \rho_{\rm damp}}\right)^{1/3} ~ = ~  \left[\frac{9 V^4(\phi_\textrm{ini})}{\MPl^8 \Gamma^2 m_\textrm{eff}^6(\phi_\textrm{ini})}\right]^{1/3}.
\end{align}

\paragraph{Start of kination (starting).}
The scalar field reaches the kination-liked equation of state when the radial field-value settles down to its final VEV $\phi \rightarrow f_a$, when the energy density is
\begin{equation}
\rho_{\textrm{KD},i} ~ = ~ \cancel{\frac{1}{2}\dot{\phi}^2} +  \frac{1}{2}\phi^2\dot{\theta}^2 + V(\phi \rightarrow f_a) \simeq ~ \frac{1}{2}\phi^2\dot{\theta}^2 \qquad \implies \qquad
\rho_{\textrm{KD},i} ~ \simeq ~ \frac{1}{2}f^2 m_\textrm{eff}^2(f),
\end{equation}
where the radial kinetic energy vanishes after $H < \Gamma$ and where we assume the vacuum energy to vanish at the minimum.
The duration of the matter era reads
\begin{align}
 \frac{a_{\textrm{KD},i}}{\textrm{min}(a_\textrm{\rm dom}, ~a_{\rm damp})} ~ = ~ \left(\frac{ \rho_{\rm damp}}{\rho_\textrm{KD}}\right)^{1/3}.
\end{align}

\paragraph{End of kination.}
The period of kination lasts until the energy density of the scalar field drops below that of thermal bath.
The value of the radiation energy density depends on whether entropy has been injected during the damping of the radial mode. This depends on whether radial damping has occured after domination $\rho_{\rm damp}<
\rho_{\rm dom}$, or not. Its energy density at the end of kination era follows the scaling law
\begin{align}
\rho_{\textrm{rad}} ~ = ~ \rho_{\textrm{KD},f} ~ = ~ 
\begin{cases}
\rho_{\rm dom}\left(\frac{a_{\rm dom}}{a_{\textrm{KD},f}}\right)^4,~~~\qquad \qquad \text{if}~\rho_{\rm damp} > \rho_{\rm dom},\\[0.75em]
\rho_{\rm damp}\left(\frac{a_{\rm damp}}{a_{\textrm{KD},f}}\right)^4,  \qquad \qquad \text{if}~\rho_{\rm damp} < \rho_{\rm dom}.\\
\end{cases}
\end{align}
Similarly, the scalar field which has undergone a period of matter folowed by a period of kination has the energy density
\begin{align}
\rho_{\Phi} ~ =  ~ \rho_{\textrm{KD},f} ~ = ~ 
\begin{cases}
\rho_{\rm dom}\left(\frac{a_{\rm dom}}{a_{\textrm{KD},i}}\right)^3 \left(\frac{a_{\textrm{KD},i}}{a_{\textrm{KD},f}}\right)^6, ~~~\qquad \qquad \text{if}~\rho_{\rm damp} > \rho_{\rm dom},\\[0.75em]
\rho_{\rm damp}\left(\frac{a_{\rm damp}}{a_{\textrm{KD},i}}\right)^3 \left(\frac{a_{\textrm{KD},i}}{a_{\textrm{KD},f}}\right)^6, \qquad  \qquad \text{if}~\rho_{\rm damp} < \rho_{\rm dom}.\\
\end{cases}
\end{align}
From the above two equations, we deduce the energy density at the end of kination $\rho_{\textrm{KD},f}$ 
\begin{equation}
 \rho_{\textrm{KD},f} ~ = ~ 
\begin{cases}
\frac{\rho_{\rm KD, i}^2}{\rho_{\rm dom}}, ~\qquad \qquad \text{if}~\rho_{\rm damp} > \rho_{\rm dom},\\[0.75em]
\frac{\rho_{\rm KD, i}^2}{\rho_{\rm damp}}, \qquad \qquad \text{if}~\rho_{\rm damp} < \rho_{\rm dom},\\
\end{cases}
\end{equation}
and the duration of the kination era
\begin{align}
\frac{a_{\textrm{KD},f}}{a_{\textrm{KD},i}} ~ = ~ 
\begin{cases}
\left(\frac{ \rho_{\rm dom}}{\rho_{\textrm{KD},i}}\right)^{1/6}, ~\qquad \qquad \text{if}~\rho_{\rm damp} > \rho_{\rm dom},\\[0.75em]
\left(\frac{ \rho_{\rm damp}}{\rho_{\textrm{KD},i}}\right)^{1/6},  \qquad \qquad \text{if}~\rho_{\rm damp} < \rho_{\rm dom}.
\end{cases}
\end{align}

\small
\bibliographystyle{JHEP}
\bibliography{inter_kination}
\end{document}